\documentclass[3p,12pt,eqsecnum]{article}

\usepackage{etoolbox}
\pdfoutput=1

\usepackage{lmodern}
\usepackage[T2A,T1]{fontenc}
\usepackage[utf8]{inputenc}
\usepackage{bm}
\usepackage{graphicx}
\usepackage{amsmath}
\usepackage{amssymb}
\usepackage{color}
\usepackage{xifthen}
\usepackage{shuffle}
\usepackage{multirow}
\usepackage{subcaption}
\usepackage[titletoc]{appendix}
\usepackage{cancel}
\usepackage{hyperref}
\usepackage[margin=1in]{geometry}
\usepackage[numbers,sort&compress]{natbib}
\usepackage{textcomp}
\usepackage{array}  

\usepackage{afterpage}
\usepackage{longtable}
\usepackage{fancyhdr}
\usepackage{caption}
\usepackage{calc}
\def\@oddfoot{\footnotesize\itshape {}}

\usepackage{amsmath}
\numberwithin{equation}{section}
\newcounter{homework}[section]  
\newenvironment{homework}[1][]{\refstepcounter{homework}\par\bigskip
   \noindent\text{{\underline{{\rm Exercise}~\thesection.\thehomework:}} #1} \rmfamily}{\bigskip}

\newlength\imheight
\newlength\imwidth	 
\def\diagsizetwo{0.15\textwidth} 
\def\diagsize{0.19\textwidth} 

\newcommand{\num}[1]{
	\settoheight\imheight{\includegraphics{#1}}
	\settowidth\imwidth{\includegraphics{#1}}
	c \ifthenelse{ \lengthtest{\imheight < 0.9 \imwidth}}
	{  \left( 
		\parbox{\widthof{\scalebox{.96}{\includegraphics[width=\diagsize]{#1}}}}{\includegraphics[width=\diagsize]{#1}}
	}
	{ \left(		\parbox{\widthof{\scalebox{.96}{\includegraphics[height=\diagsizetwo]{#1}}}}{\includegraphics[height=\diagsizetwo]{#1}}		
	} 
	\right)}

\newcommand{\col}[1]{
	\settoheight\imheight{\includegraphics{#1}}
	\settowidth\imwidth{\includegraphics{#1}}
	c \ifthenelse{ \lengthtest{\imheight < 0.9 \imwidth}}
	{  \left( 
		\parbox{\widthof{\scalebox{.96}{\includegraphics[width=\diagsize]{#1}}}}{\includegraphics[width=\diagsize]{#1}}
		}
	{ \left(		\parbox{\widthof{\scalebox{.96}{\includegraphics[height=\diagsizetwo]{#1}}}}{\includegraphics[height=\diagsizetwo]{#1}}	
	} 
	\right)}

\def\draftnote#1{{}}  

\def\P{{\rm P}}
\def\NP{{\rm NP}}

\def\NeqFour{{{\cal N} = 4}}
\def\NeqFive{{{\cal N} = 5}}

\def\NeqEight{{{\cal N} = 8}}
\def\oneloop{{\rm 1\hbox{-}loop}}
\def\twoloop{{\rm 2\hbox{-}loop}}

\def\tree{{\rm tree}}

\def\Ord{{\cal O}}
\def\pol{\varepsilon}
\def\e{\epsilon}
\def\eps{\epsilon}
\def\Tr{\, {\rm Tr}}

\def\Li{\mathop{\rm Li}\nolimits}

\def\A{A^{\tree}_4}
\def\calA{{\cal A}^{\tree}_4}
\def\M{M^{\tree}_4}
\def\cN{{\mathcal N}}
\def\n{{\tilde n}}
\def\f{\tilde f}
\def\T{T}
\def\Tadj{\f}

\def\ck{CK }
\def\ckDash{CK}
\def\cknospace{CK}

\def\calM{{\cal M}^{\tree}_4}
\def\calMthree{{\cal M}^{\tree}_3}
\def\calMfour{{\cal M}^{\tree}_4}
\def\calMfive{{\cal M}^{\tree}_5}
\def\calMsix{{\cal M}^{\tree}_6}

\def\Afive{A^{\tree}_5}

\def\be{\begin{equation}}
\def\ee{\end{equation}}
\def\eea{\end{eqnarray}}
\def\bea{\begin{eqnarray}}
\def\nn{\nonumber}
\def\scut{{s\hbox{-}\rm cut}}
\def\tcut{{t\hbox{-}\rm cut}}
\def\sect#1{Sec.~{\ref{#1}}}
\def\Sect#1{Sec.~{\ref{#1}}}
\def\sects#1#2{Secs.~{\ref{#1}} and {\ref{#2}}}
\def\app#1{Appendix~{\ref{#1}}}
\def\App#1{Appendix~{\ref{#1}}}

\def\fig#1{Fig.~{\ref{#1}}}

\def\figs#1#2{Figs.~{\ref{#1}} and {\ref{#2}}}

\def\eqn#1{Eq.~(\ref{#1})}
\def\Eqn#1{Equation~(\ref{#1})}
\def\eqns#1#2{Eqs.~(\ref{#1}) and~(\ref{#2})}

\def\tab#1{Tab.~{\ref{#1}}}

\def\N#1{{N$^{#1}$MC}}

\def\spa#1.#2{\left\langle#1\,#2\right\rangle}
\def\spb#1.#2{\left[#1\,#2\right]}
\def\spash#1.#2{\spa{\smash{#1}}.{\smash{#2}}}
\def\spbsh#1.#2{\spb{\smash{#1}}.{\smash{#2}}}
\def\sand#1.#2.#3{%
\left\langle\smash{#1}{\vphantom1}^{-}\right|{#2}%
\left|\smash{#3}{\vphantom1}^{-}\right\rangle}
\def\sandpp#1.#2.#3{%
\left\langle\smash{#1}{\vphantom1}^{+}\right|{#2}%
\left|\smash{#3}{\vphantom1}^{+}\right\rangle}
\def\sandpm#1.#2.#3{%
\left\langle\smash{#1}{\vphantom1}^{+}\right|{#2}%
\left|\smash{#3}{\vphantom1}^{-}\right\rangle}
\def\sandmp#1.#2.#3{%
\left\langle\smash{#1}{\vphantom1}^{-}\right|{#2}%
\left|\smash{#3}{\vphantom1}^{+}\right\rangle}
\def\sand#1.#2.#3{%
  \left\langle\smash{#1}{\vphantom1}\right|{#2}%
  \left|\smash{#3}{\vphantom1}\right\rangle}
\def\sandp#1.#2.#3{%
  \left\langle\smash{#1}{\vphantom1}^{-}\right|{#2}%
  \left|\smash{#3}{\vphantom1}^{+}\right\rangle}
\def\sandpp#1.#2.#3{%
  \left\langle\smash{#1}{\vphantom1}^{+}\right|{#2}%
  \left|\smash{#3}{\vphantom1}^{+}\right\rangle}
\def\sandmm#1.#2.#3{%
  \left\langle\smash{#1}{\vphantom1}^{-}\right|{#2}%
  \left|\smash{#3}{\vphantom1}^{-}\right\rangle}
\def\sandpm#1.#2.#3{%
  \left\langle\smash{#1}{\vphantom1}^{+}\right|{#2}%
  \left|\smash{#3}{\vphantom1}^{-}\right\rangle}
\def\sandmp#1.#2.#3{%
  \left\langle\smash{#1}{\vphantom1}^{-}\right|{#2}%
  \left|\smash{#3}{\vphantom1}^{+}\right\rangle}

\DeclareMathAlphabet\mathbfcal{OMS}{cmsy}{b}{n}

\newbox\charbox
\newbox\slabox
\def\s#1{{      
        \setbox\charbox=\hbox{$#1$}
        \setbox\slabox=\hbox{$/$}
        \dimen\charbox=\ht\slabox
        \advance\dimen\charbox by -\dp\slabox
        \advance\dimen\charbox by -\ht\charbox
        \advance\dimen\charbox by \dp\charbox
        \divide\dimen\charbox by 2
        \raise-\dimen\charbox\hbox to \wd\charbox{\hss/\hss}
        \llap{$#1$} }}

\usepackage{afterpage}
\usepackage{longtable}
\usepackage{fancyhdr}
\def\MHVbar{$\overline{\hbox{MHV}}$}
\def\ie{i.e.\ }
\def\eg{e.g.\ }

\begin{document}

	
	\thispagestyle{empty}
\noindent
{\small UCLA/TEP/2019/104 \hfill CERN-TH-2019-135   \hfill NUHEP-TH/19-11 } \\
{\small UUITP-35/19  \hfill  NORDITA 2019-079} 
$\null$\\

	\begin{center}

		{ \bf 
\LARGE The Duality Between Color and Kinematics and its Applications
			
			\vspace{0.25cm}
			
		}
		\bigskip\vspace{0.4cm}{
			{\large 
		Zvi Bern,${}^{ab}$ John Joseph Carrasco,${}^{cd}$ \\ \medskip
		Marco Chiodaroli,${}^e$ Henrik Johansson,${}^{ef}$ Radu Roiban ${}^g$
		}
		} \\[7mm]
		{\small \it  
			${}^a$Mani L. Bhaumik Institute for Theoretical Physics, \\[-1mm]
			  Department of Physics and Astronomy, UCLA, Los Angeles, CA 90095 \\ [1mm]
                         ${}^b$Theoretical Physics Department, CERN, \\[-1mm] 
                         1211 Geneva 23, Switzerland \\ [1mm]
                         ${}^c$Department of Physics and Astronomy\\[-1mm]
                               Northwestern University, Evanston, IL 60208, USA\\[1mm]
                         ${}^d$Institute of Theoretical Physics (IPhT), \\[-1mm]
                                    CEA/CNRS-Saclay and University of Paris-Saclay\\[-1mm]
                                    F-91191 Gif-sur-Yvette cedex, France \\[1mm]
                         ${}^e$Department of Physics and Astronomy,\\[-1mm]
                                   Uppsala University, 75108 Uppsala, Sweden\\[1mm]
                         ${}^f$Nordita, Stockholm University and  KTH Royal Institute of Technology, \\[-1mm]
                                   Roslagstullsbacken 23, 10691 Stockholm, Sweden\\[1mm]
			${}^g$Institute for Gravitation and the Cosmos, \\[-1mm]
			  Pennsylvania State University, University Park, PA 16802, USA \\ [1mm]
		} 
                  
	\end{center}
	
	\medskip


\section*{Abstract}
This review describes the duality between color and kinematics and its
applications, with the aim of gaining a deeper understanding of the
perturbative structure of gauge and gravity theories.  We emphasize,
in particular, applications to loop-level calculations, the broad web
of theories linked by the duality and the associated double-copy
structure, and the issue of extending the duality and double copy
beyond scattering amplitudes.  The review is aimed at doctoral
students and junior researchers both inside and outside the field of
amplitudes and is accompanied by various exercises.

\newpage
\tableofcontents

\newpage


\section{Introduction}
\label{IntroductionSection}

Gauge and gravity theories play a crucial role in our understanding of physical phenomena.
Yet, they appear to be distinct.
The weak, strong and electromagnetic interactions are manifestations of gauge theories, 
while gravity shapes the macroscopic evolution of the universe and spacetime itself. 
Finding a unified framework which seamlessly combines these two classes of theories 
constitutes, arguably, the most important open problem in theoretical physics.  
It is by now clear that realizing this unification requires a departure from conventional approaches 
through new principles or novel symmetries. 
The double-copy perspective reviewed here offers a
radically different way to interpret gravity. Its relation to the
other forces through color/kinematics duality~\cite{BCJ,BCJLoop} leads to remarkable new insights and powerful computational tools.

Despite their clear differences, 
gauge and gravity theories are already known to share  many features, supporting 
the existence of an underlying unified framework, such as string theory.  
While many of these similarities are not
apparent from a standard Lagrangian or Hamiltonian standpoint,
the study of objects closely related to observable quantities, such as
scattering amplitudes, reveals deep and highly-nontrivial connections.
This is most apparent in their perturbative expansions, which make it clear that 
the dynamics of these two classes of theories are governed by the same kinematical 
building blocks, even when their physical properties
are strikingly different.

The developments which exposed these features were systematized by the
introduction of the duality between color and kinematics and of the
double-copy construction.
The scattering amplitudes of many perturbative quantum field theories
(QFTs) exhibit a double-copy structure. It is central to our ability
to carry out calculations to very high loop orders and a property of
all supergravities whose amplitudes have been analyzed in detail.
This leads to the natural question whether all (super)gravity theories
are double copies of suitably-chosen matter-coupled gauge theories.
Perhaps more importantly, the double copy
realizes a unification of gauge and gravity theories in the sense of
providing a framework where calculations in both theories can be
carried out using an identical set of building blocks, yielding vast
simplifications.  

The primary purpose of this review is to offer an introduction to the
duality between color and kinematics---also referred to as
color/kinematics (\cknospace) duality and Bern-Carrasco-Johansson
(BCJ) duality---and the associated double-copy relation in the hope of
stimulating new progress both inside and outside of the fairly
well-understood setting of scattering amplitudes.  Beyond gauge and
gravity theories, double-copy relations also provide a new perspective
on QFT, generating a surprisingly wide web of theories through
building blocks obeying the same algebraic relations.  
  
The duality essentially states that scattering amplitudes in gauge
theories---and, more generally, in theories with some Lie-algebra
symmetry---can be rearranged so that kinematic building blocks obey the
same generic algebraic relations as their color factors.  Via the
duality, we can not only constrain the kinematic dependence of each
graph, but we can also convert gauge-theory scattering amplitudes to
gravity ones through the simple replacement
\begin{equation}
\hbox{color} \Rightarrow \hbox{kinematics}\,.
\label{ColorToKinematics}
\end{equation}
Evidence provided by explicit calculations suggests that \ck duality
and the double-copy construction hold for a wide class of theories at
loop level~\cite{BCJLoop, Neq44np, FivePointN4BCJ, WhiteIRBCJ,
  SimplifyingBCJ, Du:2012mt, Yuan:2012rg, FourLoopFormFactor,
  Boels:2013bi, Bjerrum-Bohr:2013iza, OneTwoLoopPureYMBCJ,
  Ochirov:2013xba, MafraSchlottererTwoLoop, BCJDifficulty,
  HeMonteiroSchlottererBCJNumer, FiveLoopFormFactor, BoelsFourLoop,
  HeSchlottererZhangOneLoopBCJ, JohanssonTwoLoopSusyQCD,
  Jurado:2017xut, Boels:2017ftb, Faller:2018vdz}.
Formal proofs, using a variety of methods~\cite{BjerrumMomKernel,
  MafraExplicitBCJNumerators, BjerrumManifestingBCJ, DuTengBCJ,
  delaCruz:2017zqr, Bridges:2019siz}, have been constructed for only
tree-level scattering amplitudes in these theories.
The duality also gave novel descriptions for tree-level amplitudes in
bosonic and supersymmetric string theories, as well as in various
effective field theories related to spontaneous symmetry breaking, and
more. It has also been observed that, in the presence of
adjoint-representation fermions, the duality implies
supersymmetry~\cite{Chiodaroli2013upa}.

The schematic rule~\eqref{ColorToKinematics} has served as a powerful
guide for many studies in perturbative gravity and supergravity,
especially on their loop-level ultraviolet (UV) properties (see \eg
Refs.~\cite{BCJLoop, SimplifyingBCJ,N46Sugra, N46Sugra2, Bern:2012cd,
  Bern:2012gh, Boels:2012sy, Bern:2013qca, Bern:2014lha, UVFiveLoops,
  Herrmann:2016qea, HerrmannTrnkaUVGrav}), showing a surprisingly tame
behavior.
For many supergravity theories, the physical degrees of freedom are obtained by the
substitution (\ref{ColorToKinematics}).
In others, such as pure Einstein gravity, the desired spectrum can only be obtained
after a subset of the double-copy states are projected out.
As we describe in some detail in
\sect{ZoologySection1}, \ck duality and the associated double-copy
properties hold for a remarkably large web of theories.  

Given the success at exploiting the double-copy structure for
scattering amplitudes, it is natural to wonder whether it also carries over
to other areas of gravitational physics, especially for
understanding and simplifying generic classical solutions.  Scattering
amplitudes have an important property that makes transparent the 
duality and  double-copy structure: they are independent of 
the choice of gauge and field-variables.  Generic classical solutions, on the other hand, 
do depend on these choices, making the problem of relating gauge and gravity classical solutions 
inherently more involved.  
Nevertheless, the prospect of solving problems in gravity by recycling gauge-theory
solutions is especially alluring.  While the differences with
scattering amplitudes are significant and make it a nontrivial
challenge to implement this program, there has been significant
progress in unraveling both the underlying principles of \ck
duality~\cite{Square, WeinzierlBCJLagrangian, Monteiro2011pc,
  OConnellAlgebras, Monteiro:2013rya, Ho:2015bia, Fu:2016plh,
  BjerrumManifestingBCJ, Fu:2018hpu,Chen:2019ywi} and finding explicit
examples of classical solutions related by the double-copy
property~\cite{Saotome2012vy, Monteiro2014cda, Luna2015paa,
  Ridgway2015fdl, Luna2016due, White2016jzc, Cardoso2016amd,
  Goldberger2016iau, Luna2016hge, Goldberger2017frp, 
  Adamo2017nia, DeSmet2017rve, BahjatAbbas2017htu, 
  CarrilloGonzalez2017iyj, Goldberger2017ogt, Li2018qap,
  Ilderton:2018lsf, Lee:2018gxc, Plefka:2018dpa, ShenWorldLine, 
  Berman:2018hwd, Gurses:2018ckx,  Adamo:2018mpq, 
 Bahjat-Abbas:2018vgo, Luna:2018dpt,
Farrow:2018yni, CarrilloGonzalez:2019gof, PV:2019uuv}. 
One of the most promising applications of the double copy beyond scattering amplitudes 
relates to gravitational-wave physics, as highlighted by Refs.~\cite{Goldberger2016iau, ShenWorldLine, CheungPM, Kosower:2018adc, 3PM, Buananno3PMCheck,3PMLong}.

The origins of the double copy can be traced back to the dawn of string theory, with the
observation of a curious connection between the Veneziano scattering
amplitude~\cite{VenezianoAmplitude}, $A(s,t)$, (later identified as an open-string scattering amplitude)  
and the Virasoro-Shapiro amplitude~\cite{VirasoroAmplitude, Shapiro}, $M(s,t, u)$, (later identified as a
closed-string amplitude).  With an appropriate normalization, these  two amplitudes are related as~\cite{KLT}
\begin{equation}
M(s,t, u) = \frac{\sin (\pi \alpha' s)}{\pi \alpha'} A(s,t) A(s,u) \,,
\label{EarlyDoubleCopy}
\end{equation}
where 
$\alpha'$ is the inverse string tension.  The arguments are the kinematic (Mandelstam) invariants 
of a four-point scattering process,
\begin{equation}
s = (p_1 + p_2)^2\,, \hskip 1.5 cm 
t = (p_2 + p_3)^2\,, \hskip 1.5 cm 
u = (p_1 + p_3)^2\,.
\label{STUDef}
\end{equation}
\Eqn{EarlyDoubleCopy} carries over to all string states, including
the gluons of the open string and the gravitons in the closed string.
In the low-energy limit, when string theory reduces to field theory,
it yields a relation between scattering amplitudes in Einstein
gravity and those of Yang-Mills (YM) theory~\cite{GSWBook},
\begin{equation}
\calMfour(1,2,3,4) = \Bigl(\frac{\kappa}{2} \Bigr)^2
    s \A(1,2,3,4) \A(1,2,4,3) \,,
\label{FourPointKLT}
\end{equation}
where $\A(1,2,3,4)$ is a color-ordered gauge-theory four-gluon partial
scattering amplitude, $\calMfour(1,2,3,4)$ is a four-graviton tree
amplitude and $\kappa$ is the gravitational coupling to related to
Newton's constant via $\kappa^2 = 32 \pi^2 G_N$ and, for reasons 
that will become clear shortly, the polarization vectors of gluons on the right-hand side of \eqn{FourPointKLT}
are taken to be null.
We will suppress the gravitational coupling by setting $\kappa = 2$ throughout this review.  The
color-ordered partial tree amplitudes are the coefficients of basis elements once the amplitude's
color factors are expressed in the trace color basis, and the coupling $g$ is set to unity.
They are gauge invariant---see \eg
Refs.~\cite{ManganoParkeReview, TasiLance, BDKUniarityReview, 
 ElvangHuangReview, Cheung:2017pzi}
for further details.  \Eqn{FourPointKLT} is rather striking, asserting
that tree-level four-graviton scattering is described completely by gauge-theory
four-gluon scattering, bypassing the usual machinery of general
relativity.  Similar relations were later derived for higher-point
string-theory tree-level amplitudes~\cite{KLT}, and generalized in the
field-theory limit to an arbitrary number of external
particles~\cite{MultiLegOneLoopGravity}.  Besides the remarkable
implication that the detailed dynamics of the gravitational field can
be described in terms of the dynamics of gauge fields,
\eqn{FourPointKLT} has other surprising features not visible in
standard Lagrangian formulations.  For example, \eqn{FourPointKLT}
implies that the four-graviton amplitude can be re-arranged so that
Lorentz indices factorize~\cite{BernGrant,CheungRemmen} into ``left''
indices belonging to one gauge-theory amplitude and ``right'' indices
belonging to another gauge theory.

\subsection{Motivation: Complexity of gravity versus gauge theory}

\begin{figure}[tb]
\begin{center}
\includegraphics[scale=.5]{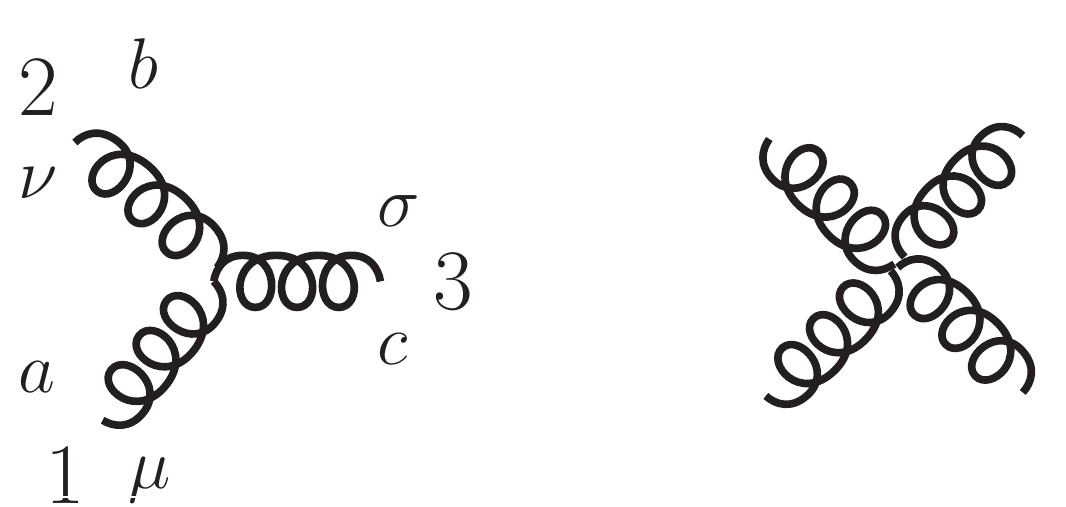}
\end{center}
\caption[a]{\small Gauge theories have three- and four-point vertices
in a Feynman diagrammatic description.}
\label{YMVertFigure}
\end{figure}

\begin{figure}[tb]
\begin{center}
\includegraphics[scale=.45]{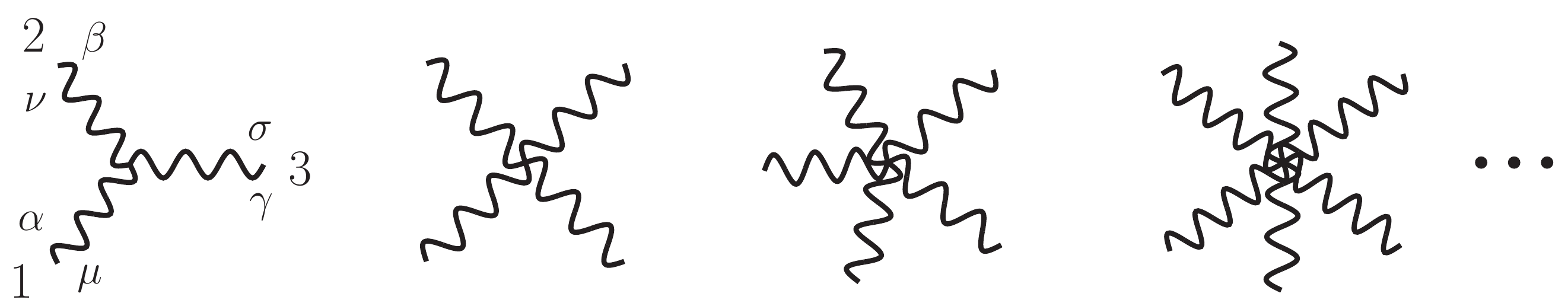}
\end{center}
\caption[a]{\small Gravity theories have an infinite number of
higher-point contact interactions in a Feynman diagrammatic description.}
\label{GravVertFigure}
\end{figure}

It is interesting to contrast the remarkable simplicity encoded in
the relation (\ref{FourPointKLT}) with the much more complicated expressions 
that arise from standard Lagrangian methods.
Scattering amplitudes for gauge and gravity theories can be obtained using the Feynman rules 
derived from their respective  Lagrangians
\begin{equation}
{\cal L}_{\rm YM} = - \frac{1}{4} F^a_{\mu\nu} F^{a\, \mu\nu}
 \,, \hskip 2 cm
{\cal L}_{\rm EH} =\frac{2}{\kappa^2} \sqrt{-g} R \,.
\label{Lagrangians}
\end{equation}
Here $F_{\mu\nu}^a$ is the usual YM field strength and $R$ the Ricci scalar. 

Following standard Feynman-diagrammatic methods, we gauge-fix and then
extract the propagator(s) and the three- and higher-point vertices.  
For gravity we also expand around flat spacetime, taking the metric to be $g_{\mu\nu} =
\eta_{\mu\nu} + \kappa h_{\mu\nu}$ where $\eta_{\mu\nu}$ is the
Minkowski metric and $h_{\mu\nu}$ is the graviton field.  As
illustrated in \figs{YMVertFigure}{GravVertFigure}, with standard
gauge choices,  gauge theory has  only three- and four-point
vertices, while  gravity has an infinite number of vertices of arbitrary multiplicity.
The complexity of each individual interaction
term is perhaps more striking than their infinite number.  Consider,
for example, the three-graviton interaction.  In the standard de~Donder
gauge, $\partial_\nu h^{\nu}_{\ \mu} = {1 \over 2} \partial_\mu
h^{\nu}_{\ \nu}$, the corresponding vertex is~\cite{DeWitt:1967uc, SannanVertex},
\begin{align}
& \hskip -.7 cm
G_{3\,\mu\rho,\nu\lambda,\sigma\tau}(p_1,p_2,p_3)\nn \\
&\null \hskip .5 cm
 =  i {\rm Sym}\Bigl[ - \frac{1}{2} P_3(p_1\cdot p_2\eta_{\mu\rho}\eta_{\nu\lambda}
\eta_{\sigma\tau}) - \frac{1}{2} P_6 (p_{1\nu}p_{1\lambda}
\eta_{\mu\rho}\eta_{\sigma\tau}) + \frac{1}{2}
P_3 (p_1\cdot p_2 \eta_{\mu\nu}\eta_{\rho\lambda}\eta_{\sigma\tau}) \nn\\
&\null \hskip 1.8cm
+ P_6(p_1\cdot p_2 \eta_{\mu\rho}\eta_{\nu\sigma}\eta_{\lambda\tau})
+2P_3(p_{1\nu}p_{1\tau}\eta_{\mu\rho}\eta_{\lambda\sigma})
-P_3(p_{1\lambda}p_{2\mu}\eta_{\rho\nu}\eta_{\sigma\tau}) \nn\\
&\null \hskip 1.8cm
 +P_3(p_{1\sigma}p_{2\tau}\eta_{\mu\nu}\eta_{\rho\lambda})
+P_6(p_{1\sigma}p_{1\tau}\eta_{\mu\nu}\eta_{\rho\lambda})
+2P_6(p_{1\nu}p_{2\tau}\eta_{\lambda\mu}\eta_{\rho\sigma}) \nn \\
&\null \hskip 1.8cm
+2P_3(p_{1\nu}p_{2\mu}\eta_{\lambda\sigma}\eta_{\tau\rho})
-2P_3(p_1\cdot p_2 \eta_{\rho\nu}\eta_{\lambda\sigma}\eta_{\tau\mu})\Bigr]
\,,
\label{deDonderVertex}
\end{align}
where we set $\kappa = 2$, $p_i$ are the momenta of the
three gravitons, $\eta_{\mu \nu}$ is the flat metric, ``Sym'' implies a symmetrization in each pair of graviton
Lorentz indices $\mu\leftrightarrow\rho$, $\nu\leftrightarrow\lambda$
and $\sigma\leftrightarrow\tau$, and $P_3$ and $P_6$ signify a
symmetrization over the three graviton legs, generating three or six
terms respectively.  The symmetrization over the three external legs
ensures the Bose symmetry of the vertex.  In total, the vertex has of
the order of 100 terms. This generally undercounts the number of terms, because 
within a diagram each vertex momentum is a linear combination of the
independent momenta of that diagram.  

We may contrast this to the three-gluon vertex in Feynman gauge,
\begin{equation}
V^{abc}_{3\,\mu\nu\sigma}(p_1,p_2,p_3) =
g f^{abc} \Bigl[ (p_1 - p_2)_\sigma \eta_{\mu\nu} + \hbox{cyclic} \Bigr] \,.
\label{FeynmanVertex}
\end{equation}
which does not appear to bear any obvious relation to the
corresponding three-graviton vertex~\eqref{deDonderVertex}.  
These considerations seemingly suggest that gravity is much more complicated than gauge theory.  
Moreover, the three-graviton vertex immediately appears to
conflict with the simple factorization of Lorentz indices into left and
right sets visible in \eqn{FourPointKLT}.  The first term in
\eqn{deDonderVertex}, for example, contains a factor
$\eta_{\mu \rho}$ which explicitly contracts a left graviton index
with a right one.

The reason why the three-graviton vertex is so complicated is that it is gauge-dependent.\footnote{While 
somewhat less complicated than the three-graviton vertex, the three-gluon vertex is also gauge-dependent.}  
With special gauge choices and appropriate field redefinitions~\cite{vandeVen,BernGrant,CheungRemmen,CheungSimple},
it is possible to considerably simplifying the Feynman rules.  Still, 
direct perturbative gravity calculations in a Feynman diagram approach
are rather nontrivial, especially beyond leading order, even with modern computers.
To eliminate the gauge dependence we should instead focus on the three-vertex with on-shell conditions imposed on 
external legs, by demanding that the vertex is contracted into physical states that satisfy,
\begin{equation}
\pol^{\mu\rho} = \pol^{\rho\mu}\,, \hskip 1cm 
p_{\mu} \pol^{\mu\rho} = 0\,, \hskip 1cm 
p_{\rho} \pol^{\mu\rho} = 0\,,  \hskip 1cm
\pol_{\mu}{}^{\mu}  \equiv \eta^{\mu\nu}\pol_{\mu\nu}= 0\,,
\end{equation}
where $p$ is a graviton momentum and $\pol^{\mu\nu}$ the associated
graviton polarization tensor.  This removes all trace and longitudinal terms,
reducing the vertex to a simple form,
\begin{equation}
G_{3\,\mu\rho,\nu\lambda,\sigma\tau}(p_1,p_2,p_3)\
= -i \Bigl[ (p_1 - p_2)_\sigma \eta_{\mu\nu} + \hbox{cyclic} \Bigr] 
          \Bigl[ (p_1 - p_2)_\tau \eta_{\rho\lambda} + \hbox{cyclic} \Bigr]\,,
\end{equation}
exposing its simple relation to the three-gluon vertex of gauge theory.  This is a
hint that there should be much better ways to organize the perturbative
expansion of gravity.  We now turn to four-graviton scattering amplitude, which
is a better example as it corresponds directly to a physical
process.

\subsection{Invitation: four-point example}

\begin{figure}[tb]
\begin{center}
\includegraphics[scale=.5]{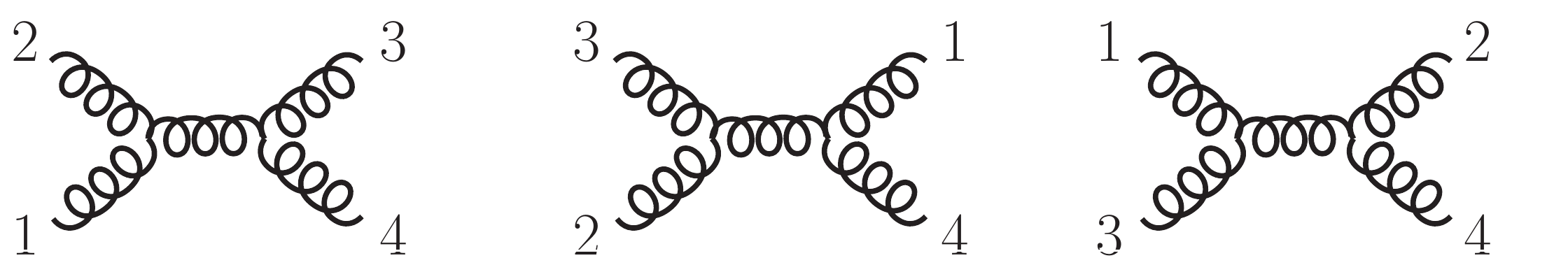}
\end{center}
\vskip -.3 cm 
\caption[a]{\small The three Feynman diagrams corresponding to the $s$, $t$ and $u$ channels.}
\label{FourPtGluonsFigure}
\end{figure}

Consider the full four-gluon tree amplitude in YM theory, which can be obtained, 
for example, by 
following textbook Feynman rules~\cite{PeskinSchroeder}.
We write it as a sum over three channels
corresponding to the three diagrams in \fig{FourPtGluonsFigure}
\begin{equation}
i \calA =  g^2 \Bigl(\frac{n_s c_s}{s}+ \frac{n_t c_t}{t}+ \frac{n_u c_u}{u}\Bigr)\,,
\label{YMFourAmplitude}
\end{equation}
where the Mandelstam variables are defined in \eqn{STUDef}. 
The $s$-channel color factor, normalized to be compatible with the scattering amplitudes
literature~\cite{ManganoParkeReview}, is
\begin{equation}
c_s= - 2 f^{a_1 a_2 b}f^{b a_3 a_4}\,,
\end{equation}
where the color-group structure constants $f_{abc}$ are the standard textbook
ones~\cite{PeskinSchroeder}.
With this normalization, the $s$-channel kinematic numerator, $n_s$, is
\begin{align}
n_s & = -\frac{1}{2}\Bigl\{ \Big[(\varepsilon_1 \cdot \varepsilon_2) p_1^\mu+2 
(\varepsilon_1 \cdot p_2) \varepsilon_2^\mu-(1 \leftrightarrow 2)\Big] 
\Big[ (\varepsilon_3 \cdot \varepsilon_4) p_{3\mu}+2(\varepsilon_3 \cdot p_4) 
\varepsilon_{4\mu}-(3 \leftrightarrow 4)\Big] \nn  \\
& \null \hskip 2 cm 
+ s \Bigl[ (\varepsilon_1 \cdot \varepsilon_3)(\varepsilon_2 \cdot \varepsilon_4)
- (\varepsilon_1 \cdot \varepsilon_4)(\varepsilon_2 \cdot \varepsilon_3)\Bigr] \Bigr\}\,,
\label{s_num}
\end{align}
 where the momenta and polarization vectors satisfy on-shell
conditions $p_i^2=\varepsilon_i \cdot p_i=0$.  The other color factors
and numerators are obtained by cyclic permutations of the particle labels~$(1,2,3)$:
\begin{equation}
c_t n_t =c_s n_s\big|_{1\rightarrow 2\rightarrow
  3\rightarrow 1}\,,  \hskip 2 cm 
c_u n_u =c_s n_s\big|_{1\rightarrow
  3\rightarrow 2\rightarrow 1}\,. 
\label{OtherChannels}
\end{equation}
Feynman rules for gluons contain a four-gluon vertex, as in \fig{YMVertFigure}.  
Here we have absorbed its contribution into
the three diagrams in \fig{FourPtGluonsFigure} according to the color
factors, by multiplying and dividing by an appropriate propagator.  
This is the origin of the term on the second line of \eqn{s_num}.

A key property of the gauge-theory scattering amplitude \eqref{YMFourAmplitude} is
its linearized gauge invariance.  To check this, we need to verify that the amplitude vanishes with the replacement $\varepsilon_4 \rightarrow p_4$. 
Upon doing this replacement for the $s$-channel numerator we get, after some algebra, the nonzero result
\begin{equation}
n_s\big|_{\varepsilon_4 \rightarrow p_4} 
= - \frac{s}{2} \Big[(\varepsilon_1 \cdot \varepsilon_2)  \big( (\varepsilon_3  \cdot p_2)-(\varepsilon_3  \cdot p_1) \big) 
+ {\rm cyclic}(1,2,3)\Big] \equiv s \, \alpha(\varepsilon,p) \,,
\label{Ngaugetr}
\end{equation}
which is no surprise since individual diagrams are, in general, gauge dependent. The function $\alpha(\varepsilon, p)$ 
is clearly invariant under cyclic permutations of the labels $(1,2,3)$. For the full amplitude we get therefore
\begin{equation}
\frac{n_s c_s}{s} + \frac{n_t c_t}{t} + \frac{n_u c_u}{u}\Big|_{\varepsilon_4 \rightarrow p_4}
= (c_s+c_t+c_u) \,  \alpha(\varepsilon,p) \, ,
\end{equation}
where $\alpha(\varepsilon,p)$ is the expression in \eqn{Ngaugetr}. 
Hence the amplitude is gauge invariant if $c_s+c_t+c_u$ vanish, \ie
\begin{equation}
c_s+c_t+c_u = -2 (f^{a_1 a_2 b}f^{b a_3 a_4}+f^{a_2 a_3 b}f^{b a_1 a_4}+f^{a_3 a_1 b}f^{b a_2 a_4}) = 0\,.
\label{BCJFourPtcolor}
\end{equation}
This is the standard Jacobi identity, which indeed is satisfied by
the group-theory structure constants in a gauge theory.

Consider the three-term sum over kinematic numerators in \eqns{s_num}{OtherChannels}, $n_s+n_t+n_u$, 
analogous to the sum over  color factors on the left-hand side of \eqn{BCJFourPtcolor}.
Remarkably,  this combination vanishes when the on-shell conditions are applied, 
\begin{equation}
 n_s+n_t+n_u=0\, .
\label{BCJFourPt}
\end{equation}
We will refer to this relation as a \emph{kinematic Jacobi identity}. 
This was originally noticed some time ago for four-point amplitudes, as a curiosity
related to radiation zeros in four-point amplitudes~\cite{EarlyDualJacobi, Goebel:1980es, Harland-Lang:2015faa}. 
Generic representations of four-point amplitudes in terms of diagrams with only cubic vertices obey these identities, 
but at higher points nontrivial rearrangements are needed. The significance of the identity \eqn{BCJFourPt} and its 
generality was understood later~\cite{BCJ,BCJLoop}.  
We  refer to kinematic identities that are analogous to generic color-factor identities
as a duality between color and kinematics. It turns out that they constitute an ubiquitous, 
yet hidden, structure not only of gauge theories, but also of an ever-increasing web of theories, as described in
\sect{ZoologySection1}.

\begin{homework}
\label{HomeworkJacobi}
Use \eqns{s_num}{OtherChannels} to verify the numerator Jacobi
identity \eqref{BCJFourPt}.  Redefine the numerators by eliminating
$c_u$ in favor of $c_s$ and $c_t$, defining new numerators $n_s'$ and
$n_t'$ as the coefficient of $c_s/s$ and $c_t/t$.  The numerator
$n_u'$ vanishes by construction.  Show that the kinematic Jacobi
identity still holds for these redefined numerators.
\end{homework}

The fact that the kinematic factors satisfy the same relations as the
color factors suggests that they are mutually exchangeable. Indeed, we
can swap color factors for kinematic factors in the YM four-point
amplitude \eqref{YMFourAmplitude}, which gives a new gauge-invariant object that, as we will discuss momentarily,
is a four-graviton  amplitude,
\begin{equation}
i \calA \Big|_{
 \begin{matrix}
\! c_i \! \rightarrow  \tilde n_i \atop
\, g\rightarrow \kappa/{2}~
\end{matrix}}   
\equiv i \calM = \Bigl(\frac{\kappa}{2} \Bigr)^2 \biggl(
\frac{n_s^2}{s}+ \frac{n_t^2}{t}+ \frac{n_u^2}{u}\biggr)\,.
\label{cnrepl}
\end{equation}
The new amplitude $\calM$ doubles up the kinematic numerators, and so we refer
to it as a double copy. (The $i$ in front of the $\calM$ is a phase convention.)
The expression in Eq.~\eqref{cnrepl} has the following properties:
the external states are captured by symmetric polarization tensors
$\pol^{\mu \nu} = \pol^\mu \pol^\nu$, the
interactions are of the two-derivative type, and the amplitude is
invariant under linearized diffeomorphism transformations. By choosing 
the polarization vectors to be null $\pol^2=0$ (corresponding to circular polarization), 
implying that $\pol^{\mu \nu}$ is traceless, this amplitude should describe the scattering of four gravitons in
Einstein's general relativity, up to an overall normalization. 
There are a number of ways to prove that this is the case, including using on-shell recursion 
relations~\cite{Square} and ordinary gravity Feynman rules~\cite{BernGrant}; here we will show that  \eqn{cnrepl}
reproduces the Kawai-Lewellen-Tye (KLT) form of gravity amplitudes~\cite{KLT}, derived using the low-energy 
limit of string theory.

The diffeomorphism invariance of the amplitude requires some elaboration. Consider a linearized diffeomorphism of the asymptotic (weak) graviton field $h_{\mu \nu}$. The diffeomorphism is parametrized by the function $\xi_\mu$ and take the simple form
\be
\delta h_{\mu \nu} = \partial_\mu \xi_\nu+\partial_\nu \xi_\mu\,. 
\ee
Translating this to momentum space implies that a diffeomorphism-invariant amplitude should vanish upon replacing a polarization tensor as: $\varepsilon^{\mu \nu} \rightarrow p^\mu \varepsilon^\nu+ p^\nu \varepsilon^\mu$. 
Applying this to leg 4 of the amplitude, we find 
\be
 \frac{n_s^2}{s}+ \frac{n_t^2}{t}+ \frac{n_u^2}{u}\Big|_{\varepsilon_4^{\mu \nu} \rightarrow p^\mu_4 \varepsilon^\nu_4+ p^\nu_4 \varepsilon^\mu_4}=2 (n_s+n_t+n_u) \,  \alpha(\varepsilon,p) = 0\,.
\ee
Thus, we see that the kinematic Jacobi identity needs to be satisfied for the amplitude to be invariant under linearized diffeomorphism transformations, in complete analogy to the color Jacobi identity in the gauge-theory amplitude.

Returning to the YM amplitude, we note that the amplitude can be written in a manifestly gauge-invariant form if 
we solve the Jacobi relation by choosing $c_t=-c_u-c_s$,
\begin{align}
i \calA & = g^2 \Big(\frac{n_s c_s}{s} + \frac{n_t c_t}{t} 
  + \frac{n_u c_u}{u}\Bigr) \nn\\
& = g^2\Bigl(\Big(\frac{n_s}{s}- \frac{n_t}{t}  \Big) c_s - \Big(\frac{n_t}{t} 
  -\frac{n_u}{u} \Big) c_u \Bigr)\nn  \\
& \equiv  i g^2 \A(1,2,3,4) c_s - i g^2 \A(1,3,2,4) c_u \,.
\end{align}
The partial amplitudes $\A(1,2,3,4)$ are gauge invariant because the color-dressed amplitude $\calA$ is now 
decomposed in a basis of independent color factors, with elements $c_s$ and $c_u$, and thus the gauge invariance 
of $\calA$ implies the gauge invariance of the individual terms of this decomposition.

It is not difficult to show that the partial amplitude can be written as
\be
 \A(1,2,3,4) = -i \frac{ t_8F^4}{s t}\,,
\label{4ptExplicit}
\ee
where 
\begin{equation}
t_8F^4 \equiv \big[ 4{\rm Tr}(F_1F_2F_3F_4) - {\rm Tr}(F_1 F_2)  {\rm Tr}(F_3 F_4) + {\rm cyclic}(1,2,3)\big]
\label{t8F4}
\end{equation}
contains various Lorentz traces over four linearized Fourier transformed field strengths, 
\begin{equation}
F^{\mu\nu}_i \equiv p^{\mu}_i \varepsilon^{\nu}_i- \varepsilon^{\mu}_i p^{\nu}_i\,,
\label{FieldStrength}
\end{equation}
where the fields
are replaced with polarization vectors.  These are manifestly invariant under linearized gauge transformations.

We can also solve the kinematic Jacobi relation \eqref{BCJFourPt} by choosing $n_t=-n_u-n_s$. The partial amplitudes
then become
\bea
i \A(1,2,3,4) &=&\frac{n_s}{s}- \frac{n_t}{t} = n_s \Big(\frac{1}{s}+\frac{1}{t}\Big)+ \frac{n_u}{t} \,,  \nn \\
i \A(1,3,2,4) &=& \frac{n_t}{t}- \frac{n_u}{u}= -n_u \Big(\frac{1}{u}+\frac{1}{t}\Big) - \frac{n_s}{t}\, ,
\label{pAmp}
\eea
which may also be organized as a matrix relation
\be
i \begin{pmatrix} 
\A(1,2,3,4) \\
\A(1,3,2,4)
\end{pmatrix}
= 
\begin{pmatrix} 
\frac{1}{s}+\frac{1}{t} &  \frac{1}{t}  \\
-\frac{1}{t} & -\frac{1}{u} -\frac{1}{t}  \\
\end{pmatrix}
\begin{pmatrix} 
n_s \\
n_u
\end{pmatrix} \, .
\label{matrixform}
\ee 
It might seem that it is possible to solve for the numerators in terms of the
partial amplitudes by inverting the two-by-two matrix of propagators. Existence of
a solution would contradict, however, the fact that on the one hand numerators are gauge-dependent
and on the other partial amplitudes are gauge-invariant.
Indeed, the matrix of propagators has no inverse as its determinant is proportional to
$s+t+u=0$.  At best, we can solve for one of the numerators, say, $n_u$, 
\begin{equation}
n_u = i t \A(1,2,3,4)+u \frac{n_s}{s}\,.  
\end{equation}
Replacing this into $\A(1,3,2,4)$ in \eqn{pAmp}, the dependence on the undetermined kinematic numerator 
$n_s$ cancels out, and we obtain the gauge-invariant relation
\begin{equation}
 \A(1,3,2,4)=   \frac{s}{u} \A(1,2,3,4)\,.
\label{4PtBCJAmplitude}
\end{equation}
Given the vanishing of the determinant of the above matrix of propagators, it is
not surprising to find that the two partial amplitudes are linearly dependent.  
In fact, one may phrase \eqn{4PtBCJAmplitude} as the orthogonality condition of the left-hand side
of \eqn{matrixform} onto the null eigenvector of the matrix of propagators.

The existence of relations between partial amplitudes is a
general feature. Such \emph{BCJ amplitude relations} exist
whenever the duality between color and kinematics and gauge invariance
conspire to prevent the relation between partial amplitudes and
numerators to be inverted. These relations have been demonstrated in a
variety of ways, including using both string theory~\cite{Monodromy,
  Stieberger:2009hq, Sondergaard:2009za, Jia:2010nz, Mafra:2010gj,
  MafraBCJAmplString, Ma:2011um, Grassi:2011ky,Barreiro:2013dpa} and
field theory methods~\cite{amplituderelationProof,
  PureSpinorsBCJAmplProof, CachazoBCJProof, Naculich:2014naa, Weinzierl:2014ava, 
  delaCruz:2015dpa}.

In string theory, one finds similar identities that follow from
world-sheet monodromy relations.  For massless vector amplitudes of
the open string, from world-sheet monodromy relations~\cite{Monodromy, Stieberger:2009hq}
one finds
\begin{equation}
 \A(1,3,2,4)=   \frac{\sin(\pi \alpha'  s)}{\sin(\pi\alpha' u)} \A(1,2,3,4)\,.
\end{equation}
where $\alpha'$ is the inverse string tension.

We can also use the two relations, $n_t=-n_u-n_s$ and $n_u = t \A(1,2,3,4)+u \,n_s/s$, in  \eqn{cnrepl}.
The result is
\begin{equation}
\calMfour(1,2,3,4) = -i\Bigl[ \frac{n_s^2}{s}+ \frac{n_t^2}{t}+ \frac{n_u^2}{u} \Bigr] 
= - i\frac{s t}{u} \Big[\A(1,2,3,4)\Big]^2 \,,
\label{4ptKLT}
\end{equation}
where, as usual, we have suppressed the gravitational coupling setting $\kappa=2$.
As for the gauge-theory case, $n_s$ drops out; as in that case, this is to be expected as it 
would otherwise lead to a relation between gauge invariant and gauge-dependent quantities, $\calMfour$ and $n_s$ respectively.
We can put this equation into a more standard form using a relabeling identity \eqref{4PtBCJAmplitude},
\begin{equation}
\calMfour(1,2,3,4) = - i s \A(1,2,3,4) \A(1,2,4,3)\,,
\label{KLTFourPt}
\end{equation}
which is the simplest of the \emph{KLT relations} between gravity and gauge-theory amplitudes.
We derived it here as a consequence of \ck duality and gauge-invariance
constraints, but the original derivation~\cite{KLT} comes from string theory. It is worth noting that
these relations are not unique given amplitude relations such as \eqn{4PtBCJAmplitude}.

Replacing the four-point YM amplitude in the from \eqn{4ptExplicit} into
the KLT relation \eqref{KLTFourPt}, we obtain an explicit form for the four-graviton
amplitude
\begin{equation}
\calMfour(1,2,3,4) = - i \frac{ t_{16} R^4}{stu}\,,
\end{equation}
where we define $t_{16} R^4$ in terms of $t_8 F^4$ in \eqn{t8F4} as
\begin{equation}
t_{16} R^4 \equiv  \big(t_8 F^4\big)^2 \, .
\end{equation} 
As the notation suggest, $t_{16} R^4$ can also be written as a contraction between a
rank-16 tensor $t_{16}$ and four linearized Riemann tensors, using the
relationship to linearized gauge-theory field strengths in \eqn{FieldStrength},
\begin{equation}
R_i^{\mu \nu \rho \sigma} = F_i^{\mu\nu}
F_i^{\rho\sigma}=(p^{\mu}_i \varepsilon_i^{\nu} - \varepsilon_i^{\mu} p^{\nu}_i
)(p^{\rho}_i \varepsilon_i^{\sigma} - \varepsilon_i^{\rho} p^{\sigma}_i )\,.
\end{equation}

\subsection{Outline of topics}

\begin{figure}[tb]
\begin{center}
\includegraphics[scale=.45]{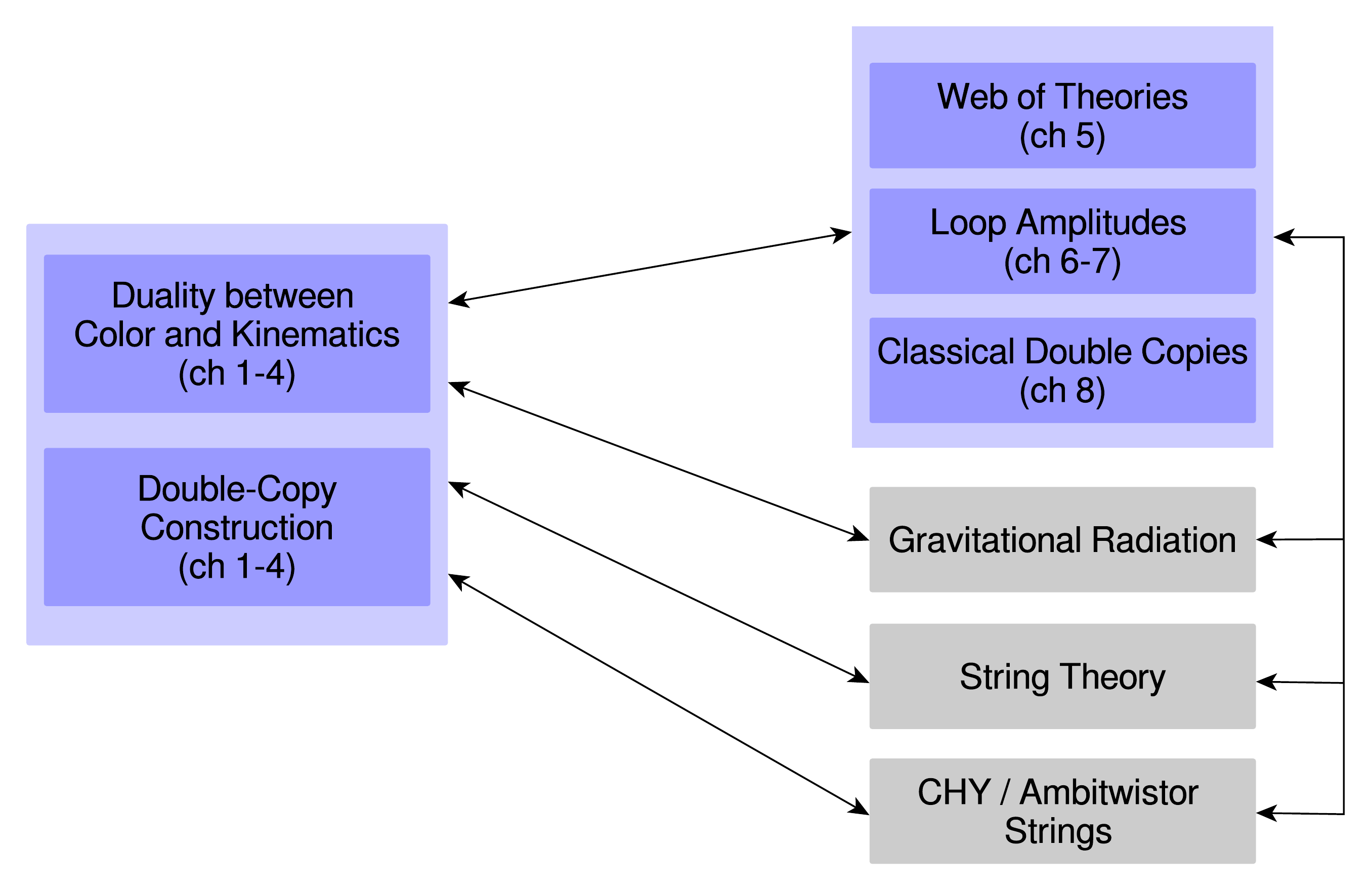}
\end{center}
\vskip -.3 cm
\caption{Connections of \ck duality to various
  topics.  This review will discuss in some detail 
the connection of \ck duality to the topics in 
the upper right (with the main chapters indicated) and less so to the topics on the
lower right. The various topics are intertwined with each other as well. }
\label{ConnectionsFigure}
\end{figure}

In this review, we will describe the duality between color and
kinematics and the double copy, as proposed in the original
work~\cite{BCJ, BCJLoop}, and later refined through various extensions
and applications.  As indicated in \fig{ConnectionsFigure}, \ck
duality and double copy are intertwined with the topics of several
vigorous research fields. The areas that the review will mainly focus
on include the web of theories, loop amplitudes and the classical
double copy.  The web of theories allude to the large classes of known
double-copy constructions and their underlying single-copy theories,
whose existence became clear after important theories, such as
Chern-Simons~\cite{Bargheer2012gv},
Yang-Mills-Einstein~\cite{Chiodaroli2014xia},
Maxwell-Einstein~\cite{Chiodaroli2014xia,Chiodaroli2015wal},
spontaneously-broken theories~\cite{Chiodaroli2015rdg} and gauged
supergravities~\cite{Chiodaroli2017ehv} were observed to fit into the
general framework. In addition, from the Cachazo, He and Yuan (CHY)
formulation~\cite{Cachazo:2013hca}, it was
observed~\cite{Cachazo2014xea} that also effective field theories such
as the non-linear-sigma-model (NLSM)~\cite{Chen2013fya},
(Dirac)-Born-Infeld (DBI) and special-Galileon theory played a central
role.

The usefulness of the duality and the double copy for loop amplitudes
becames clear once the framework was applied to obtain compact
integrands for the three-~\cite{BCJLoop} and
four-loop~\cite{SimplifyingBCJ} amplitudes in $\cN=4$ SYM and in
$\cN=8$ supergravity. By now it is clear that loop amplitudes in many
other theories can be obtained using the duality and double copy.

When the double copy was shown to be applicable to problems of
classical gravity, such as the Schwarzschild and Kerr
metrics~\cite{Monteiro2014cda} as well as other perturbatively
constructable metrics~\cite{Luna2016hge}, it opened the door to
further applications relevant to gravitational physics. With the
discovery of gravitational waves from merging binary black holes and
neutron stars~\cite{Abbott:2016blz,TheLIGOScientific:2017qsa}, it is
becoming increasingly important to find better ways to accurately
calculate classical observables in general relativity. The double-copy
approach is still in its infancy, but it bears the promise of
drastically changing the way we think of carrying out computations in gravity.

In order to keep the discussion manageable, we
will not discuss in much detail the challenges of understanding
gravitational radiation and potentials (see \eg
Refs.~\cite{BlanchetReview, Buonanno:2014aza, Porto:2016pyg,
  LeviReview} for reviews and Refs.~\cite{CheungPM,3PM,3PMLong} for a
state-of-the-art application of the double copy). Nor will we be
thorough in describing the connections to string theory (see \eg
Refs.~\cite{Monodromy,Stieberger:2009hq, Stieberger2016lng,
  SchlottererEYMHeterotic}), the CHY
construction~\cite{Cachazo:2013gna, Cachazo:2013iaa, Cachazo:2013hca,
  Cachazo:2013iea} and ambitwistor strings~\cite{Mason:2013sva,
  Adamo:2013tsa, Casali2015vta, He:2015yua, Geyer:2015jch,
  Geyer:2015bja, Cachazo:2015aol, Azevedo:2017lkz, Geyer:2017ela,
  Geyer:2018xwu, Geyer:2019hnn}, all of which have interesting
connections to \ck duality and the double-copy construction.

The outline of topics in each section is as follows: In
\sect{DualitySection}, we describe the duality in some detail and give
various examples, and show how the double copy implies diffeomorphism
invariance of gravity.  In \sect{TreeSection}, we give a way to
visualize how the duality can be thought of as specifying amplitudes
in terms of boundary data on a graph of graphs and on making use of
relabeling invariance.  Then, in \sect{GravitySymmetriesSection}, we
discuss the inheritance of symmetries in the double-copy theories from
their component theories. \Sect{ZoologySection1} gives a detailed
description of the web of double-copy constructible theories,
emphasizing the widespread applicability of these ideas.  In
\sect{ExamplesSection}, we give loop-level examples of the duality
between color and kinematics.  In \sect{GeneralizedDoubleCopySection},
we explain a generalized double-copy procedure that does not require
loop integrands to manifest the duality.
\Sect{ClassicalDoubleCopySection} discusses the important issue of
extending the double-copy procedure to solutions of the classical
equations of motion.  Conclusions and prospects for the future are
given in \sect{ConclusionSection}. In \app{NotationSection}, we
collect acronyms and notation used throughout the review.
\App{SpinorSuperspaceSection} summarizes spinor helicity and on-shell
supersymmetry, which will be useful in various sections. Finally,
\app{GeneralizedUnitaritySection} briefly describes generalized
unitarity, used in
\sects{ExamplesSection}{GeneralizedDoubleCopySection}.


\section{The duality between color and kinematics}
\label{DualitySection}

The duality between color and kinematics is by now an extensive topic
with a variety of perspectives and applications.  However,
it is not always clear from the literature what rules govern this
framework. In this section, the central aspects of \ck
duality will be described, with the aim of clarifying the reason for
imposing various requirements as well as providing an understanding of when
they can be relaxed.

\subsection{What is the duality between color and kinematics? \label{sec2.1}}

\ck duality in its original formulation states that
it is possible to reorganize the perturbative expansion of tree-level
amplitudes in $D$-dimensional pure YM theory with a general
gauge group $G$ in terms of cubic diagrams where the kinematic
numerators obey the same Jacobi relations and
symmetry properties as their color factors~\cite{BCJ,
  BCJLoop}. While it is not a priori obvious why such a reorganization
is possible or even desirable, from a Lagrangian perspective this is a
highly nontrivial statement about YM theory.  The associated
double-copy construction however, does
make it clear that the duality is worth understanding because of the
way it connects gravity to gauge theory.  While there are
tree-level proofs of the duality from the amplitudes
perspective~\cite{KiermaierTalk, BjerrumMomKernel,
  MafraExplicitBCJNumerators}, at present, only a partial Lagrangian-level 
understanding has been achieved~\cite{Square, WeinzierlBCJLagrangian, Vaman:2014iwa,
  Mastrolia:2015maa}.

More generally, \ck duality refers to the statement that
in many gauge theories, extending well beyond  YM
theories with or without matter, it should be possible to reorganize the
perturbative expansion so
that there is a one-to-one map between the Lie-algebra identities of
the color factors carried by certain diagrams (with cubic or
higher-point vertices) and the identities of the kinematic numerators
of the same diagrams. In the broad class of general gauge
theories, one can think of \ck duality as a constraint
that can be imposed on fields, gauge-group representations,
interactions and operators, such that the theories give amplitudes
that exhibit the duality structure. These constraints often result in
theories with properties that are interesting for reasons not directly
related to the duality~\cite{Chiodaroli2014xia,Chiodaroli2015rdg, Chiodaroli2015wal,
  Johansson2017srf, JohanssonConformal}. 

In generalizing beyond gauge theories, one can consider  matter theories that are
comprised of spin $< 1$ states that transform nontrivially under a
semi-simple global group. In this case, \ck duality refers to the one-to-one map
between the Lie-algebra relations of this global group and the
relations satisfied by the corresponding kinematic numerators of the
diagrams. It is convenient to still refer to the global group as the color group
since such theories can often be regarded as the matter sector of a
gauge theory. Such matter theories can have amplitudes that
nontrivially obey the duality (as in the case of the nonlinear sigma model (NLSM)~\cite{Chen2013fya}
discussed in \sects{subsecNLSM}{sec-zoo-nongrav}), thus mimicking the
intricate kinematic structure of gauge theories, or they can be
completely trivial manifestations of the duality (e.g.\ bi-adjoint
$\phi^3$ theory \cite{Du2011js, OConnellAlgebras}).  The most
remarkable aspect of \ck duality is that it naturally leads to
scattering amplitudes in double-copy theories.  \Sect{ZoologySection1}
describes a remarkable web of theories that are connected by the
duality and the  double copy.

Finally, for amplitudes that are not obtained from the standard
QFT framework involving Feynman diagrams, such as
string-theory amplitudes, it is convenient to define \ck duality to
mean that these amplitudes obey the same relations as if they were
generated by a duality-satisfying diagrammatic expansion of the
gauge-theory type.  For example, the single-trace vector-amplitude
sector of the heterotic string obeys the same relations as that of
YM theory~\cite{Stieberger2014hba}. Hence, we can write heterotic
string amplitudes as a sum over cubic diagrams with duality-satisfying
kinematic numerators, even if this might not seem completely natural
from a string-theory perspective.

\subsection{General statement of the duality and the double copy for gauge theories}

Consider scattering amplitudes in a nonabelian
gauge theory with the following properties: there is a gauge-group
$G$ under which all fields transform nontrivially; particles of
different mass are assigned to various representations of the gauge
group; the interactions are controlled by a gauge coupling constant
$g$ and a set of elementary color tensors ${\cal C}=\big\{ f^{abc},
(t^a)_i^{~j}, \ldots \big\}$. The set of elementary color tensors may include higher-rank tensors as indicated by the ellipsis.  

An $L$-loop $m$-point scattering amplitude in this $D$-dimensional
gauge theory can then be organized as\footnote{Our conventions for the overall phase in the representations of gauge-theory and gravity amplitudes follow the one in \cite{Chiodaroli2017ngp} rather than the original BCJ papers \cite{BCJ,BCJLoop}.}
\begin{equation}
{\cal A}^{(L)}_{m} = i^{L-1} g^{m-2+2L} \sum_{i}\, 
 \int \frac{d^{LD}\ell}{(2\pi)^{LD}} \frac{1}{S_i} \frac{c_{i} n_i}{D_i} \, ,
 \label{gaugeAmp}
 \end{equation}
where the sum runs over the distinct $L$-loop $m$-point diagrams that
can be constructed by contracting the elements of  ${\cal C}$ in various allowed ways 
(consistent with the choice of external
particle representations, and where the valency of each vertex is determined by the tensor rank). We take each such diagram to correspond to a
unique color factor $c_i$. Each diagram has an associated denominator
factor $D_i$ which is constructed by taking a product of the
denominators of the Feynman propagators $\sim 1/(p^2-m_j^2)$ of each
internal line of the diagram. For simplicity of notation, we assume that the color
representation of the line uniquely specifies the mass $m_j$ of the
propagator.  Cases with differing masses, but the
same color representation, are easily taken into account by setting
appropriate masses and representations equal at the end.  The adjoint
representation is by default massless and is associated to gluons 
(and, in some cases, additional fields).
The remaining nontrivial kinematic dependence is collected in the
kinematic numerator $n_i$ associated with each diagram. The numerators
$n_i$ are in general gauge-dependent functions that depend on external
momenta $p_j$, loop momenta $\ell_l$, polarizations $\varepsilon_j$,
spinors, flavor, etc., everything except for the color degrees of
freedom. The integral measure is defined as $d^{LD}\ell=\prod_{l=1}^L d^D\ell_l$.
Finally, $S_i$ are standard symmetry factors that remove
internal overcount of loop diagrams; they can be computed by counting
the number of discrete symmetries of each diagram with fixed external
legs.
 
The color factors $c_i$ are in general not independent. They satisfy
linear relations that are inherited from the Lie algebra structure,
such as the Jacobi identity and the defining commutation relation,
\begin{align}
f^{dae}f^{ebc}-f^{dbe}f^{eac}&=f^{abe}f^{ecd}\,, \nn \\
(t^a)_i^{~k}(t^b)_k^{~j}-(t^b)_i^{~k}(t^a)_k^{~j} 
  &= i f^{abc} (t^c)_i^{~j}\,,
\label{ColorID}
\end{align}
and similar identities for other color tensors that might appear
in the theory. In \eqn{ColorID} we follow the standard textbook normalization
of color generators~\cite{PeskinSchroeder},
\begin{equation}
\Tr(t^a t^b) = \frac{\delta^{ab}}{2}\,.
\label{TextBookTraceNorm}
\end{equation}
Such Lie-algebra relations are directly tied to 
gauge invariance of amplitudes.  

\begin{figure}[t]
\centering
\includegraphics[scale=01.0,trim=0 0 0 0,clip=true]{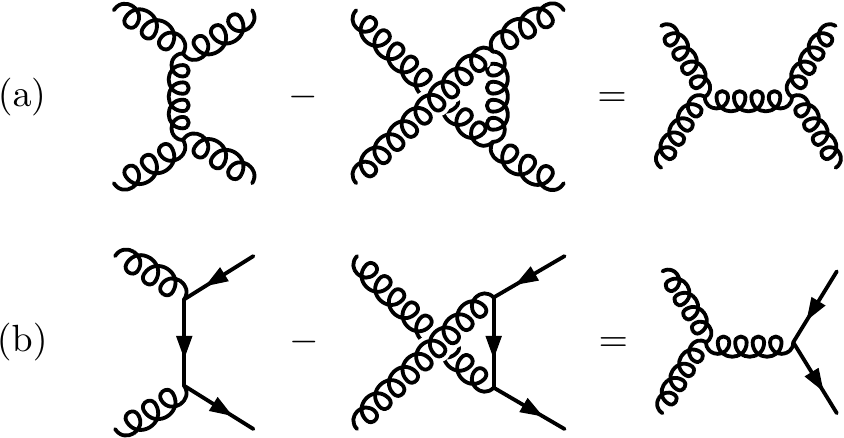}
\vspace{-3pt}
\caption{\small Color-algebra relations in the adjoint~(a) and
  fundamental representation~(b).  The curly lines represent adjoint
  representation states and the straight lines fundamental
  representation. The vertices correspond to the color matrices in 
  \eqn{ftildeNorm}. }
\label{fig:fundjacobi}
\end{figure}

In the amplitudes community, color generators
differ from the textbook definition by a $\sqrt{2}$ factor absorbed into each generator~\cite{ManganoParkeReview}.
It is also useful to rescale the group-theory structure constants,
\begin{equation}
\T^{a} \equiv \sqrt{2} t^a \,, \hskip 1.5 cm 
\f^{abc} \equiv i \sqrt{2} f^{abc} \,,
\label{ftildeNorm}
\end{equation}
so that we have the identity 
\begin{equation}
{\rm Tr}(\T^{a}\T^{b})=\delta^{ab}\,,
\label{TraceNorm}
\end{equation}
With these changes in normalization the defining commutation relations 
are,
\begin{align}
\f^{dae} \f^{ebc}- \f^{dbe} \f^{eac}&= \f^{abe} \f^{ecd}\,, \nn \\
(\T^a)_i^{~k}(\T^b)_k^{~j}-(\T^b)_i^{~k}(\T^a)_k^{~j}
  &=  \f^{abc} (\T^c)_i^{~j}\,,
\label{ColorIDRecale}
\end{align}
as illustrated in \fig{fig:fundjacobi}.  These identities imply that
there exist relations between triplets of color factors
$\{c_i,c_j,c_k\}$ which take, for example, the form $c_i - c_j=c_k$.

The scattering amplitude (\ref{gaugeAmp}) is said to obey
\ck duality if the kinematic numerator factors obey the
same general algebraic relations as the color factors do, e.g.\
\begin{equation}
 n_i - n_j = n_k \quad \Leftrightarrow \quad c_i - c_j = c_k\, ,
\label{duality}
\end{equation}
which is a generalization of the kinematic Jacobi identity in
\eqn{BCJFourPt}.  The relative signs between the terms depend on
choices in defining the color factors for each diagram.  The essential
point regarding the signs is that whatever choice is made for the
color factors are inherited by the corresponding numerator factors.
Another form of the duality in terms of color traces has also been
found~\cite{Bern:2011ia, Du:2013sha, Fu:2013qna, Du:2014uua, Naculich:2014rta, 
  Fu:2018hpu}, but the most natural form is in terms of color factors
of diagrams as described above.

It is a nontrivial task to find duality-satisfying
numerators since standard methods such as Feynman rules, on-shell
recursion~\cite{BCFW}, or generalized
unitarity~\cite{UnitarityMethod,Fusing,TripleCuteeJets,BCFUnitarity},
generally do not automatically gives such numerators.  A
straightforward but somewhat tedious way to find such representations
is to use an ansatz constrained to match the amplitude and manifest
the duality~\cite{FivePointN4BCJ, SimplifyingBCJ}.  Constructive ways
to obtain numerators have also been devised~\cite{Mafra:2009bz,
  BjerrumMomKernel, MafraExplicitBCJNumerators, Mafra:2014oia, BjerrumManifestingBCJ,Carrasco2016ldy,
  DuTengBCJ, delaCruz:2017zqr, Bridges:2019siz}.  Aside from amplitudes, the duality
has also been found to hold for currents with one off-shell
leg~\cite{FourLoopFormFactor, FiveLoopFormFactor, PureSpinorsBCJAmplProof,
  Mafra:2016ltu, PureSpinorsBCJAmplProof, Mafra2016mcc,
  SchlottererBGCurrent,Jurado:2017xut, Boels:2017ftb}.  A natural way
for making the duality valid for general off-shell quantities would be
to find a Lagrangian that generates Feynman rules whose diagrams
manifest the duality. At present, such Lagrangian is only known to a few orders
in perturbation
theory~\cite{Square,WeinzierlBCJLagrangian,Vaman:2014iwa,
  Mastrolia:2015maa};  an important problem is to find a closed form
of such a Lagrangian valid to all orders.

The color relations (\ref{ColorIDRecale}) have important implications
for kinematic numerators of diagrams.
If we start with a set of numerators that
satisfy the duality \eqref{duality}, and shift the numerators,
\begin{equation}
 n_i = n_i'- \Delta_i  \, .
\label{GaugeShift}
\end{equation}
subject to the constraint,
\begin{equation}
\sum_{i}\,
 \int \frac{d^{LD}\ell}{(2\pi)^{LD}} \frac{1}{S_i} \frac{c_{i} \Delta_i}{D_i}  = 0 \, ,
\end{equation}
the amplitude is unchanged. Because the color factors are not
independent, nontrivial shifts of the kinematic numerators
can be carried out.
 In this way, without changing the
amplitude, we can rewrite the amplitude in terms of a set of
numerators $n_i'$ not obeying the duality relations \eqref{duality}
starting from ones that do obey it.  The $\Delta_i$ are pure gauge
functions, i.e. they drop out of the amplitude. 

When  we have numerators $n_i$  that obey the same
algebraic relations as the color factors $c_i$, we can obtain 
sensible objects by formally replacing color factors  by kinematic
numerators as
\begin{equation} c_i \rightarrow n_i\,,
\label{c_n_replace}
\end{equation}
in any given formula or amplitude. Given the algebraic properties are
the same, this replacement is consistent with gauge-invariance
properties inherited from the gauge theory.  As we discuss below, this
color-to-kinematics replacement---or double-copy construction---gives
us gravity amplitudes with remarkable ease.

Consider two amplitudes ${\cal A}^{(L)}_{m}$ and ${\widetilde {\cal
    A}}^{(L)}_{m}$, and organize them as in
\eqn{gaugeAmp}. Furthermore, take the color factors to be the same in
the two amplitudes, and label the two sets of numerators as $n_i$ and
$\tilde n_i$, respectively. If at least one of the amplitudes, say
${\widetilde {\cal A}}^{(L)}_{m}$, manifests \ck
duality, we may now replace the color factors of the first amplitude
with the duality-satisfying numerators $\tilde n_i$ of the second
one. This gives the double-copy formula for gravitational scattering
amplitudes \cite{BCJ,BCJLoop},
\begin{equation}
{\cal M}^{(L)}_{m} ~ = ~   {\cal A}^{(L)}_{m} \Big|_{c_i \! \rightarrow  \tilde n_i 
                            \atop g\rightarrow  {\kappa}/{2}}
  = ~ i^{L-1}\;\!\Big(\frac{\kappa}{2}\Big)^{m-2+2L} \sum_{i}\, \int \frac{d^{LD}\ell}{(2\pi)^{LD}} \frac{1}{S_i} \frac{n_i \tilde{n}_i}{D_i} \,,
\label{DCformula}
\end{equation}
where the gravitational coupling $\kappa/2$ which
compensates for the change of engineering dimension when replacing
color factors with kinematic numerators.  In general we will 
omit the factors of $\kappa/2$ by taking $\kappa = 2$.

For the replacement $c_i \rightarrow \tilde n_i$ to be valid under the
integration symbol, it is important that the color factors are not
explicitly evaluated by summing over the contracted indices. At least
one contracted index per loop should not be explicitly summed over;
this is required so that the duality is not spoiled by treating color
and kinematics differently.  The numerators depend on loop momenta
$n_i=n_i(\ell)$ that is not yet integrated over, thus analogously the
color factors should be thought of as depending on the unevaluated
internal indices. If this subtlety is ignored, it may happen that
color factors explicitly vanish when combining the color sum with
symmetries of particular color factors, and this vanishing behavior
should not be imposed on the un-integrated numerators.  Stated
differently, we do not wish to impose any specific color-factor
properties on the numerator factors, only generic ones.

As the notation suggests, the two sets of numerators $n_i$, $\tilde n_i$ can differ in several ways: (1) they can describe different gauge choices for the same scattering process, (2) they can describe different external states in the same theory, and (3) they can originate from two different gauge theories. The first case allows us to work with numerators where only one set obeys the duality manifestly. The second case allows us to describe gravitational states that are not built out of a symmetric-tensor product
\begin{equation}
{(\rm gravity\,\,state)}  = {(\rm gauge\,\,state)} \otimes {(\rm \widetilde{gauge\,\,state})} \,.
\end{equation}
The third case allows us to describe gravitational theories that are not left-right symmetric double copies of gauge theories
\begin{equation}
{(\rm gravity\,\,theory)}  = {(\rm gauge\,\,theory)} \otimes {(\rm \widetilde{gauge\,\,theory})} \,.
\end{equation}
In \sect{ZoologySection1}, we will see  that this latter case is crucial for probing the web of double-copy-constructible theories. 

When two different gauge theories are considered in the double-copy
formula, it is important that both, in principle, can be put into a
form displaying \ck duality, even if this property needs only to be
explicit in one of the amplitudes. 
This ensures
that the generalized unitarity cuts of the loop-level double-copy
formula will be unique and gauge invariant. The link between gauge
invariance and BCJ amplitude relations has been
explored in Refs.~\cite{Arkani-Hamed:2016rak, RodinaGaugeInv,
  Du:2018khm, Plefka:2018zwm, Hou:2018bwm}. The amplitude relations
can also be understood in terms of a symmetry that act as
momentum-dependent shifts on the color factors~\cite{Brown:2016mrh,
  Brown:2016hck}.  Note that the precise form of the BCJ amplitude
relations depends on the details of the gauge-group representations
and elementary color tensors. The standard BCJ amplitude
relations~\cite{BCJ}, for example, follow from considering theories
with only adjoint particles that interact via $f^{abc}$ color tensors.

We will come back to the double-copy constructions of different
theories in later sections, but for now we will focus on illustrating
the details of \ck duality on some familiar gauge theories.

\subsection{Example 1: Tree level amplitudes with adjoint-only particles}

Consider pure YM theory in $D$ spacetime dimensions, consisting of gluons transforming in the adjoint representation of a gauge group $G$, with Lagrangian
\be
{\cal L}_{\rm YM}= -\frac{1}{4}(F_{\mu \nu}^a)^2\,,~~~~\text{where}~~~~F_{\mu \nu}^a= \partial_\mu A_\nu^a - \partial_\nu A_\mu^a + g f^{abc} A_\mu^b A_\nu^c\,.
\ee
Next consider $m$-point tree-level amplitudes. We know that the only color structure that appears are contractions of $f^{abc}$ structure constants, thus the color factors must be in one-to-one correspondence with all possible cubic diagrams with $m$ external legs. 

Cubic diagrams at multiplicity $m=j+1$ can be built recursively by attaching a new leg to every possible edge of a multiplicity-$j$ diagram. There are $(2j-3)$ edges of a given $j$-point diagram, hence the recursion gives:
\be
\text{number of cubic diagrams} =1\times 3 \times 5 \times 7 \times \cdots \times \big( 2j-3\big) = (2m-5)!! \,.
\ee 

We organize the tree amplitude in terms of all such propagator-distinct diagrams with 
only cubic vertices, 
\be
{\cal A}^\tree_m \equiv {\cal A}^{(0)}_m = -i  g^{m-2} \sum_{i=1}^{(2m-5)!!} \frac{c_i n_i}{D_i}\,,
\label{TreeAmpMpt}
\ee 
where $c_i$ are the color factors that are straightforwardly
obtained from the $i$-th diagram.  Similarly, the $D_i$ denote the
denominators of the propagators that correspond to the diagram lines.
The $n_i$ are the corresponding kinematic
numerators.  Depending on the context we will
alternate between using the diagram weights $n_i,c_i, D_i$ with
subscripts indexed by a diagram-id number, as well as a
functional maps from graph to their respective weights: $n_i \equiv
n(g_i)$, $c_i\equiv c(g_i)$, and $D_i\equiv D(g_i)$ where $g_i$ is the graph corresponding 
to the index $i$.

It is useful to first clarify what we mean by independent diagrams. The
least redundancy occurs when we insist on only one instance of a diagram
with the same propagator contribution.  This is distinct from the
number of unique diagram topologies.  Let us take a concrete example at
four-points.  We have discussed in \sect{IntroductionSection} that we
need $s$, $t$, and $u$ diagrams at four point. They have same graphical
topology, but different external labels, which results in different
generic propagator contributions.  

As a trivial example, at four points for each distinct 
propagator structure we can relabel the external 
legs without altering the propagators but flipping
the signs of the color.  For example, consider the $s$-channel
diagram in \fig{FourPtGluonsFigure} which we can label as $g_{s:1}$.
Taking the graph $g_{s:2}$ to be $g_{s:1}$ but with legs $1$ and $2$ swapped,
we obtain the same propagator but the color factors are different:
\begin{align}
c(g_{s:1}) &= \f^{a_1 a_2 b} \f^{b a_3 a_4} \,, \nonumber\\
c(g_{s:2}) &= \f^{a_2 a_1 b} \f^{b a_3 a_4} \,,
\label{4ptcolorfactors}
\end{align}
where we use the normalization in \eqn{ftildeNorm}.
The color
factors, while distinct, are related by a negative sign
 inherited by the antisymmetry of the
structure generators: $c(g_{s:2})=-c(g_{s:1})$.  For the purpose of describing scattering
amplitudes in terms of functions of diagrams, we will always take the
kinematic weights of the diagrams to obey the same antisymmetry:
$n(g_{s:2})=-n(g_{s:1})$, whether or not we are discussing a
\cknospace-satisfying representation.  This means that for any
multiplicity and loop order we will have in mind a canonical layout of
distinct diagrams which determine the color factor and numerator signs.
These signs cancel from color-dressed amplitudes because the numerator
sign are correlated with the color signs.  However, they will affect the
signs appearing in the relation between color-ordered partial 
amplitudes and kinematic numerators, as well as the 
relative signs between terms in the Jacobi identities.

\begin{figure}[tb]
\begin{center}
\includegraphics[width=6.in]{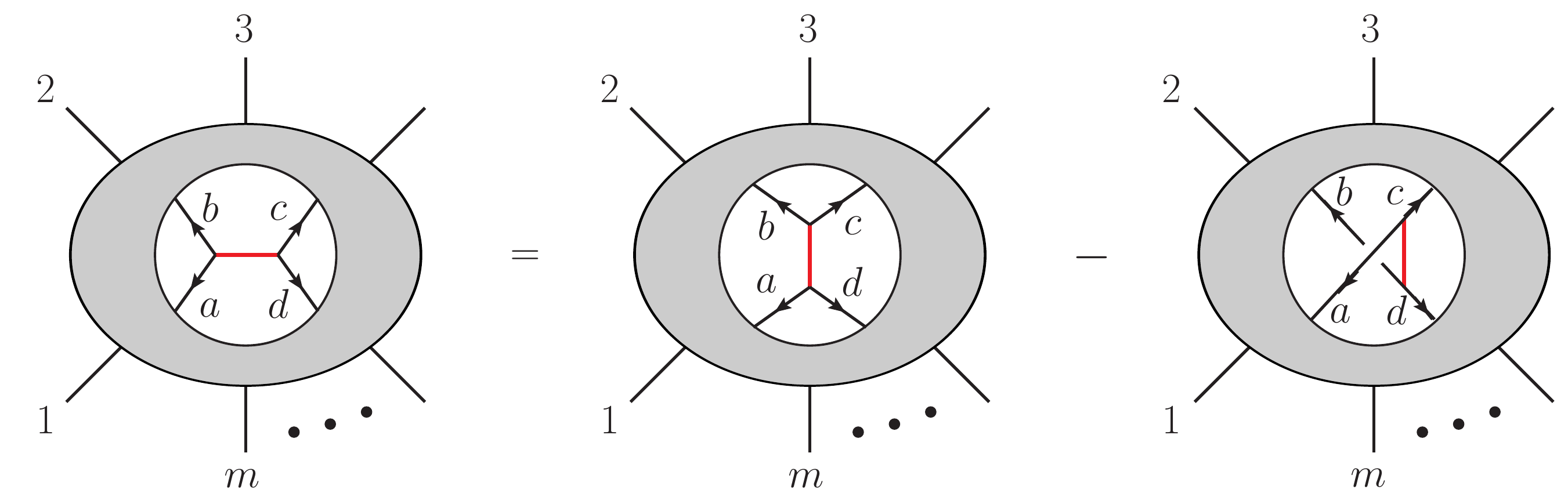}
\end{center}
\vskip -.4 cm
\caption{A Jacobi identity embedded in a generic diagram. The diagram can be either 
at tree level or at loop level. The arrows indicate that the lines are oriented the same way.}
\label{GeneralJacobiFigure}
\end{figure}

To be more explicit, as illustrated in \fig{GeneralJacobiFigure},
triplets of diagrams $(i,j,k)$ satisfy Jacobi
relations of the form
\begin{equation}
c_i-c_j+c_k = \big(\f^{d a  e} \f^{e b c} - \f^{a  b e} \f^{e c d} + \f^{d b e} \f^{e ca}\big) C^{abcd} =0\,.
\label{jacTwo}
\end{equation}
where the last factor $C^{abcd}$ is a color tensor that is common to
the diagrams in the triplet (external adjoint indices $a_1,\ldots, a_m$
are suppressed).  
As noted above, the relative signs are simply due
to choices in the ordering of the color indices in the $\f^{abc}$s.
While these relative sign choices are arbitrary, these signs are the
same as for the corresponding kinematic Jacobi identities.

More generally, the $(2m-5)!!$ color factors in (\ref{TreeAmpMpt}) are
related by Jacobi identities.  In total, at every multiplicity
$m$ there are $\frac{1}{3}(m-3)(2m-5)!!$ such Jacobi relations;
however, only $(2m-5)!!-(m-2)!$ of them are independent equations
because we can formally solve all Jacobi relations by mapping the color
factors to a $(m-2)!$ basis.

Writing the adjoint generator matrices as $(\Tadj^a)_{bc} \equiv \f^{bac}$, defined in \eqn{ftildeNorm}, we can
write any color factor as products of $\Tadj^{a_i}$'s, possibly involving
commutators of the adjoint generators. For example, pick a cubic tree
diagram and find the unique path through the diagram that connect leg $1$
and leg $m$. For each cubic vertex along this path, write down the
corresponding commutator of $\Tadj^{a_i}$'s that describes the subdiagram
that attaches this vertex. The product of these factors give $c_i$ for
the full diagram. For example, consider the color factor of the
following diagram
\def\usegraph#1#2{\includegraphics[scale=0.45,trim=0 #1 0 0]{#2}}
\be
c\Bigg(\hskip -.05 cm \raisebox{-.4 cm }{\usegraph{10}{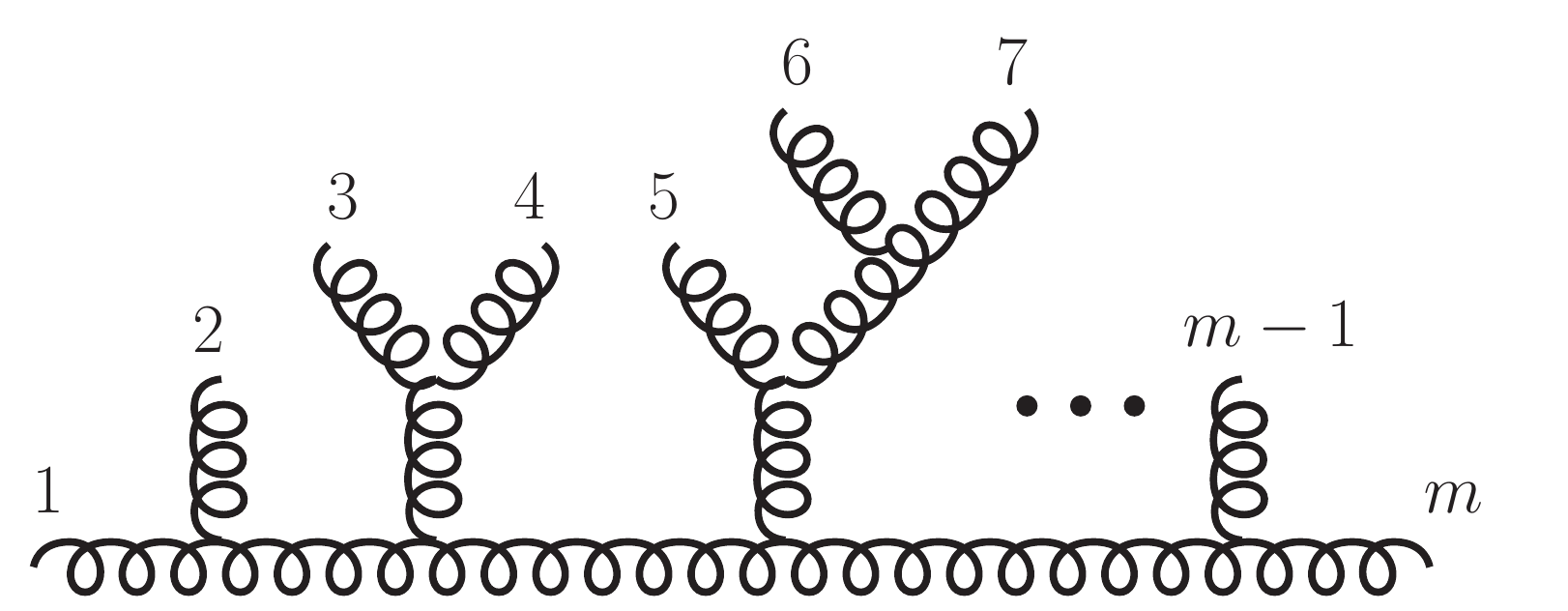}} \hskip -.3 cm\Bigg) \! 
 = (\Tadj^{a_2}[\Tadj^{a_3}\!,  \Tadj^{a_4}] 
    [\Tadj^{a_5}\!, [\Tadj^{a_6}\!, \Tadj^{a_7}  ] ]  \cdots  \Tadj^{a_{m-1}})_{a_1 a_{m}}\,,
\ee
where the adjoint indices of leg $1$ and $m$ correspond to the
external matrix indices of the adjoint representation. The commutators
arise from systematically eliminating subdiagrams involving $\f^{bac}$
using the standard Lie-algebra identity $\f^{abc} \Tadj^c = [\Tadj^a,
  \Tadj^b]$. Once only commutators of $\Tadj^a_i$'s remain, they can of be
written out as differences and sums of generators in different orders.

In summary, any color factor can in general be written as
\begin{equation}
c_i ~= ~ \sum_{\sigma \in S_{m-2}}  b_{i\sigma} \, \big(\Tadj^{a_{\sigma(2)}} \Tadj^{a_{\sigma(3)}} \Tadj^{a_{\sigma(4)}}  \cdots   \Tadj^{a_{\sigma(m-1)}}\big)_{a_1 a_{m}}\,,  
\end{equation}
where $b_{i \sigma} \in \{0,\pm 1\}$ are  coefficients that
depend on the permutation and on the specific color factor. They can
be evaluated case by case, but their explicit values are not important
here for our purposes. The main result is that color factors $c_i$ in
\eqn{TreeAmpMpt} can be eliminated in favor of expressing the
gauge-theory tree amplitude in terms of a sum over the possible
products of adjoint generators $\Tadj^{a_i}$, where the first and $m$-th
leg is kept fixed.  This gives a so-called  Del Duca-Dixon-Maltoni (DDM) color 
decomposition~\cite{DixonMaltoni} of the gauge-theory tree amplitude,
\begin{equation}
{\cal A}_m^\tree = g^{m-2} \sum_{\sigma \in S_{m-2}}   
A_m^\tree \big(1, \sigma(2), \sigma(3), \ldots, \sigma(m-1),m\big) 
\big(\Tadj^{a_{\sigma(2)}} \, \Tadj^{a_{\sigma(3)}} \cdots   \Tadj^{a_{\sigma(m-1)}}\big)_{a_1 a_{m}} \,,
\label{DDMdecomp} 
\end{equation}
where the sum runs over $(m-2)!$ permutations. The kinematic
coefficients multiplying the color factors define a basis of $(m-2)!$
partial amplitudes, which we indicate as $A^{\text{tree}}_m\big(1, \sigma(2), \sigma(3), \ldots,
\sigma(m-1),m\big)$. This is usually called the Kleiss-Kuijf (KK)
basis~\cite{KleissKuijf}.

The partial tree amplitudes in YM theory, $A_m^{\text{tree}}\big(1, 2,
\ldots, m\big)$, have a number of useful
properties~\cite{ManganoParkeReview}:
\begin{itemize}
\item They are functions of kinematic variables only, $(\varepsilon_i,
  p_i)$; the color dependence is only reflected by the ordering of the
  external particle labels.
\item They at most have poles in planar channels, i.e. when
  consecutive momenta add up to a null momentum $\big(\sum_{j \le
    i\le k} p_i\big)^2=0$ (mod $m$).
\item The amplitudes are invariant under cyclic permutations:
  \begin{equation}
  A_m^\tree\big(1, 2, \ldots, m\big)= A^{\text{tree}}_m\big(2, \ldots, m, 1\big) \,.
\end{equation}
\item Under reversal of the ordering, they at most change by a sign flip: 
\begin{equation}
A_m^\tree\big(m, \ldots, 2, 1\big)= (-1)^{m} A^{\text{tree}}_m\big(1, 2, \ldots, m\big)\,.
\end{equation}
\item They satisfy a photon-decoupling identity: 
\be
\sum_{\rm \sigma \in cyclic}  A_m^\tree\big(1, \sigma(2), \ldots,\sigma(m) \big)=0\,,
\ee
 where cyclic permutations of all but one leg are summed over.
\item They satisfy KK relations~\cite{KleissKuijf}: 
\be
A_m^\tree(1, \alpha, m, \beta) = (-1)^{|\beta|} \sum_{\sigma \in \alpha \shuffle {\beta^T} } A^{\text{tree}}_m(1, \sigma, m)\,, 
\ee
where $\alpha$ and $\beta$ are arbitrary-sized lists of the external legs, $\beta^T$ is used to represent the reverse ordering of the list $\beta$, and $ \alpha \shuffle \beta^T$ is the shuffle product of these lists (i.e. permutations that separately maintain the order of the individual elements belonging to each list). $|\beta|$ denotes the number of elements in the list $\beta$.
\item They obey BCJ relations, which in the simplest incarnation take the form~\cite{BCJ}:
\be
\sum_{i=2}^{m-1} p_1 \cdot (p_2+\ldots + p_i) \, A_m^\tree(2, \ldots , i, 1, i+1,  \ldots, m)=0\,. \label{BCJrels}
\ee
\item  After considering all permutations of the above BCJ relation, there are  only $(m-3)!$ independent partial tree amplitudes~\cite{BCJ}. The position of three consecutive legs can be fixed in the cyclic ordering; for example, $A^{\text{tree}}_m\big(1, 2, \sigma(3), \ldots, \sigma(m-1),m\big)$ can be chosen as the independent BCJ basis. 
\end{itemize}
The first property is obvious from our definition of the partial amplitudes; however, the remaining ones require some explanation. 

The fact that the partial tree amplitudes are invariant under cyclic
permutations of their arguments is most easily seen after a basis
change of the color factors. 
 We rewrite the color factors in
terms of traces of generators, $\T^a$. 
From \eqn{ColorIDRecale}
\be
\f^{abc} \equiv i \sqrt{2} f^{abc} = {\rm Tr}([\T^a,\T^b] \T^c) =  {\rm Tr}(\T^a\T^b \T^c)- {\rm Tr}(\T^b\T^a \T^c)\,,
\ee
which follows from the identity (\ref{ColorIDRecale}) after
multiplying both sides with $T^{c'}$, tracing over the fundamental
indices, and using \eqn{TraceNorm}.  This basis change implicitly
assumes that we have specialized to a gauge group were we can use 't~Hooft's
double-line notation~\cite{tHooft:1973alw}, say $G=U(N_c)$.

The generators of $U(N_c)$ obey the completeness relation
$(\T^{a})_{i}^j (\T^{a})_{k}^l =\delta_i^{l}\delta_k^{j}$, implying that
products of several $f^{abc}$ can be expressed by merging several
traces
\begin{align}
(\Tadj^{a_{2}} \Tadj^{a_{3}} \cdots   \Tadj^{a_{m-1}})_{a_1 a_{m}} 
&=   \f^{a_1 a_2 b_1} \f^{b_1 a_3 b_2} \cdots \f^{b_{m-3} a_{m-1} a_m} 
\label{trMap} \\ 
&=  {\rm Tr}(\T^{a_1}\T^{a_2}\T^{a_3}\cdots \T^{a_m}) + (-1)^m  {\rm Tr}(\T^{a_m}\cdots \T^{a_3}\T^{a_2} \T^{a_1}) + \ldots\,
\nn 
\end{align}
where on the last line the suppressed terms corresponds to
$2^{m-1}$ distinct permutations of the trace over $m$ generators. Out
of all the permutations that appear, only the two displayed terms have
the property that the generators $\T^{a_1}$ and $\T^{a_m}$ are adjacent
(in the cyclic sense). This implies that, after replacing the DDM color
factors with the trace-basis color factors in \eqn{DDMdecomp}, we can
uniquely identify the location of, say, the ${\rm
  Tr}(\T^{a_1}\T^{a_2}\T^{a_3}\cdots \T^{a_m})$ factor. It appears only
once in $(\Tadj^{a_{2}} \Tadj^{a_{3}} \cdots \Tadj^{a_{m-1}})_{a_1 a_{m}}$,
which uniquely multiplies the partial tree amplitude $A^{\text{tree}}_m(1,2,3,
\ldots, m)$. Hence, $A^{\text{tree}}_m(1,2,3, \ldots, m)$ must be the kinematic
coefficient of ${\rm Tr}(\T^{a_1}\T^{a_2}\T^{a_3}\cdots \T^{a_m})$ in the
trace-basis decomposition of the YM tree amplitude.

By crossing symmetry in the trace basis, the decomposition into partial
amplitudes has the form
\begin{equation}
{\cal A}_m^\tree= g^{m-2} \sum_{\sigma \in S_{m-1}}   
A_m^\tree\big(1, \sigma(2), \sigma(3), \ldots, \sigma(m)\big) \, 
 {\rm Tr}(\T^{a_1}\T^{a_{\sigma(2)}}\T^{a_{\sigma(3)}}\cdots \T^{a_{\sigma(m)}}) \,,
\label{TraceAmp} 
\end{equation}
which can be straightforwardly verified starting from \eqn{DDMdecomp}.
Crossing symmetry requires the summation over $(m-1)!$ terms, since we
can fix the location of one leg, say leg 1, by the cyclic property of
the trace. $(m-1)!$ is significantly larger than the $(m-2)!$
terms in \eqn{DDMdecomp}. Where did the extra terms come from? In
fact, they are the terms we suppressed in \eqn{trMap}, which have
combined in various ways to complete the formula (\ref{TraceAmp}).
Finally, since the partial amplitudes in the DDM decomposition are the
same partial amplitudes that appear in the trace decomposition, it
follows that the partial amplitudes inherit the cyclic invariance of
the trace. Further details of the trace basis and partial amplitudes
may be found in Refs.~\cite{ManganoParkeReview}.

A number of the other properties discussed above also follow from the
exercise of mapping between the DDM and trace basis. The reversal
(anti-)symmetry follows from observing that the term $ (-1)^m  {\rm
  Tr}(\T^{a_m}\cdots \T^{a_3}\T^{a_2} \T^{a_1})$ in \eqn{trMap} always goes together with ${\rm
  Tr}(\T^{a_1}\T^{a_2}\T^{a_3}\cdots \T^{a_m})$. The photon-decoupling
identity follows from realizing that we can replace one generator in
\eqn{TraceAmp} by the $U(1)$ ``photon'' generator $T_{U(1)}=1$, which
naturally belongs to the gauge group $U(N_c)=SU(N_c)\times
U(1)$. However, gluons  do not couple directly
to photons since the latter have no charges.  This can be seen directly
by looking at the structure constants $\f^{ab U(1)} = {\rm  Tr}([\T^a,\T^b]\, 1 ) = 0$.  
Hence, the photon-decoupling identity follows from the vanishing of the amplitude with
one photon.

The KK relations are explained by the fact that there are
two different decompositions of the tree amplitude, the DDM
(\ref{DDMdecomp}) and the trace (\ref{TraceAmp}) decomposition, which
use a different number of partial amplitudes, $(m-2)!$ and $(m-1)!$,
respectively. The only way that this can be consistent is if there
exist relations that map the $(m-1)!$ partial amplitudes into a
$(m-2)!$ basis. This is precisely what the KK relations
do. Recall that it was because the color factors are built only out of
$\f^{abc}$'s, which obey the Jacobi relations, that we could
find the $(m-2)!$ basis. Thus, any theory where all fields transform in
the adjoint representation and the whose amplitudes depend only color tensor is $f^{abc}$ will obey the
KK relations.

The BCJ amplitude relations are a consequence of \ck duality,
specifically of its interplay with gauge invariance. Consider the
$(m-2)!$ partial amplitudes expressed in terms of numerators, they
take the form
\begin{equation}
A_m^\tree(1,\sigma(2),\ldots,\sigma(m-1),m) = - i \sum_{i \in {\rm planar}}  b_{i \sigma} \frac{n_i}{D_i}\,,
\label{partialAmplitudeForm}
\end{equation}
where $n_i$ are the kinematic numerator weights of the diagrams
canonical to some ordering layout, $D_i$ are the propagators of the
diagram, $b_{i \sigma} \in \{0,\pm 1\}$ are coefficients that
depend on the ordering $\sigma$, and the sum is only nonvanishing for
planar diagrams with respect to the ordering $\sigma$. 

We can impose kinematic Jacobi identities on the numerators, expressing all
diagram numerators in terms of $(m-2)!$ independent {master numerators}.
We can then imagine attempting to invert the matrix between color-ordered amplitudes and these master numerators.  One will find a
remarkable surprise---only $(m-3)!$ master numerators can be solved for in
terms of some $(m-3)!$ color-ordered amplitudes---the rest of the
master numerators contribute only as unfixed parameters representing a
kind of generalization of gauge freedom.  The remaining equations
relate color-ordered amplitudes directly to simple functions
of the ordered amplitudes with $(m-3)!$ legs fixed with no dependence
on numerator choice.  For example,  following the original presentation~\cite{BCJ}, one can express the entirety  of the
KK $(m-2)!$ basis amplitudes in terms of $(m-3)!$
amplitudes as follows: 
\newcommand\AtreeCO[1]{{ A}^{\rm tree}_#1}
\begin{equation}
\AtreeCO{m}(1,2,\{\alpha\},3,\{\beta\})=
\!\!\!\!\!\!\!\!\! \sum_{ \sigma  \in {\rm POP}(\{\alpha\},\{\beta\})}  \!\!\!\!\!\!\!\!\! 
\AtreeCO{m}(1,2,3,\sigma )  \prod_{k=4}^{|\{\alpha\}|+3}  
{\frac{{\cal F}_k(3,\sigma ,1)}{ s_{24\ldots k} }} \,,
\label{allnBCJ}
\end{equation}
where $|\{\alpha\}|$ is the length of the list $\{\alpha\}$, 
and the sum runs over {partially ordered permutations} (POP) of
the merged $\{\alpha\}$ and $\{\beta\}$ sets.  To be clear we are referring to leg labels,  e.g. in $s_{24\ldots k}$,  with labels $4$ through $k$ as the first $(k-3)$ entries of the ordered list $\{\alpha,\beta\}$. \Eqn{allnBCJ} gives all
permutations of $\{\alpha\} \bigcup \{\beta\}$ consistent with the
order of the $\{\beta\}$ elements. Either $\alpha$ or $\beta$ may be empty, trivially so for the $\alpha$ case.
The  function ${\cal F}_k$ associated with leg $k$ is given by,
\begin{align}
{\cal F}_k(\{ \rho\} ) &=
\left\{ 
\begin{array}{ll}
          \sum_{l=t_k}^{m-1} {\cal S}_{k,\rho_l} & \mbox{if $t_{k-1} < t_{k}$}\\
        - \sum_{l=1}^{t_k} {\cal S}_{k,\rho_l} & \mbox{if $t_{k-1} > t_{k}$}
\end{array} \right\} \nn \\&
 \null +
\left\{ 
\begin{array}{ll}
         s_{2 4\ldots k} & \mbox{if $t_{k-1} < t_{k}< t_{k+1}$}\\
        -s_{2 4\ldots k} & \mbox{if $t_{k-1} > t_{k} > t_{k+1}$}\\
         0 & \mbox{otherwise}
\end{array} \right\} \,,
\hskip 1 cm 
\end{align}
where 
\begin{equation}
 s_{i j  \ldots k} \equiv (p_i + p_j + \cdots + p_k)^2\,,
\end{equation}
and $t_k$ is the position of leg $k$ in the set $\{\rho\}$,
except for $t_3$ and $t_{|\{\alpha\}|+4}$ which are always defined to
be,
\begin{equation}
t_3\equiv t_5 \,, \hskip 2cm t_{|\{\alpha\}|+4} \equiv 0 \,.
\end{equation}
For $|\{\alpha\}|=1$ this means that $t_3=t_5=t_{|\{\alpha\}|+4}=0$.
The expression ${\cal S}_{i,j}$ is given by,
\begin{equation}
\label{bcjFinal}
{\cal S}_{i,j}=\left\{ 
\begin{array}{ll}
        s_{ij}  & \mbox{if $i< j$ or $j=1$ or $j=3$}\\
        0& \mbox{otherwise}
\end{array} \right\} .
\end{equation}
The so-called {\em fundamental} BCJ relations (\eqn{BCJrels}) occur when the $|\{\alpha\}|= 1$. These amplitude relations were first identified in Ref.~\cite{BCJ}, and then
proven, first as a low-energy limit of string-theory
relations~\cite{Monodromy, Stieberger:2009hq}, and then directly using the Britto-Cachazo-Feng-Witten (BCFW) recursion relations
in field theory~\cite{amplituderelationProof,CachazoBCJProof}.

\begin{figure}[tb]
\begin{center}
\includegraphics[width=5.7in]{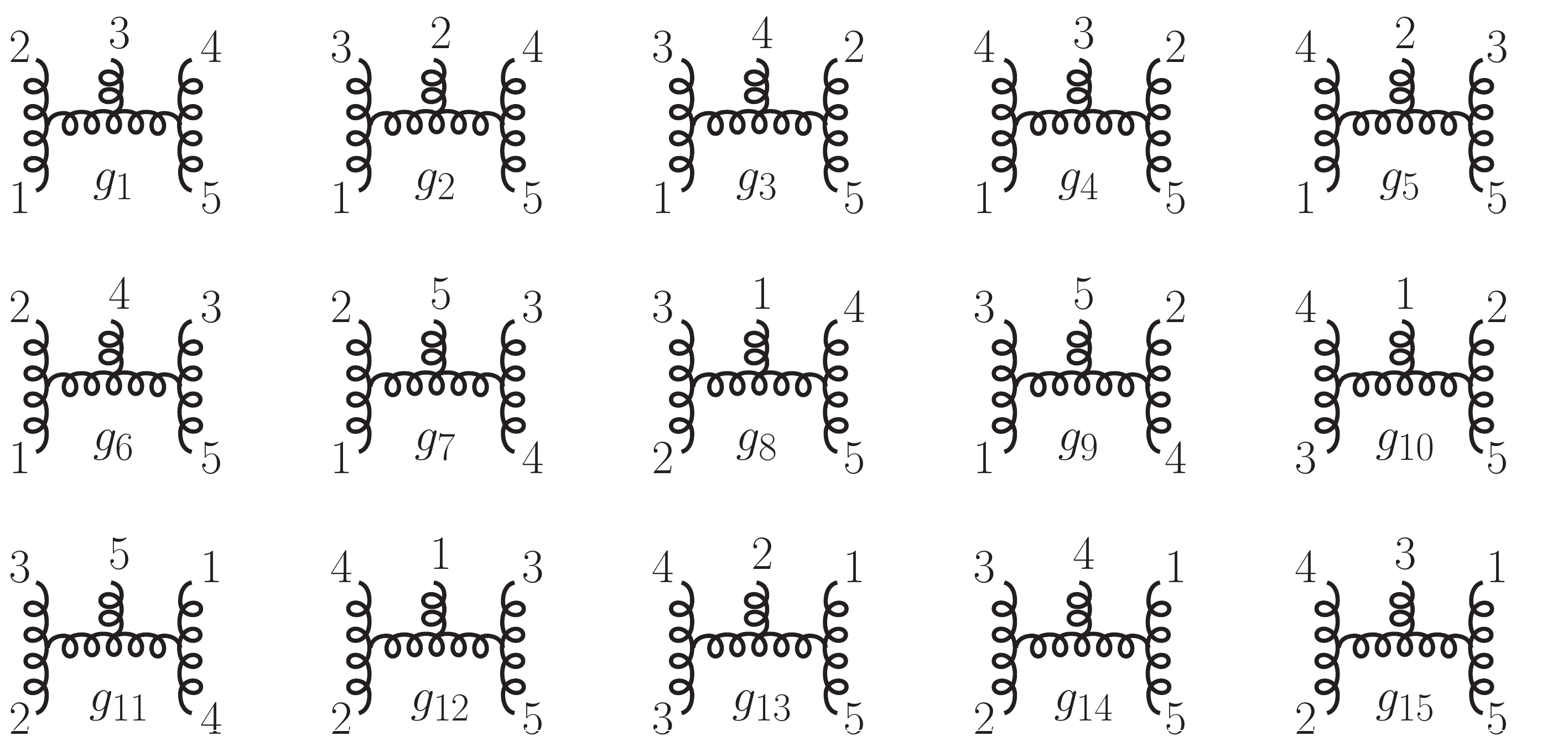}
\end{center}
\vskip -.3 cm
\caption{The color-dressed tree-level five-point amplitude an be organized using these fifteen graphs with only cubic vertices. }
\label{fivePointGraphs}
\end{figure}

Consider the five-point amplitude (e.g. governing two-to-three scattering), which
offers a first nontrivial example.  In this case, 15
distinct cubic diagrams contribute, as illustrated in 
\fig{fivePointGraphs}. Only five of these contribute to
a given color-ordered partial amplitude. Let us consider diagram nine
from \fig{fivePointGraphs}.  To see which color-orderings (and with which
signs) this diagram can contribute, we expand its canonical color-factor
in the trace basis. The color factors follow from dressing with the
structure functions $\f^{abc}$. Going to a trace basis we see that the
color weight associated with diagram nine is:
\begin{align}
c_9  =\ & \text{Tr}\left[\T^{a_ 1} \T^{a_ 2} \T^{a_ 4} \T^{a_ 5} \T^{a_ 3}\right]
-\text{Tr}\left[\T^{a_ 1} \T^{a_ 3} \T^{a_ 2} \T^{a_ 4} \T^{a_ 5}\right]
+\text{Tr}\left[\T^{a_ 1} \T^{a_ 3} \T^{a_ 4} \T^{a_ 2} \T^{a_ 5}\right] \nn \\
& 
+\text{Tr}\left[\T^{a_ 1} \T^{a_ 3} \T^{a_ 5} \T^{a_ 2} \T^{a_ 4}\right]
-\text{Tr}\left[\T^{a_ 1} \T^{a_ 3} \T^{a_ 5} \T^{a_ 4} \T^{a_ 2}\right]
-\text{Tr}\left[\T^{a_ 1} \T^{a_ 4} \T^{a_ 2} \T^{a_ 5} \T^{a_ 3}\right] \nn \\
&
-\text{Tr}\left[\T^{a_ 1} \T^{a_ 5} \T^{a_ 2} \T^{a_ 4} \T^{a_ 3}\right]
+\text{Tr}\left[\T^{a_ 1} \T^{a_ 5} \T^{a_ 4} \T^{a_ 2} \T^{a_ 3}\right] \,.
\end{align}
This implies that diagram nine will contribute to multiple
color-ordered partial amplitudes, defined as the coefficient of each
color trace in the full amplitude, with a variety of signs.  
The signs associated with each diagram in a partial amplitude are
easily determined for a given color ordering by reordering the legs of
each diagram to match the color ordering without allowing lines to
cross, and keeping track of the signs from permuting the ordering of
legs in each vertex.

Taking the layout as depicted in \fig{fivePointGraphs},
each diagram contributes to a color-ordered partial amplitude according
to whether we can flip the legs at each vertex (with a minus sign for each flip)
so that the cyclic ordering of legs matches the ordering of the arguments of the partial 
amplitudes.  For example, we have,
\be
iA_5^\tree(1,3,5,4,2)=\frac{n_ 1}{D_ 1} + \frac{n_ 2}{D_ 2} + 
  \frac{n_ 6}{D_ 6}-\frac{n_9}{D_ 9}-\frac{n_{12}}{D_{12}}\,,
\label{a13542}
\ee
as well as
\be
iA_5^\tree(1,3,5,2,4)=\frac{n_ 3}{D_ 3}+\frac{n_ 4}{D_ 4}+\frac{n_ 5}{D_ 5}
+\frac{n_9}{D_ 9}+\frac{n_{12}}{D_{12}}\,,
\label{a13524}
\ee 
where the $n_i$ are the kinematic numerators and the $1/D_i$ are
the products of Feynman propagators that can be read of from 
graph $g_i$ in \fig{fivePointGraphs}.

Jacobi relations imply that the $n_i$ of the diagrams in \fig{fivePointGraphs} are given
as linear functions of numerators
$\{n_1,n_2,n_3,n_4,n_5,n_6\}$, which we take as the master numerators.
In total there are nine independent Jacobi relations,
\begin{align}
& \hskip -.7 cm 
n_7  = n_6 - n_1 \,, \hskip .5 cm
n_8  = n_2 - n_1 \,, \hskip .5 cm
n_9  = n_3- n_2 \,, \hskip .5 cm
n_{10} = n_4 - n_3  \,, \hskip .5 cm
n_{11} = n_5 - n_4  \,,  \nn \\
& \hskip -.7 cm 
n_{12} = n_5 - n_6  \,, \hskip .5 cm
n_{13} = n_{10} - n_7  \,, \hskip .5 cm
n_{14} = n_{11} + n_8 \,, \hskip .5 cm
n_{15} = n_{12} - n_9 \,.
\end{align}
Solving this system in terms of the six  master numerators gives
\begin{align}
& \hskip -.7 cm 
n_7=-n_1 + n_6\,,\hskip.5cm
n_8=-n_1 + n_2\,,\hskip.5cm
n_9=-n_2 + n_3\,,\hskip.5cm
n_{10}=-n_3 + n_4\,,\hskip.5cm \nn\\
& \hskip -.7 cm
n_{11}=-n_4 + n_5\,,\hskip.5cm
n_{12}=n_5 - n_6\,,\hskip.5cm
n_{13}=n_1 - n_3 + n_4 - n_6\,,\hskip.5cm \nn\\
& \hskip -.7 cm
n_{14}=-n_1 + n_2 - n_4 + n_5\,,\hskip.5cm
n_{15}=n_2 - n_3 + n_5 - n_6\,.
\label{niSol}
\end{align}

Remarkably, by using \eqn{niSol}, we can show that the partial
amplitudes~\eqref{a13542} and \eqref{a13524} contain all information
necessary to describe all other ordered amplitudes at five point.  For
the sake of argument, solving \eqns{a13542}{a13524} for $n_1$ and
$n_4$ gives:
\begin{align}
n_1 &=i D_1  A_5^\tree(1,3,5,4,2) - n_2 \frac{D_1 }{D_2} - n_6\frac{D_1}{D_6} + (n_3-n_2)\frac{D_1}{D_9} + (n_5-n_6)\frac{D_1}{D_{12}} \,,  \nn\\ 
n_4 &= i D_4 A_5^\tree(1,3,5,2,4) - n_3\frac{D_4}{D_3} -  n_5 \frac{D_4}{D_5} +(n_2-n_3)\frac{D_4}{D_9} +(n_6-n_5)\frac{D_4}{D_{12}}\,, 
\label{n14_5pt_sol}
\end{align}
where we have replaced $n_9$ and $n_{12}$ with the master numerators,
using \eqn{niSol}.  Using this, we express any other partial amplitude in terms of
$A_5^\tree(1,3,5,4,2)$ and $A_5^\tree(1,3,5,2,4)$ by plugging in the solution \eqref{niSol}
for these expressions for $n_1$ and $n_4$.
Consider, for example, the partial amplitude:  
\be
iA_5^\tree(1,3, 2,5,4) =-\frac{n_2}{D_2}-\frac{n_3}{D_3}-\frac{n_4}{D_4}-\frac{n_8}{D_8}+\frac{n_{11}}{D_{11}} \,.
\ee
Jacobi relations constrain $n_{8}=-n_{1}+n_{2}$, and $n_{11}=-n_4 + n_5$.
Replacing all non-master numerators with master numerators using \eqn{niSol}, 
in conjunction with \eqn{n14_5pt_sol}, we find:
\begin{align}
iA_5^\tree(1,3,2,5,4) &= iA_5^\tree(1,3,5,4,2) \frac{ D_1}{D_8}+ iA_5^\tree(1,3,5,2,4) \biggl(-1-\frac{D_4}{D_{11}}\biggr) \nn\\
~&~ - \biggl(\frac{1}{D_2}+\frac{1}{D_8}+\frac{D_1}{D_2 D_8}+\frac{1}{D_9}+\frac{D_1}{D_8
 D_9}+\frac{D_4}{D_9 D_{11}}\biggr) n_2 \nonumber \\
~&~+\biggl(\frac{1}{D_9}+\frac{D_1}{D_8 D_9}+\frac{D_4}{D_3 D_{11}}+\frac{D_4}{D_9 D_{11}}\biggr) n_3\nonumber\\
~&~+\biggl(\frac{1}{D_5}+\frac{1}{D_{11}}+\frac{D_4}{D_5
D_{11}}+\frac{1}{D_{12}}+\frac{D_1}{D_8 D_{12}}+\frac{D_4}{D_{11} D_{12}}\biggr) n_5\nonumber\\
~&~ -\biggl(\frac{D_1}{D_6 D_8}+\frac{1}{D_{12}}+\frac{D_1}{D_8 D_{12}}+\frac{D_4}{D_{11}
D_{12}}\biggr) n_6 \,. \hskip 1 cm 
\end{align}
Using the explicit value of the propagators,
a dramatic cancellation occurs under
momentum conservation:
\begin{align}
A_5^\tree(1,3,2,5,4) = \null & - A_5^\tree(1,3,5,2,4) \left(1+\frac{s_{25}}{s_{23}}\right)
+ A_5^\tree(1,3,5,4,2) \frac{ s_{345}}{s_{23}} \nn\\
~&~ - n_6 \frac{\left(s_{23}+s_{24}+s_{25}+s_{345}\right)}{s_{23}
s_{24} s_{35}} + n_3 \frac{ \left(s_{23}+s_{24}+s_{25}+s_{345}\right)}{s_{23} s_{24} s_{245}}\nonumber\\
~&~ +n_5\frac{ \left(s_{24} \
\left(s_{25}+s_{35}\right)+s_{23}
\left(s_{24}+s_{235}\right)+s_{235} \left(s_{25}+s_{345}\right)\right)}{s_{23} s_{24} s_{35} s_{235}}\nonumber\\
~&~ - n_2\frac{ \left(s_{23} \left(s_{24}+s_{45}\right)+s_{45}
\left(s_{25}+s_{345}\right)+s_{24} \left(s_{245}+s_{345}\right)\right)}{s_{23} s_{24} s_{45} s_{245}}\nonumber\\
&= - A_5^\tree(1,3,5,2,4) \biggl(1+\frac{s_{25}}{s_{23}}\biggr) + A_5^\tree(1,3,5,4,2) \frac{ s_{345}}{s_{23}}\,. \hskip 1 cm 
\end{align}
All the coefficients in front of the remaining explicit numerators vanish, giving
$A_5^\tree(1,3,2,5,4)$ solely in terms of a basis of partial amplitudes
$A_5^\tree(1,3,5,4,2)$ and $A_5^\tree(1,3,5,2,4)$.  Indeed, this occurs for every
partial amplitude, yielding, the BCJ and KK amplitude
relations~\cite{BCJ}.

An interesting corollary of the independence of the remaining
five-point partial amplitudes on $n_2, n_3, n_5, n_6$, once $n_1$ and
$n_4$ are chosen as in \eqn{n14_5pt_sol}, is that we can choose to set
the former numerators to zero since they have no effect on any of the
partial amplitudes. This forces $n_1$ and $n_4$ to be nonlocal since
they must absorb the propagators of the diagrams whose numerators are
set to zero.

\subsubsection{KLT formula and proof of tree-level adjoint \ck duality  \label{secKLT}}

In this section we briefly review the Kawai-Lewellen-Tye (KLT) formulae for gravity
tree-level amplitudes~\cite{KLT}, first derived using string theory. We will show how they are intimately tied to the BCJ double copy in \eqn{DCformula}, and how they can be used to directly construct duality-satisfying numerators in purely-adjoint gauge theories. 

 Let us start by quoting the explicit KLT relations at three-, four-,
five- and six-points,
\begin{align}
\calMthree(1,2,3) = \null &   i A_3^\tree(1,2,3) \widetilde A_3^\tree(1,2,3)\,, \nn \\
\calMfour(1,2,3,4) = \null &  - i  s_{12} A_4^\tree(1,2,3,4) \widetilde A_4^\tree(1,2,4,3)\,, \nn \\
\calMfive(1,2,3,4,5) =
  \null &  i\, s_{12} s_{45} \Afive(1,2,3,4,5)  \widetilde A_5^\tree(1,3,5,4,2)  \nn\\
  & \hskip 1.2 cm
   + i \, s_{14} s_{25}\Afive(1,4,3,2,5) \widetilde A_5^\tree(1,3,5,2,4) \,, \nn \\
\calMsix(1,2,3,4,5,6) = \null & -i s_{12} s_{45} A_6^\tree(1,2,3,4,5,6) \bigl(s_{35} \widetilde A_6^\tree(2,1,5,3,4,6) \nn \\
&\hskip 1.2 cm  
  + \, (s_{34} + s_{35} ) \widetilde A_6^\tree(2,1,5,4,3,6) \bigr)  + {\cal P}(2,3,4) \,,
\label{KLT456}
\end{align}
where the ${\cal M}_n^\tree$ are tree-level gravity amplitudes and the
$A_n^\tree$ are color-ordered gauge-theory partial amplitudes, 
and ${\cal P}(2,3,4)$ represents a sum
over all permutations of leg labels $2, 3,$ and $4$. 

If $A$ and ${\tilde A}$ are the tree-level amplitudes of $D$-dimensional pure YM theory, then the map between the two sets of 
on-shell gluon polarization vectors $\varepsilon^i_\mu$,
with $SO(D-2)$ little-group indices $i$, 
and those of the double-copy 
fields can be made explicit,
\bea
(\varepsilon^h)^{ij}_{\mu \nu} &=&\varepsilon^{((i}_{\mu}  \varepsilon^{j))}_{\nu} ~~~~~~~~~ (\text{graviton})\,, \nn \\
(\varepsilon^{B})^{ij}_{\mu \nu} &=&\varepsilon^{[i}_{\mu}  \varepsilon^{j]}_{\nu}  ~~~~~~~~~~~ (B\text{-field})\,,     \label{polarizationdoublecopy} \\
(\varepsilon^{\phi})_{\mu \nu} &=&  \frac{\varepsilon^i_\mu   \varepsilon^j_\nu \delta_{ij}}{D-2}  ~~~~~~~\; (\text{dilaton}) \, .\nn
\eea
On the first line the gluon polarizations are multiplied in symmetric-traceless combinations corresponding to the $\frac{1}{2}(D-2)(D-1)-1$ states 
of a graviton. On the second line they are antisymmetrized corresponding to the $\frac{1}{2}(D-2)(D-3)$ states of an antisymmetric tensor field. 
The completeness of the set of gluon polarization vectors implies that the right-hand side of the third line of Eq.~\eqref{polarizationdoublecopy} 
is proportional to $\eta_{\mu\nu}$ up to momentum-dependent terms, so $(\varepsilon^{\phi})_{\mu \nu}$ describes a single state.
Adding them all up we find the $(D-2)^2$ states in the tensor product of two massless vectors, as we should.  

The double copy of $D$-dimensional pure YM theory gives gravity amplitudes ${\cal M}^\text{tree}$ that follow from the Lagrangian~\cite{Scherk:1974ca,Gross:1986mw}
\begin{equation}
S=\int d^D x\sqrt{-g}\left[-\frac{1}{2}R
+\frac{1}{2(D-2)}\partial^\mu\phi \partial_\mu \phi
+\frac{1}{6}e^{-4\phi/{(D-2)}}H^{\lambda\mu\nu}
H_{\lambda\mu\nu}\right] ,
\label{SN=0}
\end{equation}
where $H_{\lambda\mu\nu}$ is the field strength of the two-index antisymmetric tensor $B_{\mu\nu}$ and 
the non-canonical normalization of the 
dilaton quadratic term is chosen to avoid non-rational dependence on the spacetime dimension $D$.
The $\mathbb{Z}_2$ symmetry $B_{\mu\nu}\rightarrow -B_{\mu\nu}$ generates a consistent truncation of this Lagrangian 
to Einstein gravity coupled to $\phi$. The further $\mathbb{Z}_2$ symmetry of this truncation, $\phi\rightarrow -\phi$,
allows a further consistent truncation to Einstein gravity. The double copy analog of this truncation is realized by
choosing gluon polarizations in symmetric-traceless combinations, as for the graviton polarizations in \eqn{polarizationdoublecopy}.

To show the connection to the BCJ double copy, consider, for example, the five-point tree amplitude.
The double-copy amplitude in terms of Jacobi-satisfying numerators is,
\begin{align}
\calMfive(1,2,3,4,5) &= -i \sum_{i=1}^{15} \frac{n_i \tilde{n}_i}{D_i} \nn \\
& \null = -i \,\tilde{n}_1
   \left(\frac{n_1}{D_1}+\frac{n_1}{D_7}+\frac{n_1}{D_8}+\frac{n_1+n_4}{D_{13}}+\frac{n_1+n_4}{D_{14}}\right) \nn\\
   &~~~-i \, \tilde{n}_4
   \left(\frac{n_4}{D_4}+\frac{n_4}{D_{10}}+\frac{n_4}{D_{11}}+\frac{n_1+n_4}{D_{13}}+\frac{n_1+n_4}{D_{14}}\right) ,
\end{align}
where we used the solution \eqref{niSol} and the propagators $1/D_i$
can be read off from the diagrams in \fig{fivePointGraphs}.  As usual,
where we suppress an overall factor of $(\kappa/2)^3$. Remarkably,
after using the solution \eqref{niSol} for both the $n_i$ and $\n_i$,
the result depends only on the numerators $n_1,n_4$ and $\n_1,\n_4$.
Using \eqn{n14_5pt_sol}, we have,
\begin{align}
\calMfive(1,2,3,4,5) 
   &= - \, \left( \tilde{n}_1 \, \Afive(1,2,3,4,5) +\tilde{n}_4 \, \Afive(1,4,3,2,5) \right ) \nn\\
   &=  i\, s_{12} s_{45} \Afive(1,2,3,4,5)  \widetilde A_5^\tree(1,3,5,4,2)  \nn\\
   &~~~+
   i \, s_{14} s_{25}\Afive(1,4,3,2,5) \widetilde A_5^\tree(1,3,5,2,4) \,.
\end{align}
This implies that we can express the double copy in terms of partial tree amplitudes of the two gauge theories.

This structures applies also at higher points, and is captured by the $m$-point formula~\cite{KLT}:
\begin{align}
{\cal M}_m^\tree = -i
\!\!\!\!\!\!\!\!\!\!\!\!  \sum_{\sigma,\rho \in S_{m-3}(2,\dots,m-2)}\!\!\!\!\!\! \!\!\!\!\!\! 
A_m^\tree(1,\sigma,m-1,m)S[\sigma|\rho]\widetilde{A}_m^\tree(1,\rho,m,m-1)\,,
\label{KLT}
\end{align}
where we suppress an overall factor of $(\kappa/2)^{m-2}$.
The formula makes use of a matrix $S[\sigma|\rho]$ known as the field-theory KLT kernel.
This is an
$(m-3)! \times (m-3)!$ matrix of kinematic polynomials that acts on
the color-ordered amplitudes for $(m-3)!$ permutations of the external
legs~\cite{MultiLegOneLoopGravity,BjerrumBohr2010ta,BjerrumBohr2010yc,BjerrumMomKernel}:
\begin{align}
 S[\sigma|\rho]=\prod_{i=2}^{m-2}\biggl[
2p_1\cdot p_{\sigma_i}+\sum_{j=2}^{i}2p_{\sigma_i}\cdot p_{\sigma_j}\theta(\sigma_j,\sigma_i)_\rho\biggr]\,,
 \label{momKernel}
\end{align}
where $\theta(\sigma_j,\sigma_i)_\rho=1$ if $\sigma_j$ is before $\sigma_i$ in the permutation $\rho$, and zero otherwise. 
A compact definition, which reproduces Eq.~\eqref{momKernel} upon use of momentum conservation and on-shell conditions, can be given recursively 
\footnote{This recursive presentation of the KLT kernel has a string theory origin~\cite{BjerrumMomKernel}.} as~\cite{Carrasco2016ldy}, 
\be
S[A,j|B,j,C]_1 = 2 (p_1+p_B)\cdot p_j  \, S[A|B,C]_1  \,, \ \ \ \ \ \ S[2|2]_1 = s_{12} \, ,
\label{2.6}
\ee
where multiparticle labels $B=(b_1,b_2,\ldots,b_p)$ involve multiple external legs, and  we use the notation $p_B = p_{b_1}+p_{b_2}+\ldots+p_{b_p}$.
Using the recursive formula, we can obtain four-, five- and six-point KLT relations as particular cases.

\begin{figure}[tb]
\begin{center}
\includegraphics[width=2.6in]{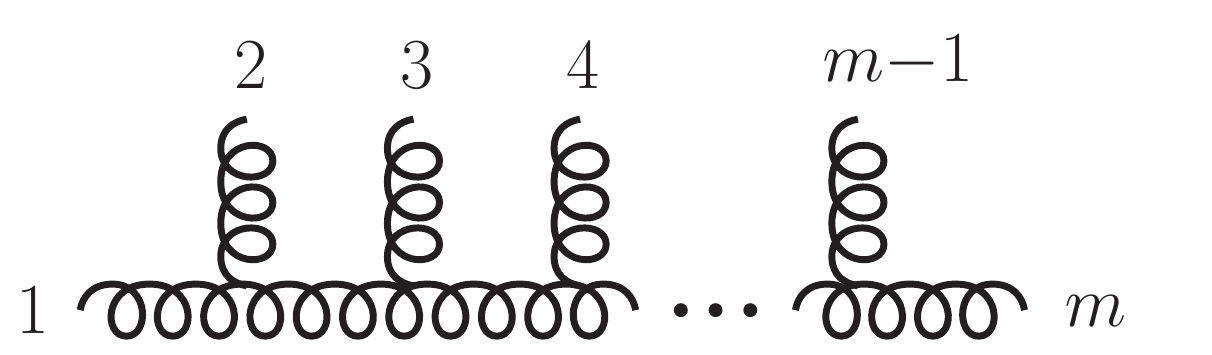}
\end{center}
\caption{An $m$-point half-ladder tree diagram.  
}
\label{HalfLadderFigure}
\end{figure}

In addition, the field-theory KLT kernel allows us to find explicit expressions for duality-satisfying tree-level numerators in the purely-adjoint case.  The construction that we give  here was independently worked out in refs.~\cite{KiermaierTalk,BjerrumMomKernel}.  The idea is to define the numerators for a subset of diagrams called half-ladder (or multi-peripheral) diagrams, whose structure is illustrated in \fig{HalfLadderFigure}.
Corresponding to permutations of these half-ladder diagrams we specify $(m-2)!$ master numerators via,
\bea
n(1,\sigma(2,\dots, m-2), m-1,m) &=& -i \sum_{\rho\in S_{m-3}} S[\sigma | \rho] \widetilde A_m^{\tree}(1,\rho,m, m-1)\,, \nn \\
n(1,\tau(2,\dots, m-1),m)\Big|_{\tau(m-1) \ne m-1} \!\!\!\!\!\!\!\!\!\!\!\!\!\!\!\!&=& 0 \,,
 \label{nonlocalNum}
\eea
and take the remaining $(2m-5)!!-(m-2)!$ numerators to be determined by the Jacobi relations. By definition these numerators satisfy all the kinematic Jacobi relations and hence they obey the CK duality. However, to conclude that they define valid numerators we must also prove that they give correct amplitudes, both in gauge theory and in gravity. 

Consider the DDM decomposition introduced in \eqn{DDMdecomp} for gauge-theory amplitudes with only adjoint particles. We use CK duality to replace the color factors in that formula with the above defined numerators,
\bea
{\cal M}_m^\tree = {\cal A}_m^\tree \Big|_{c_i \rightarrow n_i} \!  \! \!\! \!&=& \!\! \!   \sum_{\tau \in S_{m-2}}   
A_m^\tree \big(1, \tau(2,\ldots, m-1),m) \, 
n(1,\tau(2,\ldots, m-1),m)  \nn \\
\!  \!  \!\! \!&=& \!\! \! -i \!  \! \! \sum_{\sigma, \rho \in S_{m-3}}   \! \! \! 
 A_m^\tree \big(1, \sigma, m-1,m\big) 
S[\sigma | \rho] \widetilde A_m^\tree(1,\rho, m, m-1) \,.~~ 
\label{GravityDDMdecomp} 
\eea
On the first line we have a DDM decomposition for gravity amplitudes, where the half-ladder numerators play the same role as the half-ladder color factors in \eqn{DDMdecomp}. On the second line we have plugged in the explicit numerators, and used the fact that only $(m-3)!$ of them are non-vanishing. As is obvious, the KLT formula (\ref{KLT}) is reproduced. This implies that we get correct gravity amplitudes, given that both $A_m^\tree$ and $\widetilde A_m^\tree$ are gauge-theory amplitudes. If we take $A_m^\tree$ to be amplitudes in bi-adjoint $\phi^3$-theory and $\widetilde A_m^\tree$ are YM amplitudes, then the above KLT formula gives back YM amplitudes. Hence the numerators in \eqn{nonlocalNum} give correct amplitudes. 



This completes the constructive proof showing that \ck duality can be
satisfied for gauge theories with adjoint particles, given that all
the BCJ amplitude relations~(\ref{allnBCJ}) hold, which implies that the KLT formula hold. This argument relies on the availability of a DDM representation of the amplitude and on the existence of the field-theory KLT kernel.  Pure
YM theory, or ${\cal N}=1,2,4$ super-Yang-Mills (SYM) theory are
examples where tree-level \ck duality is proven by this argument.
Note, however, that the above numerators are nonlocal functions and
the crossing symmetry of the amplitude does not follow automatically
from relabeling the numerators. Hence it often desirable to find
other representations of tree-level numerators. In specific cases, we
can find representations of an amplitude with desired properties by
imposing these properties on an ansatz whose coefficients are
determined by requiring that it match the amplitude, a strategy also tremendously useful 
at loop level---see Secs.~\ref{subsecNLSM} and \ref{ExamplesSection}.

Similar considerations hold for amplitudes with multiple distinguishable adjoint
scalars, although it may be necessary to introduce four-scalar
interactions for the duality to hold~\cite{Johansson:2013nsa}.  As we
discuss in \sect{Sec-adjoint-ferm}, imposing the duality on
fermionic amplitudes implies supersymmetry.

\subsection{Example 2: Matter in fundamental representation \label{Sec-Fund-Fermions}}

\def\tf{f}

\begin{figure}[t]
\centering
$\f^{abc} =  c\! \left(\! \includegraphics[scale=0.45,trim=0 40 0 0]{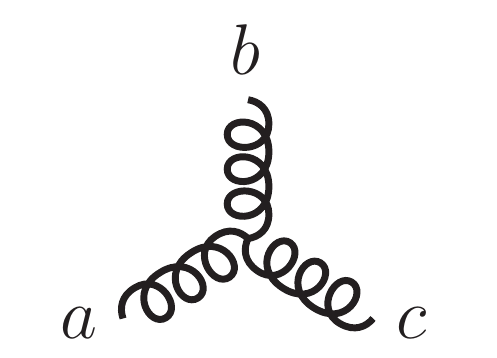} \! \! \!\right)
\hskip0.4cm
(\T^{b})_{i}^{\ j} =  c\!\left(\! \includegraphics[scale=0.45,trim=1 40 0 0]{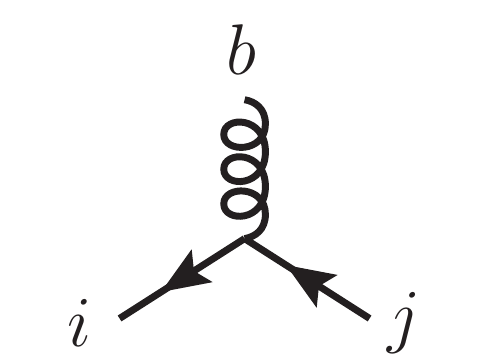} \! \! \!\right)
\hskip0.4cm
(\T^{b})^{ j}_{\ i} 
\equiv  c\!\left(\! \includegraphics[scale=0.45,trim=0 40 0 0]{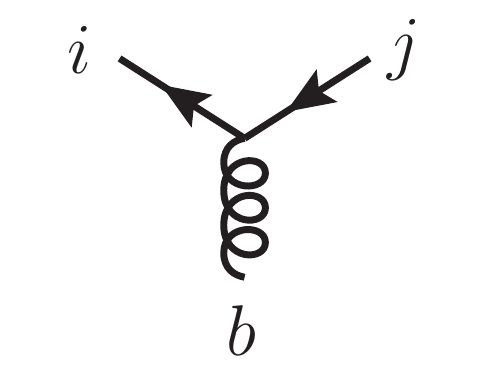} \! \!\! \right)= -(\T^{b})_{i}^{\ j}
$
\vspace{-5pt}
\caption{\small Color vertices with planar ordering
         consistent with the color-ordered Feynman rules.
}
\label{fig:colorvertices}
\end{figure}

We now generalize the discussion in the previous subsection by introducing matter in the fundamental representation, as  it appears in Quantum Chromodynamics (QCD)~\cite{Johansson2014zca, Johansson:2015oia, Johansson:2019dnu}. To be specific, let us consider YM theory with gauge group $G$ and with $N_{\!f}$ fundamental fermions.\footnote{A similar example of YM theory with scalars in matter representations will be discussed in \Sect{sec-zoo-tols}.}  For simplicity we call this theory QCD, given that it precisely matches QCD once we specify the
gauge group to be $SU(3)$ and the number of quark flavors to be six; its Lagrangian is
\begin{equation}
{\cal L}_{\rm QCD}= -\frac{1}{4}(F_{\mu \nu}^a)^2+ \overline{q}^\alpha (i \cancel{D}-M^{\ \beta}_{ \alpha} ) q_\beta \,,~~~~\text{where}~~~~D_\mu= \partial_\mu -i g A_\mu^a t^a \,, 
\end{equation}
where $\alpha,\beta=1,\ldots, N_f$ are flavor indices, and spinor
indices and fundamental gauge group indices are suppressed. The mass
matrix $M_\alpha^{\ \beta}$ is taken to be diagonal.  The only color
tensors are in this case $f^{abc}$ and $(t^a)_{i}^{ \ j}$ which
both have three free indices. Thus,  all color factors will
again correspond to cubic diagrams. A difference with the pure-adjoint case is
that we now need to decorate the lines of the diagrams with the
appropriate representation: adjoint,  fundamental or anti-fundamental. This is
illustrated in \fig{fig:colorvertices}.  A general color decomposition
of tree-level amplitudes with matter representations may be found in
Ref.~\cite{Johansson:2015oia} (see also Refs.~\cite{Melia:2013bta,Melia:2015ika,Ochirov:2019mtf}).

Without loss of generality we write the QCD $m$-point tree amplitude in terms of diagrams with cubic vertices,
\be
   {\cal A}^{\text{tree}}_{m,k} = - i g^{m-2}\!\!\!
      \sum_{i \in \text{cubic diag.}}^{\nu(m,k)}  \frac{c_i n_i}{D_i} \,,
\label{BCJformYM}
\ee
where $c_i$ are color factors, $n_i$ are kinematic numerators, and
$D_i$ are denominators encoding the propagator structure of the cubic diagrams.
The denominators (and numerators) may in principle contain masses,
corresponding to massive quark propagators. 
For $k$ quark-antiquark pairs and $(m-2k)>0$ gluons, we may count the number of cubic diagrams. 
Assuming that the quarks are all of distinct flavor,  one can then show that the number of nonzero diagrams is $\nu(m,k)=\frac{(2 m-5)!!}{(2 k-1)!!}$~\cite{Johansson:2015oia}.  As exemplified in Table~\ref{tab:CubicGraphsDistinguishable}, the numbers grow modestly with the number of  quarks. 

\begin{table*}[bt]
\centering
\begin{tabular}{c||c|c|c|c|c|c}
$k \setminus m$ & 3 & 4 & 5 & 6 & 7 & 8 \\
\hline
\hline
0 & 1 & 3 & 15 & 105 & 945 & 10395 \\
1 & 1 & 3 & 15 & 105 & 945 & 10395 \\
2 & - & 1 &  5 &  35 & 315 &  3465 \\
3 & - & - &  - &   7 &  63 &   693 \\
4 & - & - &  - &   - &   - &    99 \\
\hline
\end{tabular}
\\ \vspace{8pt}
\caption{\small Number of cubic diagrams, $\nu(m,k)$,
         in the full $m$-point amplitude with $k$ distinguishable quark-antiquark pairs
         and $(m-2k)$ gluons.}
\label{tab:CubicGraphsDistinguishable}
\end{table*}

Amplitudes with multiple quarks of the same flavor and mass can be
obtained from distinct-flavor amplitudes by setting masses to be equal
and summing over permutations of quarks with appropriate fermionic
signs.  Therefore, we do not lose generality by taking all $k$
quark-antiquark pairs to have distinct flavor and mass.  To be
explicit, in \tab{tab:CubicGraphsDistinguishable} we provide total
counts $\nu(m,k)$ of cubic diagrams for different amplitudes up to eight
particles and four quark pairs. It agrees with the usual counting of
standard QCD Feynman diagrams restricted to those diagrams that only
have trivalent vertices.

The color factors $c_i$ in \eqn{BCJformYM} are constructed
from the cubic diagrams using only two building blocks:
the structure constants $\f^{abc}$ for three-gluon vertices
and generators $(\T^a)_{i}^{ \ j}$ for quark-gluon vertices,
as shown in \fig{fig:colorvertices}.
When separating color from kinematics,
the diagrammatic crossing symmetry only holds up to signs
dependent on the permutation of legs.
These signs are apparent in the total antisymmetry of $\f^{abc}$. For a uniform treatment of the fundamental representation, 
it convenient to introduce a similar antisymmetry for the fundamental generators,
\be
 (\T^a)_{\ i}^{j}  \equiv - (\T^a)_{i}^{\ j} ~~~~ \Leftrightarrow~~~~\f^{cab} = - \f^{bac} \,.
\label{signflip}
\ee
This allows us to introduce a similar antisymmetry 
in color-ordered kinematic vertices, so that they are effectively the same as for the adjoint representation. 
As noted in \eqn{ColorIDRecale} the color factors obey Jacobi and commutation identities.
They both imply color-algebraic relations of the form given in
\eqn{duality}, and differ only by the subdiagrams as drawn in
\fig{GeneralJacobiFigure}, but otherwise have common diagram structure.
The interdependence among the color factors~$c_i$ means that the
corresponding kinematic coefficients~$n_i/D_i$ are in general not
unique, as reflected by the underlying gauge dependence of the
numerators.

\begin{figure}[t]
\centering
\includegraphics[scale=0.52,trim=0 0 0 0,clip=true]{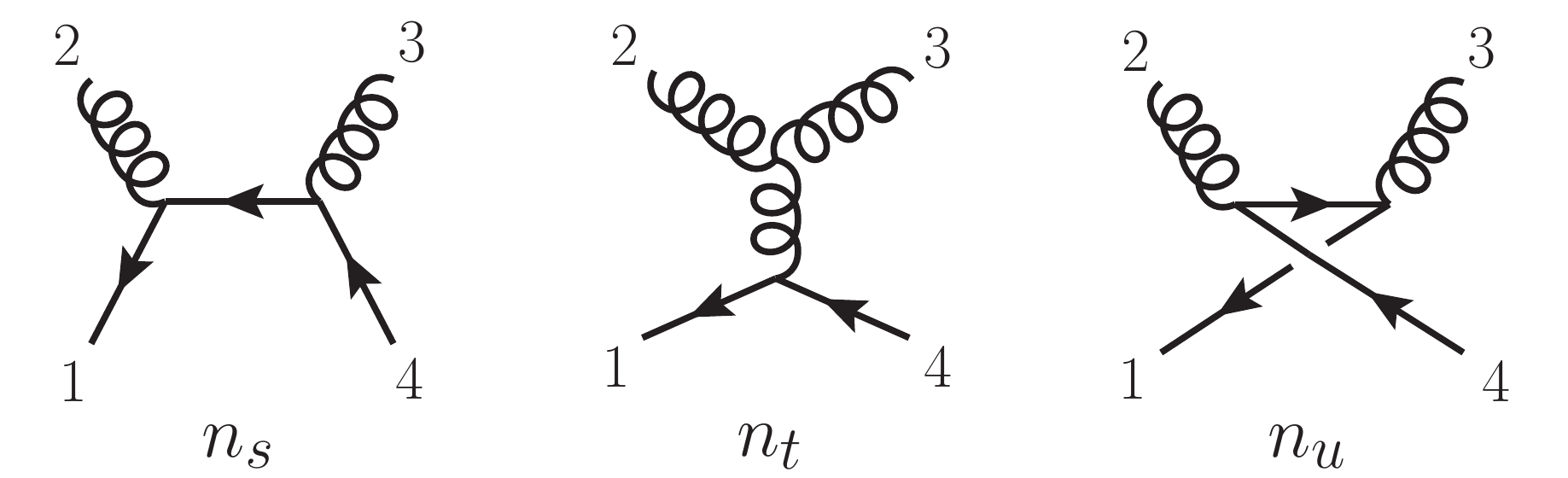}
\vspace{-3pt}
\caption{\small The diagrams contributing to the two-quark two-gluon amplitude.  
}
\label{fig:4ptAdjFund}
\end{figure}

A first interesting example of an amplitude is the four-point amplitude for two gluons and a quark-antiquark pair displayed in \fig{fig:4ptAdjFund},\footnote{When useful, we use the slightly-nonstandard notation ${\cal A}_n\big(1 \Phi_1 , \dots  , n \Phi_n \big)$ to display explicitly the external states in an amplitude.}
\be
{\cal A}^\tree_{4,1}(1{\bar q},2{g},3{g},4q)= -i \left( \frac{n_s c_s}{s-m_q^2} + \frac{n_t c_t}{t}+ \frac{n_u c_u}{u-m_q^2} \right) ,
\ee
where the numerators are
\be
n_s= {1 \over 2} \bar{u}_1 \s \varepsilon_2   (\s p_{12} \delta_{\alpha_4}^{\ \alpha_1} - M_{\alpha_4}^{\ \alpha_1})  \s \varepsilon_3  v_4  \,,~~~~ 
n_u= {1 \over 2} \bar{u}_1 \s \varepsilon_3   (\s p_{13} \delta_{\alpha_4}^{\ \alpha_1} - M_{\alpha_4}^{\ \alpha_1}) \s \varepsilon_2  v_4 \,,~~~~ n_t=n_u-n_s \, ,
\ee
and the color factors
\be
c_s= (T^{a_3}T^{a_2})_{i_4}^{\ i_1} \,,~~~~ c_u= (T^{a_2}T^{a_3})_{i_4}^{\ i_1} \,,~~~~ c_t=c_u-c_s = \f^{a_2 a_3 b}(T^{b})_{i_4}^{\ i_1}  \, .
\ee
Here the Greek indices $\alpha_1,\alpha_2$ are global (flavor) indices carried by the fermions.

Following the same steps as in pure YM theory one can show that the numerator relation together with the kinematics constraints at four points imposes massive BCJ relations for the partial amplitudes,
\be
(s-m_q^2) A^\tree_{4,1}(1{\bar q},2{g},3{g},4q)  =  (u-m_q^2)  A^\tree_{4,1}(1{\bar q},3{g},2{g},4q) \,.
\ee
More generally, one can understand this relation as a consequence of gauge redundancy. We have two independent numerators, which are not invariant under gauge transformations. We can thus at most build one gauge-invariant quantity out of these, and hence all partial amplitudes must be related.  

At general multiplicity $m$, the BCJ amplitude relations in their simplest incarnation take the form,
\be
\sum_{i=2}^{m-1} p_1 \cdot (p_2+\ldots + p_i) \, A_{m,k}^\tree(2, \ldots , i, 1g, i+1,  \ldots, m)=0\,,
\ee
where leg 1 must be a massless gluon in the adjoint. Unlike \eqn{BCJrels}, here the particles $2,\dots,n$
may have any spin, mass, and gauge-group representation. The partial amplitude is constructed as a sum over planar Feynman graphs in the same fashion 
as for the purely adjoint case; however, the color decomposition for these mixed adjoint-generic-representation amplitudes is quite different. See Refs.~\cite{Johansson:2015oia, Melia:2013bta,Melia:2015ika,Ochirov:2019mtf}) for details.

\begin{figure}[t]
\centering
\includegraphics[scale=0.52,trim=0 0 0 0,clip=true]{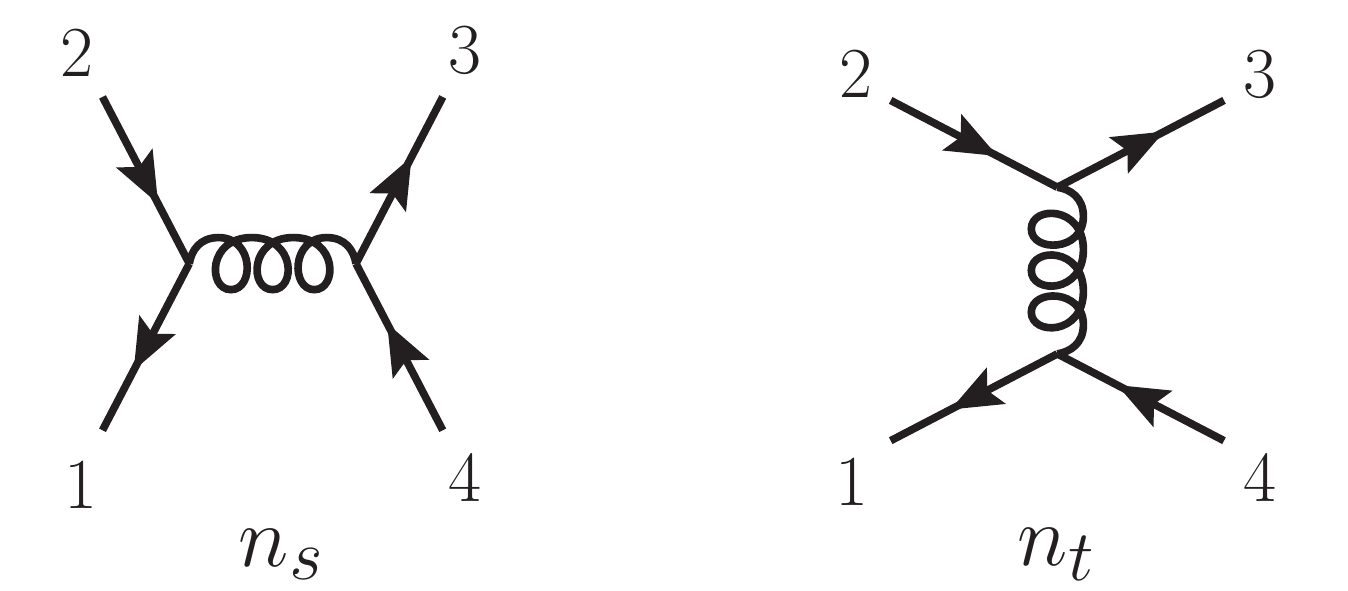}
\vspace{-3pt}
\caption{\small The diagrams contributing to the four-point pure-quark amplitude.  
}
\label{fig:4ptFundamental}
\end{figure}

As a further nontrivial example at four points, consider the fundamental representation four-quark amplitude displayed in \fig{fig:4ptFundamental}.
This amplitude is given as a sum over two displayed diagrams,
\begin{equation}
{\cal A}_{4,2}^\tree(1{\bar q},2{q},3{\bar q},4{q})= -i g^2 \left( \frac{n_s c_s}{s} + \frac{n_t c_t}{t} \right) ,
\end{equation}
where the color factors are 
\be
c_s =  (\T^{a})_{ i_2}^{\ i_1}  (\T^{a})_{ i_4}^{\ i_3}\,,  \hskip 1.5 cm 
c_t =  (\T^{a})_{ i_4}^{\ i_1}  (\T^{a})_{ i_2}^{\ i_3} \,,
\ee
and the kinematic factors are
\be
n_s=- {1 \over 2} (\bar{u}_1 \gamma_\mu v_2) (\bar{u}_3 \gamma^\mu v_4) \delta_{\alpha_2}^{\ \alpha_1}  \delta_{\alpha_4}^{\ \alpha_3} \,,  \hskip 1.5 cm 
n_t=- {1 \over 2} (\bar{u}_1 \gamma_\mu v_4) (\bar{u}_3 \gamma^\mu v_2) \delta_{\alpha_4}^{\ \alpha_1}  \delta_{\alpha_2}^{\ \alpha_3} \,.
\ee
For this amplitude, neither the color nor the kinematic factors
satisfy any relations among themselves, hence \ck duality is trivially
satisfied. Indeed, each kinematic numerator is gauge invariant by
itself and thus the amplitude representation is necessarily unique.
We will however see in \Sect{ZoologySection1} that in some cases it is possible or even necessary to impose additional numerator relations for matter amplitudes without external gluons.

We now look more in detail at the theory obtained from the double-copy formula with two sets of  QCD numerators. It will consist of gravity coupled to a single massless complex scalar, 
as well as a set of massive photons and scalars. Massless and massive fields in this theory originate from the double copy of adjoint and fundamental gauge-theory fields, respectively.  
As an example, we give the amplitude between four massive (complex) photons $\gamma$,
\be
{\cal M}_4^\tree(1{ \bar \gamma},2{\gamma},3{ \bar \gamma},4{\gamma})= - i \left(
\frac{n_s \,(n_s |_{N_{\! f}\rightarrow 1})}{s} + 
\frac{n_t \, (n_t |_{N_{\! f}\rightarrow 1})}{t} \right) \,,
\ee
where we trivialize the number of flavors on one side in order to avoid a redundant description with a factorized flavor group in the gravitational theory. 

Writing out the expression we have
\be
{\cal M}_4^\tree(1{ \bar \gamma},2{\gamma},3{ \bar \gamma},4{\gamma})= - i \left\{  \frac{\big[(\bar{u}_1 \gamma_\mu v_2) (\bar{u}_3 \gamma^\mu v_4)\big]^2}{4s}  \delta_{\alpha_2}^{\ \alpha_1}  \delta_{\alpha_4}^{\ \alpha_3}+ \frac{\big[ (\bar{u}_1 \gamma_\mu v_4) (\bar{u}_3 \gamma^\mu v_2)\big]^2}{4t} \delta_{\alpha_4}^{\ \alpha_1}  \delta_{\alpha_2}^{\ \alpha_3} \right\} .
\ee
The square can be upgraded to a tensor product since the external spinors can be chosen differently for the two numerator copies.

In order to better understand the double-copy amplitude, we may write it in terms of chiral spinors and explicitly write out the little group indices. For example,  using the massive spinor-helicity variables reviewed in \app{SpinorSuperspaceSection}, we simplify the above expression to obtain
\begin{align}
{\cal M}_4^\tree(1{ \bar \gamma}^{aa'},2{\gamma}^{bb'},3{ \bar \gamma}^{cc'},4{\gamma}^{dd'})&=
 -i \left\{ \vphantom{\frac{\delta}{s}} 
\big(\langle{1^a}\,{3^c}\rangle  [{2^b}\,{4^d}]+ [{1^a}\,{3^c}]  \langle{2^b}\,{4^d}\rangle \right.  \\
& \qquad \quad 
\left. + \langle{1^a}\,{4^d}\rangle  [{2^b}\,{3^c}] + [{1^a}\,{4^d}]  \langle{2^b}\,{3^c}\rangle\big)^2 \frac{\delta_{\alpha_2}^{ \alpha_1}  \delta_{\alpha_4}^{ \alpha_3}}{s}  + \big(2 \leftrightarrow 4 \big) \right\} .  \nn
\end{align}

\subsection{Double copy implies diffeomorphism symmetry\label{diffeoSubsection}}

Why does the double copy of gauge-theory
amplitudes yield amplitudes of some gravity theory?
A minimal criterion is that the expression obtained from the
double-copy method be invariant under linearized diffeomorphisms. 
Here we show that invariance of the double-copy amplitudes
under linearized diffeomorphisms is a direct consequence of color-kinematics duality and gauge invariance of the two gauge-theory factors entering the construction.

We start from a general linearized gauge transformation acting on a single external gluon with momentum $p$. Its polarization vector transforms as: 
 $\varepsilon_\mu(p) \rightarrow \varepsilon_\mu(p)+ p_\mu$. 
Gauge invariance of the amplitude implies that every diagram numerator should shift as
\begin{equation}
 n_i \rightarrow  n_i + \delta_i \, , \qquad \quad  \delta_i = n_i \Big|_{\varepsilon\rightarrow p} \, .
\end{equation}
Then, the entire amplitude is unaffected provided that the shifts $\delta_i$ obey 
\begin{equation}
 \sum_{i}\,  \frac{c_{i} \delta_i}{D_i} = 0 \, ,\label{gaugevariation}
\end{equation}
which must hold since by assumption the gauge-theory amplitude is gauge invariant. 

Aside from the explicit expressions for the numerator factors, the above equation must rely exclusively on the generic algebraic properties of the color factors $c_i$, namely antisymmetry and Jacobi identities. This means that, if we have \cknospace-duality-satisfying numerators $\tilde{n}_i$ in some gauge theory and consider their double copy with another set of gauge numerators $n_i$, then any linearized gauge transform of the $n_i$ will leave the double-copy amplitude invariant:
\begin{equation}
 \sum_{i}\,  \frac{n_{i} \delta_i}{D_i} = 0 \, .\label{gravvariation}
\end{equation}
We now analyze in more detail the significance of this transformation. A general coordinate transformation can be used to  impose  both transversality and tracelessness on the   on-shell asymptotic states that enter the definition of a scattering amplitude. This, in turn, results in imposing the conditions $\varepsilon_{\mu \nu}(p) p^{\nu}_{\vphantom{\mu\nu}} = 0 = \varepsilon_{\mu \nu}(p)\eta^{\mu\nu}$ on the graviton's polarization tensor. After this choice of gauge, 
amplitudes will still be invariant under the subset of linearized diffeomorphisms that do not modify the above conditions. These will act as 
\begin{equation}
\varepsilon{}_{\mu\nu}(p)  \rightarrow \varepsilon{}_{\mu\nu} (p) + p_{(\mu} q_{\nu)} \,,
\label{gaugediffeo}
\end{equation}
where $q$ is a reference vector that obeys $p \cdot q =0$, but is otherwise generic. The parenthesis denote symmetrization of spacetime indices.

The first step of formulating a double-copy construction is to establish a map between gravity asymptotic states and pairs of gauge-theory states. In general, the double-copy graviton will be obtained by taking the symmetric-traceless part of the product of the two gauge-theory gluons, i.e. its polarization tensor will be obtained from the gluon's polarizations as  $\varepsilon_{\mu \nu}= \varepsilon_{((\mu} \tilde \varepsilon_{\nu ))}$, where the double brackets indicate the symmetric-traceless part.\footnote{ 
We also note that the antisymmetric and trace parts of the product of the two gauge-theory gluon polarizations are identified with an antisymmetric tensor field and the dilaton.
These two field are generically present in amplitudes from the double copy unless additional steps are taken to ensure their removal, as we will see in  \Sect{secpureSG}.} 
 
We now study tree-level amplitudes obtained from the double-copy method. We take a set of duality-satisfying numerators $n_i$ only for one of the gauge-theory factors. 
The other set of numerators is taken in the form
\begin{equation}
\tilde n'_i = \tilde n_i + \tilde \Delta_i \, , \qquad \qquad
\sum_{i}  {\tilde \Delta_i c_i \over D_i} = 0 \, . 
\label{relGen}
\end{equation}
While the numerators $\tilde n'_i$ can violate \ck duality, they can
be obtained from a set of duality-satisfying numerators $\tilde n_i$
with a transformation of the form (\ref{GaugeShift}) with parameters
$\tilde \Delta_i$. Hence, we are assuming that there exists an
amplitude presentation for which the duality is satisfied also for the
second gauge theory. However, in the double-copy method, we use a set
of numerators with different properties for one of the theories, a
fact that will be advantageous in practical calculations.

Starting from the double-copy gravity amplitude in \eqn{DCformula} at tree level,
a tree amplitude can then be expressed as
\begin{equation}
{\cal M}_n = - i  \left\{ \sum_{i} {n_i \tilde n_i  \over D_i} 
  + \sum_{i } {n_i \tilde \Delta_i  \over D_i}\right\} =  - i \sum_{i } {n_i \tilde n_i  \over D_i}  \,, 
\label{DCformula2}
\end{equation}
where we have not included the overall $(\kappa/2)^{n-2}$.  Because
numerator factors $n_i$ obey the same algebraic relations as the color
factors $c_i$, equation (\ref{relGen}) implies the last equality
above.  Using \eqn{DCformula2}, the variation of the double-copy
amplitude under a linearized diffeomorphism of the form
(\ref{gaugediffeo}) becomes
\begin{equation}
{\cal M }_n \rightarrow {\cal M}_n - i \left\{  \sum_{i} { \delta_i \ \tilde n_i \big|_{\tilde \varepsilon\rightarrow q} \over D_i}
+ \sum_{i} { n_i \big|_{\varepsilon\rightarrow q} \tilde \delta_i \over D_i} \right\} .
\label{Mvariation}
\end{equation} 
The two terms are of the form (\ref{gravvariation}) and hence vanish
because of \ck duality. We then conclude that invariance of the
amplitude under linearized diffeomorphisms at tree level follows from
gauge invariance of the gauge theories entering the double-copy
construction provided that \ck duality is obeyed.  Diffeomorphism
invariance of the amplitudes at loop level can also be established
through generalized unitarity~\cite{BDDPR}. We will see in the next
subsection and in \sects{GravitySymmetriesSection}{ZoologySection1} that the double copy can also be used to
engineer amplitudes which are invariant under other symmetries,
including supersymmetry and gauge symmetry. In fact, one can think of
the double copy as a clever procedure to write down amplitudes that
obey a prescribed set of on-shell Ward identities starting from gauge-theory
data.  
By construction, these amplitudes also obey standard
factorization properties as well as crossing symmetry.  The basic
intuition is that gauge invariance together with mild assumptions
on the singularity structure are sufficient to fix the form of
amplitudes~\cite{Arkani-Hamed:2016rak,RodinaGaugeInv}.

\subsection{Adjoint fermions $+$ duality $\Rightarrow$ supersymmetry 
\label{Sec-adjoint-ferm}}

In \sect{Sec-Fund-Fermions} we discussed \ck duality in the context of
YM theory with matter fermions.  We now look at the case of
adjoint fermions in arbitrary dimension.  In this case, we will see
that the duality is equivalent to the
existence of supersymmetry, as argued in Ref.~\cite{Chiodaroli2013upa} (see also \cite{Weinzierl:2014ava} for a related discussion).  For
concreteness, we specialize to $D$-dimensional YM theory
minimally coupled to a single adjoint Majorana fermion, described by the Lagrangian
\begin{equation}
{\cal L}={\rm Tr} \Big[-\frac{1}{4}F_{\mu\nu}F^{\mu\nu} + 
   {i\over 2}\bar{\psi} \cancel{D}  \psi\Big] \, .
\label{gluon_1fermion}
\end{equation}
In all dimensions, the four-gluon and two-gluon-two-fermion amplitudes respect the  duality between color and kinematics 
without any further constraint.   
However, four-fermion amplitudes leads to an interesting constraint.  This amplitude is given by
\bea
\mathcal{A}^\text{tree}_4(1 \psi,2 \psi,3 \psi,4 \psi)
\!\!\!&=&\!\!\! i \left( \frac{(\bar{u}_1\gamma_{\mu}v_2)(\bar{u}_3\gamma^{\mu}v_4)c_s}{2s}
+\frac{(\bar{u}_2\gamma_{\mu}v_3)(\bar{u}_1\gamma^{\mu}v_4)c_t}{2t} +\frac{(\bar{u}_3\gamma_{\mu}v_1)(\bar{u}_2\gamma^{\mu}v_4)c_u}{2u} \right) , \nn  \\
\eea
where the $\bar{u}_i$ and $v_i$ are spinor external states which obey 
$\bar{u}_i \gamma^\mu v_j =  \bar{u}_j \gamma^\mu v_i$ due to the Majorana condition,
$\bar{u}_i = v_i^T {\cal C}$. 
In dimensions in which a Weyl representation can be chosen one of the terms above vanishes.

The requirement that 
$\mathcal{A}^\text{tree}_4(1 \psi,2 \psi,3 \psi, 4 \psi)$ obeys \ck duality forces the gamma matrices to obey the relation
\begin{equation}
\label{ck10}
(\bar{u}_1\gamma_{\mu}v_2)(\bar{u}_3\gamma^{\mu}v_4)+(\bar{u}_2\gamma_{\mu}v_3)(\bar{u}_1\gamma^{\mu}v_4)
+(\bar{u}_3\gamma_{\mu}v_1)(\bar{u}_2\gamma^{\mu}v_4)=0 \,.
\end{equation}
Eq.~(\ref{ck10}) is the equivalent to the Fierz identity  that appears in the supersymmetry transformation of the Lagrangian \eqref{gluon_1fermion}. This analysis can be repeated for pseudo-Majorana spinors with analogous results. Overall, an identity of this form can be 
satisfied only for $D=3, 4, 6, 10$, i.e. the dimensions for which the theory \eqref{gluon_1fermion} is supersymmetric.

The relation between \ck duality with adjoint fermions and
supersymmetry should not seem surprising in hindsight. In principle,
adjoint fermions can be combined with gluons with the double-copy procedure, resulting in a
gravity theory which includes spin-$3/2$ fields. However, it is known
that local supersymmetry is required to have a consistent theory of
interacting spin-$3/2$ fields. This is another example of the duality
between color and kinematics underpinning the consistency of the
gravity theory from the double copy. We shall see more examples along
this line in the following sections.

By repeating the discussion in \sect{diffeoSubsection}, 
it is straightforward to see that the double copy of a gauge theory with another gauge theory 
that has global supersymmetry leads to a theory that exhibits local supersymmetry. 
Indeed, a linearized local supersymmetry transformation of a gravitino polarization vector-spinor $u{}_{\mu}^\alpha(p)$ is
\be
u{}_{\mu}^\alpha(p)  \rightarrow u{}_{\mu}^\alpha(p) + p_{\mu} \xi_{\alpha} \,,
\ee
where $\xi_{\alpha}$ is the transformation parameter which, in order to preserve the $\gamma$-tracelessness of the gravitino 
wave function must obey the massless Dirac equation, $p\llap/ \xi = 0$. Then, the transformation of a double-copy amplitude
(after a discussion similar to the one that led to \eqn{Mvariation}) is
\be
{\cal M}_m^\tree\rightarrow {\cal M}_n^\tree + \sum_i \frac{\delta_i {\tilde n}_i\Big|_{{\tilde u}^\alpha\rightarrow \xi^\alpha} }{D_i} \,,
\ee
where ${\tilde u}^\alpha$ is the spinor that, through the double copy, generates the gravitino under consideration.
Since the parameter $\xi$ of the supersymmetry transformation has the
same properties as the original spinor ${\tilde u}^\alpha$ it
replaces, the factor ${\tilde n}_i\Big|_{{\tilde u}^\alpha\rightarrow
  \xi^\alpha} $ has the same properties as ${\tilde n}_i$, in
particular it obeys Jacobi relations. Thus, the variation of the
double-copy amplitude under (linearized) supersymmetry transformations
vanishes, implying that the double-copy theory exhibits local
supersymmetry.

More generally, we may expect that, under the right circumstances, the
double copy of a gauge theory with a theory that exhibits a global
symmetry leads to a theory where the global symmetry is promoted to a
local symmetry. We shall return to this point in
\sect{GravitySymmetriesSection}.

The emergence of supersymmetry from \ck duality offers a novel
perspective on the maximal number of gravitini that can consistently
enter a supergravity theory. As we have seen, a gauge theory
coupled to fermions can exhibit \ck duality in at most ten dimensions
dimensions. Thus, this is the highest dimension which a supergravity
theory can be given a double-copy interpretation in the sense
described here. Taking two such theories gives therefore the largest
number of supersymmetries, which is two in ten dimensions or, upon dimensional
reduction, eight in four dimensions.
This observation recovers the usual bound following from the
requirement that the exist multiplets of supersymmetry algebra
containing fields of spin $s\le 2$.

\subsection{General lessons from applying \ck duality}

In the previous subsections, we presented various concrete
examples of theories which obey \ck duality. Statements about \ck
duality often depend on the details of the theories under
consideration and on what observables are being studied. Since the
duality is often used as a shortcut for computing gravitational
amplitudes, one can restrict to gauge theories suitable for
giving broad classes of consistent gravitational
theories once the numerators are assembled via the double-copy
method. Expanding on the examples discussed earlier, in the rest of
this review we will focus mostly on theories with the following
general features:
\begin{itemize}
\item There exists (at most) one massless gauge field, the gluon, that
  transform in the adjoint of a gauge group $G$, and all fields of the
  gauge theory are charged under this group.  We will see in 
  \Sect{ZoologySection1} that this requirement translates to the
  equivalence principle in gravity.

\item The gauge group $G$ is a completely general Lie group in the
  sense that no assumptions on its rank need be imposed on it. Note
  that throughout this section the only properties of the gauge group
  we have utilized are the Jacobi relations of its structure constants
  and the commutation relations of its representation matrices, which
  do not require to spell out our choice of Lie group.

\item Amplitudes involving adjoint fields (gluons or adjoint matter)
  should admit perturbative expansions where the kinematic numerators
  obey the same Lie algebra relations (e.g.\ Jacobi identities) as the
  corresponding adjoint-valued color factors. We have seen in
  \Sect{diffeoSubsection} that this property is essential for obtaining
  a gravitational theory after the double copy.  This condition also
  implies the universality of gravitational self-interactions.

\item Amplitudes involving fields in generic representations of the
  gauge group should admit perturbative expansions where the kinematic
  numerators obey the same Lie algebra relations as the generators of
  those representations. The simplest example of non-purely-adjoint
  theory has been discussed in  \Sect{Sec-Fund-Fermions}.
\end{itemize}
These general properties guarantee that every diagram in the
perturbative expansion of an amplitude has a unique nontrivial color
factor, which obeys the minimal constrains imposed by the Lie algebra
of the gauge group, and furthermore that the coupling to the unique
gluon is universally controlled by the gauge-group
representations. The kinematic factors can then be constrained to obey
the duality by enforcing the one-to-one map between color and
kinematic identities.  Along these lines, in  \Sect{sec-zoo-CK}
we will articulate a more precise set of working rules which will
define the properties of the gauge theories employed for obtaining a
web of double-copy-constructible theories.


\section{Geometric organization}
\label{TreeSection}

The dual Jacobi identities give nontrivial relations between diagram
numerators.  Here we describe the systematics of these relations and
how they can be used to express amplitudes' integrands in terms of the
contributions of a small set of master diagrams.  This is
generally very helpful at higher perturbative orders because it allows
us to express an integrand in terms of a (small) subset of all of its
terms.  To this end, we will first describe a useful geometric
organization via a graph of graphs\footnote{See
  Ref.~\cite{Arkani-Hamed:2017mur} for a related application of such
  an approach towards identifying \emph{scattering forms} of amplitudes.}
that offers insight into the information flow of the duality
identities.  We will then illustrate the general case through
some examples.

\subsection{Amplitudes in terms of boundary data}

The duality between color and kinematics provides a set of relations
between diagrams.  \Sect{DualitySection} frames the discussion
of the duality in terms of vector and matrix operations between linear
spaces of numerators of diagrams and linear spaces of scattering
amplitudes.  
Here we give an alternative perspective, using the language of graphs~\cite{JJHenrikReview,jjmcTASI2014}.
This offers a useful way to visualize how a small set of graphs is
sufficient to describe the entire amplitude.  We shall see that the
minimal set of graphs whose numerators need to be specified can be
thought of as boundary data on the graph of graphs describing the
amplitude.

\begin{figure}[tb]
\begin{center}
\includegraphics[height=3in]{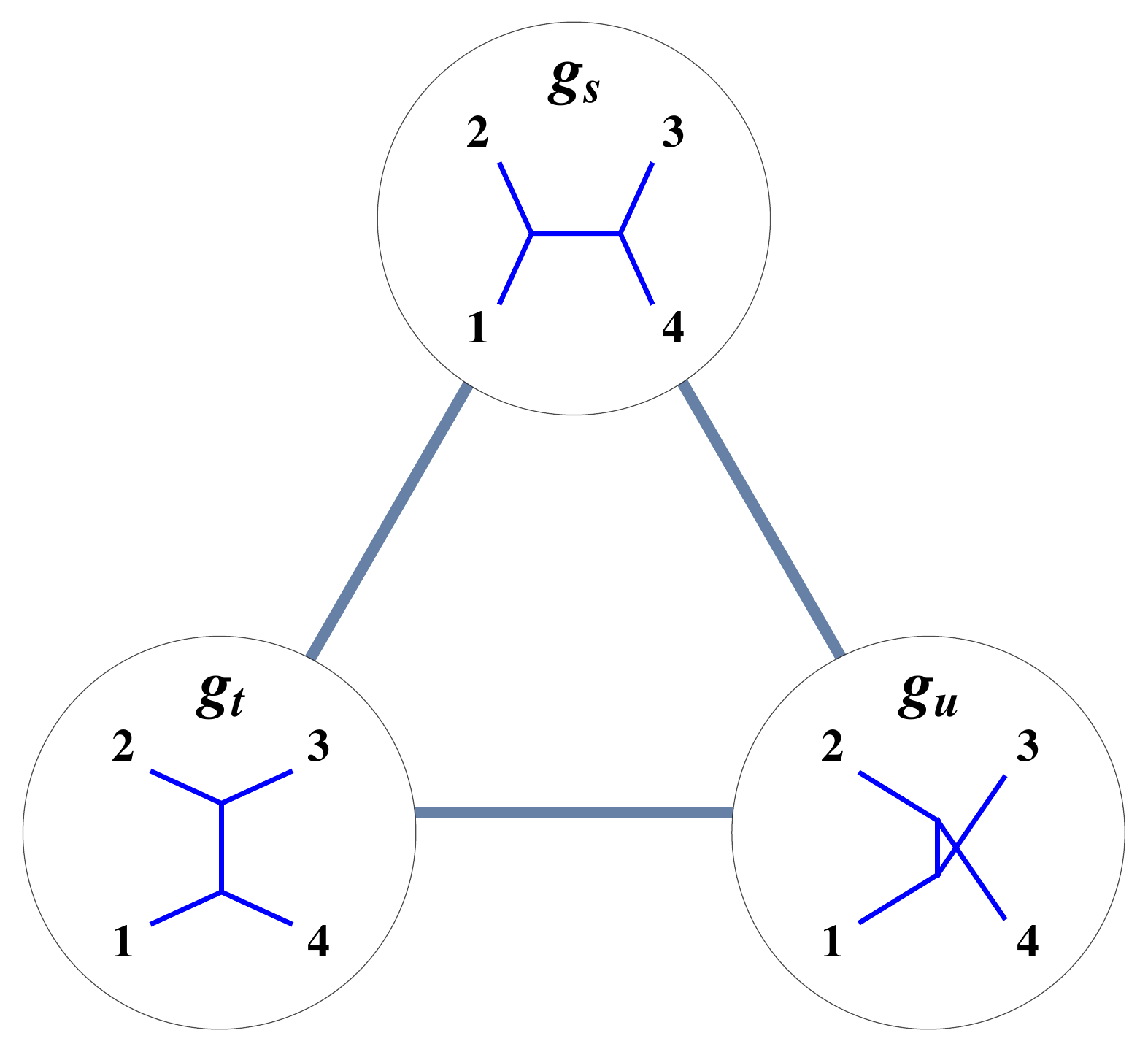}
\end{center}
\caption{Graph of graphs relevant to four-point tree-level scattering 
forms a triangle.  Each vertex represents one of the three graphs in \fig{FourPtGluonsFigure}, and every edge represents a Jacobi or Whitehead move (c.f. \fig{TUOperators}).  Every triangle in the graph of graphs represents a Jacobi relation that can be used to constrain a dressing of one vertex graph in terms of dressings of the  other two vertex graphs.
 }
\label{graphOfGraphs4ptFigure}
\end{figure}

\begin{figure}[tb]
\begin{center}
\includegraphics[height=3.3in]{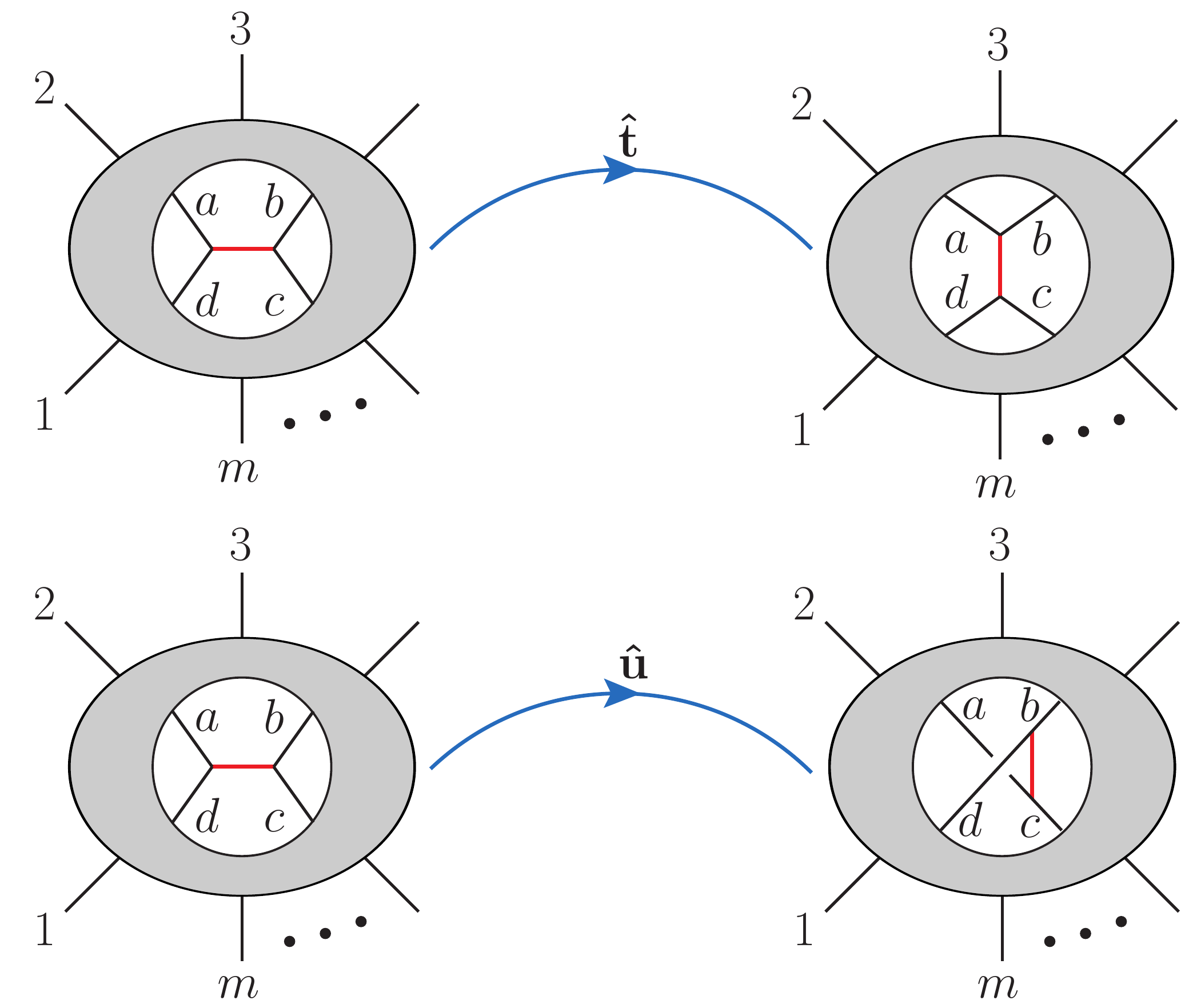}
\end{center}
\caption{Operations that relate edges of one graph to edges of another.
 }
\label{TUOperators}
\end{figure}

As a simple example, consider the four-gluon tree amplitude.  
After absorbing any contact terms into graphs with only
cubic vertices, this amplitude can be described by the three graphs 
in \fig{FourPtGluonsFigure}, corresponding to the $s$-channel,  $t$-channel, and
$u$-channel.   Both the kinematic and color numerators of these three graphs satisfy the 
(dual) Jacobi identities in \eqn{BCJFourPtcolor} and \eqn{BCJFourPt},
$c_s+c_t+c_u=0$ and $n_s + n_t + n_u = 0$.  
These equations are equivalent to the statement that any two ``single-copy'' numerator dressings determine the third one.
We can draw this relationship as a graph of graphs, where each vertex 
represents a specific graph participating in the Jacobi relation and the edges connect 
to the other two vertices (i.e. graphs) that determine the first 
vertex. 

These relations can be summarized in the triangle shown in \fig{graphOfGraphs4ptFigure}. 
Each vertex or
node in \fig{graphOfGraphs4ptFigure} is one of the three cubic graphs 
from \fig{FourPtGluonsFigure} that would contribute to the four-point
amplitude.  Every edge in this graph of graphs corresponds to a Jacobi move 
on the internal propagator of one of the vertex graphs that transmutes
it into another.  
In the mathematics literature these moves are known as Whitehead 
moves~\cite{rafi2013}.  
The basic moves for
acting on graphs represented by the edges in \fig{graphOfGraphs4ptFigure}
are denoted $\hat t$ and $\hat u$ and  are shown in \fig{TUOperators}.  The
first move, which we call $\hat{t}$, takes the $s$-channel graph in \fig{FourPtGluonsFigure}
and converts it to the $t$-channel. 
Similarly, we call $\hat{u}$ the move that converts the $s$-channel graph to the $u$-channel 
one.  
The move that takes the $t$-channel graph to the $u$-channel graph can be
understood as the composition $\hat{u} \circ \hat{t}$.  
Alternatively, we can view it as one of the same basic
operations as on the $s$-channel graph but acting on a graph with permuted labels.
Strictly speaking, one should associate a direction with each move,
but we will ignore this distinction because, up to relabelling,
the reverse operation is identical to the forward one.  
It is not difficult to see that each edge in the triangle graph of
graphs contains the graphs contributing to a particular four-point
color-ordered partial amplitude, and the entire triangle itself
represents an occasion for Jacobi to be satisfied by a dressing of the
graphs (whether color or kinematic).

\begin{figure}[tb]
\begin{center}
\includegraphics[height=3.5in]{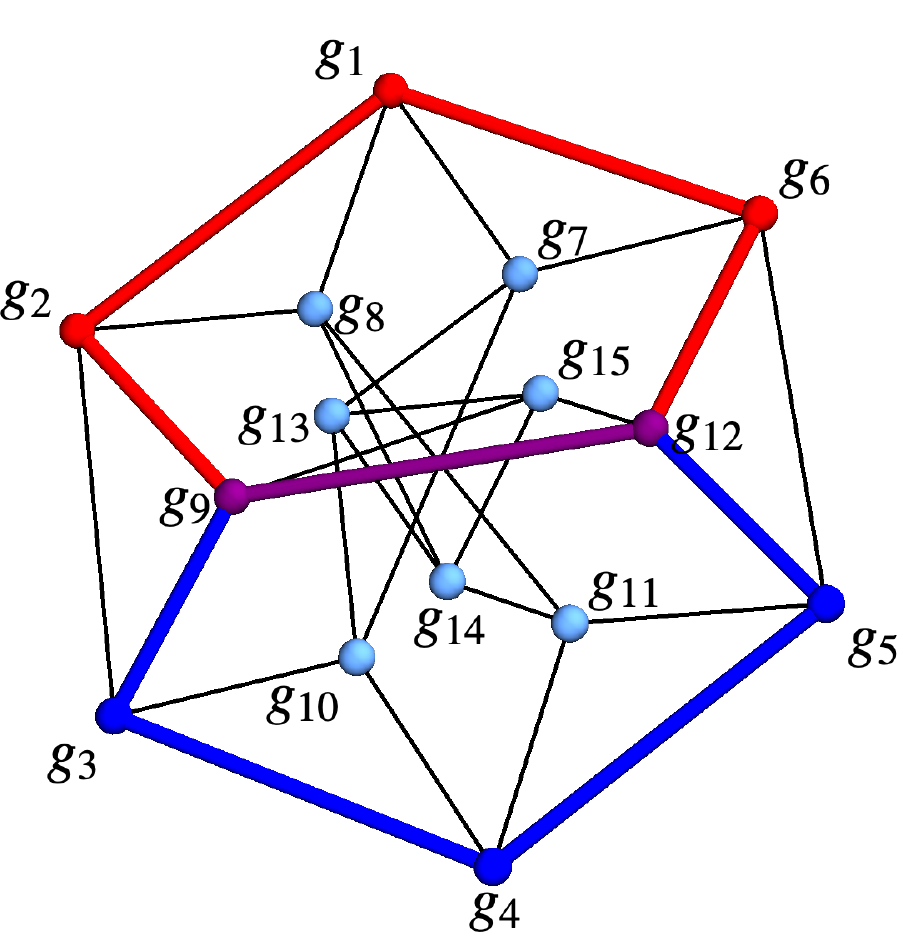}
\end{center}
\caption{Graph of graphs for the full color-dressed five-point
    partial amplitude, with vertex labels corresponding to the graphs
    given in \fig{fivePointGraphs}. Two color-ordered partial amplitude
    graphs are highlighted, corresponding to $A^\tree_5(1,3,5,4,2)$ and
    $A^\tree_5(1,3,5,2,4)$, c.f. \eqn{a13542} and \eqn{a13524} respectively,
     as well as \fig{assoc5pt}. 
 }
\label{full5pt}
\end{figure}

\begin{figure}[tb]
\begin{center}
\includegraphics[width=4in]{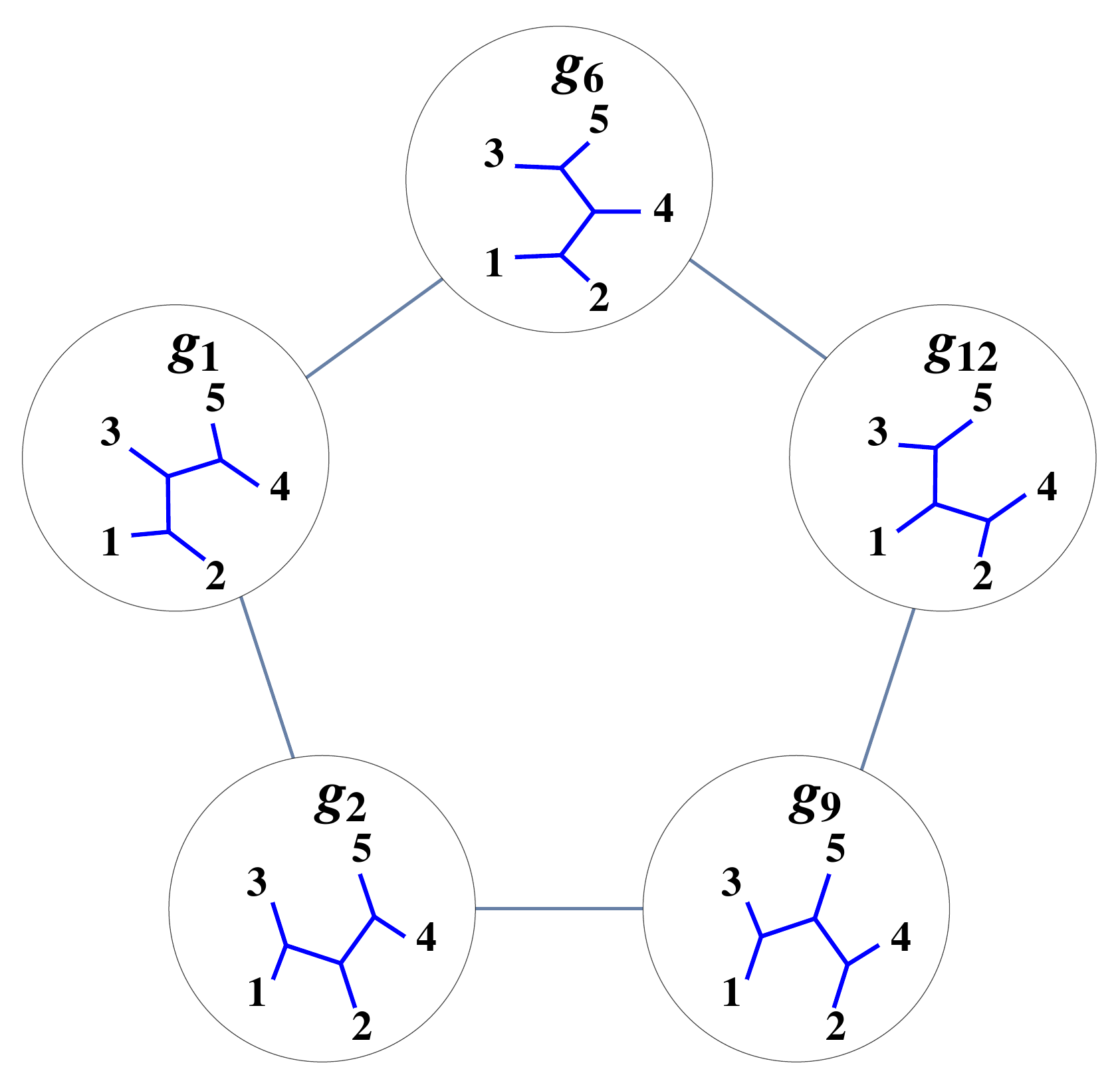}
\end{center}
\caption{ Graph of graphs relevant to the color-ordered
  five-point partial amplitude $A^\tree_5(1,3,5,4,2)$.  The vertices correspond to the indicated five-point graphs given in
  \fig{fivePointGraphs}, with the understanding that each labeled graph
  contributes to this color order with signs determined by its
  color-weight as in \eqn{a13524}.  The edges in the color-ordered graph of graph
  represent application of ${\hat{t}}$ on vertex-graph edges until closure. }
\label{assoc5pt}
\end{figure}

This basic structure generalizes straightforwardly to higher-point
amplitudes.  Consider it at five points: in total there are 15
distinct cubic graphs (constructible by starting with one five point
cubic graph and applying $\hat{t}$ and $\hat{u}$ on each of it's
internal edges, and repeating on new graphs until closure)
contributing to the full color-dressed integrand as displayed in
\fig{full5pt}.

\begin{homework}
Draw individual five-point graphs for each vertex in \fig{full5pt}. 
Label the edge operations to get from graph to graph.
\label{drawFivePointHW}
\end{homework}

Let us focus on the subset  of  five  graphs comprising  the
ordered partial amplitude  $A^\tree_5(1,3,5,2,4)$, as indicated in \eqn{a13524}.
As shown  in \fig{assoc5pt}, all five graphs can  be found by starting with any  
graph that has the relevant external ordering and  applying $\hat{t}$ to its two internal edges to
find and connect two other graphs with the same color order.  It is interesting to note that each edge in the graph represents a shared factorization channel for the internal propagator not mutated between the two connected graphs.  We keep applying $\hat{t}$ until closure. 
This procedure of repeated $\hat{t}$ application, building a graph of all the graphs of a given color-order, carves out the skeletal graph\footnote{Typically called a one-skeleton (cf.~Ref.~\cite{polytopeSkeletons}).} of a polytope known as the 
\emph{associahedron}.
(cf.~\cite{StasheffPolytope,Tonks95}).  Associahedra are often also called Stasheff polytopes.  As $\hat{t}$ preserves color order, each Stasheff skeletal graph is
composed of all the graphs that contribute  to a fixed color order~\cite{jjmcTASI2014}.  Indeed every Stasheff subgraph of the full graph of graphs represents the contributions of a particular ordered partial amplitude to the full amplitude. This is exemplified by the two highlighted pentagonal subgraphs of  \fig{full5pt} which represent the ordered amplitudes  $A^\tree_5(1,3,5,2,4)$ and $A^\tree_5(1,3,5,4,2)$.

\begin{homework}
Find another pentagonal subgraph in \fig{full5pt} besides the two highlighted ones.  Which color-ordered amplitude does it represent?
\end{homework}

Of course, the full five-point amplitude requires all 15
cubic graphs (cf. \eqn{TreeAmpMpt}), as displayed in the complete
five-point graph of graphs \fig{full5pt}.  Every triangle subgraph represents a
Jacobi identity that single-copy numerator dressings could satisfy.

One question when presented with the graph of graphs, is whether it is easy to see how many ordered amplitude (Stasheff) subgraphs are
required to specify a full color-dressed amplitude.   A natural conclusion is $(m-2)!$ of them, because this is the minimal number of ordered amplitude subgraphs 
whose union of vertex-graphs includes all $(2m-5)!!$ vertex-graphs that contribute to the full amplitude.
In terms of our five-point graph of graphs, drawn in \fig{full5pt}, it would be necessary to identify six different pentagonal subgraphs for every 
vertex-graph to be included at least once.  This corresponds to the 
KK~\cite{KleissKuijf} basis, which gives the full amplitude in terms of partial amplitudes~\cite{DixonMaltoni} as per \eqn{DDMdecomp}.

\begin{homework}
Find six color-ordered partial amplitudes (pentagonal subgraphs) in the five-point graph of graphs shown in \fig{full5pt} that together include every node at least once.  What color-ordered amplitudes do they correspond to?  Is it a KK basis?
\end{homework}

\begin{figure}[tb]
\begin{center}
\includegraphics[height=2.5in]{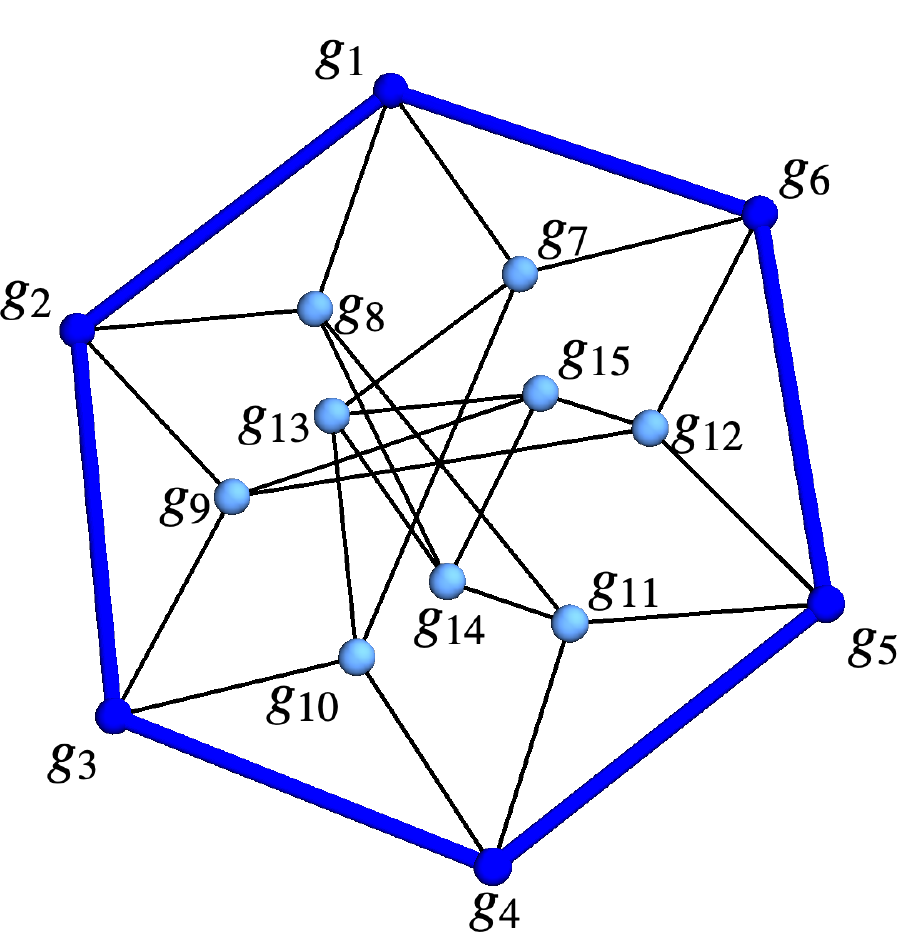}
\end{center}
\caption{Due to the nine Jacobi relations (six
  outer triangles, and three inner triangles), only the six outer
  boundary graphs are needed to specify bulk data via Jacobi
  relations. 
  }
\label{triagFivePoint}
\end{figure}

The counting works differently for an amplitude where both the kinematic 
numerators and the color factors obey Jacobi identities.
Consider the external boundary graphs of \fig{triagFivePoint}.  As
noted above, Jacobi identities are represented by triangles in the
graph of graphs.  By working inwards using Jacobi identities in \fig{triagFivePoint}, we see 
that each pair of neighboring graphs in the six external boundary graphs 
completely specify all other graphs.  
So, for theories that can satisfy the dual Jacobi identities, we need
only specify data on this external boundary.  More generally, for an
$m$-point amplitude, it turns out that only  $(m-2)!$ boundary graph numerators are
sufficient to specify all other graph numerators via Jacobi and $(m-3)!$  ordered
amplitudes are independent. For example, in the case of the five-point amplitudes, 
the two highlighted partial amplitudes in \fig{full5pt} are sufficient to generate all others.

The spanning boundary graphs at tree level, sometimes referred to as the master graphs, have a geometric association as well.  
Consider the half-ladder graph shown in~\fig{HalfLadderFigure}. A half-ladder graph
(also called a multiperipheral graph) is a cubic tree-graph where
all vertices but two connect two internal edges.  They are called
half-ladders because they can be drawn to resemble
one half of a ladder split down all rungs.  
For any multiplicity, a choice of master graphs can be obtained by taking any 
half-ladder graph and acting with the $\hat{u}$ move on all internal edges until no
new graphs are generated.  All
graphs so generated will remain half-ladders with different labels and
their corresponding graph of graphs forms the one-skeleton of a polytope known as a \emph{permutahedron}~\cite{Tonks95}.
\begin{homework}
 Why is at least one half-ladder graph required at any multiplicity in the set of Jacobi master-graphs?
\end{homework}

As discussed in \sect{DualitySection}, the mismatch between the number of independent gauge-theory amplitudes and the number of independent numerators leads to a gauge
freedom that allows some numerators to take on arbitrary values,   neither altering the amplitudes 
nor the BCJ amplitude relations.  
At five points, four of the numerators can be arbitrarily chosen, \ie even set to zero, 
making the remaining numerators nonlocal. 
In addition, the fact that there are only $(m-3)!$ independent gauge-theory amplitudes is directly related to the fact that in the KLT
formula \eqref{KLT} only $(m-3)!$ independent partial amplitudes
appear for each of the two gauge theories.

As typical with graphical organization of amplitudes, the total number of
independent graphs will increase factorially,  going as $(m-2)!$ for multiplicity $m$.  
While this is a reduction over the total $(2 m - 5)!!$ cubic graphs that contribute to 
the complete amplitude, a more useful question is the minimal information required
to build the complete amplitude (at tree level, and, more generally, the amplitude's integrand at loop level).  
Remarkably, as we now show, by imposing diagram
symmetry, we can specify only a single half-ladder
diagram, which then determines all other diagrams at any given
multiplicity at tree level. This allows us to avoid specifying a factorially-growing
number of diagrams.

\subsection{Applying relabeling invariance at tree-level \label{subsecNLSM}}

As a warm-up before turning to gauge theory, we consider the NLSM \cite{Gell-Mann1960}, as defined by the Lagrangian in the Cayley
parameterization~\cite{macfarlane1968,Kampf:2013vha,Carrasco2016ldy},
\begin{equation} 
\label{NLSMaction}
{\cal L}_{{\rm NLSM}} = {1\over 2} {\rm Tr} \bigg\{ \partial_\mu \varphi \, 
{1\over 1- \lambda  \varphi^2} \, 
  \partial^\mu \varphi \, {1\over 1 - \lambda\varphi^2} \bigg\} \,,
\end{equation}
where $\varphi$ is a Lie-algebra valued Goldstone-boson scalar field
in the adjoint representation. Here the color symmetry is global.
Although this theory has only even-point interactions, we can assign
its data to graphs with only cubic vertices by multiplying and
dividing by appropriate inverse propagators.  In terms of diagrams
with only cubic vertices, dimensional analysis dictates that each
vertex effectively carries two powers of momentum.  The first nonvanishing
amplitude is at four points.  We will see that, by imposing Jacobi
identities and relabeling symmetry on the kinematic numerators of the
diagrams, we can obtain the four-point scattering amplitude of
this theory.  At higher points, we will also either need to impose additional conditions
to uniquely fix the amplitudes to this theory.  It is sufficient to impose the manifestation of only quartic poles in amplitudes~\cite{Carrasco:2019qwr}, which in combination with color-kinematics encodes
 the necessary~\cite{CheungSoft, LowSoft, RodinaSoft} vanishing soft-scalar limits, or \emph{Adler zero} conditions~\cite{AdlerZero}.
Higher-derivative deformations of the NLSM and their compatibility with the KK and the BCJ amplitudes relations have been discussed in~\cite{Carrillo-Gonzalez:2019aao}

To start the construction of the four-point amplitude, consider the four-point half-ladder graph.
Since we require the dimensions of the numerator to match that following from the NLSM Lagrangian and
therefore carry four powers of momentum, we take a numerator ansatz:
\begin{equation}
n(a,b,c,d)\equiv \, n \left( \hskip -.1 cm 
  \parbox{2.6 cm}{\includegraphics[width=0.17\textwidth]{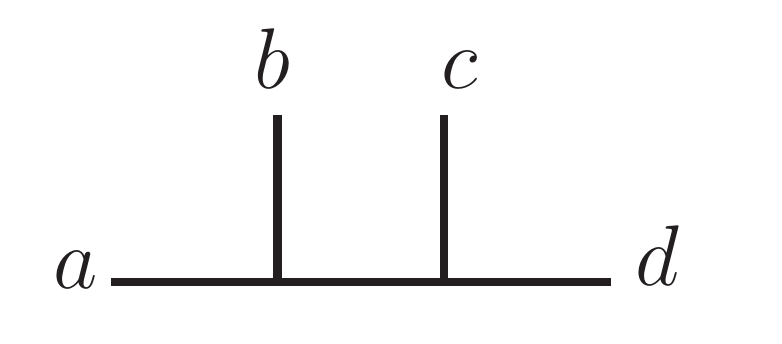}}\!\!
 \right) =  s_{ab} (\alpha s_{ab} + \beta s_{bc}),
\end{equation}
where $s_{ab} = (p_a + p_b)^2$ and $\alpha$ and $\beta$ are to be
constrained by symmetry and the kinematic Jacobi identities.  
The other kinematic invariant is $s_{ac}=-s_{ab} - s_{bc}$, and it
is not independent.

Imposing the Jacobi constraints, 
\begin{equation}
n(a,b,c,d) = n(c,a,d,b) + n(d,a,b,c)\,,
\label{NumeratorJacobi4Pt}
\end{equation}
relates $\alpha$ and $\beta$ according to
\begin{align}
0 &=  \alpha  s_{ab}^2-\alpha  s_{bc}^2 - \beta  s_{bc} s_{ac} 
               - \alpha  s_{ac}^2 \nonumber \\
  &= -2 \alpha  s_{ab} s_{bc} + \beta s_{ab} s_{bc} 
       - 2 \alpha  s_{bc}^2 +  \beta  s_{bc}^2 \,,
\end{align}
where we used $s_{ac}=-s_{ab}-s_{bc}$.  Given that the
Mandelstam invariants $s_{ab}$  and $s_{bc}$ are independent, we find 
$\beta = 2\alpha$; thus, the numerator is uniquely fixed up to its overall scale, which may be identified as the coupling of the model:
\begin{equation}
 n(a,b,c,d)  \propto s_{ab} (s_{ab} + 2 s_{bc})\,.
\end{equation}
One can verify that this expression satisfies all necessary antisymmetry constraints,
\eg it changes sign with $a\leftrightarrow b$ or $c\leftrightarrow d$. 
The full amplitude, up to overall normalization is then 
\begin{equation}
A^\tree_{\rm NLSM} \propto \frac{c(1,2,3,4) n(1,2,3,4)}{s_{12}} + 
                     \frac{c(3,1,4,2) n(3,1,4,2)}{s_{13}} + 
                     \frac{c(4,1,2,3) n(4,1,2,3)}{s_{14}} \,,
\end{equation}
where the color factors are obtained by dressing each vertex with a structure constant
$f^{abc}$.  The key lesson is that, by taking the
numerators to be functions of the graph labels, only a single 
numerator needs to be specified.

\begin{homework}
Repeat the above analysis assuming  only degree-one monomials in the Mandelstam
invariants.  What theory could this construction correspond to?
\end{homework}

Next, consider the case of YM theory at four points.  In this case the numerator ansatz is 
constructed out of external momenta and polarization vectors $\{ \varepsilon_1,\dots,\varepsilon_4 \}$,
subject to the requirements that every $\varepsilon_i$ appears 
once in each term and that every term has exactly two momenta. These constraints guarantee 
consistency with the structure of Feynman rules.
The possible third-degree monomials are constructed  from the following independent Lorentz invariants:
\begin{align}
&\Big\{s_{12}, s_{13} ,
   \left(p_1\cdot \varepsilon_2\right), \left(p_1\cdot \varepsilon _3\right), 
   \left(p_2\cdot \varepsilon_1\right), \left(p_2\cdot \varepsilon _3\right), \left(p_2\cdot \varepsilon_4\right),
   \left(p_3\cdot \varepsilon _1\right), ~~~~ \nonumber \\
&   \left(p_3\cdot \varepsilon_2\right), \left(p_3\cdot \varepsilon _4\right),
   \left(\varepsilon _1\cdot \varepsilon
   _2\right),\left(\varepsilon _1\cdot \varepsilon _3\right),
   \left(\varepsilon _1\cdot
   \varepsilon _4\right),\left(\varepsilon _2\cdot \varepsilon
   _3\right),\left(\varepsilon _2\cdot \varepsilon _4\right),
   \left(\varepsilon _3\cdot \varepsilon _4\right)\Big\} \,.
\label{YMFourPtMonomial}
\end{align}
There are 30 possible combinations, leading to an ansatz with an equal number of parameters.
Besides constraining it with the kinematic Jacobi identity \eqref{NumeratorJacobi4Pt}, 
we  also impose the antisymmetry constraints at the two vertices,\footnote{See \sect{sec2.1}, \eqn{4ptcolorfactors} and discussion below it.}
\begin{equation}
n(a,b,c,d) =- n(a,b,d,c)=-n(b,a,c,d)\, ,
\end{equation}
matching the antisymmetry of the color factors.

Applying these constraints on the ansatz built from the monomials in
\eqn{YMFourPtMonomial} fixes all but five of the
ansatz' coefficients. Further imposing gauge invariance on one external leg then fixes the
form of the numerator.  In fact, it is sufficient to impose gauge
invariance on one leg when the amplitude is factorized on the pole of a given channel, i.e.
for $s_{ab} \to 0$,
\begin{equation}
n(a,b,c,d)|_{s_{ab} \to 0, \; \hbox{and} \; \varepsilon_a \to p_a} \to 0\, .
\end{equation}
This gives
\begin{align}
n(a,b,c,d) & \propto \Bigl\{ \Big[(\varepsilon_a \cdot \varepsilon_b) 
p_a^\mu+2(\varepsilon_a \cdot p_b) \varepsilon_b^\mu-(a \leftrightarrow b)\Big]
\Big[ (\varepsilon_c \cdot \varepsilon_d) p_{c\mu}+2(\varepsilon_c \cdot p_d)
\varepsilon_{d\mu}-(c \leftrightarrow d)\Big] \nn  \\
& \null \hskip 2 cm
+ s_{ab} \Bigl[ (\varepsilon_a \cdot \varepsilon_c)(\varepsilon_b \cdot \varepsilon_d)
- (\varepsilon_a \cdot \varepsilon_d)(\varepsilon_b \cdot \varepsilon_c)\Bigr] \Bigr\}\,,
\end{align}
in agreement with \eqn{s_num}.  We note that, as explained in
Refs.~\cite{Arkani-Hamed:2016rak,RodinaGaugeInv}, one can also
determine the amplitude using other constraints, in particular from
gauge invariance and mild assumptions on the singularity structure.

\begin{homework}
Verify explicitly that the four-point YM numerator given above
satisfies the Jacobi constraint.
\end{homework}

Emboldened by the success to obtain four-point amplitude by imposing
dual-Jacobi relations, we continue to the next multiplicity for the
NLSM. As for four points, the duality involves not
only imposing kinematic Jacobi relations, but also the same antisymmetry
carried by color factors. For the half-ladder numerators, 
\begin{equation}
n(a,b,c,d,e)\equiv  
 \, n \left( \hskip -.055 cm 
  \parbox{2.55 cm}{\includegraphics[width=0.20\textwidth]{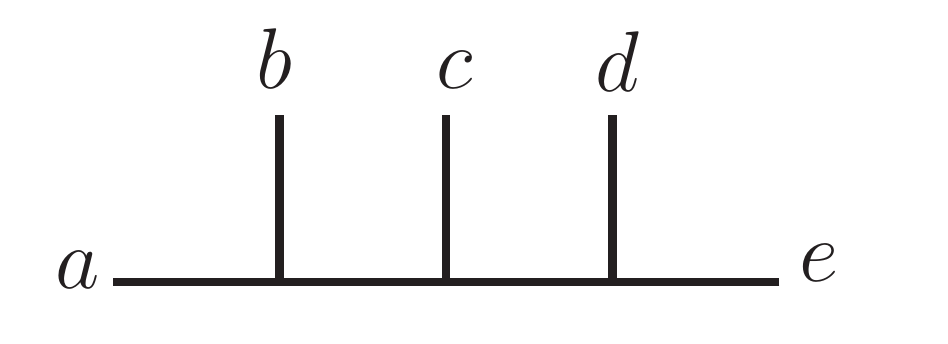}}~~\; 
 \right) ,
\end{equation}
these antisymmetry constraints read:
\begin{align}
 n(a,b,c,d,e)& =-n(a,b,c,e,d) = -n(b,a,c,d,e) = -n(d,e,c,a,b) \, .
\label{FivePointAntisym}
\end{align}
The Jacobi identities corresponding to the two independent propagators of the five-point half-ladder graphs are:
\begin{align}
 n(a,b,c,d,e) & =n(a,c,b,d,e) + n(c,b,a,d,e) \,, \nonumber \\
 n(a,b,c,d,e) & =n(a,b,d,c,e) + n(b,a,e,c,d) \,.
\label{FivePointBCJ}
\end{align}
One can immediately see the need to impose two Jacobi relations from the two
triangles that touch every vertex in the graph of graphs for the
five-point tree as drawn in \fig{triagFivePoint}.

At five points, we have a 35-parameter ansatz 
comprised of all degree-three monomials with factors from
\begin{equation}
\left\{ s_{ab}, s_{ac}, s_{ad}, s_{bc}, s_{bd} \right\} \,.
\end{equation}
The constraints in \eqns{FivePointAntisym}{FivePointBCJ}
fix 34 parameters, leaving us with a unique expression,
up to an overall coefficient, 
\begin{equation}
n(a,b,c,d,e) \propto 
\left( s_{ac}+s_{bc}\right) 
\left( s_{ad}  \left(s_{bd}+s_{be} \right) 
  - \{ a\leftrightarrow b\} \right) .
\label{NLSM5Pt}
\end{equation}
Remarkably, the five-point amplitudes obtained from these numerators
actually vanish, in line with the fact that odd-point amplitudes vanish in
the NLSM.
Moreover, this amplitude vanishes without having to impose the requirement that the underlying theory
 has no three-point vertex (\ie that there is no two-particle factorization channel).
This in turn suggests that there do not exist \cknospace-satisfying scalar two-derivative 
theories with only fields in the adjoint representation that are not the NLSM.

\begin{homework}
Verify that the above numerator satisfies the two independent Jacobi relations at five points.
\end{homework}

\begin{homework}
Verify explicitly that a color-ordered amplitude, say $A^\tree_5(1,2,3,4,5)$,
expressed in terms of its cubic graphs using the diagram numerator in \eqn{NLSM5Pt} 
vanishes.
\end{homework}

One can continue in this way, systematically building up higher-point amplitudes.  It is also
useful to impose other physical constraints such as the vanishing of all factorization limits 
where at least one factor is an odd-point amplitude:
\begin{equation}
\lim_{s_{1\dots 2k}\rightarrow 0} s_{1\dots 2k} A^\tree_n(1,\dots, n) = 0 \,,
\end{equation}
or, equivalently, the vanishing of the residue of the simple pole in $s_{1\dots 2k}$:
\begin{equation}
\sum_{\text{states}} A^\tree_{2k+1}(1,\dots 2k, p) A^\tree_{n-2k+1}(-p, 2k+1, n)\big|_{p^2 = 0} = 0  \,.
\end{equation}
These conditions guarantee recursively consistency with the vanishing of the odd-point tree amplitudes 
with multiplicity smaller than $n$. Let us illustrate this for the six-point amplitude.
For our scalar theory, we have nine independent external momentum invariants; from them we can construct 
495 degree-four monomials thus obtaining a 495-parameter ansatz for the half-ladder graph.
\ck duality alone constrains all but 23.  
Imposing the vanishing of the color-ordered factorization
\begin{equation}
\sum_{\text{states}}A^{\rm tree}_3 \left(1,2,p\right)\, 
  A^{\rm tree}_5\left(-p,3,4,5,6\right)  = 0 \,,
\end{equation}
leaves six unconstrained parameters.  
Although individual diagrams depend on them, these parameters always appear in
the same linear combination in front every color-ordered partial amplitude, which  indeed 
reproduce the six-point partial NLSM amplitudes, up to the overall normalization.
Other factorization limits, \eg $0=\text{Res}_{ s_{23} = 0}(A_6^\text{tree}(1,2,3,4,5,6)) = \sum_{\text{states}}A^{\rm tree}_3 \left(2,3,p\right)\, 
  A^{\rm tree}_5\left(-p,4,5,6,1\right)$, do not constrain them any further, implying that it should be possible to remove five of the remaining six parameters by
  a local generalized gauge transformation.

When combined, relabeling symmetry and the dual Jacobi
relations are extremely constraining and can be used to determine scattering 
amplitudes in the NLSM and gauge theory.  
As the number of legs and loops increases, this process of constraining an ansatz becomes increasingly more
tedious.  However, there are now a variety of constructive approaches for building tree-level and low-loop numerators that satisfy
the kinematic Jacobi identities~\cite{KiermaierTalk,BjerrumMomKernel,MafraExplicitBCJNumerators,
  MafraExplicitBCJOneLoop, MafraSchlottererTwoLoop, delaCruz2016gnm, 
  BjerrumManifestingBCJ, FuDuFengEYM, Chiodaroli2017ngp, TengFengBCJNumerators, Chen:2017bug, Du2017gnh, delaCruz:2017zqr, Bridges:2019siz}. 

In general, higher-loop integrands is a more involved problem, perhaps more in gauge theories than for the NLSM. 
Direct approaches based on constraining ans\"atze have proven an effective means of 
generating gauge-theory loop integrands~\cite{SimplifyingBCJ, FiveLoop}. We will see explicit examples in \sect{ExamplesSection}. It turns out, however, 
that it can be difficult to find gauge-theory numerators that manifest the
duality between color and kinematics thus complicating the construction of
corresponding gravity integrands. 
Nevertheless, a generalized double-copy procedure outlined in \sect{GeneralizedDoubleCopySection} can be
used to convert gauge-theory integrands in generic representations to integrands in gravity theories; for example, this procedure 
was used to obtain the five-loop four-point integrand of $\NeqEight$ supergravity and determine its
UV behavior~\cite{GeneralizedDoubleCopyFiveLoops, UVFiveLoops}.


\section{Gravity symmetries and their consequences}
\label{GravitySymmetriesSection}

Symmetries are essential for understanding the properties of gauge and
gravity theories. In the context of the framework provided by \ck
duality and the double copy, which relates QFTs order-by-order in
perturbation theory, it is hence interesting to explore how symmetries
originate and transfer.
Not all symmetries of double-copy theories are currently
well-understood from this perspective; likewise, the consequences of
certain symmetries of single-copy parent theories have yet to be
properly understood.\footnote{Moreover, symmetries of sectors of a
single-copy parent theories that relate to the gauge group---such as
symmetries of the planar sector--seem difficult to capture because of
``contamination" from other sectors.}
In this section we review the current status of the relation between
the symmetries of the single- and double-copy theories. We begin by
outlining which (part) of the symmetries of a Lagrangian can be
identified and analyzed through scattering-amplitude techniques
emphasizing that, while the linearized part of symmetries can be
manifest, nonlinear symmetries affect only special momentum
configurations of scattering amplitudes.
We then proceed to discuss the linearly-realized global symmetries and
to extend the diffeomorphism and local-supersymmetry discussion
in \sect{DualitySection} to also include nonabelian gauge
symmetry. All these symmetries are inherited from the symmetries of
their single-copy parents.
We then discuss certain enhanced symmetries, \ie symmetries which,
while unrelated to any of the single-copy symmetries, act linearly on
the double-copy asymptotic states.
The ability to efficiently compute amplitudes and analyze them for
special momentum configurations is essential to explore the emergence
of nonlinear symmetries in the double copy.

\subsection{Symmetries: Lagrangian vs. scattering amplitudes \label{lagrangian_vs_symmetries}}

Lagrangians exhibiting nonlinear symmetries---such as supersymmetry in
a formulation without auxiliary fields, or nonabelian gauge
symmetries---are usually constructed through an iterative Noether
procedure. One starts with the free-field theory with the desired
spectrum, which is invariant under the linearized form of the desired
symmetries and simultaneously deforms the action and the
transformation rules such that the resulting action is invariant
{off-shell} under the deformed transformations.  The resulting
symmetry algebra closes up to the equations of motion.
Thus, this approach leads to actions and transformation rules of the form
\begin{align}
S&= S_2 + S_3 + S_4 + ...  \ ,
\cr
\delta &= \delta^{(0)}  +   \delta^{(1)} +  \delta^{(2)}+ ... \ ,
\end{align}
where the $n$-field term $S_n$ in the action determines the $(n-2)$-field term $\delta^{(n-2)}$ in the transformation rules. For example, 
\begin{align}
\delta^{(0)} S_3 + \delta^{(1)} S_2 &= 0 \,, \nn
\\
\delta^{(0)} S_4 + \delta^{(1)} S_3 + \delta^{(2)} S_2 &= 0 \, .
\end{align}
The first relation implies that the cubic term is invariant under the 
undeformed transformations up to terms proportional to the free equations of motion.

Quantum mechanically, symmetries are realized through Ward identities, which relate time-ordered correlation functions 
of the fundamental fields of the theory. Nonlinear transformation rules imply that the relevant Ward identities contain 
correlation functions of different multiplicities. For example, for a transformation $\delta\phi\propto \phi^k$, an $n$-point Green's function 
is related to $(n+k-1)$-point Green's functions. Moreover, locality of the transformation rules imply that, for $k\ge 2$,  these $k$ fields are 
at the same spacetime point(s).

Upon Lehmann-Symanzik-Zimmermann (LSZ) reduction, Ward identities simplify considerably.  For asymptotic states with momenta $p_1, \dots,p_n$, the amputation leading 
to the $n$-point amplitude selects the most singular term, proportional\footnote{Here we assume that external states have generic masses $m_i$. 
Assuming from the outset that external states are massless does not alter the conclusion.} to $\prod_{i=1}^n (p_i^2-m_i^2)^{-1}$.
For an  $(n+k-1)$-point  ($k=2,3,...$) Green's function resulting from a nonlinear term in a (symmetry) transformation,  
momentum conservation requires that it has a different pole structure. Thus, all such terms are amputated away  and 
all effects of nonlinear terms in the symmetry  transformations that underlie the structure of off-shell Ward identities 
are projected out by the LSZ reduction.
The resulting on-shell Ward identities imply that, for generic momenta, S-matrix elements are invariant only under 
the linearized symmetry transformations.
This argument fails when the additional fields appearing in nonlinear symmetry transformations all carry vanishing momenta; indeed, 
in this case, the off-shell $(n+k)$-point Green's function develops a pole $\prod_{i=1}^n (p_i^2-m_i^2)^{-1}$ and gives a nonvanishing contribution
after the LSZ $n$-point 
amputation. 
We shall return in \sect{soft_limits} to this  special momentum configuration and interpret it as  the soft limit of a higher-point 
scattering amplitude.

It is not difficult to identify these features in nonabelian YM theory: the amplitudes vanish if the polarization vector of a gluon  $\varepsilon_\mu(p)$  is replaced by the momentum\footnote{It is worth mentioning that, from the perspective of the gauge-fixed theory, the transformation 
$\pol^\mu \rightarrow \pol^\mu + \Lambda(p) p^\mu$ can also be interpreted as $\pol^\mu$ not being a proper 
Lorentz vector~\cite{Arkani-Hamed:2016rak}. Indeed, the polarization vector is constrained to obey $p\cdot \pol = 0$ so, on shell, 
any transformation of $\pol$ can include a shift by $p^\mu$. }
\be 
\delta^{(0)} A_\mu = \partial_\mu\Lambda 
~~ \longrightarrow~~
\delta \varepsilon_\mu(p)  =  p_\mu\, \Lambda(p) \,.
\ee
They are also invariant under the  global part of the gauge group, which is the only remnant of the nonlinearity of the gauge 
transformation. 

Not all symmetry transformations have a linearized approximation. An outstanding class of examples are the U-duality symmetries 
of  extended supergravity theories, such as the $E_{7(7)}$ duality group of ${\cal N}=8$ supergravity. It turns out that only their maximal 
compact subgroup, which is isomorphic to the on-shell  $R$-symmetry group, has  such an approximation.
It is therefore an interesting question whether on-shell methods can probe symmetry transformations which are inherently nonlinear.

A possible approach, put forward in Ref.~\cite{Kallosh2016qvo} and further explored in Refs.~\cite{Kallosh:2016lwj, Karlsson:2017bfv},
 effectively amounts to constructing the quantum one-particle irreducible (1PI) effective action and studying its symmetries. Indeed, the quantum 1PI effective action 
is determined\footnote{One may use other methods, such as those outlined in Ref.~\cite{Kallosh2016qvo}, to construct the effective action.} 
by the S-matrix of the theory up to terms proportional to the free equations of motion. Consequently,  up to the 
corresponding contact terms and assuming absence of anomalies, the off-shell Ward identities of all symmetries---in particular 
of the nonlinear ones---should hold. In this formalism, anomalies appear as violations of the Ward identities of the corresponding 
symmetries which cannot be removed by the addition of finite local counterterms to the (effective) action. These counterterms 
may be simultaneously interpreted 
both as part of the definition of the theory and as an ambiguity in the construction of the effective action from the S-matrix.

Another approach geared towards the exploration of nonlinearly-realized symmetries was first described in Ref.~\cite{ArkaniHamed2008gz} 
for the $E_{7(7)}$ symmetry of ${\cal N}=8$ supergravity in four dimensions. 
It amounts to (1) the vanishing of scattering amplitudes in the limit in which momenta of one scalar field vanish and (2) 
the identification/extraction of the structure constants of the nonlinearly-realized part of the symmetry group from the limit 
in which two scalar fields have vanishing momenta. 
In \sect{soft_limits} we will outline this approach and summarize some of its many generalizations to nonlinearly-realized (Volkov-Akulov) supersymmetry~\cite{Volkov:1972jx,Chen2014xoa, Kallosh:2016lwj}, Bondi-Metzner-Sachs (BMS) symmetry~\cite{Conde:2016rom, H:2018ktv, Distler2018rwu}, anomalous symmetries~\cite{Huang2015sla}, effective theories~\cite{Elvang:2016qvq}, string theory~\cite{DiVecchia:2015oba, DiVecchia:2015bfa, DiVecchia:2016szw} and theories with spontaneously-broken conformal invariance~\cite{DiVecchia:2015jaq}.
For discussions of soft theorems at the quantum level see Refs.~\cite{Bern:2014oka, Bern:2014vva, Guerrieri:2017ujb,DiVecchia:2018dob,DiVecchia:2019kle}. 
While neither of these two approaches is specifically tied to the double-copy construction, they may provide strategies to understanding  
aspects of symmetries of the double-copy theories and their relation to their single-copy parents.

\subsection{Global symmetries; on-shell $R$ symmetry \label{Rsym}}

In the absence of anomalies, the scattering amplitudes of a theory exhibit its off-shell 
symmetries  to all orders in perturbation theory.\footnote{This assumes the existence of a regulator 
that preserves these symmetries. }  Below, we shall review how the double copy expresses this property. 

As we reviewed at length in previous chapters, at tree level the KLT relations build gravity scattering amplitudes 
from gauge-theory amplitudes. More generally, for all double-copy theories (including the non-gravitational ones), 
there exist analogous relations that build their scattering amplitudes in terms of data the single-copy parent theories.
It is therefore clear that, at tree level, the global symmetry group $G$ of a double-copy theory is at least as large as the 
product of the global symmetry groups $G_{1,2}$ of the parent theories:
\be
G \supset G_1\otimes G_2 \,.
\label{groups}
\ee

In a Feynman-diagram approach to the construction of scattering
amplitudes, one can arrange that each diagram exhibits all off-shell
global symmetries of the classical Lagrangian. The construction of
tree-level \cknospace-satisfying numerators in terms of tree-level
amplitudes~\cite{KiermaierTalk,BjerrumMomKernel} implies that, at tree
level, the same is true for each single-copy parent theory if one also
demands that the amplitude obey \ck duality. Thus, \eqn{groups} also
holds in this approach.

In the presence of a symmetry-preserving regulator, generalized unitarity then guarantees that the regularized cuts 
of higher-loop amplitudes of the double-copy theory also inherit all the global symmetries of the single-copy parent theories.

\homework{Explore if there is a general statement that can be made about anomalous global symmetries, i.e. whether all anomalous global symmetries of the single-copy parent theories remain anomalous in the double-copy theory. To this end, consider the example
of a four-dimensional gauge theory with chiral fermions and construct examples of the double-copy amplitudes that involve 
scattering amplitudes of this theory that are sensitive to the chiral anomaly.}

Not all global symmetries of a double-copy theory are inherited; in fact, inheritance of some symmetries demands that others 
be enhanced. Consider, for example, the case of the double copy of two theories with ${\cal N}_1$ and ${\cal N}_2$-extended
supersymmetry, respectively.  Their supersymmetry algebras have $SU(N_1)$ and $SU(N_2)$  $R$ symmetry (perhaps with additional decoupled $U(1)$ factors) and, according to the previous discussion, the double-copy theory will be invariant under 
at least $SU(N_1)\times SU(N_2)$ transformations.
However, the $({\cal N}_1+{\cal N}_2)$-extended supersymmetry algebra that is expected based on the number of supercharges 
has a larger $R$ symmetry, $SU(N_1+N_2)$. Thus, to extend $R_1\otimes R_2$ to the complete $R$ symmetry group, it is 
necessary to identify further $2N_1 N_2 +1$ generators.  
The first $2 N_1 N_2$ generators were constructed in Refs.~\cite{Anastasiou2015vba, Anastasiou2017nsz} in terms of the supersymmetry generators
of the two single-copy parent theories. In four dimensions, they are
\be
G_{I {\tilde J}} = Q{}_+{}_{I} {\tilde Q}{}_- {}_{{\tilde J}}
\quad\text{and}\quad
G^{I {\tilde J}} =  Q{}_-{}^{I} {\tilde Q}{}_+ {}^{{\tilde J}} \,, 
\label{off_diagonal_generators}
\ee
where  $Q$ and ${\tilde Q}$ are the supersymmetry generators of the two single-copy parent theories, respectively, and the $\pm$ indices represent their 
helicity.
These generators have vanishing total helicity. From their structure it is clear that they change the helicities 
of the two single-copy components in opposite ways, such that the helicity of the double-copy state is unchanged. For the case of 
${\cal N}=8$ supergravity, their action on states is given in fig.~\ref{Neq8withArrows}.

\begin{table}[t]
\begin{center}
\includegraphics[width=4.6in]{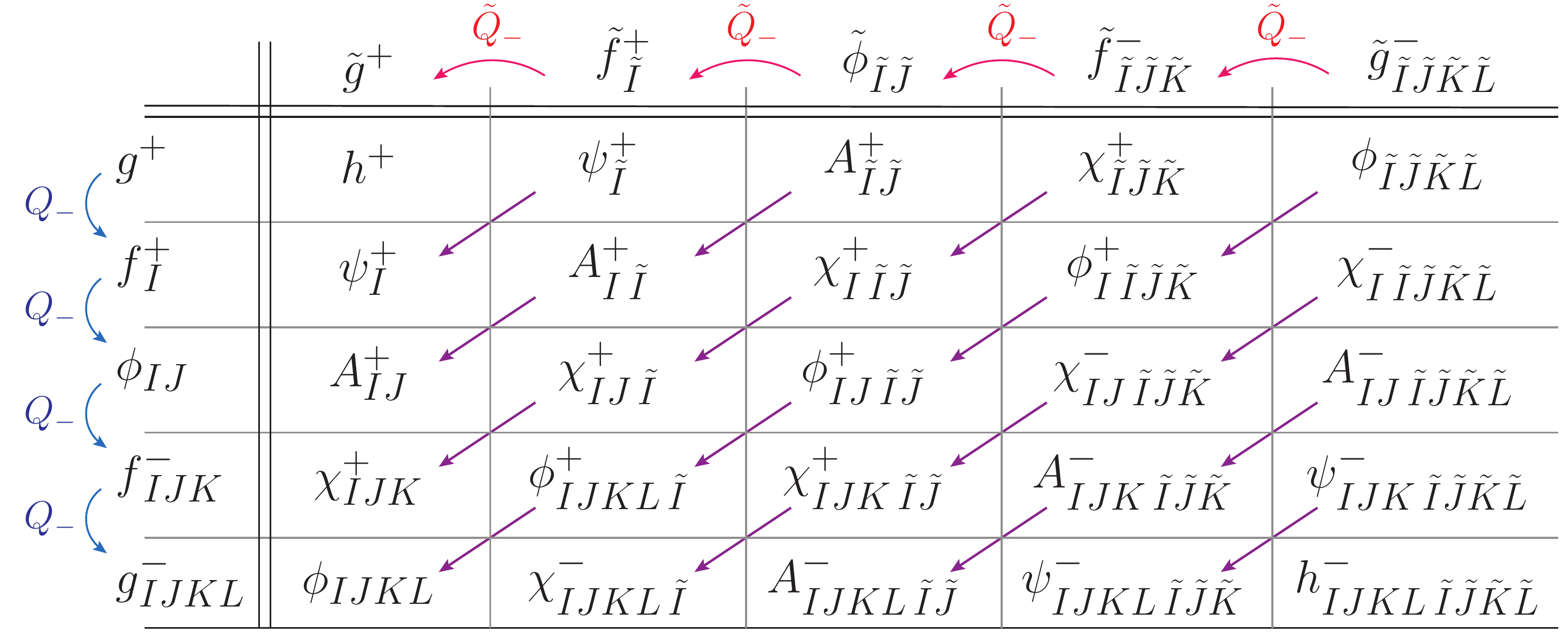}
\end{center}
\caption{Action of the $G^{I {\tilde J}}$ 
generator defined in \eqn{off_diagonal_generators} on the states 
of ${\cal N}=8$ supergravity. The action of the generators $G_{I {\tilde J}}$ is obtained by reversing the direction of the arrows.
}
\label{Neq8withArrows}
\end{table}

The remaining (Cartan) generator which is necessary to recover the complete (and expected) $R$-symmetry group may in principle 
be obtained from the closure of the off-diagonal $G_{I{\tilde J}}$ and $G^{I {\tilde J}}$. In \sect{DualitySubSection}, we shall 
review another way of identifying it,
as well as its physical interpretation.

An interesting feature which has been observed in explicit examples, some of which are described in \sect{ZoologySection1}, is that certain 
gravity theories have two distinct double-copy realizations. In these cases, each version of the construction exhibits different manifest symmetries and, while following the pattern above,
the details of the symmetry enhancement are different.
An example discussed in Ref.~\cite{OneLoopSusy} from a double-copy perspective and in 
\cite{Tourkine:2012vx} from a string-theory point of view, is ${\cal N}=4$ supergravity with two vector multiplets, which can be realized 
both as $({\cal N}=4$~SYM$)\times$(YM$+2$\,scalars$)$ and $({\cal N}=2$~SYM)$\times$$({\cal N}=2$~SYM$)$. 
While in the former construction the complete $SU(4)$ $R$ symmetry of supergravity is manifest, the latter only has a manifest $SU(2)\times SU(2)$
symmetry. 
A more dramatic example is provided by three-dimensional ${\cal N}\ge 8$ supergravities, which can be realized~\cite{Huang2012wr,Huang:2013kca} 
either in terms of two three-dimensional 
SYM theories or in terms of two Chern-Simons-matter theories~\cite{Bagger:2007jr, Gustavsson:2007vu, Bagger:2008se, Bagger:2010zq, Chen:2009cwa, Chen:2010xj} (see also Ref.~\cite{Bargheer2012gv} for the double-copy realization of maximally supersymmetric three-dimensional supergravity~\cite{Marcus:1983hb}).

To study the origin of the symmetries of a gravitational theory from
the double-copy factors, it is sometimes convenient to introduce a
manifestly covariant formulation by defining the action of the
double copy on the off-shell linearized (super)fields, following an
approach introduced in Refs.~\cite{Borsten2013bp, Borsten2013bp,
Anastasiou2014qba, Nagy:2014jza, Anastasiou2015vba, Cardoso2016amd,
Cardoso2016ngt, Anastasiou:2018rdx}.  To give an explicit example, we
consider the double copy of two vector fields, which in this language
is written as
\be 
 H_{\mu \nu} = h_{\mu \nu} + B_{\mu \nu} +\phi\eta_{\mu\nu}= A_\mu^a \star  \Phi^{-1}_{a a'} \star {\tilde A}_\nu^{a'}\,,
 \label{fatgraviton}
\ee
where $A_\mu^a $ and  ${\tilde A}{}_\mu^{a'} $ are the fields in the left and right gauge theory, respectively. The resulting double-copy field $H_{\mu \nu}$ 
(sometimes referred to as the ``fat graviton'' \cite{Luna2016hge}, see also \sect{ClassicalDoubleCopySection}) needs to be decomposed in irreducible representations of the Lorentz group, giving the graviton field, the dilaton and an antisymmetric tensor $B_{\mu \nu}$. This version of the construction is formulated in position space and hence relies on the convolution among linearized superfields, which is defined as 
\be 
[f \star g](x) = \int d^4 y f(y)g(x-y) \,.
\ee
Crucially, $\Phi_{a a'}$ is a bi-adjoint scalar field which is
employed to contract the gauge indices of the left and right   
fields. The action of a symmetry transformation on left and right
fields is then written as
\begin{align}
\delta A_\mu^a & = \partial_\mu \Lambda^a + f^{a}{}_{bc} A_\mu^b   \theta^c + \hat \delta A_\mu^a   \,,
 \nn\\
\delta {\tilde A}_\mu^a &= \partial_\mu {\tilde \Lambda}^a + f^{a}{}_{bc} {\tilde A}_\mu^b {\tilde\theta}^c + \hat \delta  {\tilde A}_\mu^a \,, 
\end{align}
where $\Lambda^a$, ${\tilde \Lambda}^a$, $\theta^a$ and ${\tilde\theta}^a$ are the parameters of local abelian and global nonabelian gauge transformations, while $\hat \delta$  indicates a global transformation under the (super)Poincar\'{e} group. The bi-adjoint scalars are designed to offset the left and right gauge transformations, and transform as \cite{Anastasiou2014qba}
\be 
\delta \Phi^{-1}_{a a'} = - f^{b}{}_{a c} \Phi^{-1}_{b a'} \theta^c - f^{b'}{}_{a' c'} \Phi^{-1}_{ab'} {\tilde \theta}^{c'} + \hat \delta \Phi^{-1}_{aa'} \,.
\label{biadjointscalar}
\ee
While this approach treats the action of the gauge-theory symmetries in an elegant way, the full dictionary is known only at the linearized level.
As it was shown in Ref.~\cite{Anastasiou:2018rdx}, the linearized gravitational equations of motion can be obtained from the linearized gauge-theory ones.
It remains an open question how to include interactions in this formalism (which are naturally incorporated from the perspective of scattering amplitudes).   

\homework{Use Eq.~\eqref{biadjointscalar} to show that the fat graviton defined in Eq.~\eqref{fatgraviton} is inert under nonabelian gauge transformations. Moreover, show that its local transformation rules are a linear combination of linearized diffeomorphisms and gauge transformations 
of a two-index antisymmetric tensor field.}

\subsection{Local symmetries}

In \sect{DualitySection}, we discussed in detail the emergence of diffeomorphism invariance  
and local supersymmetry in  gravity scattering amplitudes obtained from the double-copy construction. The former is a direct consequence of the gauge invariance
of the two single-copy gauge theories and manifest \cknospace-satisfying form for at least one of the two 
gauge theories~\cite{Arkani-Hamed:2016rak,RodinaGaugeInv, Chiodaroli2017ngp}. The latter is a consequence of the gauge invariance
of one of the single-copy gauge theories, supersymmetry of the second, and manifest \ckDash-satisfying form for at least one of them.
The on-shell supersymmetry Ward identities of the double-copy theory follow from those of the single-copy parents.
In this section we review how similar mechanisms lead to other local symmetries in double-copy theories.

As discussed in the beginning of this section, scattering amplitudes in theories with local symmetries that act nonlinearly 
on fields are invariant under the global part of the symmetry group (if it acts linearly) as well as under its linearized local transformations. 
The converse, however, does not necessarily hold: scattering amplitudes that are invariant under some global symmetry group 
$G$ and under abelian local transformations do not necessarily describe a QFT with a local $G$ symmetry. 
For example, they may correspond to a field theory with ${\rm dim}(G)$ abelian vector fields.
It is of course not difficult to distinguish between these two possibilities by inspecting the scaling dimension of certain scattering 
amplitudes, which is different according to whether the theory involves abelian and nonabelian vector fields.

From the discussion in \sect{DualitySection}, it is clear that, in any double-copy theory,  each vector field whose asymptotic states 
are realized as a product of a scalar- and a vector-field asymptotic states  exhibits a Maxwell gauge symmetry, which is a consequence 
of the corresponding gauge symmetry of the vector field in the single-copy parent theory.\footnote{We note that the single-copy origin of
Maxwell gauge symmetry is not obvious if the vectors are realized as in terms of two fermions.}
In order to associate these vector fields to  a local nonabelian symmetry, the corresponding amplitudes must exhibit several properties:
(1) be invariant under the adjoint action of some a global nonabelian symmetry group on the asymptotic states of vector fields and
(2) have the correct dimension to be consistent with minimal coupling.
The second property demands that a three-vector amplitude have  unit dimension,
\be
[{A}^{(0)}_3] = 1 \,,
\ee
as in a standard nonabelian gauge theory. 
To obtain such amplitudes through double copy, at least one of the single-copy parent theories
must have amplitudes of dimension zero. Lorentz invariance and locality imply then that the corresponding single-copy fields labeling 
such amplitudes must be scalars. To satisfy property (1), the corresponding amplitude must be momentum-independent and 
coming from a Lagrangian of the type
\be
{\cal L} = \dots + f^{abc} F^{ABC} \phi^{aA}\phi^{bB}\phi^{cC}+\dots \,,
\ee
where the ellipsis stand for other interactions. 
This reasoning led to the double-copy realization of Yang-Mills-Einstein (YME) theories, as described in Ref.~\cite{Chiodaroli2014xia}.
It was also used in Ref.~\cite{Bern1999bx} to obtain amplitudes in the same theory through a KLT-like construction. 
This procedure can be extended 
to give a double-copy realization of spontaneous breaking of YM gauge symmetry of supergravity theory \cite{Chiodaroli2015rdg}; 
spontaneous breaking of this symmetry is related to explicit breaking of a global symmetry of one of the single-copy parent theories.
We shall review its applications more thoroughly in \sect{Higgsed}. 

The same analysis implies that the double-copy fields that are realized as products of single-copy fields with nonzero spin
cannot couple directly to the nonabelian vector potential and can couple only to its field strength. Indeed, in a conventional gauge
theory, any three-point amplitude with at least one field with nonzero spin has unit dimension.  Thus, the corresponding double-copy 
three-point amplitude has dimension 2 and cannot be given by a minimal-coupling term.

The analysis above can be extended to interactions of gravitini with abelian or nonabelian gauge potentials. Such interactions
are the tell-tale of gauged supergravities---that is, supergravities in which part of the $R$ symmetry is gauged. 
The gravitino minimal coupling around Minkowski space is
\be
{\cal L}_3 \sim {\bar\psi}_\mu \gamma^{\mu\nu\rho}D_\nu\psi_\rho \,,
\label{gravitino3pt}
\ee
and, thus, as for the case of lower-spin fields, the two-gravitini-vector amplitudes have again unit dimension.  Since all three-point amplitudes 
in conventional gauge theories that could have this spin content in their product have at least unit dimension, it follows that their double copy can only 
describe the coupling of gravitini and field strengths and thus, a more refined argument is needed to accommodate minimal couplings of gravitini.  

The observation that sidesteps the difficulty  exposed above \cite{Chiodaroli2017ehv} is that, assuming that the theory has a Minkowski ground 
state,  the three-point amplitude following from the minimal coupling of a gravitino with a vector field spontaneously breaks supersymmetry.  
Consequently, some of the gravitini must be massive and therefore their double-copy realization must involve a single-copy theory with massive 
vector fields and another one with massive fermions. The former must therefore be a spontaneously-broken gauge parent theory while the latter 
turns out to exhibit explicit supersymmetry breaking (this construction will be illustrated in detail in \sect{gaugedsugras}).
The general pattern is that, through the double copy, explicit breaking of a global symmetry can be promoted to spontaneous breaking of the 
local version of the same symmetry.
%


\subsection{Dualities\label{DualitySubSection}}

We have seen in \sect{Rsym} that double-copy theories inherit all the global symmetries of their single-copy parents and that some of the single-copy symmetries combine in nontrivial ways (\eg two supersymmetry generators combine to 
become a bosonic $R$-symmetry generator) to enhance the inherited symmetries.
Lagrangian-based supergravity considerations suggest the existence of
much larger symmetries --- the U-duality symmetries --- which are
typically noncompact. In pure supergravities in various dimensions,
these symmetries were originally discussed in Refs.~\cite{Cremmer:1979up,Julia:1980gr}; their maximal
compact subgroup is isomorphic to the on-shell $R$-symmetry group of
the theory. Depending on the dimension, they are either symmetries of the
equations of motion (in four dimensions) or symmetries of the Lagrangian
(\eg in five dimensions).
In four dimensions, the field strengths and their duals form an {irreducible} representation of the U-duality group, which therefore 
contains electric/magnetic duality as one of its generators.
While the general understanding U-duality symmetries from the double-copy perspective is currently an open problem, 
their dimension-dependent properties suggest that their realization should involve transformations that are not off-shell symmetries in the single-copy parent theories.  

Ref.~\cite{CarrascoN4Anomaly} showed that a universal generator of the U-duality groups of four-dimensional supergravities
can be realized as the difference of the little-group generators (helicity) of the two single-copy parent theories;  
the charges of the double-copy fields under this generator
are
\begin{equation}
Q = q(h - {\tilde h}) \,.
\label{U1charges}
\end{equation}
Note that this transformation acts on the positive and negative helicity vector fields in the double-copy theory with opposite phases. 
Because of this property, the above transformation can be identified   as an electric/magnetic duality transformation acting on vector fields, combined with additional 
transformations of other fields  \cite{Boels2008fc, Roiban2012gi}.
It is not { a priori} clear why this transformations should be a symmetry of the double-copy theory at tree level. One can nonetheless check that 
for  ${\cal N}\ge 5$ it is part of the on-shell $R$ symmetry of the theory and thus part of the maximal compact subgroup of the U-duality. For example, 
decomposing the positive-helicity (denoted with the index $``+"$ below) and scalar states of ${\cal N}=8$ supergravity in representations of the 
$(SU(4),{\widetilde {SU(4)}})^{U(1)}$ subgroup of the $SU(8)$ $R$ symmetry (of which only $SU(4)\times{ {SU(4)}}$  is manifest in the double copy)
one finds
\def\rep#1{{\bf #1}}
\def\no{\nonumber}
\bea
\rep{1}^+&=&(\rep{1},{\rep1})^0 \ ,
\no\\
\rep{8}^+&=&(\rep{4},{\rep1})^q\oplus (\rep{1},\rep{4})^{-q} \ ,
\no\\
\rep{28}^+&=&(\rep{6},{\rep1})^{2q}\oplus (\rep{1},\rep{6})^{-2q}\oplus (\rep{4},\rep{4})^{0}
\label{state_decomposition} \ ,
 \\
\rep{56}^+&=&({\bar{\rep{4}}},{\rep1})^{3q}\oplus (\rep{1},{\bar{\rep{4}}})^{-3q}\oplus (\rep{6},\rep{4})^{q}
\oplus (\rep{4},\rep{6})^{-q} \ ,
\no\\
\rep{70}^{\hphantom{+}}&=&(\rep{1},{\rep1})^{4q}\oplus (\rep{1},\rep{1})^{-4q}
\oplus ({\bar{\rep{4}}},\rep{4})^{2q}\oplus (\rep{4},{\bar{\rep{4}}})^{-2q}
\oplus(\rep{6},\rep{6})^{0} \,.
\no
 \eea
As pointed out in Ref.~\cite{CarrascoN4Anomaly}, the $U(1)$ charges resulting from this decomposition are exactly given by eq.~\eqref{U1charges}.
The decomposition of the negative helicity states is obtained by conjugating the first four lines of \eqn{state_decomposition};
the $U(1)$ charge changes sign under conjugation and may be also identified as being proportional to the net number of indices
of the states in Table~\ref{Neq8withArrows}. From this table, we also see that supersymmetry generators change the $U(1)$ charge
by $1/2$ unit, while the off-diagonal $R$-symmetry generators enhancing $SU(4)\times {\tilde {SU(4)}}\rightarrow SU(8)$ 
change the $U(1)$ charge by one unit.
While we illustrated here its relevance for ${\cal N}=8$ supergravity, the $U(1)$ symmetry described by eq.~\eqref{U1charges}
is required for obtaining the complete on-shell $R$-symmetry for all ${\cal N}\ge 5$ supergravities, as follows from the fact that the latter
theories can be obtained as consistent truncations of the former.

For $1\le {\cal N}\le 4$, this symmetry is present at tree level but it
is anomalous \cite{MarcusAnomaly, CarrascoN4Anomaly}.  This anomaly
sources certain loop-level amplitudes which vanish at tree level. 
In the realization of these theories as double copies with one non-supersymmetric gauge-theory factor, these anomalous 
amplitudes  \cite{CarrascoN4Anomaly} can be traced to a self-duality anomaly of YM theory \cite{Cangemi:1996rx}.
When realized as double copies of supersymmetric gauge theories, the identification of anomalous amplitudes is more subtle: 
they arise from $\mu$-terms\footnote{$\mu$-terms are numerator terms proportional to the extra-dimensional parts of loop 
momenta. Such terms vanish identically if the integrand is evaluated in four dimensions.} which, in 
each gauge theory, give only ${\cal O}(\epsilon)$ terms but give finite terms only after the double copy  \cite{JohanssonTwoLoopSusyQCD}.
It turns out~\cite{BPRAnomalyCancel} that, at least for ${\cal N}=4$
supergravity, the anomalous amplitudes can be canceled at one loop by the
addition of a finite local counterterm to the classical action; this
counterterm restores the $U(1)$ symmetry at the expense of breaking
other symmetries\footnote{They are the two generators that, together
with $U(1)$, form the $SU(1,1)$ classical U-duality group of the
theory.} that do not appear to impose any obvious selection rules on
amplitudes. 
The same counterterms also cancels the two-loop anomalous amplitudes \cite{Bern:2019isl}.
The full consequences of these cancellations remain to be
explored.\footnote{For ${\cal N}=0$ supergravity, defined as the double copy of two pure YM theories,  
this symmetry is also present; it represents the $U(1)$ rephasing of the dilaton-axion. It
survives at the quantum level because there are no fields that can contribute to its anomaly.}

It is instructive to consider the $U(1)$ transformation with charges \eqref{U1charges} {vis \`a vis} the observation 
discussed in \sect{Rsym} that the same supergravity theory may have (two or perhaps more) different double-copy realizations.
While a general analysis is yet to be carried out, it is not difficult to see on a case-by-case basis that this symmetry may play 
different roles. To this end, consider ${\cal N}=4$ supergravity realized as $({\cal N}=4$~SYM$)\times$(YM$+2$ scalars$)$ 
and $({\cal N}=2$~SYM)$\times$$({\cal N}=2$~SYM$)$, both of which can be obtained as different (orbifold) 
truncations\footnote{See \sect{SecOrbifold} for details on field-theory orbifolds in this and related contexts.}
of the double-copy constructions of ${\cal N}=8$ supergravity.
In the former, the $SU(4)$ $R$ symmetry in manifest and the $U(1)$
symmetry is part of the $SU(1,1)$ duality group of ${\cal N}=4$
supergravity.  In the latter, only $SU(2)\times SU(2)\subset SU(4)$ is
manifest and the $U(1)$ symmetry is required to enhance it to the
complete $SU(4)$ $R$ symmetry. In this formulation the origin of
$U(1)\subset SU(1,1)$ is not clear.
Similarly, the further enhancement to the $SO(6,2)\times SU(1,1)$ complete U-duality group (see \sect{ZoologySection1}) is currently 
an open problem, on the same footing as $SU(8)\rightarrow E_{7(7)}$ in ${\cal N}=8$ supergravity.\footnote{For supergravity 
theories for which the scalar fields parametrize the locally-homogeneous space $G/H$ with $H$ being the maximal compact subgroup of $G$, the 
noncompact part of $G$ can be identified once $H$ and its representations carried by scalars are determined \cite{Freedman:2012zz}; 
see also Refs.~\cite{Anastasiou:2013hba, Anastasiou2015vba,Anastasiou2017nsz} for further details from the double-copy perspective at the
noninteracting level. It is nevertheless not clear how to construct the noncompact $G$-generators 
in terms of operators in the single-copy theories.}

The definition of this universal $U(1)$ symmetry in \eqref{U1charges}, does not single out supergravities as the only double-copy 
theories that exhibit this 
symmetry. There exist many non-gravitational four-dimensional field theories exhibiting electric/magnetic duality; for
all those that have a double-copy realization, the transformations of the asymptotic states under duality have the same form \ref{U1charges}. 
An example is the Born-Infeld theory; in this case duality implies that only split-helicity amplitudes (\ie amplitudes with 
an equal number of positive and negative vector fields) are nonvanishing. While this can be proven to all multiplicities through
various techniques \cite{Boels2008fc, Roiban2012gi}, it would be interesting to understand this property from the perspective of the 
double-copy construction.
Quite generally, it remains an interesting open question to understand the consequences of duality from the perspective of the 
single-copy parent theories. 

The one-loop all-multiplicity all-plus and single-minus amplitudes of the Born-Infeld theory were constructed using $D$-dimensional unitarity and 
supersymmetric decomposition in \cite{Elvang:2019twd} and integrated using dimension-shifting relations in \cite{BernMorgan}.  
The amplitudes in both classes turn out to be nonvanishing, implying that, similarly to ${\cal N}=4$ supergravity, duality appears to be anomalous in 
the non-supersymmetric  Born-Infeld theory in this regularization scheme. It remains an open question \cite{Elvang:2019twd}  whether 
the anomaly is physical or whether it can be removed by a finite local counterterm at the expense of other symmetries.
Arguments presented in \cite{Roiban2012gi} suggest that it should be possible to restore duality with a counterterm that breaks Lorentz invariance.

\subsection{Soft theorems as tests of enhanced global symmetries \label{soft_limits}}

As described at length in \sect{lagrangian_vs_symmetries}, supergravity considerations suggest the existence of a much larger symmetry group then the one manifestly realized on  on-shell scattering amplitudes. Part of this (U-duality) group acts nonlinearly and thus does not impose standard selection rules on scattering amplitudes; consequently, the corresponding generators cannot be realized manifestly (\ie linearly) on scattering amplitudes simultaneously with supersymmetry and Lorentz  invariance.\footnote{We note that a Lagrangian formulation of ${\cal N}=8$ that has manifest $E_{7(7)}$ symmetry was 
constructed in Ref.~\cite{Hillmann:2009zf} and further explored in Ref.~\cite{Bossard:2010dq}. This formulation, however, breaks 
manifest Lorentz invariance. Moreover, diffeomorphism transformations on vector fields are realized in a nonstandard way. 
It would be interesting to explore the scattering amplitudes of ${\cal N}=8$ supergravity in this formulation and compare them to the standard form.}
Because of these features, it is not currently known how to identify the single-copy origin of the noncompact U-duality 
transformations. 
The discussion in \sect{lagrangian_vs_symmetries} and the ability to compute scattering amplitudes efficiently (both at tree 
and loop level) gives us an alternative route to probe the existence of these symmetries, borrowing from the supergravity knowledge 
that scalar fields of the theory parametrize  coset space of the form $G/H$, where $G$ is the U-duality group and $H$ its maximal compact subgroup.
In this section, departing from the philosophy in the rest of this review, we shall assume that supergravity amplitudes are 
available (through the double-copy or by some other means) and describe how to identify the hidden existence of the 
noncompact U-duality symmetries.

This problem was first discussed in detail in Ref.~\cite{ArkaniHamed2008gz}, where it was shown that the existence of nonlinearly 
realized symmetries of this type can be identified through the vanishing of single-soft-scalar limit of scattering amplitudes, while the
precise group structure can be inferred from the limit in which the momenta of two scalar fields become simultaneously soft.
We review this construction, which was also extended to other nonlinearly-realized or spontaneously-broken symmetries, 
as well as to fields with nontrivial Lorentz-transformation properties, in 
\cite{Chen2014xoa, Huang2015sla, Guerrieri:2017ujb, Kallosh:2016lwj, DiVecchia:2015bfa}.
A thorough analysis of the soft limits in effective field theories was carried out in Refs.~\cite{Brodel2009hu, Elvang:2016qvq}.

Consider, following Ref.~\cite{ArkaniHamed2008gz}, a symmetry group $G$ with generators falling into two sets, $T$ and $X$, broken to 
the subgroup $H$ generated by $T$. Schematically, the commutation relations are 
\be
[T,T]\sim T\,,
\qquad
[T, X]\sim X \,,
\qquad
[X,X]\sim T \,.
\ee
From a Lagrangian point of view (if one is available), there exists a (Nambu-Goldstone) scalar for each of the broken generators $X$. 
In general, this Lagrangian has many degenerate vacua; moving from one to another amounts to giving nonzero vacuum expectation values (VEVs) 
to the Nambu-Goldstone scalars. From the perspective of scattering amplitudes, a vacuum-expectation value of a field 
corresponds to a condensate of the zero-momentum mode. Thus, exploring the change in vacuum state is equivalent to exploring the 
properties of scattering amplitudes in the zero-momentum limit for some of the scalars.

A similar conclusion may be reached by revisiting the argument in \sect{lagrangian_vs_symmetries} showing that, for generic momentum
configurations, LSZ reduction renders scattering amplitudes  insensitive to nonlinear field transformations. Assuming a generic transformation
rule $\delta\phi \sim \phi^{k\ge 2}$, the same argument implies that, if all but one of the fields on the right-hand side of the transformation carry 
vanishing momentum, then the transformed Green's function has the same poles as the original one and therefore survives the LSZ reduction.
Thus, the nonlinear parts of a symmetry transformation should have a reflection on higher-multiplicity scattering amplitudes in which the additional
asymptotic states have vanishing momenta.

Starting with some vacuum state $|0\rangle$, a neighboring one is obtained through a $G$ transformation with parameters given by the 
VEVs of the old scalars in the new vacuum:
\be
|0\rangle_\theta = e^{iX^\alpha\theta_\alpha } |0\rangle  \,.
\ee
Since the $G$-symmetry requires that amplitudes around the two vacua be the same, the conclusion is therefore that the scattering 
amplitudes with at least one zero-momentum scalar field vanish identically.  For a single soft scalar, this reproduces the 
celebrated Adler zero~\cite{AdlerZero}. 
One may turn the single-soft-scalar limit argument around and infer  \cite{Chiodaroli2015wal} that, in a theory that has vanishing single-soft-scalar limits,  the scalar fields belong to a locally-homogeneous space (\ie a space that has a transitive local group action).

A more involved argument \cite{ArkaniHamed2008gz} extracts the structure constants of the broken symmetry group from the double-soft-scalar  limit of scattering amplitudes:
\be
{\cal M}_{n+2}(1,2,3,\dots,n+2)\stackrel{p_1, p_2\rightarrow 0}{-\!\!\!\!-\!\!\!\!-\!\!\!\!-\!\!\!\!-\!\!\!\!-\!\!\!\!-\!\!\!\!\longrightarrow}
\frac{1}{2}\sum_{i=3}^{n+2} \frac{p_i\cdot (p_2-p_1)}{p_i\cdot (p_2+p_1)} T{\cal M}_{n}(3,\dots,n+2) \,,
\ee
where $T$ is the $G$-generator given by the commutator of the $X$ generators corresponding to the two soft scalars.
The momenta of the two scalars should be taken soft at the same rate.  

\homework{
As we discussed in \sect{lagrangian_vs_symmetries} we have seen that scattering amplitudes with generic momenta are 
insensitive to nonlinear terms in symmetry transformations because the LSZ reduction projects out their contribution. An interesting
unexplored problem is the contribution of terms with special momentum configurations. 
Consider a nonlinear symmetry transformation whose nonlinear parts contains bilinears and cubic terms. Assuming that only one
of the fields in the nonlinear terms carries nonzero momentum, explore the features of the single- and double-soft-scalar limits of amplitudes
by applying LSZ construction to Green's functions acted upon by such special transformations.\\
}

The construction reviewed above does not refer to any specific order in perturbation theory and thus relies on absence of  U-duality anomalies. 
Its conclusions have been used to constrain and characterize possible counterterms of ${\cal N}=8$ supergravity, which should be such that 
their contributions to  scattering amplitudes have soft limits following the same pattern. 
Through this reasoning it was shown that a suggested three-loop $R^4$ counterterm is inconsistent with  the $E_{7(7)}$ 
symmetry of ${\cal N}=8$ supergravity~\cite{Brodel2009hu}. Along the same lines, Ref.~\cite{Beisert2010jx} argued that the first deformation
of ${\cal N}=8$ supergravity that is consistent with the soft-scalar behavior required by $E_{7(7)}$ symmetry can appear at seven loops
and corresponds to a supersymmetric completion of a $D^8R^4$ operator.

Generic diffeomorphism transformations are nonlinear. As we discussed
in the beginning of this section and in \sect{DualitySubSection},
infinitesimal/linearized diffeomorphisms are symmetries of scattering
amplitudes: shifting the graviton polarization tensor
$\pol(p)^{\mu\nu}
\mapsto\pol(p)^{\mu\nu}+p^{(\mu}\Lambda^{\nu)}$ with $\Lambda^\mu$ being the parameter of the transformation, leaves amplitudes invariant.
By definition, large diffeomorphisms do not have a linearized
approximation; the BMS transformations (named after Bondi, van der
Burg, Metzner and Sachs~\cite{Bondi1962px, Sachs1962wk, Sachs1962zza})
arise, in a certain gauge, as residual diffeomorphism symmetries of
asymptotically-flat spacetimes which do not fall off at infinity.
It was argued in Refs.~\cite{Strominger2013jfa, He2014laa,
Kapec2014opa} that the Ward identities of these symmetries imply the
tree-level single-soft-graviton behavior of scattering
amplitudes. Quantum corrections have been discussed in
Refs.~\cite{Bern:2014vva, Larkoski:2014bxa}, with the conclusion that
they affect the linear order in the small momentum if all other
momenta are generic.
The identification of the BMS algebra in the double-soft-graviton limit of scattering amplitudes was discussed in Ref.~\cite{Distler2018rwu}.

Other symmetries can also be probed through double-soft limits. For
example, by explicitly inspecting the tree-level amplitudes of a
certain Akulov-Volkov theory~\cite{Volkov:1972jx},
Ref.~\cite{Chen2014xoa} showed that the double-soft-goldstino limit
yields the supersymmetry algebra.  Moreover, for $4\le {\cal N}\le 8$
supergravities in four dimensions and for ${\cal N}=16$ supergravity
in three dimensions, tree-level scattering amplitudes have a universal
behavior in the double-soft-fermion limit which is analogous to the
scalar one.
The photon and graviton soft theorems were discussed from an
effective-field-theory standpoint in Ref.~\cite{Elvang:2016qvq}, where
a complete classification of local operators responsible for
modifications of soft theorems at subleading order for photons and
subsubleading order for gravitons was derived.

The original discussion~\cite{ArkaniHamed2008gz} of the U-duality
symmetries in supergravity and its subsequent generalizations assumed
absence of anomalies of the spontaneously-broken symmetry.  Possible
anomalies have been included in this framework in
Ref.~\cite{Huang2015sla}, from the perspective of the effective
action; the conclusion of the analysis is that, while the single-soft
limits receive corrections signaling the anomalous breaking of the
symmetry, double-soft limits are unaffected. This is { probably} a
reflection of the anomaly (defined as the nonvanishing of the
divergence of the symmetry current) being invariant under the
classical symmetry.


\section{A web of double-copy-constructible theories}
\label{ZoologySection1}

\def\no{\nonumber}
\def\ha{\hat a}
\def\hb{\hat b}
\def\hc{\hat c}
\def\hd{\hat d}
\def\he{\hat e}
\def\hA{\hat A}
\def\hB{\hat B}
\def\hC{\hat C}
\def\hD{\hat D}
\def\hE{\hat E}

\def\tR{t_{\cal R}}

\newcommand{\co}{\ , \ \ \ \ \ \ }
\newcommand{\dd}{\mathrm{d}}
\newcommand{\te}{\textrm}
\newcommand{\al}{\alpha}
\newcommand{\la}{\lambda}
\newcommand{\vph}{\varphi}
\newcommand{\ap}{{\alpha'}}

\newcommand{\haa}{{\hat \alpha}}
\newcommand{\hbb}{{\hat \beta}}
\newcommand{\hgg}{{\hat \gamma}}
\newcommand{\hdd}{{\hat \delta}}
\newcommand{\hee}{{\hat \epsilon}}

\def\cN{{\cal N}}
\def\wone{0.18\textwidth}
\def\wtwo{0.315\textwidth}
\def\wfour{0.13\textwidth}
\def\wthree{0.3\textwidth}
\def\wfive{0.4\textwidth}

\setlength{\LTcapwidth}{\textwidth}

\begin{figure}[t]
\begin{center} 
  \includegraphics[width=1.01\textwidth]{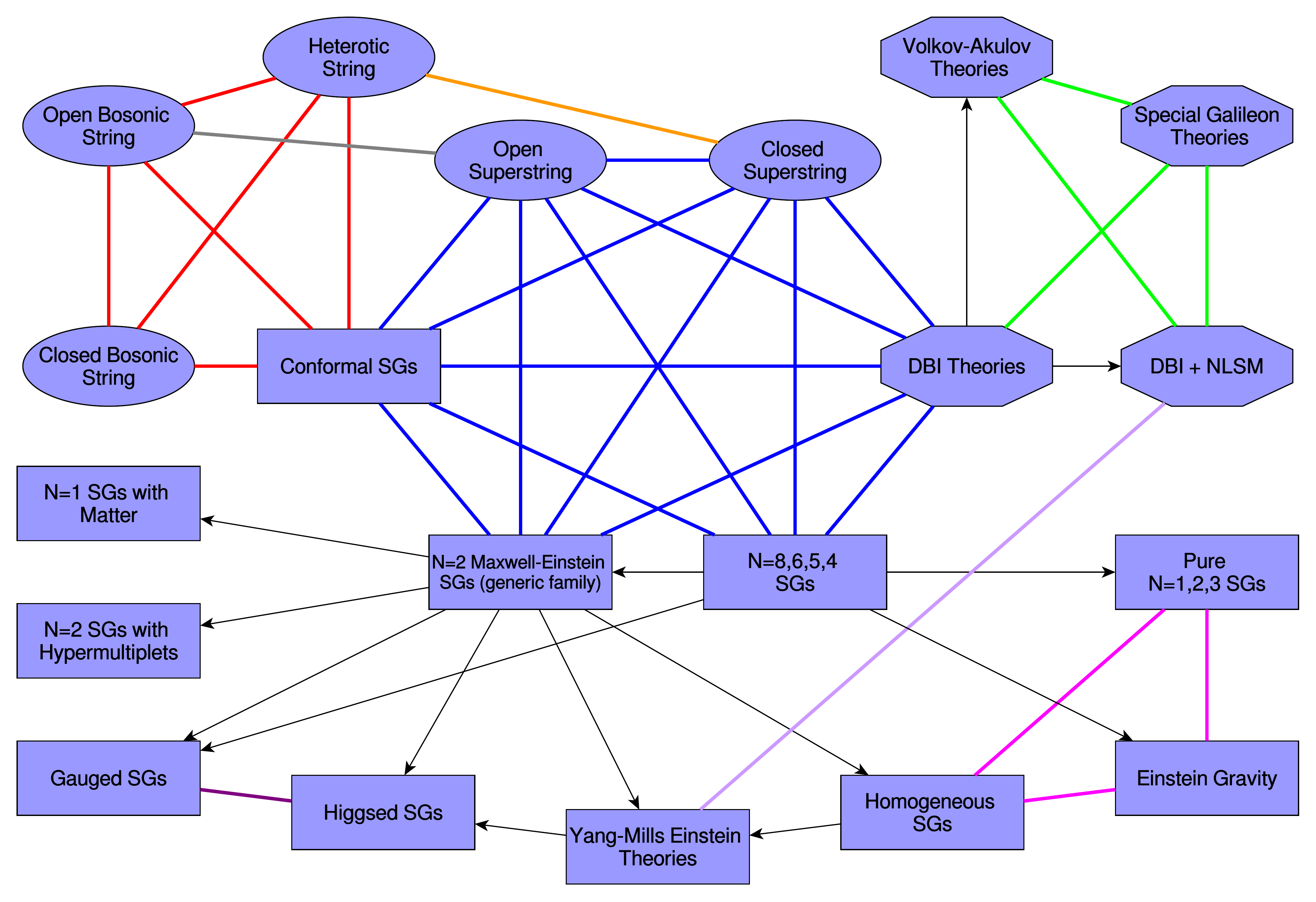} \caption{Schematic
 rendition of the web of theories. Nodes represent the main
 double-copy-constructible theories discussed in this section, which
 include gravitational theories (rectangular nodes), string theories
 (oval nodes) and non-gravitational theories (octagonal
 nodes). Undirected links are drawn between theories that have a
 common gauge-theory factor in their construction (different
 gauge-theory factors correspond to different colors). Directed links
 connect theories obtained by modifying/deforming both gauge-theory
 factors (e.g. adding matter, assigning VEVs). Details are given
 throughout \sect{sec-zoo-examples}. \label{FigWeb}}
\end{center}
\end{figure}

As we have seen in the previous sections, the duality between color
and kinematics and the double-copy construction express amplitudes of
gravitational theories in terms of simpler building blocks from gauge
theory. It has become clear that this property is not an accident of
few very special theories, but extends to large classes of
gravitational and non-gravitational theories. Seemingly unrelated
theories have been shown to share---and thus be connected by---the
same set of building blocks, yielding a ``web of theories'' which can
be analyzed with double-copy methods (see \fig{FigWeb}).
In this section, we aim to probe this web  more in detail. Particularly prominent results will be the classification of homogeneous $\cN=2$ Maxwell-Einstein supergravities~\cite{deWit1991nm}, which can be reproduced and streamlined by double-copy methods,  the double-copy construction for YME \cite{Bern1999bx,Chiodaroli2014xia,Cachazo2014nsa,Cachazo2014xea} and gauged supergravities \cite{Chiodaroli2017ehv,Chiodaroli2018dbu}, and the construction for Dirac-Born-Infeld (DBI) theories \cite{Cachazo2014xea,CheungUnifyingRelations}. We will also see that some of the building blocks which appear, for example, in the double-copy construction for conformal supergravities play a role in a family of ``stringy'' double-copy constructions.   
Similar webs of theories have appeared, for example,  in the contexts of the scattering equations formalism~\cite{Cachazo2014xea}, amplitude transmutation~\cite{CheungUnifyingRelations}, and soft limits~\cite{Cheung:2016drk}.

The simplest examples of double-copy-constructible theories we have discussed so far include $\cN\geq  4$ supergravity and  Einstein gravity 
coupled to a dilaton and two-form field. Once a double-copy structure has been established for a 
given gravitational theory, it is relatively straightforward to obtain the tree-level amplitudes of its consistent truncations. 
In this way, we can study  amplitudes in a handful of additional theories. 
At the same time, it is well-known that supergravity theories with $\cN<4$ 
have a very rich structure which goes beyond the few theories that can be understood as truncations of more supersymmetric gravities. 
Ungauged supergravities with $\cN \geq 5$ and two-derivative actions are unique. Starting from $\cN=4$, it becomes possible to have various matter contents.
While $\cN=3,4$ supergravities are completely specified by the number of vector multiplets, additional information on interactions needs to be provided for 
theories with $\cN=2$ supersymmetry. Supergravity theories generically involve scalar fields, which can be regarded 
as the coordinates of a manifold. While extended $\cN>2$ supersymmetry allows only a discrete set of symmetric scalar manifolds, supersymmetry poses less stringent constraints when $\cN\leq 2$. Specifically, in four dimensions, supergravities with vector multiplets possess special-K\"{a}hler scalar manifolds, 
while the geometry is  quaternionic-K\"{a}hler in the case of supergravities with hypermultiplets~\cite{Supergravity}. 

\begin{table}[t]
	\small
	\begin{tabular}{@{}c@{}|c|@{}c}
		\bf Supergravities & \bf Free Parameters	 & \bf Scalar geometry \\
		\hline
		\hline
		\parbox{0.3\textwidth}{ \centering \medskip $\cN>4$ supergravities  } & none &  \parbox{0.33\textwidth}{\centering \medskip symmetric spaces\\ \medskip}  \\
		\parbox{0.3\textwidth}{\centering \medskip $\cN=4$ supergravity} & \parbox{0.3\textwidth}{number of vector multiplets} & \parbox{0.33\textwidth}{\medskip \centering symmetric space \\ \medskip} \\
		\parbox{0.3\textwidth}{\centering \medskip $\cN=3$ supergravity} & \parbox{0.3\textwidth}{number of vector multiplets} & \parbox{0.33\textwidth}{\medskip \centering symmetric space \\ \medskip} \\
		\parbox{0.3\textwidth}{\centering \medskip $\cN=2$, vector multiplets, \\ 5D uplift} &  \parbox{0.3\textwidth}{\centering $C_{IJK}$-tensor} & \parbox{0.33\textwidth}{\medskip \centering very-special K\"{a}hler geometry\\ \medskip} \\
		\parbox{0.3\textwidth}{\centering \medskip $\cN=2$, vector multiplets, \\ 4D only \\ \medskip} &  \parbox{0.3\textwidth}{\centering  free degree-two holomorphic function (prepotential)} & \parbox{0.35\textwidth}{\centering special K\"{a}hler geometry} \\
		\parbox{0.3\textwidth}{\medskip \centering $\cN=2$, hypermultiplets, \\ from $c$-map \\ \medskip} & \parbox{0.3\textwidth}{\centering $C_{IJK}$-tensor or prepotential} & \parbox{0.33\textwidth}{\centering special/very special quaternionic K\"{a}hler geometry} \\	
		\parbox{0.3\textwidth}{\medskip \centering $\cN=2$, hypermultiplets, \\ general \\ \medskip} & See text  &  \parbox{0.33\textwidth}{\centering quaternionic K\"{a}hler geometry} \\
		\hline
	\end{tabular}
	\caption{Freedom in specifying the two-derivative action in extended (ungauged) supergravities with $2\leq\cN\leq 8$ in four dimensions.
\label{specify}}
\end{table}

In Table~\ref{specify}, we list the information which needs to be provided to specify unambiguously ungauged supergravity theories with $2 \leq \cN \leq 8$, 
together with the corresponding geometries. 
It should be noted that theories with $\cN=2$ have different geometrical properties depending on whether or not they have a five-dimensional uplift. 
Theories with vectors multiplets which can be lifted to five dimensions are uniquely specified by a symmetric constant tensor $C_{IJK}$ whose indices 
run over the total number of vector fields. Since this tensor can be obtained from inspecting specific three-point interactions, supergravities of this sort 
have the pleasant property of being entirely constructible from their three-point amplitudes, a property that we will utilize extensively later in this section.
Intrinsically-four-dimensional theories are significantly less constrained. 
They are fully specified by a homogeneous degree-two holomorphic function---the prepotential---which is otherwise arbitrary. Theories with hypermultiplets 
possess even more freedom: $\cN=2$ supersymmetry only require the hypermultiplet scalar manifold be  quaternionic-K\"{a}hler (that is, to admit an hermitian 
metric and three complex structures which satisfy the quaternionic algebra). At the same time, a subset of these theories can be regarded as the image of 
supergravities with vector multiplets under an operation known as $c$-map; specifying these theories requires the same information as their vector counterparts.
When studying supergravities with reduced supersymmetry it is important to keep in mind how this freedom is reflected in the gauge-theory data entering 
the double-copy construction.

\def\spacebefore{\vskip 0.35cm} 
\def\spaceafter{\vskip 0.35cm}
\def\indentspace{\hskip 0.12cm}

\afterpage{
\begin{longtable}{@{}c@{}|c|c|c}
	{\bf Gravity}	& {\bf Gauge theories} & {\bf Refs.} & {\bf Variants and notes} \\
	\hline
	\hline
	\parbox{\wone}{ \footnotesize \spacebefore
		$\cN>4$ \\ supergravity  \medskip } &
	\parbox{\wtwo}{ \footnotesize \spacebefore
		$\bullet$ $\cN=4$ SYM theory \\
		$\bullet$ SYM theory ($\cN=1,2,4$)  \spaceafter
	} & \parbox{\wfour}{\footnotesize \spacebefore
\cite{BCJ,BCJLoop,N46Sugra,GravityFour,N5GravFourLoop}} &
	\parbox{\wthree}{} \\
	\hline
	\parbox{\wone}{ \footnotesize \spacebefore
		$\cN=4$ \\ supergravity with \\ vector  multiplets   \spaceafter } &
	\parbox{\wtwo}{ \footnotesize \spacebefore
		$\bullet$ $\cN=4$ SYM theory \\
		$\bullet$ YM-scalar theory from dim. \\ $\vphantom{x}$ \indentspace reduction  \spaceafter
	} & \parbox{\wfour}{\footnotesize \spacebefore
	\cite{BCJ,BCJLoop,N46Sugra,N4GravFourLoop}} &
	\parbox{\wthree}{ \footnotesize \spacebefore 
		$\bullet$	$\cN=2 \times \cN=2$ construction \\ $\vphantom{x}$ \indentspace is also possible 
		 \spaceafter }\\
	\hline
	\parbox{\wone}{ \footnotesize \spacebefore
		pure $\cN<4$ \\ supergravity  \medskip } &
	\parbox{\wtwo}{ \footnotesize \spacebefore
		$\bullet$  (S)YM theory with matter \\
		$\bullet$ (S)YM theory with ghosts    \spaceafter 
	} & \footnotesize \cite{Johansson2014zca} &
	\parbox{\wthree}{ \footnotesize \spacebefore 
		$\bullet$ ghost fields in fundamental rep  \spaceafter
		\smallskip }\\
	\hline
		\parbox{\wone}{ \footnotesize \spacebefore
		Einstein gravity \medskip } &
		\parbox{\wtwo}{ \footnotesize \spacebefore
			$\bullet$  YM theory with matter \\
			$\bullet$  YM theory with ghosts    \spaceafter 
		} & \footnotesize \cite{Johansson2014zca} &
		\parbox{\wthree}{ \footnotesize \spacebefore 
			$\bullet$ ghost/matter fields in \\ $\vphantom{x}$ \indentspace fundamental rep  \spaceafter
			\smallskip }\\
		\hline
	\parbox{\wone}{ \footnotesize \spacebefore
		$\cN=2$ \\ Maxwell-Einstein \\ supergravities \\ (generic family)  \spaceafter } &
	\parbox{\wtwo}{ \footnotesize \spacebefore
		$\bullet$ $\cN=2$ SYM theory \\
		$\bullet$ YM-scalar theory from dim. \\ $\vphantom{x}$ \indentspace reduction  \spaceafter
	} & \footnotesize \cite{Chiodaroli2014xia} &
	\parbox{\wthree}{\spacebefore \footnotesize 
		$\bullet$ truncations to $\cN=1,0$ \\
		$\bullet$ only adjoint fields  \spaceafter} \\
	\hline
	\parbox{\wone}{ \footnotesize \spacebefore
		$\cN=2$ \\ Maxwell-Einstein \\ supergravities \\ (homogeneous\\ theories)  \spaceafter } &
	\parbox{\wtwo}{ \footnotesize \spacebefore
		$\bullet$ $\cN=2$ SYM theory with half \\ $\vphantom{x}$ \indentspace hypermultiplet \\
		$\bullet$ YM-scalar theory from dim. \\ $\vphantom{x}$ \indentspace reduction 
		with matter fermions  \spaceafter 
	} & \footnotesize \cite{Chiodaroli2015wal,Ben-Shahar:2018uie} &
	\parbox{\wthree}{ \footnotesize \spacebefore $\bullet$ fields in pseudo-real reps \\
		$\bullet$ include Magical Supergravities  \spaceafter
		} \\
	\hline		
	\parbox{\wone}{ \footnotesize \spacebefore
		$\cN=2$ \\ supergravities with \\ hypermultiplets  \spaceafter } &
	\parbox{\wtwo}{ \footnotesize \spacebefore
		$\bullet$ $\cN=2$ SYM theory with half \\ $\vphantom{x}$ \indentspace hypermultiplet \\
		$\bullet$ YM-scalar theory from dim. \\ $\vphantom{x}$ \indentspace red.   
		with extra matter scalars  \spaceafter 
	} & \footnotesize \cite{Chiodaroli2015wal,Anastasiou2017nsz} &
	\parbox{\wthree}{ \footnotesize \spacebefore $\bullet$ fields in matter representations\\
		$\bullet$ construction known in \\ $\vphantom{x}$ \indentspace particular cases  \spaceafter 
		} \\
	\hline		
	\parbox{\wone}{\footnotesize \spacebefore
		$\cN=2$ \\ supergravities \\  with  vector/ \\ hypermultiplets  \spaceafter } &
	\parbox{\wtwo}{ \footnotesize  \spacebefore
		$\bullet$ $\cN=1$ SYM theory with chiral \\ $\vphantom{x}$ \indentspace multiplets \\
		$\bullet$ $\cN=1$ SYM theory with chiral \\ $\vphantom{x}$ \indentspace multiplets  \spaceafter 
	} & 
    \parbox{\wfour}{\footnotesize
	\cite{OneLoopSusy,Damgaard2012fb,Anastasiou2015vba}  \spaceafter }
     &
	\parbox{\wthree}{\smallskip \footnotesize $\bullet$ construction known in \\ $\vphantom{x}$ \indentspace particular cases  \spaceafter } \\
	\hline	
	
	\parbox{\wone}{ \footnotesize \spacebefore
		$\cN=1$ \\ supergravities  with \\  vector multiplets  \spaceafter } &
	\parbox{\wtwo}{ \footnotesize \spacebefore
		$\bullet$ $\cN=1$ SYM theory with chiral \\ $\vphantom{x}$ \indentspace multiplets \\
		$\bullet$ YM-scalar theory with fermions  \spaceafter 
	} & 
    \parbox{\wfour}{\footnotesize \spacebefore
    	\cite{OneLoopSusy,Damgaard2012fb,Anastasiou2015vba,Johansson2014zca}  \spaceafter }
     &
	\parbox{\wthree}{\footnotesize \spacebefore $\bullet$ fields in matter reps \\
		 $\bullet$ construction known in \\ $\vphantom{x}$ \indentspace particular cases  \spaceafter } \\
	\hline		
	\parbox{\wone}{ \footnotesize \spacebefore
		$\cN=1$ \\ supergravities with \\ chiral multiplets  \spaceafter } &
	\parbox{\wtwo}{ \footnotesize \spacebefore
		$\bullet$ $\cN=1$ SYM theory with chiral \\ $\vphantom{x}$ \indentspace multiplets \\
		$\bullet$ YM-scalar with extra matter \\ $\vphantom{x}$ \indentspace scalars  \spaceafter
	} & 
    \parbox{\wfour}{\footnotesize \spacebefore
    	\cite{OneLoopSusy,Damgaard2012fb,Anastasiou2015vba,Johansson2014zca} \spaceafter }
     &
	\parbox{\wthree}{ \footnotesize \spacebefore $\bullet$ fields in matter reps\\
		$\bullet$ construction known in \\ $\vphantom{x}$ \indentspace particular cases  \spaceafter } \\
	\hline		
	\parbox{\wone}{ \footnotesize \spacebefore
		Einstein gravity \\ with matter  \spaceafter } &
	\parbox{\wtwo}{ \footnotesize \spacebefore
		$\bullet$ YM theory with matter \\
		$\bullet$ YM theory with matter  \spaceafter
	} & \footnotesize \cite{BCJ,Johansson2014zca}  &
	\parbox{\wthree}{\footnotesize \spacebefore $\bullet$ construction known in \\ $\vphantom{x}$ \indentspace particular cases  \spaceafter} \\
	\hline		
	\parbox{\wone}{\footnotesize \spacebefore
		 $R$ + $\phi R^2$ + $R^3$ \\ gravity  \spaceafter } &
	\parbox{\wtwo}{ \footnotesize \spacebefore
		$\bullet$ YM theory + $F^3$ + $F^4$ + $\ldots$ \\
		$\bullet$ YM theory + $F^3$ + $F^4$ + $\ldots$  \spaceafter 
	} & \footnotesize \cite{Broedel2012rc} &
	\parbox{\wthree}{ \footnotesize \spacebefore $\bullet$ extension to $\cN\leq 4$ replacing
		one of the factors by  undeformed SYM theory \spaceafter } \\
	\hline		
	\parbox{\wone}{ \footnotesize \spacebefore
		Conformal \\ (super)gravity  \spaceafter } &
	\parbox{\wtwo}{ \footnotesize \spacebefore
		$\bullet$ $DF^2$ theory  \\
		$\bullet$ (S)YM theory   \spaceafter 
	} & \footnotesize \cite{Johansson2017srf,JohanssonConformal} &
	\parbox{\wthree}{\footnotesize \spacebefore $\bullet$ $\cN \leq 4$ \\
		$\bullet$	involves specific  gauge theory \\ $\vphantom{x}$ \indentspace with dimension-six operators   \spaceafter }\\
	\hline
		\parbox{\wone}{ \footnotesize \spacebefore
			$3D$ maximal \\ supergravity  \spaceafter} &
		\parbox{\wtwo}{ \footnotesize \spacebefore
			$\bullet$ BLG theory  \\
			$\bullet$ BLG theory   \spaceafter 
		} & 
	 \footnotesize
	 	\cite{Huang2012wr,Bargheer2012gv,AllicABJM}
	  &
		\parbox{\wthree}{\footnotesize \spacebefore $\bullet$ $3D$ only  \spaceafter}\\
		\hline 
\multicolumn{3}{c}{}\\
	\caption{ 	\small  Non-inclusive list of ungauged gravities and supergravities for which a double-copy construction 
		is presently known. Theories are given in four dimensions unless otherwise stated. \label{table-zoology-ungauged}}
\end{longtable}}

Supergravities studied in the double-copy context have thus far been mostly theories of the Maxwell-Einstein class, \ie theories in which all 
vector fields are abelian and there are no charged matter fields. 
From a Lagrangian perspective, supergravities with nonabelian gauge interactions have also been studied, see~\cite{Supergravity,Samtleben2008pe}
for reviews.
They can be further divided into YME theories and proper gauged supergravities. In the former class, a nonabelian subgroup of the isometry 
group of the scalar manifold is promoted to a gauge symmetry. In the latter case, part of the $R$ symmetry is promoted to gauge symmetry. This procedure, 
customarily referred to as gauging, does not introduce additional vector fields. It minimally couples some of the existing vector fields while also giving them nonabelian self-interactions  and extending the resulting theory so that it is invariant under the required number of supercharges.
In an amplitude context, YME theories have been studied from a variety of perspectives, including scattering equations \cite{Cachazo2014nsa,Cachazo2014xea,Nandan2016pya}, collinear limits of gauge theory amplitudes~\cite{Stieberger:2015qja}, BCFW recursion \cite{TengFengBCJNumerators,Du2017gnh},  string theory~\cite{Stieberger2016lng,SchlottererEYMHeterotic}, ambitwistor strings \cite{Casali2015vta} and, of course, the double-copy construction~\cite{Chiodaroli2014xia}. Through this work, it has become clear that amplitudes in such theories may be written as linear combinations
of (color-ordered) amplitudes or ordinary YM theory~\cite{Stieberger2016lng,Nandan2016pya,TengFengBCJNumerators,Du2017gnh}. 
We will see later in this section that the above property has a very straightforward double-copy interpretation. 
Gauged supergravities display a considerably more involved structure. Once a subset of the $R$ symmetry is gauged (\ie some of the $R$-symmetry generators appear in the covariant derivatives), supersymmetry requires a scalar potential to appear in the theory. According to whether the potential vanishes or not at a critical point, the theory admits Minkowski, Anti-de Sitter or de Sitter vacua. Minkowski vacua break supersymmetry spontaneously (partly or completely), resulting in massive gravitini.   
The study of gauged supergravities in the double-copy framework is still in the early stages, 
but encouraging results are available which will be reviewed later in this section.

A growing body of work seems to suggest that the existence of a double-copy structure is not merely an accidental feature of highly-supersymmetric  theories, but a generic property of very large classes of gravities. To determine whether the double-copy property is a hidden structure of 
gravitational interactions it is necessary to  identify the gauge-theory counterparts of all data required to specify a generic gravity theory, whether it be ungauged, YME or gauged.
While this program has not yet been completed, important progress has been made in formulating double-copy constructions for  theories 
which include, among others, pure supergravities, homogeneous $\cN=2$ Maxwell-Einstein supergravities, homogeneous $\cN=2$ theories with hypermultiplets, large classes of YME or gauged theories, and conformal supergravities. A list of ungauged and gauged theories for which a double-copy construction is currently known can be found in Table~\ref{table-zoology-ungauged} and Table~\ref{table-zoology-gauged}, respectively. 
Gauge theories with fields in various matter (non-adjoint) representations of the gauge group are a rather common building block for this class of extended constructions. Useful tools for treating matter representations in a way that makes manifest color and numerator relations will be introduced 
in  \sect{sec-zoo-tols}. We will then discuss systematics of the process of identifying the gravity theory given, through double copy, by a pair of 
gauge theories and study several examples in \sect{sec-zoo-examples}.

Double-copy constructibility is a property that goes beyond gravitational theories. Various theories without a graviton, most prominently some variants of the DBI theory have also been shown to possess this property (see Table \ref{table-zoology-nongrav}). We shall briefly review their construction in \sect{sec-zoo-nongrav}.

\subsection{The rules of the game \label{sec-zoo-CK}}

To capture as many gravities as possible, we need to consider gauge theories which are more general 
than the ones discussed at length in previous sections. At the same time, having in mind a double-copy construction which leads 
to a sensible gravity theory with desirable basic properties, it makes sense to impose some requirements on the gauge theories under consideration. 
Some additional requirements will also be imposed for simplicity reasons; in both cases, 
one can contemplate generalizations in which some of the stated rules of the game  bent or broken.

First of all, for simplicity, we choose to focus on theories for which amplitudes can be organized exclusively in terms of cubic graphs. 
This is a natural generalization of the gauge theories from the previous sections, which possess this property, 
and is a natural choice for describing gravities that are entirely specified by their three-point interactions. Hence, we restrict the space of gauge theories 
under consideration according to the following rule:

\begin{center}
	\parbox{0.9\textwidth}{
		{\bf  Working Rule 1:} Consider gauge theories with only cubic invariant tensors or, alternatively, theories for which amplitudes can be organized in terms of cubic graphs. 
	}
\end{center}

\noindent
Allowed invariant tensors will include, for example, structure constants, representation matrices and cubic Clebsch-Gordan coefficients. It should be emphasized that the gauge theories under consideration can and will possess quartic vertices. Our requirement constrains higher-point 
interaction vertices to be made of color building blocks which are cubic. If this property is satisfied, amplitudes can be expressed in terms of cubic graphs by including a suitable number of inverse propagators in the numerator factors.  
While this rule is quite desirable for the sake of simplicity, it can in principle be broken.
A notable violation are the the Bagger-Lambert-Gustavsson (BLG) and Aharony-Bergman-Jefferis-Maldacena (ABJM) theories, which are most naturally organized in terms of quartic graphs~\cite{Huang2012wr,Bargheer2012gv,AllicABJM}.

Within the class of cubic theories, however, we need to consider cases which are as general as possible. This motivates the second rule:

\begin{center}
	\parbox{0.9\textwidth}{
		{\bf Working Rule 2:} The gauge theories will include matter fields transforming in general (not necessarily irreducible) representations of the gauge group (which is not necessarily semisimple). 
		Only one adjoint representation will be allowed.  }
\end{center}

\begin{table}[t]
	\begin{tabular}{@{}c@{}|c|c|c}
		{\bf Gravity}	& {\bf Gauge theories} & {\bf Refs.} & {\bf Notes} \\
		\hline
		\hline
		\parbox{\wone}{ \footnotesize \spacebefore
			YME \\ supergravities \spaceafter } &
		\parbox{\wtwo}{ \footnotesize \spacebefore
			$\bullet$ SYM theory \\
			$\bullet$ YM + $\phi^3$ theory   \spaceafter
		} & \parbox{\wfour}{\footnotesize \spacebefore
		\cite{Bern1999bx,Chiodaroli2014xia,Cachazo2014nsa,Cachazo2014xea,Nandan2016pya,TengFengBCJNumerators,Du2017gnh,Stieberger2016lng,Casali2015vta,CheungUnifyingRelations,SchlottererEYMHeterotic} } &
	\parbox{\wthree}{ \footnotesize \spacebefore $\bullet$ trilinear scalar couplings 
		\\ 
		$\bullet$ $\cN=0,1,2,4$ possible  \spaceafter } \\
	\hline		
	\parbox{\wone}{ \footnotesize \spacebefore
		Higgsed \\ supergravities  \spaceafter } &
	\parbox{\wtwo}{ \footnotesize \spacebefore
		$\bullet$ SYM theory (Coulomb branch)\\
		$\bullet$ YM + $\phi^3$ theory with extra \\  $\vphantom{x}$ \indentspace massive scalars  \spaceafter
	} & \footnotesize \cite{Chiodaroli2015rdg} &
	\parbox{\wthree}{ \footnotesize \spacebefore $\bullet$ $\cN=0,1,2,4$ possible\\
		$\bullet$ massive fields in supergravity  \spaceafter } \\
	\hline
	\parbox{\wone}{ \footnotesize \spacebefore
		$U(1)_R$ gauged \\ supergravities \spaceafter} &
	\parbox{\wtwo}{ \footnotesize \spacebefore
		$\bullet$ SYM theory (Coulomb branch)\\
		$\bullet$ YM theory with SUSY broken \\
		$\vphantom{x}$ \indentspace by fermion masses \spaceafter
	} & \footnotesize \cite{Chiodaroli2017ehv} &
	\parbox{\wthree}{ \footnotesize \spacebefore $\bullet$ $0\leq \cN \leq 8$ possible \\
		$\bullet$  SUSY is spontaneously broken  \\
		$\bullet$ only theories with Minkowski \\ $\vphantom{x}$ \indentspace vacua  \spaceafter
	} \\
	\hline

	\parbox{\wone}{ \footnotesize \spacebefore
		gauged \\ supergravities \\ (nonabelian) \spaceafter} &
	\parbox{\wtwo}{ \footnotesize \spacebefore
		$\bullet$ SYM theory (Coulomb branch)\\
		$\bullet$ YM + $\phi^3$ theory with massive \\ $\vphantom{x}$ \indentspace fermions
		\spaceafter}
	& \footnotesize \cite{Chiodaroli2018dbu}  &
	\parbox{\wthree}{ \footnotesize  \spacebefore
		$\bullet$  SUSY is spontaneously broken \\
		$\bullet$  only theories with Minkowski \\ $\vphantom{x}$ \indentspace vacua \spaceafter } \\
	\hline 
\end{tabular}
\caption{Gauged/YME gravities and supergravities for which a double-copy construction 
	is presently known.  \label{table-zoology-gauged}}  	
\end{table}

\begin{table}[t]
	\begin{center}
		\begin{tabular}{@{}c@{}|c|c|c}
			{\bf Double copy}	& {\bf Starting theories} & {\bf Refs.}  & {\bf Variants and notes} \\
			\hline
			\hline
			\parbox{\wone}{ \footnotesize \spacebefore
				DBI \\ theory} &
			\parbox{\wtwo}{ \footnotesize \spacebefore
				$\bullet$ NLSM  \\
				$\bullet$ (S)YM theory  \\ } 
			&\footnotesize 	\parbox{\wfour}{  \cite{Cachazo2014xea,CheungUnifyingRelations,Chen2013fya,Du2016tbc,Chen2014dfa,Chen2016zwe,Cheung:2017yef}}
			& \parbox{\wthree}{ \footnotesize \spacebefore
				$\bullet$ $\cN \leq 4$ possible  \\
				$\bullet$ also obtained as $\alpha' \rightarrow 0$ limit \\ $\vphantom{x}$ \indentspace
				of abelian Z-theory   \spaceafter } 	    
			\\
			\hline		
			\parbox{\wone}{ \footnotesize \spacebefore
				Volkov-Akulov  \\ theory} &
			\parbox{\wtwo}{ \footnotesize \spacebefore
				$\bullet$ NLSM  \\
				$\bullet$ SYM theory (external fermions)  \\ } 
			&   		\parbox{\wfour}{\centering \footnotesize \cite{Bergshoeff:1996tu,Bergshoeff:1997kr,Kallosh:1997aw,Bergshoeff:2013pia,Cachazo2014xea, Cachazo2016njl, He2016mzd, Elvang:2018dco} }
			& \parbox{\wthree}{ \footnotesize \spacebefore
				$\bullet$ restriction to external fermions \\$\vphantom{x}$ \indentspace from supersymmetric DBI   \spaceafter } 
			\\
			\hline		
			\parbox{\wone}{ \footnotesize \spacebefore
				Special Galileon \\ theory \spaceafter} &
			\parbox{\wtwo}{ \footnotesize \spacebefore
				$\bullet$ NLSM  \\
				$\bullet$ NLSM  \\ 
			} & 	\parbox{\wfour}{ \footnotesize \spacebefore \cite{Cachazo2014xea,CheungUnifyingRelations,Cheung2016prv,Cachazo2016njl,Cheung:2017yef} \spaceafter }
			& \parbox{\wthree}{ \footnotesize \spacebefore $\bullet$ theory is also characterized by \\ $\vphantom{x}$ \indentspace its soft limits  
				\spaceafter } 
			\\
			\hline
			\parbox{\wone}{ \footnotesize \spacebefore
				DBI  + (S)YM   \\  theory } &
			\parbox{\wtwo}{ \footnotesize \spacebefore
				$\bullet$ NLSM + $\phi^3$  \\
				$\bullet$ (S)YM theory   \\ 
			} &  		\parbox{\wfour}{ 	\footnotesize \cite{Chiodaroli2017ngp,Cachazo2014xea,CheungUnifyingRelations,Chen2013fya,Du2016tbc,Chen2014dfa,Chen2016zwe,Cachazo2016njl,Carrasco2016ygv}} 
			& \parbox{\wthree}{ \footnotesize \spacebefore
				$\bullet$ $\cN \leq 4$ possible  \\
				$\bullet$ also obtained as $\alpha'\rightarrow 0$ 
				limit \\ $\vphantom{x}$ \indentspace
				of semi-abelianized  Z-theory \spaceafter } 
			\\
			\hline
			\parbox{\wone}{ \footnotesize \spacebefore
				DBI + NLSM  \\   theory } &
			\parbox{\wtwo}{ \footnotesize \spacebefore
				$\bullet$ NLSM   \\
				$\bullet$ YM + $\phi^3$ theory   \\ 
			} &  		\parbox{\wfour}{ 	\footnotesize \cite{Chiodaroli2017ngp,Cachazo2014xea,CheungUnifyingRelations,Chen2013fya,Du2016tbc,Chen2014dfa,Chen2016zwe}} 
			& \parbox{\wthree}{ \footnotesize \spacebefore \spaceafter } 
			\\
			\hline			
		\end{tabular}
		\caption{List of non-gravitational theories constructed as double copies.  
 \label{table-zoology-nongrav}}
	\end{center}
\end{table}

\noindent
Considering general gauge groups and representations will allow us to capture very large families of (super)gravities which would not 
otherwise be accessible through double-copy methods. The main observation is that there is nothing in the double-copy construction that 
requires that representations be divided into irreducible blocks. At the same time, we want to obtain theories with a single graviton. 
This forces us to combine all gauge-theory gluons in a single adjoint representation, even when the gauge group is the product of several factors each possessing its own adjoint representation.
In case of more than one semi-simple factor in the gauge group, we need to take all gauge coupling constants to be the same. Since all fields in the gauge theory
have canonical couplings with gluons, our second rule can also be regarded as the double-copy incarnation of the Equivalence Principle. 

Additionally, massive fields are typically assigned to non-adjoint representations such that all the fields in a given representation have the same mass. This will be accompanied by mass-matching conditions of the spectrum of the two sides of the double copy.

Combining the first two rules, we obtain a generic amplitude structure  that involves cubic graphs in which internal and external legs carry definite representations of the gauge group. Cubic vertices between three representations are allowed only when it is possible to extract a gauge singlet in their tensor product (or, alternatively, there exist a nonvanishing invariant tensor with the three corresponding indices). Whenever a vertex involves two lines carrying the same representation, its symmetry or antisymmetry will be dictated by the representations under consideration (real representations will imply antisymmetry, pseudo-real representation will imply symmetry). Additionally, color factors will obey three-term identities following from the Jacobi relations, the generators' commutation relations and additional algebraic relations which may also involve the Clebsch-Gordan coefficients. Consequently, the duality between color and kinematics must to be imposed in the following way:

\begin{center}
	\parbox{0.9\textwidth}{
		{\bf Working Rule 3:} Numerator factors in a duality-satisfying presentation of an amplitude need to have the same algebraic properties as the color factors. This includes symmetry properties as well as obeying two- and three-term identities.}
\end{center}
\noindent As discussed in \sect{DualitySection}, this rule ensures that the gravity theory obtained through double copy is 
invariant under linearized diffeomorphisms. 
If there exists (massive) vector fields that transform in non-adjoint representations, additional gauge-group Lie algebra relations are needed 
to guarantee that gauge invariance aside from Jacobi and commutation relations. The same relations should be imposed on the kinematic numerators 
for all fields that transform in the same representations as the vectors. For some classes of constructions, it will be convenient to consider a slight variant of Working Rule 
3 which instructs to impose the algebraic properties of the color factors of one theory on the numerators of the other theory entering the double-copy construction (and vice versa).
\\

Finally, we  need a procedure for consistently pairing representations in the two gauge theories 
when we substitute color factors with numerator factors following the double-copy prescription. 
A priori, several choices are possible. However, the following criterion is convenient, elegant and easy to implement:

\begin{center}
	\parbox{0.9\textwidth}{
		{\bf Working Rule 4:} Each state in the double-copy (gravitational) theory corresponds to a gauge-invariant bilinear of gauge-theory states. 
		For this to be possible, we will identify the  gauge groups of the two theories entering the construction.}
\end{center}

\noindent A concrete consequence of this rule is that gauge-theory states in the adjoint representation will double copy among themselves, but not with states in matter non-adjoint representations. Similarly, states in two matter representations will be combined only when the tensoring of the representations includes a singlet. Considering general graphs, two numerators will be combined only
when Working Rule 4 is satisfied by each internal and external line. We will see that this requirement is essential for preventing the gravity from the double copy from having too many gravitini. 

The space of all possible gauge theories is quite vast (though perhaps not quite as vast as that of gravitational theories).
The purpose of the working rules we laid out is to restrict this space to a subset which is sufficiently large to capture a considerable 
number of theories and yet sufficiently small to allow a thorough analysis.
It is not difficult to enlarge it by relaxing some of the rules. We emphasize that many  gauge theories which might be naively rejected as unphysical, 
such as theories with ghost fields, may be admissible---even in some sense necessary---from a double-copy perspective.
This is because, through double copy,  gauge-theory data is deconstructed and reassembled in a highly-nontrivial way and undesirable features
of gauge theories can be rendered harmless by this process.

The rules stated in this section should be slightly modified when constructing not gravitational. In this case, 
the gauge group should be replaced by a global symmetry group in the theories entering the construction.     

\subsection{Tools for extensions \label{sec-zoo-tols}}

Having established the general rules of the game, we will now analyze particular examples. We start by considering a YM-scalar theory with only adjoint fields and trilinear cubic couplings \cite{Chiodaroli2014xia}. Its Lagrangian can be written as
\begin{eqnarray}
{\cal L}_{{\rm YM}+\phi^3} &=&-\frac{1}{4}F_{\mu\nu}^{\ha}F^{\mu\nu \ha}+\frac{1}{2}(D_\mu\phi^{A})^{\ha} (D^\mu\phi^{A})^{\ha}
- \frac{g^2}{4} f^{\ha \hb \he} f^{\he \hc \hd}\phi^{A \ha}\phi^{B \hb}\phi^{A \hc}\phi^{B \hd}
\no \\
&&
\null + \frac{1}{3!}\lambda g F^{ABC} f^{\ha \hb \hc} \phi^{A \ha}\phi^{B \hb}\phi^{C \hc} \,.
\label{YMscalar}
\end{eqnarray}
The indices $\ha,\hb,\hc$ are gauge-group adjoint indices. $A,B,C= 1, \ldots, n$ are global indices carried by the scalars.\footnote{In this section, we frequently use hatted indices for gauge-group indices of the gauge theories that enter the double copy, to help distinguish them from global indices and gauge indices that appears in gravitational theories. } The theory has a $SO(n)$ global symmetry which 
is broken by the trilinear couplings to the subgroup preserved by the $F^{ABC}$ tensor. 
Field strengths and covariant derivatives are
\begin{eqnarray}
F_{\mu\nu}^{\hat a}&=& \partial_\mu A_\nu^{\hat a} - \partial_\nu A_\mu^{\hat a} 
+ g  f^{\hat a \hat b \hat c}A^{\hat b}_\mu A^{\hat c}_\nu \no \,,\\
(D_\mu\phi^{A})^{\hat a} &=& \partial_\mu \phi^{ { A \hat a}}  
+ g  f^{\hat a \hat b \hat c}A^{\hat b}_\mu \phi^{{ A} \hat c}\,. 
\end{eqnarray}
To understand the constraints imposed by \ck duality on the parameters of this theory we first analyze the four-scalar amplitudes. 
There is a clean separation between the contribution from the trilinear scalar coupling and the one from gluon exchange (the latter including the contact term).
After a short calculation, the $s$-channel numerator can be written as
\begin{equation}
n_s = \delta^{AB} \delta^{CD} (t-u) - (\delta^{AC} \delta^{BD}-\delta^{AD} \delta^{BC} ) s - \lambda^2 F^{ABE}F^{ECD}\,, 
\end{equation}
while the other numerators can be obtained by relabeling the external lines.
The three corresponding color factors obey standard Jacobi relations; imposing the duality between color and kinematics then results in the condition
\begin{equation}
\lambda^2 (F^{ABE}F^{ECD} + F^{BCE}F^{EAD} + F^{CAE}F^{EBD} ) = 0  \,.
\label{globalJacobiId}
\end{equation}
The $\lambda^0$ part of the numerator factors satisfies the duality automatically. This follows from the YM-scalar theory with $\lambda =0$ 
being the dimensional reduction of a pure YM theory in higher dimension, which is known to satisfy the duality at arbitrary multiplicity. 
At order $\lambda^2$,  the kinematic Jacobi relations imply that $F^{ABC}$-tensors must themselves obey Jacobi relations. 
This implies that they can be regarded as the structure constants of some global group which is unrelated to the gauge group.
What remains to be done is to consider amplitudes involving vectors and amplitudes at higher points. 
It turns out that no further constraint on the theory appears. 
It has been explicitly checked that the Lagrangian obeys \ck duality up to at least six points \cite{Chiodaroli2014xia}.

A second example which we review in detail is YM theory with complex scalars in a matter representation \cite{Chiodaroli2015rdg}; an analogous 
example involving matter fermions was discussed in \sect{DualitySection}. For such a field content, trilinear couplings are forbidden by gauge symmetry;
the first possible scalar self-interaction is quartic, so the Lagrangian is
\begin{equation}
{\cal L}_{\text{scalar}} =  D_{\mu} \overline{\varphi} D^{\mu}\varphi - a {g^2 \over 2} (\overline{\varphi} t^{\ha} \varphi)(\overline{\varphi} t^{\ha} \varphi) \,, \label{L-matterscalar}
\end{equation}
where $\varphi$ is a complex scalar, $t^{\ha}$ are representation matrices and $a$ is a constant. 
Products are understood in the sense of matrix 
multiplication, as representation indices are not displayed explicitly to avoid cluttering the expression. 
The two-scalar two-gluon amplitude in this theory is\footnote{We use the slightly-nonstandard notation, \eg ${\cal A}_n\big(1 \Phi_1 , \dots  , n \Phi_n \big)$, which displays explicitly the external states.}
\begin{eqnarray}
{\cal A}_4\big(1\overline{\varphi}^{{\hat \imath}},2\varphi_{{\hat \jmath}},3 A^{\ha}, 4 A^{\hb} \big) &=& i g^2 
\left\{ \left( {4 (\varepsilon_4 \cdot k_1) (\varepsilon_3 \cdot k_2) + t (\varepsilon_3 \cdot \varepsilon_4) \over t} (t^{\ha} t^{\hb})_{\hat \jmath}^{\ {\hat \imath}} + (3\leftrightarrow 4) \right) \right. +  \no \\
&& \hskip -65pt \left.  i { 4 (\varepsilon_3 \cdot k_1) (\varepsilon_4 \cdot k_2)- 4 (\varepsilon_4 \cdot k_1) (\varepsilon_3 \cdot k_2)  + (u-t) (\varepsilon_3 \cdot \varepsilon_4) \over s} f^{\ha \hb \hc} (t^{\hc })_{\hat \jmath}^{ \ {\hat \imath}} \right\}   , 
\end{eqnarray}
where we have displayed explicitly the gauge representation indices ${\hat \imath} , {\hat \jmath}$.
As a consequence of the commutation relation for the group generators, the color factors obey a three-term identity,
\begin{equation}
[t^{\ha}, t^{\hb}] = i f^{\ha \hb \hc} t^{\hc} \qquad \rightarrow \qquad c_t-c_u = c_s \,.
\end{equation}
It is easy to verify that the same identity is automatically satisfied by the numerators in the above amplitude,
\begin{equation}
n_t-n_u= n_s \,.
\end{equation}
This is possibly the simplest nontrivial example of the duality between color and kinematics for theories with non-adjoint fields. 
While \ck duality for the two-scalar two-gluon amplitude is a rather straightforward generalization of the case of amplitudes with fields in the adjoint representation, the four-scalar amplitude exposes new subtleties. This amplitude is
\begin{eqnarray}
{\cal A}_4\big(1\overline{\varphi}^{{\hat \imath}}, 2\overline{\varphi}^{{\hat \jmath}},3\varphi_{\hat k}, 4\varphi_{\hat l}  \big) &=& 
i g^2 \left\{ {s-u-a t \over t} (t^{\ha})^{\ \hat \imath}_{ \hat l} (t^{\ha})^{\ \hat \jmath}_{ \hat k} + (3\leftrightarrow 4) \right\}  \,.
\end{eqnarray}
Due to the scalar being complex, the amplitude involves only two terms; in principle, we may consider imposing the extra identity
\begin{equation}
(t^{\ha})^{\ \hat \imath}_{  \hat l} (t^{\ha})^{\ \hat \jmath}_{  \hat k} - (t^{\ha})^{\ \hat \jmath}_{  \hat l} (t^{\ha})^{\ \hat \imath}_{ \hat k} = 0 \,. \label{2term-identity}
\end{equation} 
However, this identity is not satisfied except for special gauge groups and representations, so it would seem that our Working Rule 3 does not compel us to impose (\ref{2term-identity}) in the general case.
At the same time, the numerator factors can easily obey the corresponding two-term kinematic identity if we fix $a=1$. 
Whether of not this choice should be made depends on the situation in which the kinematic numerators are used.
For example, we might choose to use this theory in a double-copy construction that involves massive $W$ fields in a matter representation 
(we will see that this is required, for example, for constructing Higgsed supergravities). 
In these cases, the spontaneously-broken gauge symmetry results in Ward identities that can be satisfied only if the massive vectors belong 
to specific representations for which color factors obey additional relations which are of the form (\ref{2term-identity}). 
Hence, it will be appropriate to impose two-term identities on the numerators of the {second} gauge theory entering the double copy. 
In contrast, whenever (\ref{2term-identity}) is not necessary for deriving some Ward identity in the gravity theory, there is no particular 
reason for imposing the corresponding numerator identity. 

So far, we have presented two examples of theories which obey \ck duality. In some cases, it is sufficient to write down simple gauge theories, 
verify that they obey the duality up to at least a certain multiplicity, and feed the corresponding numerators in the double-copy 
apparatus. However, this approach quickly becomes inconvenient as the number of matter representations increases. 
Hence, we would like to have systematic tools for obtaining more general theories which obey the duality from simpler ones. 
These tools will be reviewed in the next three subsections.

\subsubsection{Breaking representations into pieces \label{break}}

A first step for generating theories 
with fields transforming in matter representations in a way that preserves the duality is to start 
from the adjoint representation of a larger gauge group and decompose it into representations of a subgroup. 
This amounts to splitting the adjoint index of the larger groups $\hA$ as
\begin{equation}
\hA  \; \rightarrow \; (\ha, \haa_1 , \ldots , \haa_p) \,,
\end{equation}
where $\ha$ is the adjoint index of the smaller subgroup and $\haa_1, \ldots, \haa_p$ are indices of other representations. 
While it is always possible to choose them to correspond to irreducible representations, we will not do so here.
The structure constants of the original gauge group are broken down as follows:
\begin{eqnarray}
\{ f^{\hA \hB \hC} \} \; \rightarrow \; \{f^{\ha \hb \hc}, f^{\ha\haa_i \hbb_i}, f^{\haa_i \hbb_j \hgg_k} \} \,.
\end{eqnarray}
Here $f^{\ha \hb \hc}$ are the structure constants of the unbroken subgroup, $f^{\ha \haa_i \hbb_i}$ give the representation matrices of the $i$-th 
matter representation 
and    $f^{\haa_i \hbb_j \hgg_k}$ give Clebsch-Gordan coefficients for representations $i,j$, and $k$. We note that $f^{\ha \hb \haa}= 0$ 
from closure of the algebra of the unbroken gauge group. The notation above suggests that we have assumed the matter representations 
above to be real; the complex case can be treated analogously by introducing a pairing between some representations $i,j,k$ and their 
conjugate, denoted as $\bar \imath, \bar \jmath, \bar k$. The breaking of the adjoint representation acts in the following way on color 
and numerator factors:

\begin{figure}[tb]
	\begin{center}
		\def\widthfig{0.17\textwidth}
		\begin{eqnarray*}
			\col{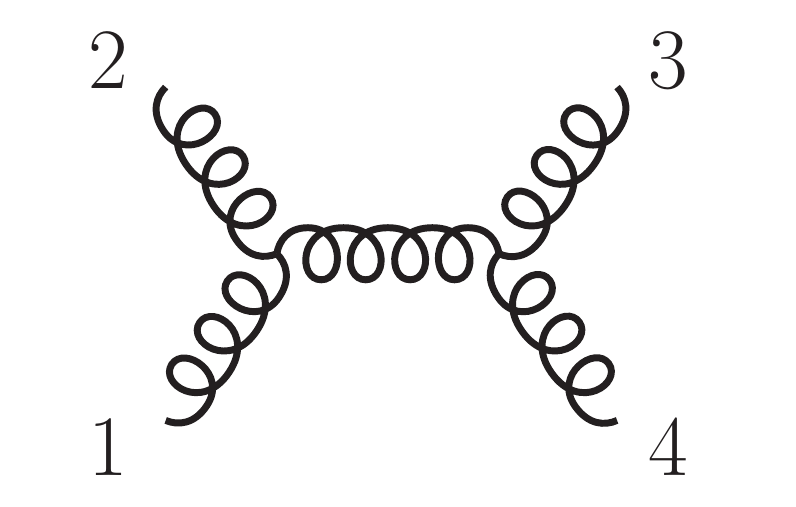}
			& \quad \rightarrow \quad &
			\left\{  \col{figs/diagbreak1}  \,, \ 
			\col{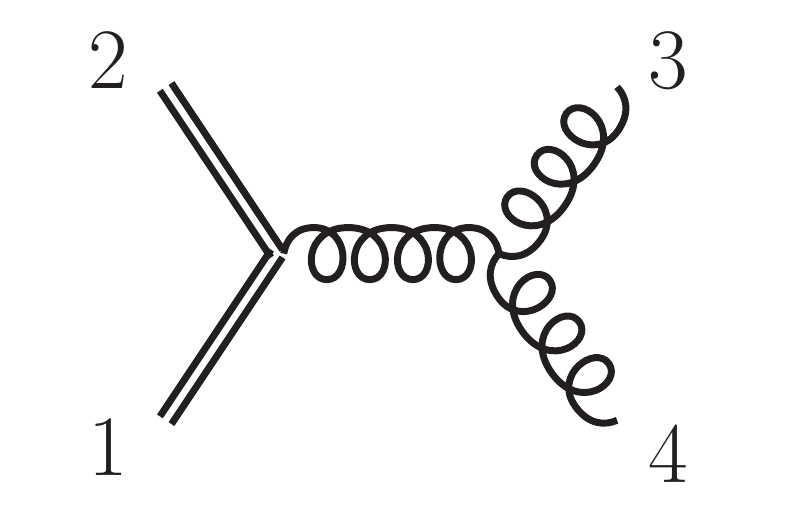} \right. \,, \ \\
			&&  \left. \col{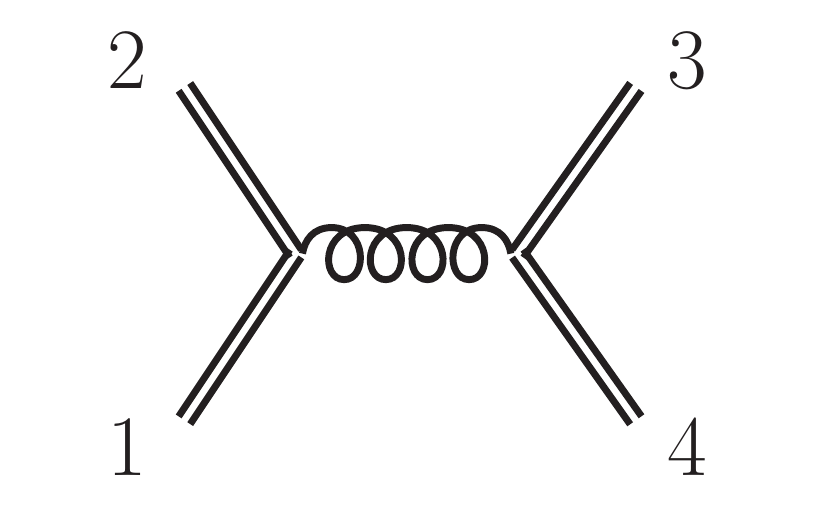} \,, \ \col{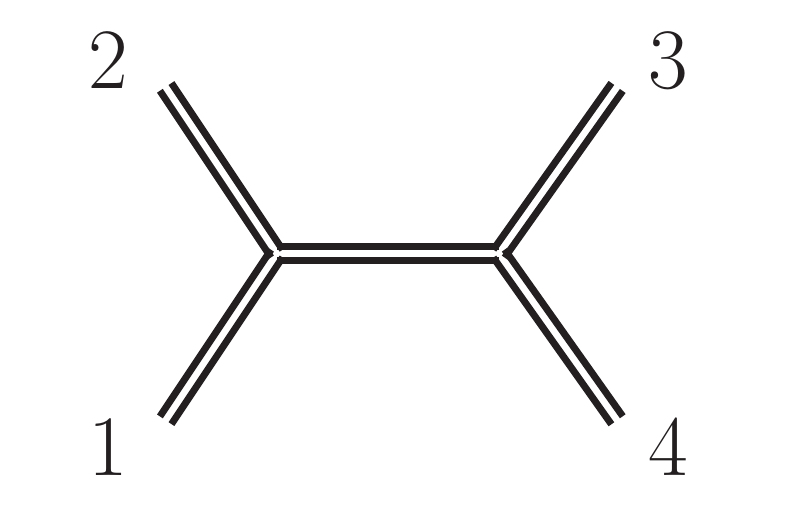} \,,  \cdots    \right\}
		\end{eqnarray*}	
	\end{center}
	\caption{Breaking of an adjoint representation into representations of a smaller subgroup. Curly lines denote the adjoint representation of the smaller group; 
		double lines denote matter representations.
 \label{figbreak}}
\end{figure}

\begin{itemize}
	\item Color factors are split into different pieces according to the representations carried by internal and external lines (see \fig{figbreak}); 
	color identities are preserved by this operation, but one needs to take into account that some color factor may vanish upon direct evaluation. \\
	
	\item Numerator factors are unchanged. Graphs with the same topology but  different representation labels will inherit the same 
	numerator factors as the original graphs of the unbroken theory. Whenever numerators obey a three-term identity in the unbroken theory, 
	the identity will be inherited by the broken theory. 	
\end{itemize} 
Two-term identities of the form (\ref{2term-identity}) deserve a more detailed discussion. 
Before decomposition into representations of a subgroup, color factors obey the standard Jacobi relations, which at four points can be written as 
\def\widthfig{0.18\textwidth}
\begin{equation}
\col{figs/diagbreak1} - \col{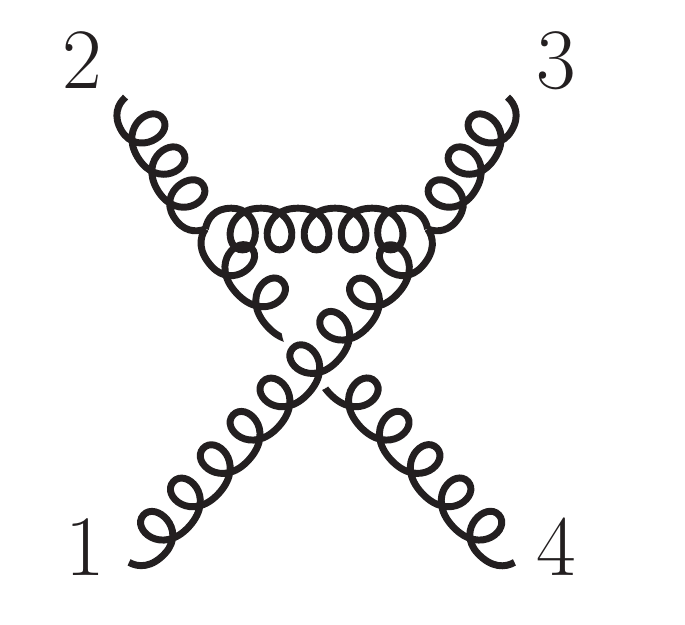}  =
\, \col{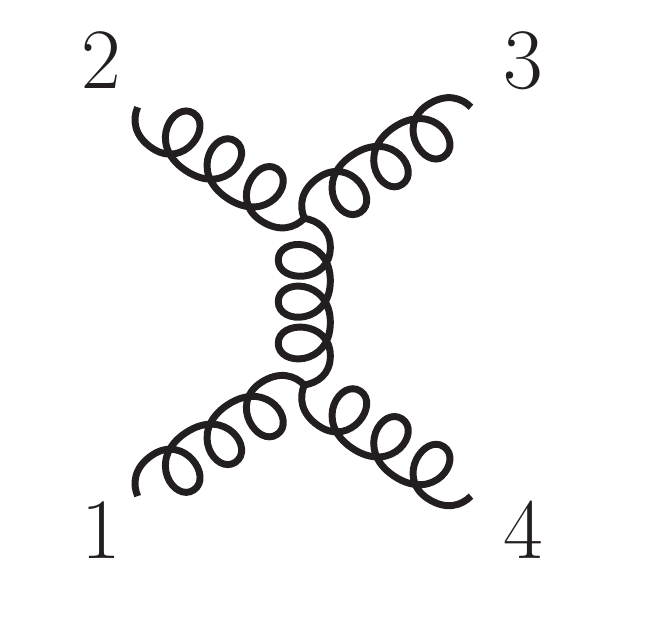} .
\end{equation}
If we consider the case in which the initial adjoint representation is broken into three pieces (adjoint, a single complex matter representation, its conjugate), 
the corresponding identity for external matter  
is
\begin{equation}
\col{figs/diagbreak3}  =
\col{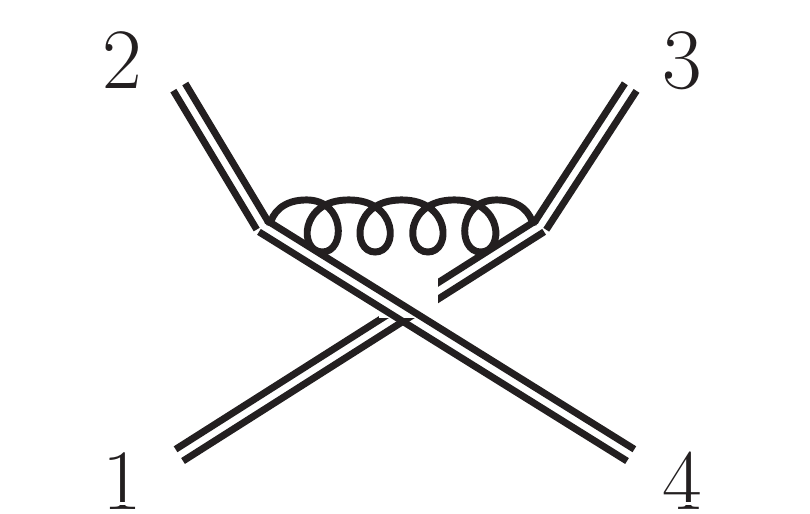}   ,
\end{equation}
because the structure constant does not contain a component with two indices in the same complex representation.
Hence, the original three-term color identity has collapsed into a two-term identity.
However, the decomposition of color factors with respect to a subgroup does not affect the kinematic numerators so, if the original theory 
obeys \ck duality, the numerators still obey three-term kinematic identities after the decomposition.
In other words, a nonvanishing numerator is associated to a vanishing color factor. 
We will see that this will require extra care, \eg in the construction of Higgsed supergravities in \sect{Higgsed}.
Note that a nonzero kinematic numerator can be associated with a vanishing color factor also in theories with only adjoint fields, as we shall see 
in \sect{ExamplesSection}.

\subsubsection{Field-theory orbifolds\label{SecOrbifold}}

If we consider a gauge theory which possesses a certain number of global/flavor symmetries (which may include the $R$ symmetry), 
it is always possible to  truncate it to its sector which is invariant under the combined action of 
some elements of the global and  gauge groups. Specifically, a generic 
adjoint field $\Phi$ of the original theory will transform as 
\begin{equation}
\Phi \rightarrow R F g \Phi g^\dagger \,,
\label{fieldaction}
\end{equation} 
where \textit{}$g$ is the gauge-group element and $R,F$ are the corresponding elements of the $R$-symmetry and global-flavor group ($R$-symmetry and global indices are not explicitly displayed). It is convenient to consider elements $(g,R,F)$ which belong to a discrete subgroup $\Gamma$ of the symmetry group of the theory we are considering, 
that is we have elements $g,R,F$ such that $g^k=I, F^k=I, R^k=I$ for some $k$.\footnote{Other subgroups can also be considered.} 
Theories obtained with this construction  are referred to  as field-theory orbifolds in the literature \cite{Bershadsky1998cb}.   
Given $(g,R,F)$ above, we can immediately write a projector
\begin{equation}
{\cal P}_\Gamma \Phi = {1 \over |\Gamma|} \sum_{(g,R,F) \in \Gamma} R F g \Phi g^\dagger \,,\label{projectorOrbifold}
\end{equation} 
where $|\Gamma|$ denotes the rank of $\Gamma$.
It is easy to verify that this projector sets to zero all components of $\Phi$ which are not invariant under $\Gamma$.

To give a simple example, we start from $\cN=4$ SYM theory with $SU(2N)$ gauge group and consider an $\Gamma={\mathbb{Z}}_2$ orbifold with generators
\begin{equation}
r = \text{diag} \big(1,1,-1,-1 \big) \,, \qquad  g = \left( \begin{array}{cc} I_N & 0  \\ 0 & -I_N \end{array} \right) .
\label{Z2Action}
\end{equation}
The matrix $r$ gives the action of the unique nontrivial generator of $\mathbb{Z}_2$ on the fundamental $R$-symmetry indices; the action 
of $\mathbb{Z}_2$ on other representation of the $R$ symmetry can be obtained by taking tensor products of $r$.
In this case, it is convenient to represent the action 
of the projector on the components of a on-shell $\cN=4$ superfield ${\cal V}^{\hat A}_{\cN=4}$ which is written as \eqref{Neq4multiplet}.
The part of this superfield which  survives the 
orbifold projection \eqref{projectorOrbifold} is
\begin{eqnarray}
{\cal V}^{\hA}_{\cN=4} &\rightarrow& A^{\ha}_+ + \eta_i \lambda^{\ha i}_+ + \eta_r \lambda^{\haa r}_+  
+ \eta_1 \eta_2 \phi^{\ha 12} + \eta_i \eta_r \phi^{\haa ir} \no \\
 &&
\null +  \eta_3 \eta_4 \phi^{\ha 34}
+ \eta_1 \eta_2 \eta_r \lambda^{\haa r}_- +  \eta_i \eta_3 \eta_4 \lambda^{\ha i}_- + \eta_1 \eta_2\eta_3\eta_4 A^{\ha}_-  \,,
\end{eqnarray} 
where $i,j=1,2$ and $r,s=3,4$. The gauge-group indices $\ha,\hb$ and $\haa, \hbb$ run over the (reducible) $SU(N)\times SU(N) \times U(1)$ 
adjoint representation and the bi-fundamental representation, respectively.
This result can be organized in $\cN=2$ on-shell superfields as
\begin{equation}
{\cal V}^{\hA}_{\cN=4} \rightarrow {\cal V}^{\ha}_{\cN =2} + \eta_r {\Phi}_{\cN=2}^{\haa r}  
+ \eta_3\eta_4 \overline{\cal V}_{\cN=2}^{\ha} \,, \qquad r=3,4 ,
\end{equation}
where ${\Phi}_{\cN=2}$ is the on-shell hypermultiplet superfield. Hence, we see that the theory resulting 
from the orbifold projection \eqref{projectorOrbifold} with $\Gamma=\mathbb{Z}_2$ acting as \eqref{Z2Action} is an $\cN=2$ SYM theory with gauge group 
$SU(N)\times SU(N) \times U(1)$ and one matter hypermultiplet in the bi-fundamental representation.

\begin{homework}
	Work out spectrum and on-shell superfield organization for the $\mathbb{Z}_2$ orbifold projection of $\cN=4$ SYM theory with generators
	\begin{equation}
	r = \text{diag} \big(-1,-1,-1,-1 \big) \,, \qquad  g = \left( \begin{array}{cc} I_N & 0  \\ 0 & -I_N \end{array} \right) .
	\end{equation} 
	What is the residual supersymmetry?
\end{homework}

\begin{homework}
	Formulate an orbifold projection of $\cN=4$ SYM preserving $\cN=1$ supersymmetry. Work out the multiplet structure of the 
	on-shell superfields.
\end{homework}

Field-theory-orbifold amplitudes are constructed through a set of Feynman rules which are obtained directly by taking the Feynman 
rules of the parent theory and dressing both internal and external lines with projectors of the form (\ref{projectorOrbifold}). 
For a tree-level amplitude, one can use invariance of the propagators and vertices under global and gauge symmetries to move all projectors from internal to external lines. 
The result is that all tree-level amplitudes of a theory constructed as an orbifold can be obtained from the amplitudes of the parent theory by 
inserting projectors on the external legs or, alternatively, by ensuring that the asymptotic states 
are invariant under the orbifold group. Particularly relevant to us, 
this property has the consequence that all numerator relations of the parent theory are preserved by the orbifold construction \cite{Chiodaroli2013upa}.

\begin{homework}
	Consider $\cN=4$ SYM theory with $SU(2N+1)$ gauge group. Show that the projection with orbifold group generators
	\begin{equation}
	r = \text{diag} \big(1,1,-1,-1 \big) \,, \qquad  g = \left( \begin{array}{cc} -I_{2N} & 0  \\ 0 & 1 \end{array} \right) ,
	\end{equation} 	 
	yields a $\cN=2$ theory with a hypermultiplet in the fundamental representation.
\end{homework}

The reader may wonder whether there is a straightforward way to extend this result to loop level. At one loop, using the symmetries of propagators and vertices, projectors can be removed from all but one internal line (which can be chosen freely). Additionally, particular classes of loop-level amplitudes (for example, planar amplitudes in the large-$N$ limit) are inherited from the parent theory (for a subset of so-called regular orbifolds) \cite{Bershadsky:1998mb,Bershadsky1998cb}. 
Loop-level amplitudes can of course be constructed from tree-level ones with unitarity methods. For general amplitudes and choices of orbifold groups, the properties 
at loop-level will not be directly related to the ones of the parent theory and the orbifold construction will be used to obtain tree-level building blocks to be employed with unitarity methods. This construction has been instrumental, for example, in the study of one-loop amplitudes for supergravities that can be embedded in the $\cN=8$ maximal theory \cite{OneLoopSusy,Chiodaroli2013upa}.

\subsubsection{Masses as compact momenta \label{masses}}

Once representations of a larger gauge group are broken into smaller pieces, it is in principle possible to introduce nonzero masses for some of the fields. At the same time, if we intend to consider more general theories of gravity coming from the double copy, 
we need some procedure for generating massive states (for which the most natural choice  is the Higgs mechanism).    
If we consider gauge theories that can be written in higher dimension, a straightforward way to create mass terms from an amplitude perspective 
consists of assigning to some of the fields momenta in one of the extra (compact) dimensions.

Our starting point is to consider adjoint fields in a higher-dimensional theory which are written as 
\begin{eqnarray}
A^{ \mu \hA }(\vec{x},x_{D+1}) \Big|_{D+1}&=& \big(e^{i x_{D+1} m}\big)^{\! \hA \hB} A^{  \mu \hB} (\vec{x}) \,, \no \\
\phi^{a \hA }(\vec{x},x_{D+1}) \Big|_{D+1}&=& \big(e^{i x_{D+1} m}\big)^{\! \hA \hB} \phi^{  a \hB} (\vec{x}) \,, \quad a = 1,\ldots,n \, ,  \label{compactmom}
\end{eqnarray} 
where $m^{\hA \hB}$ is a mass matrix with adjoint indices $\hA, \hB$, $\vec{p}$ is the $D$-dimensional momentum and $p_{D+1}$  is the momentum in the $D+1$ internal direction.
If the mass matrix vanishes, this is equivalent to ordinary dimensional reduction. The condition above can also be implemented in position space through the differential equation
\begin{equation}
\partial_{D+1} 
\left( \begin{array}{c} A^{\mu \hat A} \\  \phi^{a \hat A} \end{array}  \right) \Big|_{D+1}
=  i \,m^{\hat A \hat B} \left( \begin{array}{c} A^{\mu \hat B} \\  \phi^{a \hat B} \end{array}  \right).
\label{diffeqn}
\end{equation}
Introducing this mass term has the effect of breaking the adjoint representation of the gauge group into various representations with respect to which $m^{\hA \hB}$ is block-diagonal. We choose $m^{\hA \hB}$ to be given by 
\begin{equation}
m^{\hA \hB} = ig V f^{\hat 0 \hA \hB} \,.
\end{equation}
Fields that commute with the gauge-group generator $t^{\hat 0}$ will not have a mass since that implies that $f^{0 \hat A\hat B}$ vanish. 
We can now explicitly show that the kinetic term of the scalars in $(D+1)$ dimensions is identical to a kinetic term in $D$ dimensions plus a $\phi^4$-term in which a scalar acquires a VEV:
\begin{eqnarray}
\frac{1}{2}\big(  {\cal D}_\mu \phi^{a \hA} \big)^2\Big|_{D+1} & \rightarrow &\frac{1}{2}\big(  {\cal D}_\mu \phi^{a \hA} \big)^2-\frac{1}{2}\big(i \, m^{\hA \hB} \phi^{a \hB}+ g f^{\hA \hB \hC}\phi^{0 \hB} \phi^{a \hC} \big)^2 \nn \\ 
&&= \frac{1}{2}\big(  {\cal D}_\mu \phi^{a \hA} \big)^2+\frac{g^2}{2}{\rm tr}\big( [V t^0 + \phi^0, \phi^a]^2 \big)\,,
\end{eqnarray}
where we have renamed the gauge field in the internal direction, $A^{\hat A}_{D+1} \rightarrow \phi^{0 \hat A}$ (the global index $a$ does not include $a=0$). 
We then inspect  the $(D+1)$-dimensional vector-field kinetic term, 
\begin{eqnarray}
-\frac{1}{4}\big({\cal F}^{\hat A}_{\mu \nu}\big)^2 \Big|_{D+1}& \rightarrow &-\frac{1}{4}\big({\cal F}^{\hat A}_{\mu \nu}\big)^2+\frac{1}{2}\big(\partial_{\mu}\phi^{0 \hat A}- i\,m^{\hat A \hat B} A_\mu^{\hat B}+g f^{\hat A \hat B \hat C} A_\mu^{\hat B}\phi^{0\hat C}\big)^2 \nn \\
&&=-\frac{1}{4}\big({\cal F}^{\hat A}_{\mu \nu}\big)^2+\frac{1}{2}\big(({\cal D}_{\mu}\phi^{0 })^{\hat A}- i\, m^{\hat A \hat B} A_\mu^{\hat B}\big)^2 \nn \\
&&=-\frac{1}{4}\big({\cal F}^{\hat A}_{\mu \nu}\big)^2+\frac{1}{2}\big(({\cal D}_{\mu}\phi^{0}+{\cal D}_{\mu}\langle \phi^0 \rangle)^{\hat A}\big)^2\,.
\end{eqnarray}
This term is identical to the $D$-dimensional vector-field kinetic term plus the kinetic term for $\phi^{0}$ in the presence of a VEV
\begin{equation}
\langle \phi^0 \rangle =V t^{\hat 0}\,. \label{vev}
\end{equation} 
Adding the quartic potential terms for the  scalars, one sees that the $(D+1)$-dimensional massless (S)YM Lagrangian in the presence of a compact momentum of the form (\ref{compactmom}) is indeed equivalent to a  spontaneously-broken $D$-dimensional SYM Lagrangian with VEV given by (\ref{vev}). 
Strictly speaking, we have shown that this procedure works only in the presence of a 
quartic potential generated by dimensional reduction of a higher-dimensional pure (S)YM theory. 
The case of more general scalar potentials need to be considered separately.
The argument in this subsection gives a prescription for finding amplitudes of theories with fields 
becoming though Higgs mechanism in terms of higher-dimensional massless amplitudes. If the higher-dimensional theory obeys \ck duality, the massive amplitudes will inherit the same algebraic properties \cite{Chiodaroli2015rdg}.  
BCJ amplitude relations with massive particles were also derived in \cite{Naculich:2014naa} using the CHY formalism.

\subsubsection{Identifying the right supergravity\label{sec-zoo-id}}

At this point, we have developed some basic techniques to start from the amplitudes of an arbitrary
theory which is known or can be shown to obey the duality between color and kinematics 
and generate the amplitudes of more involved theories, which may include fields in non-adjoint representations and mass terms from the Higgs mechanism, 
in a way that preserves numerator relations.
In principle, we can use the numerators from various theories  obtained with this procedure for producing amplitudes through the double-copy technique. 
As discussed in \sects{DualitySection}{GravitySymmetriesSection}, these
amplitudes will obey the Ward identities related to invariance under linearized diffeomorphisms and hence should be the amplitudes from some gravitational theory. 
Identifying the precise theory, however, is not always straightforward. 
In principle, one could consider a generic Lagrangian involving the Einstein-Hilbert term (or, in case of conformal gravity, some Weyl${}^2$ gravitational action) and arbitrary matter interactions. Up to terms which vanish due to the equations of motion, this Lagrangian can be fixed order by order by comparing its amplitudes  with the ones from the double-copy method. 
In practice, the implementation of this program is limited by one's desire to evaluate higher-point tree amplitudes.
For many theories however, minimal information about symmetries and lower-point interactions 
can be sufficient for identifying the theory completely and, in principle, for writing down its Lagrangian (with some help from the relevant supergravity literature).  
More specifically:
\begin{itemize}
	\item Symmetry considerations are sufficient for identifying supergravities with extended $\cN \geq 4$ supersymmetry and particular theories with 
	reduced supersymmetry that can be viewed as truncations of more supersymmetric theories \cite{N46Sugra,OneLoopSusy,Anastasiou2015vba}. 
	Such considerations can also be sufficient to formulate constructions for theories with homogeneous scalar manifolds, with some residual freedom 
	that needs to be fixed with minimal information on their interactions \cite{Chiodaroli2015wal,Anastasiou2017nsz}.
	\item Very broad classes of Maxwell-Einstein supergravities with $\cN=2$ supersymmetry which can be lifted up to at least five spacetime dimensions can be uniquely specified 
	by their three-point interactions (specifically, three-vector amplitudes in five-dimensions). In a sense, these theories constitute a natural testing ground for double-copy constructions with reduced supersymmetry \cite{Chiodaroli2014xia,Chiodaroli2015wal}.
	\item More generally, there exist amplitudes which capture physical features of the desired supergravity theory. For example, YME theories or gauged supergravities with nonabelian gauge group will possess non vanishing three-point amplitudes between three gluons \cite{Chiodaroli2014xia}. Gauged supergravities will have nonvanishing amplitudes between two gravitini and one vector \cite{Chiodaroli2017ehv}. Knowledge of these amplitudes can either allow identification of the theory or point to the general class to which the theory belongs.
	\item Some theories are characterized in terms of their soft limits. These include e.g. theories with homogeneous target spaces \cite{ArkaniHamed2008gz,Chiodaroli2015wal}, the NLSM and some of its extensions \cite{Cachazo2014xea,CheungUnifyingRelations,Cachazo2016njl}, and the special Galileon theory \cite{Cheung2016prv}.  \\	
\end{itemize}

\subsection{Examples \label{sec-zoo-examples}}

We now proceed to discussing some examples. A list of the main double-copy constructible theories at the time 
of this writing can be found in Tables \ref{table-zoology-ungauged}, \ref{table-zoology-gauged}, and \ref{table-zoology-nongrav}.

\subsubsection{Theories with $\cN\geq 4$ supersymmetry}

Pure supergravities with ${\cal N}= 4$ and $8$ have been originally formulated from a Lagrangian perspective in \cite{Cremmer:1977tt, Das:1977uy} 
and \cite{deWit:1977fk, Cremmer:1979up}, respectively, while the ${\cal N}=5$ and ${\cal N}=6$  Lagrangians were obtained by 
truncation~\cite{Fischler:1979yk} from that of the ${\cal N}=8$ supergravity.
Amplitudes of theories with extended $\cN \geq 4$ supersymmetry  will be given by a double copy involving $\cN=4$ SYM theory 
together with a YM  or SYM theory. 
The possibilities are the following \cite{N46Sugra}:
\begin{eqnarray}
\cN = 8  \text{ supergravity}  &:& \quad  (\cN = 4 \text{ SYM}) \otimes (\cN = 4 \text{ SYM}), \no \\
\cN = 6  \text{ supergravity}  &:& \quad  (\cN = 4 \text{ SYM}) \otimes (\cN = 2 \text{ SYM}), \no \\
\cN = 5  \text{ supergravity}  &:& \quad  (\cN = 4 \text{ SYM}) \otimes (\cN = 1 \text{ SYM}), \no \\
\cN = 4  \text{ supergravity}  &:& \quad  (\cN = 4 \text{ SYM}) \otimes (\cN = 0 \text{ YM}). 
\label{DoubleCopySusy}
\end{eqnarray} 
All double copies above involve gauge theories with only adjoint fields. These are cases in which the symmetries of the desired supergravity single 
out the correct construction,
without any free parameters. Supergravities with $\cN>4$ are unique. For $\cN=4$ supergravities one can 
add matter in the form of $\cN=4$ vector multiplets, which correspond to adding adjoint scalars in the non-supersymmetric gauge theory.

Perturbative mass spectra and on-shell superfield structure of the theories listed above can be straightforwardly obtained 
from the on-shell superfields of the gauge theories. Alternatively, all theories with $\cN>4$ and some examples of $\cN=4$ theories 
can be seen as truncations of $\cN=8$ supergravity using a field-theory orbifold construction. 

$\cN=4$ supergravity has an alternative double-copy construction, in terms of two $\cN=2$ SYM theories coupled to hypermultiplets in matter 
representations. Apart from the mass spectra, it has been verified that tree-level and four-point one-loop amplitudes in the two realizations are the same, including anomalous amplitudes \cite{OneLoopSusy,JohanssonTwoLoopSusyQCD}.

We now look more in detail at the pure $\cN=4$ supergravity in four dimensions. The theory involves one complex scalar, whose asymptotic states are 
obtained by taking the double 
copy of  gauge-theory gluons with opposite polarizations. Geometrically, the scalar can be regarded as the complex coordinate of the coset space
\be 
\boldsymbol{\cal M}_{4D} =  {SU(1,1) \over U(1)} \,. 
\label{cosetNeq4}
\ee 
As we will discuss more in detail later, the fact that the scalar lives in an homogeneous space can be confirmed by checking the vanishing 
of the scalar soft limits at tree level.\footnote{In $\cN=4$ supergravity the single-soft-scalar limit no longer vanishes at one loop \cite{CarrascoN4Anomaly} 
due to an anomaly of the $U(1)$ symmetry in Eq.~\eqref{cosetNeq4}. Finite local counterterms can be used to restore this symmetry (at the expense of 
the other $SU(1,1)$ generators) \cite{BPRAnomalyCancel}. This counterterm also restores the vanishing single-soft-scalar limit.}
More explicitly, the bosonic part of the Lagrangian for pure $\cN=4$ supergravity has a relatively simple form,
\bea
e^{-1}{\cal L} &=& -\frac{R}{2} + {1 \over 4} {\partial_\mu \tau \partial^\mu \bar \tau \over ({\rm Im}\, \tau)^2} - {1 \over 4} {\rm Im}\,  \tau F^I_{\mu \nu} F^{I \mu \nu} - {1 \over 8} {\rm Re} \, \tau e^{-1}\, \epsilon^{\mu \nu \rho \sigma} F^I_{\mu \nu} F^I_{\rho \sigma} \no \\
& =& -\frac{R}{2} + {1 \over 4} \left( {\partial_\mu \tau \partial^\mu \bar \tau \over ({\rm Im}\, \tau)^2} + i \tau (F^{+I}_{\mu \nu})^2 - i \bar \tau (F^{-I}_{\mu \nu})^2  \right) ,
\label{Lag-N4}\eea
where $\tau=i e^{-\phi} +\chi$ is the dilaton-axion  scalar, $\tilde F^{I}_{\mu \nu}= (i/2) e \epsilon_{\mu \nu \rho \sigma}  F^{I\rho \sigma}$, $F^{\pm I}_{\mu \nu}=(F^{ I}_{\mu \nu} \pm \tilde F^{I}_{\mu \nu})/2$,  and $I=1,\ldots,6$ is an index running over the vector fields in the theory. Alternatively, the kinetic term for the scalars 
can be written with a Cayley parameterization of the form \eqref{NLSMaction}. 

In contrast to more supersymmetric settings, the double-copy construction for $\cN=4$ supergravities can be easily modified by adding extra adjoint scalars 
in the non-supersymmetric  gauge theory.  This can be done by considering a YM theory coupled to $N$ scalars, which is the reduction to four dimensions of
$D=(N+4)$ pure YM theory.\footnote{Recall that dimensional reduction is an operation which is known to preserve \ck duality. These theories will sometimes be denoted as YM${}_{\text{DR}}$.}
This theory is invariant under an $SO(N)$ symmetry, which is the subgroup of the $D$-dimensional Lorentz group transverse to four dimensions, $SO(1,3+N)\rightarrow SO(1,3)\times SO(N)$. Under this symmetry, the vector fields are inert while the scalars transform in the vector (fundamental) representation.
Since these scalars transform in the adjoint representation of the gauge group, they can be double-copied with the $\cN=4$ vector multiplet to yield 
$N$ vector multiplets in the supergravity theory. Their scalars transform in the $({\boldsymbol N}, {\boldsymbol 6})$ representation of $SO(N) \times SO(6)$, where latter factor is the $R$-symmetry group.\footnote{Vector fields in this theory are of two types: graviphotons, which are part of the gravition multiplet and transform in 
$({\boldsymbol 1}, {\boldsymbol 6})$ and vectors which are part of the additional vector multiplets, which transform as $({\boldsymbol N}, {\boldsymbol 1})$.}

In $\cN=4$ supergravity scalars fields outside the graviton multiplet parametrize a homogeneous space of the form 
\be 
\boldsymbol{\cal M}_{4D} = {G\over H} \,, 
\ee 
where the stabilizer group $H$ is the symmetry which is linearly realized and thus visible in amplitudes involving scalars. 
Thus, the symmetry which is manifest in the double-copy construction cannot be larger than $H$. In our case, we have 
$SO(N) \times SO(6) \subseteq H$. This suggests that the $6N$ vector multiplet scalars parametrize $SO(6,N)/(SO(6) \times SO(N))$
and, together with Eq.~\eqref{cosetNeq4}, that the $(6N+2)$ real scalars in the four-dimensional $\cN=4$ supergravity theory 
parametrize the symmetric space
\be
\boldsymbol{\cal M}_{4D}= {SO(6,N) \over SO(6) \times SO(N)} \times {SU(1,1) \over U(1)} \,.
\ee  

The double-copy construction for $\cN \geq 4$ supergravities can be
used to find expressions for amplitudes at one loop, which are
discussed in \sect{ExamplesSection}.  Beyond one loop, amplitudes in
extended supergravity theories have been the subject of intense
investigation, especially on their UV properties.  Lore has it that
all supergravity must diverge at a sufficiently high loop order. Is
this actually true or might there be surprises?  A variety of
multiloop calculations for $\cN \ge 4$ supergravity have been
carried out to analyze UV properties:
\begin{itemize}
\item Four-point amplitudes for pure $\cN=4$ supergravity have been shown to be UV-finite 
 at three loops and UV-divergent at four loops in four
 dimensions \cite{Bern:2012cd,Bern:2013qca,Bern:2014lha,N4GravFourLoop}.
 The four-loop UV-divergence appears to be related to a $U(1)$
 anomaly \cite{MarcusAnomaly,CarrascoN4Anomaly,BPRAnomalyCancel,Bern:2019isl}.
 Full one- and two-loop four-point amplitudes in $\cN=4$ supergravity are
 given in Refs.~\cite{N46Sugra, N46Sugra2}.
\item Four-point amplitudes for $\cN=5$ supergravity are finite at least through 
 four loops in four dimensions \cite{N5GravFourLoop}.  Despite various
 attempts, there is no standard symmetry
 explanation for the ``enhanced cancellations'' that lead to this
 improved UV behavior~\cite{NewKalloshFreedmanN5}.  
 See, however, Refs.~\cite{Kallosh:2018wzz, Gunaydin:2018kdz} for arguments suggesting that U-duality invariance may be ultimately  responsible.
 It is of
 considerable interest to settle the origin of these cancellations,
 and to know whether they continue to higher orders.
\item The complete two-loop four-point amplitude of $\cN=6$ supergravity may
   be found in Ref.~\cite{N46Sugra2}.  As yet there have not been any
   direct studies of the critical dimension of this theory at high loop
   orders, altough it follow from the calculations in $\cN=5$ supergravity that 
   divergences cannot appear before five loops. Standard symmetry considerations imply that divergences
   are delayed until at least five loops~\cite{Bossard:2010bd, NewKalloshFreedmanN5}.
\item UV properties of four-point amplitudes in $\cN=8$ supergravity have been analyzed in detail
    through five loops~\cite{UVFiveLoops}.  In contrast to the case of
    $\cN=5$ supergravity, $\cN=8$ supergravity at five loops does not
    appear have enhanced cancellations, but it is possible that this
    is an artifact of the fact that the analysis is carried out in the
    fractional critical dimension $D=24/5$, where from various
    considerations~\cite{Bjornsson:2010wm,Vanhove:2010nf} divergences
    are first expected to appear.  A proper study of this issue in the
    most interesting dimension $D=4$ requires a seven-loop
    computation, as suggested by symmetry
    considerations~\cite{Howe:1980th, Kallosh:1980fi, Beisert2010jx,
    Green:2010sp, Vanhove:2010nf, Bjornsson:2010wm,
    Bossard:2011tq}. The complete three-loop four-point and two-loop
    five-point amplitudes of $\NeqEight$ supergravity have
    been obtained~\cite{Chicherin:2019xeg, Henn:2019rgj}, starting
    from integrands constructed via the double copy~\cite{BCJLoop,
    FivePointN4BCJ}.  The construction of $\cN = 8$ one-, two- and
    three-loop integrands via the double copy is described
    in \sect{ExamplesSection}.
\end{itemize}

\subsubsection{Maxwell-Einstein theories with $\cN=2$ supersymmetry \label{jordan}}

In this section we discuss amplitudes in theories with  $\cN=2$ supersymmetry in four dimensions (eight supercharges). 
Theories of this type are no longer specified solely by 
their matter content. Hence, we need a strategy to conveniently classify the interactions consistent with $\cN=2$ supersymmetry.   
An efficient approach is  to focus on theories that can be uplifted to five dimensions. 
The Lagrangians for these theories have long been known explicitly \cite{Gunaydin1983bi,Gunaydin1984ak,Gunaydin1984nt,Gunaydin1986fg}.
Here we will write only the bosonic part of the Lagrangian:\footnote{For consistency with the rest of this review, this Lagrangian is written with a metric of mostly-minus signature, in contrast to most of the supergravity literature.}
\begin{equation}
e^{-1}\mathcal{L} \! = \!-\frac{1}{2}R- \frac{1}{4}
{\stackrel{\circ}{a}}_{IJ} F_{\mu\nu}^{I}F^{J\mu\nu} \!\! + \! \frac{1}{2}
g_{xy}\partial_{\mu}\phi^{x} \partial^{\mu}
\phi^{y} + \!  \frac{e^{-1}}{6\sqrt{6}}C_{IJK}
\varepsilon^{\mu\nu\rho\sigma\lambda}F_{\mu\nu}^{I}
F_{\rho\sigma}^{J}A_{\lambda}^{K} .  \label{Lbossugra}
\end{equation}
All vectors in the Lagrangian are taken to be abelian (YME theories will be discussed in \sect{YMESec}). The index $I=0,1,\ldots, n$ runs over the number of vectors in the theory with $I=0$ corresponding to the graviphoton.
$F_{\mu\nu}^{I}$ are the  
field strengths, while ${\stackrel{\circ}{a}}_{IJ}$ and $g_{xy}$ are functions of the physical scalars $\phi^x$ ($x=1,\ldots,n$). 
The key insight is that the  symmetric constant tensor $C_{IJK}$ 
is sufficient to specify the theory completely, \ie to fix all functions appearing in the two-derivative Lagrangian.
The formalism manifesting this feature introduces an auxiliary ambient space with coordinates $\xi^I$ and dimension 
equal to the number of vectors  in the theory which, together with the $C_{IJK}$-tensor, are used to define a cubic polynomial 
\begin{equation}
\mathcal{V}(\xi)\equiv C_{IJK}\xi^{I} \xi^{J} \xi^{K}\, .
\end{equation}
In turn, this is used to define a metric on the ambient space:
\begin{equation}\label{aij}
a_{IJ}(\xi)\equiv -\frac{1}{3}\frac{\partial}{\partial \xi^{I}}
\frac{\partial}{\partial \xi^{J}} \ln \mathcal{V}(\xi)\, .
\end{equation}
The scalar manifold $\boldsymbol{\cal M}_{5D}$  is defined as the hypersurface obeying the equation
\begin{equation}
{\cal V} (h)=C_{IJK}h^{I}h^{J}h^{K}=1\,, \qquad  \quad h^I = \sqrt{2 \over 3} \xi^I .
\label{prepotentialJordan}
\end{equation}
The functions ${\stackrel{\circ}{a}}_{IJ}(\phi)$ and $g_{xy}(\phi)$ which appear in the Lagrangian    are given by the  restriction of the ambient-space metric to $\boldsymbol{\cal M}_{5D}$ and the pullback to that surface of the ambient space metric, respectively:
\begin{equation}
{\stackrel{\circ}{a}}_{IJ}(\phi)=a_{IJ}\big|_{{\cal V}(h)=1}\; ; \qquad
g_{xy}(\phi) = \frac{3}{2}\left. \frac{\partial \xi^I}{\partial \phi^x}
\frac{\partial \xi^J}{\partial \phi^y} a_{IJ} \right|_{ {\cal V}(h) = 1} \,. \end{equation}
The functions appearing in the fermionic part of the Lagrangian can also be expressed in terms of the $C_{IJK}$-tensor.
Since the $C_{IJK}$-tensor can be obtained by inspecting three-point amplitudes,
$\cN=2$ Maxwell-Einstein theories in five dimensions are uniquely specified by their three-point interactions. 
This is in contrast to Maxwell-Einstein theories that only exist in four dimensions as well as theories with hypermultiplets.
It is in principle possible to compute amplitudes from the Lagrangian \eqref{Lbossugra} using Feynman rules. To this end one should first 
expand around a scalar-base point (\ie some background values for the scalar fields)  at which the scalar and vector kinetic terms are positive-definite. 
The quadratic terms should then be diagonalized in order to find the spectrum, the propagators, and the vertices.
For practical calculations, it is often convenient to reduce the theory to four dimensions and use the spinor-helicity formalism.

To identify the simplest supergravity theories we will utilize  symmetry considerations  together with 
minimal information on the trilinear interaction terms. 
A natural starting point is to consider a double copy of the form 
\be 
\cN = 2  \text{ supergravity}  : \quad  (\cN = 2 \text{ SYM}) \otimes (\cN = 0 \text{ YM}) \,, 
\no 
\ee
in which the non-supersymmetric theory is a pure $(4+n)$-dimensional YM theory reduced to four dimensions. Bosonic asymptotic states 
from the double copy are identified with those from the supergravity Lagrangian as follows~\cite{Chiodaroli2014xia}:\footnote{
The phase in the map between the asymptotic states from the supergravity Lagrangian and the ones from the double copy was chosen to match the phase conventions in the supergravity literature, see \eg \cite{Chiodaroli2015wal, Gunaydin1983bi, Gunaydin1986fg}. 	
Note that  $z^0$ is the same scalar as  $\tau$ from the previous subsection.}
\begin{align}
A^{-1}_- &=  \bar \phi \otimes A_- \,,    &  h_- &= A_- \otimes A_- \,, &  A^{-1}_+ &=  \phi \otimes A_+ \,, &   h_+ &= A_+ \otimes A_+ \,,  \no \\
A^0_- &= \phi \otimes A_-  \,,  &  i \bar z^0 &= A_+ \otimes A_- \,,  &  A^0_+ &= \bar\phi \otimes A_+  \,, &  -i  z^0 &= A_- \otimes A_+ \,, \no \\
A^A_- &= A_- \otimes \phi^A  \,,  &  i \bar z^A &= \bar \phi \otimes \phi^A \,, &   A^A_+ &= A_+ \otimes \phi^A  \,, &  -i z^A &= \phi \otimes \phi^A \,.  
\label{map1}
\end{align}
The index $A$ above has range $A=1,2,\ldots,n$.
Note that, in four dimensions, an extra vector field ($A^{-1}_\mu$) is present.
$\phi$ denotes the single complex scalar in the $\cN=2$ SYM theory.
Since all gauge-theory fields are in the adjoint representations, each field bilinear is associated to a supergravity state.
Overall, the construction produces a supergravity with the following properties:
\begin{enumerate}
	\item   has $\cN=2$ supersymmetry in four dimensions;
	\item   has $(n+1)$ vector multiplets in four dimensions. Scalars obtained as $\phi\otimes \phi^A$   
	transform under a $U(1) \times SO(n)$ symmetry; 
	\item uplifts to five dimensions whenever $n>0$;
	\item has vanishing single-soft limits at tree level (this can be checked explicitly);
	\item the supergravity can be seen as a truncation of $\cN=4$ supergravity with the same number of vector multiplets.
\end{enumerate}
Putting together all available information, the scalar manifold of the resulting theory turns out to be
\begin{equation}
\boldsymbol{\cal M}_{4D} = {SO(n,2) \over SO(n) \times SO(2)} \times {SU(1,1) \over U(1)}  \,.
\end{equation}
This infinite family of theories is known in the supergravity literature as the generic Jordan family of $\cN=2$ 
Maxwell-Einstein supergravities. The corresponding cubic polynomial in the natural basis is \cite{Gunaydin1983bi}:\footnote{The detailed 
form of the $C$-tensor may be changed by field redefinitions without changing the scattering amplitudes. See Ref.~\cite{Gunaydin1983bi} for 
a discussion of the {\em canonical} and {\em natural} basis.}
\begin{equation} 
{\cal V}(\xi) = \sqrt{2}\big( \xi^0 (\xi^1)^2 - \xi^0 (\xi^i)^2 \big) \,, \qquad  i=2,3,\ldots, n  \,.
\end{equation}
\begin{homework}
	Calculate explicitly ${\stackrel{\circ}{a}}_{IJ}$ and $g_{xy}$ corresponding to the cubic polynomial above.
\end{homework}	

\begin{homework}
	Calculate the three point amplitude ${\cal M}^{\tree}_3 \big( 1 A^0_- , 2 A^A_- , 3 \bar z^B   \big)$ using the double-copy prescription and the
	map~\eqref{map1}.   
\end{homework}

\subsubsection{Homogeneous $\cN=2$ Maxwell-Einstein supergravities \label{homogeneousSection}}

We now want to consider more general theories with $\cN=2$ supersymmetry and homogeneous scalar manifolds.  
A scalar manifold is said to be homogeneous if it admits a transitive group of isometries. 
From an amplitude perspective, not all these isometries will linearly realized, \ie some of them correspond go constant shifts of 
the scalars which modify the vacuum of the theory. 
Hence, in the homogeneous case, 
all coordinates of the manifold are Goldstone bosons and, consequently all single-soft limits of scalar amplitudes 
vanish (see for example~\cite{ArkaniHamed2008gz}). 
In short, drawing from the discussion in \sect{soft_limits}, we  have the following criterion: 

\begin{center}
	\parbox{0.9\textwidth}{ A necessary condition for a theory to possess a (locally) homogeneous scalar manifold is that all single-soft limits
		of scalar amplitudes vanish.}
\end{center}
\noindent
We also note that double-soft limits can be used to identify the particular homogeneous space under consideration (\ie $G$ in $G/H$)~\cite{ArkaniHamed2008gz}. More generally, 
each independent vanishing single-soft scalar limit will correspond to an isometry of the scalar manifold. 

We now return to the double-copy construction outlined in \sect{jordan}. A natural extension  consists of adding some matter fields 
in both gauge theories.  Hypermultiplets are the only available matter that can be coupled to $\cN=2$ SYM theory. 
One hypermultiplet consists of four real scalars and two Majorana fermions and is an irreducible representation of the $\cN=2$ supersymmetry algebra. 
For the construction described below, we will need to assign matter representations of the gauge  group to hypermultiplets.
If the gauge-group representation is pseudo-real, an additional option becomes available: we may consider a half-hypermultiplet instead of a full one. 
A single half-hypermultiplet is by itself a representation of the supersymmetry algebra,  but one is forced to include the Charge-Parity-Time reversal (CPT)-conjugate states 
unless its gauge-group representation is pseudo-real. This leads to a full hypermultiplet. 
Taking a single half-hypermultiplet, \ie choosing the smallest representations of supersymmetry algebra, amounts to introducing the minimal possible number of states and, in principle, allows us to manifest a larger 
global symmetry in the non-supersymmetric theory.  

If the desired supergravity theory is of the Maxwell-Einstein class, the non-supersymmetric gauge theory needs to be
a YM-scalar theory with extra fermions, so that additional vector multiplets are obtained as double copies involving one gauge-theory 
hypermultiplet and one fermion. Because of Working Rule 4, we will take the additional fermions to transform in the same pseudo-real representation ${\cal R}$
used for the supersymmetric theory. The Lagrangian is then written as  
\begin{eqnarray} 
{\cal L} \!\!&= \!\!& - {1 \over 4} F^{\ha}_{\mu \nu} F^{\ha \mu \nu} + {1 \over 2} (D_\mu \phi^a)^{\ha} (D^\mu \phi^a)^{\ha}  \no 
+{i \over 2}  \overline{\lambda}^\alpha  D_{\mu} \gamma^\mu \lambda_\alpha \ \no \\ &&
\!\!\! \!\!\! \null  + {g \over 2} 
\phi^{a\ha} \Gamma^{a \ \beta}_{\alpha} \overline{\lambda}^\alpha \gamma_5 \tR^{\ha} \lambda_\beta 
- {g^2 \over 4} f^{\ha \hb \he} f^{\hc \hd \ha} \phi^{a \ha} \phi^{b \hb} \phi^{a \hc} \phi^{b \hd} \,, 
\ \ \label{Lfermion} 
\end{eqnarray}
where $\ha, \hb$ are adjoint indices of the gauge group and $\alpha, \beta=1,\ldots, n_F$ and $a,b=1, \ldots, (D-4)$ are global indices. 
The  matrices $\Gamma^{a \ \beta}_{\alpha}$ in the global indices need to be constrained by imposing the duality between color and kinematics at four points.

\begin{table}[t]
	\centering
	\begin{tabular}{c@{$\quad$}|@{$\quad$}c@{$\quad$}|@{$\quad$}c@{$\quad$}|@{$\quad$}c@{$\quad$}|@{$\quad$}c}
		$D$ &  ${\cal D}_D$ & $4D$ fermions $n_F(D,P,\dot{P})$ & conditions & flavor group \\  
		\hline
		\hline & & &  \\[-8pt]
		$4$&  $1$ & $P$ &  R or W &   $SU(P)$   \\[3pt]
		$5$&  $1$ & $P$ &  R &   $SO(P)$   \\[3pt] 
		$6$ & $1$ & $P \! + \! \dot P$ & RW   & $SO(P)\! \! \times \! SO(\dot{P})$  \\[3pt]
		$7$  &  $2$ & $2P$ & R & $SO(P)$  \\[3pt]
		$8$  & $4$ & $4P$ & R or W  & $U(P)$ \\[3pt]
		$9$ &  $8$ & $8P$ & PR  & $USp(2P)$  \\[3pt]
		$10$  & $8$ & $8P \! + \! 8\dot P$ & PRW   & $USp(2P)\! \! \times \! USp(2\dot{P})$  \\[3pt]
		$11$  & $16$ & $16P$ & PR  & $USp(2P)$  \\[3pt]
		$12$  & $16$ & $16P$ & R or W &   $U(P)$  \\[3pt]
		$k \!+\! 8$  & $16 \, {\cal D}_k$ &  $16 \, r(k,P,\dot{P}) $ & as for $k$ & as for $k$     \\[3pt]
		\hline
	\end{tabular}
	
	\medskip
	
	\caption{\small Parameters in the double-copy construction for homogeneous supergravities \cite{Chiodaroli2015wal,Anastasiou2017nsz}. 
		$n_F(D,P,\dot{P})$ is the number of $4D$ irreducible spinors
		in the non-supersymmetric gauge theory,
		which can obey a reality (R), pseudo-reality (PR) or Weyl (W) conditions. Note that the pattern repeats itself with periodicity $8$. \label{tab1}} 
\end{table}

Imposing the duality on  amplitudes between two adjoint scalars and two matter fermions 
gives the constraint \cite{Chiodaroli2015wal}
\begin{equation}
\{ \Gamma^a , \Gamma^b \} = - 2 \delta^{ab} \,,  
\label{clifford}
\end{equation}
that is, the  matrices $\Gamma^a$ are gamma matrices which belong to a $(D-4)$-dimensional Euclidean Clifford algebra. 
Because of this relation, the non-supersymmetric theory can be regarded  as the dimensional reduction of a YM theory coupled to fermions in $D$ dimensions. A second parameter, $P$, will count the number of irreducible fermions in $D$ dimensions. 

\begin{homework}
	Show that imposing \ck duality on amplitudes the two-scalars two-fermion amplitudes given by the Lagrangian~\eqref{Lfermion} 
	yields the relation (\ref{clifford}).	
\end{homework}

An important difference with the standard treatment of $D$-dimensional spinors is the fact that fermions 
transform in pseudo-real representations of the gauge group. To obtain irreducible spinors, a case-by-case analysis is necessary. 
Depending on the value of the parameter $D$, one can impose  reality (R) or pseudo-reality (PR) conditions~\cite{Chiodaroli2015wal}
\begin{equation} 
\overline{\lambda} =  \lambda^t  {\cal C}_4 C V \,, \qquad
{\rm R:} \ \ C =   {\cal C}_{D-4}   \,, 
\qquad {\rm PR:} \ \ C =   {\cal C}_{D-4}  \Omega \    , \label{RPRcond}
\end{equation}
where ${\cal C}_{D-4}$  and ${\cal C}_{4}$ are the internal and spacetime charge-conjugation matrices, respectively. They obey  the relations
${\cal C}_{D-4} \Gamma^a {\cal C}_{D-4}^{-1} = - \zeta (\Gamma^a)^t$, ${\cal C}_4 \gamma^\mu {\cal C}_4^{-1} = - \zeta (\gamma^\mu)^t$, $\zeta = \pm1$.
$V$ is the unitary antisymmetric matrix entering the pseudo-reality condition for the gauge-group representation matrices, $V \tR^{\hat a} V^\dagger = - (\tR^{\hat a})^*$.
$\Omega$ is an antisymmetric real matrix acting on  indices which run over the number $P$ of irreducible spinors.
Alternatively, if $D$ is even, one can impose Weyl conditions.
If we have more than one  irreducible spinor, an extra flavor symmetry is present  (either $U(P)$, $SO(P)$ or $USp(P)$, depending on 
whether Weyl, Reality or pseudo-Reality conditions were employed).  
A separate treatment is needed when $D=6,10$ (mod $8$). In these dimensions, there are two inequivalent irreducible 
spinors with different chirality and one needs to introduce parameters $P, \dot P$ which  count the number of each. 

Explicit computations reveal that  soft-scalar limits vanish for amplitudes constructed by the double copy \cite{Chiodaroli2015wal}. Hence, this generalized construction yields supergravities with homogeneous scalar manifolds. The dimension-by-dimension analysis is given in Table \ref{tab1}.
The number of vector multiplets in the four-dimensional supergravity is equal to $(D - 3 + n_F)$, where $n_F(D,P,\dot P)$ is the number of $4D$ fermions in the non-supersymmetric gauge theory; the supergravity bosonic states are obtained as double copies in the following way \cite{Chiodaroli2015wal}:
\begin{align}
A^{-1}_- &=  \bar \phi \otimes    A_- \,,               & h_- &= A_- \otimes A_- \,,                 &   A^{-1}_+ & =  \phi \otimes A_+ \,,         & h_+ & = A_+ \otimes A_+ \,,  \no \\
A^0_- &= \phi \otimes A_-  \,,                          & i \bar z^0 &= A_+ \otimes A_- \,,          &   A^0_+ & = \bar\phi \otimes A_+  \,,        &  -i  z^0 &= A_- \otimes A_+ \,, \no \\
A^A_- &= A_- \otimes \phi^A  \,,                        & i \bar z^A &= \bar \phi \otimes \phi^A \,, &   A^A_+ &= A_+ \otimes \phi^A  \,,           &  -i z^A &= \phi \otimes \phi^A \,, \no \\ 
A_{\alpha -} &= \chi_-  \otimes \lambda_{\alpha-}  \,,  & i \bar z_{\alpha} &= \chi_+ \otimes  \lambda_{\alpha-}  \,,  &    A^{\alpha}_+ &= \chi_+  \otimes \lambda_+^{\alpha}  \,, & -i z^{\alpha} &= \chi_+ \otimes  \lambda_-^{\alpha} \,. 
\label{map}
\end{align}

\begin{homework}
	Show that the amplitude ${\cal M}_5^\tree\big(z^0, \bar z^0, z^{\alpha}, \bar z_{\beta}, z^0\big)$
	has vanishing single-soft limits for all external scalars.
\end{homework}	

\begin{homework}
	Show that the amplitude ${\cal M}_3^\tree\big( 1 A^a_-,	 2A^\alpha_-, 3\bar z^\beta \big) $ can be expressed as 
	\begin{equation}
	{\cal M}_3^\tree\big( 1 A^a_-,	 2A^\alpha_-, 3\bar z^\beta \big) = {\kappa \over 2 \sqrt{2}} \langle 12 \rangle^2 \big( U^t \Gamma^a C^{-1} \big)^{\alpha \beta}\,.
	\end{equation}
\end{homework}

A remarkable result is that the theories listed in Table~\ref{tab1} reproduce the complete classification of homogeneous 
supergravities by de Wit and van Proeyen \cite{deWit1991nm}. Theories obtained with this construction include some classic examples. 
Specifically, for $P=1$ and $D=7,8,10,14$, we find the so-called Magical Supergravities. 
These theories exhibit additional symmetry enhancement, which results in the corresponding scalar manifolds being symmetric spaces. 
Their scalar manifolds are:
\begin{equation}
\boldsymbol{\cal M}^{\mathbb{R}}_{4D} = {Sp(6,\mathbb{R}) \over U(3)}, \quad \boldsymbol{\cal M}_{4D}^{\mathbb{C}} = {SU(3,3) \over S(U(3)\times U(3))}, \quad \boldsymbol{\cal M}_{4D}^{\mathbb{H}} = {SO^*(12) \over U(6)}, \quad 
\boldsymbol{\cal M}^{\mathbb{O}}_{4D} = {E_{7(-25)} \over E_6 \times U(1)} \,. 
\end{equation}
An important property is that Magical theories are unified, that is there exists a symmetry with respect to which all 
vector fields transform in a single irreducible representation. Physically, this implies that fields from different matter multiplets have the same properties. 
In contrast, vectors in generic homogeneous theories typically have different interactions according to whether they are obtained as vector-scalar or as
fermion-fermion from a double-copy perspective.
The construction of these theories from a supergravity perspective relies on degree-three Jordan algebras which have as elements $3 \times 3$ matrices with entries in the four division algebras ($\mathbb{R},\mathbb{C},\mathbb{H},\mathbb{O}$). Reviewing the supergravity construction is beyond the scope of this review; we refer the reader to Ref.~\cite{Gunaydin1983bi} for details.

Aside from the Magical Supergravities, there is another class of examples of unified theory in four dimensions. They are obtained by choosing
$D=4$  and $P$ is arbitrary. It is not difficult to see that the construction exhibits a global $U(P)$ flavor symmetry, as well as that the 
resulting supergravity theory will have $(P+1)$ complex scalars in its spectrum. Putting together this information results in the scalar manifold \cite{Chiodaroli2015wal,Anastasiou2017nsz}
\begin{equation}
\boldsymbol{\cal M}_{4D} = {U(P+1,1) \over U(P+1) \times U(1) } \,,
\end{equation}  
which is the complex projective space $\mathbb{CP}^{P+1}$. Theories in this family are also referred to as minimally-coupled or the Luciani model.  The analysis can be repeated in dimensions different from four. In five and six dimensions we find exactly one infinite-dimensional family of unified theories: 
\begin{eqnarray}
5D:  \qquad &\text{Generic non-Jordan family} & \boldsymbol{\cal M}_{5D} = { SO(P+1,1) \over SO(P+1)} \,, \\
6D:  \qquad &\text{Generic Jordan family} & {\boldsymbol{\cal M}}_{6D} = {SO(P+1,1) \over SO(P+1)} \,.
\end{eqnarray}
In both cases, the double-copy construction is similar to the one in four dimensions: the non-supersymmetric gauge theory is a YM theory in the 
appropriate dimension with an arbitrary number of fermions and no additional scalars. 
While the parameter $P$ is by construction non-negative, it should be noted that pure supergravities in dimensions $4,5,6$ can be obtained as particular cases by setting $P=-1$. 
This observation will be consequential in formulating double-copy constructions for pure supergravities with $\cN=2$ in various dimensions. 
The construction outlined in this section has been used to compute one-loop matter amplitudes in these theories and to analyze their UV properties at 
that order, see Ref.~\cite{Ben-Shahar:2018uie}.

\subsubsection{Pure supergravities\label{secpureSG}}

Pure supergravities with ${\cal N}=1, 2, 3$ have been originally formulated from a Lagrangian perspective in Refs.~\cite{Freedman:1976xh, Freedman:1976py, Deser:1976eh},
\cite{Ferrara:1976fu} and~\cite{Freedman:1976nf, Ferrara:1976vf}, respectively.
Regardless of the number of supercharges which are manifest in the construction, a double-copy  gravity theory in four dimensions 
contains a complex scalar which is obtained from the product of gluons of opposite polarizations. For extended $\cN\geq 4$ supersymmetry, 
this scalar still belongs to the gravity multiplet.
For $\cN<4$, however, the gravity multiplet does not contain any scalar field so 
the complex scalar under consideration belongs to a matter multiplet.
Hence, to obtain pure supergravities with $\cN<4$, one needs to modify the construction and remove the contributions to amplitudes of the unwanted scalar.     
At tree level, one can always project out the unwanted scalars from the amplitudes by judiciously choosing the asymptotic states. 
Special care is however necessary for loops.  

\begin{table}
	\begin{center}
		\begin{tabular}{c|c|c} \small
			$\cN$  & tensoring vector states &  ghosts = matter $\otimes$ $\overline{\text{matter}}$ \\ 
			\hline
			\hline 
			&& \\[-8pt]
			$0+0$ & $A_\mu \otimes A_\nu = h_{\mu \nu} \oplus \phi \oplus a$ & $(\psi_+ \otimes \psi_-) \oplus (\psi_- \otimes \psi_+) = \phi \oplus a$ \\
			$1+0$& $ {\cal V}_{\cN=1} \otimes A_\mu = {\cal H}_{\cN=1} \oplus \Phi_{\cN=2}$ & $(\Phi_{\cN=1}\otimes \psi_-)\oplus(\bar \Phi_{\cN=1}\otimes \psi_+) = \Phi_{\cN=2}$ \\
			$2+0$& $ {\cal V}_{\cN=2} \otimes A_\mu = {\cal H}_{\cN=2} \oplus {\cal V}_{\cN=2}$ & $(\Phi_{\cN=2}\otimes \psi_-)\oplus(\bar \Phi_{\cN=1}\otimes \psi_+) = {\cal V}_{\cN=2}$ \\
			$1+1$& $ {\cal V}_{\cN=1} \otimes {\cal V}_{\cN=1} = {\cal H}_{\cN=2} \oplus 2 {\Phi}_{\cN=2}$ & \! $(\Phi_{\cN=1}\otimes \bar \Phi_{\cN=1})\oplus(\bar \Phi_{\cN=1}\otimes \Phi_{\cN=1}) = 2 {\Phi}_{\cN=2}$ \\
			$2+1$& $ {\cal V}_{\cN=2} \otimes {\cal V}_{\cN=1} = {\cal H}_{\cN=3} \oplus {\cal V}_{\cN=4}$ & \! $(\Phi_{\cN=2}\otimes \bar \Phi_{\cN=1})\oplus(\bar \Phi_{\cN=2}\otimes \Phi_{\cN=1}) =  {\cal V}_{\cN=4}$ \\
			$2+2$& $ {\cal V}_{\cN=2} \otimes {\cal V}_{\cN=2} = {\cal H}_{\cN=4} \oplus 2{\cal V}_{\cN=4}$ & \! $(\Phi_{\cN=2}\otimes \bar  \Phi_{\cN=2})\oplus( \bar \Phi_{\cN=2}\otimes  \Phi_{\cN=2}) = 2 {\cal V}_{\cN=4}$ \\[-8pt] && \\ \hline
\end{tabular}
		\caption{ Pure gravities constructed as double copies \cite{Johansson2014zca}. The construction necessitates
			ghosts from matter-antimatter double copies. Barred multiplets transform in the anti-fundamental representation. 
			For compactness, graviton and vector supermultiplets ${\cal H}, {\cal V}_{\cN<4}$ include the CPT-conjugate states. 
			Pairs of chiral/antichiral $\cN=1$ supermultiplets are grouped in $\cN=2$ hypermultiplets, denoted as $\Phi_{\cN=2}$.  \label{tabpure}}
	\end{center}
\end{table}

A solution to the problem was first outlined in Ref.~\cite{Johansson2014zca}. 
The first step is to introduce an additional matter representation (the fundamental representation, without any loss of generality) in both gauge theories. The precise map  depends on the desired amount of supersymmetry and is listed in Table \ref{tabpure}.
Since the various graphs contributing to the amplitude carry representation information associated to all internal and external lines, 
we can organize the graphs with no external matter  according to the number of matter loops. 
We can then treat the additional matter as a ghost multiplet by associating an extra minus sign to each matter loop 
(as with Faddeev-Popov ghosts).
More explicitly, loop-level pure-supergravity amplitudes are constructed using the prescription
\begin{equation}
{\cal M}^{(L)}_m = i^{L-1}\Big( {\kappa \over 2} \Big)^{m+2L-2} \sum_{i \in \text{cubic}}\int {d^{LD}\ell \over (2\pi)^{LD}}
\frac{({-}1)^{|i|}}{S_i} { \frac{n_i \tilde n_i }{ D_i} } \,,
\end{equation}
where $|i|$ denotes the number of matter loops in the $i$-th graph. 
It has been shown by explicit calculation through two loops and argued to all loop orders 
that amplitudes obtained with this prescription have the same unitarity cuts as the ones of the pure supergravities listed in Table \ref{tabpure}.
It is interesting to note that $\cN=2$ ghost multiplets are constructed as double copies involving fermions in the non-supersymmetric theory.
One can in principle consider an analogous construction
involving scalars, but amplitudes constructed in this way would have unitarity cuts which are different from those of pure supergravities. 
This observation provides a clue on the meaning of the construction:
formally, the prescription above is equivalent to considering one of the unified infinite families of supergravities described at the end of the last subsection and setting $P=-1$.  

Explicit calculations show that pure Einstein gravity is
UV-divergent at two loops\footnote{The interpretation of the
divergence is rather subtle because of its dependence on evanescent operators and choice of
fields~\cite{PureGravityTwoLoops,Bern:2017puu}.}~\cite{Goroff:1985th,
vandeVen:1991gw}, although it is finite at one loop because the candidate
counterterm is a total derivative in four
dimensions~\cite{tHooft:1974toh}.  For $\cN=1,2, 3$ pure
supergravities, ultraviolet divergences cannot appear before three
loops~\cite{Grisaru:1976nn,Tomboulis:1977wd}; the relevant explicit calculations
at this loop order, probing the appearance of
divergences, have as yet not been carried out.

\subsubsection{Theories with hypermultiplets and supergravities with $\cN<2$}

An alternative option is to couple matter hypermultiplets with $\cN=2$
supergravity.  In general, supergravities with hypermultiplets are
less constrained than theories with vector multiplets. A subset of
such theories, however, appears to be closely related to theories with
vector multiplets through a procedure known as
$c$-map \cite{Cecotti1988qn}.

Starting from a Maxwell-Einstein theory in four dimensions, one first reduces the theory to three dimensions. After dualization of the vector field, each supermultiplet in the three-dimensional theory contains four real scalars and four Majorana fermions, which is the field content corresponding to a hypermultiplet. Since the hypermultiplet action is the same in any dimension up to six, the three-dimensional theory obtained with this procedure can be uplifted to higher dimension, leading to the image of the original Maxwell-Einstein theory under the $c$-map.

The basic double-copy construction for theories with hypermultiplets was mentioned in Ref.~\cite{Chiodaroli2015wal} and further detailed in Ref.~\cite{Anastasiou2017nsz}. 
It relies on taking as one copy  $\cN=2$ SYM theory with matter hypermultiplets and, as the second copy, a YM theory with extra matter scalars. 
The simplest realization of this construction involves  
a SYM theory with a single  hypermultiplet in a real representation and 
a YM theory with $m$ real scalars. 
A Lagrangian for the latter theory is:\footnote{In principle, it is possible to choose a different coefficient for the quartic scalar coupling while preserving \ck duality. Indeed the scalar sector of this theory is the same as the theory discussed at the end of \sect{sec-zoo-tols}. The theory given here can also be constructed as a field-theory orbifold of an adjoint YM theory in higher dimension. See also Ref.~\cite{Chiodaroli2015rdg} for a similar discussion.}
\be 
{\cal L} = -{1 \over 4} F^{\hat a}_{\mu \nu} F^{\hat a \mu \nu} + {1 \over 2} D_\mu  \varphi^I D^\mu \varphi^I + {g^2 \over 4} \big( \varphi^{[I} \tR^{\hat a} \varphi^{J]} \big) \big( \varphi^{[I} \tR^{\hat a} \varphi^{J]} \big) \,. 
\ee
Scalar fields $\varphi^I$ are labeled by flavor indices $I,J=1,\ldots,m$, which refer to the global $SO(m)$ symmetry, and gauge-group 
representation indices for some real representation ${\cal R}$, which we do not display; $\tR^{\hat a}$ are the gauge-group generators in this representation 
and $\hat a, \hat b$  are adjoint indices.
One can verify that this theory obeys \ck duality at four points. 
The computation is identical to the one for a higher-dimensional YM theory reduced to four dimensions, with the only 
difference being related to representation of the scalar fields.

Based on the symmetry $SO(m)\times SO(4)$ which is manifest in the construction, the $4m$ real hypermultiplet scalars in the theory 
together with the universal dilaton-axion parametrize the scalar manifold 
\be 
\boldsymbol{\cal M}_{4D} = {SU(1,1) \over U(1)} \times {SO(m,4)\over SO(m) \times SO(4)} \,. 
\ee
The second term in the product manifold is the special quaternionic-K\"{a}hler manifold which is the image of the generic Jordan family 
scalar manifold under the $c$-map. 
From the point of view of scattering amplitudes, the relation between the two classes of theories is a consequence of the fact that 
the kinematic numerator factors from the non-supersymmetric gauge theory are identical in the two constructions. The differences
relate to the pairing between the kinematic numerators of the two gauge theories, which is now different because of the different
gauge-group representations and  color factors.

Several additional constructions for ungauged supergravities with various matter contents deserve mention:
\begin{itemize}
	\item Various $(\cN=1) \times (\cN=1)$ double copies were studied in Refs.~\cite{OneLoopSusy,Schreiber2016sss,Damgaard2012fb,Anastasiou2015vba}. 
	In this case, at least one hypermultiplet is present. Since $\cN=1$ gauge theories do not generically uplift to higher dimensions, the construction 
	does not manifestly give a supergravity which can be written in five dimensions and, hence, three-point amplitudes cannot be used to specify 
	the theory completely. Instead, the identification relies on symmetry consideration and on the possibility of embedding the theory into a supergravity 
	with extended supersymmetry.  
	\item Several examples of $\cN=1$ supergravities constructed as double copies are known. The known examples can often 
	be seen as truncations of theories with a larger number of supersymmetries~\cite{OneLoopSusy,Schreiber2016sss,Damgaard2012fb,Anastasiou2015vba}. 
	\item Various examples of non-supersymmetric gravities constructed as double copies are known \cite{BCJ,Johansson2014zca}. Among these, the simplest example is Einstein gravity with a scalar and antisymmetric tensor, which we have already encountered in \sect{DualitySection}. This theory is constructed as the square of YM theory.
	In four-dimensions, the antisymmetric tensor is dual to an axion. The scalar-sector Lagrangian is then identical to the one in \eqref{Lag-N4}.  
	\item An interesting version of the construction applies to the so-called twin supergravities~\cite{Anastasiou2016csv}. 
	These are pairs of supergravities with different amounts of supersymmetries which share the same bosonic Lagrangian but have different fermionic field 
	content and interactions.
\end{itemize}

\begin{homework}
	Consider two supergravities constructed as field theory orbifolds of $\cN=8$ supergravity with the following generators:
	\begin{eqnarray}
	\text{Theory 1}: && R=\text{diag}\big( 1, e^{2\pi i\over 3}, e^{2\pi i\over 3}, e^{2\pi i\over 3}, 1, 1, 1, 1 \big) \,, \qquad R'=\text{diag}\big( 1, 1, 1, 1, -1, -1, -1, -1 \big) \,, \nonumber \\
	\text{Theory 2}: && R=\text{diag}\big( 1, i, i, i, -1, -1, -1, -i \big)  \nonumber \,.
	\end{eqnarray}	
	Find the corresponding spectra and, using the manifest symmetries of the construction, find a candidate for the scalar manifolds.
\end{homework}

\subsubsection{Yang-Mills-Einstein theories \label{YMESec}}

YME theories are supergravities that involve nonabelian gauge interactions among (some of) the vector fields. 
Surprisingly, they admit a very simple double-copy realization, which relies on the following principle \cite{Chiodaroli2014xia}: 

\begin{center}
	\parbox{0.9\textwidth}{To introduce nonabelian gauge interactions in a gravitational 
		theory from the double copy, it is sufficient to add trilinear couplings among adjoint scalar fields in one of the gauge theories entering the construction. }
\end{center}
\noindent

\begin{table}[t]
	\centering
	\begin{tabular}{l|c|c}
		Gravity coupled to YM & Gauge theory 1  &  Gauge theory 2
		\\
		\hline \\[-13pt]
		\hline
		${\cal N}=4$ YME supergravity  $\raisebox{.3 cm}{ \vphantom{|} }  $  & ${\cal N}=4$ SYM & YM~+~$\phi^3$
		\\
		${\cal N}=2$ YME supergravity (gen.Jordan)   & ${\cal N}=2$ SYM & YM~+~$\phi^3$
		\\
		${\cal N}=1$ YME supergravity  & ${\cal N}=1$ SYM & YM~+~$\phi^3$ 
		\\
		${\cal N}=0$ YME + dilaton + $B^{\mu\nu}$   &  YM & YM~+~$\phi^3$
		\\
		${\cal N}=0$ ${\rm YM}_{\rm DR}$-E + dilaton + $B^{\mu\nu}$   &  ${\rm YM}_{\rm DR}$ & YM~+~$\phi^3$
		\\
		\hline
	\end{tabular}
	\small \caption[a]{\small
		Amplitudes in YME gravity theories for different number of supersymmetries, corresponding to different choices for the left gauge-theory
		factor entering the double copy  \cite{Chiodaroli2014xia}. YM${}_{\text{DR}}$ denotes the YM-scalar theory obtained from dimensional reduction.  \label{TheoryConstructions1}}
\end{table}

\noindent
The relevant Lagrangian was introduced in \eqref{YMscalar}. The effect of the trilinear coupling is to introduce nonzero supergravity 
amplitudes of the form
\bea
{\cal M}_3 (1 A^A_-, 2A^B_-,3A^C_+) &=& i {A}_3 (1 A_-, 2A_-,3A_+)  {A}_3 (1 \phi^A, 2 \phi^B,3 \phi^C) \nn \\
&=& 
- {\kappa \over 2\sqrt{2}} \lambda F^{ABC} {\langle 12 \rangle^3 \over \langle 23 \rangle \langle 31 \rangle } = 
i {\kappa \over 4} \lambda \tilde F^{ABC} {\langle 12 \rangle^3 \over \langle 23 \rangle \langle 31 \rangle } \,, 
\eea
\ie amplitudes between three spin-1 fields which are proportional to an antisymmetric tensor obeying Jacobi relations 
and have the same momentum dependence as the three-gluon amplitudes from the supergravity Lagrangian. In particular, 
the supergravity gauge coupling  constant $g_s$ is related to the parameter $\lambda$ in (\ref{YMscalar}) as 
\begin{equation}
g_s= \left( {\kappa \over 4} \right) \lambda  \, , \label{YMEcouplingidentification}
\end{equation}
where we have temporarily re-introduced $\kappa$.
In this construction, the global-symmetry tensor $F^{ABC}$, which obeys the Jacobi identity \eqref{globalJacobiId}, 
is identified with the structure constants of  the supergravity gauge group. 
Hence, a global symmetry in one of the two gauge theories becomes a local symmetry in the resulting double-copy gravity theory.

We note that this approach gives, by construction, gauge groups which are subgroups of the manifest isometry group of the 
 corresponding Maxwell-Einstein theory. These groups are necessarily compact. 
Gauging a subgroup of the $R$ symmetry, a construction which results in the so-called gauged supergravities, requires a more 
involved procedure which will be outlined in \sect{gaugedsugras}.
The double-copy construction for YME theories can be adapted to supergravities with various amounts of supersymmetry, 
which are listed in Table \ref{TheoryConstructions1} \cite{Chiodaroli2014xia}.

Aside from spelling out the construction at the level of the gauge-theory Lagrangian, it is interesting to consider the implications of the double-copy structure on YME amplitudes.
We start from the double copy
\begin{equation}
\big( \text{YM}+\phi^3 \big) = \big( \text{YM}+\phi^3 \big) \otimes \big( \phi^3 \text{ theory}  \big)  ,
\end{equation}
i.e. we note that the double copy between the YM+$\phi^3$ theory and the bi-adjoint $\phi^3$ theory gives amplitudes from the YM+$\phi^3$ theory itself.
By choosing numerator factors corresponding to the DDM basis \cite{DixonMaltoni}, we then write a color-ordered tree amplitude between $k$ gluons and $m\ge 2$ scalars in the YM+$\phi^3$ theory as follows
\be
{\cal A}^{\text{YM}+\phi^3}_{k,m}(1, \ldots, k \, |\, k+1,\ldots,k + m)   =  - i  \sum_{\, w \in \sigma_{12\ldots k}} 
N_k(w) A^{\phi^3}_{k+m}(w) ~ + ~ \text{Perm}(1, \ldots, k)  \, .
\label{scalaramp2}
\ee
$A^{\phi^3}_{k+m}(w)$ are amplitudes  in bi-adjoint $\phi^3$ theory that are color-ordered only with respect to one of the two colors. In the above formula we are summing over all color orderings $w$ that belong to the set $\sigma_{123\cdots k}$; which is explicitly constructed using a shuffle product $\shuffle$, as
\bea
\sigma_{123\cdots k} &=&  \Big\{\{k+1, \gamma, k+m\} \,\Big| \, \gamma \in \alpha \shuffle \beta \Big\}\,, ~~\text{where} \no  \\
\alpha &=&  \{1,2,3, \ldots, k\}\, ~~\text{and} ~~ \beta = \{k+2, \ldots, k+m-1\}\,. \label{shuffle}
\eea
The set $\sigma_{123\cdots k}$ contains all shuffles of the gluon ($\alpha$) and scalar ($\beta$) sets that respect the ordering within each set, with the additional constraint that the first and last scalars are held fixed. We will refer to the elements of this set as ``words'' $w$. A remarkable observation is that gauge invariance is sufficient to fix the numerators $N_k(w)$ in the expression above.

\begin{figure}
	\begin{center}
		\includegraphics[width=0.8\textwidth]{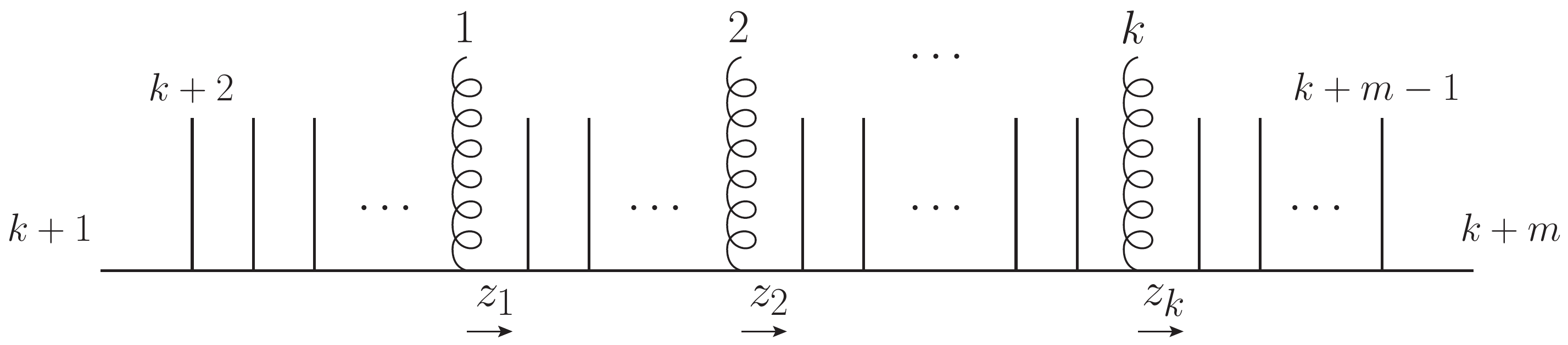}
		\caption{Half-ladder graph for the YM+$\phi^3$ theory. The gluons are denoted with $1,2,\ldots, k$ and the remaining particles are scalars. $z_i$ denote the momentum of the internal scalar to the right of gluon $i$. \label{figmulti2}}
	\end{center}
\end{figure}

Color-ordered single-trace YME amplitudes are obtained by replacing the $\phi^3$ partial amplitudes with partial amplitudes belonging to 
pure YM theory (or its supersymmetric relatives, depending on the target gravitational theory) \cite{Chiodaroli2017ngp},
\bea
\hskip -.4 cm 
M^{\text{YME(SG)}}_{k,m}(1,\ldots,k \, | \, k+1, \ldots, k+m) \! &=& \!\! \!\! \sum_{w \in \sigma_{12\ldots k}} \!\! \!\!
N_k(w) A^{\text{(S)YM}}_{k+m}(w) +
\text{Perm}(1, \ldots,k)\,. \quad \quad
\label{ST_YME}
\eea
Since the partial amplitudes $A^{\text{(S)YM}}_{k+m}(w)$ obey the same relations as $A^{\phi^3}_{k+m}(w)$ (including in particular the BCJ relations), the YME amplitudes given by this formula will be by construction gauge invariant. The numerators $N_k(w)$ are obtained by imposing gauge invariance on (\ref{scalaramp2}), that is by imposing that 
the amplitude vanishes after the replacement 
\begin{equation}
\varepsilon_i \rightarrow p_i \, . \label{gauge}
\end{equation}
Along the same lines of \sect{subsecNLSM}, we construct the numerators $N_k(w)$  from the following
independent Lorentz invariants:
\begin{equation}
~~~~~~ \big\{ \, (\varepsilon_i \cdot  z_i) \, , \ \
(p_i \cdot  z_i) \, , \ \
(\varepsilon_i \cdot \varepsilon_j) \, , \ \
(\varepsilon_i \cdot  p_j) \, , \ \
(p_i \cdot  p_j) \, \big\} \, ,  ~~~~~~~~ (i,j= 1,\ldots, k) 
\end{equation}
where $p_i$ denote only the momenta of the gluons. The momenta of the scalars will only appear implicitly through the region momenta $z_i=z_i(w)$ that we define as
\begin{equation}
z_i(w)= \mathop{\sum_{1 \le j \le l}}_{w_l=i} p_{w_j}\,,
\end{equation}
which give the sum of the momenta of all the particles to the left of the $i$-th gluon in the multiperipheral graph corresponding 
to the word $w$ (including the gluon momentum  $p_i$, see \fig{figmulti2}).

We consider the case of one external gluon ($k=1$). By dimensional analysis, each term in $N(w)$ will need to contain a 
single factor of momentum. $\sigma_1$ is the set of external-leg orderings in which the order of the scalars is preserved and 
the single external gluon is inserted in different positions (leaving a scalar as the first and last entry). 
A natural guess for the numerator is given by  
\begin{equation}
N_1 =  2 (\varepsilon_1 \cdot z_1) \,. 
\label{YME1grav}
\end{equation}
We can check gauge invariance with the replacement (\ref{gauge}); the gauge variation of the amplitude becomes
\begin{equation}
\sum_{w \in \sigma_{1}} (p_1 \cdot z_1) A^{\text{(S)YM}}_{m+1}(w)  = 0 \, ,
\label{FundBCJrel}
\end{equation}
which is zero as a consequence of the BCJ relations. Indeed, (\ref{FundBCJrel}) is precisely the fundamental BCJ relation~\eqref{BCJrels} \cite{BCJ,amplituderelationProof}. That this BCJ relation can be obtained from YME amplitudes with a single graviton, using the numerator (\ref{YME1grav}), was first shown in Ref.~\cite{Stieberger2016lng}. 

The next-simplest example has two external gluons. Now $\sigma_{12}$ will be the set of external-leg orderings in which the two gluons are inserted in different positions while leaving the order of scalars and gluons unchanged (and keeping scalars as the first and last entries). The last term in (\ref{ST_YME}) is obtained by exchanging the two external gluons. The numerators are
\begin{equation}
N_2 = 4  (\varepsilon_1 \cdot z_1)  (\varepsilon_2  \cdot z_2) + 2 (p_2 \cdot z_2) (\varepsilon_1 \cdot \varepsilon_2) \,,
\end{equation}
where the last (contact) term is fixed by imposing gauge invariance. Taking a gauge variation $\varepsilon_1 \rightarrow p_1$ of the amplitude we obtain:
\begin{eqnarray}
&& 4 \Big( \sum_{w \in \sigma_{12}} + \sum_{w \in \sigma_{21}} \Big) (p_1 \cdot z_1) (\varepsilon_2 \cdot z_2) A^{\text{(S)YM}}_{m+2}(w) \no \\
&& \qquad 
\null + 2 (\varepsilon_2 \cdot p_1) \Big(  \sum_{w \in \sigma_{12}} (p_2 \cdot z_2) A^{\text{(S)YM}}_{m+2}(w) + \sum_{w \in \sigma_{21}}  (p_1 \cdot z_1)  A^{\text{(S)YM}}_{m+2}(w)
\Big)  \,. \label{variationk2-0}
\end{eqnarray}
Using the fundamental BCJ relation (\ref{BCJrels}), the reader can verify that the above gauge variation reduces to
\begin{eqnarray}
(\varepsilon_2 \cdot p_1) \sum_{w \in \sigma_{12}} \big( (p_1 \cdot z_1) + (p_2 \cdot z_2) \big) A^{\text{(S)YM}}_{m+2}(w) = 0 \,. \label{variationk2}
\end{eqnarray} 
This is equivalent to the sum of two BCJ relations and thus vanishes for all amplitudes that satisfy \ck duality. 

\begin{homework}
Verify 	\eqref{variationk2} starting from \eqref{variationk2-0} and using the fundamental BCJ relation.
 \end{homework}

Semi-recursive expressions for YME amplitudes with up to five external gravitons were given in Ref.~\cite{Chiodaroli2017ngp}. General expressions for any multiplicity based on BCFW recursion were given in Ref.~\cite{TengFengBCJNumerators} in the single-trace case and in Ref.~\cite{Du2017gnh} in the multi-trace case. 
The reader may also consult \cite{Nandan2016pya,delaCruz2016gnm} for alternative expressions obtained through the CHY formalism and \cite{He2016mzd,Nandan2018ody} for loop-level amplitudes in YME theory.

\subsubsection{Higgsed supergravities\label{Higgsed}}

A key feature of the double-copy construction for YME theories is that it can be generalized to cases in which the nonabelian gauge supersymmetry of the supergravity theory is spontaneously broken. Since YME theories possess a moduli space in which the unbroken-gauge phase is given by a single isolated point, the fact that the double-copy construction admits an extension of this sort gives a strong hint of its applicability for generic gravity theories. 
At the same time, the construction we review here is one of the simplest examples in which some of the fields are massive.
The double-copy construction for a Higgsed supergravity has the schematic form  
\begin{equation}
\Big( \text{Higgsed YME SG} \Big) = \Big( \text{Coulomb-branch SYM theory} \Big) \otimes \Big( \text{YM + massive scalars} \Big) \,.
\end{equation}
Schematically, amplitudes for the first gauge-theory factor can be obtained with a two-step process: (1) break the gauge-group down to a 
subgroup (see \sect{break}); (2) assign masses (seen as compact momenta) to fields transforming in matter representations of the 
unbroken subgroup (see \sect{masses}).
We have seen in \sect{YMESec} that, with the appropriate choices of
gauge theories, the global symmetry in one of the gauge-theory factors becomes a
nonabelian gauge symmetry in (super)gravity. The scenario discussed
here extends this property by showing that an explicitly-broken
symmetry in one of the two gauge theories becomes, through the double
copy, a spontaneously-broken gauge symmetry in (super)gravity.

To avoid a notationally-heavy discussion, we will review here the
simplest example of the Higgsed double-copy construction.  We start from a $\cN=2$ SYM
theory with $SU(N+M)$ gauge group and decompose it with respect to the
$SU(M)\times SU(N)\times U(1)$ subgroup.  The direct sum of the
adjoint representations of the unbroken gauge-group factors is denoted
as ${\cal G}$; the corresponding fields are left massless.  In
addition, there will be two vector multiplets transforming in the
bifundamental ${\cal R}=(\boldsymbol M,\overline{\boldsymbol N})$ and
anti-bifundamental $\overline{\cal R}=(\overline{\boldsymbol
M}, \boldsymbol N)$ representations. These fields are made massive by
a suitable assignment of momenta along one single compact dimension. This relies, implicitly, on
the fact that the theory can be uplifted to higher dimension. The
resulting bosonic states are given in Table~\ref{TabExHiggs} (while
fermionic states are not displayed).
%
\begin{table}
	\begin{center}	
		\begin{tabular}{c@{$\quad$}|@{$\quad$}c@{$\quad$}|@{$\quad$}c}
			Fields & Representation & Mass \\
			\hline \hline && \\[-8pt]
			$(A_\mu, \phi, \bar \phi )$ & $\cal G$ &  $0$ \\
			$(W_\mu, \varphi )$ & $\cal R$ &  $m$ \\
			$(\overline{W}_\mu, \overline{\varphi})$ & $\overline{\cal R}$ &  $-m$ \\
           \hline
		\end{tabular}
		\hskip .5in
		\begin{tabular}{c@{$\quad$}|@{$\quad$}c@{$\quad$}|@{$\quad$}c}
			Fields & Representation & Mass \\
			\hline \hline && \\[-8pt]
			$(A_\mu, \phi, \bar \phi )$ & $\cal G$ &  $0$ \\
			
			$\varphi $ & $\overline{\cal R}$ &  $m$ \\
			
			$\overline{\varphi}$ & $\cal R$ &  $-m$ \\
           \hline
		\end{tabular}
		\caption{Bosonic fields in gauge-theory factors for the example of double-copy construction for Higgsed supergravities.\label{TabExHiggs} }
	\end{center}
\end{table}
%
As discussed in \sect{masses}, this is equivalent to giving a scalar VEV 
\begin{equation}
\langle \phi \rangle  = V t^{\hat 0} \,,
\label{symbreakingEG}
\end{equation}
where $t^{\hat 0}=$diag$\big({1 \over M} I_M,- {1 \over N} I_N\big)$ and $V$ is a real parameter (since our assignment of compact momenta only involves a single compact dimension).

At this stage, we need to examine the constraints coming from the duality between color and kinematics at four points.
\ck duality for amplitudes between two adjoint and two matter fields is automatically satisfied. This is a consequence of the fact that the theory can be obtained by assigning compact momenta to a higher-dimensional massless theory, as explained in  \sect{masses}. Alternatively, one could adopt a bottom-up approach and start from a Lagrangian involving massive vectors and scalars and leave free parameters in the interaction terms. 
Imposing \ck duality would fix the interaction terms to be the ones of the Coulomb-branch theory. 

Amplitudes involving four matter fields require a more detailed analysis. In particular, scattering amplitude of four massive scalars can be cast in the form
\be 
{\cal A}_4 \big( 1 \varphi_{\haa}, 2 \varphi_{\hbb}, 3 \overline{\varphi}^{\hgg}, 4 \overline{\varphi}^{\hdd} \big) = 
-i g^2 \left( {n_t c_t \over D_t } + {n_u c_u \over D_u } + {n_s c_s \over D_s }  \right) , 
\label{4massform_gauge} 
\ee
where the color factors are\footnote{As discussed in \sect{DualitySection}, 
it is convenient to write the color factors in terms of $\tilde f^{ABC} = \sqrt{2} i f^{ABC}$.}
\bea 
c_t = \tilde f^{\ha \  \hdd}_{\ \haa} \tilde f^{\ha \  \hgg }_{ \ \hbb} \,,  
\qquad  c_u = \tilde f^{\ha \ \hgg}_{\  \haa} \tilde f^{\ha \ \hdd}_{ \ \hbb} \,,  \qquad c_s = \tilde f^{\hgg  \hdd}_{\  \ \hee} \tilde f^{\hee}_{\ \haa \hbb} \,, 
\eea
while the inverse propagators are
\bea 
D_t = (p_1+p_4)^2 \, ,  \qquad 
D_u = (p_1+p_3)^2 \,  , \qquad  
D_s = (p_1+p_2)^2 - (2m)^2 \,.  
\label{4massprops} 
\eea
To understand the mass $(2m)$ in the massive channel it is useful to recall that
masses have been assigned as momenta in some additional dimensions. Because of this, masses are conserved at each vertex.
Since the color factor $c_s$ contains two fields of the same complex representation of masses $m$ meeting at a vertex, the third field 
must necessarily have mass $2m$. 
The kinematic numerators are given by:
\bea 
&&  n_t =   - p_1 \cdot p_2    + p_1 \cdot p_3 + 2 m^2   \,, \qquad 
      n_u =  - 2 p_1 \cdot p_2 + m^2  - p_1 \cdot p_3   \,, \no \\
&&  n_s = p_1 \cdot p_2 + 2 p_1 \cdot p_3 +  m^2  \,. 
\label{numstilde}
\eea
These numerators can be obtained from (\ref{s_num}) by assigning momenta along a single compact dimension or, alternatively, from the YM-scalar Lagrangian (\ref{L-matterscalar}) with $a=0$.

\begin{homework}
	Modify the example discussed above by introducing a VEV that corresponds to compact momenta along two compact dimensions. Write the numerators for four-scalar amplitudes. 	
\end{homework}

We note that the $s$-channel color factor is zero because there does not exist an invariant gauge-group object with two bifundamental 
and one anti-bifundamental indices.\footnote{In a standard formulation of the Higgs mechanism, 
this channel does not appear in the amplitude because the necessary vertices are absent from the Lagrangian.}
However, the corresponding numerator factor is nonzero. Alternatively stated, the color factors obey 
two-term algebraic relations, while numerator factors obey three-term relations. This observation affects the choice of the second gauge-theory 
factor entering the double-copy construction, which must have an identically-vanishing $s$-channel numerator.

We choose a non-supersymmetric theory with one complex massive scalar transforming in the representation conjugate to the one of the 
Coulomb-branch theory (see Table \ref{TabExHiggs}). Its Lagrangian is 
\bea 
{\cal L} & \! \! \! \! =& \! \! \! \! - {1 \over 4} \tilde F^{\ha}_{\mu \nu} \tilde F^{\ha \mu \nu} + {1 \over 2} D_\mu \phi^{a\ha} D^{\mu} \phi^{a\ha} 
+  D_\mu \overline{\varphi} D^{\mu} \varphi - m^2 \overline{\varphi} \varphi  -{g^2 \over 4} f^{\ha\hb\he}f^{\hc\hd\he} \phi^{a\ha} \phi^{b\hb}  \phi^{a\hc} \phi^{b\hd} \no \\
&& \null - {g^2 \over 2} (\overline{\varphi} \tR^{\ha} \varphi) (\overline{\varphi} \tR^{\ha} \varphi) +  
{g^2 } \phi^{a\ha} \phi^{a\hb} \overline{\varphi} \tR^{\ha} \tR^{\hb} \varphi + g \lambda \phi^{2\hat a} \overline{\varphi} \tR^{\ha} \varphi \,, 
\eea 
where $a,b=1,2$, $\tR$ are gauge-group representation matrices for the massive scalars, and only $\phi^{2\ha}$ enters the trilinear scalar couplings. 
The reason for this latter choice is that we want a 
construction which manifestly uplifts to five dimensions. 
Without any loss of generality we can rotate the other scalars which appear in the trilinear couplings into $\phi^{2\ha}$. 
One can check that numerators of this theory obey a two-term relation.

Putting all together, the spectrum of the resulting supergravity theory is given by \cite{Chiodaroli2015rdg}:
\begin{align}
A^{-1}_- &=  \bar \phi \otimes A_- \,,  &  h_- &= A_- \otimes A_- \,,   &   A^{-1}_+ &=  \phi \otimes A_+ \,, &  h_+ &= A_+ \otimes A_+ \,,  \no \\
A^0_- &= \phi \otimes A_-  \,,  & i \bar z^0 &= A_+ \otimes A_- \,,  &  A^0_+ &= \bar\phi \otimes A_+  \,, &  -i  z^0 &= A_- \otimes A_+ \,, \no \\
A^A_- &= A_- \otimes \phi^A  \,, & i \bar z^A &= \bar \phi \otimes \phi^A \,, &    A^A_+ &= A_+ \otimes \phi^A  \,, &  -i z^A &= \phi \otimes \phi^A \,, \no \\  
W_i &= W_i \otimes \varphi \,, &\varphi &= \varphi \otimes \varphi \,. & & & &
\end{align}
with massive fields given by ${\cal R} \otimes \overline{\cal R}$ bilinears (the index $i$ runs over the massive-vector three physical polarizations). Note that this construction has two free parameters: the mass $m$ and the constant $\lambda$ in the trilinear scalar couplings. Comparison with amplitudes from the Higgsed supergravity Lagrangian leads to the identification (\ref{YMEcouplingidentification}).\footnote{To obtain a Higgsed supergravity, we take $\lambda>0$ strictly.}
The masses of supergravity fields are the same as those of the gauge-theory fields from which they are constructed.
In turn, this determines the choice of scalar base-point for the supergravity perturbative expansion that matches the result of the double copy.
Given the presence of two massive $W$ bosons in the supergravity spectrum, the supergravity gauge-symmetry breaking is $SU(2)\rightarrow U(1)$.

This is arguably the most straightforward example of Higgsed supergravity constructed as double copy.
In the general case, we need to consider a generic breaking of the Coulomb-branch theory.
The structure constants, generators and Clebsch-Gordan coefficients obey relations inherited from the Jacobi relations of the original gauge group. A first set of relations is
\bea
f^{\hd \ha \hc} f^{\hc \hb \he} - f^{\hd \hb \hc} f^{\hc \ha \he}  &=& f^{\ha \hb \hc} f^{\hd \hc \he} \,, \nn \\
f^{\ha \ \hbb}_{\ \hgg } f^{\hb \ \hgg }_{\ \haa}-  f^{\hb \ \hbb}_{\ \hgg }f^{\ha \ \hgg }_{\ \haa}  &=& f^{\ha \hb \hc} f^{\hc \ \hbb }_{\ \haa} \, , \no \\
f^{\ha \ \hgg}_{\ \hee} f^{\hee \ \hbb}_{\ \hdd} -
f^{\ha \ \hbb}_{\ \hee} f^{\hee \ \hgg }_{\ \hdd} &=& 
f^{\ha \ \hee}_{\ \hdd} f^{\hgg \ \hbb }_{\ \hee}  \, . 
\label{fck}
\eea
These relations are necessary to ensure gauge invariance. The Clebsch-Gordan coefficients $f^{\hgg \ \hbb }_{\ \hee}$ need to obey additional identities: 
\bea
f^{\haa \ \hgg}_{\ \hee} f^{\hee \ \hbb}_{\ \hdd}  -
f^{\haa \ \hbb}_{\ \hee} f^{\hee \ \hgg }_{\ \hdd}  &=& 
f^{\haa \ \hee}_{\ \hdd} \, f^{\hgg \ \hbb}_{\ \hee} \,, \nn\\ 
\Big(f^{\hbb \ \hee}_{\ \hgg} f_{\hee \ \hdd}^{\ \haa}
+ f^{\haa \ \hee}_{\ \hdd} f_{\hee \ \hgg}^{\ \hbb} 
+ f^{\ha \ \hbb}_{\ \hgg} f^{\ha \ \haa}_{\ \hdd}\Big)- (\haa\leftrightarrow\hbb)
&=& f^{\haa \ \hbb}_{\ \hee} f_{\hdd \ \hgg}^{\ \hee}  \, . 
\label{extraf} 
\eea
The seven-term identity is to be thought of as a compact way of writing a set of three- and two-term identities. The general construction for Higgsed supergravities proceeds as follows~\cite{Chiodaroli2015rdg}:
\begin{itemize}
	\item One introduces a non-supersymmetric gauge theory with massive scalars and imposes the identities \eqref{fck} and \eqref{extraf} 
	on its numerator factors. Note that the numerators of the Coulomb-branch theory need not obey the same identities. 
	\item Masses need to be matched on both gauge-theory factors. For gaugings that uplift to five dimensions, the Higgs mechanism 
	requires that the Coulomb-branch theory masses be proportional to a preferred $U(1)$ gauge generator (given by the direction of 
	the VEV). Imposing \ck duality results in demanding than the masses in the explicitly-broken massive-scalar theory also be 
	proportional to a preferred $U(1)$ global generator (in our example, the $U(1)$ acting as a phase rotation on the complex scalars).
	\item In general, the symmetry-breaking pattern (number of factors in the gauge group, number of matter representation, 
	existence of Clebsch-Gordan coefficients corresponding to a given triplet of representations) from the Coulomb-branch gauge theory 
	matches both that of the explicitly-broken theory and that of the supergravity theory.
	\item Identification of the supergravity relies on the unbroken limit (setting all masses to zero), as well as on the symmetry breaking information encoded in the trilinear scalar couplings. 
\end{itemize}
A list of constructions for Higgsed supergravities with various amounts of supersymmetry can be found in Table \ref{TheoryConstructions2}. 

\begin{table}[t]
	\centering
	\begin{tabular}{l@{$\quad$}|@{$\quad$}c@{$\quad$}|@{$\quad$}c}
		Gravity coupled to \cancel{YM} & Left gauge theory  & Right gauge theory  
		\\
		\hline & &  \\[-13pt]
		\hline 
		$\raisebox{.3 cm}{ \vphantom{|} } {\cal N}=4$ \cancel{YM}E supergravity & ${\cal N}=4$ S\cancel{YM}  & YM + $\cancel{\phi^3}$ 
		\\
		${\cal N}=2$ \cancel{YM}E supergravity (gen.Jordan) & ${\cal N}=2$ S\cancel{YM}  & YM + $\cancel{\phi^3}$ 
		\\
		${\cal N}=0$ $\cancel{\rm YM}_{\rm DR}$-E + dilaton + $B^{\mu\nu}$ &  $\cancel{\rm YM}_{\rm DR}$  & YM + $\cancel{\phi^3}$ 
		\\
		\hline
	\end{tabular}
	\caption{\small New double-copy constructions corresponding to spontaneously-broken YME gravity theories for different 
		amounts of supersymmetry  \cite{Chiodaroli2015rdg}. The dimensionally-reduced YM$_{\rm DR}$ theory must have at least one scalar to provide the 
		VEV responsible for spontaneous symmetry breaking. }
	\label{TheoryConstructions2}
\end{table}

\begin{homework}
	What would happen if we attempted to double copy two Coulomb-branch theories realized both in terms of compact momenta? 
	Find out as much information as possible on the resulting gravity theory. 	
\end{homework}

\subsubsection{Gauged supergravities\label{gaugedsugras}}

\begin{table}
	\begin{center}	
		\begin{tabular}{c@{$\quad$}|@{$\quad$}c@{$\quad$}|@{$\quad$}c}
			Fields & Representation & Mass \\
			\hline
			\hline
			$\raisebox{.3 cm}{ \vphantom{|}} (A_\mu, \, \bar \phi^a )$ & $\cal G$ &  $0$ \\
			
			$(W_\mu, \varphi^s )$ & $\cal R$ &  $m$ \\
			
			$(\overline{W}_\mu, \overline{\varphi}^s)$ & $\overline{\cal R}$ &  $-m$ \\
	\hline
		\end{tabular}
		\hskip 0.5in
		\begin{tabular}{c@{$\quad$}|@{$\quad$}c@{$\quad$}|@{$\quad$}c}
			Fields & Representation & Mass \\
			\hline	
			\hline	
			$\raisebox{.3 cm}{ \vphantom{|}} (A_\mu, \varphi^\alpha )$ & $\cal G$ &  $0$ \\
			
			$\chi$ & $\overline{\cal R}$ &  $m$ \\
			
			$\overline{\chi}$ & $\cal R$ &  $-m$ \\
			\hline
		\end{tabular}
		\caption{Fields in gauge-theory factors for a simple example of a double-copy construction for $\cN=2$ gauged supergravities.\label{TabExGauged} }
	\end{center}
\end{table}

In this subsection, we consider an important variant of the construction for Higgsed supergravities. In a sense, the construction outlined in the previous 
subsection can be regarded as the simplest double-copy prescription which produces a gravity with massive vector fields. 
Various details of the construction can then be traced back to the requirement that such massive vectors obey the relevant
Ward identities corresponding to spontaneous symmetry breaking.   

Along similar lines, we might want to consider double copies leading to massive spin-$3/2$ fields.
It turns out that the construction will lead to gauged supergravities---supergravities in which a subgroup 
of the $R$ symmetry is promoted to a gauge symmetry under which gravitini are charged. 
In a gauged supergravity  with a  Minkowski vacuum, minimal coupling between gravitini and gauge vector produces a nonzero amplitude of the form
\begin{equation}
{\cal M}_3 \big(1 \overline{\psi}_i , 2\psi_j, 3 A^a \big) = i g_{{R}} t^a_{ij} \bar v_1^\mu \cancel{\varepsilon}_3 v_{2\mu} + {\cal O}\big((g_{R})^0 \big)
\label{gravitinoamp} \, .
\end{equation}  
$g_{{R}}$ is the coupling constant and $v_{l\mu}~(l=1,2)$ are the gravitini's polarization spinor-vectors. The matrices $t^a_{ij}$ 
generate the gauged $R$-symmetry subgroup acting nontrivially on the gravitini.
The above amplitude does not vanish with the replacement 
\begin{equation} v_{l\mu} \rightarrow v_{l\mu} + k_{l\mu} \epsilon \,,\qquad \qquad \cancel{k}_l \epsilon=0 \, .
\end{equation}
Since this replacement correspond to an linearized supersymmetry transformation, the presence of a nonzero amplitude of the form (\ref{gravitinoamp})
signifies that supersymmetry is spontaneously broken. Indeed, $R$-symmetry gauging and spontaneous supersymmetry breaking go hand in hand for
supergravities which admit Minkowski vacua. In turn, the fact that supersymmetry is spontaneously broken results in (some) massive gravitini. 
This can be understood by comparing the number of physical polarizations of our gravitini; because some of the supersymmetry generators are broken, they cannot be used to eliminate components of gravitini. Some of the gravitini will have four physical polarizations and must therefore become massive.

This observation provides a hint on how to find a double-copy construction for amplitudes of gauged supergravities with Minkowski vacua. 
In analogy with the previous subsection, we start by seeking a construction that has the following two properties:
\begin{enumerate}
	\item contains massive spin-$3/2$ fields, realized as the double copies of a massive $W$ bosons in one gauge theory with  massive fermions in the other;
	\item reduces to the construction of the corresponding ungauged supergravity in the massless limit.
\end{enumerate}
The simplest realization with these properties has the schematic form
\begin{eqnarray}
\Big(\text{Gauged Supergravity} \Big) = \Big(  {\text{Coulomb-branch YM}} \Big) \otimes \Big(\text{s}\cancel{\text{upe}\vphantom{YM}}\text{r}   \  \text{YM}  \Big) \, ,  
\end{eqnarray}
where the second factor stands for a theory with explicit supersymmetry breaking and massive fermions.

Next, we discuss the two gauge theories separately, focusing on the particular case of $\cN=2$ supersymmetry and using the toolbox introduced in \sects{SecOrbifold}{masses}. Unlike the case of Higgsed YME theories, the Coulomb-branch theory is non-supersymmetric; we will take it  
to be a pure YM theory coupled with $n$ scalars, obtained from dimensional reduction from $D=(n+4)$ dimensions. 
The corresponding VEV will be parameterized by a $n$-dimensional vector which we will denote as $V^a$.
The theory with supersymmetry explicitly broken by fermion masses is obtained by starting with four-dimensional $SU(N+M)$ ${\cal N}=2$ SYM theory 
and spontaneously breaking the gauge group to  $G=SU(N)\times SU(M) \times U(1)$ by introducing  a VEV
	\begin{equation}
	\langle \varphi_\alpha \rangle = \widetilde V_\alpha \times \text{Diag}\Big(\frac{1}{N}I_N, -\frac{1}{M}I_M\Big) \, . 
	\label{vev2}
	\end{equation} 
We then orbifold the theory by a $\mathbb{Z}_2$   generated by $\gamma = \text{diag}(I_N, -I_M)$:
	\be
	A_\mu \mapsto \gamma A_\mu \gamma^{-1} \,,
	\qquad
	\chi\mapsto -  \gamma \chi \gamma^{-1} \,,
	\qquad
	\varphi\mapsto \gamma \varphi \gamma^{-1}
	\,.
	\label{orbifold}
	\ee
	Note that, as explained in  \sect{SecOrbifold}, this operation preserves \ck duality.
The VEVs in both theories are chosen to have the same magnitude $(V^a)^2 = (\widetilde V_\alpha)^2$, so that the two 
theories have common mass spectra. The explicitly-broken theory has Lagrangian
\begin{eqnarray}
{\cal L}_{\cN \! \cancel{\, =\, }  2} \! = \! {1 \over g^2}\Tr\Big[ \! - \! \frac{1}{4}F_{\mu \nu}F^{\mu \nu} \! - \! \frac{1}{2}D_\mu \varphi_\alpha D^\mu \varphi_\alpha \! + \! \frac{1}{4} [\varphi_\alpha, \varphi_\beta]^2 \! + \! \frac{i}{2}{\overline\chi} \Gamma^\mu D_\mu \chi \! + \!  \frac{1}{2} {\overline\chi} \Gamma^\alpha [\varphi_\alpha \! + \! \langle \varphi_\alpha \rangle , \chi] \Big] \no \,,\\
\label{LNeq2}              
\end{eqnarray}
where $\chi$ is a six-dimensional Weyl fermion and $\alpha,\beta=5,6$.  
The fields in the gauge-theory factors are listed in Table \ref{TabExGauged}.
Denoting with $\xi_\mu$ the massive gravitino field on the supergravity side, the fermionic states have the following double-copy origin:
\begin{eqnarray}
&&  \xi_\mu = W_\mu  \otimes \chi  ~  - ~ 
W_\nu \otimes \Big({\gamma_\mu \over 3}  -  {ip_\mu\over 3 m} \Big)\gamma^\nu \chi \,, \no \\   
&& 
\xi =  W_\nu \otimes \gamma^\nu \chi
\,, \qquad \quad
(U \lambda)^s =  \varphi^s \otimes \chi   \,.  \quad
\end{eqnarray}
The combination on the first line is manifestly transverse and $\gamma$-traceless. $U$ is a unitary matrix diagonalizing the spin-$1/2$ mass terms.
Last, the $U(1)_R$ gauge vector is:
\begin{eqnarray}
A^{U(1)_R}_+ &=& - A_+ \otimes {\varphi^6 } \,\, \pm \,\,  \phi^2 \otimes A_+   \,.
\end{eqnarray}
We note that the massless limit leads an ungauged theory belonging to the Generic Jordan family discussed in \sect{jordan}.
The freedom of choosing the $U(1)_R$ gauge group corresponds to the choice of VEVs in the two gauge theories entering the construction.
As for Higgsed supergravities, this is the simplest example of the double-copy construction. However, it is immediate to generalize 
the construction reviewed here to $U(1)_R$ gaugings of $\cN=4,6,8$ supergravity by adjusting the supersymmetry of the gauge-theory factors.

\begin{homework}
	Introduce massive spinor-helicity notation by splitting massive momenta as	
	$p_i = p_i^\perp- {m^2 \over 2 p_i \cdot q} q $.  Here $q$ is a reference momentum and $p_i^\perp, q$ are both massless.
	Write massive spinor polarizations as 
	$v^t_+ =\big(  |i^\perp] , \  m {|q \rangle  / \langle i^\perp q \rangle} \big)$ and $
	v^t_- = \big( m {|q ]  / [ i^\perp q ] } , \ |i^\perp\rangle  \big)$.
	Show that an  amplitudes involving massive gravitini with $\pm$ polarizations and the $A^{-1}$ vector field can be written as
	\begin{eqnarray}
	& \!\!\! & \!\!\! {\cal M}^\tree_3 \big( 1 \bar \xi_+ , 2 \xi_-, 3 A^{-1}_+ \big) =  -\sqrt{2} i  m \Omega  {\langle   2^\perp q \rangle \over \langle  1^\perp q \rangle} \,, 
\qquad \Omega={[3^\perp 1^\perp]^3 \over [1^\perp 2^\perp] [2^\perp  3^\perp ]} \,.
	\end{eqnarray}
\end{homework}

The construction outlined above can be generalized to allow gauging of  nonabelian subgroups of the $R$ symmetry.
To do so, we need to consider double copies that \cite{Chiodaroli2018dbu}:
\begin{enumerate}
	\item contain massive spin-$3/2$ fields;
	\item give the suitable  ungauged supergravity in the massless limit;
	\item involve gauge theories with trilinear scalar couplings.
\end{enumerate}
The first two requirements parallel the abelian example discussed at the beginning of this section, while the last property 
reflects the fact that cubic couplings involving gauge-theory scalars result in nonabelian interactions in the theory from the double copy, as seen in the example of the construction for YME theories. 
 As before, the gauge theories entering the double copy are obtained from higher dimension  with a combination of Higgsing and orbifolding. 
 We specialize to the case of gaugings of $\cN=8$ supergravity and start by writing both copies of $\cN=4$ SYM as the dimensional reduction of SYM theories in ten dimensions. 
 For the left gauge-theory factor, we choose undeformed $\cN=4$ SYM theory on the Coulomb branch.
 In the right gauge-theory factor, we  
 introduce a massive deformation  which involves trilinear scalar couplings, 
\bea
{\cal L} \!\!\!\! &=& \!\!\!\! - {1 \over 4} (F^{\hat a}_{\mu \nu})^2 + {1 \over 2} (D_\mu \phi^{\hat a I})^2 
-{1\over 2 } m^2_{IJ}  \phi^{\hat a I}  \phi^{\hat a J} \!\!
- {g^2 \over 4} f^{\hat a \hat b \hat e} f^{\hat c \hat d \hat e} \phi^{\hat a I} \phi^{\hat b J} \phi^{\hat c I} \phi^{\hat d J} - {g\lambda \over 3!} f^{\hat a \hat b \hat c} F^{IJK} \phi^{\hat a I}  \phi^{\hat b J}  \phi^{\hat c K}  \no \\
&& \null + {i \over 2} \bar \psi \cancel{D} \psi - {1 \over 2}  \bar \psi M \psi  + {g \over 2} \phi^{\hat a I} \bar \psi \Gamma^{I} \tR^{\hat a} \psi \, .
\label{actionGSGgt}
\eea
\ck duality of the two-scalar-two-fermion four-point amplitude demands that the fermionic mass matrix $M$ obey the relation
\be 
\big[ \Gamma^I , \big\{ \Gamma^J , M  \big\} \big]  + i \lambda F^{IJK} \Gamma^K  = 0 \,, \label{masterintro} 
\ee
where $\Gamma^I$ are the Dirac matrices in higher dimensions and $F^{IJK}$ are related to the structure constants of the supergravity gauge group. 
The right gauge theory is then Higgsed and orbifolded, following the same strategy outlined in the $\cN=2$ example. 
Double copies involving  theories obtained with this prescription need however to satisfy additional consistency requirements. 

Referring to the literature for the general construction~\cite{Chiodaroli2018dbu}, we consider the action \eqref{actionGSGgt} with an $SU(3N)$ gauge group  and the deformation 
\be 
M = i {g \over 4}  \Gamma^{789}  \,, \qquad \lambda F^{789} = g 
\label{solution-constraint}    \,. 
\ee
This deformation breaks ten-dimensional Lorentz invariance to $SO(3) \times SO(6,1)$ and can be uplifted to seven dimensions. 
Starting from $D=7$, we take a a $\mathbb{Z}_5$ orbifold which acts as
\begin{equation}
\psi \rightarrow e^{{2\pi \over 5}\Gamma_{56}} g^\dagger \psi g \,, \qquad  \phi^I \rightarrow R^{IJ}\big( {4 \pi \over 5} \big)  g^\dagger \phi^J g, \ \  \qquad g = \text{diag}\big( I_N, e^{i{2\pi \over 5}} I_N, e^{i{4\pi \over 5}} I_N \big)
\end{equation}
where $I,J=5,6$ and $R_{56}$ generates a rotation in the 5-6 plane.  We also
take the scalar mass-matrix to be
\begin{equation}
m_{55}= m = m_{66} \,, \qquad {m_{IJ}=0} \ \ \text{ otherwise} \,.
\end{equation}
After the projection, the fields of the theory are organized schematically as:
\begin{equation}
\left( \begin{array}{ccc} A_\mu , \phi^i & \psi^r & \phi^+ \\
\tilde \psi^{r'} & A_\mu, \phi^i & \psi^r \\
\phi^- & \tilde \psi^{r'} & A_\mu, \phi^i   \end{array} \right) \,,
\end{equation}
where $i=4,7,8,9$, $r=1,2$, $r'=3,4$, and $\phi^\pm= \phi^5\pm i \phi^6$. In the above equation, we represent the fields surviving the projection 
as entries the in $3N\times 3N$ matrices originating from the parent theory; each entry is an $N\times N$ block. 
To obtain a number of states that reproduces the spectrum of $\cN=8$ supergravity, we need to combine the representations $(\boldsymbol N,\boldsymbol{\bar N}, \boldsymbol 1)$ 
with $(\boldsymbol 1, \boldsymbol N,\boldsymbol{\bar N})$ and the representation $ (\boldsymbol{\bar N}, \boldsymbol N, \boldsymbol 1)$ with $(\boldsymbol 1, \boldsymbol{\bar N}, \boldsymbol N)$ into a (reducible) representation which is denoted as ${\cal R}_1$.
This can be realized by rewriting the Lagrangian in a way that only representation matrices for ${\cal R}_1$ appear explicitly. 

In the left theory,  we take a 
a VEV of the form
\begin{equation}
\langle \phi^4 \rangle = \text{diag} \big( u_1 I_N, u_2 I_N, u_3 I_N \big)  \,, \qquad u_1+u_2+u_3=0 \,.
\end{equation}  
Since the two irreducible representations that have been combined into ${\cal R}_1$ need to have the same mass, we get a condition involving the VEV parameters,
\begin{equation}
u_1 - u_2 = u_2 - u_3 \quad \rightarrow \quad u_2 = {u_1+u_3\over 2} = 0 \,.
\end{equation} 
In addition, we get the following conditions by matching the mass spectra of the two theories:
\begin{eqnarray}
M^2 = - u_1^2  &\qquad & m^2  = 4 u_1^2 \,.
\end{eqnarray} 
We list the fields from the double copy with their respective mass spectra in Table \ref{Tab-N0}.

\begin{table}[tb]
	\begin{center}
		\begin{tabular}{c@{$\qquad$}|@{$\qquad$}c@{$\qquad$}|@{$\qquad$}c@{$\qquad$}|@{$\qquad$}c@{$\qquad$}|@{$\qquad$}c}
			Rep. & R & L & Supergravity fields & mass${}^2$  \\	
			\hline \hline 
			$\raisebox{.3 cm}{ \vphantom{|}} {\cal G}$  & ${\cal V}^0_{\cN=4}$ & $A_\mu \oplus \phi^i$ & ${\cal H}_{\cN=4} \oplus 4 {\cal V}^0_{\cN=4}$  & $0$ \\
			${\cal R}_1$  & ${\cal V}^m_{\cN=4}$ & $\psi^r$ & $2{\Psi}^m_{\cN=4}$ & $u_1^2$  \\  
			$\bar {\cal R}_1$ &  ${\cal V}^m_{\cN=4}$ & $\tilde \psi^{r'}$ & $2{\Psi}^m_{\cN=4}$ & $u_1^2$  \\  
			${\cal R}_2$ &  ${\cal V}^m_{\cN=4}$& $\phi^+$ & ${\cal V}^m_{\cN=4}$ & $4u_1^2$  \\  
			$\bar {\cal R}_2$  & ${\cal V}^m_{\cN=4}$ & $\phi^-$ & ${\cal V}^m_{\cN=4}$ & $4u_1^2$  \\  
\hline
		\end{tabular}
		\caption{Fields and mass spectra for gauging of $\cN=8$ supergravity with $\cN=4$ residual supersymmetry \cite{Chiodaroli2018dbu}.\label{Tab-N0}}
	\end{center}
\end{table}	
The vacuum of this theory has an unbroken $SU(2)\times U(1)$ gauge group which is reflected by the $F^{IJK}$ tensors in (\ref{solution-constraint}).  $\cN=4$ unbroken supersymmetry is inherited from the Coulomb-branch gauge-theory factor. Many additional examples can be worked out along similar lines. A complete classification of double-copy-constructible gaugings is currently an open problem.

\subsubsection{Conformal supergravity}

A double-copy construction for conformal gravity was set forth in Ref.~\cite{Johansson2017srf} and further investigated in Ref.~\cite{JohanssonConformal}. Before we get into the details of that construction, let us review some general properties of conformal gravity. The simplest model is that of Weyl gravity, which has the four-derivative action
\be
S = -\frac{1}{\varkappa^2} \int d^4x \sqrt{-g} \, (W_{\mu \nu \rho \sigma})^2 \ ,
\ee 
where $W_{\mu \nu \rho \sigma}$ is the Weyl curvature tensor, and $\varkappa$ is a dimensionless coupling. The action is invariant under local rescaling of the metric, $g_{\mu \nu} \rightarrow \Omega(x) g_{\mu \nu}$; more generally the theory possesses local conformal symmetry at the classical level. The symmetry can be extended to local superconformal symmetry by considering supergravity formulations of the Weyl theory. It is believed that $\cN=4$ is the maximum allowed supersymmetry. In contrast to expectations from SYM and ordinary two-derivative supergravity, the maximally supersymmetric theory is not unique, in fact it has an infinite number of free parameters~\cite{Butter:2016mtk}. The free parameters are encoded in a free holomorphic function that multiplies the square of the Weyl tensor,\footnote{Note that compared to Ref.~\cite{Johansson2017srf} we are using a convention where we have swapped $i\bar\tau$ with $-i\tau$. This changes the sign of the axion field, which is physically unobservable.} 
\be
- \varkappa^2 \sqrt{-g}^{-1}{\cal L}_{\cN =4} = f(\tau) (W_{\mu \nu \rho \sigma}^{+})^2 + \overline{f(\tau)} (W_{\mu \nu \rho \sigma}^{-})^2 + \ldots\,,
\ee
where $W_{\mu \nu \rho \sigma}^{\pm} =W_{\mu \nu \rho \sigma}/2 \pm  (i \sqrt{-g}/4) W_{\mu \nu}^{\ \ \lambda \kappa} \epsilon_ {\lambda \kappa \rho \sigma}$ is the (anti-)selfdual Weyl tensor and the complex scalar $\tau=i e^{-\phi}+\chi$ is the dilaton-axion field. The ellipsis denotes additional terms that are fully constrained by the superconformal symmetry. The choice $f(\tau)=1$ corresponds to the supersymmetrization of the Weyl theory, and it is usually called {\it minimal} conformal supergravity. When $f(\tau)$ is not constant, the theory corresponds to {\it non-minimal} conformal supergravity. The double-copy constructions that we will consider corresponds to the two cases~\cite{JohanssonConformal}: 
\bea
f(\tau)&=&-i \tau\,~~~~~  (\cN=4~\text{Berkovits-Witten theory})\,, \nn \\ 
f(\tau)&=&1 \,~~~~~~~~~ (\cN=4~\text{minimal conformal supergravity})\,. 
\eea
These two cases are special. The Berkovits-Witten
theory~\cite{Berkovits:2004jj} corresponds to the unique conformal
supergravity theory that has an uplift to 10
dimensions~\cite{Johansson2017srf,JohanssonConformal,Azevedo:2017lkz,
deRoo:1991at}. At tree level, the minimal theory has the same
$SU(1,1)$ electromagnetic duality symmetry as $\cN=4$ supergravity,
and certain all-multiplicity tree-level amplitudes are the same as in
that theory. All $\cN=4$ conformal supergravities are expected to be
anomalous at loop level unless they are coupled to four vector
multiplets~\cite{Fradkin:1983tg,Tseytlin:2017qfd}.

For reasons of conciseness, we will restrict the discussion in this section to scattering amplitudes where the external states are plane waves. As is well known, the four-derivative action of conformal gravity also permits other types of asymptotic states, see e.g. Refs.~\cite{Adamo:2018srx, JohanssonConformal} for further details. The double copy that gives amplitudes in the Berkovits-Witten conformal supergravity theory has the schematic form
\be
\big( \text{Berkovits-Witten CSG} \big)=\big({\rm SYM}\big) \otimes \big(\textrm{$(DF)^2$-theory}\big)\, ,
\label{CG_DC}
\ee
where SYM is the maximally supersymmetric Yang-Mills theory, and the $(DF)^2$ theory is a bosonic gauge theory with dimension-six operators which has the following Lagrangian~\cite{Johansson2017srf}:
\begin{align}
{\cal L}_{(DF)^2}&= \frac{1}{2}(D_{\mu} F^{a\, \mu \nu})^2  - \frac{g}{3} \,   F^3+ \frac{1}{2}(D_{\mu} \varphi^{\alpha})^2  + \frac{g}{2}  \,  C^{\alpha ab}  \varphi^{ \alpha}   F_{\mu \nu}^a F^{b\, \mu \nu }  +  \frac{g}{3!}  \, d^{\alpha \beta \gamma}   \varphi^{ \alpha}  \varphi^{ \beta} \varphi^{ \gamma} \notag  \,.
\end{align}
The vector $A_{\mu}^a$ transforms in the adjoint representation of a gauge group $G$ with indices $a,b,c$. $\varphi^{\alpha}$ are additional scalars transforming in a real representation for which  $C^{\alpha ab}$ and $d^{\alpha \beta \gamma}$ are invariant tensors. We have used the short-hand notation $F^3 = f^{ a  b  c} F^{a \nu}_\mu  F^{b \lambda}_\nu F^{c \mu}_\lambda$.
It should be noted that $C^{\alpha ab}, T^a_{\cal R}$ and $d^{\alpha \beta \gamma}$ are implicitly defined through the two relations:
\bea
&&C^{\alpha ab}C^{\alpha cd} = f^{ace}f^{edb}+ f^{ade}f^{ecb}\,, \label{ClebschGordan}  \\
&&C^{\alpha ab}d^{\alpha \beta \gamma}= (T_{\cal R}^a)^{\beta \alpha} (T_{\cal R}^b)^{\alpha \gamma}+ C^{\beta ac} C^{\gamma cb} + (a \leftrightarrow b)\,, \nn
\eea
which are sufficient relations for expressing tree-level gluon amplitudes in terms only $f^{abc}$ tensors.

A massive deformation of this theory was also introduced in Ref.~\cite{Johansson2017srf} and is defined by the Lagrangian:
\begin{align}
	{\cal L}_{(DF)^2 + {\rm YM}}&= \frac{1}{2}(D_{\mu} F^{a\, \mu \nu})^2  - \frac{g}{3} \,   F^3+ \frac{1}{2}(D_{\mu} \varphi^{\alpha})^2  + \frac{g}{2}  \,  C^{\alpha ab}  \varphi^{ \alpha}   F_{\mu \nu}^a F^{b\, \mu \nu }  +  \frac{g}{3!}  \, d^{\alpha \beta \gamma}   \varphi^{ \alpha}  \varphi^{ \beta} \varphi^{ \gamma} \notag \\ 
	& \ \ \ \   -   \frac{1}{2} m^2 (\varphi^{\alpha})^2- \frac{1}{4} m^2 (F^a_{\mu \nu})^2\,.
	\label{massdefL}
\end{align}
This  theory interpolates between the $(DF)^2$ theory  and a pure YM theory and has  the mass as a free parameter.
Along similar lines, the theory (\ref{massdefL}) can be further augmented by introducing adjoint scalars $\phi^{aA}$ which  
are also charged under a global group and appear in trilinear couplings which are analogous to the ones introduced for YME theories 
and nonabelian gauged supergravities:
\begin{align}
{\cal L}_{(DF)^2+{\rm YM}+\phi^3} &=  \frac{1}{2}(D_{\mu} F^{a\, \mu \nu})^2  - \frac{g}{3} \, F^3+ \frac{1}{2}(D_{\mu} \varphi^{\alpha})^2  
+ \frac{g}{2}  \,  C^{\alpha ab}  \varphi^{ \alpha}   F_{\mu \nu}^a F^{b\, \mu \nu } 
+  \frac{g}{3!}  \, d^{\alpha \beta \gamma}   \varphi^{ \alpha}  \varphi^{ \beta} \varphi^{ \gamma} \notag \\
& \ \ \     -   \frac{1}{2} m^2 (\varphi^{\alpha})^2  - \frac{1}{4} m^2 (F^a_{\mu \nu})^2+ \frac{1}{2} (D_{\mu} \phi^{aA})^2 
+ \frac{g}{2} C^{\al ab} \vph^\al  \phi^{aA} \phi^{bA} 
\label{4.7} \\
& \ \ \ + \frac{g\lambda}{3!} f^{abc} F^{ABC} \phi^{aA} \phi^{bB} \phi^{cC} \,.  \notag 
\end{align}
These deformations of the $(DF)^2$ theory will also play an important role for double-copy constructions involving various string theories, 
which are reviewed in the next subsections. 
We also note that the $(DF)^2$ theory is just a representative of a large class of gauge theories with higher-dimension operators. 
An investigation of their amplitudes in the general case is an open problem; we refer the reader to~\cite{Broedel2012rc} for 
a similar construction of supergravities with higher-dimension operators and to~\cite{Azevedo:2017lkz} for a study of the $(DF)^2$ 
theory from the point of view of ambitwistor strings.

\begin{homework}
Show that three- and four-gluon color-ordered amplitudes in the $(DF)^2$ theories have the expressions
\begin{align}
	A_{(DF)^2}(1,2,3) &=  -4 (\varepsilon_1 \cdot p_2)(\varepsilon_2 \cdot p_3)(\varepsilon_3\cdot p_1) \,,
	\label{2.91}
	\\
	A_{(DF)^2}(1,2,3,4) &=    4\frac{ s^2_{12}s^2_{23}}{ s_{13} } \!
	\Big( \frac{ p_{4} {\cdot} \varepsilon_1}{s_{23}} - \frac{ p_2 {\cdot} \varepsilon_1 }{s_{12}} \Big)\!
	\Big( \frac{ p_{1} {\cdot} \varepsilon_2}{s_{12}} - \frac{ p_3 {\cdot} \varepsilon_2 }{s_{23}} \Big)\!
	\Big( \frac{ p_{2} {\cdot} \varepsilon_3}{s_{23}} - \frac{ p_4 {\cdot} \varepsilon_3 }{s_{12}} \Big)\!
	\Big( \frac{ p_{3} {\cdot} \varepsilon_4}{s_{12}} - \frac{ p_1 {\cdot} \varepsilon_4 }{s_{23}} \Big) \,,
	\notag
\end{align} 
and that they obey color-kinematics duality. Note that the products between polarization vectors, $\varepsilon_i \cdot \varepsilon_j $,  always cancel out (this is a special property of the $(DF)^2$ theory). 
\end{homework}

Finally, we will consider amplitudes in the minimal $\cN=4$ conformal supergravity theory.  For external plane waves at tree level, the relation is
\be
\big( \text{minimal CSG} \big)=\big({\rm SYM}\big) \otimes \big(\textrm{minimal  $(DF)^2$-theory}\big)\, ,
\label{CG_DC_min}
\ee
where we have truncated the bosonic gauge theory to a ``minimal'' version,
\begin{align}
{\cal L}_{{\rm min.}\,(DF)^2}&= \frac{1}{2}(D_{\mu} F^{a\, \mu \nu})^2 \,.
\end{align}
However, as the reader may confirm, the all tree-level plane-wave amplitudes in this theory vanish---a property that is also true of minimal conformal supergravity. In order to have something nonvanishing to compare with, we must deform the two theories by a mass term,
\begin{align}
{\cal L}_{{\rm min.}\,(DF)^2+{\rm YM}}&= \frac{1}{2}(D_{\mu} F^{a\, \mu \nu})^2- \frac{1}{4} m^2 (F^a_{\mu \nu})^2 \,.
\end{align}
The resulting double copy
\be
\big( \text{mass-deformed minimal CSG} \big)=\big({\rm SYM}\big) \otimes \big(\textrm{minimal  $(DF)^2$ + YM}\big)\, ,
\label{CG_DC_massdef}
\ee
gives amplitudes in a mass-deformed minimal $\cN=4$ theory that interpolates between (Weyl)${}^2$ and a Ricci scalar term
\be
- \varkappa^2 \sqrt{-g}^{-1}{\cal L}_{\cN =4} = (W_{\mu \nu \rho \sigma})^2 - 2m^2 R + \ldots
\ee
where the ellipsis are additional terms fixed by supersymmetry. The tree amplitudes, for external plane waves, in the mass-deformed theories, are proportional to the corresponding amplitudes in ordinary YM and supergravity~\cite{JohanssonConformal},
\bea
A_{{\rm min.}\,(DF)^2+{\rm YM}} &= & m^2 A_{{\rm YM}}\,, \nn \\
M_{\text{mass-def.\,min.\,CSG}} &= & m^2 M_{{\rm SG}}\,.
\eea

In addition to considering $\cN=4$ conformal supergravity, the corresponding theories with reduced supersymmetry $\cN=0,1,2$ can be obtained by replacing the $\cN=4$ SYM factor in the double copies~(\ref{CG_DC}), (\ref{CG_DC_min}) and (\ref{CG_DC_massdef}) by $\cN=0,1,2$ (S)YM. The $\cN=0,1,2$ conformal (super)gravity theories will not be pure, as they will inherit a dilaton-axion multiplet from the  $\cN=4$ theory, in close analogy to the case of ordinary two-derivative supergravity theories.

\subsubsection{Perturbative string theories}

In refs. \cite{MafraBCJAmplString,Broedel2013tta}, disk integrals that appear in open-string amplitudes were organized in terms of building 
blocks 
\be
Z_\sigma(\rho(1,2,\ldots,n)) = (2 \ap)^{n-3} \! \! \! \! \! \! \! \! \! \! \! \! \! \! \! \! \! \! \! \!  \! \! \! \!\int \limits_{\sigma \, \{  - \infty \leq z_{1}\leq z_{2} \leq \ldots \leq z_{n}\leq \infty \}} \! \! \! \! \! \! \! \! \! \! \! \! \! \! \! \! \! \! \! \! \frac{d z_1 \, d z_2 \, \ldots\, d z_{n}}{{\rm vol}({\rm SL}(2,\mathbb{R}))} \ 
\frac{ \prod_{i<j}^{n} |z_{ij}|^{\alpha' s_{ij}}  }{ \rho \, \{ z_{12} z_{23} \cdots z_{n-1,n} z_{n,1} \} } \, .
\ee
Here we use the notation $z_{ij}=z_i-z_j$ and ${\rm vol}({\rm SL}(2,\mathbb R))$ refers to fixing three punctures on the disk to $z_i,z_j,z_k\rightarrow (0,1,\infty) $ while introducing a Jacobian $|z_{ij}z_{ik} z_{jk}|$. Such building blocks depend on two permutations $\sigma,\rho \in S_n$ and obey field-theory  BCJ relations with respect to~$\rho$,
\be 
\sum_{j=2}^{n-1}  (p_1\cdot p_{23\ldots j})Z_\sigma (2,3,\ldots,j,1,j+1, \ldots, n) = 0 \,,
\label{ftarel}
\ee
and string-theory monodromy relations~\cite{Monodromy,Stieberger:2009hq} with respect to $\sigma$,
\be \sum_{j=1}^{n-1}  e^{ 2 i \pi \alpha' p_1\cdot p_{23\ldots j}}Z_{(2,3,\ldots,j,1,j+1, \ldots, n)}(\rho) = 0 \,.
\ee
One may therefore think of $Z_\sigma(1,\rho(2,\dots,n-2), n-1, n)$ as partial amplitudes ordered with respect to two symmetry groups. One of them corresponds to dressing $Z$ with traces built out of Chan-Paton factors following the permutation $\sigma$ and the other corresponds to dressing it with trace color factors, unrelated to the Chan-Paton factors, 
following the permutation~$\rho$. 

With these building blocks, the open-superstring amplitudes with massless external states color-ordered, with respect to the Chan-Paton factors, can be expressed directly in terms of Yang-Mills scattering amplitudes~\cite{MafraBCJAmplString}, and written in terms of a field theoretic double-copy factorization in~\cite{Broedel2013tta},
\begin{equation}
A^{\tree}_{\rm OS}(\sigma(1,2,3,\ldots,n)) ~ =\!\!\!\!\!\!\!\!\!  \sum_{\tau,\rho \in S_{n-3}(2,...,n-2)}\!\!\!\!\!\!\!\!\!  Z_\sigma (1,\tau,n,n{-}1)  S[\tau | \rho]  A_{\rm SYM}(1,\rho,n{-}1,n) \,,
\label{2.2cOLD} 
\end{equation}
where the  $(n-3)!\times(n-3)!$ matrix  $S[\tau|\rho] = S[\tau(2,\ldots,n{-}2) |\rho (2,\ldots,n{-}2)]$ is the field-theory KLT kernel\footnote{Note that the $\alpha'$-dependent KLT kernel, given in Ref.~\cite{BjerrumMomKernel} (and its inverse in Ref.~\cite{Mizera:2016jhj}), needs not feature in the factorization of tree-level  string amplitudes, cf.~\eqn{CSvsOS}.
}  introduced in \sect{secKLT}. It is fascinating to note that a suggestive hint of this type of field-theoretic double-copy factorization was identified in Ref.~\cite{Stieberger:2013hza}.

Rather than focusing on the partially-ordered open string amplitudes \eqref{2.2cOLD}, consider instead the content of the full Chan-Paton-dressed open supersymmetric string amplitude.  Dressing $Z_{\sigma}$ with all relevant $(n-1)!$ traces built out of Chan-Paton factors, yields a singly ordered function, 
\be
{\mathbf Z}^\tree (1,...,n) ~ \equiv \!\!\!\!\!\!\! \sum_{\sigma \in S_{n-1}(2, \ldots, n)}  \!\!\!\!\!\!\!  \text{Tr}\left[{\T^{a_ 1} \T^{a_{\sigma(2)}} \cdots  \T^{a_{\sigma(n-1)}} \T^{a_{\sigma(n)}}}\right] \sum_{\rho \in S_{n-3}}  Z_\sigma( \, 1,\rho,n-1,n)\,,
\ee
which obeys only the field-theory amplitude relations (\ie Eq.~\eqref{ftarel} with the replacement $Z_\sigma \rightarrow {\mathbf Z}^\tree$).
The full Chan-Paton-dressed open superstring amplitude, 
\be
{\cal{A}}_{\rm OS}=  \sum_{\sigma \in S_{n-1}}  \text{Tr}\left[{\T^{a_ 1} \T^{a_{\sigma(2)}} \cdots  \T^{a_{\sigma(n-1)}} \T^{a_{\sigma(n)}}}\right]  A^{\tree}_{\rm OS}(1,\sigma)\,,
\ee
can also be written entirely as a field-theory double copy
\be
{\cal{A}}_{\rm OS}=  \sum_{\tau,\rho \in S_{n-3}}  {\mathbf Z}^\tree( 1,\tau,n,n-1) S[\tau | \rho]  A_{\rm SYM}(1,\rho,n{-}1,n) \, .
\label{CP-dressedSOS} 
\ee

An interesting open problem is the physical interpretation of the above building blocks.  Given the adjoint field-theory relations obeyed by the ordered ${\mathbf Z}(\rho)$, it is natural to consider the orderless-functions resulting by dressing the $\rho$ ordering with adjoint $f^{abc}$ structure constants as per a DDM basis. 
This yields a fully dressed function that can be expressed in terms of cubic graphs dressed with two factors that both satisfy Jacobi identities and antisymmetry:
\be
{\mathbfcal{Z}} =  \sum_{i} \frac{ {\mathbf z}_i c_i}{D_i} =  \sum_{\rho \in S_{n-2}} c_{1|\rho|n} \, \mathbf{Z}^\tree(1,\rho,2) \,,
\ee
such that:
\be
{\mathbf Z}^\tree(1,\rho(2),\ldots \rho(m-1),m) = - i \sum_{i \in {\rm planar}}  b_{i \, \rho} \frac{\mathbf{z}_i}{D_i}\,,   
\ee
where $\mathbf{z}_i$ are Jacobi satisfying functions of both higher-derivative scalar kinematics and string Chan-Paton factors, $D_i$ are the propagators of the graph,
and  $b_{i \, \rho} \in \{0,\pm 1\}$ are integer coefficients that depend on the ordering $\rho$.
Both $\mathbfcal{Z}$ and ${\mathbf Z}$ can be  be derived as the tree-level amplitudes, color-dressed and ordered respectively, of an effective field theory of 
double-colored scalar fields in which the scalars obey an equation of motion of the schematic form~\cite{Mafra2016mcc}
\begin{equation}
\Box \Phi = \Phi^2 + {\alpha'}^2 \zeta_2 \big( \partial^2 \Phi^3 + \Phi^4  \big) +
{\alpha'}^3 \zeta_3 \big(  \partial^4 \Phi^3 + \partial^2 \Phi^4 + \Phi^5  \big) + {\cal O}({\alpha'}^4) \, .
\end{equation}
 This theory was named Z-theory in refs.
\cite{Carrasco2016ldy, Mafra2016mcc, Carrasco2016ygv}.  It is worth noting that the color structure of the leading term in the equation of motion is the same as the bi-adjoint $\phi^3$ theory.  
 
The entire tower of higher derivative operators relevant to the open-string are encoded in this effective scalar theory, whose double copy with the supersymmetric gauge theory yields the supersymmetric open string. Schematically, the formula (\ref{CP-dressedSOS}) can be rewritten with the short-hand notation 
 \be
 (\text{massless open superstring}) = (\text{Z-theory}) \otimes  ({\rm SYM})\, .
 \label{OS_DC}
 \ee

The simplest set of Z-theory amplitudes arise when one trivializes the string Chan-Paton factors, taking all the generators to be the identity, corresponding to a $U(1)$ group. This operation on the Chan-Paton dressed open string results in a symmetrization over all orders referred to as the abelian or photonic open-string whose low-energy limit yields amplitudes in maximally supersymmetric DBI theory, where the fermionic sector is of Volkov-Akulov type~\cite{Fradkin:1985qd,Metsaev:1987qp,
Aganagic:1996pe, 
Aganagic:1996nn,
Bergshoeff:1996tu,Bergshoeff:1997kr,Kallosh:1997aw,Rocek:1997hi,Bergshoeff:2013pia}. Abelian ${\cal Z}$ amplitudes yield in the low-energy limit NLSM amplitudes\footnote{See Eqn.~\ref{NLSMaction} for one representation of the action.} in the $\alpha'\rightarrow 0$~\cite{Carrasco2016ldy}.  This is consistent with the realization that the NLSM double-copies with ${\cal N}=4$ SYM in four dimensions to generate DBI-VA amplitudes~\cite{Cachazo2014xea, He2016mzd}.

A closed-string  version of the Z-theory amplitudes involves integrals over the Moduli space of punctured Riemann spheres \cite{Stieberger2014hba,Schlotterer2018zce,Vanhove:2018elu,Brown2018omk}
\be
{\rm sv} \, Z(\tau  | \sigma) =  \left( \frac{2\ap}{\pi } \right)^{n-3}  \! \! \!  \int  \frac{d^2 z_1 \, d^2 z_2 \, \ldots\, d^2 z_{n}}{{\rm vol}({\rm SL}(2,\mathbb{C}))} \ 
\frac{ \prod_{i<j}^{n} |z_{ij}|^{2 \alpha' s_{ij}}  }{ \tau \, \{ \bar{z}_{12} \bar{z}_{23} \cdots \bar{z}_{n-1,n} \bar{z}_{n,1} \}  \sigma \, \{ z_{12} z_{23} \cdots z_{n-1,n} z_{n,1} \} } \,.
\label{2.2bcl}
\ee
The notation ${\rm sv} Z$ refers to the so-called single-valued projection  of multiple zeta values (MZVs), which can be regarded 
as a formal operation acting on the   building blocks which arise in the construction for tree amplitudes of massless open-superstring states in the low-energy expansion \cite{Schnetz2013hqa, Brown2013gia}. Here, we use (\ref{2.2bcl}) as the definition of ${\rm sv} \, Z(\tau  | \sigma)$. 
Amplitudes in the Z-theory, together with their closed-string counterparts ${\rm sv} \, Z(\tau  | \sigma)$, enter a particular class of tree-level double-copy constructions which 
combine the amplitudes of a string theory with the amplitudes of a gauge theory. For example, amplitudes in the closed superstring 
with massless asymptotic states can be obtained with the construction \cite{Schlotterer2012ny,Stieberger:2013wea}
\be
(\text{closed superstring})= ({\rm SYM}) \otimes  {\rm sv}\big(\textrm{open superstring} \big) \, .
\label{CSvsOS}
\ee
Remarkably, various incarnations of the $(DF)^2$ theory introduced in the previous subsection in a completely different context enter these double-copy constructions for string amplitudes~\cite{Azevedo2018dgo}:
\bea
(\text{open bosonic string})&=& \big(\textrm{Z-theory}\big) \otimes \big((DF)^2+{\rm YM}\big) \,,\label{BS_DC} \\
(\text{closed bosonic string})&=&\big((DF)^2+{\rm YM}\big)\otimes  {\rm sv}\big(\textrm{open bosonic string} \big) \, , \\ 
(\text{heterotic string}) &=& \big((DF)^2+{\rm YM}+\phi^3\big)\otimes  {\rm sv}\big(\textrm{open superstring} \big)  \, .
\label{HET_DC3}
\eea
We should note that these constructions are of the generic form \eqref{2.2cOLD}, \ie they involve the field-theory KLT kernel, and apply at tree level and with massless external states. Remarkably, the free mass parameter in the $(DF)^2+$YM  theory is related to the inverse string tension $\alpha'$ as
\be
m^2= -\frac{1}{\alpha'}\,.
\label{massIdentification}
\ee
Various relations between Z-theory and string amplitudes are summarized in Table \ref{overview}. Some extensions to loop level can be found in refs. \cite{MafraSchlotterOneLoopString,Mafra:2012kh,Mafra2018nla,Mafra2018pll,Mafra2018qqe,Mafra:2019ddf,Mafra:2019xms}. Additionally, a double-copy construction
for string amplitude in terms of field-theory amplitudes in the CHY formalism was obtained in Refs.~\cite{He:2018pol,He:2019drm}.

\begin{table}[tb]\small
	\be \! \! \!
	{
		\begin{array}{c|c|c|c}
			{\rm string}\otimes {\rm QFT}  &\te{SYM} &(DF)^2 \,{+}\, \te{YM}&\ (DF)^2\, {+}\, \te{YM}\,{+}\,\phi^3 \\\hline \hline
			\te{Z-theory} \ & \ \te{open superstring} \ \, & \, \te{open bosonic string} \, &{ \begin{array}{c} 
					\te{compactified open} \\
					\te{bosonic string}
				\end{array}}  \\
				\te{sv}(\te{open superstring})  \ & \ \te{closed superstring}  \ \, &\ \te{heterotic (gravity)} \ \, &\, \te{heterotic\,(gauge/gravity)}  \\
				\te{sv}(\te{open\,bosonic\,string})  \, &  \, \te{heterotic\,(gravity)} \, & \, \te{closed\,bosonic\,string} \, &{ \begin{array}{c} 
						\te{compactified closed} \\
						\te{bosonic string}
					\end{array}}
				\end{array}} \nonumber
				\ee
				\caption{Various known double-copy constructions of string amplitudes \cite{Azevedo2018dgo}. The single-valued projection sv($\bullet$) converts disk to sphere integrals. }
				\label{overview}
			\end{table}  

\subsubsection{Other theories \label{sec-zoo-nongrav}}
We conclude the section by listing further examples of double-copy constructions. 
\begin{itemize}

	\item The non-gravitational (supersymmetric) DBI theory was constructed in Ref.~\cite{Cachazo2014xea} using the scattering 
	equation formalism (see also \cite{CheungUnifyingRelations,Cheung:2017yef}). It can be regarded as the double copy of (S)YM theory 
	and the NLSM. It should be noted that the NLSM can be obtained in the $\alpha' \rightarrow 0$ limit of 
	abelian Z-theory~\cite{Carrasco2016ldy}. A further interesting feature of the NLSM is that it admits a Lagrangian in which the duality between 
	color and kinematics is manifest~\cite{Cheung2016prv}.

	\item Similarly, the (supersymmetric) DBI theory coupled to (S)YM theory can be constructed as a double copy 
	involving (S)YM theory and the NLSM coupled to bi-adjoint $\phi^3$ theory~\cite{Cachazo2016njl}. The latter gauge-theory 
	factor can be obtained from the $\alpha'\rightarrow 0$ limit of partially-Abelianized Z-theory~\cite{Carrasco2016ygv}.  
	
	\item The DBI theory coupled to the NLSM can be constructed as a double copy involving YM coupled 
	to bi-adjoint $\phi^3$ theory and the NLSM \cite{Chiodaroli2017ngp}.  
	
	\item Volkov-Akulov theory has tree-level amplitudes that can be obtained from supersymmetric DBI by restricting the external states to be fermions. Since DBI only has nonvanishing even-point amplitudes, and internal bosons would require tree-level factorization with an odd number of particles (2$\times$fermions + 1\,boson), this restriction gives a consistent truncation of the theory. The double copy for Volkov-Akulov theory can thus be inferred to be a product between NLSM and SYM with only external fermions. 
	
	\item Two copies of the NLSM give the so-called special-Galileon theory~\cite{Cachazo2014xea,Cheung:2017yef}.
				
	\item In three dimensions, two copies of the BLG theory~\cite{Bagger:2007jr, Gustavsson:2007vu} yield an alternative construction for maximal three-dimensional supergravity \cite{Huang2012wr,Bargheer2012gv,AllicABJM,Bianchi:2013pfa,Huang:2013kca}. The 
	three-dimensional version of \ck duality relevant to this construction is based on a so-called three-algebra. The three-algebra for BLG theory is introduced formally using a totally antisymmetric triple product $[X,Y,Z]$. Using a basis of generators the triple product can be expressed using rank-four structure constants,
	\be 
	\big[ T^a, T^b, T^{ c} \big] = f^{ab c}_{\ \ \ d} T^d \,.
	\ee 
  Consistency of the algebra requires that the  structure constants satisfy the four-term identity
	\be
	0= f^{abc}_{\ \ \ l} f^{dl e g} + f^{bae}_{\ \ \  l} f^{d  lc g} + f^{ c e b}_{\ \ \ \ l}f^{ d a l g}
        + f^{ ec a}_{\ \ \ \  l} f^{ d b l g } \,, 
        \ee
	which plays the same role as the standard Jacobi identity for a Lie two-algebra. It turns out that the only nontrivial compact three-algebra is SO(4)~\cite{Gustavsson:2008dy},  where $f^{abc}_{\ \ \  d}=\epsilon^{abcd}$. However, for color-kinematics duality to work, it is sufficient to impose the four-term identity, whereas identities specific to SO(4) should be ignored. Finally, we may note that the closely-related ABJM theory~\cite{Aharony:2008ug} appears to not have similarly nice properties under color-kinematics duality. While the tree-level ABJM amplitudes up to six points obey the duality and their double copy gives three-dimensional maximal supergravity, at eight points the double copy does not reproduce the corresponding amplitude in maximal supergravity~\cite{Huang2012wr,Huang:2013kca}. Since there is no dynamical graviton in three dimension, this mismatch is not forbidden by the diffeomorphism symmetry argument in \sect{diffeoSubsection}. 
\end{itemize}
Additional theories for which a double-copy construction has been proposed involve massive higher-spin ${\cal N}=7$
W-supergravity theories \cite{Ferrara:2018iko,Borsten:2018jjm}; this amount of supersymmetry has not been accessible through different constructions. 
Chiral higher-spin theories have been shown to obey generalized 
BCJ relations in Ref.~\cite{Ponomarev:2017nrr}. Theories with gravitationally-coupled fermions have been discussed in Ref.~\cite{delaCruz:2016wbr}.
A construction of the free spectrum of $D = 3$ supergravities in terms of SYM theories with fields valued in the four division algebras 
was given in~\cite{Anastasiou:2017taf}. 
Further examples of constructions in higher dimensions include half-maximal supergravity in six dimensions \cite{Berg:2016fui} and  
the so-called $(4,0)$ theory in six dimensions \cite{Hull:2000zn,Anastasiou:2013hba}, which can be seen as the double copy of two $(2,0)$ theories, at least 
at the level of the free spectrum \cite{Chiodaroli:2011pp,Borsten:2017jpt}.


\section{BCJ duality at loop level}
\label{ExamplesSection}

In this section, we describe loop-level examples of BCJ duality and
the associated double-copy construction.  Whenever a gauge-theory
integrand can be found in a form that manifests the duality between
color and kinematics, corresponding gravity integrands can be
immediately written down via the double-copy procedure.  This
procedure enormously simplifies the construction of gravity loop
integrands and has been successful for carrying out a variety of
loop-level studies in perturbative quantum gravity theories (see
e.g. Refs.~\cite{N46Sugra, N46Sugra2, Bern:2012cd, Bern:2013qca,
N4GravFourLoop, BCJDifficulty, N5GravFourLoop, BoelsFourLoop,
FiveLoopFormFactor,Faller:2018vdz}). As explained
in \sect{ZoologySection1}, the precise gravity theory to which the
integrands belong depends on the choice of input gauge theories.
We start by briefly recalling the definition and the main points of
the duality and of the double-copy construction, discussed at length
in \sect{DualitySection}.  With the appropriate separation of
diagrams' symmetry factors and judicious choice of loop momenta, they
are essentially the same as at tree level.

Similarly to tree-level amplitudes, loop-level amplitudes in a gauge
theory coupled to matter fields can be organized as a sum over
diagrams with only cubic (trivalent) vertices by multiplying and
dividing by appropriate propagators to absorb contact diagrams into
diagrams with only cubic vertices.  If all fields are in the adjoint
representation of the gauge group, this rearrangement puts the
amplitude in a form equivalent to Eq.~\eqref{gaugeAmp},
\begin{equation}
\mathcal{A}^{L\hbox{-}\mathrm{loop}}_{m}=i^{L-1} g^{m-2+2 L}
\sum_{\mathcal{S}_{m}}\sum_{j}\int\prod_{l=1}^{L}\frac{d^{D}\ell_{l}}{(2\pi)^{D}}
\frac{1}{S_{j}}\frac{c_{j} n_{j}(\ell)}{D_{j}}\,,
\label{CubicRepLoop}
\end{equation}
where the $c_i$ are color factors obtained by assigning structure constant factors
$\tilde f^{abc} = i \sqrt{2} f^{abc}$ to each cubic vertex.  
The first sum runs over the set $\mathcal{S}_{m}$ of $m!$ permutations of the
external legs. The second sum runs over the distinct $L$-loop
$m$-point diagrams with only cubic vertices. As at tree level, by
multiplying and dividing by propagators, it is trivial to absorb
contribution from higher-than-three-point
vertices into numerators of diagrams with only cubic vertices.  The
symmetry factor $S_{j}$ counts the number of automorphisms of the
labeled diagram $j$ from both the permutation sum and from any internal
automorphism symmetries.\footnote{Note that this symmetry factor is different from the symmetry factor in Eq.~\eqref{gaugeAmp}, 
where $S_j$ counts the automorphisms of graphs with fixed external legs.}  This symmetry factor should not be included
in the kinematic numerator.

\begin{figure}[t]
\begin{center}
\begin{eqnarray*}
\hbox{\large $n$} \left(\parbox{2.9 cm}
    {\includegraphics[scale=.4] {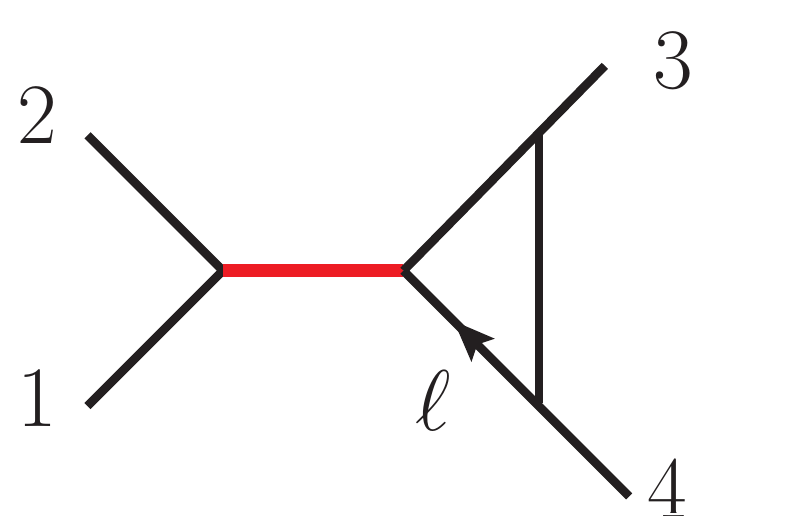}}
                 \right) \hbox{ \Large $=$ }
\hbox{\large $n$} \left(\parbox{2.3 cm}
    {\includegraphics[scale=.4] {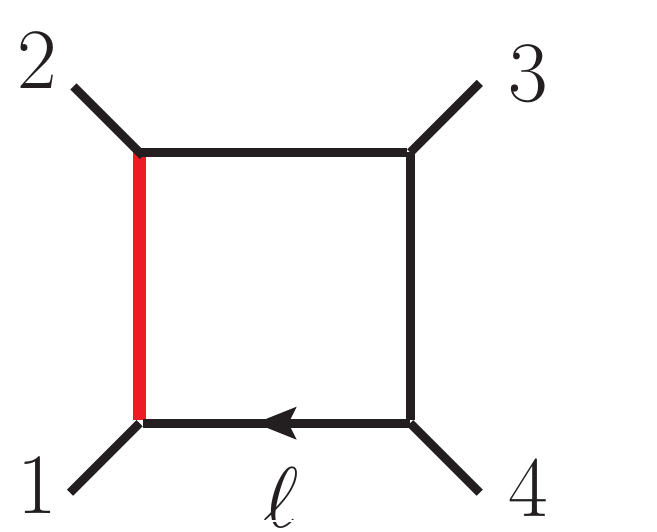}} 
       \right) \hbox{ \Large $-$ } 
\hbox{\large $n$} \left(\parbox{2.6 cm}
   {\includegraphics[scale=.4]  {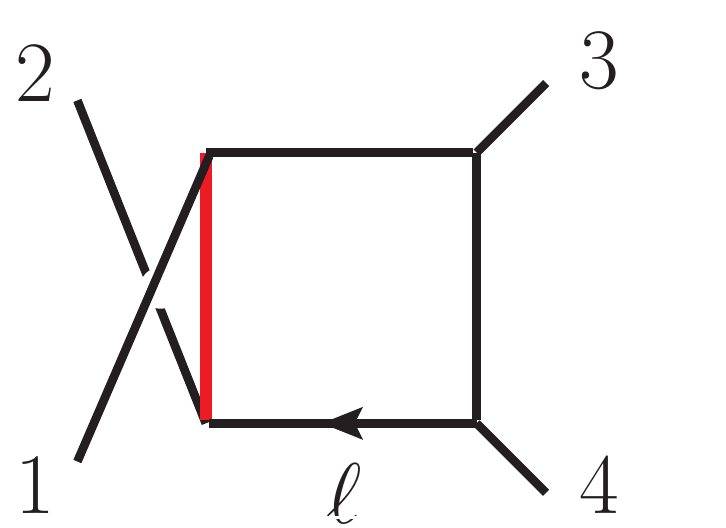}} \right) 
\end{eqnarray*}
\end{center}
\caption{A BCJ kinematic numerator relation at one loop. When the
  external particles are gluons this holds just as well for adjoint or
  fundamental representation particles circulating in the loop. 
  The shaded (red) line differs between the diagrams, but the others 
  are identical.
}
\label{OneLoopBCJFigure}
\end{figure}

The nontrivial conjecture is that, as at tree level, for every loop-level color Jacobi identity there is a matching kinematic numerator identity~\eqref{duality}. 
\begin{equation}
 c_i - c_j=c_k  \quad \Leftrightarrow \quad  n_i(\ell) - n_j(\ell) = n_k(\ell) \, .
\label{BCJDualityLoop}
\end{equation}
However, unlike at tree level, one has to be cautious with the
treatment of degrees of freedom that are not fixed by the external
states. This includes proper accounting of the loop momenta of the
numerators, generically called $\ell$, as well as being careful to not
set to zero color factors that vanish when summing over internal indices.

We can change the signs of the color factors using the antisymmetry of the $f^{abc}$s, but any relative signs between color factors in the Jacobi relation are then inherited by the corresponding 
relation between the kinematic numerator factors.  
A simple example of such loop-level relations is illustrated
in \fig{OneLoopBCJFigure} for the case of a one-loop amplitude.  At
loop-level, the duality between color and kinematics (\ref{duality})
remains a conjecture~\cite{BCJLoop}, although evidence in its favor
continues to accumulate~\cite{FivePointN4BCJ, WhiteIRBCJ,
SimplifyingBCJ, FourLoopFormFactor, Boels:2013bi,
Bjerrum-Bohr:2013iza, OneTwoLoopPureYMBCJ, Ochirov:2013xba,
MafraSchlottererTwoLoop, BCJDifficulty, HeMonteiroSchlottererBCJNumer,
FiveLoopFormFactor, Chiodaroli2017ngp, BoelsFourLoop, HeSchlottererZhangOneLoopBCJ, JohanssonTwoLoopSusyQCD, 
Jurado:2017xut, Boels:2017ftb,Faller:2018vdz}.

Just as for tree-level numerators, once gauge-theory numerator factors
which satisfy the duality are available, replacing the color factors by
the corresponding numerator factors, $c_i \rightarrow n_i$ yields the
double-copy form of gravity loop integrands~\eqref{DCformula},
\begin{align}
\mathcal{M}^{L\hbox{-}\mathrm{loop}}_{m} =
i^{L-1} \left(\frac{\kappa}{2}\right)^{m-2+2 L}
\sum_{\mathcal{S}_{m}}\sum_{j}\int\prod_{l=1}^{L}\frac{d^{D}\ell_{l}}{(2\pi)^{D}}
\frac{1}{S_{j}}\frac{\tilde{n}_{j}(\ell)n_{j}(\ell)}{D_{j}}
\,,
\label{DoubleCopyLoop}
\end{align} 
where $\tilde{n}_j$ and $n_j$ are gauge-theory numerator factors,
which can come from distinct gauge theories and $\kappa$ is the gravitational
coupling defined below \eqn{Lagrangians}.  The duality needs to be
manifest in only one of the two gauge-theory amplitudes for the double-copy formula to hold.

\subsection{One-loop examples of BCJ duality: $\cN=4$ SYM theory}

The simplest example that illustrates \ck duality at
loop level is the one-loop four-point superamplitude of $\NeqFour$ SYM
theory.  These amplitudes are remarkably simple, making them very useful 
for this purpose.

\begin{figure}[t]
\centerline{\includegraphics[width=3.5cm]{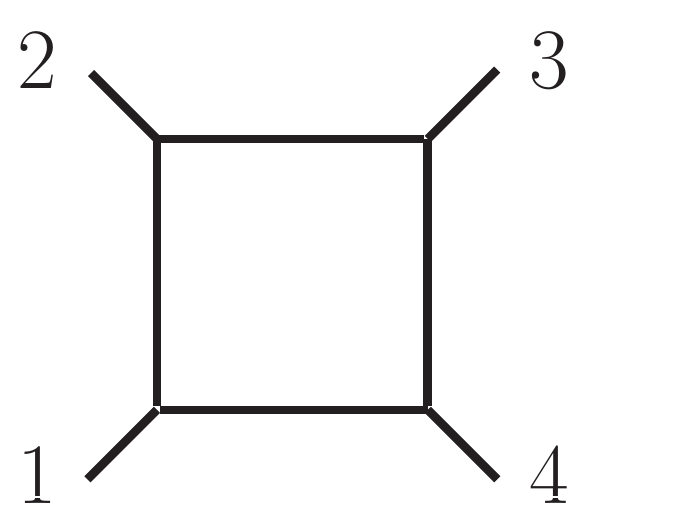}}
\caption{A one-loop box integral, $I_4(s,t)$, appearing in the one-loop four-point
  $\NeqFour$ SYM and $\NeqEight$ supergravity amplitudes. The three independent 
relabelings of external legs appear in the amplitudes.}
\label{OneLoopBoxFigure}
\end{figure}

The Jacobi identity obeyed by the structure constants of any Lie algebra guarantees that,  in any gauge theory with all fields in 
the adjoint representation of the gauge group, any one-loop four-point amplitude can be organized as
\begin{equation}
{\cal A}^{\oneloop}_4(1,2,3,4)= g^4 
  \Bigl(c_{1234} A^\oneloop_4(1,2,3,4) + c_{1243} A^\oneloop_4(1,2,4,3) + c_{1423} A^\oneloop_4(1,4,2,3) \Bigr) \, ,
\label{YMFourPointColorDecomp}
\end{equation}
where the color factor $c_{1234}$ in \eqn{YMFourPointColorDecomp} corresponds to the
one of the box diagram in \fig{OneLoopBoxFigure} and is given by 
dressing each three-point vertex with an $f^{abc}$ structure constant, and summing over all repeated indices,
\begin{equation}
c_{1234}= 4 f^{ba_1c} f^{ca_2d} f^{da_3e}f^{ea_4b} \, .
\end{equation}
The other two color factors are obtained by relabeling and we normalized $c_{1234}$ following standard
conventions~\cite{ManganoParkeReview}. 
Passing to a trace basis for the color factors identifies $A^\oneloop(1,2,3,4)$ with the one-loop four-point color-ordered amplitudes. 
The form \eqref{YMFourPointColorDecomp} can be obtained by applying the color Jacobi
identity to the color factors of any valid representation (e.g. Feynman diagrams) of the amplitude
to trade other color factors in favor of the three box ones~\cite{DixonMaltoni}. 
Similar manipulations, together with use of the defining commutation relations of the Lie algebra, can be used to map the 
color factors of all one-loop four-point amplitudes in a theory with fields in any representation to the color factors of a box diagram;
in this subsection we we will  however restrict ourselves to theories with fields in the adjoint representation.


\begin{homework}
Prove \eqn{YMFourPointColorDecomp} by starting from standard Feynman
diagrams and then applying color Jacobi identities to express all
color factors in terms of the color factors of the box diagrams.
\end{homework}

Consider now the one-loop four-point superamplitude of $\NeqFour$
SYM theory.  Each color-ordered superamplitude in \eqn{YMFourPointColorDecomp}
is especially simple and given by~\cite{GSB} 
\begin{equation}
A^{\oneloop}_{\NeqFour}(1,2,3,4)= i  s t A_{\NeqFour}^\tree(1,2,3,4) I_4(s,t) \,,
\label{NeqFourYMFourPoint}
\end{equation}
where $I_4(s,t)$ is the box integral illustrated
in \fig{OneLoopBoxFigure}, $s = (p_1 + p_2)^2$ and $t = (p_2 + p_3)^2$
are the usual Mandelstam invariants and $A_\NeqFour^\tree(1,2,3,4)$,
standing for the $n=4$ case of Eq.~\eqref{MHVSuperAmplitude}, is the
color-ordered four-point tree superamplitude.

Since the diagram structure of the kinematic propagators in the three
color-ordered amplitudes entering \eqn{YMFourPointColorDecomp} matches
that of their color factors, the kinematic numerators of the
representation \eqref{CubicRepLoop} of the one-loop amplitude can be
straightforwardly identified.
The combination $s t A_{\NeqFour}^\tree(1,2,3,4)$ is fully
crossing-symmetric, as a consequence of the BCJ four-point tree-level
amplitude relations \eqref{4PtBCJAmplitude}, so all three numerators
are the same,
\begin{equation}
n_{1234} = n_{1243} = n_{1423} = i st A_{\NeqFour}^\tree(1,2,3,4) 
= \frac{[12][34]}{\langle 12\rangle \langle 34\rangle} 
\delta^{(8)}(\sum_{i=1}^4 \lambda_i \eta_i^I)\,,
\label{FourFourNumeratorCrossing}
\end{equation}
where we have specialized to four-dimensional external kinematics in the last equality.

Because triangle and bubble diagrams do not appear in the
$\NeqFour$ SYM amplitude \eqref{NeqFourYMFourPoint} (or, alternatively, they enter with vanishing numerators), 
it is straightforward to check, using \eqn{FourFourNumeratorCrossing}, that the BCJ duality relation
illustrated in \fig{OneLoopBCJFigure} holds.
The remaining kinematic Jacobi relations are also satisfied for similar reasons. 

The corresponding $\NeqEight$ supergravity amplitude follows
immediately from the basic double-copy
substitution~\eqref{c_n_replace}, replacing color factors by
numerators and compensating for the change in coupling. This gives
\begin{equation}
{\cal M}_{\NeqEight}^{\oneloop}(1,2,3,4)= 
 - i stu {\cal M}_{\NeqEight}^\tree(1,2,3,4) 
\Bigl(I_4(s,t) + I_4(s,u) +  I_4(t,u) \Bigr) \,,
\label{NeqEightGravityFourPoint}
\end{equation}
where we used \eqref{FourPointKLT},
\begin{equation}
\Bigl(\frac{\kappa}{2} \Bigr)^4  \bigl(st A^\tree_{\NeqFour\, \rm SYM}(1,2,3,4) \bigr)^2 = stu {\cal M}^\tree_{\NeqEight} (1,2,3,4) \,,
\end{equation}
to replace the square of the $\NeqFour$ SYM four-point tree-level amplitude with the
$\NeqEight$ supergravity four-point tree-level amplitude.  This is a consequence of the KLT
relations \eqref{KLTFourPt} and the BCJ amplitude relation \eqref{4PtBCJAmplitude}.  The
amplitude in \eqn{NeqEightGravityFourPoint} reproduces the known
$\NeqEight$ supergravity four-point tree-level amplitude~\cite{GSB,BDDPR}.

The explicit value of the massless scalar box integral $I_4(s,t)$ appearing in both the ${\cal N}=4$ SYM and 
${\cal N}=8$ supergravity one-loop four-point amplitudes is
\begin{equation} 
I_4(s,t) = \int \frac{d^D \ell}{ (2 \pi)^D} \frac{1} 
{\ell^2 (\ell-p_1)^2 (\ell - p_1-p_2)^2 (\ell + p_4)^2 } \,,
\label{4PtIntegral}
\end{equation}
where the $p_i$'s are the external momenta and the Feynman $ i
\varepsilon$ prescription, not included explicitly, is used to define the propagators.  In
dimensional regularization, we take $D= 4 -2 \eps$ with $\epsilon$
small.  The explicit functional form of $I_{4}(s,t)$ is (see
e.g.\ Refs.~\cite{BKStringLimit,PentagonIntegral})
 \begin{equation}
I_{4}(s,t) = i \frac{c_{\Gamma}}{st}
\left[ \frac{2}{\e^2} \Big((-s)^{-\e}+(-t)^{-\e} \Big) -
\ln^2 \left(\frac{-s}{-t} \right)-\pi^2 \right] + \cal{O}(\epsilon) ~,
\label{ZeroMassBoxFullExpand}
\end{equation}
with
\begin{equation}
c_{\Gamma} = \frac{(4\pi)^{\epsilon}}{16\pi^2}\frac{\Gamma(1+\epsilon)
\Gamma(1-\epsilon)^2}{\Gamma(1-2\epsilon)} \,.
\label{cGamma}
\end{equation}
The other box integrals can be obtained from this one by relabeling. Using these explicit expressions
one can verify general properties of (super)gravity amplitudes, such as existence of only soft infrared (IR) divergences.

We can use \eqn{YMFourPointColorDecomp}, together with the
$\NeqFour$ SYM numerators \eqref{FourFourNumeratorCrossing}, to
immediately obtain the four-point amplitudes of any $4 \le {\cal N}
\le 8$ supergravity after integration.  Because the duality satisfying
$\NeqFour$ four-point SYM kinematic numerators
(\ref{FourFourNumeratorCrossing}) are independent of the loop
momentum, they come out of the integral as in
\eqn{NeqEightGravityFourPoint} and behave essentially the same way as
color factors.  Thus, to obtain results for ${\cal N} \ge 4$ supergravity, we can start with \eqn{YMFourPointColorDecomp} 
evaluated for ${\cal N}\le 4$ (S)YM theory and replace the color factors with the $\NeqFour$ SYM numerators 
in \eqn{FourFourNumeratorCrossing}.
This gives us a general representation of the four-point amplitudes of all  ${\cal N} \ge 4$ supergravities:
\begin{align}
 \mathcal{M}_{ {\cal N}+4\,\,{\rm susy}}^{\oneloop} (1,2,3,4) &=
\Bigl( \frac{\kappa}{2} \Bigr)^4 i s t A^{\tree}_4 (1, 2, 3, 4)
 \Bigl(A_{ {\cal N}\,{\rm susy}}^{\oneloop} (1, 2, 3, 4)
 +  A_{ {\cal N}\,{\rm susy}}^{\oneloop} (1, 2, 4, 3) \nn \\
& \hskip 4.5 cm \null
 +  A_{ {\cal N}\,{\rm susy}}^{\oneloop} (1, 4, 2, 3)\Bigr)\,.
\label{NeqM}
 \end{align}
As explained above, $A_{ {\cal N}\,{\rm susy}}^{\text{1-loop}}$ are one-loop color-ordered gauge-theory amplitudes 
after loop integration for a theory with ${\cal N}$ (including zero) supersymmetries (c.f. \eqn{YMFourPointColorDecomp}). 
This expression applies just as well for external matter multiplets in $\NeqFour$ supergravity. The needed integrated gauge-theory 
amplitudes may be found in Refs.~\cite{BKStringLimit,BernMorgan}.

The double copy of amplitudes of gauge theories with ${\cal N} <4$ supersymmetry is  less straightforward because
 the required gauge-theory numerators are in general not independent of loop momenta. Because of this,  although the double-copy 
 construction holds at the integrand level, one cannot simply carry over the integrated results from gauge to gravity theories. 
It is nevertheless remarkable that there is such a simple relation between these two different theories.

To illustrate \ck duality and the double-copy construction at one loop,
we consider the one-loop  identical-helicity four-gluon amplitude in QCD with $N_{f}$ quark flavors in the fundamental representation, 
originally constructed in Ref.~\cite{BernMorgan}. It is\footnote{It may also be obtained via the dimension-shifting relation \cite{Bern:1996ja} 
from the four-gluon superamplitude in ${\cal N}=4$ SYM theory~\eqref{YMFourPointColorDecomp},\eqref{NeqFourYMFourPoint}.}
\begin{align}
\label{QCDFourPointOneLoop}
&{\cal A}^{\oneloop}_{\rm QCD}(1^+,2^+,3^+,4^+)=  2  g^4  \frac{\spb1.2 \spb3.4}{\spa1.2 \spa3.4} 
\Bigl( (c_{1234} -  N_{\!f} c^f_{1234}) I_4(s,t)[\mu^4]
        \\
& \null \hskip 3.5 cm 
 +
       (c_{1243} -  {N_f} c^f_{1234}) I_4(s,u)[\mu^4] + (c_{1423} -  {N_f} c^f_{1234}) I_4(t,u)[\mu^4] \Bigr) \,,
       \nn
\end{align}
where the color factor associated with the quark loop is 
\begin{equation}
c^f_{1234} = \Tr[T^{a_1} T^{a_2} T^{a_3} T^{a_4}] + \Tr[T^{a_4} T^{a_3} T^{a_2} T^{a_1}] \,.
\end{equation}
For simplicity, we have assumed that the quarks are massless.  Here $\mu$ is the $(-2\eps)$-dimensional
component of loop momentum, so 
\begin{equation}
\ell = \ell^{(4)} + \mu\,,  \hskip 2 cm 
\ell^2 = (\ell^{(4)})^2  - \mu^2\, ,
\end{equation}
and $I_4(s,t)[\mu^4]$  is the integral corresponding to the diagram in \fig{OneLoopBoxFigure} with a $\mu^4$ numerator factor.
As required by Bose symmetry, the prefactor is fully cross symmetric, \ie
\begin{equation}
 \frac{\spb1.2 \spb3.4}{\spa1.2 \spa3.4} = \frac{\spb2.3 \spb4.1}{\spa2.3 \spa4.1} =  \frac{\spb1.3 \spb2.4}{\spa1.3 \spa2.4} \,,
\label{PrefactorCrossing}
\end{equation}
and, up to the supermomentum conservation delta function, it is the same as in Eq.~\eqref{FourFourNumeratorCrossing}.

\begin{homework}
Show the prefactor in \eqn{PrefactorCrossing} is crossing
symmetric. Spinor properties may be found in
Appendix \ref{SpinorSuperspaceSection} and in various
reviews~\cite{ManganoParkeReview,TasiLance,ElvangHuangReview}.
\end{homework}

It is not difficult to check that the amplitude in \eqn{QCDFourPointOneLoop} obeys \ck duality.
Consider the duality relation in \fig{OneLoopBCJFigure}: because the triangle diagrams have vanishing numerators 
in \eqn{QCDFourPointOneLoop}, the duality requires the different box integrals to have an identical numerator, which follows 
from \eqn{PrefactorCrossing} and the integrals' numerators being crossing symmetric.

\begin{homework}
Make the quarks massive.  For the identical helicity case, the
integral numerator is obtained with the replacement $\mu^4 \rightarrow (\mu^2 +
m_q)^2$~\cite{BernMorgan} while the loop propagators become massive
with mass $m_q$. Do the  BCJ relations hold?  What does the double-copy theory correspond to?
\end{homework}

Consider now the double-copy construction with one of the two amplitude factors being Eq.~\eqref{QCDFourPointOneLoop} with $N_f = 0$.  
Taking the second amplitude to be the four-gluon superamplitude of ${\cal N}=4$ SYM theory given in Eqs.~\eqref{YMFourPointColorDecomp},\eqref{NeqFourYMFourPoint} leads  to an anomalous superamplitude in ${\cal N}=4$ supergravity \cite{CarrascoN4Anomaly}:
\begin{align}
{\cal M}^{\oneloop}(1,2,3,4)_{{\cal N}=4}&= 2  \Bigl(\frac{\kappa}{2}\Bigr)^4 
   \biggr( \frac{\spb1.2 \spb3.4}{\spa1.2 \spa3.4} \biggr)^2
  \delta^{(8)}(\sum_{i=1}^4\lambda_i\eta_i^I) 
 \nn\\
 &\hskip1cm \times \Bigl( I_4(s,t)[\mu^4] + I_4(s,u)[\mu^4] 
 + I_4(t,u)[\mu^4] \Bigr)  \, .
 \label{Neq4anomalous4}
\end{align}
As outlined in \sect{GravitySymmetriesSection}, 
this amplitude breaks  the $U(1)$ duality symmetry of this theory \cite{CarrascoN4Anomaly} and is the amplitude-level manifestation
of the duality anomaly identified in~\cite{MarcusAnomaly} from a Lagrangian perspective. 

Another example is the double copy in which both amplitudes are given by Eq.~\eqref{QCDFourPointOneLoop}.
In $D$ dimensions, the double copy of a gluon has a total for $(D-2)^2$ states, corresponding to
a graviton ($D(D-3)/2$ states), antisymmetric tensor  ($(D-2)(D-3)/2$ states) and 
dilaton (1 state).   Taking both amplitudes in the double copy to be given by Eq.~\eqref{QCDFourPointOneLoop} with $N_f = 0$
leads to the four-graviton amplitude in a theory with a dilaton and antisymmetric tensor,
\begin{equation}
{\cal M}^{\oneloop}(1^+,2^+,3^+,4^+)= 4 \Bigl(\frac{\kappa}{2}\Bigr)^4 
   \biggr( \frac{\spb1.2 \spb3.4}{\spa1.2 \spa3.4} \biggr)^2
\Bigl( I_4(s,t)[\mu^8] + I_4(s,u)[\mu^8] 
 + I_4(t,u)[\mu^8] \Bigr) \, .
\label{DoubleQCDFourPointOneLoop}
\end{equation}
The polarization vectors in the spinor-helicity basis used in Eq.~\eqref{QCDFourPointOneLoop} project out the dilaton and antisymmetric 
tensor asymptotic states from this amplitude.
BCJ duality and the double copy for general helicity have been
described in Refs.~\cite{OneTwoLoopPureYMBCJ,OneTwoLoopMatterYMBCJ}.  For a
theory with only gravitons and no anti-symmetric tensor or dilaton,
the result for the identical helicity four-graviton amplitude is the
same as in \eqn{DoubleQCDFourPointOneLoop}, except that the overall
factor of 4 becomes a factor of 2.  This can be proven by inserting
graviton physical-state projectors into the unitarity cuts, as described
in \app{GeneralizedUnitaritySection}.

The integrals in the gauge-theory and gravity amplitudes in Eqs.~\eqref{QCDFourPointOneLoop}, \eqref{Neq4anomalous4} and \eqref{DoubleQCDFourPointOneLoop},
\begin{align}
 I_4(s,t)[\mu^{4k}] & = \int \frac{d^D \ell}{ (2 \pi)^D} \frac{\mu^{4k}}
{\ell^2 (\ell-p_1)^2 (\ell - p_1-p_2)^2 (\ell + p_4)^2 }  \,,
\label{muintegrals}
\end{align}
evaluate to
\begin{align}
 I_4(s,t)[\mu^{4}] & = - \frac{i}{(4 \pi)^2} \frac{1}{6} + \Ord(\eps) \,,
 \hskip 1.5 cm 
I_4(s,t)[\mu^{8}]  = - \frac{i}{(4 \pi)^2} \frac{1}{840} (2 s^2 + 2 t^2 + st) \,.
\end{align}
Their finite values arise due to a cancellation of the ${\cal O}(\epsilon)$ numerator 
factors and ${\cal O}(\epsilon^{-1})$ IR divergences. From this perspective, the nonvanishing amplitude \eqref{QCDFourPointOneLoop}
may be interpreted as a self-duality anomaly~\cite{Cangemi:1996rx}. The integrals~\eqref{muintegrals} 
may also be interpreted in terms of higher-dimensional 
integrals~\cite{Bern:1996ja}.

\begin{homework}
Consider the double copy of amplitudes in QCD with $N_f > 0$ flavors of quarks in the adjoint representation. 
Write down the spectrum of the resulting gravity theory. Construct the corresponding four-graviton amplitude.
Would you expect that this theory is consistent quantum 
mechanically for any value of $N_f$?
Answer the same questions if the  $N_f > 0$ flavors of quarks are in the fundamental representation. 
\end{homework}

\begin{figure*}[tb] 
\begin{center}
\includegraphics[scale=0.5]{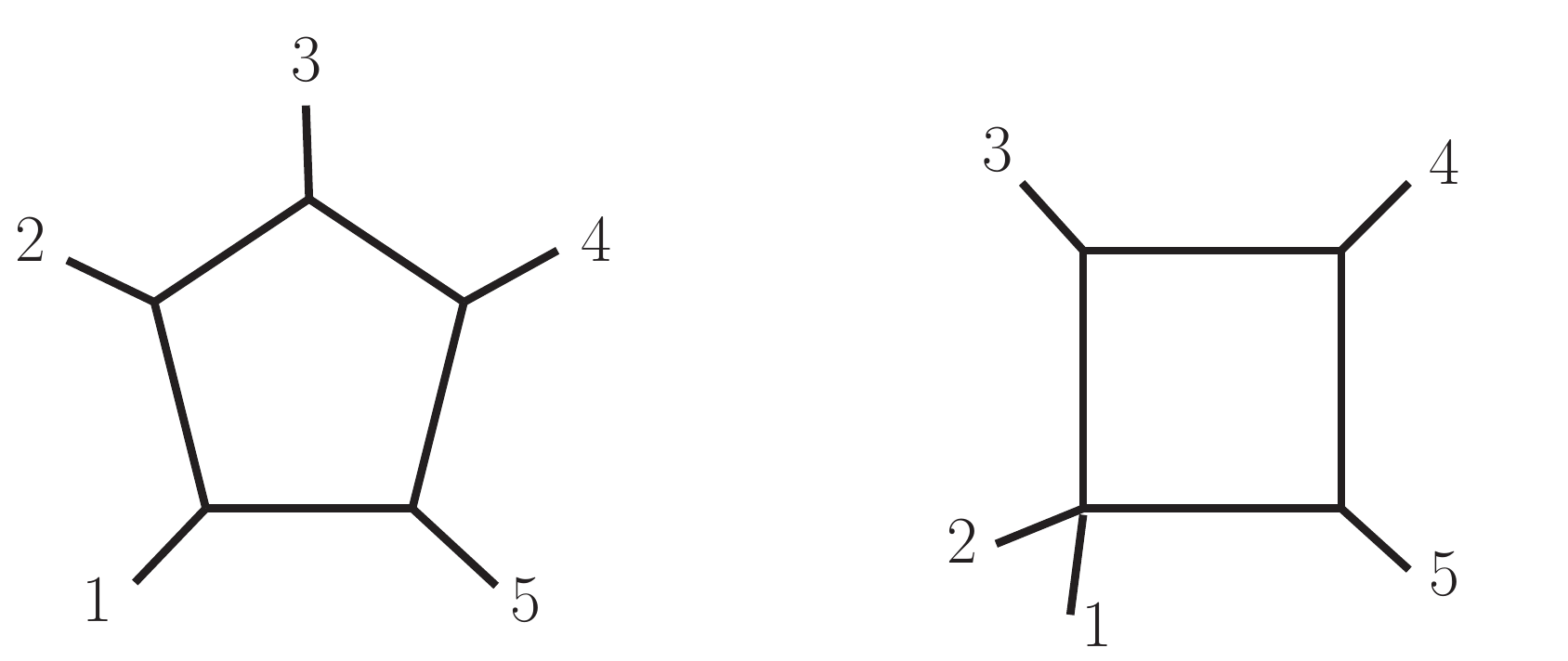}
\end{center}
\caption[a]{\small Pentagon and box integrals appearing in the 
$\NeqFour$ SYM five-point one-loop amplitudes.  The
complete set of such integrals is generated by permuting external legs
and removing overcounts.}
\label{PentagonBoxFigure}
\end{figure*}
%

As a more sophisticated example, consider the one-loop five-gluon amplitude. We will eventually restrict to the $\NeqFour$ SYM theory, but for now 
the discussion is quite general.  We only need to discuss 
the maximally-helicity-violating (MHV) amplitude, as the only other nonvanishing one, the \MHVbar{} amplitude, can be obtained by 
hermitian conjugation. Five-point amplitudes with other external states can be obtained through a suitable sequence of supersymmetry 
transformations.
This amplitude was constructed in Refs.~\cite{FiveYM,UnitarityMethod} in a color-trace basis.  Here we
rearrange it slightly and write it in the structure-constant basis,
\begin{equation}
{\cal A}^{\oneloop}_5(1,2,3,4,5) = g^5 
\sum_{S_5/(\mathbb{Z}_5\times \mathbb{Z}_2)} \!\! c_{12345}\, A^{\oneloop}_5(1,2,3,4,5)\, ,
\label{YMFivePtColor}
\end{equation}
where $A^{\oneloop}_5$ on the right-hand side are the five-point color-ordered partial amplitudes.
The sum runs over the distinct permutations of the external legs: this is the set of all $5!$ permutations, 
$S_5$, but with cyclic, $\mathbb{Z}_5$, and reflection symmetries, $\mathbb{Z}_2$, removed, leaving 12 distinct permutations.  
The color factor $c_{12345}$ is the one obtained from  the  pentagon diagram shown in \fig{PentagonBoxFigure},
with legs following the cyclic ordering, by dressing each vertex with an $\f^{abc}$.
This color decomposition holds for any gauge-theory amplitude with only adjoint-representation particles and can be reached
by starting from a generic color decomposition in terms of products of structure constants and repeatedly using the Jacobi identity
to favor structure constants with a single external color index.

\begin{homework}
By starting from Feynman diagrams, apply color Jacobi identities to
express all color factors in terms of those of pentagon diagrams.
What is the generalization for an arbitrary number of external legs?
(Feynman diagrams can helpful proving various properties, even if not
useful for high-multiplicity explicit calculations.)
\end{homework}

\begin{homework}
Generalize \eqn{YMFivePtColor} to include quarks in the fundamental representation  in the loop. (See
\eqn{QCDFourPointOneLoop} at four points).
\end{homework}

\begin{figure}[t]
\begin{center}
\begin{eqnarray*}
\hbox{\large $n$} \left(\parbox{2.9 cm}{
   \includegraphics[scale=.4]    {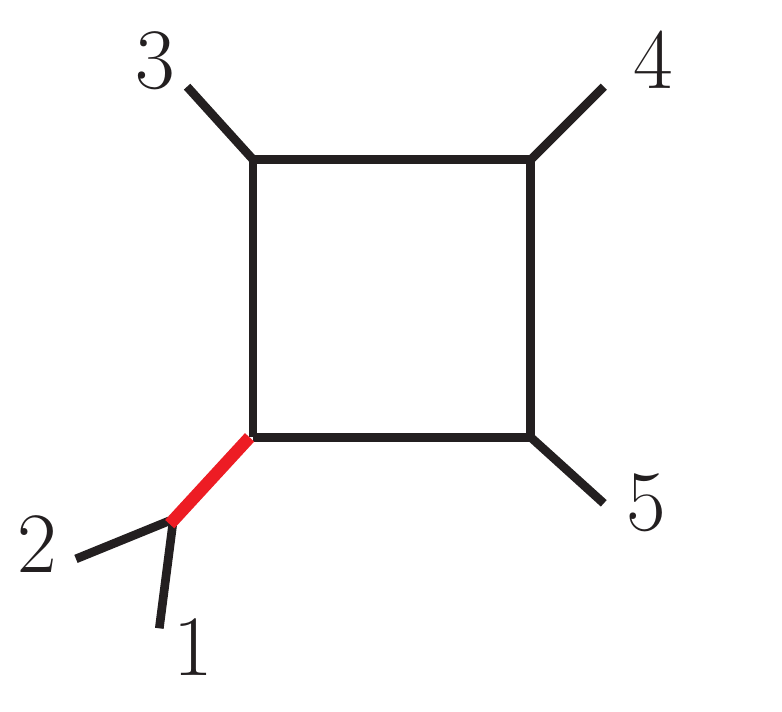}} 
          \right) \hbox{ \Large $=$ }
\hbox{\large $n$} \left(\parbox{2.7 cm}
   {\includegraphics[scale=.4] {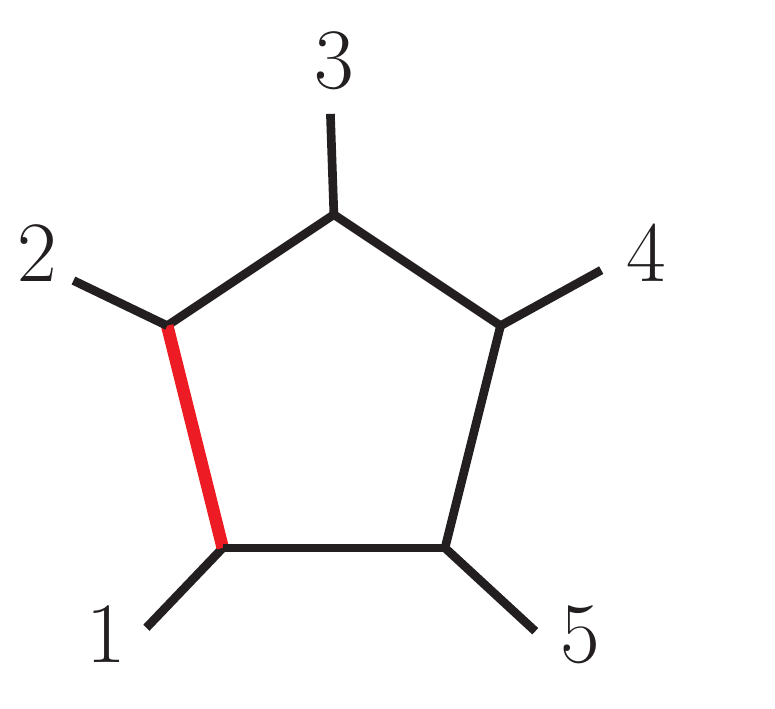}} 
          \right) \hbox{ \Large $-$ } 
\hbox{\large $n$} \left(\parbox{2.7 cm}
   {\includegraphics[scale=.4] {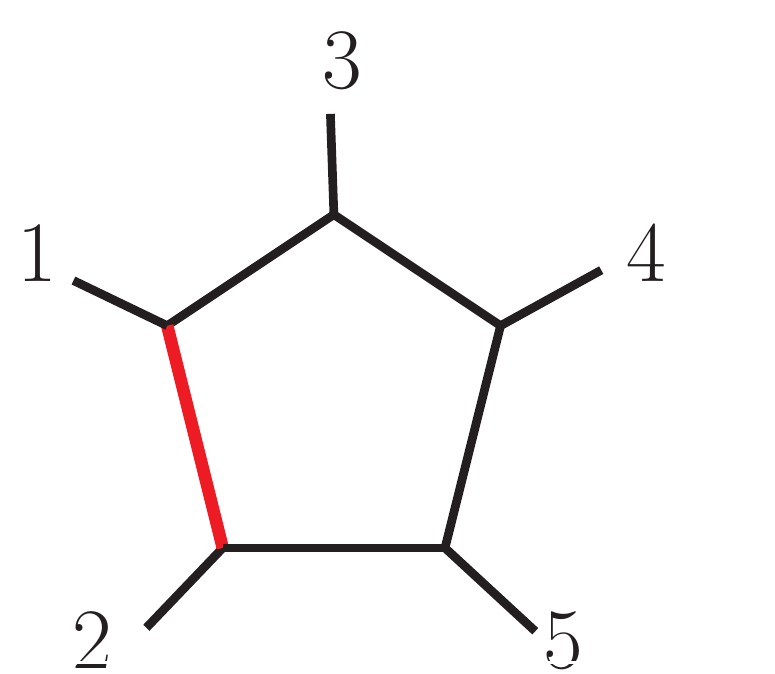}}
          \right) 
\end{eqnarray*}
\end{center}
\vskip -.2 cm 
\caption{A BCJ kinematic numerator relation between a diagram containing a box integral and
   two pentagon diagrams.  The shaded (red) line differs between the
  diagrams, but the others are identical.  }
\label{OneLoopPentagonBCJFigure}
\end{figure}

For $\NeqFour$ SYM theory, the color-ordered one-loop five-point amplitudes 
in \eqn{YMFivePtColor} are~\cite{FiveYM,UnitarityMethod},
\begin{align}
A_{\NeqFour}^{\oneloop}(1,2,3,4,5) &=\frac{1}{2}A^\tree_5(1,2,3,4,5)
\Bigl(s_{34}s_{45} I_4^{(12)345}+s_{45}s_{15} I_4^{1(23)45}+
s_{12}s_{15} I_4^{12(34)5}\nn \\
&\hskip3.3cm \null
+s_{12}s_{23} I_4^{123(45)}+s_{23}s_{34} I_4^{234(51)} \Bigr) + \Ord(\eps)\,,
\hskip 1 cm 
\label{FiveMHV}
\end{align}
where $A^\tree_5(1,2,3,4,5)$ is the color-ordered MHV tree-level amplitude. We may obtain the entire one-loop five-point MHV superamplitude 
by replacing $A^\tree_5(1,2,3,4,5)$ with the five-point tree-level MHV superamplitude in Eq.~\eqref{MHVSuperAmplitude}. 
The external kinematic invariants are $s_{ij} = (p_i + p_j)^2$.  The
$I_4^{abc(de)}$ are scalar box integrals where the legs in parenthesis
connect to the same vertex, { e.g.} $I_4^{(12)345}$ is the box
diagram in \fig{PentagonBoxFigure}.  
This representation (\ref{FiveMHV}) of the
amplitude does not manifestly satisfy the duality.
An alternative representation of the MHV superamplitude, which manifests the duality between color and kinematics, is~\cite{FivePointN4BCJ}:
\begin{equation} 
\mathcal{A}_{\NeqFour}^{\oneloop}(1,2,3,4,5) = g^5
\Big(\sum_{S_5/(\mathbb{Z}_5\times \mathbb{Z}_2)}\hskip -.2 cm 
 c_{12345} n_{12345}  I_5^{12345} 
  + \sum_{S_5/\mathbb{Z}_2^2} c_{[12] 345} n_{[12]345} \frac{1}{s_{12}}I_4^{(12)345} \Big)
\,.
\hskip 1 cm 
\label{FivePtMHV}
\end{equation}
Each of the two sums runs over the distinct permutations of the
external legs of the integrals.  For $I_5^{12345}$, the set
$S_5/(\mathbb{Z}_5\times \mathbb{Z}_2)$ denotes all permutations but with cyclic and
reflection symmetries removed, leaving 12 distinct permutations.  For
$I_4^{(12)345}$ the set $S_5/\mathbb{Z}_2^2$ denotes all permutations but with
the two symmetries of the one-mass box removed, leaving 30 distinct
permutations.  
The pentagon numerator for this representation of the superamplitude is
\begin{equation}
n_{12345} = - \delta^{(8)}(Q) \frac{ \spb1.2 \spb2.3\spb3.4 \spb4.5 \spb5.1}
 {4 i \epsilon(1,2,3,4)}\,,
\label{pentnumer}
\end{equation}
where $ 4 i \epsilon(1,2,3,4)=
4 i \epsilon_{\mu \nu \rho \sigma}k_1^{\mu}k_2^{\nu}k_3^{\rho}k_4^{\sigma}
=\spb{1}.{2}\spa{2}.{3}\spb{3}.{4}\spa{4}.{1}-\spa{1}.{2}\spb{2}.{3}\spa{3}.{4}\spb{4}.{1}$.
With this pentagon numerator, the box numerators that manifest the kinematic Jacobi relations illustrated
in \fig{OneLoopPentagonBCJFigure} are
\begin{equation}
n_{[12]345} = n_{12345} - n_{21345}\, .
\label{boxnumer}
 \end{equation}   
Other box numerators are obtained by relabeling.
It is not difficult to see that the diagrams with triangle or bubble integrals have vanishing numerators.  For example, the numerator of the
triangle diagram with momenta $p_1+p_2$ at one vertex and $p_4 + p_5$ at another is
\begin{align}
\label{TriangleDiagramN4Numerator}
n_{[12]345} - n_{[12]354} & = n_{12345} - n_{21345} - n_{12354} + n_{21354} \\
& = 
- \frac{\delta^{(8)}(Q)}{4 i \epsilon(1,2,3,4)}  \Big\{
 \spb1.2 \spb2.3\spb3.4 \spb4.5 \spb5.1 
+ \spb2.1 \spb1.3\spb3.4 \spb4.5 \spb5.2   \nn \\
& \hskip 3.0 cm \null  
+ \spb1.2 \spb2.3\spb3.5 \spb5.4 \spb4.1
+ \spb2.1 \spb1.3\spb3.5 \spb5.4 \spb4.2  \Big\} \,,
\nn
\end{align}
where we used momentum conservation to relate all Levi-Civita symbols contracted with four external momenta.
Upon use of the Schouten identities,
\begin{align}
\spb5.1 \spb2.3 - \spb5.2\spb1.3 = \spb1.2 \spb3.5 \, ~ ,
\quad
\spb2.3 \spb4.1 - \spb1.2\spb3.4 = \spb1.3 \spb4.2 \, ,
\label{schouten}
\end{align}
the term in brackets in \eqn{TriangleDiagramN4Numerator} vanishes, so
\begin{eqnarray}
n_{[12]345} - n_{[12]354} = n_{12345} - n_{21345} - n_{12354} + n_{21354} = 0 \,.
\end{eqnarray}
The first of the identities \eqref{schouten} is used to combine the first two terms
in \eqn{TriangleDiagramN4Numerator} and the second identity shows that the remaining term cancel. 
\begin{homework}
Show that all kinematic numerator relations hold for the amplitude given in \eqn{FivePtMHV}.
\end{homework}

A nice feature of this representation is that the numerator factors of both the pentagon and box 
integrals do not depend on loop momentum. This greatly simplifies the construction of the corresponding 
supergravity amplitudes.

Given that the duality holds for the representation \eqref{FivePtMHV} of the five-point one-loop MHV  $\NeqFour$ SYM
superamplitude,  we can immediately obtain the corresponding $\NeqEight$ amplitude. We replace the color factors with a 
numerator factor \eqref{c_n_replace},
\begin{equation}
c_{12345} \rightarrow n_{12345}\,, 
\hskip 2 cm 
c_{[12] 345} \rightarrow n_{[12] 345}\,, 
\end{equation}
as well as the gauge coupling with the gravitational one.
The resulting five-graviton one-loop MHV superamplitude in ${\cal N}=8$ supergravity reads \eqref{DCformula}
\begin{equation} 
{\cal M}^{\text{1-loop}}_{\NeqEight}(1, 2, 3, 4, 5) =  
\Bigl({\kappa \over 2} \Bigr)^5  
\Big(\sum_{S_5/(\mathbb{Z}_5\times \mathbb{Z}_2)} \hskip -.3 cm  (n_{12345})^2  I_5^{12345} + 
   \sum_{S_5/\mathbb{Z}_2^2} (n_{[12]345})^2 \frac{1}{s_{12}}I_4^{(12)345} \Big) \,,
\label{FiveGrav}
\end{equation}
where the sums run over the same permutations as in \eqn{FivePtMHV} and, as discussed in \sect{GravitySymmetriesSection}, 
the $\delta^{(16)}(Q)$ should be understood as containing eight different $\eta$ parameters for each external particle.

The scalar pentagon integral and the one external-mass box integral have been computed in Ref.~\cite{,PentagonIntegral}.
We include them here for convenience:
\begin{align}
I_4^{(12)345} = &  -\frac{2ic_{\Gamma}}{s_{34}s_{45}} \Bigg \{ -\frac{1}{\epsilon^2}
\Bigl[ (-s_{34})^{-\e} + (-s_{45})^{-\e} - (-s_{12}^2)^{-\e} \Bigr]\nn \\
 &\null 
  + \Li_2\left(1-{s_{12}\over s_{34}}\right)
  +\Li_2\left(1-{s_{12}\over s_{45}}\right)
   + \frac{1}{2}\ln^2\left({s_{34}\over s_{45}}\right) + \frac{\pi^2}{6} \Bigg \}
+ \Ord(\e) \,, \\
I_5^{12345} =& \sum_{\mathbb{Z}_5}{ -ic_{\Gamma} (-s_{51})^\eps (-s_{12})^\eps
   \over (-s_{23})^{1+\eps} (-s_{34} )^{1+\eps} (-s_{45})^{1+\eps} }
  \left[ {1\over \eps^2} + 2\Li_2\Bigl(1- {s_{23} \over s_{51} }\Bigr)
  + 2\Li_2\Bigl( 1- {s_{45}\over s_{12}} \Bigr)
  - {\pi^2 \over 6} \right]
 \nn \\
& \null \hskip 5 cm
+ \Ord(\eps)\,,
\label{FivePtN4Integrals}
\end{align}
where $\Li_2(x)$ is the dilogarithm function and $c_\Gamma$ is defined in Eq.~\eqref{cGamma}.

\begin{homework}
Show that the double copy of the one-loop five-point amplitude where one copy is of an MHV amplitude and the second an \MHVbar{} amplitude 
vanishes. The \MHVbar{} amplitude is obtained from the MHV one by parity which amounts to replacing $\spa{a}.{b} \leftrightarrow \spb{a}.{b}$
and flipping the overall sign of the amplitude.  Do you expect a similar property to hold at $n$ points or at higher loops?
\end{homework}

As discussed above, the double-copy construction works even if the
duality is manifest in only one gauge-theory factor.  Starting with the color
decomposition in \eqn{YMFivePtColor} and using the fact that for the one-loop
five-point $\NeqFour$ SYM amplitude the pentagon numerators
are independent of loop momentum, we immediately obtain  five-point superamplitudes for $({\cal N}+4)$-extended supergravities. 
By taking the second copy to be any pure SYM theory, with  color-ordered one-loop five-point amplitudes $A^\oneloop_{\cal N}(1,2,3,4,5)$, 
we find
\begin{align}
{\cal M}^\oneloop_{\cal N}(1,2,3,4,5) =& \Bigl({\kappa \over 2} \Bigr)^5
\sum_{S_5/(\mathbb{Z}_5\times \mathbb{Z}_2)} \hskip -.3 cm 
n_{12345}\, A^\oneloop_{\cal N}(1,2,3,4,5)\, . \hskip 1cm 
\label{NeqFourSixMatter}
\end{align}
Here $n_{12345}$ is given in \eqn{pentnumer} and the sums run, as in the case of the ${\cal N}=4$ amplitude, over all the permutations which 
are not related to each other by cyclic permutations or reflections.

\subsection{One-loop examples of BCJ duality:  SYM theories with reduced supersymmetry}

Gauge theories with reduced supersymmetry provide an opportunity to discuss the construction of duality-satisfying (loop-level) scattering amplitudes
with fields in representations other than the adjoint. A simple example, which we will review here in some detail, is the one-loop four-matter-field 
superamplitude in  $\cN=2$ SYM theory with a single hypermultiplet in a complex representation ${\cal R}$ \cite{Chiodaroli2013upa,Ben-Shahar:2018uie}. 
The color factors $c_j$ in Eq.~\eqref{CubicRepLoop} are now constructed by dressing every vertex of every diagram with a gauge-group generator in the appropriate 
representation. This more complicated color structure is a consequence of reduced supersymmetry, which allows for matter fields in non-adjoint representations.
To keep supersymmetry manifest, we organize the hypermultiplet asymptotic states as on-shell superfields and their CPT-conjugates, which are treated as distinct:
\begin{equation}
{\Phi}_{\cN=2}{}_{\hat\alpha} = \chi_{+ \hat \alpha} + \eta^i \varphi_{i \hat \alpha} + \eta^1 \eta^2 \tilde \chi_{- \hat \alpha} 
\qquad
\overline{\Phi}_{\cN=2}{}^{\hat\alpha}  = \tilde \chi_{+}^{\hat \alpha} + \eta^i \overline{\varphi}{}_{i}^{\hat \alpha} + \eta^1 \eta^2  \chi_{-}^{\hat \alpha} .
\label{Neq2superfields}
\end{equation}
The lower and upper $\hat \alpha$  is the ${\cal R}$ and ${\bar {\cal R}}$ representation indices, respectively.
As outlined in \sect{SecOrbifold}, such superfields with reduced supersymmetry can in principle be obtained from the ones of $\cN=4$ SYM 
theory by an orbifold truncation.

At one loop, a duality-satisfying representation of the four-hypermultiplet superamplitude can be constructed in terms of two master numerators, which 
can be chosen to belong to two box diagrams. 
Adopting the standard notation for theories with matter (super)fields, we denote the adjoint vector multiplet with a curly  line and the 
complex-representation hypermultiplet with a solid line with an arrow. The master numerator factors and the corresponding diagrams are \cite{Chiodaroli2013upa,Ben-Shahar:2018uie}:
\begin{eqnarray}
n \left( \parbox{2.5 cm}{\includegraphics[width=0.17\textwidth]{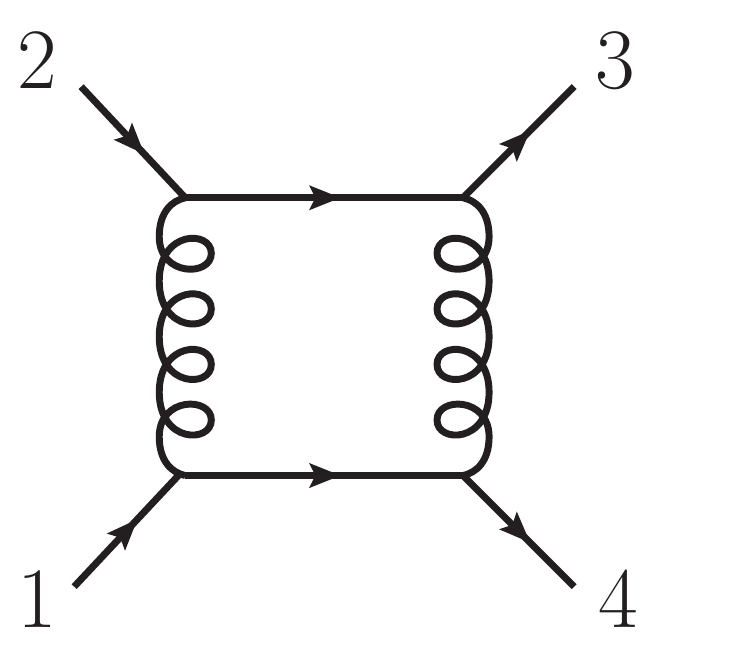}} \right)
   =  {s^2 \over \langle 12 \rangle \langle 34 \rangle} \delta^{(4)} \big( Q \big) && 
n \left( \parbox{2.5 cm}{\includegraphics[width=0.17\textwidth]{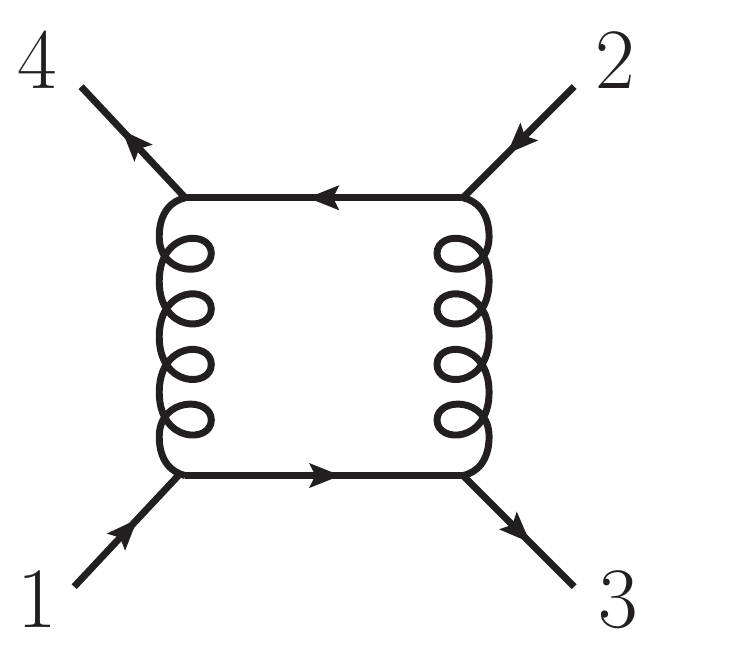}} \right)
=  - {st \over \langle 12 \rangle \langle 34 \rangle} \delta^{(4)} \big( Q \big)  \,, \qquad     
\end{eqnarray} 
where $\delta^{(4)} \big( Q \big) = \delta^{(4)} \big( \sum_n \eta_n^i \lambda_n \big)$. 
The other box integrals can then be obtained by permutation, keeping in mind that the overall superamplitude possesses a $\mathbb{Z}_2 \times \mathbb{Z}_2$ Fermi symmetry under the exchange  of hypermultiplet superfields. 
For example, a third box-integral numerator is
\begin{equation}
n \left( \parbox{2.5 cm}{\includegraphics[width=0.17\textwidth]{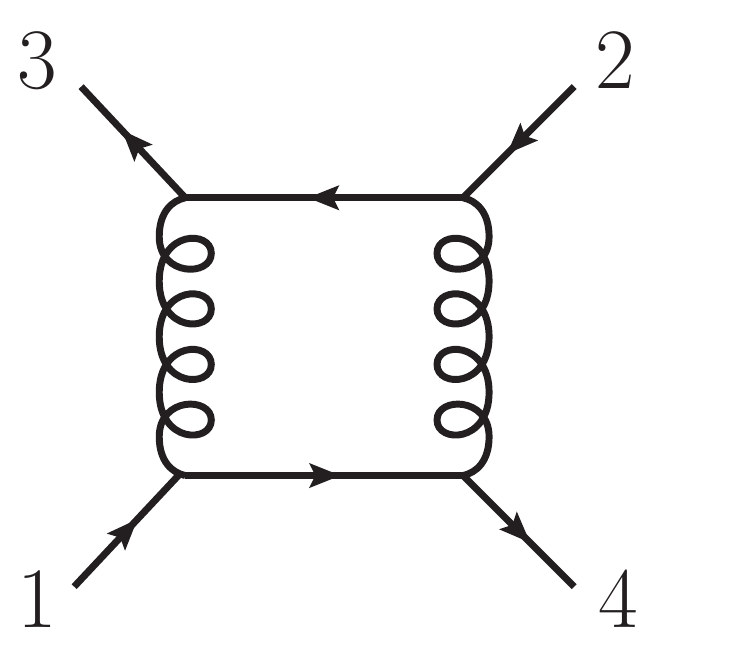}} \right)
= - {su \over \langle 12 \rangle \langle 34 \rangle} \delta^{(4)} \big( Q \big) \,.
\end{equation}

Numerators for the triangle and bubble diagrams can be obtained via the kinematic numerator relations. They can be organized in two distinct sets:
(1) those that mirror relations between color factors which are a consequence of the defining commutation relations of the color Lie algebra 
and
(2) those that corresponding to color relations that hold only for certain groups and representations but are nonetheless required for the consistency of the 
double copy of a hypermultiplet with a vector multiplet.
An example of numerator relations from the first group is
\begin{equation}
n \left( \parbox{2.5 cm}{\includegraphics[width=0.17\textwidth]{figs/Box1review}} \right) -
n \left( \parbox{2.5 cm}{\includegraphics[width=0.17\textwidth]{figs/Box2breview}} \right) 
=   
n \left( \parbox{2.1 cm}{\includegraphics[width=0.15\textwidth]{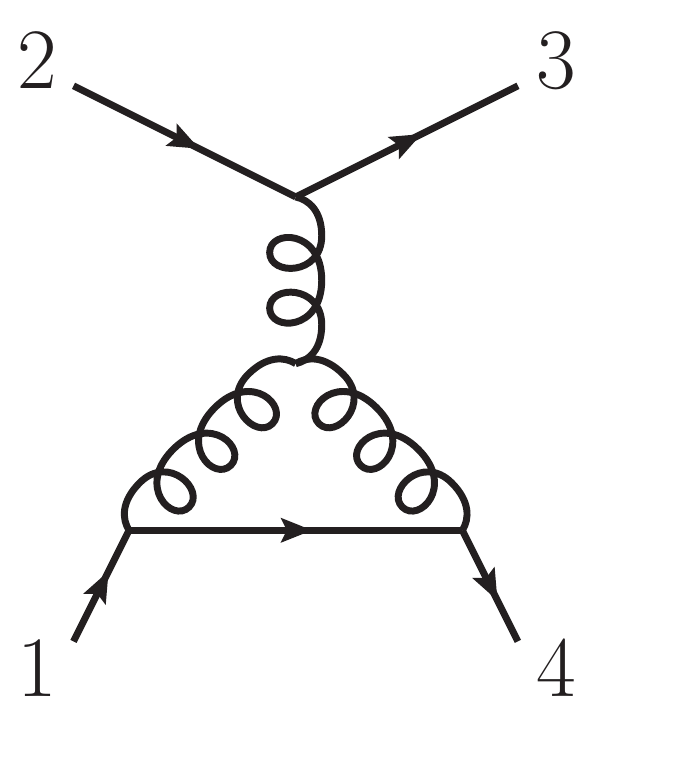}} \right) . \label{numrel1}
\end{equation}
From the double-copy perspective, following the argument presented in \sect{diffeoSubsection}, these relations are required for obtaining gravity 
amplitudes invariant under linearized diffeomorphisms.
An example of color relations that hold only for certain groups and representations is
\begin{equation}
T^{\hat a \ \hat \gamma}_{\ \hat \alpha} T^{\hat a \ \hat \delta}_{\ \hat \beta} = T^{\hat a \ \hat \delta}_{\ \hat \alpha} T^{\hat a \ \hat \gamma}_{\ \hat \beta} \,.
\end{equation}
The corresponding numerator relations include, for example,
\begin{equation}
n \left( \parbox{2.5 cm}{\includegraphics[width=0.17\textwidth]{figs/Box2areview}} \right) = 
n \left( \parbox{2.8 cm}{\includegraphics[width=0.19\textwidth]{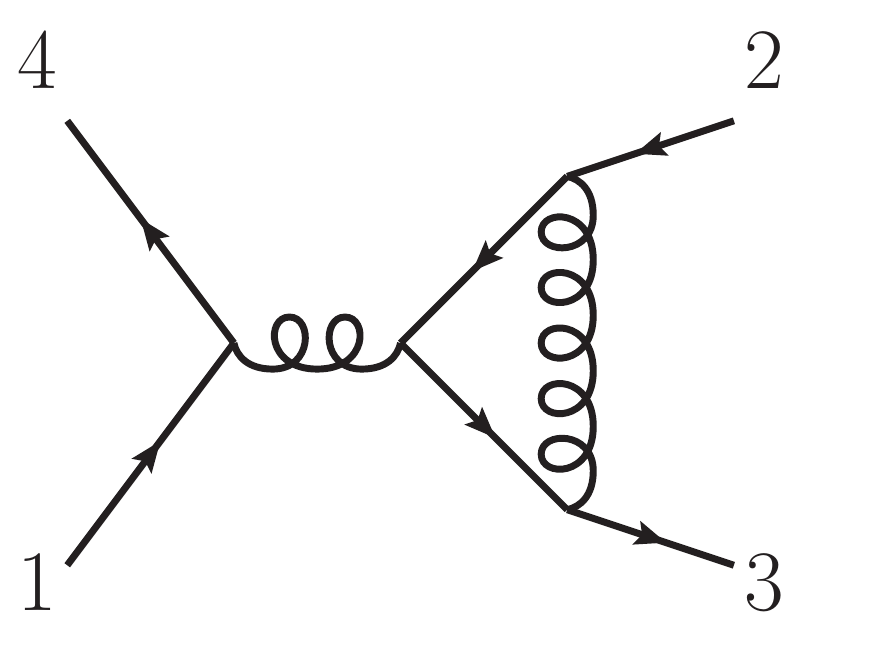}} \right) .  \label{numrel2} 
\end{equation}
While these color relations are satisfied only for certain choices of gauge group and representations, the fact that the form \eqref{CubicRepLoop}
is independent of such choices suggests that one may choose, as we do here, to always impose the corresponding numerator relations.
One may easily convince oneself that these numerator relations are required by consistency of the double-copy construction in case 
the hypermultiplet fields are combined with spin one fields in the conjugate matter representation. In other cases they may be regarded 
as ``bonus''  relations; it is not clear { a priori} that there exist solutions to the numerator relations in the second group even when solutions to 
the numerator relations in the first group do.

In  \sect{homogeneousSection}, we have reviewed the double-copy construction for homogeneous supergravities, and showed that it reproduces the 
existing classification of such theories. An important ingredient of the construction are matter fields in pseudo-real representations. It is therefore instructive 
to see how our one-loop numerators described above are modified in this case.  
To enforce the pseudo-reality of the gauge group representation (\ie the equivalence of the upper and lower ${\hat\alpha}$ indices in Eq.~\eqref{Neq2superfields}), the on-shell superfields $\Phi_{\cN=2}$ and $\overline{\Phi}_{\cN=2}$ are identified. Consequently, the superamplitude needs to acquire a complete Fermi symmetry for all of its external legs. Drawing from this observation, the 
numerators for a theory with pseudo-real half-hypermultiplets can be obtained as the unique set of numerators which are both invariant under the permutation 
of all external legs and reduce to the numerators for the complex case whenever the corresponding color factors are nonzero.  
More concretely, in the pseudo-real case we have only one master numerator: 
\begin{eqnarray}
n \left( \parbox{2.5 cm}{\includegraphics[width=0.17\textwidth]{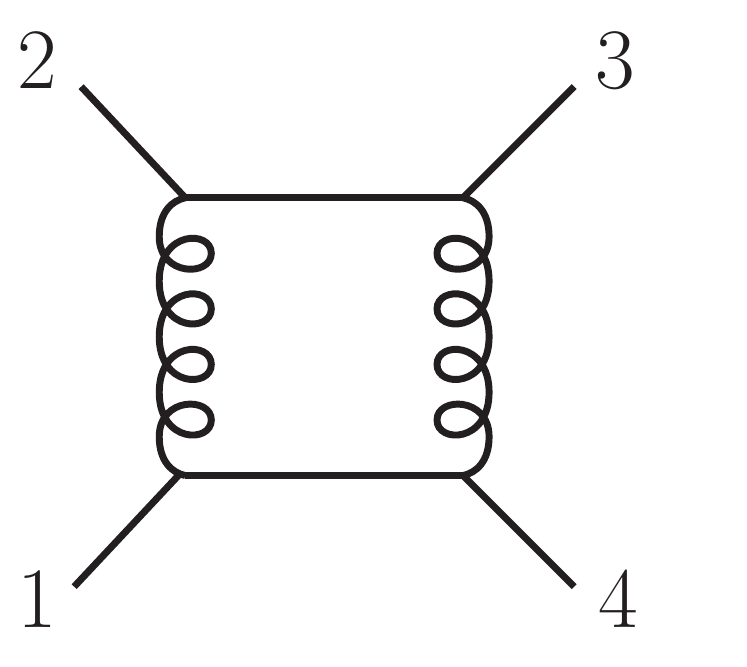}} \right)
=  {s^2 \over \langle 12 \rangle \langle 34 \rangle} \delta^{(4)} \big( Q \big)    \, ,
\label{master-pr}
\end{eqnarray} 
and all the other numerators are obtained either from permutation symmetry or from the numerator relations (\ref{numrel1}) and (\ref{numrel2}). 
For half-hypermultiplets in pseudo-real representations, solid lines no longer carry an arrow since the matter 
half-hypermultiplets are CPT-self-conjugate. 

\begin{homework} \label{ex-pr}
Given the master numerator (\ref{master-pr}), use numerator relations to generate all nonzero numerators (up to permutation symmetry).	
\end{homework}

We emphasize that, as in the case of the one-loop four- and five-point superamplitudes of ${\cal N}=4$ SYM theory, the duality-satisfying 
kinematic numerators of the superamplitude reviewed here are independent of the loop momentum. Consequently, the physical properties of 
the double-copy supergravity theory can be directly 
related to properties of the other gauge theory entering the construction. 
Consider, for example,  the construction for homogeneous Maxwell-Einstein supergravities explained in  \sect{homogeneousSection}.
We can relate the one-loop divergences of supergravity amplitudes with four vector superfields constructed as hypermultiplet~$\times$~fermion 
to a linear combination of various parts of the one-loop beta function of the non-supersymmetric gauge theory, 
\begin{eqnarray}
{\cal M}^{\text{1-loop}}\Big|_{\text{div}} \! \! \! \! \! \! &=& \! \! \! { -i \over (4 \pi)^2} {   s \; \delta^{(4)}(Q) \over \langle 12 \rangle \langle 34 \rangle   } \Big({\kappa \over 2}\Big)^4 
\! \!  \left\{ s A^{\text{tree}}_{s,\phi} \Big( \beta_\phi \big|_{T(G)} \! - \! { \beta_\phi \over 2} \Big|_{T(R)} \Big) 
\! + \! s A^{\text{tree}}_{s,A}  \Big( \beta_A \big|_{T(G)} \! - \! { \beta_A \over 2} \Big|_{T(R)} \Big) \! \!
\right\} {1 \over \epsilon} \no \\ && \hskip 10cm + \text{Perms} \, .
\label{betafunc}
\end{eqnarray}
Here $\beta_\phi, \beta_A$ are the beta-functions for the gauge coupling and the Yukawa interactions. 
We use the notation $\beta\big|_{T(G), T(R)}$ to label the parts of the relevant beta functions that are proportional 
to the index of the adjoint, $T(G)$,and pseudo-real, $T(R)$, matter representations. 
$A^{\text{tree}}_{s,A}$ and $A^{\text{tree}}_{s,\phi}$ are, respectively, the $s$-channel gluon and scalar exchange parts of the 
gauge theory tree-level amplitudes. 
Finally, it should be noted that compact expressions for two-loop amplitudes for $\cN=2$ gauge theories with matter can be found in Ref.~\cite{KalinN2TwoLoop}.  

\begin{homework}
Use the result of Exercise 7.9 to verify equation (\ref{betafunc}).	
\end{homework}

\subsection{Two-loop examples}

\begin{figure*}[tb]
\begin{center}
\includegraphics[scale=0.4]{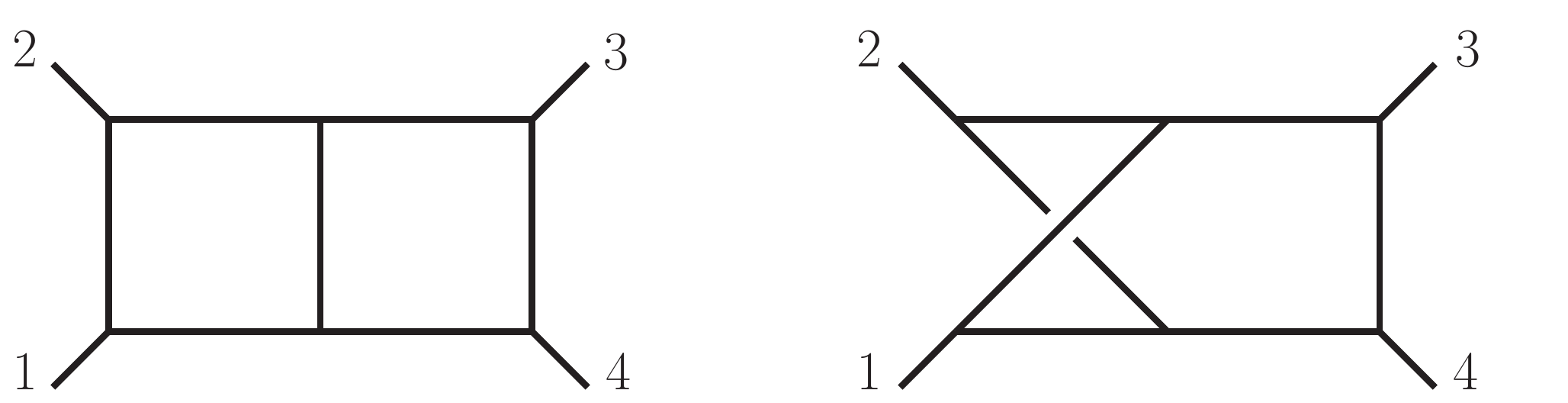}
\end{center}
\vskip -.2 cm
\caption[a]{\small Diagrams for the two-loop integrals appearing in the
two-loop four-point $\NeqFour$  and $\NeqEight$ supergravity amplitudes.
}
\label{TwoloopDiagramFigure}
\end{figure*}

If the duality between color and kinematics holds at tree-level in $D$ dimensions, then
it also holds on all $D$-dimensional generalized cuts that decompose a
loop amplitude into a sum of products of tree amplitudes. Thus,
barring anomalies, it is expected to hold beyond one-loop level. As an illustrative
example, consider the two-loop four-point
amplitude of $\NeqFour$ SYM theory. This amplitude,
originally constructed in Refs.~\cite{BRY,BDDPR}, is
\begin{align}
{\cal A}_4^{\twoloop}(1,2,3,4) &=  - g^6 st  \,
              A_4^{\rm tree}(1,2,3,4) \Bigl(
    c^{\P}_{1234} \, s \, I_4^{\twoloop,\P}(s, t)
  + c^{\P}_{3421} \, s \, I_4^{\twoloop,\P}(s, u)
\nn \\
 & \null \hskip  2.5 truecm
  + c^{\NP}_{1234} \, s\, I_4^{\twoloop , \NP}(s,t)
  + c^{\NP}_{3421} \, s\, I_4^{\twoloop , \NP}(s,u)
+  {\rm cyclic} \Bigr)\,, 
\label{TwoLoopYM}
\end{align} 
where ``$+$~cyclic'' indicates that one should add the two cyclic permutations of $(2,3,4)$.  
The integrals correspond to the scalar planar and nonplanar double-box diagrams shown in \fig{TwoloopDiagramFigure}.
As at one loop, the color factor of each diagram is obtained by dressing each cubic vertex with an $\f^{abc}$ factor. 

\begin{figure}[t]
\begin{center}
\begin{eqnarray*}
&& \hbox{\large $n$} \left(\parbox{2.5 cm}{\includegraphics[scale=.43]{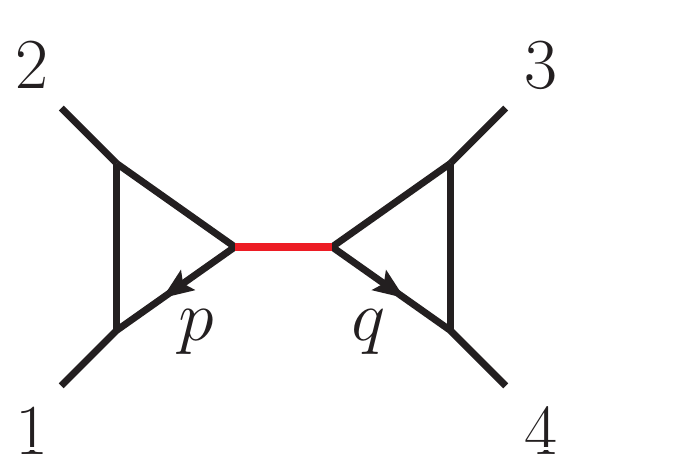}} \right) \hbox{\Large $=$ }
\hbox{\large $n$} \left(\parbox{2.5  cm}{\includegraphics[scale=.43]{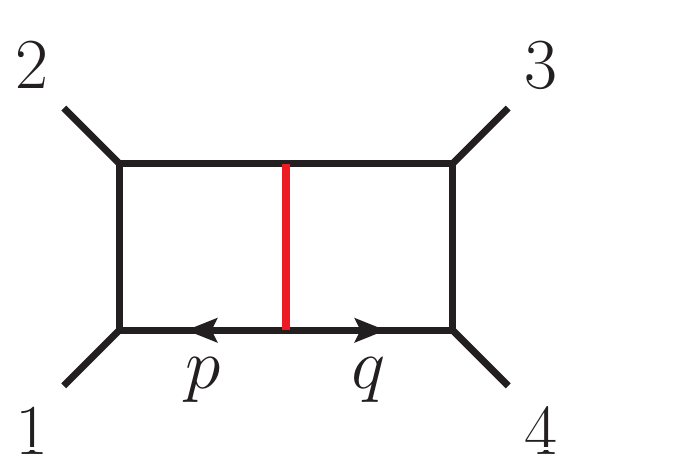}} \right) \hbox{\Large $-$ }
\hbox{\large $n$} \left(\parbox{2.8cm}{\includegraphics[scale=.43]{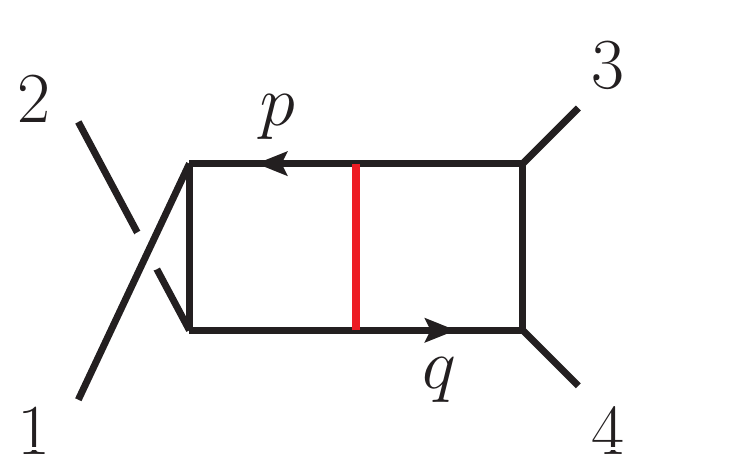}} \right) \\
&&
\hbox{\large $n$} \left(\parbox{2.5 cm}{\includegraphics[scale=.43]{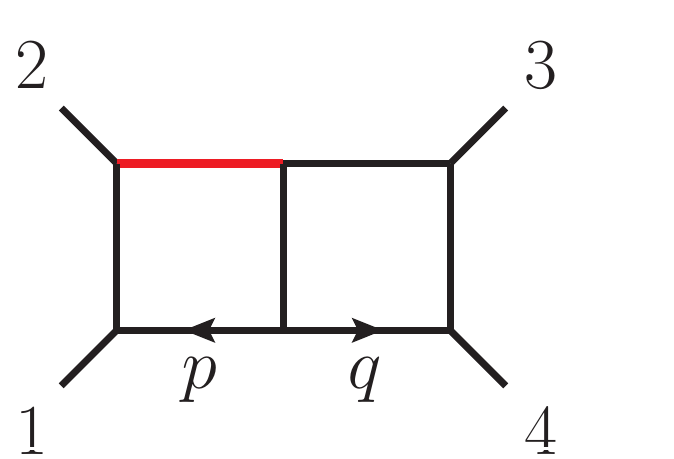}} \right) \hbox{\Large $=$ }
\hbox{\large $n$} \left(\parbox{2.  cm}{\includegraphics[scale=.43]{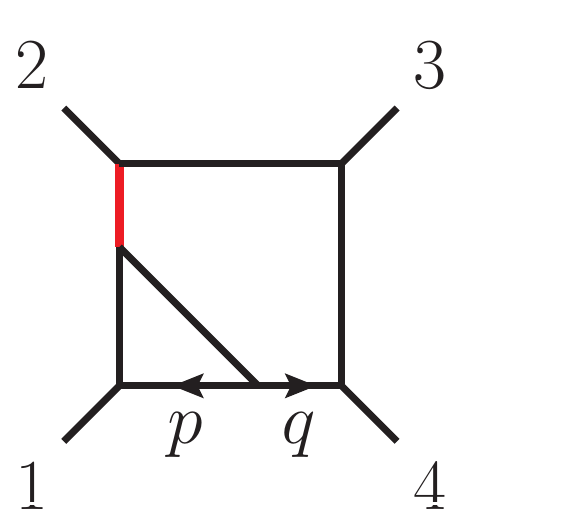}} \right) \hbox{\Large $-$ }
\hbox{\large $n$} \left(\parbox{2.5 cm}{\includegraphics[scale=.43]{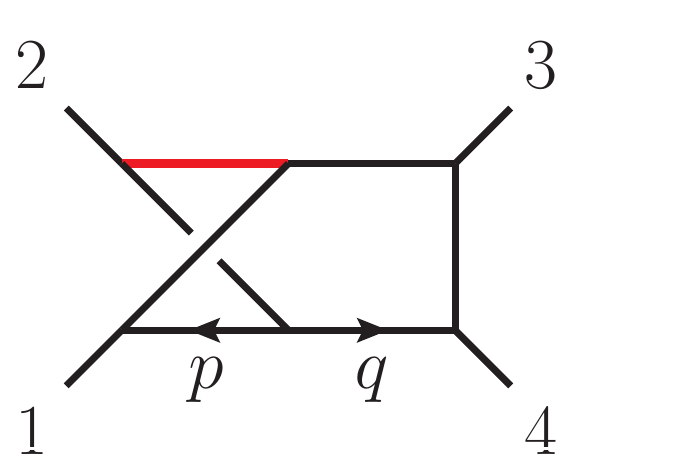}} \right) \\
&& 
\hbox{\large $n$} \left(\parbox{2.5 cm}{\includegraphics[scale=.43]{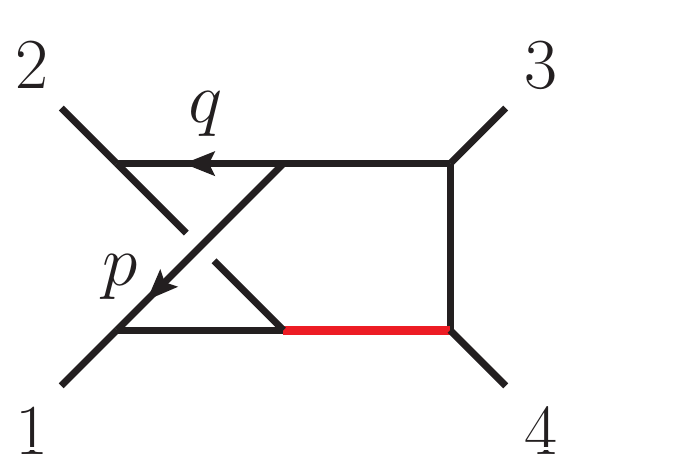}} \right) \hbox{\Large $=$ }
\hbox{\large $n$} \left(\parbox{2.5  cm}{\includegraphics[scale=.43]{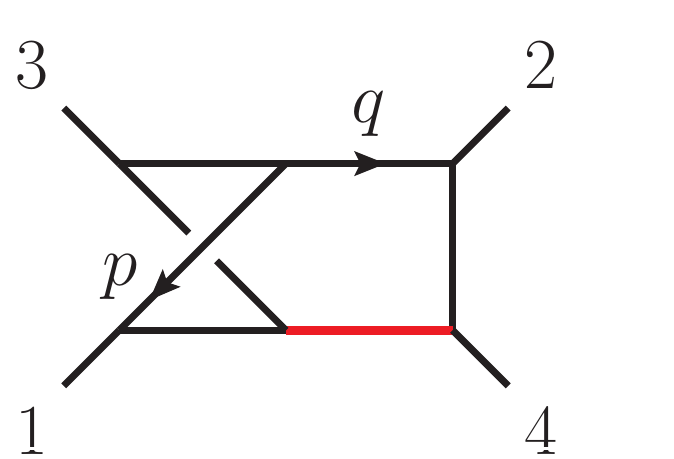}} \right) \hbox{\Large $-$ }
\hbox{\large $n$} \left(\parbox{2.5 cm}{\includegraphics[scale=.43]{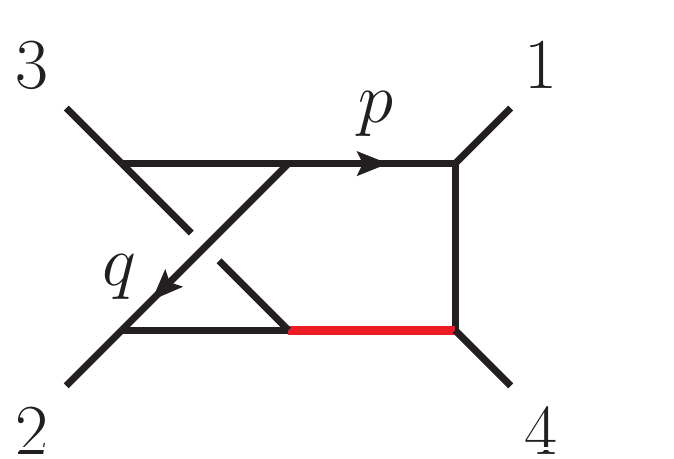}} \right)
\end{eqnarray*}

\end{center}
\caption{Examples of BCJ kinematic numerator relations at two loops.}
\label{TwoLoopBCJFigure}
\end{figure}

As the diagrams appearing in the amplitude are already cubic, we can read off the kinematic numerators for each diagram.
They are:
\begin{align}
n^\P_{1234} = n^\NP_{1234} = i s^2 t A^\tree_4(1,2,3,4)\,, \hskip 1 cm 
n^\P_{3412} = n^\NP_{3412} = i s^2 t A^\tree_4(1,2,3,4)\,, \nn \\
n^\P_{1342} = n^\NP_{1342} = i u s t A^\tree_4(1,2,3,4)\,, \hskip 1 cm 
n^\P_{4213} = n^\NP_{4213} = i u s t A^\tree_4(1,2,3,4)\,, \nn \\
n^\P_{1423} = n^\NP_{1423} = i s t^2 A^\tree_4(1,2,3,4)\,, \hskip 1 cm 
n^\P_{2314} = n^\NP_{2314} = i s t^2 A^\tree_4(1,2,3,4)\, .
\label{Neq42loopNumerators}
\end{align}
The factor $s t A^\tree_4(1,2,3,4)$, being crossing symmetric, remains as overall factor 
for the complete amplitude, after all cyclic permutations of $(2,3,4)$ are added.

Because of the limited set of nonvanishing diagrams, it is straightforward to check that this amplitude satisfies all duality relations.  
Three of them are shown in \fig{TwoLoopBCJFigure}.  
The complete set may be obtained by starting with the diagrams in \fig{TwoloopDiagramFigure} and systematically generating the duality relations.

Following  the double-copy prescription (\ref{c_n_replace}), we obtain the corresponding $\NeqEight$ supergravity amplitude by replacing 
the color factor with a numerator factor,
\begin{equation}
 c^{\P}_{1234} \rightarrow  i s^2 t A^\tree(1,2,3,4)\,,  \hskip 2 cm
 c^{\NP}_{1234} \rightarrow i s^2 t A^\tree(1,2,3,4) \,,
\label{ColorReplacement}
\end{equation}
including relabelings and then swapping the gauge coupling for the gravitational one.  
Indeed, this gives the correct $\NeqEight$ supergravity amplitude, as first noted in Ref.~\cite{BDDPR} which also verified it against the direct 
construction from unitarity cuts.

As mentioned in \sect{DualitySection}, generalized gauge invariance implies that only one of the two copies must be in a form
manifestly satisfying the duality (\ref{duality}); for the second copy, such a form should exist but its use is not required. 
The color Jacobi identity allows us to express any four-point color factor of an adjoint representation in terms of the ones in
\fig{TwoloopDiagramFigure}~\cite{DixonMaltoni}. If the duality and double-copy properties hold and because of the independence of the 
momentum of the ${\cal N}=4$ SYM numerator factors \eqref{Neq42loopNumerators}, it is possible to obtain
integrated ${\cal N} \ge 4$ supergravity amplitudes starting from ${\cal N} \le 4$ SYM theory and applying the 
replacement rule~(\ref{ColorReplacement}) \cite{N46Sugra2}.  

\begin{figure}[t]
\centerline{\includegraphics[width=13cm]{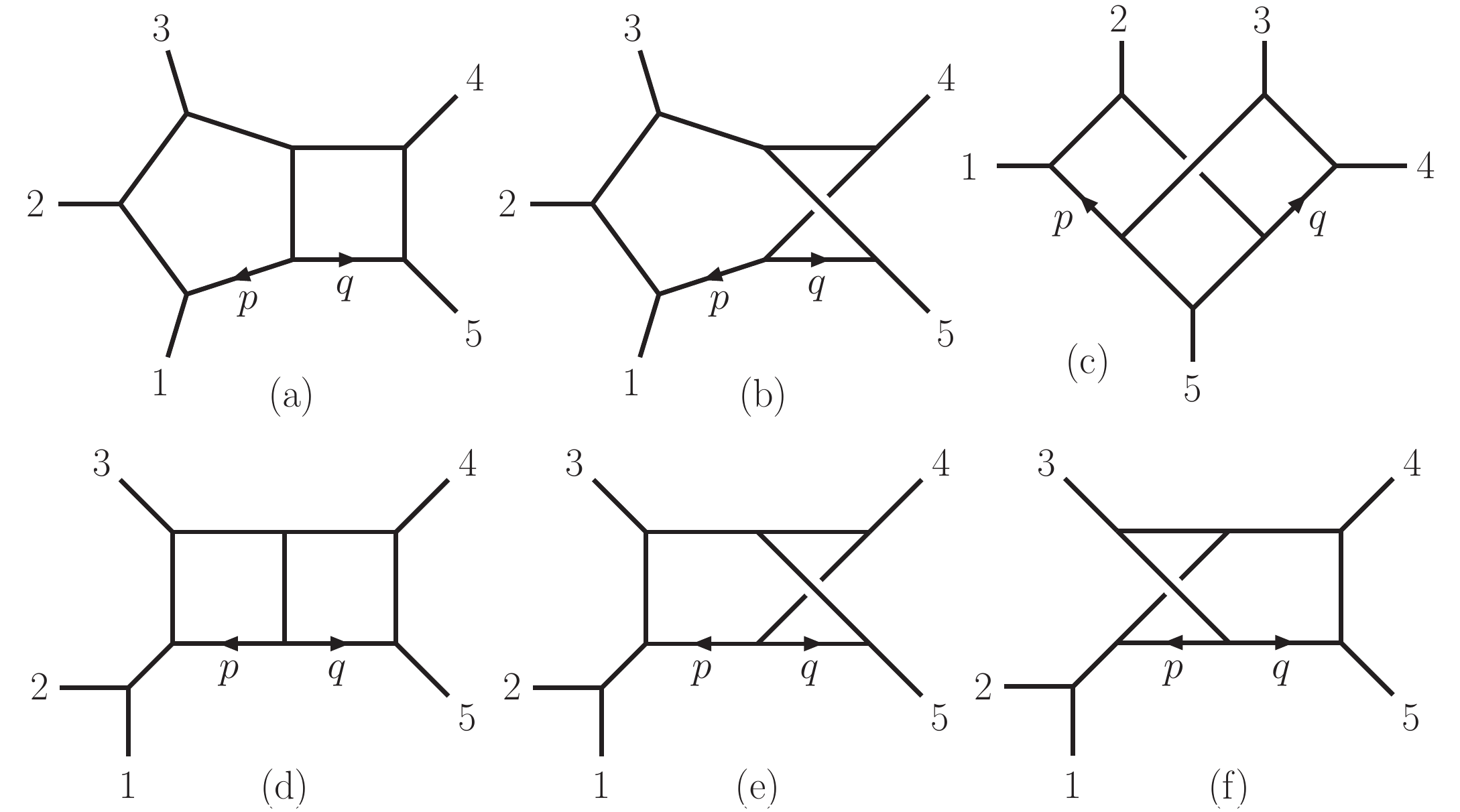}}
\caption{The six nonzero diagrams that contribute to two-loop five-point amplitude in $\NeqFour$ SYM and $\NeqEight$ supergravity.}
\label{TwoloopfivepointDiagramFigure}
\end{figure}

A further interesting and nontrivial example  is the five-point amplitude in $\cN=4$ SYM~\cite{FivePointN4BCJ}. Due to the high degree of supersymmetry, one can express the amplitude in terms of only six contributing diagrams shown in \fig{TwoloopfivepointDiagramFigure}. To be concise, we will only quote the duality-satisfying numerator of diagram (a); it is
\bea
n^{\rm (a)}_{12345}(p,q)&=&\frac{1}{4}\Big(\gamma_{12}(2s_{45} - s_{12} + \tau_{2p} - \tau_{1p}) + \gamma_{23}(s_{45} + 2s_{12} - \tau_{2p} + \tau_{3p}) \nn \\ 
&&\null + 2\gamma_{45}(\tau_{5p} - \tau_{4p}) + \gamma_{13}(s_{12} + s_{45} - \tau_{1p} + \tau_{3p})\Big)\,,
\eea
where the two independent loop momenta are called $p$ and $q$. The Lorentz invariants are $\tau_{ip}=2p_i\cdot p$, $\tau_{iq}=2p_i\cdot q$ and $s_{ij}=(p_i+ p_j)^2$.  The external state dependence for the MHV amplitude is captured by the $\gamma_{ij}$, where
\be
\gamma_{12} \equiv n_{[12]345} = \delta^{(8)}(Q)\frac{\spb{1}.{2}^2 \spb{3}.{4} \spb{4}.{5} \spb{5}.{3}}{4 i \epsilon(1,2,3,4)}
\ee
is the one-loop box numerator given in \eqn{boxnumer}, and the other $\gamma_{ij}$ are given by $S_5$ permutations of this expression. Note that the $\gamma_{ij}$ satisfy the relations
\be
\gamma_{ij}= - \gamma_{ji}\,, ~~~~ \sum_{i=1}^5 \gamma_{ij} =0\,,
\ee
from which it follows that there are only six independent variables of this type. The diagram numerators of the $\overline{\rm MHV}$ amplitude are obtained by replacing $\gamma_{ij}$ by their CPT conjugates.

\begin{homework}
Show that the kinematic numerators corresponding to diagrams (b)--(f) in \fig{TwoloopfivepointDiagramFigure} can be obtained from $n^{\rm (a)}_{12345}(p,q)$ using kinematic Jacobi relations. Which numerators happens to be independent of loop momenta? Which numerators are identical to each other (due to the Jacobi relation collapsing to a two-term identity)?  
\end{homework}

The two-loop $\cN=4$ SYM amplitude is given by the sum over the six diagrams (a)--(f) in \fig{TwoloopfivepointDiagramFigure}, together with the sum  over the ${\cal S}_5$ permutations over the external legs,
\begin{equation}
{\cal A}_5^\twoloop = i g^{7}
\sum_{\mathcal{S}_{5}}\sum_{j\in \{\rm a, \dots f\}}\int \frac{d^{D}p\, d^{D}q}{(2\pi)^{2D}}
\frac{1}{S_{j}}\frac{c^{(j)}_{12345} n^{(j)}_{12345}(p,q) }{\prod_{\alpha_{j}}p^{2}_{\alpha_{j}}}\,.
\label{2L5ptAmplitude}
\end{equation}
The corresponding  $\cN=8$ supergravity amplitude is obtained by the double-copy replacements $c^{(j)}_{12345} \rightarrow n^{(j)}_{12345}(p,q)$ and $g \rightarrow \kappa/2$.

\begin{homework}
By inspecting the diagrams in \fig{TwoloopfivepointDiagramFigure}, compute the symmetry factors $S_j$ that appear in \eqn{2L5ptAmplitude}.
\end{homework}

The two-loop five-point amplitudes of both $\cN=4$ SYM and $\cN=8$ supergravity, as presented above,  were integrated in  $D=4-2\epsilon$ dimensions in refs.~\cite{Abreu:2018aqd,Chicherin:2018yne,Chicherin:2019xeg,Abreu:2019rpt}.

\subsection{Three-loop example}

So far, we illustrated various one- and two-loop amplitudes that manifest the color-kinematics duality. To be more concrete, in this subsection we go through in detail how to construct duality-satisfying amplitudes when the system of numerators is quite large.   As a sophisticated example---though still quite manageable---consider the
three-loop four-point amplitude of the $\NeqFour$ SYM and $\NeqEight$
supergravity theories~\cite{BCJLoop}.

Apart from the duality and unitarity constraints, it is beneficial to
systematically impose various other constraints which become more important as
the complexity of the problem increases.  Although not required, such
auxiliary constraints, when appropriately chosen, can greatly facilitate 
the construction.  If a constraint is too strong and
leads to an inconsistency with unitarity, then one may relax or modify it as needed.  
This strategy is especially effective for
theories with high degrees of supersymmetry, because of their
restricted power counting.  For the three-loop four-point $\NeqFour$
SYM amplitudes, a natural set of constraints is as follows.
\begin{enumerate}
\item One-loop tadpole, bubble and triangle subdiagrams do not appear
in any diagram~\cite{BernNoTriangle, BjerrumNoTriangle, ArkaniHamed2008gz}.
\item A one-loop $n$-gon subdiagram carries no more than $n-4$ powers
  of loop momentum for that loop.
\item After extracting an overall factor of $s t A^\tree_4$, the
numerators are polynomials in $D$-dimensional Lorentz scalar products
of the independent loop and external momenta.
\item Numerators carry the same relabeling symmetries as the diagrams 
(cf. discussion in \sect{TreeSection}).
\end{enumerate}

In general, the choice of auxiliary constraints depends on the problem
at hand.  For example, the third constraint above is specific to the
four-point amplitude, and should be modified for higher-point
amplitudes because of their more complicated external-state structure.
As described in the previous subsection, a relatively simple
generalization has been found for the five-point
(super)amplitude~\cite{FivePointN4BCJ}, involving prefactors that are
proportional~\cite{virtuousTrees, Naculich:2014rta, Fu:2014pya} to
linear combinations of five-point color-ordered tree-amplitudes.  For
amplitudes in less supersymmetric theories, all but the fourth
condition must also be relaxed, because their power counting is such
that one-loop triangle and bubble subdiagrams do appear; this is
related to \eg the running of their couplings.  The above constraints
also work well for the four-loop four-point amplitudes of $\cN =4$
SYM~\cite{SimplifyingBCJ}, but fail at five loops. A procedure which
works for this case is described in \sect{ClassicalDoubleCopySection}.

Because the duality imposes stringent relations between diagrams'
numerators, a remarkably small subset of generalized unitarity cuts is
then sufficient to completely determine the integrand. Of course, to
confirm that it is correct, it is necessary to verify that it
reproduces correctly a spanning set of unitarity cuts that fully
determine the amplitude.
Quite generally, one expects that a problem with a generalized cut can be addressed by relaxing some of the auxiliary constraints.

\begin{figure}[t]
\centerline{\includegraphics[scale=0.5]{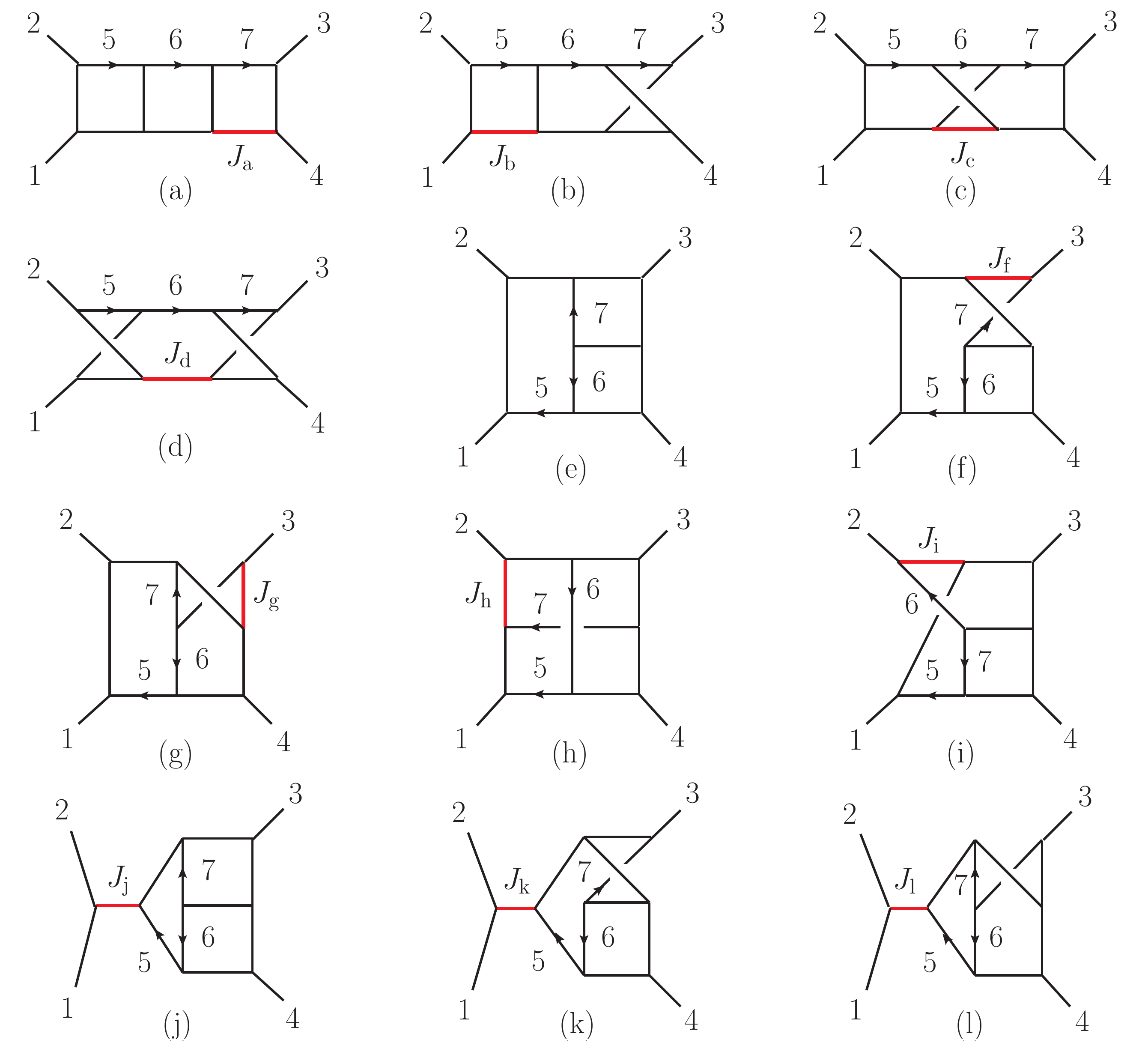}}
\caption{The diagrams for constructing the 
  $\NeqFour$ SYM and $\NeqEight$ supergravity three-loop four-point
  amplitudes.  The shaded (red) lines indicate the application of the
  duality relation.  The
  external momenta are outgoing and the arrows indicate the directions of
  the labeled loop momenta. Diagram (e) is the master diagram. }
\label{ThreeLoopDiagramsFigure}
\end{figure}

\begin{figure}
\centering
        \begin{subfigure}[t]{0.2 \textwidth}
        \begin{equation*}
      \hbox{$n$} 
 \left( \hskip -.1 cm \parbox{3.2cm} {\includegraphics[width=1.\textwidth]{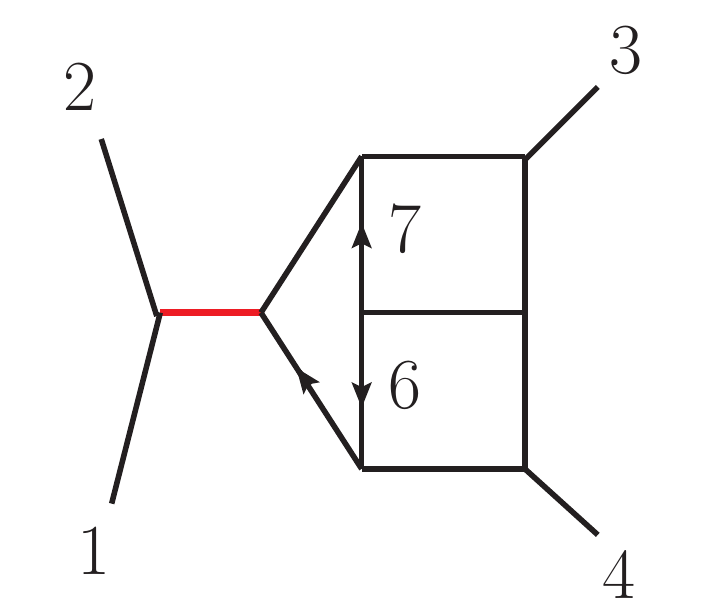}} \right)
        \end{equation*}
\vskip - .7 cm  
               \caption*{\hskip 1.1  cm \hphantom{(j)}\vphantom{(j)}}  
        \end{subfigure}
  \begin{subfigure}[t]{0.105 \textwidth}
        \begin{eqnarray*}
\raisebox{-1.1 cm}{\hskip 1.2 cm \hbox{\Large $=$ }}
        \end{eqnarray*}
 \caption*{ }
 \end{subfigure}
       \begin{subfigure}[t]{0.2 \textwidth}
        \begin{eqnarray*}
  \hbox{ $n$} 
     \left(\hskip -.1 cm  \parbox{3.2 cm} {\includegraphics[width=1.\textwidth]{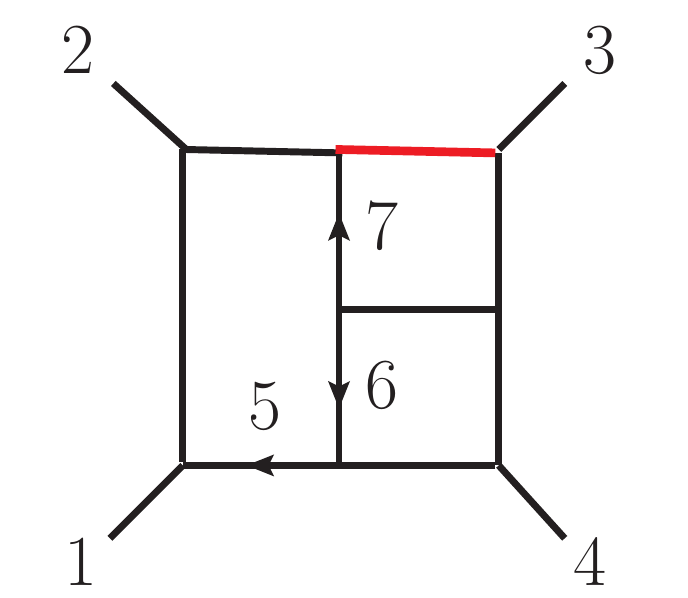}} \right)
        \end{eqnarray*}
 \vskip - .7 cm  
        \end{subfigure}
  \begin{subfigure}[t]{0.105 \textwidth}
        \begin{eqnarray*}
\raisebox{-1.1 cm}{\hskip 1.2 cm \hbox{\Large $-$ }}
        \end{eqnarray*}
 \caption*{ }
 \end{subfigure}
       \begin{subfigure}[t]{0.2 \textwidth}
        \begin{eqnarray*}
  \hbox{ $n$} 
     \left(\hskip -.1 cm  \parbox{3.2 cm} {\includegraphics[width=1.\textwidth]{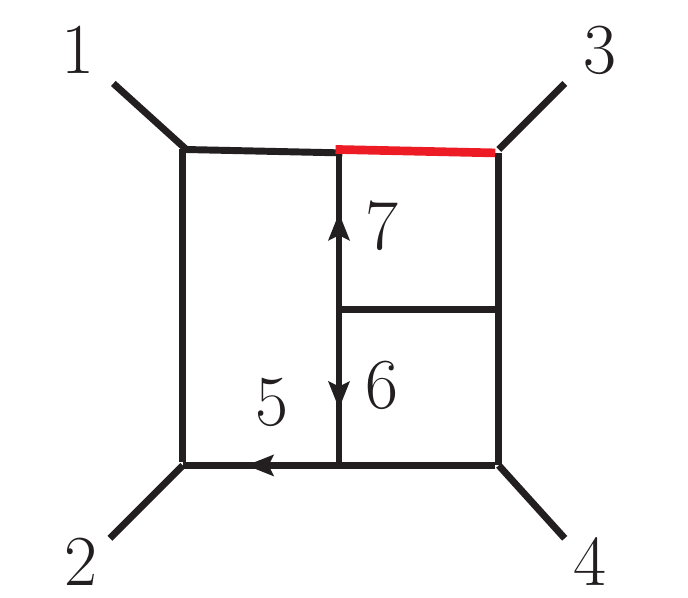}} \right)
        \end{eqnarray*}
 \vskip - .7 cm  
        \end{subfigure}
        \begin{subfigure}[t]{0.2 \textwidth}
        \begin{equation*}
      \hbox{ $n$} 
 \left( \hskip -.1 cm \parbox{3.2cm} {\includegraphics[width=1.\textwidth]{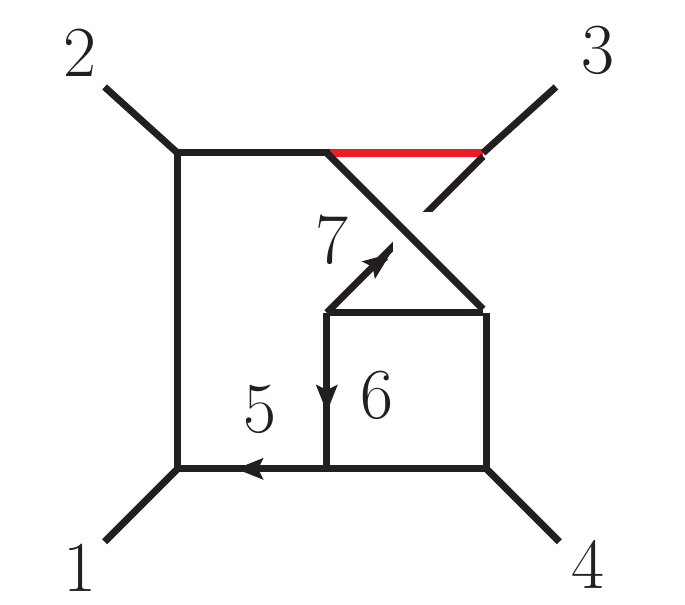}} \right)
        \end{equation*}
\vskip - .7 cm  
        \end{subfigure}
  \begin{subfigure}[t]{0.105 \textwidth}
        \begin{eqnarray*}
\raisebox{-1.1 cm}{\hskip 1.2 cm \hbox{\Large $=$ }}
        \end{eqnarray*}
 \caption*{ }
 \end{subfigure}
       \begin{subfigure}[t]{0.2 \textwidth}
        \begin{eqnarray*}
  \hbox{ $n$} 
     \left(\hskip -.1 cm  \parbox{3.2 cm} {\includegraphics[width=1.\textwidth]{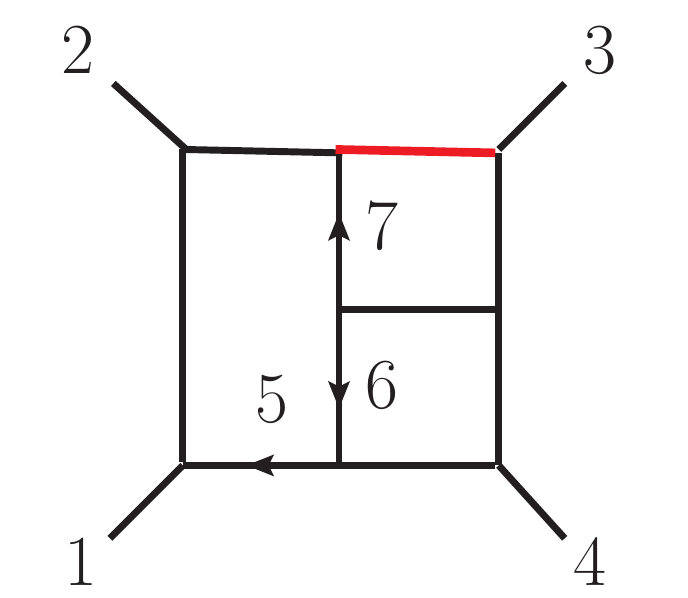}} \right)
        \end{eqnarray*}
 \vskip - .7 cm  
        \end{subfigure}
\caption{Examples of a BCJ kinematic numerator relation at three loops for $\NeqFour$ SYM theory.
In the two term relations one of the three numerators in a Jacobi
  triplet of diagrams vanishes.  }
\label{ThreeLoopBCJFigure}
\end{figure}

Let us return now to the three-loop four-point amplitudes of $\NeqFour$ SYM theory and illustrate these ideas. 
A straightforward enumeration shows that there are 17 distinct cubic
diagrams with three loops and four external legs, which do not have
one-loop triangle, bubble or tadpole subdiagrams. It turns out that the
twelve diagrams shown in \fig{ThreeLoopDiagramsFigure} are sufficient
for finding a solution to the duality and unitarity cut constraints,
as shown in Ref.~\cite{BCJLoop}. Had we kept all 17 diagrams, the
construction would be slightly more involved, with the result that the
numerators of the additional diagrams vanish identically.

The four-point amplitudes of $\NeqFour$ SYM theory are special.
Applying the third condition above we write the numerator as
\begin{equation}
n^{(x)} = -i st A_4^\tree(1,2,3,4) \, N^{(x)}\, ,
\end{equation}
where $(x)$ refers to the label for each diagram in
\fig{ThreeLoopDiagramsFigure} and
$N^{(x)}$ are scalar functions which depend on three 
independent external momenta, labeled by $p_1, p_2, p_3$, and on (at most) three independent loop
momenta, labeled by $\ell_5, \ell_6, \ell_7$, 
\begin{eqnarray}
N^{(x)} \equiv N^{(x)}(p_1,p_2,p_3,\ell_5,\ell_6,\ell_7) \, .
\end{eqnarray}
The coefficient $st A_4^\tree(1,2,3,4)$  is fully crossing symmetric, as noted in Eq.~\eqref{FourFourNumeratorCrossing}.

Next, consider the duality relations.  We need to discuss first those that allow us to express 
the complete set of numerators $N^{(x)}$ in terms of a small subset---the master numerators.  
Some of them are shown in  \fig{ThreeLoopBCJFigure}.
The remaining relations are subsequently verified once the former are solved together with 
the constraints imposed by the unitarity cuts.
For the three-loop four-point $\NeqFour$ SYM amplitude, 
a simple restricted set of duality relations is~\cite{JJHenrikReview,SimplifyingBCJ}:
\begin{align}
 N^{(\rm a)}&= N^{(\rm b)}(p_1,p_2,p_3,\ell_5,\ell_6,\ell_7)\,, \nn \\
 N^{(\rm b)}&= N^{(\rm d)}(p_1,p_2,p_3,\ell_5,\ell_6,\ell_7)\,, \nn \\
 N^{(\rm c)}&= N^{(\rm a)}(p_1,p_2,p_3,\ell_5,\ell_6,\ell_7)\,, \nn \\
 N^{(\rm d)}&= N^{(\rm h)}(p_3,p_1,p_2, \ell_7, \ell_6, p_{1,3}-\ell_5+\ell_6-\ell_7)
   + N^{(\rm h)}(p_3, p_2, p_1,\ell_7,\ell_6,  p_{2,3}+\ell_5-\ell_7)\,,\nn \\
 N^{(\rm f)}&= N^{(\rm e)}(p_1,p_2,p_3,\ell_5,\ell_6,\ell_7)\,, \nn \\
 N^{(\rm g)}&= N^{(\rm e)}(p_1,p_2,p_3,\ell_5,\ell_6,\ell_7)\,, \nn \\
 N^{(\rm h)}&= - N^{(\rm g)}(p_1, p_2, p_3, \ell_5, \ell_6, p_{1,2} - \ell_5 - \ell_7)
- N^{(\rm i)}(p_4, p_3, p_2,  \ell_6-\ell_5,  \ell_5 - \ell_6 + \ell_7-p_{1 ,2}, \ell_6)\,,\nn \\
 N^{(\rm i)}\,&= N^{(\rm e)}(p_1,p_2,p_3,\ell_5,\ell_7,\ell_6)
   - N^{(\rm e)}(p_3, p_2, p_1, -p_4 - \ell_5 - \ell_6, - \ell_6 - \ell_7, \ell_6)\,, \nn\\
 N^{(\rm j)}\,&= N^{(\rm e)}(p_1,p_2,p_3,\ell_5,\ell_6,\ell_7)
             - N^{(\rm e)}(p_2,p_1,p_3,\ell_5,\ell_6,\ell_7)\,,\nn \\
 N^{(\rm k)}&= N^{(\rm f)}(p_1,p_2,p_3,\ell_5,\ell_6,\ell_7)
             - N^{(\rm f)}(p_2,p_1,p_3,\ell_5,\ell_6,\ell_7)\,,\nn \\
 N^{(\rm l)}&= N^{(\rm g)}(p_1,p_2,p_3,\ell_5,\ell_6,\ell_7)
             - N^{(\rm g)}(p_2,p_1,p_3,\ell_5,\ell_6,\ell_7)\,,
 \label{BCJjacobi}
\end{align}
where $p_{i,j}\equiv p_i + p_j$.  To simplify the notation, we have suppressed
the canonical arguments $(p_1,p_2,p_3,\ell_5,\ell_6,\ell_7)$ of the
numerators on the left-hand side of the equations (\ref{BCJjacobi}).
Each relation specifying an $N^{(x)}$ is generated by considering the
kinematic Jacobi relations dual to the color Jacobi relations corresponding to the shaded (red) line and labeled
$J_x$ in \fig{ThreeLoopDiagramsFigure}. 
In general, duality relations relate triplets of numerators; if however one of the diagrams is not present  in \fig{ThreeLoopDiagramsFigure}, \eg because it has a 
one-loop triangle subdiagram, then we obtain a two-term relation.
Five of the equations above are of this type and they result in pairs of numerators being equal.

The system~\eqref{BCJjacobi} can be used to express any kinematic numerator factor as a combination of the 
numerator $N^{\rm (e)}$ with various different arguments. Thus, diagram (e) can be taken as the sole master diagram. This
is a convenient choice, but not the only possible one; for example, either diagram (f) or (g) can also be used as a single 
master diagram. None of the remaining nine diagrams, however, can act alone as a master diagram.

The numerator factor of diagram (e) is constructed such that the unitarity cuts are satisfied simultaneously with the duality constraints.
An expression that satisfies the maximal cuts is given by the so-called ``rung-rule'' numerator~\cite{BRY},
\begin{equation}
N_{\rm rr}^{\rm (e)} = s (\ell_5 + p_4)^2\, ,
\label{RungRuleNumerator}
\end{equation}
which follows from the general features of iterated two-particle cuts. 

We wish to find a modification $N_{\rm rr}^{\rm (e)} \rightarrow N^{\rm (e)}$ such that 
all the other numerators determined from it via \eqn{BCJjacobi} are consistent with the unitarity cuts.
We start by requiring that the maximal-cut of diagram (e) is correct (see \app{MaximalCutsSubsection}
for a description of the maximal cuts), and that the auxiliary constraints above are satisfied. That is, the departure from $N_{\rm rr}^{\rm (e)}$
vanishes on the maximal cut,  the  numerator $N^{\rm (e)}$ has mass dimension four and possesses the symmetry of the diagram; no
loop momentum for any box subdiagram in (e) appears in it (ruling out
$\ell_6$ and $\ell_7$), and $N^{\rm (e)}$ is at most quadratic in the
pentagon loop momenta $\ell_5$.   
The last condition is a little weaker than the second auxiliary condition listed earlier, which demands linearity in $\ell_5$; we relax it 
slightly to make it easier to find deformations that vanish on maximal cuts, and impose later that the $\ell_5^2$ terms cancel out. 
The symmetry condition implies that $N^{\rm (e)}$ is invariant under
\begin{equation}
\{ p_1 \leftrightarrow p_2, \ p_3 \leftrightarrow p_4, \
     \ell_5 \rightarrow p_1 + p_2 - \ell_5 \}\,.
\end{equation}
The most general polynomial consistent with these constraints is
\begin{equation}
N^{\rm (e)} =  s (\ell_5 + p_4)^2 +
         (\alpha s + \beta t) \ell_5^2 +
         (\gamma s + \delta t)(\ell_5-p_1)^2 +
         (\alpha s + \beta t) (\ell_5-p_1-p_2)^2 \,,
\label{ThreeLoopAnsatz}
\end{equation}
where the four parameters $\alpha, \beta, \gamma, \delta$ are to be determined by further constraints.  
All added terms are proportional to inverse propagators and therefore vanish on the 
maximal cut.  
Thus, given that \eqn{RungRuleNumerator} is consistent with the maximal cuts, so is \eqn{ThreeLoopAnsatz}.

The second auxiliary constraint above demands that the numerator of a pentagon subdiagram be at most linear in the 
corresponding loop momentum, $\ell_5$, not quadratic as assumed above. Therefore we impose that the coefficient of $\ell_5^2$ 
in \eqn{ThreeLoopAnsatz} vanishes. This yields the relation $\gamma = -1- 2 \alpha $ and $\delta = - 2 \beta$, which 
simplifies \eqn{ThreeLoopAnsatz} to
\begin{equation}
N^{\rm (e)} = s (\tau_{45}  +\tau_{15})
+  ( \alpha s +   \beta t) (s+\tau_{15} - \tau_{25})  \,,
\label{CleanAnsatz}
\end{equation}
where we use the notation,
\begin{equation}
\tau_{ij} \equiv 2 p_i \cdot \ell_j\,, \hskip .3 cm
(i \le 4, j \ge 5) \,.
\label{tauDef}
\end{equation}
We are therefore left with two undetermined parameters, $\alpha$ and $\beta$.

We determine the remaining parameters by imposing that the numerators of other diagrams
determined through \eqn{BCJjacobi}  are consistent with the auxiliary constraints and unitarity cuts.
A convenient starting point is the numerator of diagram (j), $N^{\rm (j)}$, which is determined in terms of $N^{\rm (e)} $
by the 9th duality constraint in \eqn{BCJjacobi}.
Inserting \eqn{CleanAnsatz} into this relation leads to
\begin{equation}
 N^{(\rm j)}
 = s (1+2\alpha -\beta) (\tau_{15} - \tau_{25})
               + \beta s (t-u) \,.
\label{Njtemp}
\end{equation}
Because the smallest loop in diagram (j) carrying $\ell_5$ is a box
subdiagram, our auxiliary constraints require that this momentum be
absent from $N^{\rm (e)}$.  Setting the first term in \eqn{Njtemp}
to zero implies that $\beta = 1+2 \alpha$, which in turn leads to 
\begin{align}
N^{\rm (e)} &= s (\tau_{45}  +\tau_{15}) +
( \alpha (t-u)  + t) (s+\tau_{15} - \tau_{25})  \,,
\label{Simplified_e}\\
N^{\rm (j)} &= (1+2 \alpha) ( t-u ) s\,,
\label{Simplified_j}
\end{align}
leaving undetermined a single parameter $\alpha$.

To obtain the value of the final parameter we use the numerator of diagram (a) expressed in terms of $N^{\rm (e)} $ 
by \eqn{BCJjacobi}.
Because every loop in diagram (a) is part of a box, the auxiliary constraint that a  one-loop box 
subdiagram cannot carry loop momentum then implies that $N^{\rm (a)}$ cannot contain loop momentum.
By solving the duality relations (\ref{BCJjacobi}),  the numerator $N^{\rm (a)}$ is given by
\begin{align}
N^{(\rm a)} =\null & 
  N^{\rm (e)}(p_1, p_2, p_4, -p_3 + \ell_5 - \ell_6 + \ell_7, \ell_5 - \ell_6, -\ell_5) \nn \\
& \null
+ N^{\rm (e)}(p_2, p_1, p_4, -p_3 - \ell_5 + \ell_7, -\ell_5, \ell_5 - \ell_6) \nn \\
& \null
- N^{\rm (e)}(p_4, p_1, p_2, \ell_6 - \ell_7, \ell_6, \ell_5 - \ell_6)
- N^{\rm (e)}(p_4, p_2, p_1, \ell_6 - \ell_7, \ell_6, -\ell_5)\nn \\
& \null
- N^{\rm (e)}(p_3, p_1, p_2, \ell_7, \ell_6, \ell_5 - \ell_6)
-  N^{\rm (e)}(p_3, p_2, p_1, \ell_7, \ell_6, -\ell_5) \,.
\label{DualityEquationsForA}
\end{align}
Plugging in the value of the numerator factor $N^{\rm (e)}$ in
\eqn{Simplified_e}, and simplifying we obtain
\begin{equation}
N^{\rm (a)} = s^2 +
 (1 + 3 \alpha) \Bigl( (\tau_{16} - \tau_{46}) s
 - 2 (\tau_{17} + \tau_{37}) s
 + (\tau_{16} - 2 \tau_{17} - \tau_{26} + 2\tau_{27}) t + 4 u t \Bigr)\,.
\end{equation}
Demanding that this expression is independent of loop
momenta, fixes the final parameter to be $\alpha = -1/3$
and completely determines the numerator of diagram (e) to be
\begin{equation}
N^{\rm (e)}= s (\tau_{45}  +\tau_{15}) +
 \frac{1}{3}(t-s) (s+\tau_{15} - \tau_{25})\,.
\label{NumeratorE}
\end{equation}

With a proposed expression for $N^{\rm (e)}$ in hand, \eqn{BCJjacobi} then determines all other numerators and thus the complete amplitude.  
The resulting numerators are collected in \tab{BCJNumeratorTable}. 
To confirm that this is indeed the correct amplitude, it is necessary to verify a complete set of unitarity cuts.  
The three-loop four-gluon amplitude in ${\cal N}=4$ SYM theory is determined only by its maximal and next-to-maximal 
cuts, so it is relatively straightforward to check them all.  As a highly-nontrivial test, one can also check the next-to-next-to-maximal cuts.
The resulting cuts  match those of previous expressions of the amplitude~\cite{GravityThree,CompactThree} on all $D$-dimensional unitarity cuts.  
Thus, the amplitude is complete.
We stress again that it is highly-nontrivial that there exists a solution to {\it all} duality relations which is consistent with all unitarity cuts and exhibits 
all the diagram symmetries.

Squaring the numerators $n^{(x)} = stA_4^{\rm{tree}}(1,2,3,4)n^{(x)} $, using \eqn{DoubleCopyLoop}, yields the numerators for the three-loop four-point $\NeqEight$ 
supergravity superamplitude.  This form has been confirmed against previous expressions~\cite{GravityThree,CompactThree} on a 
spanning set of $D$-dimensional unitarity cuts~\cite{BCJLoop}.
Using  as the second copy  the three-loop four-point numerator factors of ${\cal N}<4$ SYM theories yields the three-loop four-graviton amplitudes in 
$(4+{\cal N})$-extended supergravity theories. The case ${\cal N}=0$  was discussed at length in Refs.~\cite{Bern:2012cd, Bern:2013qca, Bern:2014lha}, where it
was used to explore the UV properties of half-maximal supergravities and demonstrate the absence of UV divergences at this loop order in four dimensions.

\begin{table}
\begin{center}
\begin{tabular}{c|c}
diagram &  $\raisebox{-.3 cm}{ \vphantom{|}} \NeqFour$ SYM ($\sqrt{\vphantom{\big|}\NeqEight~{\rm supergravity}}$) numerator  \\
\hline
\hline
(a)--(d) &  $\raisebox{.3 cm}{ \vphantom{|}} \raisebox{-.15 cm}{ \vphantom{|}} 
 s^2$   \\
\hline
(e)--(g) & $\raisebox{.3 cm}{ \vphantom{|}} \raisebox{-.3 cm}{ \vphantom{|}} 
\big(\,s \left(-\tau _ {3 5}+\tau _ {4 5}+t \right)- t \left(\tau _ {2 5}+\tau _ {4 5}\right)+                                                               
 u \left(\tau _ {2 5}+\tau _ {3 5}\right)-s^2 \, \big)/3$   \\
\hline
(h)& $ \raisebox{.3 cm}{ \vphantom{|}} 
\big(\, s \left(2 \tau _ {1 5}-\tau _ {1 6}+2 \tau _ {2 6}-\tau _ {2 7}+2 \tau _ {3 5}+\tau _ {3 6}+\tau _ {3 7}-u \right)$\\
&$\null + t \left(\tau _ {1 6}+\tau _ {2 6}-\tau _ {3 7}+2\tau _ {3 6}-2 \tau _ {1 5}-2\tau _ {2 7}-2\tau _ {3 5}-3 \tau _ {1 7}\right)+s^2  
\,\big)/3\raisebox{-.3 cm}{ \vphantom{|}} $\\
\hline
(i)& $\raisebox{.3 cm}{ \vphantom{|}}  
\big(\, s \left(-\tau _ {2 5}-\tau _ {2 6}-\tau _ {3 5}+\tau _ {3 6}+\tau _ {4 5}+2 t \right)$\\
&$ \null + t \left(\tau _ {2 6}+\tau _ {3 5}+ 2\tau _ {3 6}+2\tau _ {4 5}+3 \tau _ {4 6}\right)+ u\,\tau _ {2 5}+s^2 \,\big)/3
\raisebox{-.3 cm}{ \vphantom{|}} $\\
\hline
(j)-(l) & $\raisebox{.3 cm}{ \vphantom{|}} \raisebox{-.25 cm}{ \vphantom{|}}  
s (t-u)/3 $
 \\
\hline
\end{tabular}
\end{center}
\caption{The numerator factors for diagrams in
  \fig{ThreeLoopDiagramsFigure} \cite{BCJLoop}. The first column labels the diagram,
  the second column the relative numerator factor for $\NeqFour$
  SYM theory.  The square of this is the relative
  numerator factor for $\NeqEight$ supergravity. The momenta are
  labeled as in \fig{ThreeLoopDiagramsFigure} and the $\tau_{ij}$ are defined
  \eqn{tauDef}.
\label{BCJNumeratorTable} }
\end{table}

\begin{homework}
Work through the entries in \tab{BCJNumeratorTable} to explicitly confirm that they do indeed satisfy BCJ duality.
\end{homework}

The strategy followed above generalizes straightforwardly to the four-loop four-point~\cite{SimplifyingBCJ} and two-loop five-point amplitudes of
$\NeqFour$ SYM and supergravity.  It has also been tested in a variety of other cases, including the one- and two-loop amplitudes in various 
theories with fewer supersymmetries~\cite{OneLoopSusy}, and nonsupersymmetric gauge and gravity theories~\cite{OneTwoLoopPureYMBCJ,OneTwoLoopMatterYMBCJ}  
as well as to the construction of form factors in $\NeqFour$ SYM through five loops~\cite{FourLoopFormFactor, FiveLoopFormFactor}.  
While the application of the double-copy construction to gauge-theory form factors yields quantities consistent with the linearized diffeomorphism 
invariance of a gravity theory, their precise physical interpretation is currently an open question. 

\subsection{Other examples}

The examples described above are but a sample of the many loop-level amplitudes that have representations that manifest the duality
between color and kinematics. Among them are various examples of
supersymmetric~\cite{FivePointN4BCJ, SimplifyingBCJ, Du:2012mt,
Yuan:2012rg, Bjerrum-Bohr:2013iza, MafraSchlottererTwoLoop,
HeMonteiroSchlottererBCJNumer, HeSchlottererZhangOneLoopBCJ,
JohanssonTwoLoopSusyQCD, KalinN2TwoLoop} and
nonsupersymmetric~\cite{OneTwoLoopPureYMBCJ, BCJDifficulty,
Chiodaroli2017ngp, Nandan2018ody,Boels:2013bi} gauge-theory
amplitudes, form factors~\cite{FourLoopFormFactor, FiveLoopFormFactor,
BoelsFourLoop, Boels:2017ftb, Faller:2018vdz}, string theory amplitudes, and their
field theory limits~\cite{PierreHigherLoopMonodromy,
PierreOneLoopMonodromy,Boels:2011mn, Ochirov:2013xba,
MafraSchlotterOneLoopString}.
Additionally, a systematic method to determine BCJ numerators for one-loop amplitudes which makes 
use of the global constraints on the loop-momentum dependence of the numerators imposed by the kinematic Jacobi identities was introduced in Ref. \cite{Bjerrum-Bohr:2013iza}.

It has moreover been shown that the leading and subleading \cite{WhiteIRBCJ,Vera:2014tda} factorization theorems of gauge and gravity theories are 
consistent with the double-copy procedure to all orders in perturbation theory, thus providing some all-loop-levels evidence for this conjecture. 
In another interesting example, the duality has been applied to QCD scattering amplitudes, in order to find hidden relations between 
coefficients of loop integrals~\cite{Chester:2016ojq, Primo:2016omk}.

The multitude of nontrivial examples suggests that the duality between color and kinematics does extend to loop amplitudes, even though no proof exists as yet.  
Finding a proof  would likely provide a guide towards more
systematic constructions of representations of amplitudes that manifest the duality.

Even when they are expected to exist, the construction of duality-satisfying amplitudes representations
is not always straightforward. An alternative, discussed and illustrated on the two-loop four-point all-plus pure-YM amplitude in Ref.~\cite{Bern:2015ooa}, is to 
relax the demand that the duality be manifest off shell
and impose instead that it be manifest only on a spanning set of generalized unitarity cut.  The double-copy construction then yields an expression 
which coincides with the corresponding supergravity amplitude on a spanning set of cuts; the two must therefore be the same.
   
It has proven difficult to find representations of the five-loop
four-point $\NeqFour$ SYM amplitude for which the duality between
color and kinematics is manifest. In particular, the expected
power-counting constraints suggested by supersymmetry appear to not be
compatible with duality and $D$-dimensional unitarity cuts.
To find the corresponding $\NeqEight$ supergravity amplitude the
generalized double-copy construction provides an efficient approach,
as it uses gauge-theory amplitudes' representations that should
exhibit the duality but do not manifest it; we review it in the next
section.  The success of the generalized double copy, using the
five-loop amplitudes' representations constructed in
Refs.~\cite{GeneralizedDoubleCopy, GeneralizedDoubleCopyFiveLoops,
UVFiveLoops}, strongly suggests that it should be possible to manifest
the duality for the five-loop four-point $\NeqFour$ SYM
amplitude. Presumably, this will require integrands that relax some
the simplifying assumptions, such as locality or manifest relabeling
symmetry of the diagrams.  Using string theory to define global diagram
labels in the field-theory limit, there has been some very interesting
progress on finding integrands that manifest \ck
duality~\cite{Tourkine:2019ukp}.



\section{Generalized double copy}
\label{GeneralizedDoubleCopySection}

\def\G{{\rm GR}}
\def\tn {\tilde n}
\def\BCJ{{\rm BCJ}}
\def\E{{\cal E}}
\def\x#1{{\bullet}}
\def\tJ{{\tilde J}}

Whenever gauge-theory amplitudes are available in a form that
manifests the duality between color and kinematics, the BCJ
double-copy construction provides the most efficient means for
obtaining the corresponding gravity integrands.  However, in some
cases, such as the five-loop four-point amplitude of $\NeqEight$
supergravity, it has proven difficult to find such representations.
In other cases, such as the all-plus two-loop five-gluon amplitude in
pure-YM theory, the BCJ form of the amplitude has a superficial
power-count much worse than that of Feynman
diagrams~\cite{BCJDifficulty} and thus an analysis of UV properties of
its double copy is cumbersome at best.
It can therefore be advantageous to have a double-copy method for converting generic representations 
of gauge-theory amplitude to gravity ones, without first constructing BCJ
representations for them.
Such a procedure has been developed in
Ref.~\cite{GeneralizedDoubleCopy} and applied in
Refs.~\cite{GeneralizedDoubleCopyFiveLoops, UVFiveLoops} to construct
the five-loop four-point integrand of $\NeqEight$ supergravity and to
extract its UV properties after integration.\footnote{Another
possible method, proposed in Ref.~\cite{Bern:2015ooa} and illustrated on
the two-loop four-point pure-YM amplitude in $D$ dimensions, is to
demand that the duality between color and kinematics holds only on
unitarity cuts. This provides a straightforward construction of the
generalized unitarity cuts of the double-copy theory, which need to be subsequently
assembled into the complete gravity amplitude.}

If we start with a generic representation of a gauge-theory amplitude
where BCJ duality is not manifest and apply the double-copy
substitution rule (\ref{c_n_replace}), in general, we do not obtain a
correct gravity amplitude.  Nevertheless, this ``naive double copy''
can be systematically corrected to give the desired amplitude.  As we
summarize below, the correction terms have a regular pattern reminiscent
of the KLT tree-level amplitudes relations~\cite{KLT}, allowing us to
obtain the most complicated corrections directly from gauge theory.

\subsection{Generalities \label{general_g2c}}

To start the generalized double-copy construction we first need to
reorganize slightly the two (possibly distinct) gauge-theory
amplitudes that comprise the two sides of 
the double copy.  Starting with any local
representations of the amplitudes, which may include four- or
higher-point contact terms, we reorganize them into a format that has
only three-point vertices and the maximum number of propagators.  If a given
term has fewer propagators we multiply and divide by the propagators
needed to form diagrams with only cubic vertices that correspond to
the color factor of the given term.  Once the gauge-theory amplitudes
are written in this format the next step is to apply the double-copy
substitution \eqref{c_n_replace} to these amplitudes, despite neither
gauge theory manifesting the BCJ duality between color and kinematics.
As already mentioned, this so-constructed naive
double-copy expression is, in general, not a correct (super)gravity
amplitude.  Nonetheless, it is a good starting point for obtaining the
full gravity amplitude as, by construction, it reproduces the maximal and
next-to-maximal cuts of the desired (super)gravity amplitude.
(See \app{GeneralizedUnitaritySection} for a description of the method
of maximal cuts.)  
In maximal cuts, where all propagators are cut, the amplitude is reduced to a sum of products 
of gauge-theory three-point tree amplitudes. Because on-shell gravity three-vertices are products of 
gauge-theory ones, maximal cuts trivially satisfy the double-copy property for any representation
of the single copy amplitudes.
The next to maximal cuts, where one of the propagators are left uncut, also automatically give the correct
gravity expressions because, if present, the duality between color and kinematics is automatic for 
on-shell four-point tree amplitudes~\cite{BCJ}.

Beyond the next-to-maximal cuts, the naive double copy will generally
{\em not} give correct unitarity cuts, and nontrivial corrections are
necessary.  These required corrections can be organized into contact
terms via the method of maximal cuts described
in \app{GeneralizedUnitaritySection}.  However, for complicated
problems, such as $\NeqEight$
supergravity~\cite{GeneralizedDoubleCopyFiveLoops} at five-loops it
becomes cumbersome to use the method of maximal cuts to obtain the
missing terms.

\begin{figure}
\begin{center}
\includegraphics[scale=.41]{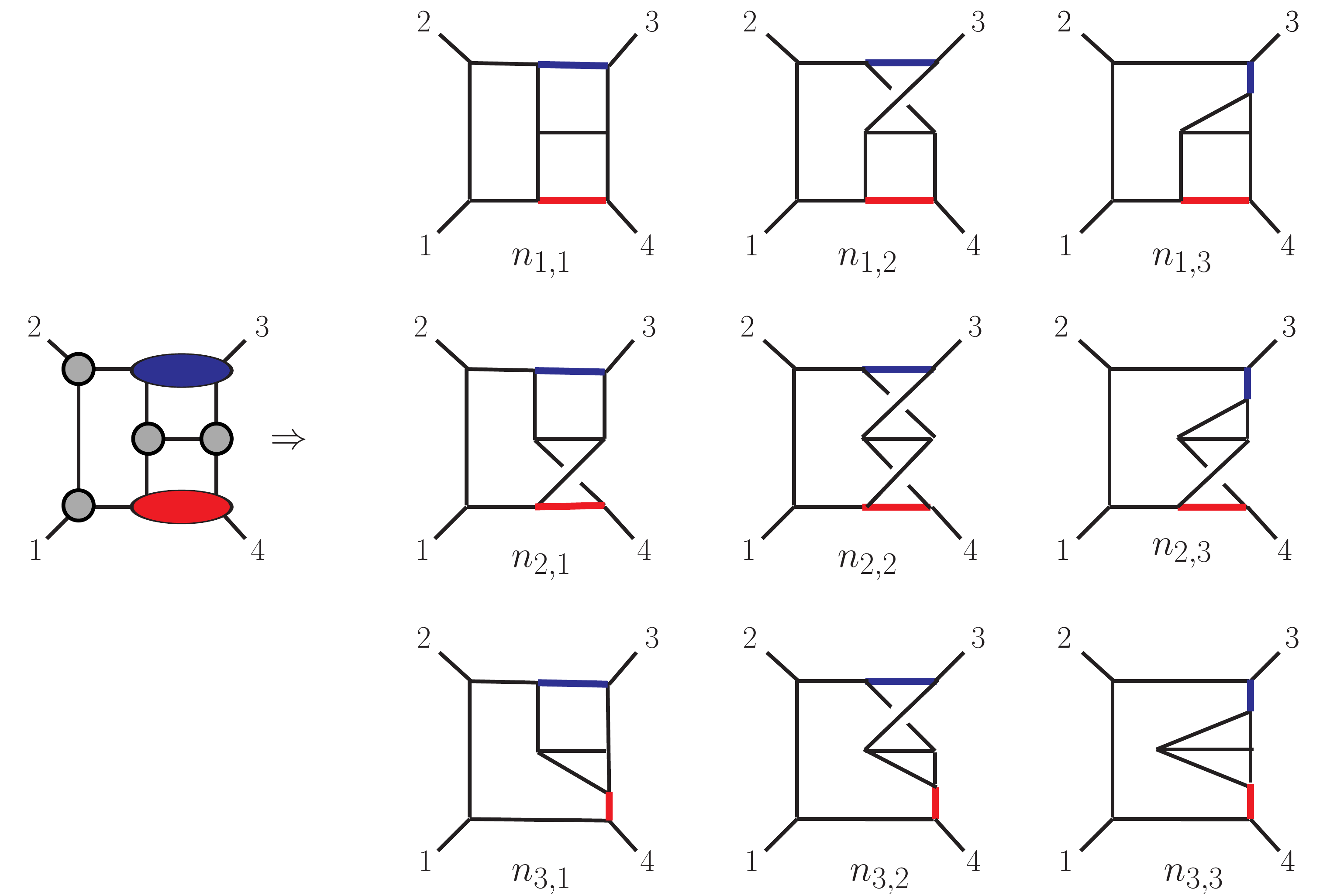}
\end{center}
\caption{\small An example illustrating the notation in \eqn{DoubleCopyCut}.
  Expanding each of the two four-point blob gives a total of nine
  diagrams.  The $n_{i,j}$ correspond to labels used in the
  generalized unitarity cut. The shaded thick (blue and red) lines are
  the propagators around which BCJ discrepancy functions are defined.
  }
\label{FourxFourBlobFigure}
\end{figure}

Instead, it turns out that it is possible to construct general
formulae that relate the necessary cut-correction terms to the
violations of the kinematic Jacobi relations~\eqref{duality} in the
gauge-theory amplitudes. The derivation of such formulae relies only
on the existence of duality-satisfying representations for all
tree-level amplitudes.

Indeed, the existence of BCJ representations at tree level implies that such 
representations should also exist for all cuts of gauge-theory amplitudes
that decompose the loop integrand into products of tree amplitudes.
This further implies that the corresponding generalized unitarity cuts 
of the gravity amplitude can be expressed in double-copy form,
\begin{equation}
{\cal C}_\G = \sum_{i_1,\dots,i_q}
\frac{n_{i_1, i_2, ...i_q}^\text{BCJ} \tn_{i_1, i_2, ...i_q}^\text{BCJ}}
{D_{i_1}^{(1)}\dots D_{i_q}^{(q)} }\,,
\label{DoubleCopyCut}
\end{equation}
where the $n^\BCJ$ and $\tn^\BCJ$ are the BCJ numerators associated
with each of the two single-copy parent theories. In this expression the cut conditions are
understood as being imposed on the numerators.  Each sum runs over the
diagrams of each tree amplitude composing the generalized cut and $D_{i_m}^{(m)}$ are the products of the uncut
propagators associated to each diagram of  $m$-th tree amplitude.
This notation is illustrated in \fig{FourxFourBlobFigure} for an \N2 at three loops.
In this figure, each of the two four-point blobs is expanded into
three diagrams, giving a total of nine diagrams.  For example, the
combination of indices $i_1 = 1$ and $i_2=1$ refers to the three-loop 
diagram obtained by taking the first diagram from each blob and connecting 
it to the three-point vertices; the result, in the ordering of diagrams chosen for each of 
the two four-point amplitude, is the first cubic diagram on the first line of  \fig{FourxFourBlobFigure}.  
The denominators in \eqn{DoubleCopyCut} correspond to the thick (colored) lines in the
diagrams.

The BCJ numerators in \eqn{DoubleCopyCut} are
related~\cite{BCJLoop,Square} to those of an arbitrary representation
by a generalized gauge transformation which shifts the numerators 
subject to the constraint that the amplitude is unchanged; the shift
parameters follow the same labeling scheme as the numerators
themselves,
\begin{equation}
n_{i_1, i_2, ...i_q} = 
 n_{i_1, i_2, ...i_q}^\text{BCJ} + \Delta_{i_1, i_2, ...i_q} \, .
\label{GeneralizedGaugeTrans}
\end{equation}
The shifts $ \Delta_{i_1, i_2, ...i_q}$ are constrained to leave the
corresponding cuts of the gauge-theory amplitude unchanged.  Using
such transformations we can reorganize a gravity cut in terms
of cuts of a naive double copy and an additional contribution,
\begin{equation}
 {\cal C}_{\G} = \sum_{i_1,\dots,i_q} \frac{n_{i_1, i_2, ...i_q} \tn_{i_1, i_2, ...i_q} }
  {D_{i_1}^{(1)}\dots D_{i_q}^{(q)}} +{\cal E}_\G(\Delta)\, ,
\label{CutNDC}
\end{equation}
where the cut conditions are imposed on the numerators.
Rather than expressing the correction ${\cal E}_\G$ in terms of the
generalized-gauge-shift parameters, it is useful to re-express the
correction terms as bilinears in the violations of the kinematic
Jacobi relations~\eqref{duality} by the generic gauge-theory
amplitude numerators. These violations are referred to as BCJ
discrepancy functions.

As an example, the generalized unitarity cut in \fig{FourxFourBlobFigure} is
composed of two four-point tree amplitudes and the rest are
three-point amplitudes. For any cut of this structure, two
four-point trees connected to any number of three-point trees,
the correction has a simple expression,
\begin{equation}
{\E}^{4\times 4}_\G =
- \frac{1}{d^{(1,1)}_{1} d^{(2,1)}_{1}}
     \Bigr(J_{\x1, 1}\tJ_{1, \x2}
        + J_{1,\x2} \tJ_{\x1,1} \Bigr) \,,
\label{Extra4x4Simple}
\end{equation}
where $d^{(b,p)}_{i}$ is the $p$-th propagator of the $i$-th diagram
inside the $b$-th 
amplitude factor\footnote{We will sometimes omit the second argument, $p$, when an amplitude factor has a single propagator.} and
\begin{align}
J_{\x1, i_2} \equiv \sum_{i_1=1}^3 n_{i_1 i_2} \,,
\hskip 1. cm
J_{i_1, \x2} \equiv \sum_{i_2=1}^3 n_{i_1 i_2}\,,
\hskip 1. cm
\tJ_{\x1, i_2} \equiv \sum_{i_1=1}^3 \tn_{i_1 i_2} \,,
\hskip 1. cm
\tJ_{i_1, \x2} \equiv \sum_{i_2=1}^3 \tn_{i_1 i_2}\,,
\end{align}
are BCJ discrepancy functions\footnote{We will sometimes denote the BCJ discrepancy function with either $\bullet$ in the position $i$ 
or by $\{i, 1\}$ when the $i$-th amplitude factor has a single propagator (\ie it is a four-point amplitude).}. Our notation is to label
the type of cut by $m_1 \times m_2 \times \cdots m_k$ where each
$m_i$ specifies the number of legs on each tree amplitude
with $m_i \ge 4$ composing the cut.
These discrepancy functions vanish whenever the numerators
involved satisfy the BCJ relations, even if the representation as a whole
does not satisfy them.
Such expressions are not unique and can be rearranged using various
relations between discrepancy functions~\cite{JConstraintsTye, JConstraintsVanhove, Vaman:2010ez,
GeneralizedDoubleCopy, GeneralizedDoubleCopyFiveLoops}.  For example,
a more symmetric version, equivalent to
\eqn{Extra4x4Simple}, is
\begin{equation}
{\E}^{4\times 4}_{\G} =
- \frac{1}{9} \sum_{i_1,i_2 =1}^3 \frac{1}{d^{(1,1)}_{i_1} d^{(2,1)}_{i_2}}
  \Bigl(J_{\x1, i_2} \tJ_{i_1, \x2}
     + J_{i_1, \x2} \tJ_{\x1, i_2} \Bigr) \,.
\label{Extra4x4Symmetric}
\end{equation}

Similarly, a cut with a single five-point tree amplitude and the
rest three-point tree amplitudes is given by
\begin{align}
{\cal C}^{5}_\G =\sum_{i=1}^{15} \frac{n_{i}  \tn_{i}}{d^{(1,1)}_{i} d^{(1,2)}_{i} }
 + {\E}^{5}_\G
 \qquad
 \text{with}
 \qquad
 {\E}^{5}_\G & = - \frac{1}{6}\sum_{i=1}^{15}
 \frac{J_{\{i,1\}} \tJ_{\{i,2\}} + J_{\{i,2\}} \tJ_{\{i,1\}} }
  { d^{(1,1)}_{i} d^{(1,2)}_{i}} \,,
\label{Extra5}
\end{align}
where $J_{\{i,1\}}$ and $J_{\{i,2\}}$ are BCJ discrepancy functions
associated with the first and second propagator of the $i$-th diagram.
(See Ref.~\cite{GeneralizedDoubleCopyFiveLoops} for further details.)

As the cut level $k$ increases, the formulae relating the amplitudes'
cuts with the cuts of the naive double copy become more intricate, but
the basic building blocks remain the BCJ discrepancy functions.
Formulas like \eqref{Extra4x4Symmetric}, \eqref{Extra5} and their generalizations 
can enormously streamline the computation of the contact
term corrections and are especially helpful at five loops at the \N2
and \N3 level, where calculating the contact terms via the maximal-cut
method can be rather involved.
Beyond this level, the contact terms become much simpler due to a
restricted dependence on loop momenta and are better dealt with using the
method of maximal cuts and KLT relations~\cite{KLT}, as described in
Ref.~\cite{GeneralizedDoubleCopyFiveLoops}.

\subsection {Three-loop example }

\begin{figure}[tb]
\begin{center}
\includegraphics[scale=.5]{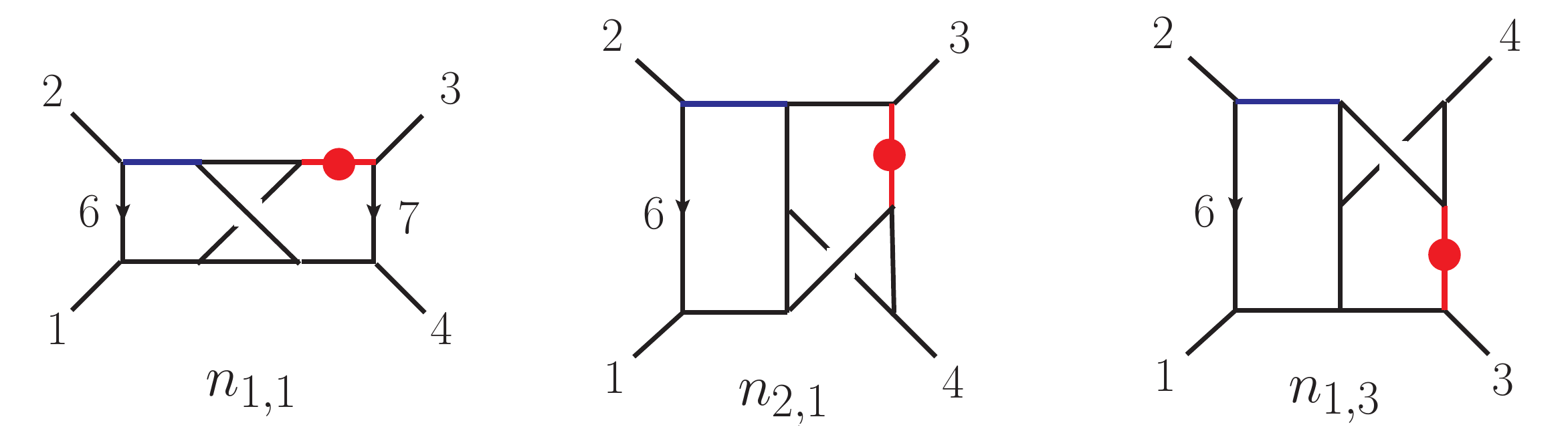}
\end{center}
\vskip -.2 cm 
\caption[a]{\small The three diagrams whose kinematic numerators
  contribute to $J_{\{1,1\},1}$.  The thick shaded (red) line marks
  the off-shell legs participating in the dual Jacobi relation.
  The shaded (red) dot indicates the off-shell leg of the second
  amplitude factor. 
 }
\label{ThreeLoopJFigure}
\end{figure}

\begin{table}[tb]
\begin{center}
\begin{tabular}{c|c}
Diagram & $\NeqFour$ SYM numerators.  \\
\hline
\hline
(a)-(d)& $\raisebox{.3 cm}{ \vphantom{|}} \raisebox{-.2 cm}{ \vphantom{|}}  s^2$\\
\hline
(e)-(g)& $\raisebox{.3 cm}{ \vphantom{|}} \raisebox{-.3 cm}{ \vphantom{|}} s (p_5^2 + \tau_{45})$\\
\hline
(h)&$\raisebox{.3 cm}{ \vphantom{|}} \raisebox{-.3 cm}{ \vphantom{|}}  s(\tau_{26} + \tau_{36}) -
      t (\tau_{17}+ \tau_{27}) + s t $ \\
\hline
(i)& $ \raisebox{.3 cm}{ \vphantom{|}} \raisebox{-.3 cm}{ \vphantom{|}}  s (p_5^2 + \tau_{45}) - t (p_5^2 + \tau_{56} + p_6^2) -
        (s - t) p_6^2/3 $\\
\hline
\end{tabular}
\end{center}
\caption{A non-BCJ form of the three-loop four-point $\NeqFour$ SYM
  diagram numerators from Ref.~\cite{CompactThree}.  
 We define $\tau_{ij} = 2 p_i \cdot p_j$, $s= (p_1
 + p_2)^2$, $t = (p_2+p_3)^2$ and $u = (p_1+p_3)^2$. 
}
\label{NonBCJNumeratorTable} 
\end{table}

To illustrate the discussion above, we now present a relatively simple though nontrivial construction 
of the three-loop four-point amplitude of $\NeqEight$ supergravity, 
which was studied in several other different approaches~\cite{GravityThree,CompactThree,BCJLoop,SimplifyingBCJ}.
As described in \sect{ExamplesSection} the most efficient  way to construct it is to first obtain 
a BCJ representation of corresponding $\NeqFour$ SYM amplitude and then apply the 
double-copy construction.
Instead, we construct it here through the generalized double copy, from
a non-BCJ form of the $\NeqFour$ SYM amplitude of Ref.~\cite{CompactThree} whose 
numerators are included in \tab{NonBCJNumeratorTable} with the momentum labeling in
\fig{ThreeLoopDiagramsFigure}(a)-(i), corresponding to the one of Ref.~\cite{BCJLoop}.  
An overall factor of $s t A_4^\tree$ is not included in \tab{NonBCJNumeratorTable}.

Following the generalized double-copy construction, the $\NeqEight$ supergravity 
numerators of diagrams (a)--(i) are squares of the corresponding $\NeqFour$ SYM ones:
\begin{equation}
N^{\NeqEight}_{(x)} =n_{(x)}^2  \,, 
\end{equation}
where $ x \in \{\rm a,\dots, i\}$.  This defines the naive double
copy.  This is not the complete supergravity amplitude given that the
gauge-theory numerators do not satisfy the BCJ
relations \eqref{duality}, as can be confirmed by checking its
generalized unitarity cuts.  To complete the supergravity amplitude we
need to find the missing contact terms.

\begin{figure}
\begin{center}
\includegraphics[scale=.44]{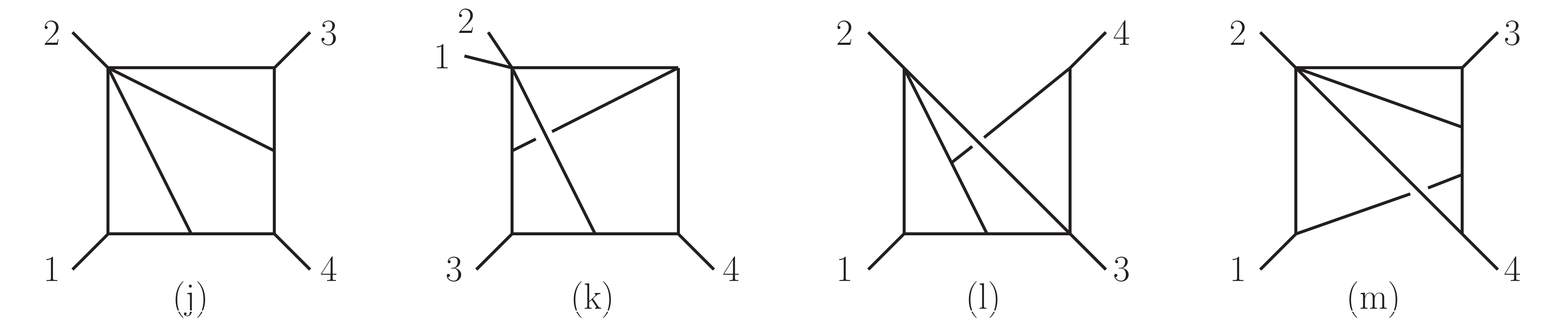}
\end{center}
\vskip -.4cm 
\caption{Nonvanishing contact terms appearing in the generalized
double copy construction of the three-loop four-point amplitude
of $\NeqEight$ supergravity. }
\label{ThreeLoopContactsFigure}
\end{figure}

Given that the \N1-level contact terms are automatically accounted for
in the naive double copy, contact terms first appear at the \N2
level.  There are a total of 62 possible independent such contact
diagram, corresponding to diagrams obtained by starting from the
first nine diagrams in \fig{ThreeLoopDiagramsFigure} and collapsing 
all pairs of propagators.
Of these, all but the four diagrams (j)-(m) in \fig{ThreeLoopContactsFigure} vanish.

 As an example,
consider the contact diagram in \fig{ThreeLoopContactsFigure}(l),
composed of two four-point vertices.  We obtain it from
\eqn{Extra4x4Simple}.
First, we identify the nine cubic diagrams that contribute to it (some
are vanishing) and pick one whose numerator we label as $n_{1,1}$; we
choose diagram (c) in \fig{ThreeLoopDiagramsFigure}. The two 
$J$-functions are calculated by relabeling the appropriate
numerators to the labels of \fig{ThreeLoopJFigure}.  For example, 
$J_{\{u_1, 1\},1}$ is obtained from the $\NeqFour$ SYM
numerators of the three diagrams shown in \fig{ThreeLoopJFigure}, 
\be
n_{1,1} = s^2, \hskip 1. cm n_{2,1} = s (t+ \tau_{26} +\tau_{36}), 
\hskip 1. cm
 n_{3,1} = s (u - \tau_{36}) \,,
\ee 
corresponding to relabeling of diagrams (c) and (g) in
\fig{ThreeLoopDiagramsFigure}.  Summing and applying momentum
conservation gives $J_{\{1, 1\}, 1} = s \tau_{26}$. Similarly, $J_{1,
  \{1, 1\}} = s \tau_{37}$.  With these labels, the two off-shell
inverse propagators are $\tau_{26}$ and $\tau_{37}$, so that from
\eqn{Extra4x4Simple} the $\NeqEight$ supergravity contact term numerator
for diagram (l) is
\begin{equation}
 N^{\NeqEight}_{\rm (l)} = -2\frac{J_{\{1, 1\}, 1} J_{1, \{1, 1\}}} 
 {\tau_{26} \tau_{37}} = -2 s^2\,.
\end{equation}
The other three independent contact terms corresponding to diagrams
(j), (k) and (m), can similarly be obtained from \eqn{Extra5}, with
the result
\be
N^{\NeqEight}_{\rm (j)} = - {\textstyle \frac{1}{9}} (s-t)^2\,, \hskip .8 cm 
N^{\NeqEight}_{\rm (k)} = N^{\NeqEight}_{\rm (m)} = -2s^2\,\!.
\end{equation}
All other nonvanishing contact terms are relabelings of these. 

\subsection{Towards general formulae}

This generalized double-copy procedure has been systematically used to
obtain the five-loop four-point integrand of $\NeqEight$
supergravity~\cite{GeneralizedDoubleCopyFiveLoops}, which was then
used to analyze the UV properties of this theory at five
loops~\cite{UVFiveLoops}.  In this case, it was sufficient to work out
formulae for the extra corrections up to the N$^3$MCs, because beyond
this the missing contact terms are simple enough to straightforwardly
obtain by numerical analysis.

As discussed before, Eqs.~\eqref{Extra4x4Simple} and \eqref{Extra5}
can be used for all N$^2$MCs in any double-copy theory. These are sufficient
to determine the three-loop four-point amplitude in $\NeqEight$
supergravity, because of its low power count. 
Beyond this order the corresponding formulae for ${\cal E}$ depend on the
detailed labeling of the corresponding cut. We include here
${\E}^{4\times 4\times 4}_\G$ and ${\E}^{5\times 4}_\G$ and 
comment on ${\E}^{6}_\G$ given as an ancillary file in
Ref.~\cite{GeneralizedDoubleCopyFiveLoops}.

The additional terms that promote a cut composed of three four-point amplitude factors of the naive double 
copy to the cut of the corresponding double-copy theory~\cite{GeneralizedDoubleCopy, GeneralizedDoubleCopyFiveLoops} are obtained by following
the steps detailed in \sect{general_g2c}. It is convenient to organize then into the contribution of 
single- and double-discrepancy functions: 
\begin{equation}
{\E}^{4\times 4\times 4}_\G = T_1 + T_2\, .
\label{Extra4x4x4}
\end{equation}
They are
\begin{align}
T_1 &= - \sum_{i_3=1}^3 \frac{J_{\x1, 1, i_3} \tJ_{1, \x2, i_3}} 
                {d^{(1)}_1 d^{(2)}_1 d^{(3)}_{i_3}} 
       - \sum_{i_2=1}^3 \frac{J_{\x1, i_2, 1} \tJ_{1, i_2, \x3}}
                  {d^{(1)}_1 d^{(2)}_{i_2} d^{(3)}_1} 
        -\sum_{i_1=1}^3 \frac{J_{i_1, \x2, 1} \tJ_{i_1, 1,  \x3}}
                 {d^{(1)}_{i_1} d^{(2)}_1 d^{(3)}_1}  + \{J \leftrightarrow \tJ\} \,, 
\nonumber \\
T_2 &=   \frac{J_{\x1, 1, 1}\tJ_{1, \x2, \x3}}{d^{(1)}_1 d^{(2)}_1 d^{(3)}_{1}}
         + \frac{J_{1, \x2, 1}\tJ_{\x1, 1, \x3}}{d^{(1)}_1 d^{(2)}_{1} d^{(3)}_{1}}
         + \frac{J_{1,1, \x3}\tJ_{\x1, \x2, 1}} 
            {d^{(1)}_1 d^{(2)}_1 d^{(3)}_{1}}  + \{J \leftrightarrow \tJ\} \, ,
\label{T1andT2}
\end{align}
where \eg $\tJ_{1, \x2, \x3}$ is defined as
\be
\tJ_{i_1, \x2, \x3} = \sum_{i_2 = 1}^3  \sum_{i_3 = 1}^3 {\tilde n}_{i_1, i_2, i_3}\,,
\ee
with $n_{i_1, i_2, i_3}$ being the numerators of the cut of the naive double copy.
As mentioned earlier, we dropped the second upper label in $d^{(b,p)}_{i}$ defined below \eqn{Extra4x4Simple}
because four-point diagrams have only a single propagator, so $b=1$ for all terms in
\eqn{T1andT2}.  

To simplify $T_2$ we used the relations
\begin{align}
\frac{J_{1, \x2, \x3}} {d^{(1)}_1}
   =\frac{J_{2, \x2, \x3}} {d^{(1)}_2}
   =\frac{J_{3, \x2, \x3}} {d^{(1)}_3} 
   ~, \quad
\frac{J_{\x1, 1, \x3}} {d^{(2)}_1}
   = \frac{J_{\x1, 2, \x3}} {d^{(2)}_2}
   = \frac{J_{\x1, 3, \x3}} {d^{(2)}_3} 
    ~, \quad
\frac{J_{\x1, \x2, 1}}{d^{(3)}_1}
   = \frac{J_{\x1, \x2, 2}}{d^{(3)}_2}=\frac{J_{\x1, \x2, 3}}{d^{(3)}_3} \, ,
\label{relations_3}
\end{align}
which identify various double-discrepancy functions.

The additional terms that promote a cut composed of one five-point and one four-point amplitude factors of the naive double 
copy to the cut of the corresponding double-copy theory can be organized as
\begin{align}
{\E}^{5 \times 4}_\G 
 &=   \sum_{i=1}^{15}\sum_{j=1}^3 \frac{1}{{d_{i}^{(1,1)}d_{i}^{(1,2)} d^{(2,1)}_{j}}} 
\biggl[
    - \frac{1}{6} {J_{\{i,1\},j} \tJ_{\{i,2\},j} }
 - \left(-\frac{1}{3}\right)\times\frac{1}{6} 
  \Bigl(  {J_{\{i,1\},j} \tJ_{\{i,2\}, \x2}} + 
     {J_{\{i,2\},j} \tJ_{\{i,1\}, \x2}} \Bigr)
\nonumber \\ 
& \null
 -  \frac{1}{5} {J_{\{i,1\}, j} \tJ_{i, \x2} }
 -  \frac{1}{5} {J_{\{i,2\}, j} \tJ_{i, \x2}}
 +  \frac{1}{30} \sum_{k\in {\cal J}_i } \sigma_{k, i} {J_{\{k, 1\}, j} \tJ_{i ,\x2} }
  +  \frac{1}{30} \sum_{k\in {\cal J}_i }\sigma_{k, i} {J_{\{k, 2\}, j} \tJ_{i, \x2}}
     \biggr]  + \{J \leftrightarrow \tJ\} \, ,
\label{Extra5x4}
\end{align}
where ${\cal J}_i $ is the set of five diagrams connected to diagram $i$ through Jacobi relations on the two propagators, including diagram $i$ which appears once,
and $\sigma_{k, i}$ are the signs with which their color factors enter in the color Jacobi relations, with the normalization that $\sigma_{i, i}=1$.\footnote{That is, the color factors of the corresponding diagrams obey the relation
$$
c_i+\sum_{k\in {\cal J}_i} \sigma_{k,i} c_k = 0 \, ,
$$
which is just the sum of the two Jacobi relations on the two propagators of diagram $i$.
}
While Eq.~\eqref{Extra5x4} is quite different from the corresponding ${\E}^{5 \times 4}_\G$ in Ref.~\cite{GeneralizedDoubleCopyFiveLoops},
the two expressions are in fact equivalent, as can be shown by reducing to a basis of BCJ discrepancy functions, or by directly evaluating 
the additional terms for a choice of representation of the five-point amplitude. 
The essential advantage of Eq.~\eqref{Extra5x4} is that it does not make reference either to a specific ordering of the diagrams of the five-point 
amplitude or  to a specific choice of order of propagators for each diagram.
These features may be the key to extending Eq.~\eqref{Extra5x4} to cuts with higher-point tree-level amplitude factors.

Similarly to  $\E_\G^{4\times 4}$ and $\E_\G^{5}$, both  $\E_\G^{4\times 4\times 4}$ and $\E_\G^{5\times 4}$ are not local.
To extract their corresponding contact terms it is necessary to subtract the contribution of the $4\times 4$- and $5$-contact 
terms which contribute to the $4\times 4\times 4$ and $5\times 4$ cuts. The strategy discussed in the previous section applies here as well, 
so we will not repeat it.

Expressions for the extra terms that promote cuts with higher-point
tree-level factors of the naive double copy to the corresponding cuts
of the double-copy theory can be obtained following the discussion
in \sect{general_g2c}.
For example, the additional terms for a cut with a single six-point factor are included 
in the ancillary file {\tt  ExtraJ\_6pt.m} of Ref.~\cite{GeneralizedDoubleCopyFiveLoops}.
Unlike $\E_\G^{4\times 4}$, $\E_\G^{5}$, $\E_\G^{4\times 4\times 4}$ and $\E_\G^{5\times 4}$ 
above however, $\E_\G^{6}$ is presented in terms of a basis of independent discrepancy 
functions, obtained by solving the constraints they obey due to their definition in terms 
of the cut kinematic numerators of the single-copy parent theories. 
The expression is also not manifestly organized in terms of the kinematic denominators 
of the 105 diagrams of the six-point tree-level diagram.
While, for these reasons, the available $\E_\G^{6}$ is not manifestly crossing symmetric,
it is sufficient for greatly simplifying the analytic structure of N$^3$MC with a single six-point tree amplitude,
compared to the direct construction of such cuts via \eg the KLT relations.
 
A feature of the nonsymmetric correction terms  $\E_\G$ expressed in terms of the some basis of BCJ discrepancy functions 
is that, when evaluated on a cut, they may lead to terms that behave as $0/0$.
These are harmless when the $0$ in the numerator is manifest, since it corresponds to an absent 
diagram.
Sometimes, however,  the $0$ in the numerator is not manifest and arises due to
a cancellation between distinct terms, that can leave behind a nontrivial finite piece. 
When this occurs, the simplest
strategy is to take advantage of the asymmetry in the formula, to relabel it to avoid such 
problematic cases. 

Generalized double-copy formulae such as those reviewed here, give the cuts of any double-copy
theory in terms of generic representations of the amplitudes of the
single-copy parent theories.
It is therefore an interesting problem to find similar general
formulae for more complicated---perhaps all---cuts at any loop order.
We can argue based on the gauge invariance of the single-copy theories
that the correction terms must be linear in the BCJ discrepancy functions of
each of the single-copy theories~\cite{GeneralizedDoubleCopyFiveLoops}.  That is,
\begin{equation}
{\cal E}  =  \sum_{i,j} M_{ij} J_{i} \tJ_{j} \,,
\label{bilinear}
\end{equation}
for some appropriate matrix $M_{ij}$ whose entries are rational
functions of the kinematic invariants of the cut.
This structure is compatible with the fact that the corrections should
all vanish if the duality between color and kinematics were manifest
in {\em either} one of the two single copies~\cite{Square}.
A further heuristic argument for the general form \eqref{bilinear} of
the correction terms ${\cal E}$ relies on an understanding of the
structure of the terms that need to be added to cuts of the naive
double copy in order to restore the linearized diffeomorphism
invariance expected of the cuts of amplitudes of a gravitational
theory.
As we saw in \sects{IntroductionSection}{DualitySection}, a gauge
transformation of tree-level amplitudes---and thus also of the cuts of
a loop amplitude---is given by a sum of terms each of which is
proportional to some linear combination of color Jacobi relations.
Consequently, a linearized diffeomorphism transformation of the naive
double copy yields a sum of terms each containing a BCJ discrepancy
function from either one of the two single copies. To restore
diffeomorphism invariance these terms must be cancelled by the
transformation of further terms that are added to the cuts of the
naive double copy.
Assuming that the structure of these terms is the same for all double
copy theories, they must be of the form \eqref{bilinear}.  See
Ref.~\cite{GeneralizedDoubleCopyFiveLoops} for more details.

While the generalized double-copy method has already been successful
for the highly nontrivial case of $\NeqEight$ supergravity at five
loops~\cite{GeneralizedDoubleCopy, GeneralizedDoubleCopyFiveLoops},
its development is only at the beginning.  Having a general tool for converting
gauge-theory amplitudes in any representation to gravity ones is clearly useful 
and important.
A good starting point would be to derive general
formulae for tree-level amplitudes~\cite{JConstraintsTye,
JConstraintsVanhove, Vaman:2010ez,GeneralizedDoubleCopyFiveLoops} in
terms of a naive double copy, plus corrections in terms of the BCJ
discrepancy functions.  At present such formulae are known only
through six points.  If an elegant solution to the tree-level problem
can be found, it should be immediately applicable to finding a general solution to the
loop-level one.  One obvious application would be towards a definitive
resolution of the UV behavior of extended supergravity
theories.  This would require calculations beyond those that have
already been carried out (see
\eg Refs.~\cite{N5GravFourLoop,UVFiveLoops}), and would likely need a
version of the generalized double copy to be practical.  $\NeqFive$
supergravity at five loops is an especially interesting case for
future study, given that at four loops it exhibits an enhanced cancellation
of UV divergences~\cite{N5GravFourLoop}. It is important to
know whether this continues at higher loops.



\section{Classical double copy}
\label{ClassicalDoubleCopySection}

As we have seen at length, the duality between color and kinematics
and the  double-copy construction are essential tools in the
construction of gauge and (super)gravity scattering amplitudes at
higher-loop orders and/or at higher multiplicity.  In close analogy
with tree-level scattering amplitudes, the perturbative construction
of solutions of the classical equations of motion of a field theory
(perhaps in the presence of sources) also exhibits an expansion in
tree-level diagrams. One may consequently expect that, with an
appropriate definition, some version of double-copy construction may
lead to a construction of solutions of Einstein's equations (perhaps
also in the presence of other fields) in terms of solutions of
YM equations of motion (perhaps also in the presence of other
fields).  If one could turn the double copy into a systematic tool for
analyzing classical solutions one could hope for new advances
analogous to the ones that have occurred for scattering amplitudes.

As we shall discuss below, such a relation between classical solutions
is not without subtleties and comes with quantifiable differences from
the case of flat-space scattering amplitudes.  Flat-space scattering
amplitudes carry an inherent simplicity in that they are completely
independent of gauge and field variable choices.  However, in contrast
to scattering amplitudes, generic classical solutions change
nontrivially under gauge transformations and, moreover, they are
sensitive to the nonlinear terms in the gauge transformations.  Thus,
to relate gauge and gravity solutions it is necessary to make
correlated gauge choices in the two theories; the principles for making
such choices are unclear.  Related to this, the
form of the equations of motion depends strongly on the choice of
field variables.  Thus, any naive extension of the
scattering-amplitudes' double copy of fields can be completely obscured by
nonlinear coupling-dependent terms that depend on some {a priori}
chosen form of the equations of motion.

As yet, no coherent set of rules for the construction of double copies
for generic classical solutions in gravity theories has been
formulated, though a variety of nontrivial tantalizing examples have
been found.  (See \eg Refs.~\cite{Saotome2012vy, Monteiro2014cda, Luna2015paa,
Ridgway2015fdl, Luna2016due, White2016jzc, Cardoso2016amd,
Goldberger2016iau, Luna2016hge, Goldberger2017frp,
Adamo2017nia, DeSmet2017rve, BahjatAbbas2017htu,
CarrilloGonzalez2017iyj, Goldberger2017ogt, Li2018qap,
Ilderton:2018lsf, Lee:2018gxc, Plefka:2018dpa, ShenWorldLine,
Berman:2018hwd, Gurses:2018ckx, Adamo:2018mpq, 
Bahjat-Abbas:2018vgo, Luna:2018dpt,
Farrow:2018yni, CarrilloGonzalez:2019gof, PV:2019uuv}.)  Ideally, any such rules should smoothly generalize
those of scattering amplitudes and reduce to them in the appropriate
limits.  The classes of examples that have been constructed and analyzed
emphasize both the similarities and the differences between classical
solutions and scattering amplitudes, and expose the subtleties that
need to be addressed in order to formulate a general framework.
Their existence, however, suggests that it may be possible to find
generic solutions of a gravity theory in terms of solutions of the two
gauge theories that give its scattering amplitudes. The most obvious
application of these ideas are towards improving calculations of as
well as calculations in post-Newtonian expansion of gravitational
interaction potentials as well as calculations potentially relevant to
gravitational-wave detection.
These type of calculations can be phrased in terms of scattering
amplitudes~\cite{Holstein2004dn, NeillRothstein,
Damour:2017zjx,Luna2017dtq, BjerrumClassical, CheungPM,
Kosower:2018adc, OConnellSpin2019} and therefore are likely to lead to useful new
results, such as the computation of the third post-Minkowskian 
contribution to the conservative two-body potential~\cite{3PM,3PMLong}.

In this section we describe the known constructions of gravity
classical solutions in terms of gauge-theory solutions, commonly referred to as
``classical double copies''. We outline their relation and
similarities with the double copy of scattering amplitudes and
summarize the examples that have been discussed in this framework.  We
start with a description of perturbative solutions in gravity before
turning to complete double copies.

\subsection{Perturbative classical solutions vs. tree-level amplitudes} 

There is a close relation between solutions of classical equations of motion of some field theory and the Green's functions of that 
theory. The classical field generated by an arbitrary source is the generating
functional for the tree-level connected Green's functions. Given a field theory of some field $\phi$ with Lagrangian ${\cal L}$, 
a solution of the equation of motion with general sources,
\begin{equation}
\frac{\delta {\cal L}}{\delta \phi} = \zeta\,,
\label{eom}
\end{equation}
is given in terms of the generating functional of connected tree-level Green's functions by  \cite{Boulware1968zz}
\begin{equation}
\phi[x, \zeta]= \frac{\delta W[\zeta]^\text{tree}}{\delta\zeta} \,,
\label{sol_from_W}
\end{equation}
and moreover 
\begin{equation}
W[\zeta] = \int d^Dx \; \big( {\cal L}[\phi[x, \zeta]] - \zeta \phi[x, \zeta] \big) \,.
\end{equation}
The relation between Green's functions and scattering amplitudes given by the LSZ reduction implies in turn that, 
by amputating the sources, $W$ becomes the generating functional of tree-level S-matrix elements. This may be realized by
taking the source to be the quadratic operator acting on an on-shell wave solution of the free equation of motion.
The solution \eqref{sol_from_W} with such sources is the generating function of Berends-Giele currents---\ie Green's functions 
of fundamental fields with exactly one leg off shell\footnote{Green's functions with two legs off shell have been constructed in gauge theories
coupled to fundamental matter in \cite{Mahlon:1992fs,Mahlon:1993si}.}; it therefore may also be interpreted as a solution of the 
Berends-Giele off-shell recursion relation \cite{Berends:1987me}.  
This idea was  used in Refs.~\cite{Selivanov1997ts,Selivanov1997aq} to construct an implicit representation (referred to as 
the ``perturbiner'') of gluon scattering amplitudes in four-dimensional YM theory and the gravitational dressing of certain 
classes of such amplitudes. Tree-level amplitudes of higher-dimensional and supersymmetric YM theories have been 
constructed using this method in Refs.~\cite{Mafra:2015gia, PureSpinorsBCJAmplProof} and in certain effective field theories and deformations 
of YM theories in Refs.~\cite{Mizera:2018jbh,SchlottererBGCurrent}.
It was also was used in Ref.~\cite{Monteiro2011pc} to construct the kinematic algebra dual to the color algebra in self-dual 
YM theory. Solutions for the supersymmetric versions of Berends-Giele current that manifest
\ck duality were given in Ref.~\cite{Lee:2015upy}.

Thus, Eqs.~\eqref{eom}~and~\eqref{sol_from_W} allow us to construct perturbative approximations of solutions with the 
appropriate source in terms of the scattering amplitudes of the theory. 
Moreover, should it be possible to resum the scattering amplitudes into a generating functional, Eq.~\eqref{sol_from_W} 
provides an exact solution of the equation of motion with the appropriate sources. 
Depending on the chosen sources, the construction can be carried out either in momentum space (if the sources are momentum eigenstates) or in position space.

\begin{figure}
	\begin{center}
	        \def\widthfigF{0.10\textwidth}
		\def\widthfig{0.15\textwidth}
		\def\widthfigT{0.12\textwidth}
		\begin{eqnarray*}
\phi[x, \zeta] =  \parbox{\widthfig}{\includegraphics[width=\widthfigF]{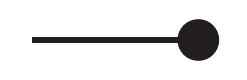}}
 \!\!\!\!\!\!\!\!\!
+ \parbox{\widthfig}{\includegraphics[width=\widthfigT]{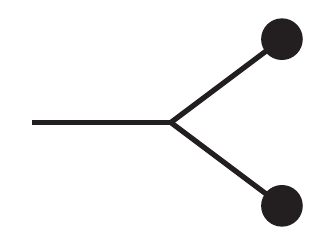}}
     +\left(\parbox{\widthfig}{\includegraphics[width=\widthfig]{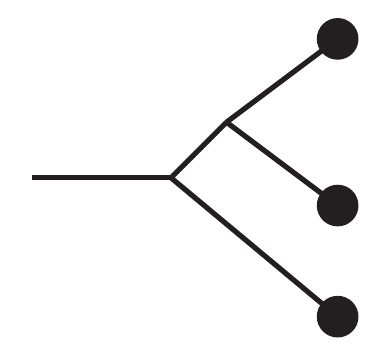}} 
            +\parbox{\widthfig}{\includegraphics[width=\widthfig]{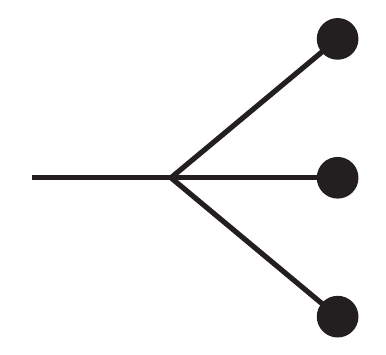}}
\right)
  +\dots \end{eqnarray*} \end{center} \caption{The first few terms in
		the expansion of a classical solution in terms of
		sources. Each heavy dot represents a source
		$\zeta$. The free end is at position $x$. The weight
		of each vertex is not specified and may contain
		derivatives acting on the propagators connecting it to
		other vertices, sources or the point
		$x$. \label{classical_sol_trees}}
\end{figure}

Introducing sources in gauge and gravity theories can be confusing for
at least two reasons. First, fixed sources coupling to vector fields
or with the graviton may break gauge invariance.  A resolution of this
would-be problem is the gauge-fixing that is necessary for any
(tree-level) computation, which already breaks gauge invariance. One
then adds sources in the gauge-fixed theory, in which the question
of gauge invariance should not arise.
Second, related, nonabelian vector fields and gravitons self-interact
and consequently they can self-source.  Examples are all solutions of
vacuum Einstein's equations as well as solutions of classical
YM equations such as the instanton. For a stable configuration
the matter stress tensor should be covariantly constant with respect
to the metric that it sources; thus, it has some knowledge of the
solution.  This implies that the perturbative construction of such
solutions requires a judicious choice of source which may itself receive
corrections order by order in perturbation theory. Examples were
discussed in \eg Refs.~\cite{Duff1973zz}
and~\cite{Sardelis1973em} for the Schwarzschild and
Reissner-Nordstr\"om black holes, respectively.

Unlike scattering amplitudes, solutions of the classical field
equations can be changed by (1) field redefinitions (2) coordinate
changes and (3) gauge transformations (if gauge symmetries are
present).\footnote{Symmetries of the equations of motion which are not
symmetries of the action, such as parts of the U-duality symmetry of
four-dimensional supergravity theories, may be used to generate
inequivalent solutions from known ones. See \eg Ref.~\cite{Gunaydin:2009pk} for a review.}
\footnote{The same choices also affect Feynman rules; however, when Feynman rules 
are combined into a scattering amplitude there is no dependence upon
these choices, although solutions of the classical equations of motion
(and also Green's functions) depend on them.}
As yet, Lagrangians that manifest \ck duality are known to only a few
perturbative orders~\cite{Square,WeinzierlBCJLagrangian,Vaman:2014iwa,
Mastrolia:2015maa}.  It is natural to expect that, if one had such a
complete Lagrangian, classical solutions constructed through a
classical double copy would solve its equations of motion. It is
natural to expect that nontrivial field redefinitions and coordinate
transformations are necessary to map such a solution to the field
variables of a more standard Lagrangian.  In fact, the perturbative
Lagrangians manifesting the double-copy properties of gravity require
this as well as elimination of auxiliary fields.

To illustrate the perturbative construction of solutions of
supergravity equations of motion we outline here the derivation of the
first terms~\cite{Sardelis1973em} of the Reissner-Nordstr\"om
solution---a charged black hole of (super)gravity coupled with a
vector field $A_\mu$ of field strength $F_{\mu\nu}$.  
The vanishing-charge limit leads to the corresponding
(first) term(s) in the Schwarzschild solution, discussed in
Ref.~\cite{Duff1973zz}.  The relevant action is
\begin{align}
S &= S_G+S_{EM}+S_\text{gauge fixing}+S_\zeta \,,
\nonumber\\
{\cal L}_G &= \frac{1}{\kappa^2} \sqrt{-g} g^{\mu\nu} R_{\mu\nu} \,,
\hskip 1.2 cm 
{\cal L}_{EM} =\frac{1}{16 \pi} \sqrt{-g} g^{\mu\rho} g^{\nu\sigma}  F_{\mu\nu}F_{\rho\sigma} \,,
\nonumber \\
{\cal L}_\zeta &=  \frac{1}{2} g^{\mu\nu}(\zeta^M_{\mu\nu} + \zeta^{EM}_{\mu\nu}) +   A_\mu \zeta^\mu
\equiv \frac{1}{2}\sqrt{-g}g^{\mu\nu}(T^M_{\mu\nu} + T^{EM}_{\mu\nu}) +  \sqrt{-g} A_\mu j^\mu \,, \nonumber
\\
{\cal L}_\text{gauge fixing}&=-\frac{1}{2\pi}(\partial^\mu A_\mu)^2 +\frac{1}{2}(\partial_\mu (\sqrt{-g} g^{\mu\nu}))^2\, .
\end{align}
To construct a perturbative solution around Minkowski space the metric
is assumed of the form\footnote{Note that this choice is different
form the one typically used for perturbative S-matrix calculations and
in later subsections, but it is useful here as it avoids nonlinear
terms involving the metric fluctuation and the sources.}
\be
g^{\mu\nu} = \eta^{\mu\nu} +\kappa h^{\mu\nu} \, .
\label{metric_expansion}
\ee

There are several sources that lead to the desired solution. One may choose, for example, the stress tensor of a charged point 
particle. Alternatively, one may choose an extended source---a sphere of radius $\epsilon$ of uniform mass density $\rho$ and 
uniform charge density $\sigma$.  The general form of the stress tensor is
\be
T_\nu{}^\mu = (\rho+p) u_\nu u^\mu + p\delta{}_\nu{}^\mu\,,
\hskip 1.5 cm 
g_{\mu\nu} u^\mu u^\nu = 1 \,,
\ee
where $\rho$ is the mass density function and $p$ the (potentially phenomenological) pressure. In the case of a ``ball of dust'' with 
uniform mass and charge densities, the components of the source turn out to be (after choosing $u=(1,0,0,0)$ and imposing
covariant constancy of the stress tensor)~\cite{Sardelis1973em}
\begin{align}
\zeta^M_{00} &= \rho \theta(\epsilon-r)= \frac{3m}{4\pi \epsilon^3}\theta(\epsilon-r) \,,
\hskip 1 cm
\zeta^M_{ij} = p^{(0)} \eta_{ij} = \frac{3Q^2}{8\pi \epsilon^6} (r^2-\epsilon^2) \delta_{ij}\theta(\epsilon-r) \,,
\cr
\zeta_\mu &= \sigma \delta_\mu^0\theta(\epsilon-r) = \frac{3Q}{4\pi \epsilon^3}\theta(\epsilon-r)  \delta_\mu^0 \,,
\label{sources_RN}
\end{align}
where $m$ and $Q$ are the total mass and charge, respectively.  The pressure $p^{(0)}$ is chosen such that this configuration of
mass and charge densities is static under Newtonian gravitational attraction and Coulomb repulsion. As the metric receives 
$\kappa^n$ corrections, so will the pressure function (hence the upper label ``$(0)$'' in $\zeta_{ij}^M$ above).
%

The first correction to the flat space metric and the electromagnetic field due to the sources \eqref{sources_RN}, given by 
the first two diagrams in \fig{classical_sol_trees}, is
\begin{align}
\langle A^\mu(x) \rangle_\zeta &=  \int d^d y \, \Delta^{\mu\nu}(x-y)\zeta_{\nu}(y)+\dots \,,
\nonumber \\
\kappa \langle h^{\mu\nu}(x)\rangle_{\zeta} &= \frac{\kappa^2}{2} \int d^d y \, 
\Delta^{\mu\nu, \rho\sigma}(x-y)\zeta^M_{\rho\sigma}(y) 
\nonumber \\
& \null \hskip .5 cm 
 +{\kappa^2}\int d^d y d^d x_1 d^d x_2 \, 
\Delta^{\mu\nu, \rho\sigma}(x-y) \gamma_{\rho\sigma, \eta\tau}(\partial_y)
\langle A^\eta(y) \rangle\langle A^\tau(y) \rangle \, .
\end{align}
Here $\Delta^{\mu\nu, \rho\sigma}$ is the graviton propagator in the
chosen de-Donder gauge (cf. ${\cal L}_\text{gauge fixing}$),
$\Delta^{\mu\nu}$ is the photon propagator in Lorentz gauge and
$\gamma_{\rho\sigma, \eta\tau}(\partial_y)$ describes the
graviton-photon three-point interaction.
We note that, due to the $\kappa$ dependence in expansion of the
metric~\eqref{metric_expansion}, the trilinear graviton-photon vertex
contributes before the three-graviton vertex.

The extended nature of the source implies that the vector potential is different for $r<\epsilon$ and $r>\epsilon$. 
Denoting by tilde  the Fourier-transform of the source, 
\be
\langle A^\mu (y) \rangle = \delta^\mu_0\int d^3 p \frac{e^{i p\cdot x}}{-p^2} {\tilde \zeta}_{0}(p) 
= \delta^\mu_0\left( \frac{Q}{r} \theta(r-\epsilon) + \left(\frac{3Q}{2\epsilon} - \frac{Q r^2}{2\epsilon^3}\right)\theta(\epsilon-r)\right) 
\equiv \delta^\mu_0 U\,,
\ee
which is just the Coulomb potential of the assumed charge distribution. Defining similarly the Newtonian potential 
of the given mass distribution,
\be
W \equiv \int d^3 p \frac{e^{i p\cdot x}}{-p^2} \tilde \zeta^M_{00}(p) =  
\frac{\rho}{4\pi r} \theta(r-\epsilon) + \left(\frac{3\rho}{8\pi\epsilon} - \frac{\rho r^2}{8\pi\epsilon^3}\right)\theta(\epsilon-r) \,,
\ee
and the action of the inverse Laplace operator on a time-independent function $F(x)$ as
\be
\frac{1}{\nabla^2}F(x) \equiv  \frac{1}{4\pi} \int d^3 y \frac{F(y)}{|x-y|} \,,
\ee
the components of the metric fluctuations around flat Minkowski space are
\begin{align}
\kappa \langle h^{00}\rangle_\zeta &= 
8\pi G\left(W+3\frac{1}{\nabla^2}p^{(0)}-\frac{\eta_{kl}}{4\pi} \frac{1}{\nabla^2} \partial^k U\partial^l U\right) \,, \nonumber
\\
\kappa \langle h^{ij}\rangle_\zeta &= 8\pi G\left(W - \frac{1}{\nabla^2}p^{(0)}
 - \frac{\eta_{kl}}{4\pi} \frac{1}{\nabla^2} \partial^k U\partial^l U\right) \delta^{ij} 
 - 4G\frac{1}{\nabla^2}  \partial^i U\partial^j U \,,
\nonumber\\
\kappa \langle h^{i0}\rangle_\zeta &= 0\,.
\end{align}

Evaluating the integrals and defining the physical mass
\be
M = m+\frac{3}{5} \frac{Q^2}{\epsilon} \,,
\label{physical_mass}
\ee
it follows~\cite{Sardelis1973em} that for $r>\epsilon$ the metric components are 
\begin{align}
g^{00} &= 1+\frac{2MG}{r} - \frac{Q^2 G}{r^2}+ {\cal O}(G^2)\,, \nn
\\
g^{ij}&= -\left(1-\frac{2MG}{r}\right)\delta^{ij} + \frac{Q^2 G}{r^4} x^i x^j + {\cal O}(G^2) \,, \nn
\\
g^{i0}&= 0 \,.
\end{align}
This matches the Reissner-Nordstr\"om solution in Cartesian
coordinates and de Donder gauge~\cite{Sardelis1973em}\footnote{While
this is different from the standard form of the Reissner-Nordstr\"om
solution, it can be mapped to it by a coordinate transformation and
field redefinition.}: 
\be
ds^2 = \frac{r^2+Q^2 G - M^2 G^2}{(r+MG)^2} dt^2 -\left(1+ \frac{M G}{r}\right)^2 (dx^i)^2 
+ \frac{(Q^2 G - M^2 G^2)(r+MG)^2}{r^4(r^2+Q^2 G - M^2 G^2)} (x_i dx^i)^2 \,.
\ee
We note that the $Q\rightarrow 0$ limit yields the Schwarzschild
solution~\cite{Duff1973zz} as well as that the size of the mass and
charge distribution do not affect the exterior solution, in agreement
with Birkhoff's theorem, which states that any spherically symmetric
solution of the vacuum field equations must be static and
asymptotically flat.  The size of the distribution enters however the
definition \eqref{physical_mass} of the physical mass $M$ for
nonvanishing electric charge.
We also note that Eq.~\eqref{physical_mass} is a reflection of the
field backreaction on sources. In fact, the redefinition
\eqref{physical_mass} is necessary for the solution to have a 
smooth limit to a point source. The relation between the physical 
mass $M$ and the ``free mass'' $m$ receives further corrections as
higher orders are included.

\subsection{Perturbative spacetimes and the double copy \label{perturbative_st}}
\def\h{\mathfrak{h}}
\def\dd#1{\frac{d^D #1}{(2\pi)^D}}
\def\del#1{(2\pi)^D\delta(#1)}

The double-copy formulation of classical gravity calculations has the
potential to streamline calculations such as those
outlined in the previous subsection by exploiting the close relation
between the tree expansion in \fig{classical_sol_trees} and that of
tree-level S-matrix elements.
Given that we do not as yet have a general framework for applying the
double copy to perturbative solutions, detailed analyses of specific
examples, as we do below, help identify the correct physical extension
of the amplitudes double-copy rules to this setting.
Before we proceed to summarize the various options and illustrate
their application to this problem, we begin with several comments
which connect it to some of the calculations above and alert the
reader to points that will arise.
 
As noted in \sect{DualitySection}, the double-copy spectrum naturally
contains a dilaton and a two-index antisymmetric tensor (or
equivalently a pseudo-scalar in four dimensions). As for tree-level
scattering amplitudes where these unwanted states can be projected out at tree level
by a suitable choice of asymptotic states, solutions of Einstein's
equation may be found by choosing gauge-theory sources such that their
double copy does not source the dilaton and/or the anti-symmetric
tensor~\cite{Luna2016hge, Goldberger2016iau}.
Choosing gauge-theory sources that are then used in the double copy appears to bypass the need for a judicious choice a matter stress tensor as source for the 
gravity solution; however, prescribed properties of supergravity solutions and their corresponding sources undoubtedly translate 
into properties of gauge-theory sources. At the time of this writing, a complete dictionary has not yet been formulated. 

\ck duality as defined for scattering amplitudes in \sect{DualitySection}, 
requires that external lines are on the free mass shell.  Thus, in the
tree expansion in \fig{classical_sol_trees} the duality can be
expected to hold only up to terms that vanish if the sources obeyed
free-field equations of motion. The discussion in
\Sect{DualitySection} then implies that such a feature leads to
breaking of linearized gauge (diffeomorphism) invariance in the double-copy 
theory due to the presence of sources. This may be interpreted as
the double-copy realization of the fact that gravity sources break
diffeomorphism invariance.
For the same reason, gravity field equations can be satisfied by a double-copy field configuration only up to terms proportional 
to the free equations of motion of the sources. Thus, for a comparison with a direct solution of supergravity equations of motion, 
such terms must be eliminated by field, coordinate and source redefinitions. 
This mirrors the backreaction of gravitational field on its source, illustrated in the previous subsection. 
It is not {a priori} obvious that gauge-theory classical solutions which differ by gauge transformations lead through the double copy 
to gravity solutions that differ by field redefinitions and coordinate transformations. 
 
Perturbative spacetimes and their relation to perturbative solutions
of the YM equations of motion were discussed in
Ref.~\cite{Luna2016hge}. Below we outline their construction. As in
the calculation of scattering amplitudes, we begin with the YM
action (see \eqn{Lagrangians}), whose equations of motion in the presence of sources are
\begin{equation}
\partial^\mu F^a_{\mu\nu} + g f^{abc} A^{b\mu} F^c_{\mu\nu} = \zeta_\mu^a \,,
\label{YM_eqs}
\end{equation}
where $g$ is the coupling constant and the field-strength tensor $F^a_{\mu\nu} $ is
\begin{equation}
F^a_{\mu\nu} = \partial_\mu A^a_\nu - \partial_\nu A^a_\mu + g f^{abc} A^b_\mu A^c_\nu \,.
\end{equation}
It is straightforward to include matter fields. Apart from their equations of motion, inclusion of the matter fields also gives specific 
expressions for the sources $\zeta$. We will not discuss this possibility any further, choosing $\zeta$ to be non-dynamical and
focusing on the gauge sector.
The goal, following Ref.~\cite{Luna2016hge}, is to solve perturbatively Eq.~\eqref{YM_eqs}, 
\begin{equation}
A^a_\mu = A^{(0)a}_\mu + g A^{(1)a}_\mu + g^2 A^{(2)a}_\mu + \cdots \,, 
\label{vector_field_expansion}
\end{equation}
and construct from it a solution of the double-copy theory.  

The action for Einstein gravity coupled with a dilaton and an antisymmetric tensor, which is the double copy of two pure $D$-dimensional 
gauge theories, is given in Eq.~\eqref{SN=0}.
For the construction of a perturbative solution of its equations of motion, the fields are expanded as
\begin{align}
h_{\mu\nu} &= h^{(0)}_{\mu\nu} + \frac{\kappa}{2} h^{(1)}_{\mu\nu} + \left(\frac{\kappa}{2}\right)^2 h^{(2)}_{\mu\nu} + \cdots \,, 
\nonumber \\
B_{\mu\nu} &= B^{(0)}_{\mu\nu} + \frac{\kappa}{2} B^{(1)}_{\mu\nu} +\left(\frac{\kappa}{2}\right)^2  B^{(2)}_{\mu\nu} + \cdots \,, 
\nonumber \\
\phi_{\mu\nu} &= \phi^{(0)}_{\mu\nu} + \frac{\kappa}{2} \phi^{(1)}_{\mu\nu} + \left(\frac{\kappa}{2}\right)^2 \phi^{(2)}_{\mu\nu} + \cdots  \,.
\label{gravity_ini_fields}
\end{align} 
We can combine these different fields into a single field $H$.
Since the asymptotic values of these fields are all obtained by projection from the tensor product of the two asymptotic gauge fields,
it is convenient to have the field $H$ which has this property at every order in $\kappa$.
That is, in its expansion in $\kappa$,
\begin{equation}
H_{\mu\nu} = H^{(0)}_{\mu\nu} + \frac{\kappa}{2} H^{(1)}_{\mu\nu} + \left(\frac{\kappa}{2}\right)^2 H^{(2)}_{\mu\nu} + \cdots \,, 
\label{FatFullGraviton}
\end{equation}
$H^{(n)}$ is the double copy of the $n$-th order term in the expansion of the gauge-theory field. 
There are no cross terms between different orders in the vector field expansion \eqref{vector_field_expansion}. This is a 
consequence of the fact that different orders are given by different tree configurations in \fig{classical_sol_trees} 
and thus do not mix in the double copy.

On shell, at the linearized level and in the appropriate 
gauges\footnote{The de Donder gauge for the graviton and the Lorentz gauge for the tensor field.} it is 
possible~\cite{Luna2016hge} to formulate the equations of motion in terms of a linear combination of the three fields:
\be
H_{\mu\nu}^{(0)} = h_{\mu\nu}^{(0)}+B_{\mu\nu}^{(0)}+ P_{\mu\nu}^q \phi^{(0)} \, .
\label{fat}
\ee
In the absence of sources they are
\be
\partial^\rho \partial_\rho  H_{\mu\nu}^{(0)}= 0 \, .
\label{fat_grav_eq}
\ee
A source modifies the right-hand side appropriately and must have the transversality and trace properties of $H_{\mu\nu}^{(0)}$. 
The field \eqref{FatFullGraviton} has been referred to in Ref.~\cite{Luna2016hge} as the ``fat graviton'', in contrast 
with the ``skinny graviton'', $h_{\mu\nu}$. 
In Eq.~\eqref{fat} $P_{\mu\nu}^q$ is a projector, which depends on a  fixed null vector $q$, defining the physical dilaton. In position space it is
\be
P_{\mu\nu}^q = \frac{1}{D-2}\left(\eta_{\mu\nu} - \frac{q_\mu\partial_\nu- q_\nu\partial_\mu}{q\cdot \partial} \right) \,.
\label{projector}
\ee
Conversely, the three physical fields can be extracted from $H^{(0)}$ by projection:
\be
\phi^{(0)} = \eta^{\mu\nu} H^{(0)}_{\mu\nu}\,,
\qquad
B_{\mu\nu}^{(0)} = \frac{1}{2}(H^{(0)}_{\mu\nu}-H^{(0)}_{\nu\mu})\,,
\qquad
h_{\mu\nu}^{(0)} = \frac{1}{2}(H^{(0)}_{\mu\nu}+H^{(0)}_{\nu\mu}) - P_{\mu\nu}^q H^{(0)}{}^\rho{}_\rho \,.
\label{projection_hbphi}
\ee

\subsubsection{Linearized solution \label{linearized_source_discussion}}

Following  Ref.~\cite{Luna2016hge}, to solve the YM equations \eqref{YM_eqs} we choose the Lorenz 
gauge, $\partial^\mu A^a_\mu = 0$, and to leading order in the coupling the equation becomes 
\begin{equation}
\partial^2 A^{(0)a}_\mu = \zeta_\mu^a \, .
\label{YM_0_w_source}
\end{equation}
Consistency with the gauge condition requires that $\zeta$ be transverse.
To start instead with a scattering state it suffices to replace $\zeta^a_\mu \rightarrow \varepsilon_\mu c^a \partial^\mu\partial_\mu \exp(i p\cdot x)$ 
where $c^a$ is some color wave function and take the limit $p^2 \rightarrow 0$ at the end of calculations. 
To cover both options simultaneously one may denote the solution to this equation by $A^{(0)a}_\mu$ while not using its specific form, though in 
specific examples it may be necessary to be more specific.

Wave solutions of the YM free-field equations, 
\be
A_\mu^{(0)a} = \sum_j c^a_j  \varepsilon^j_\mu(p) e^{ip\cdot x} \,,
\ee
with little-group indices $j$, $p\cdot \varepsilon^j =0 = q\cdot \varepsilon^j $ and $c^a_j$ color wave functions, can be straightforwardly 
double-copied to wave solutions of the free field equation of the action \eqref{SN=0}. In the absence of sources this 
is just a reorganization of the usual double copy of scattering states.
The gravity solution,
\be
H^{(0)}_{\mu\nu}(x) =h_{ij} \varepsilon^i_\mu(p) \varepsilon^j_\nu(p) e^{ip\cdot x} \, ,
\ee 
can be decomposed into the graviton, $B$-field and dilaton using \eqref{projection_hbphi}. The constant factor $h_{ij}$ is
arbitrary and can be chosen such that the gravity and gauge-theory asymptotic waves have the same normalization.
It can also be used to project out the $B$-field and dilaton and obtain a linearized solution of Einstein's equations, {\i e.g.}
\be
c^a_j = c^a a_j\,, 
\hskip 1 cm 
h_{ij} = a_i a_j\,,
\hskip 1 cm 
a\cdot a = 0 \, ,
\ee
where $a_i$ are ``kinematic gauge-theory wave functions''. 

In general, whether or not the $B$-field and dilaton can be turned off depends on the gauge-theory sources
and on their relation to gravity sources. The relevant solutions of the \eqref{YM_0_w_source}, in position and 
momentum space, for an arbitrary source is
\be
A^{(0)a}_\mu(x) \propto  \int d^D y \frac{\zeta_\mu^a(y)}{|x-y|^{D-2}} \,,
\hskip 1.5 cm 
{\cal F}[A^{(0)a}_\mu](p)  = \frac{{\cal F}[\zeta_\mu^a](p)}{p^2} \,,
\label{sol_zeta_YM}
\ee
where ${\cal F}$ is the Fourier-transform operator. The rules for constructing the corresponding linearized gravity 
solution and sources are yet to be completely clarified. Here we attempt to formalize several possibilities, while 
leaving others for future development. 

In identifying suitable relations between gauge and gravity sources it is important that the result can be interpreted 
as the linearized stress tensor of some field theory and thus that it conforms with energy conditions expected of 
such a stress tensor~\cite{Ridgway2015fdl}. Not every possible construction has this property; indeed, it was shown 
in Ref.~\cite{Ridgway2015fdl} that, while the source for Kerr-Schild solutions (whose linearized approximation is exact 
and will be discussed in some detail in Sect.~\ref{KS_sol_sect}) can be obtained by specifying the charge distribution 
sourcing the corresponding gauge-theory solutions  and imposing $\nabla_\mu T{}^\mu{}_\nu = 0$, they do not obey 
simultaneously the weak and strong energy conditions.
While discussions of energy conditions have appeared in the literature (see below for references), a thorough analysis 
is currently absent and we will refrain from attempting one here.
We emphasize that  the classical double copy is best {\em defined} so that  it yields a solution of \eqref{fat_grav_eq};
moreover, its sources should be constructed out of the gauge-theory sources such that they do not have any unphysical features. 
In general, it is necessary to verify whether the resulting source obey reasonable energy conditions before attempting 
to promote it from a non-dynamical source to a dynamical one, realized in terms of the fundamental fields of a quantum theory.
These requirements may be used to identify some of the rules of the construction.

We begin with a source of the type
\be
\zeta_\mu^a = c^a \zeta_\mu \, ,
\label{single}
\ee
with constant color factor $c^a$ and transverse $\zeta_\mu$. Even though $c^a$ need not have any particular algebraic 
properties,  it is natural to take at face value  the fact  that the solution \eqref{sol_zeta_YM} is given by the first Feynman diagram in 
\fig{classical_sol_trees} and apply the usual double-copy rules: $c^a\rightarrow {\tilde \zeta}_\mu$.
Even though nonlinear corrections to a YM solution with this source
vanish because $f^a{}_{bc}c^b c^c = 0$, nonlinear corrections to its
gravity counterpart may be present; we shall see this explicitly
in \sect{nonlinear_corrections}.  Similarly to scattering-amplitudes double copy, it is very important to {\em not} discard color
factors that vanish due to the summation over the color indices.
(Examples where this is crucial are found in
Refs.~\cite{SimplifyingBCJ, ShenWorldLine}.)

Gauge-theory sources may exhibit a less transparent separation of color and kinematics,~\eg
\be
\zeta(x)_\mu^a = \sum_i c^a_i \zeta(x)^i_\mu \,,
\ee
with several distinct independent color factors $c^a_i$ and transverse (position-dependent) $\zeta^i_\mu$.  
Similarly to the case of asymptotic scattering states, we may still apply the (color factor) $\rightarrow$ (kinematic factor) 
replacement in momentum space with the same twist as in that case (and in the case of a wave solution) of allowing for a 
constant relative rotation of sources. Formally
\be
\sum_i c^a_i {\cal F}[\zeta^i_\mu](p) \longrightarrow \sum_{i,j} h_{ij}{\cal F}[\zeta^i_\mu](p) {\cal F}[{\tilde \zeta}^i_\mu](p) \,,
\ee
where ${\cal F}$ is the Fourier-transform operator. 
In general it may be possible to allow $h_{ij}$ to be a function of momentum; Lorentz invariance demands that it should be a 
function of $p^2$ and thus it can only lead to shifts of $H^{(0)}_{\mu\nu}$  by local functions. From this perspective, 
$h_{ij}\rightarrow h(p^2)_{ij}$ should be equivalent to field and/or coordinate redefinition in the gravity theory. 

In both this case and in the simpler previous case (which may be obtained by taking the indices $i$ and $j$ to take a 
single value), the resulting space linearized solution is
\be
H_{\mu\nu}^{(0)}(p)  = \frac{\sum_{ij}h(p^2)_{ij}{\cal F}[\zeta_\mu^i](p){\cal F}[{\tilde \zeta}_\nu^j](p)}{p^2} \,.
\label{general_kernel_linearized}
\ee
Comparing this the general solution of Eq.~\eqref{fat_grav_eq} with a source, we identify the numerator 
as the Fourier-transform of that source. Transforming back to position space, it follows that the gravity source is given 
by the convolution of the two YM sources with a kernel defined by the matrix $h_{ij}$:
\be
\zeta_{\mu\nu}(x) = \int d^Dy d^D z \,  {\tilde h}(|x-y-z|)_{ij} \,  \zeta_\mu^i(y) {\tilde \zeta}_\mu^j(z) \,.
\ee
Such a relation between gauge and gravity sources was discussed in Ref.~\cite{Luna2016due} and is reminiscent of the 
off-shell definition of the linearized fat graviton in Eq.~\eqref{fatgraviton}.

Gauge transformations,  whose linearized form is $A_\mu^a\rightarrow A_\mu^a + \partial_\mu \chi^a$, can map
a solution such as \eqref{sol_zeta_YM} into one that has less straightforward identification of a momentum space 
``kinematic numerator''. 
To explore this possibility let us assume that $A_\mu^a(x)$ has the general form\footnote{Time-independent vector
potentials, of the form 
$ A_\mu^a(\vec x) =  \sum_i c^a_i \mu^i(\vec x)/|\vec x|$,
can be treated similarly. The apparent difference in the engineering dimension between the expression of $A_\mu^a(\vec x)$ here and that
in Eq.~\eqref{Ageneralform} stems from the difference in the dimension of the measure of the three-dimensional and four-dimensional (inverse) Fourier-transform operator.
}
\be
A^{(0)a}_\mu(x) = \frac{1}{x^2} \sum_i c^a_i n_\mu^i(x) \,,
\label{Ageneralform}
\ee
where $n(x)$ may contain terms that either eliminate the overall factor or introduce stronger singularities.
The Fourier transform of this vector potential can be defined formally as
\be
{\cal F}[A_\mu^{(0)a}](p) =  \sum_i c^a_i {\hat n}_\mu^i\!\left(i{\partial}/{\partial p}\right) {\cal F}\left[\frac{1}{x^2}\right](p) \,,
\ee
where the operators ${\hat n}_\mu^i\!\left(i{\partial}/{\partial p}\right)$ are obtained from $n_\mu^i(x) $ by the formal replacement 
$x^\mu \mapsto i\partial/\partial p_\mu$. 
This operation is to be understood in the sense of distributions, \ie the Fourier transform 
is taken in the presence of a test function that falls off sufficiently fast so integration by parts does not yield any boundary terms.
Interpreting the operators $n_\mu^i\!\left(i{\partial}/{\partial p}\right)$ as the kinematic numerators, the linearized double copy may 
be defined as\footnote{This construction may {\em in principle} be generalized to ${\hat n} = {\hat n}(p, i{\partial}/{\partial p})$. We leave this to the readers who read this footnote.}
\be
{\cal F}[H^{(0)}_{\mu\nu}](p) =  \sum_{ij}  
h_{ij} \,
{\hat n}_\mu^i\!\left(i{\partial}/{\partial p}\right)\, 
{\hat {\tilde n}}_\nu^j\!\left(i{\partial}/{\partial p}\right) \, 
{\cal F}\left[\frac{1}{x^2}\right](p) \,.
\ee
The commutation properties of the operators ${\hat n}$ and ${\hat {\tilde n}}$ together with the properties of $h_{ij}$ 
determine  whether or not this double copy yields a purely gravitational solution or the solution also contains nontrivial 
dilaton and/or anti-symmetric tensor. Fourier-transforming back to position space for a constant matrix $h_{ij}$ suggests
a (linearized) gravitational source (in de Donder gauge)
\begin{equation}
\zeta_{\mu\nu}(x) = \sum_{ij} h_{ij} n^i(x) {\tilde n}^j(x) \,.
\end{equation}
See Ref.~\cite{Luna2016due} for a further discussion on the relation of gauge and gravity sources in a time-dependent setting and
Ref.~\cite{CarrilloGonzalez2017iyj} for examples where symmetries help identify the appropriate sources.
The above construction is related to the position-space replacement
rules of Refs.~\cite{Monteiro2014cda, CarrilloGonzalez2017iyj}.

A non-dynamical source can also be interpreted in the spirit of a (spontaneous) breaking of the gauge 
group and thus apply the corresponding double-copy rules discussed in \sect{ZoologySection1} together with the fact 
that the linearized solution \eqref{sol_zeta_YM} is given by the first Feynman diagram in \fig{classical_sol_trees}. 
That is, the source is decomposed in irreducible representations of the unbroken 
(global part of the) gauge group and the double copy amounts to constructing gauge-invariant bilinears. 
This interpretation should also be subject to the consistency conditions discussed in \sect{ZoologySection1} regarding
the spectrum of the double-copy theory.

Ultimately gravitational sources should be dynamical (we shall review this in \sect{radiation}); as a step in this direction 
while eschewing the full dynamics of matter fields one may demand, as was done in Ref.~\cite{Goldberger2016iau}, that 
the gauge-theory source obeys covariant current conservation, 
\be
D^\mu \zeta_\mu^a = 0 \,.
\ee
Imposing it anticipates that $\zeta_\mu^a$ can be realized in terms of some other fields, in a gauge invariant Lagrangian
without settling on a specific realization.

The previous discussion and examples above refer to cases in which the sources of at least one of the two gauge 
theories are smooth functions, perhaps with compact support. If both momentum-space sources contain singular distributions 
their product requires a careful definition, especially if their product is ill-defined, such as a product of Dirac $\delta$-functions.
%
A physical perspective together with the expectation that there exists a Lagrangian that manifests the double-copy properties of 
Eq.~\eqref{SN=0} suggests a natural prescription.
Because the momenta of the two gauge theories are identified through the double copy, it is natural that constraints on 
it be imposed only once. Thus, if overlapping constraints are imposed by the gauge-theory sources, they should be included 
only once in the double copy of the source.
It is perhaps interesting that this prescription yields identical (linearized) gravity solutions from distinct (linearized) gauge-theory 
solutions---\eg two point-like sources present for all times {\em vs.} one point-like source present for all time and one instantaneous source.

To illustrate this, let us consider the field of a static point-like charge. The four-current is proportional to $u = (1,0,0,0)$
and the vector potential is~\cite{Luna2016hge}
\begin{equation}
A^{(0)a}_\mu(x) = g c^a u_\mu \frac{1}{4\pi r}\,,
 \hskip 1.5 cm 
A^{(0)a}_\mu(p) = g c^a u_\mu \frac{\delta^{(1)}(p^0)}{p^2} \,.
\label{sol_point_charge}
\end{equation}
Consequently, $H^{(0)\mu\nu}$ is
\begin{equation}
H^{(0)\mu\nu}(p) = \frac{\kappa}{2} M u^\mu u^\nu \frac{\delta^{(1)}(p^0)}{p^2} \,,
\label{H0_pp}
\end{equation}
which can be easily Fourier-transformed to position space. In writing this expression we made certain identifications between
the gauge coupling and constants in the gravity theory.
Since $H^{(0)\mu\nu}$ is symmetric, $b^{\mu\nu}=0$; it is not traceless, so there is a nontrivial dilaton
\be
\phi = H^{(0)}{}^\mu {}_\mu = +\frac{\kappa}{2} \frac{M}{4\pi r} \,.
\ee
Using Eq.~\eqref{projection_hbphi} and the projector \eqref{projector},
the correction to the metric is
\be
h^{\mu\nu} = \frac{\kappa}{2} \frac{M}{4\pi r} \left(u^\mu u^\nu + \frac{1}{2}(\eta^{\mu\nu} - q^\mu l^\nu - q^\nu l^\mu) \right) ,
\quad
\text{with}
\quad
l = \frac{1}{r+z} (0, x, y, r+z) 
\,.
\ee

Running a similar construction in the opposite direction, shock-wave
solutions of Einstein's equations which are also solutions of
linearized Einstein's equations were shown in
Ref.~\cite{Saotome2012vy} to be related, through a double-copy
procedure, to certain wave solutions of YM theory.  The relevant
gravitational source $\zeta_{\mu\nu}$ is identified such that the
scattering of some particle off a high-energy graviton is equivalent
to all orders in perturbation theory to the scattering off
$\zeta_{\mu\nu}$; the gravitational shock wave, given by the
Aichelberg-Sexl~\cite{Dray1984ha}, is the solution of (linearized)
Einstein's equation with this source.
The corresponding gauge-theory source $\zeta^a_\mu$ was similarly constructed, \ie such that the scattering of some particle 
off a high energy gluon is equivalent to all orders in perturbation theory  to the scattering off $\zeta^a_{\mu}$. The source turned 
out to be of the type \eqref{single} and the gauge-theory shock wave is the solution of (linearized) YM equations with 
this source.
The two waves are related by the usual color$\rightarrow$kinematics replacement. By construction, scattering 
off the gravitational wave can also be obtained though this replacement from  scattering 
off the gauge-theory wave, to all orders in perturbation theory.

\

\subsubsection{Nonlinear corrections \label{nonlinear_corrections}}

\

With a linearized solution in hand, nonlinear corrections can be computed directly, by evaluating increasingly higher
orders in the tree expansion in \fig{classical_sol_trees}. The goal however it to explore the realization of nonlinear corrections 
to the gravity solutions as a double copy of the nonlinear corrections to the YM solutions. 
We will review this here, loosely following Ref.~\cite{Luna2016hge}.
As we shall see, this comparison will emphasize the importance of the choice of fields, a feature that will be further discussed 
for complete solutions. 

Nonlinear corrections to a linearized solution are expressed, through the tree expansion in \fig{classical_sol_trees},
as convolutions of the linearized solution with kernels given by Feynman vertices. 
Since however, the gravity source depends on the metric it sources, one may either include explicitly such modifications 
(as ${\cal O}(\kappa^{n\ge 2})$  corrections to the source) or ignore them and obtain a solution for a choice of fields 
such that source changes are absent. 
These two perspectives have an analog in the two YM theories, where sources may either be corrected order by order in 
perturbation around the trivial solution such that they are \eg  covariantly constant, $D\cdot \zeta = 0$, or they are fixed,
respectively.

The first nonlinear correction to some solution $A^{(0)c}_\mu$ follows easily in terms of the standard three-point vertex.
To utilize the same rules as for amplitudes double copy, it is convenient to present it in momentum space:
\begin{multline}
A^{(1)a\mu}(-p_1) = \frac{i}{2p_1^2} f^{abc} \int \dd{p_2} \dd{p_3} \del{p_1 + p_2 + p_3} \\
\times \left[  (p_1 - p_2)^\gamma \eta^{\mu\beta} + (p_2 - p_3)^\mu \eta^{\beta\gamma} + (p_3 - p_1)^\beta \eta^{\gamma\mu} \right]  A^{(0)b}_\beta(p_2) A^{(0)c}_\gamma(p_3) \,.
\label{eq:correctionYM}
\end{multline}
The factor in the square parenthesis is the usual kinematic part of the off-shell three-gluon vertex and has the same antisymmetry properties as the color factor.  While this expression may be simplified somewhat by making use of the transversality of  $A^{(0)a\mu}$, we will choose not to do so.

Taking two configurations like \eqref{eq:correctionYM} and replacing the color factors of one with the kinematics of the other while leaving the propagators untouched (which, apart form using the same double-copy rules for amplitudes also includes the application of 
the results of the previous subsection  $A^{(0)a}_\mu(p) {\tilde A}{}^{(0)b}_\nu(p) \rightarrow H^{(0)}_{\mu\nu}(p)$) leads to
\begin{align}\!\!\!
H^{(1)\mu\mu'}(-p_1) = \frac{1}{4p_1^2} & \int \dd{p_2} \dd{p_3} \del{p_1 + p_2 + p_3} \nonumber\\
&\times \left[  (p_1 - p_2)^\gamma \eta^{\mu\beta} + (p_2 - p_3)^\mu \eta^{\beta\gamma} + (p_3 - p_1)^\beta \eta^{\gamma\mu} \vphantom{\eta^{\gamma'\mu'}}\right] \label{eq:H1general} \\
&\times \left[  (p_1 - p_2)^{\gamma'} \eta^{\mu'\beta'} + (p_2 - p_3)^{\mu'} \eta^{\beta'\gamma'} + (p_3 - p_1)^{\beta'} \eta^{\gamma'\mu'} \right]
H^{(0)}_{\beta\beta'}(p_2) H^{(0)}_{\gamma\gamma'}(p_3) . \nonumber
\end{align}
This expression has the same structure as the first nonlinear correction to the solutions of the equations of motion of the action
\eqref{SN=0} except that the trilinear interaction of gravitons, $B$ fields and dilatons was replaced by the factorized integrand kernel above.  This factorization is the same as that of the three-point amplitudes from Eq.~\eqref{SN=0}.
It can be seen explicitly by starting from the complete three-point vertices and using transversality {\em and} the on-shell condition 
for the external states. 
While  $H^{(0)}_{\gamma\gamma'}$ is transverse by construction, it obeys, in general, a free-field equation with a source. 
Thus, $H^{(1)\mu\mu'}(-p_1) $ given above represents the first correction to a gravity solution for the choice of a fluctuations 
such that the trilinear vertex is free of terms that vanish on the free mass shell.
This vertex is related to the one following from the expansion of the Lagrangian by a field redefinition.

There exists further freedom in the relation between $H^{(1)\mu\mu'}(-p_1)$ and the fluctuations  of the metric, $B$ field 
and dilaton. At the linearized level the later are given by the decomposition \eqref{fat}. For higher-order corrections however
this decomposition may be modified. 
As discussed in the beginning of this section, the amplitudes double copy guarantees only that the asymptotic 
states---or linearized solutions---double copy. At higher orders in $\kappa$  there may exists further terms in the relation between
gauge-theory and gravity fields which are projected out when the LSZ reduction is applied to a Green's function.
At the first nonlinear order this is
\be
H_{\mu\nu}^{(1)} = h_{\mu\nu}^{(1)}+B_{\mu\nu}^{(1)}+ P_{\mu\nu}^q \phi^{(1)} + {\cal T}_{\mu\nu}^{(1)}(h^{(0)}, b^{(0)}, \phi^{(0)})\,,
\ee
and at arbitrary order
\be
H_{\mu\nu}^{(n)} = h_{\mu\nu}^{(n)}+B_{\mu\nu}^{(n)}+ P_{\mu\nu}^q \phi^{(n)} + {\cal T}_{\mu\nu}^{(n)}(h^{(m)}, b^{(m)}, \phi^{(m)}, m<n) \,.
\ee
Such terms may be interpreted as field redefinitions connecting the initial choice of gravity fields \eqref{gravity_ini_fields} to the 
ones ``chosen'' by the double copy. They also capture various choices that can be made during the calculation, such as gauge
choices and---highlighted by their appearance in the first nonlinear correction---use of the free/lower order equations of motion 
in the definition of vertices.
Terms of this type may be eliminated by nontrivial choices of the kernel in Eq.~\eqref{general_kernel_linearized}.
These ``transformation functions''  \cite{Luna2016hge}  may be determined by comparing the perturbative solution of the 
equations of motion of the action \eqref{SN=0} with the result of the double copy. 
The main physical information they contain is that they provide the connection between the fields natural from a double-copy perspective and the natural fluctuations in the gravity Lagrangian. 
In the special case of the self-dual theory, it is known how to choose a parametrization of the metric perturbation such that the double copy is manifest~\cite{Monteiro2011pc}. For these field variables $\mathcal{T}_{\mu\nu} = 0$ to all orders in the tree diagram
expansion of self-dual spacetimes.

An example illustrating this discussion and dramatically emphasizing the relevance of the choice of field variables was given 
in Ref.~\cite{Luna2016hge} using the linearized gravity solution in Eq.~\eqref{H0_pp} and its gauge-theory counterpart in 
Eq.~\eqref{sol_point_charge}. 
This example also emphasizes the importance of {\em not} dropping
terms whose color factors vanish {\em after} summation over color
indices.
The  first nonlinear correction $H^{(1)}$ to Eq.~\eqref{H0_pp} was obtained in Ref.~\cite{Luna2016hge}; it is 
\begin{equation}
H^{(1)}_{\mu\nu}(x) = -\left(\frac{\kappa}{2}\right)^2 \frac{M^2}{4(4\pi r)^2} \hat{r}_\mu \hat{r}_\nu \,,
\label{eq:fatJNWH1}
\end{equation}
where $\hat{r}_\mu = (0, \mathbf{x}/r)$.
It turns out that a nontrivial transformation function is necessary to
turn $H = \eta + \kappa H^{(0)} + \kappa^2 H^{(1)}$ into a solution of
the equations of motion to ${\cal O}(\kappa^2)$ in the
variables \eqref{gravity_ini_fields}.
It is given by~\cite{Luna2016hge} 
\begin{equation}
\begin{split}
\mathcal{T}^{(1)\mu\nu}(-p_1)
 = \int& \dd{p_2}\dd{p_3} \del{p_1+p_2+p_3}
\frac{1}{4p_1^2}
\bigg\{
   H_{2\, \alpha\beta}^{(0)} H_3^{(0)\alpha\beta} p_1^{\mu}p_1^{\nu}
 + 8 p_{2}^{\alpha} H^{(0)}_{3\, \alpha\beta}
     H^{(0)\beta(\mu}_{2} p_1^{\nu)} \\ &
 + 8 p_2 \cdot p_3\,H_2^{(0)\mu\alpha} H^{(0)\nu}_{3\, ~~\alpha}
 - 2\eta^{\mu\nu} p_2\cdot p_3\,H_{2\, \alpha\beta}^{(0)} H_3^{(0)\alpha\beta}
 + 4\eta^{\mu\nu} p_2^{\alpha} H^{(0)}_{3\, \alpha\beta}
   H^{(0)\beta\gamma}_{2} p_{3\gamma} \\ &
 + P_q^{\mu\nu} \left[
   2(D-6) p_2\cdot p_3\,H_{2\, \alpha\beta}^{(0)} H_3^{(0)\alpha\beta}
 - 4(D-2) p_2^{\alpha} H^{(0)}_{3\, \alpha\beta}
   H^{(0)\beta\gamma}_{2} p_{3\gamma} \right]\!
\bigg\} \,,
\end{split}
\end{equation}
where we used the shorthand notation
\be
H^{(0)}_{i\, \mu\nu}\equiv  H^{(0)}_{\mu\nu}(p_i)\,,
\quad
\text{and}
\quad
p^{(\mu}q^{\nu)}\equiv \frac{1}{2}(p^\mu q^\nu + p^\nu q^\mu) \,.
\ee
This first transformation function $\mathcal{T}^{(1)\mu\nu}$ holds for all cases that have symmetric and 
transverse $H^{(0)}_{\mu\nu}$ and $h^{(0)}_{\mu\nu}$.

Since \eqn{sol_point_charge} is an exact solution of the YM equations of motion, one may wonder whether it 
is possible that it has some other, physically equivalent form which can be double-copied to an exact solution of dilaton-axion-gravity
in some field variables. To this end, it is necessary that the first correction to this equivalent form of \eqn{sol_point_charge} 
vanishes before summation over color indices. We shall see in \sect{KS_sol_sect} that this is indeed possible.

Proceeding to higher orders is in principle straightforward, but quite tedious in practice. 
The new features compared to the discussion above relates to the need of a 
representation of the  corrections to the YM equations which manifest \ck duality up to terms that are projected 
out by the LSZ reduction. Since the only difference between the asymptotic states of scattering amplitudes and $A^{a(0)}_\mu$ 
is that the latter obey an on-shell condition with sources, \ck duality can be satisfied only up to such terms.
Similarly to scattering amplitudes, a generic perturbative classical solution is related to one that exhibits the duality (in this 
restricted sense)  by generalized gauge transformations. As in that case, such transformations are not always easy to find.
As in that case, a Lagrangian whose Feynman rules lead to manifestly \ckDash-dual representation or the use of the generalized double-copy construction can alleviate this issue.

To quintic order in fields, the Lagrangian in Ref.~\cite{Square} provides the requisite Feynman rules to obtain the gauge-theory perturbative classical solution in a form that can be double copied directly. This was exploited in Ref.~\cite{Luna2016hge}, where the second nonlinear correction was discussed. As explained there, the quartic YM vertex does not contribute to a symmetric double copy and the second term in the perturbative solution of YM equations is given entirely in terms of the three-point vertex:
\begin{eqnarray}
\label{eq:NLOcorrectionYM}
A^{(2)a\mu}(-p_1) &=& \frac{i}{p_1^2} f^{abc} \int \dd p_2 \dd p_3 \del{p_1 + p_2 + p_3} \\
&&\times \left[  (p_1 - p_2)^\gamma \eta^{\mu\beta} + (p_2 - p_3)^\mu \eta^{\beta\gamma} + (p_3 - p_1)^\beta \eta^{\gamma\mu} \right]  A^{(0)b}_\beta(p_2) A^{(1)c}_\gamma(p_3) \,.
\nonumber
\end{eqnarray}
It leads to the  second correction $H^{(2)}$ in the gravitational solution 
\begin{align}
H^{(2)\mu\mu'}(-p_1) = &\,\frac{1}{2p_1^2} \int \dd p_2 \dd p_3 \del{p_1 + p_2 + p_3} \nonumber \\
&\times \left[  (p_1 - p_2)^\gamma \eta^{\mu\beta} + (p_2 - p_3)^\mu \eta^{\beta\gamma} + (p_3 - p_1)^\beta \eta^{\gamma\mu} \vphantom{\eta^{\gamma'\mu'}}   \right] \label{eq:H2} \\
&\times \left[  (p_1 - p_2)^{\gamma'} \eta^{\mu'\beta'} + (p_2 - p_3)^{\mu'} \eta^{\beta'\gamma'} + (p_3 - p_1)^{\beta'} \eta^{\gamma'\mu'} \right]
H^{(0)}_{\beta\beta'}(p_2) H^{(1)}_{\gamma\gamma'}(p_3) . \nonumber
\end{align}
The graviton, antisymmetric tensor and dilaton components can be easily extracted using the projectors; to connect 
this general expression to a solution with specific sources in specific coordinates ${\cal T}^{(2)}$ must be computed 
as well. We refer to Ref.~\cite{Luna2016hge} for details.

\begin{homework}
Explore the possibility of using a quasi-classical solution obtained by folding scattering amplitudes in BCJ representation against 
external sources to construct solutions for the gravity field equations. This is equivalent to removing terms proportional to the free-field equations from Green's functions and using the result to construct an ansatz for a classical solution. 
The resulting double-copy field configuration should be correct---for
some choice of field variables---up to terms that are proportional to
the free field equations, \ie up to field redefinitions.
\end{homework}

Steps towards the double copy of nonlinear classical solutions beyond
second order were taken in
Refs.~\cite{Mizera:2018jbh,SchlottererBGCurrent} for the special case
of perturbiners or Berends-Giele currents. Starting from the
perturbiners of certain effective field theories~\cite{Mizera:2018jbh}
and $F^3$ and $F^4$-deformed YM theory, perturbiners of the
corresponding gravity theories were constructed using the KLT
relations. Because only one leg of the Berends-Giele current is off
shell, the relation between the objects thus constructed and
the ``true'' gravitational perturbiner is simpler than in the most
general case: it consists only of a gauge transformation and involves
no field redefinition.

The need for a Lagrangian yielding \ckDash-satisfying Feynman  rules or, more generally, of Green's functions manifesting \ck duality on all of their internal lines may be circumvented through the generalized double-copy construction discussed in \sect{GeneralizedDoubleCopySection}. Generalizing slightly to Green's functions, the starting point is any general perturbative expressions for the gauge-theory solutions expressed in terms of cubic diagrams; quartic vertices, if present, are resolved in the usual way. 
Because of lack of manifest \ck duality, their double copy does not yield solutions of the equations of \eqref{SN=0} up to field redefinitions. The formulae discussed in \sect{GeneralizedDoubleCopySection} provide the correction terms.
As in the examples discussed earlier in this section, transformation functions are { probably} necessary to relate the result of the generalized double copy to a solution in some chosen coordinates. It remains an open problem to have an { a priori} understanding 
of the choice of fields in the gravitational theory that set all transformation functions to zero.

\subsection{Complete solutions; Kerr-Schild coordinates \label{KS_sol_sect}}

In the discussion of perturbative construction of gravity solutions
in \sect{perturbative_st} we encountered, following
Ref.~\cite{Luna2016hge}, linearized solutions which are exact
solutions of YM equations---such as that in
Eq.~\eqref{sol_point_charge}---which double copy to linearized
solutions of gravity which receive higher-order corrections.
While, as emphasized there, this can be understood as a consequence of
the special properties of the color factors of the YM solution, it is
important to understand whether there exists a choice of field
variables for which these contributions to not arise at all and
consequently the transformation functions vanish identically to all
orders in classical perturbation theory.
The general expectation is that if a gauge-theory solution does not receive corrections beyond $n$-th order in perturbation 
theory, then its corresponding gravity solution will also be exact beyond that order.

As pointed out in Ref.~\cite{Monteiro2014cda}, following
Ref.~\cite{Stephani2003tm}, a particular ansatz for the metric
linearizes the source-free Einstein's equations and thus can
potentially give these metrics as double copies of solutions of YM
equations which do not receive nonlinear corrections.
They are know as Kerr-Schild metrics; the ansatz is given in terms of
a scalar function $\phi$ (which is {\em not} the dilaton) and a vector
$k$ which is null and geodesic with respect to the background metric $\bar{g}_{\mu\nu}$:
\begin{equation}
g_{\mu\nu}=\bar{g}_{\mu\nu}+\kappa h_{\mu\nu} \equiv \bar{g}_{\mu\nu}+\kappa\,\phi\, k_\mu\, k_\nu
\,, \qquad
\bar{g}_{\mu\nu}\, k^{\mu}\, k^\nu=0\,,
\qquad (k\cdot {\bar \nabla})\, k_\mu=0 \, .
\label{KSdef}
\end{equation}
The background (or fiducial) metric  $\bar{g}_{\mu\nu}$ is also used to raise and lower indices on the metric fluctuation $h$ and ${\bar \nabla}^\mu$ is 
the corresponding background-covariant derivative.
One component of $k$ can be set to unity, thus absorbing its dynamics in $\phi$.
The Kerr-Schild form is special in that the metric perturbation---or the graviton---explicitly decomposes into a direct product of 
the vector $k_\mu$ with itself.
The remarkable property of this ansatz is that it linearizes the Ricci tensor and reduces Einstein's equations to a single nontrivial relation
between the function $\phi$ and the source. The components of the Ricci tensor are
\begin{equation}
R^\mu{}_\nu=\bar{R}^\mu{}_\nu + \kappa \left[-h^{\mu}{}_\rho\bar{R}^\rho{}_\nu
+\frac{1}{2} {\bar \nabla}_\rho\left({\bar \nabla}_\nu h^{\mu\rho}+{\bar \nabla}^\mu h^\rho{}_\nu
-{\bar \nabla}^\rho h^\mu{}_\nu\right)\right],
\label{Ricci}
\end{equation}
where $\bar{R}{}^\mu{}_{\nu}$ is the Ricci tensor associated with the background metric $\bar{g}_{\mu\nu}$. We emphasize  that the linear 
dependence on the metric fluctuation $h$ in Eq.~\eqref{KSdef} holds only for the index positions in Eq.~\eqref{Ricci}.

A simple choice of background metric is $\bar{g}_{\mu\nu}=\eta_{\mu\nu}$ (with a mostly-minus signature), used at length in this context in Ref.~\cite{Monteiro2014cda}.
For this choice the background-covariant derivatives become regular derivatives. Further choosing $k^0 = 1$, the components of the 
Ricci tensor are
\begin{align}
R^{0}{}_{0}& = \frac{1}{2}\partial^i\partial_i \phi \,, \nonumber \\
R^{i}{}_0 & = -\frac{1}{2}\partial_j \left[\partial^i\left(\phi k^j\right)
   - \partial^j\left(\phi k^i\right)\right], \nonumber \\
R^i{}_j & = \frac{1}{2}\partial_l\left[\partial^i\left(\phi k^l k_j\right)
   + \partial_j\left(\phi k^l k^i\right)-\partial^l\left(\phi k^i k_j\right)\right],
\nonumber \\
R&=\partial_i\partial_j\left(\phi k^i k^j\right) .
\label{Rval}
\end{align}
All Latin indices run over the space-like directions. Thus, the scalar function $\phi$ is determined by a Poisson-type equation.

A generalization of the Kerr-Schild ansatz in Eq.~\eqref{KSdef} is the
double-Kerr-Schild ansatz~\cite{Chong2004hw}, which is given in terms
of two scalar functions and two null, geodesic and mutually orthogonal
vectors:
\begin{align}
g_{\mu\nu}&=\bar{g}_{\mu\nu}+\kappa h_{\mu\nu} =\bar{g}_{\mu\nu}+\kappa\left(\phi\,k_\mu\,k_\nu+\psi\,l_\mu\,l_\nu \right) \,,
\label{doubleKS}
\\
k^2&=l^2=k\cdot l=0 \,, \hskip 1 cm (k\cdot \bar\nabla)k_\mu=0\,,
\hskip 1 cm (l\cdot \bar\nabla)l_\mu=0 \,,
\nn
\end{align}
where as before $\bar{g}_{\mu\nu}$ is a background metric and is used in all index contractions and background-covariant derivatives $\bar\nabla$. 
For these field variables the Ricci tensor is
\begin{align}
R^\mu{}_\nu&=\bar{R}^\mu{}_\nu+\kappa\left[-h^\mu_\rho\bar{R}^\rho{}_\nu
+\frac{1}{2} \bar\nabla_\rho\left(\bar\nabla_\nu h^{\mu\rho}+\bar\nabla^\mu h_\nu^\rho-\bar\nabla^\rho 
h^\mu{}_\nu\right)\right]+R^\mu{}_{\nu, {\rm non-lin.}} \,,
\label{RiccidoubleKS}
\\
R^\mu{}_{\nu, {\rm non-lin.}}&=-\frac{\kappa^2}{2}\left[\frac{1}{2}
\bar\nabla^\mu h(k)^\rho{}_\delta \bar\nabla_\nu h(l)^\delta{}_\rho
+h(l)^{\mu\delta}\bar\nabla_\rho \bar\nabla_\nu h(k)^\rho{}_\delta \right.    \notag\\
&\left.\phantom{\frac{1}{2}}+\bar\nabla_\rho\left(
h(l)^{\rho\delta}\bar\nabla_{\delta}h(k)^{\mu}{}_\nu+2h(l)^{\rho}{}_{\delta}\bar\nabla_{(\nu}h(k)^{\mu)}{}_\delta-2h(l)^{\mu\delta}
\bar\nabla^{[\rho}h(k)^{\delta]}{}_{\nu}\right)\right]+(k\leftrightarrow l),
\label{Rnonlin}
\end{align}
where
\begin{equation}
h(k)_{\mu\nu}=\phi k_\mu k_\nu,\qquad h(l)_{\mu\nu}=\psi l_\mu l_\nu \,.
\label{hkhl}
\end{equation}
The linearity of Einstein's equations in Kerr-Schild variables implies that any single Kerr-Schild metric can 
also be thought of as a double Kerr-Schild metric.

In higher dimensions further generalizations are possible, involving up to $D-2$ null, geodesic and mutually orthogonal 
vectors with the same properties as $k$ and $l$. 
Additionally, it was argued in Ref.~\cite{Chong2004hw} that in the so-called Plebansky coordinates, the nonlinear part of the Ricci tensor, $R^\mu_{\nu, {\rm non-lin.}}$, vanishes identically and solutions of the linearized Einstein's equations are  also exact solutions.

\subsubsection{Kerr-Schild exact solutions \label{KS_sol_sectExact}}

In this section we shall review the double-copy interpretation of the Schwarzschild solution, emphasizing its realization vis-\`a-vis
the discussion in the previous section. We will then summarize and comment on generalizations of this approach to other spacetimes.

The Kerr-Schild form of the Schwarzschild solution is (see e.g. Ref.~\cite{Stephani1982ac})
\begin{equation}
g_{\mu\nu} = \eta_{\mu\nu}+\frac{\kappa^2}{8\pi}\frac{M}{ r}k_\mu k_\nu \,,
\label{KS_Sch_metric}
\end{equation}
where 
$M$ is positive and the null four-vector $k$ is chosen such that the line element is rotationally invariant:
\begin{equation}
k^\mu = (1, x^i/r) \,,
\qquad
r^2 =\sum_{i=1}^3 x^i x_i \,.
\end{equation}  
The double-copy form of this solution was discussed at length in Ref.~\cite{Monteiro2014cda}. 
With the definition of the linearized double copy for singular sources discussed in \sect{linearized_source_discussion} 
and up to identification of parameters, it can be seen that the departure of the metric \eqref{KS_Sch_metric} from 
Minkowski space is given by the double copy of 
\begin{equation}
A_\mu^a = \frac{g c^a  k_\mu}{4\pi r} \,.
\label{vector_potential}
\end{equation}
It can be straightforwardly verified that $A_\mu^a$ satisfies
Maxwell's equations with a point-like source at the origin.  
Eq.~\eqref{vector_potential}, however, is not the standard potential of such a source. Rather, as shown in Ref.~\cite{Monteiro2014cda}, 
it is related to the standard potential of a point-like charge \eqref{vector_potential} by a gauge transformation with parameter $\Lambda^a$:
\begin{align}
A{}_\mu^a & = A'{}_\mu^a + \partial_\mu \Lambda^a \,, \nonumber \\
A'{}_\mu^a & = \frac{g c^a  u_\mu }{4\pi r} \,,
\qquad u_\mu = (1, 0, 0, 0) \,, 
\qquad
\Lambda^a = \frac{g c^a}{8\pi} \log r^2 \,, \\
j_\mu^a & = -g c^a u_\mu \delta^{(3)}({\bf x}) \,. 
\nonumber
\end{align}

It is interesting to contrast the two gauge-equivalent vector
potentials $A{}_\mu^a$ and $A'{}_\mu^a$. Both are proportional to a
single color vector $c^a$ and because of this they both are formally
exact solutions of the nonlinear YM equations.  For the latter one,
$A'$, the vanishing color factors of the corrections are multiplied by
nontrivial kinematic dependence and thus, as discussed
in \sect{nonlinear_corrections}, there are nonlinear corrections that
should be included which are crucial for transforming its double copy
into a solution of the full Einstein's equations.
For the former, $A$, one can check that the vanishing color factors come together with {vanishing} kinematic 
dependence.  Therefore, the corrections to the double copy of two $A$ vectors also vanish and thus the 
linearized double copy does not receive nonlinear corrections.
This underscores the importance of the gauge choice for the gauge-theory solutions that participate in the classical double copy. 

To recover the Schwarzschild solution the parameters of the two theories are replaced as
\begin{equation}
\frac{\kappa}{2}\leftrightarrow g\,, \qquad M\leftrightarrow | c | \,.
\label{replace2}
\end{equation}
We note that the norm of the color vector $c^a$, which may be identified as the charge of the source under the 
sole Cartan generator of the gauge group that is nontrivial,  corresponds to the mass of the Schwarzschild black hole.
This seems to suggest a relation between the uniqueness of the
Coulomb-like solution and Birkhoff's theorem.

It is interesting and important to note that, despite the gauge-theory
solutions being sourced by the same charge distributions and being
gauge-equivalent, their classical double copies as defined here are
inequivalent. Indeed, while the solution constructed in this section
has only a nontrivial metric, the one constructed perturbatively
in \sect{nonlinear_corrections} stating from
Eq.~\eqref{sol_point_charge} also has a nontrivial dilaton which
cannot be removed while preserving a nontrivial metric.
With the current understanding of the classical double copy, the fact
that gauge-equivalent gauge-field configurations lead to inequivalent
gravitational field configurations appears to be an unavoidable
feature.  At this juncture, it seems best to start with 
valid gravitational solutions and work backwards to gauge theory.

Considerations similar to the ones outlined above have been used to give a double-copy interpretation to the Kerr back hole, 
black brane solutions, shock-wave and plane-wave solutions~\cite{Monteiro2014cda}  and to the 
(anti) de Sitter spaces in Ref.~\cite{Luna2015paa}. In the latter cases the cosmological constant is related to the charge density 
of a uniform charge  distribution.
Gravity solutions with additional matter fields turned on, such as the Taub-NUT space, have a double Kerr-Schild form and, as 
argued in Ref.~\cite{Luna2015paa}, have a double-copy interpretation (in the same sense as discussed above) 
in terms of a dyon solution whose electric and magnetic charges are related to the mass and the NUT charge. 

Kerr-Schild solutions with time-dependent sources, describing accelerating black holes, have been discussed 
in Ref.~\cite{Luna2016due} where a relation was constructed between the electromagnetic radiation of an accelerating charge 
and the gravitational radiation of an accelerating point mass and thus represents an effective description of the complete 
vacuum solution.  The contraction of the corresponding sources with gluon and graviton polarization vector/tensor gives the
amplitude for the Bremsstrahlung process. Other gravitational wave solutions, including vacuum solutions, which are of the
double-Kerr-Schild type, were discussed in Ref.~\cite{Luna:2018dpt}.

All YM solutions that appeared in the constructions reviewed here are
also solutions of Maxwell's equations. Using the fact, discussed
in \sect{ZoologySection1}, that YM theory can be interpreted as a
double copy of itself with a theory of a bi-adjoint scalar field, more
complicated solutions can be constructed by taking the double copy of \eg a
Maxwell solution with a solution of the bi-adjoint scalar theory. This
observation was explored in Refs.~\cite{Luna2015paa,
BahjatAbbas2017htu}, while solutions of bi-adjoint scalar theory were
constructed in Refs.~\cite{White2016jzc,Bahjat-Abbas:2018vgo} and \cite{DeSmet2017rve}. The
details pertaining to the relation between the sources of various
solutions remain to be fully worked out.  This perspective also makes
contact with the off-shell Lagrangian double copy of
Refs.~\cite{Cardoso2016amd, Cardoso2016ngt}.

Following Ref.~\cite{Ridgway2015fdl}, the gravitational stress tensor
of certain double-copy Kerr-Schild solutions was expressed linearly in
terms of the current sourcing the gauge-theory solution. With this
relation, in most cases they are not stress-energy tensor of a
perfect fluid and contains shear stresses and, moreover, they do not
obey the weak-energy condition.
It is possible that other choices of coordinates and field
variables display double-copy behavior that simultaneously map YM
solutions to gravitational ones and satisfy the energy conditions.

Further generalizations, involving a nontrivial fiducial metric ${\bar g}$ in the Kerr-Schild ansatz, were discussed in 
Refs.~\cite{Luna2015paa, CarrilloGonzalez2017iyj, BahjatAbbas2017htu}.
As discussed in Refs.~\cite{Luna2015paa, BahjatAbbas2017htu}, if the fiducial metric 
is of Kerr-Schild type, then every such solution can also be interpreted as a (multiple) Kerr-Schild metric with Minkowski space 
as fiducial metric (referred to as Type A constructions in Ref.~\cite{BahjatAbbas2017htu}). Among the examples discussed are the 
de Sitter and anti de Sitter generalizations of the Schwarzschild black hole. 

Solutions with a non-Kerr-Schild background metric have been discussed in Ref.~\cite{BahjatAbbas2017htu} (referred to there 
as Type B constructions) and in Ref.~\cite{CarrilloGonzalez2017iyj}. They are realized in terms of solutions of gauge theory on a 
space with the fiducial metric.
The classical scale invariance of YM theories implies that, for a fiducial metric is conformally Minkowski, the gauge theory  
is effectively in flat space (up to a curvature-dependent scalar mass term). Examples of this type were discussed in Ref.~\cite{BahjatAbbas2017htu}.
Apart from black holes in asymptotically maximally-symmetric spaces which are also treated in this framework, Ref.~\cite{CarrilloGonzalez2017iyj}  also gives double-copy interpretations to black strings, black branes, and various types 
of gravitational waves.  
The corresponding localized sources for the YM and scalar theories, for both  stationary and time-dependent examples, 
are also identified and examples are given in terms of Kerr-Schild vectors $k$.

While a coherent picture for the classical double copy of exact gauge-theory solutions to exact (matter-coupled) gravity solutions
is still to be formulated, the examples discussed in the literature and summarized here give hope that such a relation may be 
generically possible.

\subsubsection{Good and bad coordinates: Charged black holes from higher dimensions}

To further illustrate the importance of the choice of field variables for the interpretation of the result of classical double-copy 
constructions as exact solutions of Einstein's equations (perhaps coupled to additional matter) let us briefly discuss the charged 
black hole solution in the presence of an additional scalar field. (See also Ref.~\cite{BahjatAbbas2017htu} for a discussion of charged black holes).
The equations of motion are standard\footnote{The corresponding action
is in the string frame, and may be mapped to the Einstein frame by a
rescaling of the five-dimensional metric.}
\begin{align}
R_{\mu\nu}-\frac{1}{2}g_{\mu\nu}R &= \frac{1}{2}\phi^2 (g^{\alpha\beta} F_{\mu\alpha}F_{\nu\beta} - \frac{1}{4}g_{\mu\nu}F\cdot F)
+\frac{1}{\phi} (\nabla_\mu\nabla_\nu \phi - g_{\mu\nu}\nabla\cdot\nabla \phi) \, ,
\nonumber \\
\nabla\cdot\nabla \phi &= \frac{1}{4}\phi^3 F\cdot F \, ,
\nonumber\\
\nabla^{\alpha}F_{\alpha\mu} &= -3 \frac{\nabla^{\alpha} \phi}{\phi} F_{\alpha\mu} \,, 
\label{KKeqs}
\end{align}
and can be obtained by Kaluza-Klein reduction from Einstein's equations in five dimensions through the usual ansatz
\be
g_5 = 
\begin{pmatrix}
g_4 + \phi^2 A\otimes A & \phi^2 A \cr
\phi^2 A & \phi^2
\end{pmatrix}  .
\label{KKansatz}
\ee
As we shall see, in these field variables the charged black hole solution does not a clear classical double-copy interpretation; we 
will identify the field variables in which it does, paralleling the smooth relation of double copy between theories related by dimensional reduction.

The four-dimensional charged black hole can be obtained via Kaluza-Klein reduction from a five-dimensional black string.
In Kerr-Schild form, it is 
\be
g = \eta + \varphi {\hat k}\otimes {\hat k} \,, 
\qquad
\eta^{\mu\nu} {\hat k}_\mu {\hat k}_\nu = 0 \,, 
\qquad
\varphi = \frac{M}{r_3} \,,
\ee
where $r_3$ is the radial coordinate in the three coordinates transverse to the string. A suitable solution of the constraints 
constants defining ${\hat k}$ is that it is a boost of the vector $(1, {\hat r}{}_3, 0)$ where $ {\hat r}_3$ is 
the unit vector in three dimensions orthogonal to the string:
\be
{\hat k} = (\gamma, {\hat r}_3, \beta\gamma) \equiv (k, \beta\gamma) \,,
\hskip 1 cm \gamma^2 = \frac{1}{1-\beta^2}  \,.
\ee
Using the reduction ansatz in Eq.~\eqref{KKansatz}, the four-dimensional fields are:
\begin{align}
g_4 &= \eta + \frac{\varphi}{1+\beta^2\gamma^2 \varphi}k\otimes k \,,\qquad k = (\gamma, {\hat r}_3)\, ,\qquad k^2 = 1-\gamma^2 = -\frac{\beta^2}{1-\beta^2}\,,
\nonumber \\
\phi &=\sqrt{1+\beta^2\gamma^2 \varphi}\,,
\qquad\qquad
A =  \frac{\beta\gamma\, \varphi}{1+\beta^2\gamma^2 \varphi}k \,.
\end{align}
It is not difficult to check that this field configuration is a solution of Eqs.~\eqref{KKeqs}. 

It is also not difficult to see that this field configuration departs from the Kerr-Schild ansatz in that the vector $k$ defining 
the departure  of the metric from Minkowski space is time-like rather than null. Moreover, the dependence on $\varphi$
suggests that all fields are given by a nontrivial resummation of tree diagrams. 

Another choice of field variables, 
\begin{equation}
g_5 = 
\begin{pmatrix}
\tilde g_4 &  \tilde A \cr
\tilde A & 1+ \tilde \phi
\end{pmatrix} \,,
\label{betteransatz}
\end{equation}
which is closely related to the dimensional reduction of asymptotic states of scattering amplitudes, is more suitable for a classical double-copy interpretation. 
Indeed, the four-dimensional fields are
\begin{eqnarray}
\tilde g_4 &=& \eta +{\varphi}k\otimes k\,, \qquad k = (\gamma, {\hat r}_3)\,, \qquad k^2 = 1-\gamma^2 = -\frac{\beta^2}{1-\beta^2}\,,
\nonumber\\
\tilde \phi &=& \beta^2\gamma^2 \varphi\,,
\qquad\qquad
\tilde A =  \varphi k \, ,
\label{good_coords}
\end{eqnarray}
which are related in the sense described in \sect{linearized_source_discussion} to the following solution of 
gauge theory coupled to a scalar field:
\begin{eqnarray}
A^a_\text{YM} = \varphi k \, c^a\,, 
\hskip 1.5 cm
\phi^a_\text{YM} =  {\cal N} \varphi \, c^a \, ,
\label{gt_sol}
\end{eqnarray}
where $c^a$ is some color vector.

We note that the field configuration \eqref{good_coords} is not a solution of Eq.~\eqref{KKeqs}, but it is a 
solutions of the equations obtained from them through the field redefinition mapping the fields in Eq.~\eqref{KKansatz} 
to those in Eq.~\eqref{betteransatz}. 
Moreover, while structurally similar to the Kerr-Schild ansatz, it is
not of the same type because the vector $k$ is time-like.  The
violation of the null condition is compensated by the contribution of
the vector and scalar fields.
While the relation between \eqref{gt_sol} and \eqref{good_coords} is linear, the fact that $k$ is not null allows in
principle for a nonvanishing kinematic part in the nonlinear corrections to \eqref{gt_sol} and thus to potential nonlinear 
corrections to their double copy, cf.~\sect{nonlinear_corrections}.
The fact that \eqref{good_coords} is an exact solution suggests absence of the nonlinear corrections to the scalar and 
vector fields in Eq.~\eqref{gt_sol}.
These features may allow further generalization of the classical double-copy interpretation of solutions of Kerr-Schild type.
An alternative construction of the charged dilatonic black hole solution discussed here, which uses the standard four-dimensional equations of motion \eqref{KKeqs} and a generalization of the double-Kerr-Schild ansatz \eqref{doubleKS} which also includes certain internal dimensions, was discussed in \cite{Cho:2019ype}.

\begin{homework}
Show that the first nonlinear correction to \eqref{good_coords} vanishes by evaluating it in terms of the kinematic factors of 
the corrections to the gauge-theory solution \eqref{gt_sol}.
\end{homework}

\subsection{Radiation \label{radiation}}

Earlier in this section we have reviewed and illustrated various possible definitions of the double copy of classical solutions
gauge theories to solutions of Einstein's equations coupled perhaps with additional matter and summarized the existing results.
One of the fundamental results of general relativity (and, in fact, of
any gravity theory) is the emission of gravitational waves---classical
gravitational radiation emitted in processes involving massive
astrophysical bodies such as neutron stars or black holes, perhaps
with macroscopic intrinsic angular momentum.

From the perspective of general relativity such calculations have a
long history, with numerical and perturbative results in various
approximation schemes, which we will not review here; see
Refs.~\cite{BlanchetReview, Buonanno:2014aza, Barack:2018yly} for
reviews.  They have been stunningly confirmed through the direct
experimental detection of gravitational waves by the LIGO and Virgo
collaborations \cite{Abbott2016blz}.
We expect that the double-copy approach to such calculations will lead
to important technical simplifications and bring new insight into these
problems.

The first nontrivial contribution to the radiation process involves five
particles: the two incoming and outgoing massive bodies and the
outgoing graviton. To evaluate this, it is necessary to fix a model for the
massive bodies that can be included in the double-copy construction.
In Ref.~\cite{Goldberger2016iau} they were represented in terms of
gauge fields, effectively as the linearized solutions in
Eq.~\eqref{sol_point_charge}. The classical double copy then
(effectively) yields a gravity solution whose linearized form
is \eqref{H0_pp} and thus the massive objects being scattered source
both gravitons and dilatons. It moreover appears that the double-copy
rules used in Ref.~\cite{Goldberger2016iau} assume that a Lagrangian
that manifests \ck duality is available. Indeed, the gauge-group
generators are replaced with the kinematic dependence of the off-shell
three-point vertex, thus assuming that the latter have the same
algebraic properties as the former.  While for a Lagrangian that
manifests \ck duality this replacement is, of course, equivalent to
the usual rules in \sect{DualitySection}, for a general Lagrangian,
however, further terms may be necessary.
Nevertheless, the result reproduces direct calculations \cite{Goldberger2016iau, Goldberger2017frp} in dilaton-coupled gravity.
Similar techniques have been used to obtain the corresponding results in Einstein-Yang-Mills theory~\cite{Chester2017vcz}.

In a different approach, suggested in Ref.~\cite{Luna2017dtq} based on earlier ideas of Refs.~\cite{Laenen2008gt, White2011yy}, 
incoming and outgoing spinless massive bodies are represented as double copies of minimally-coupled massive scalar $\Phi$.
The Lagrangian is 
\begin{equation}
\mathcal{L} = -\frac12 {\rm Tr} F^{\mu\nu} F_{\mu\nu} + \sum_i \left[(D_\mu \Phi_i)^\dagger (D^\mu \Phi_i) - m^2_i |\Phi|^2\right] \,,
\label{Lagrangian_YM_Phi}
\end{equation}
with the scalar field in some (complex) representation of the gauge group and $D_\mu$ the corresponding covariant derivative.
Then, a certain classical limit is taken to ensure that, as for
classical particles, masses are parametrically larger than their
spatial momenta. In this approach one can choose the couplings of
these particles such that, on the one hand, \ck duality is present and
on the other their double copy does not yield a dilaton source.
Due to Birkhoff's theorem, this model is sufficient describe the
gravitational wave emission far from the horizon of black holes, where
the large (classical) masses ensures that the linearized emission is
captured accurately. We outline the relevant calculation, following
Ref.~\cite{Luna2017dtq}.

\begin{figure}
\centering
	\begin{subfigure}[t]{0.3 \textwidth}
		\includegraphics[width=1.\textwidth]{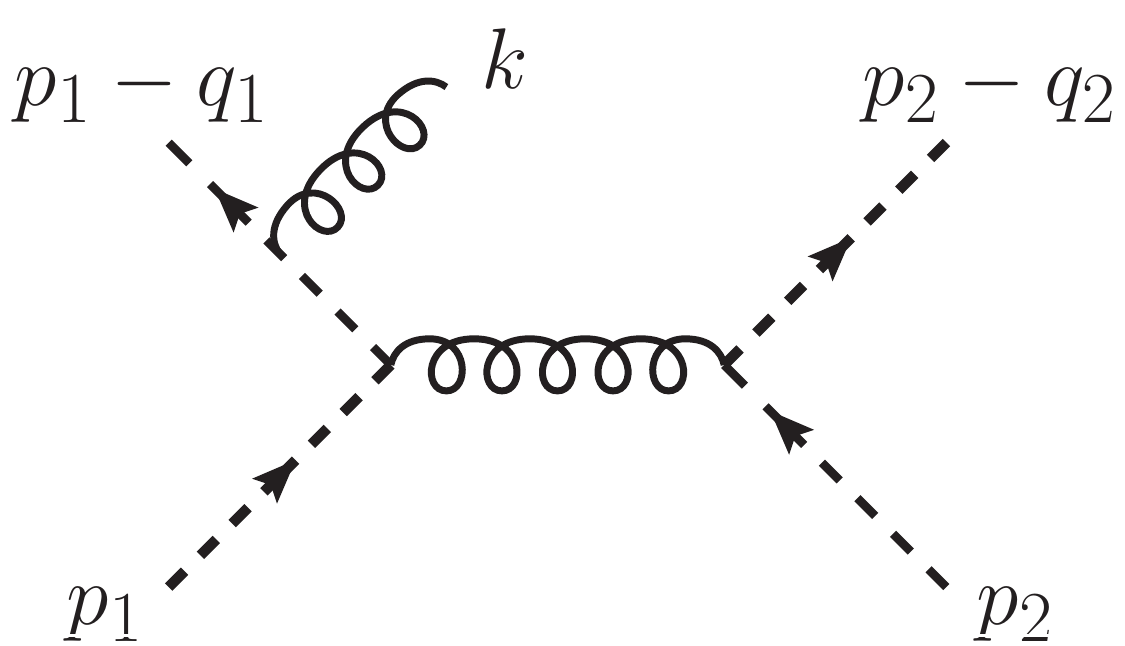}
		\caption*{(a)}
		\label{fig:scalarDiagramsA}
	\end{subfigure}
\hskip .5 cm 
	\begin{subfigure}[t]{0.3\textwidth}
		\includegraphics[width=1.\textwidth]{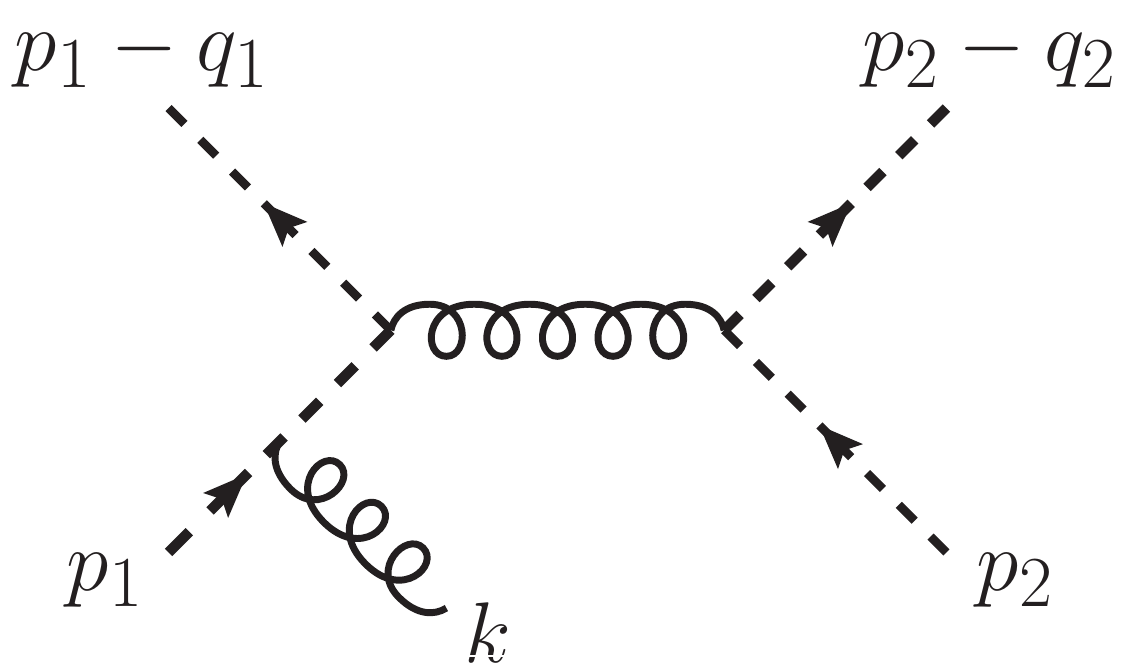}
		\caption*{(b)} 
		\label{fig:scalarDiagramsB}
	\end{subfigure}
\hskip .5 cm
	\begin{subfigure}[t]{0.3\textwidth}
		\includegraphics[width=1.\textwidth]{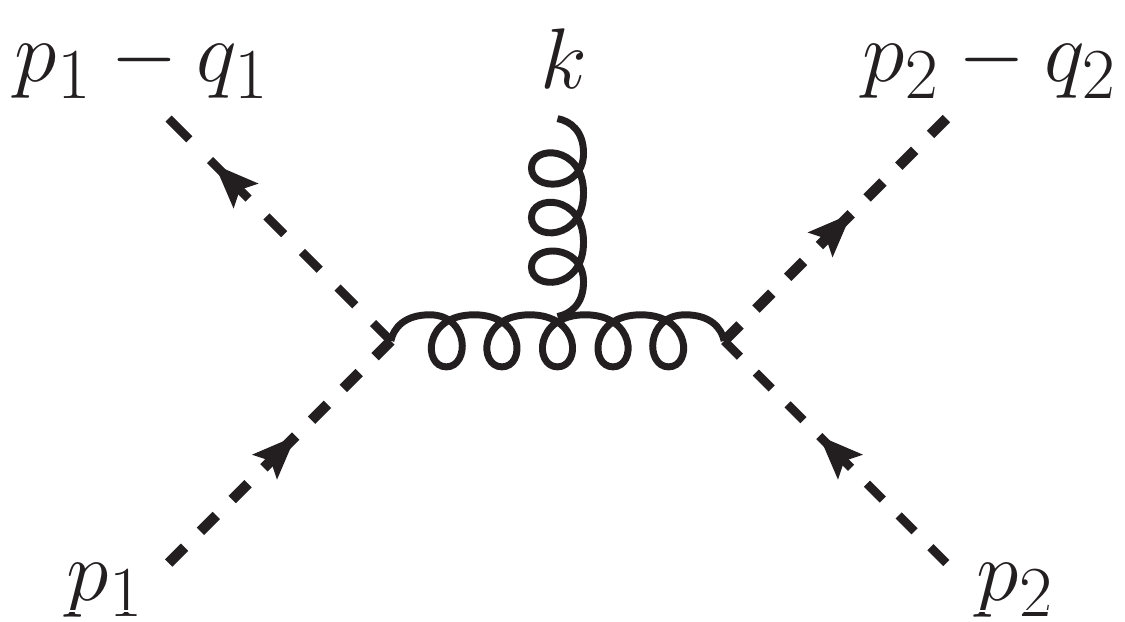}
		\caption*{(c)}
		\label{fig:scalarDiagramsC}
	\end{subfigure}
	
	\bigskip
	
	\begin{subfigure}[t]{0.3\textwidth}
		\includegraphics[width=1.\textwidth]{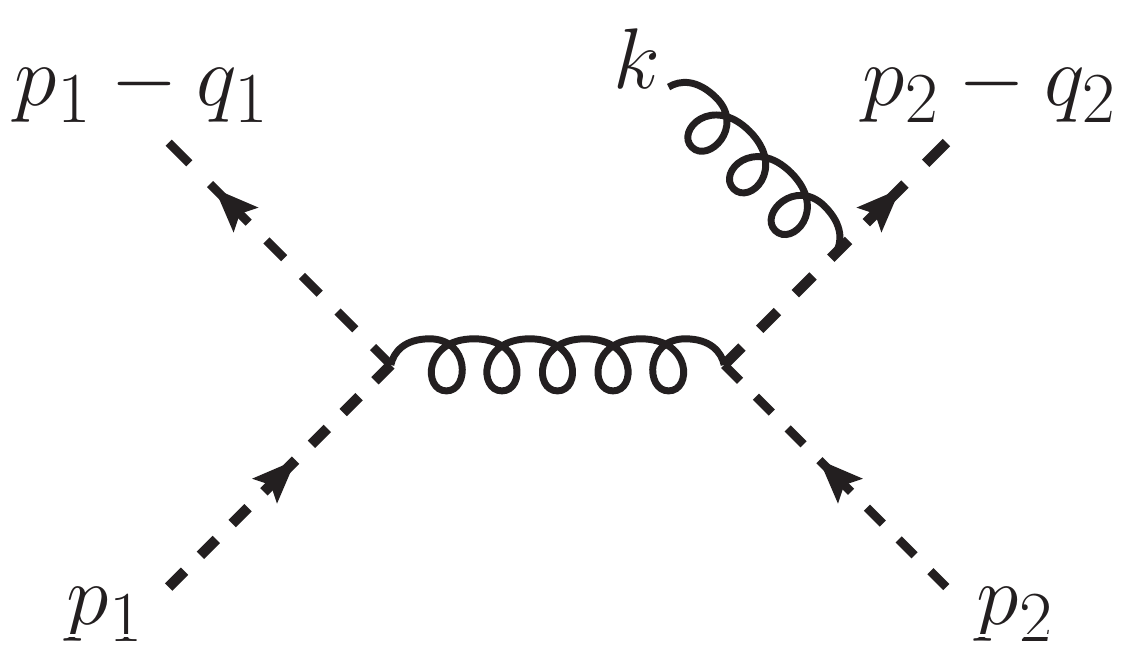}
		\caption*{(d)}
		\label{fig:scalarDiagramsD}
	\end{subfigure}
\hskip .5 cm
	\begin{subfigure}[t]{0.3\textwidth}
		\includegraphics[width=1.\textwidth]{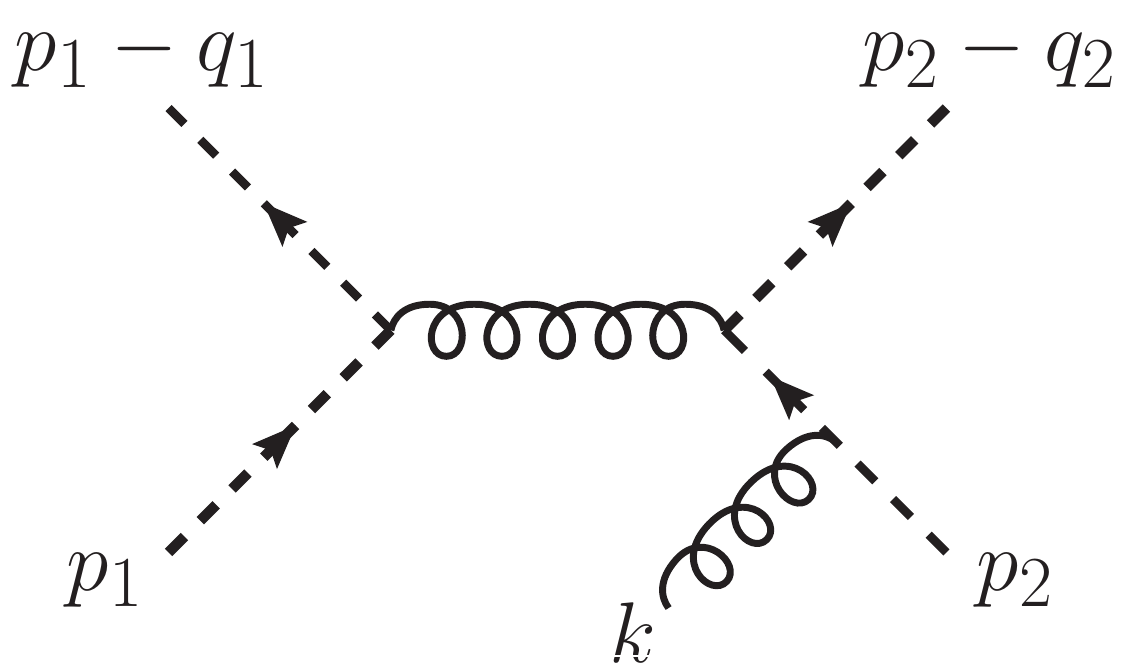}
		\caption*{(e)}
		\label{fig:scalarDiagramsE}
	\end{subfigure}
\caption{The five cubic diagrams for inelastic scalar
scattering with gluon production in gauge theory.  The legs carrying
momenta $p_1$ and $p_2$ are incoming and the remaining ones are
outgoing.  } 
\label{fig:scalarDiagrams}
\end{figure}

The five diagrams contributing to the scattering process $\Phi_i\Phi_j\rightarrow \Phi_i\Phi_jh_{\mu\nu}$ are shown in 
\fig{fig:scalarDiagrams} and the corresponding amplitude is
\begin{equation}
\mathcal{A} = -i\left(\frac{n_{\rm a} c_{\rm a}}{D_{\rm a}} + \frac{n_{\rm b} c_{\rm b}}{D_{\rm b}} + \frac{n_{\rm c} c_{\rm c}}{D_{\rm c}} 
+ \frac{n_{\rm d} c_{\rm d}}{D_{\rm d}} + \frac{n_{\rm e} c_{\rm e}}{D_{\rm e}} \right).
\label{A5pt_gt_radiation}
\end{equation}
The denominators $D_{\rm a},\ldots, D_{\rm e}$ and the color factors $c_{\rm a},\ldots, c_{\rm e}$ are easily read from the
diagrams in \fig{fig:scalarDiagrams}, taking into account that the scalar $\Phi_i$ is in some
complex representation $R_i$ with generators $T_{R_i}^a$. The
kinematic numerators follow from the Feynman rules of the
Lagrangian \eqref{Lagrangian_YM_Phi}.\footnote{We note that, as
discussed in previous sections, a quartic scalar term is not necessary
for \ck duality because the scalar fields are taken in a complex
representation of the gauge group. }
They are:
\label{eq:numerators}
\begin{align}
n_{\rm a} &= (2 p_1 + q_2) \cdot (2 p_2 - q_2) \; \varepsilon \cdot (2 p_1 + 2 q_2) - (2 p_1 \cdot q_2 + q_2^2) \; 
\varepsilon \cdot(2 p_2 - q_2) \,, \cr
n_{\rm b} &= (2 p_1 - k - q_1) \cdot (2 p_2 - q_2) \; 2 \varepsilon \cdot p_1 + 2 p_1 \cdot k \; \varepsilon \cdot(2 p_2 - q_2) \,,\cr
n_{\rm c} &= (2 p_1 - q_1)^\mu (2 p_2 - q_2)^\rho \left[(k+q_2)_\mu \eta_{\nu \rho} + (q_1 - q_2)_\nu \eta_{\rho\mu} - (k+q_1)_\rho \eta_{\mu\nu}\right] \varepsilon^\nu \,, \\
n_{\rm d} &= (2 p_1 - q_1) \cdot (2 p_2 + q_1) \; \varepsilon \cdot (2 p_2 + 2 q_1) - (2 p_2 \cdot q_1 + q_1^2) \; \varepsilon \cdot (2 p_1 - q_1)
\,,\cr
n_{\rm e} &= (2 p_1 - q_1) \cdot (2 p_2 -k - q_2) \;2 \varepsilon \cdot p_2 + 2 p_2 \cdot k\; \varepsilon \cdot (2 p_1 - q_1) \,,
\nonumber
\end{align}
where $\varepsilon$ is the gluon polarization vector.
The color identities that are important for the gauge invariance of ${\cal A}$ in Eq.~\eqref{A5pt_gt_radiation} are 
\begin{equation}
c_{\rm a} - c_{\rm b} = c_{\rm c} \quad c_{\rm d} - c_{\rm e} = c_{\rm c} \,.
\end{equation}
It can be easily checked that the numerators \eqref{eq:numerators} obey the corresponding kinematic relations.

The double-copy amplitude follows from the usual rules, see \sect{DualitySection}:
\begin{equation}
\label{eq:gravAmp}
\mathcal{M} = -i\left(\frac{n_{\rm a} n_{\rm a}}{D_{\rm a}} + \frac{n_{\rm b} n_{\rm b}}{D_{\rm b}} + \frac{n_{\rm c} n_{\rm c}}{D_{\rm c}} 
+ \frac{n_{\rm d} n_{\rm d}}{D_{\rm d}} + \frac{n_{\rm e} n_{\rm e}}{D_{\rm e}} \right).
\end{equation}
The tensor product of the two outgoing gluon polarization vectors can be projected onto a graviton state. For internal lines a more 
involved projection is necessary~\cite{Johansson2014zca}. We shall return to it shortly.

To relate the amplitude just constructed to the classical scattering of massive bodies it is necessary to focus on the 
classical kinematic regime. There exists many ``classical limits'' of a field theory and all of them involve the limit of 
vanishing Planck's constant, which must therefore be restored (on dimensional grounds) in the field theory expressions. 
The limit we are interested in is also the one in which masses and other quantum numbers, such as external momenta 
and charges, are parametrically large compared to the momenta exchanged between particles. 
Thus, the classical limit is equivalent with a large mass
expansion\footnote{Other formulations of the classical limit, leading
to the same result but with a different physical reasoning, were
discussed in \cite{CheungPM, 3PM, 3PMLong}.}~\cite{Holstein2004dn}:
\begin{align}
m_i\rightarrow \frac{m_i}{\hbar}\,,
\qquad
g\rightarrow \frac{g}{\hbar}\,,
\qquad
\hbar\rightarrow 0\,,\qquad
p_i^\mu \rightarrow m_i v_i^\mu\,,
\qquad
v_i^2 = 1 \,.
\label{classical limit}
\end{align}
Because the coupling (charges) and masses are scaled simultaneously, this limit makes parts of tree-level and loop-level diagrams 
of the same order and consequently all such contributions enter nontrivially in this classical limit~\cite{Holstein2004dn}. 

The limit \eqref{classical limit} must be taken while enforcing the exact on-shell condition for all external particles. In particular
\be
(p_i - q_i)^2 = m_i^2 - 2 m_i v_i \cdot q_i + q_i^2 = m_i^2 \Rightarrow  2 m_i v_i \cdot q_i = q_i^2 \,.
\ee
Thus, if external momenta are parametrically larger than the exchanged ones, this equation can be satisfied only if
\be
v_i \cdot q_i \sim {\cal O}(m_i^{-1}) \, .
\ee
This condition must be enforced when taking the classical limit of Eq.~\eqref{eq:gravAmp}. Defining the variables
\begin{align}
P_{12}^\mu &\equiv k \cdot v_1 \; v_2^\mu - k \cdot v_2 \; v_1^\mu, \nn\\
Q_{12}^\mu &\equiv (q_1-q_2)^\mu- \frac{q_1^2}{k \cdot v_1} v_1^\mu+ \frac{q_2^2}{k \cdot v_2} v_2^\mu,
\label{eq:PQ}
\end{align}
this classical limit of the amplitude \eqref{eq:gravAmp} is
\begin{multline}
\mathcal{M}_\textrm{cl} = -16 i m_1^2 m_2^2 \; \varepsilon_{\mu \nu} \left[
4 \frac{P_{12}^\mu P_{12}^\nu}{q_1^2 q_2^2} +2 \frac{v_1 \cdot v_2}{q_1^2 q_2^2} \left( Q_{12}^\mu P_{12}^\nu + Q_{12}^\nu P_{12}^\mu\right) \right.  \\
\left. + (v_1 \cdot v_2)^2 \left(\frac{Q_{12}^\mu Q_{12}^\nu}{q_1^2 q_2^2} - \frac{P_{12}^\mu P_{12}^\nu}{(k \cdot v_1)^2 (k\cdot v_2)^2} \right) 
\right]  ,
\label{eq:gravAmplitude}
\end{multline}
where $\varepsilon_{\mu\nu}\equiv \varepsilon_\mu\varepsilon_\nu$. 
As pointed out in Ref.~\cite{Luna2017dtq},  there is a close relation between this amplitude and the metric perturbation (\ie 
radiation field) constructed in Ref.~\cite{Goldberger2016iau}. The metric perturbation is given by the Fourier transform to position space 
of  $\mathcal{M}_\textrm{cl}$ with respect to the incoming scalar momenta, subject to the constraints imposed by the on-shell 
conditions for all external momenta. 
We note that the mass of the particles enters only as an overall factor in the amplitude \eqref{eq:gravAmplitude} and, consequently, in 
the associated metric perturbation. This property, implying that the features of the metric are essentially independent of a (spinless) source,
may be interpreted physically as a reflection of Birkhoff's theorem.

To obtain the analogous results in Einstein's gravity theory it is necessary to project out the dilaton and antisymmetric tensor field 
from all diagrams. As reviewed in \sect{ZoologySection1} following Ref.~\cite{Johansson2014zca}, this can be done by introducing 
further ``ghost'' fields whose couplings are adjusted such that they remove the (un)desired degrees of freedom.\footnote{An alternative possibility is to carry out the double copy while keeping track of the helicity of internal fields and making sure that 
only graviton modes appear on all internal lines of diagrams~\cite{3PM, 3PMLong}.}
For the case at hand the relevant Lagrangian is~\cite{Luna2017dtq}
\begin{equation}
\mathcal{L} = -\frac 12 {\rm Tr} F_{\mu\nu} F^{\mu\nu} +{\rm Tr} D^\mu \chi D_\mu \chi + \sum_i \left[(D_\mu \Phi_i)^\dagger D^\mu \Phi_i - m_i^2 \Phi_i^\dagger \Phi_i  - 2 X m_i \Phi_i^\dagger \chi \Phi_i \right] ,
\label{eq:extendedGauge}
\end{equation}
where $D_\mu$ is the gauge-covariant derivative, $\chi$ is the adjoint ghost, $X$ is its coupling to be determined. The mass factors are included such that the ghost
field has canonical dimension and $X$ is dimensionless. While in the adjoint representation, the ghost field is allowed to double
copy only with itself. 

The unknown coupling $X$ can be determined by comparing the $2\rightarrow 2$ 
massive scalar scattering obtained through double copy from the 
Lagrangian \eqref{eq:extendedGauge} with the massive scalar scattering in scalar-coupled general relativity. 
The result is
\be
X^4=\frac{1}{D-2} \, ,
\ee
where $D$ is the spacetime dimension.
The Lagrangian \eqref{eq:extendedGauge} can then be used to evaluate the additional Feynman diagrams that remove the
dilaton and axion contribution to Eq.~\eqref{eq:gravAmplitude}. The complete amplitude in scalar-coupled pure gravity is 
\begin{multline}
\label{fullGS1grav}
\mathcal{M}_{\textrm{\tiny{GR}}} = -16 i m_1^2 m_2^2 \; \varepsilon_{\mu \nu} \left[
4 \frac{P_{12}^\mu P_{12}^\nu}{q_1^2 q_2^2} +2 \frac{v_1 \cdot v_2}{q_1^2 q_2^2} \left( Q_{12}^\mu P_{12}^\nu + Q_{12}^\nu P_{12}^\mu\right) \right.  \\
\left. + \left((v_1 \cdot v_2)^2 - \frac{1}{D-2} \right) \left(\frac{Q_{12}^\mu Q_{12}^\nu}{q_1^2 q_2^2} - \frac{P_{12}^\mu P_{12}^\nu}{(k \cdot v_1)^2 (k\cdot v_2)^2} \right) 
\right] ,
\end{multline}
where the polarization tensor $\varepsilon_{\mu \nu}$ now includes only graviton degrees of freedom.
As shown in Ref.~\cite{Luna2017dtq}, this reproduces the result of the
(far more complicated) direct computation in general relativity
coupled to point particles.

It is interesting to note that, repeating the calculation in
Refs.~\cite{Goldberger2016iau, Goldberger2017frp} such that the
sources correspond to massive spinless bodies and removing the dilaton
and axion contribution through a procedure similar to the one
described above appears to lead~\cite{Luna2017dtq} to a different
result than \eqref{fullGS1grav}. While the origin of the difference
is not clear, it is possible that they are due to the
unusual double-copy rules employed in Refs.~\cite{Goldberger2016iau,
Goldberger2017frp}.

\begin{homework}
Evaluate the amplitude for the graviton production in the scattering of massive charged spinless bodies YME theory and compare the result with that of Ref.~\cite{Chester2017vcz}. 
\end{homework}

\subsection{Further comments}

The close relation between Green's functions and scattering amplitudes of quantum field theories suggests that relations between 
scattering amplitudes of different theories may translate, in particular gauges and for special choices of field variables, into relations 
between classical solutions of the corresponding equations of motion.
In this section we reviewed at length examples in which this expectation is realized and certain solutions of gauge theories
can be used to construct, through a classical double copy, certain solutions of gravity theories. 
Important points---such as the relation between the gauge choice for the gauge-theory solution and the properties of the 
corresponding gravity solution, or the identification of the best choice of gravity field variables such that no transformation functions
are present---remain to be fully understood and the complete rules of the classical double-copy construction to be spelled out.
The examples we discussed, as well as the additional ones that may be
found in the literature, show that such an approach can have useful
applications to current problems in gravitational physics. Chief among
them is precision predictions of gravitational waves; as we saw
in \sect{radiation}, the (classical) double-copy construction may help
streamline the evaluation of the expected signal from the relevant
astronomical events.

Further applications of the double copy to gravitational wave physics,
which we did not discuss in detail, relate to the calculation of
gravitational interaction potential in the post-Newtonian expansion.
While standard methods, using the gravitation Lagrangian, are
well-developed and results through fourth post-Newtonian order are
available~\cite{Damour:2014jta, Jaranowski:2015lha,
Foffa2016rgu,Foffa:2019yfl}, double-copy calculations such as in
Refs.~\cite{Donoghue_lightbending, Donoghue_lightscattering,
Donoghue_morelightbending} may bring a novel perspective to this
problem.  Indeed, advances based on the double copy and new
developments~\cite{CheungPM} in the effective field
approach~\cite{Goldberger:2004jt,NeillRothstein} resulted in a new
state of the art result at the third order in Newton's constant~\cite{3PM}.

A common feature of the classical solutions constructed to date
through such methods is that, in the appropriate field variables,
Einstein's equations become linear. This includes the case of the Kerr
black hole which was shown in~\cite{Arkani-Hamed:2019ymq} to be
related to a certain complex deformation of the Coulomb potential. It
is also shown that the change in momentum in a scattering event,
known as ``the impulse'', can be described via a double copy of a
point charge.
Progress towards further understanding some of the rules of the classical double copy may 
follow from finding examples where nonlinear contributions are nonvanishing. Perhaps the easiest approach to exploring such 
cases is the analysis of a Kerr-Schild solution for another choice of field variables; an example would be to repeat the calculations
in Refs.~\cite{Duff1973zz, Sardelis1973em} using modern approaches.

The construction of a gravitational Lagrangian whose fields are
explicitly constructed in terms of those of the two single-copy gauge
theories may also lead to new ways of relating the solutions of the
two theories. The linearized Lagrangians of certain ${\cal N}=2$
supergravities have been organized in this fashion in
Refs.~\cite{Cardoso2016amd, Cardoso2016ngt}.

Another (notoriously difficult) problem which may benefit from the
existence of a systematic classical double copy is gravitational
perturbation theory around a curved ground state such as the (anti) de
Sitter space, or the Schwarzschild black hole. The double copy will
likely relate it to gauge-theory perturbation theory around a
nontrivial classical solution of YM equations of motion.
Attempts in this direction have been discussed in
Ref.~\cite{Adamo2017nia} and \cite{Adamo:2018mpq} where, respectively, the three- and four-point 
amplitudes of gravitons
around a gravitational plane wave were expressed in terms of
three- and four-gluon amplitudes around a particular gauge-theory plane wave.
While developing general methods for such calculations is an interesting problem 
in its own right, it is likely that their main applications 
will be to gauge/string duality.

The study of the classical double copy is in its infancy and many
avenues remain to be explored; we expect that the resulting methods
will yield important new progress in gravitational physics, especially
on the problem of gravitational radiation.


\section{Conclusions}
\label{ConclusionSection}

The duality between color and kinematics and the double-copy
construction offer a radically-different perspective on gravity
theories compared to traditional Lagrangian or Hamiltonian approaches.
For the well-studied case of scattering amplitudes, the duality
provides powerful means for converting results in gauge theory to
those of gravity. This has led to progress in studying the behavior of various gravitational theories 
at high perturbative orders, such as  the UV
behavior of extended supergravity at four and five
loops~\cite{GravityFour, N4GravFourLoop, N5GravFourLoop, UVFiveLoops}
and  the third post-Minkowskian corrections to the classical Hamiltonian
for compact binaries~\cite{3PM,3PMLong}. At present there are no other
means to evaluate such high orders.

Remarkably, the idea of \ck duality and of the double-copy structure extends
to theories with no obvious connection to gauge or gravity theories,
as reviewed in \sect{ZoologySection1}.    The fact that the
scattering amplitudes of theories whose Lagrangians seem to have
little to do with each other contain the same kinematical objects is
rather striking and points to new nontrivial constraints shared by
consistent theories.  Additionally, by now the duality and double copy
have been established for a large number of examples
of classical solutions~\cite{Saotome2012vy, Monteiro2014cda, Luna2015paa,
  Ridgway2015fdl, Luna2016due, White2016jzc, Cardoso2016amd,
  Goldberger2016iau, Luna2016hge, Goldberger2017frp, 
  Adamo2017nia, DeSmet2017rve, BahjatAbbas2017htu, 
  CarrilloGonzalez2017iyj, Goldberger2017ogt, Li2018qap,
  Ilderton:2018lsf, Lee:2018gxc, Plefka:2018dpa, ShenWorldLine, 
  Berman:2018hwd, Gurses:2018ckx,  Adamo:2018mpq, 
 Bahjat-Abbas:2018vgo, Luna:2018dpt,
Farrow:2018yni, CarrilloGonzalez:2019gof, PV:2019uuv}.

There are several areas where further progress would be
welcomed.  For example, it is not at the moment clear how far the notion of \ck duality and the
double copy can be carried beyond scattering amplitudes.  Many of the
examples of classical solutions that display the duality make use of
special properties, such as the existence of Kerr-Schild forms of the
metric.  It would be very important to find more general examples.  
Classical solutions are inherently more
difficult to study because they depend on coordinate and gauge
choices and, without the appropriate choices, the double-copy structure is
obscured. This may be contrasted with scattering amplitudes, which are
independent of the choice of gauge and, to a large extent,  field
variables, making it much easier to  formulate
double-copy relations.
To avoid carrying out complicated case-by-case analyses, a key step is to 
find underlying principles for choosing gauges and field
variables in both single- and double-copy theories that make it more
straightforward to identify relations between off-shell quantities.
It would also be interesting to see if the more invariant color-trace-based formulation of the duality~\cite{Bern:2011ia, Du:2013sha,
Fu:2013qna, Du:2014uua, Naculich:2014rta, Fu:2018hpu} might shed light
on extensions of \ck duality beyond scattering amplitudes.

A possible path to unraveling the principles for choosing gauges and
field variables may be the study of correlation functions. The
computation of correlation functions of gauge-invariant operators in
gauge theories may be approached through generalized
unitarity \cite{Engelund:2012re}, which relates it to the construction
of tree-level scattering amplitudes and form factors of the
same gauge-invariant operators.  
In this respect, \ck duality has been formulated and used for the form factors of
certain operators in four-
and five-loop calculations in $\cN=4$ SYM
theory \cite{FourLoopFormFactor, BoelsFourLoop, Boels:2017ftb,
FiveLoopFormFactor}. Therefore, it seems plausible to extend \ck duality to the
correlation functions of these operators.
However, a puzzle arises if one considers the natural step of constructing the
double copy of such correlation functions.
Since correlation functions of gauge-invariant operators in gauge
theories are gauge invariant, one may conclude following the
discussion in \sect{DualitySection}
and \sect{GravitySymmetriesSection} that the corresponding
gravitational correlation functions are automatically diffeomorphism-invariant.
It is well-known however that local diffeomorphism-invariant operators
do not exist in gravitational theories.
Since correlators of gauge-dependent operators depend on choices of
field variables, it appears that the gravitational
correlation functions obtained though the double copy should be understood
as being given for a particular choice of field variables and perhaps
also for particular choice of gauge.
A further puzzle originates from contrasting the results of the
double copy for conformal gauge theories that admit a string-theory
dual to the results of the corresponding string theory in
anti-de-Sitter space. On the one hand, gauge-theory correlation
functions are given by string-theory correlators with prescribed
boundary conditions in anti-de-Sitter space; on the other, the
gauge-theory correlators can be used to construct correlators in a
gravitational theory ({\it not} a string theory) in a Minkowski vacuum
and with the same amount of supersymmetry as the AdS one. While
technically difficult, it would clearly
be interesting to understand the implications of such a relation.

Apart from formal developments such as the ones described above,
perturbative calculations in curved spacetime are playing an
increasingly-important role in the current development of our
understanding of the universe. Initial attempts to use the double-copy
construction in this context, involving calculations in certain plane
wave spacetimes, have been discussed in~\cite{Adamo2017nia,
BahjatAbbas2017htu, Adamo:2018mpq,Adamo:2019zmk, Sachs:2019wrk}.  As
in flat space, the double-copy construction may help by relating such
calculations with simpler ones in gauge theory, especially for
spacetimes which are themselves classical double copies.  It is
obviously nontrivial to extend the insights of flat-spacetime
scattering to the many conceptual and technical challenges posed by
cosmological correlators in de Sitter, yet there is already a
developing program~\cite{Raju:2012zr, Raju:2012zs, Maldacena:2011nz,
Arkani-Hamed:2015bza, Arkani-Hamed:2018bjr, Arkani-Hamed:2018kmz}
leveraging the identification of S-matrix elements emerging as
residues of well-defined singularities of such quantities.

While \ck duality and the associated double copy have been crucial for
uncovering the UV properties of various supergravities~\cite{Bern:2012cd,
Bern:2012gh, Bern:2013qca, N4GravFourLoop, 
GeneralizedDoubleCopyFiveLoops, UVFiveLoops} and for identifying a new set of
nontrivial enhanced UV cancellations~\cite{N5GravFourLoop}, to move forward it is essential
to gain a thorough grasp on the structures or symmetries that are
responsible for the appearance of the latter.  Progress in this direction has
been reported in~\cite{Bern:2012gh,HerrmannTrnkaUVGrav,
Bourjaily:2018omh}, but much more remains to be done to have a
satisfactory understanding. Presumably, the duality and double copy
play a key role in these cancellations.

Another important topic is to expand the web of theories related by
the duality and double-copy construction.  As illustrated
in \fig{FigWeb} of \sect{ZoologySection1}, theories that may appear to
be unrelated are bound together by double-copy relations.  In many of
these cases, the connection is rather obscure from a Lagrangian
perspective, \eg that Dirac-Born-Infeld theory has a relation to the
special Galileon theory, by sharing the NLSM as a
composite theory via the double copy.  A crucial open question is whether it is possible to get a
complete classification of all double-copy-constructible theories.  
An equally important question is whether all supergravity theories can
be expressed in a double-copy
format~\cite{Chiodaroli2015wal,Anastasiou2017nsz}. In this review, we discuss a
large number of examples, which are collected in
Tables~\ref{table-zoology-ungauged} and~\ref{table-zoology-gauged}.  
It is a
surprising fact that the only known unitary UV completions of
gravity and higher-dimensional YM, the closed and open
superstring, require their constituent effective field theories to be
compatible with the field-theory adjoint double-copy to all orders of
$\alpha'$ at
tree-level~\cite{Broedel2013tta,Carrasco2016ldy,Mafra2016mcc,Carrasco2016ygv}.

Another  basic research direction is to find the underlying algebra behind the
duality between color and kinematics.  A natural expectation is that
the kinematic Jacobi identities are due to an infinite-dimensional Lie
algebra~\cite{Monteiro2011pc, Monteiro:2013rya, Fu:2016plh,
Chen:2019ywi}. Indeed, for the case of
self-dual field configurations, corresponding to amplitudes with
identical helicity, the algebra has been identified as that of the area-preserving 
diffeomorphisms in one lightcone and one transverse direction~\cite{Monteiro2011pc}. 
However, extending this observation to general helicity or field configurations 
has proven to be challenging.

Constructing a Lagrangian that automatically generates
Feynman rules that manifest the duality would greatly help with
finding double-copy relations between classical solutions.  However,
at present, only perturbative order-by-order constructions of such
Lagrangians are known~\cite{Square,WeinzierlBCJLagrangian,Vaman:2014iwa,
Mastrolia:2015maa}.  
From the perspective of gravity theories, Lagrangians that display
the required factorization of Lorentz indices~\cite{BernGrant} have
been obtained to all orders~\cite{Hohm:2011dz,CheungRemmen}.  However, as
yet, it is unclear which all-orders gauge-theory Lagrangians can reproduce them though double copy.
One difficulty is that such  Lagrangians would likely contain an infinite number of auxiliary 
fields to make them local.

To further streamline higher-loop computations would be particularly desirable. 
 While there has been enormous progress in carrying out such
computations to relatively high orders (see e.g.,
Refs.~\cite{SimplifyingBCJ, N4GravFourLoop,
EnhancedCancellationsIntegrals,UVFiveLoops} for four and five-loop
calculations) we should always strive to go further.  
At high orders, it can be nontrivial to find representations of loop
integrands that manifest \ck duality~\cite{Bern:2015ooa,
BCJDifficulty}.  As described in \sect{GeneralizedDoubleCopySection},
these difficulties can be bypassed via a generalized double
copy~\cite{GeneralizedDoubleCopy, GeneralizedDoubleCopyFiveLoops} that
can be used to convert any representation of gauge-theory amplitudes
to corresponding gravity ones, relying only on the proven existence of
the duality at tree level.  Finding generalizations of this procedure 
for any number of loops or legs would be important.

Strengthening connections between the double copy and other advances in scattering amplitudes
would also be advantageous.  In particular, the
amplituhedron~\cite{Arkani-Hamed:2013jha} gives novel geometric
descriptions of amplitudes.  A detailed formulation has been given for
the planar sector of ${\cal N} = 4$ SYM theory.  Making contact
with \ck duality requires extending these results to the nonplanar
sector.  Evidence suggests that this may be possible~\cite{Bern:2015ple,
Arkani-Hamed:2017mur}.

The double copy seems to hint at some kind of interpretation of
gravitons as composed of spin-1 particles.  Of course, these cannot be
any kind of naive bound states, which are forbidden by the
Witten-Weinberg theorem~\cite{Weinberg:1980kq}.  Still, the double copy strongly
suggests that gravitons and gluons ultimately belong together,
presumably along the lines realized by string theory.  (See
Refs.~\cite{Siegel:1993sk, Siegel:2003vt, Lee:2003qw} for steps in
this direction.) Understanding any fundamental physical implications
of the way gauge and gravity theories are intertwined by the double
copy is a key problem that deserves further attention.

The application of the double copy to gravitational-wave physics~\cite{Abbott2016blz}  is currently the subject of intense investigation, 
specifically regarding the
post-Newtonian~\cite{Einstein:1938yz} and
post-Minkowskian~\cite{PMBertotti1956,PMKerr1959} approaches to the
inspiral phase of binary mergers (see the following reviews for details
and references~\cite{BlanchetReview, Buonanno:2014aza, Porto:2016pyg,
LeviReview}).  A nontrivial application of \ck duality to the study of gravitational
radiation has been discussed  with a worldline formulation in
Ref.~\cite{Goldberger2016iau}, where the duality has been established
through next-to-leading order~\cite{ShenWorldLine}.
Related progress was also reported in
Refs.~\cite{Goldberger2017frp, Chester2017vcz, Goldberger2017ogt, Li2018qap}.
Other investigations related to gravitation-wave physics that directly draw
from scattering-amplitudes methods can be found in
Refs.~\cite{NeillRothstein, Bjerrum-Bohr:2013bxa, Luna2016due,
Guevara:2017csg, Luna2017dtq, BjerrumClassical,
Damour:2017zjx,Kosower:2018adc, Guevara:2018wpp, OConnellSpin2019, Guevara:2019fsj}.  A systematic and
scalable approach for obtaining high-order corrections to conservative
two-body potentials in the post-Minkowskian framework was 
presented in Ref.~\cite{CheungPM}.  This has been successfully used
to find the third post-Minkowskian corrections~\cite{3PM,3PMLong}, starting
from two-loop amplitudes obtained via the double copy.  It is
noteworthy that this is one order beyond previous
calculations~\cite{PMWestfahl,Damour:2016gwp,Damour:2017zjx}.  While their
impact on improving templates for LIGO/Virgo is currently under
study~\cite{Buananno3PMCheck}, these  results should also
offer new insights into the general structure of high-order two-body
Hamiltonians.


\vskip .3 cm 

\subsection*{Acknowledgments}

We thank T.~Adamo, L.~Borsten, E.~Bjerrum-Bohr, J.~Bourjaily,
L.~Dixon, M.~Duff, A.~Edison, S.~Ferrara, M.~G\"unaydin, S.~He,
E.~Herrmann, Y.-t.~Huang, G.~K\"{a}lin, R.~Kallosh, D.~Kosower,
A.~Luna, D.~L\"ust, C.~Mafra, G.~Mogull, R.~Monteiro, S.~Nagy,
H.~Nicolai, A.~Ochirov, D.~O'Connell, J.~Parra-Martinez, L.~Rodina,
O.~Schlotterer, C.-H.~Shen, S.~Stieberger, F.~Teng, J.~Trnka,
A.~Tseytlin, P.~Vanhove, I.~Vazquez-Holm, C.~White and S.~Zekioglu for
helpful discussions, collaboration and comments during the course of
writing this review.
Z.B. is supported by the U.S. Department of Energy (DOE) under grant
no.~DE-SC0009937.  JJMC is grateful for the support of Northwestern
University, CEA/CNRS-Saclay, and the European Research Council under
ERC-STG-639729, {\it Strategic Predictions for Quantum Field
Theories\/}. R.R.~is supported by the U.S. Department of Energy (DOE)
under grant no.~DE-SC0013699.  The research of M.C. and H.J. is
supported by the Knut and Alice Wallenberg Foundation under grants KAW
2013.0235 and KAW 2018.0116 - {\it From Scattering Amplitudes to
Gravitational Waves}, the Ragnar S\"{o}derberg Foundation (Swedish
Foundations' Starting Grant), and the Swedish Research Council under
grant 621-2014-5722.  This
work was performed in part at the
Munich Institute for Astro- and Particle Physics (MIAPP) of the DFG
cluster of excellence ``Origin and Structure of the Universe'' and in part at the Aspen Center for Physics, which is
supported by National Science Foundation grant PHY-1607611.  In
addition, Z.B.~thanks Mani L. Bhaumik for generous support over the
years.


\newpage

\begin{appendices}

\section{Notation and list of acronyms}
\label{NotationSection}

In this appendix, we summarize our notation  and conventions for the reader's convenience. 

Throughout the review, we denote with ${\cal A}_n(1,\cdots,n)$ gauge-theory color-dressed amplitudes, while $A_n(1,\cdots,n)$ is used to indicate color-stripped partial amplitudes. In some sections, 
where theories containing different fields are discussed, it is convenient to adopt the notation 
\begin{equation}
{\cal A}_n\big(1 \Phi_1 , \dots  , n \Phi_n \big) \ ,
\end{equation}
which makes explicit which field is associated to each external leg of the amplitude. Superamplitudes are denoted with the same symbol as the corresponding amplitudes, i.e. it should be clear from the context whether $A_n(1,\cdots,n)$ and ${\cal A}_n(1,\cdots,n)$ refer to an amplitude or to the corresponding superamplitude. 
Writing the $S$-matrix as $S= 1 + i T$, our amplitudes correspond to the $iT$ term, i.e. they give the output of the Feynman-diagram calculation. 
Amplitude in a gravitational theory are denoted as  ${\cal M}_n(1,\cdots,n)$. For notational simplicity, we set $\kappa = 2$ in most formulas, where  $\kappa$ is the gravitational coupling constant.   

Our conventions for the phase in the BCJ representation of gauge-theory and gravity amplitudes follows the one in Ref.~\cite{Chiodaroli2017ngp} and departs from the original BCJ papers~\cite{BCJ,BCJLoop}. With this choice, YM numerators are real when written in terms of polarization vectors.

Calculations presented in this review involve a spacetime metric of mostly-minus signature. Our spinor-helicity conventions are obtained from the ones of Ref.~\cite{ElvangHuangReview} by the minimal replacement 
\begin{equation}
(\eta_{\mu\nu})_{\rm E\&H} \rightarrow - (\eta_{\mu\nu})_{\text{our}} \ .
\end{equation}
In particular, angle and square brackets are the same as in Ref.~\cite{ElvangHuangReview}.   

Gauge-group fundamental fields are represented with high indices and anti-fundamental fields are represented with low indices. For example, the generator for the fundamental representation is written as
\begin{equation}
 (t^a)_i^{\ j} \equiv t^a_{i \bar \jmath} \ .
\end{equation}  
Generators are normalized as 
\begin{equation}
\Tr(t^a t^b) = \frac{\delta^{ab}}{2}\ ,
\end{equation}
and obey commutation relations of the form
\begin{equation}
[t^a, t^b] = i f^{abc} t^c \ .
\end{equation}
In amplitude calculations it is convenient to rescale the group-theory generators and structure constants as
\begin{equation}
\T^{a} \equiv \sqrt{2} t^a \,, \hskip 1.5 cm 
\f^{abc} \equiv i \sqrt{2} f^{abc} \,,
\end{equation}
so that we have the identity 
\begin{equation}
{\rm Tr}(\T^{a}\T^{b})=\delta^{ab}\,.
\end{equation}
In particular, color factors entering the formula \eqref{gaugeAmp} are written in terms of $T^a$s and $\f^{abc}$s, i.e. are written in terms of hermitian objects carrying a factor of $\sqrt{2}$ with respect to the Feynman-rule normalization. When we encounter fields in a matter (non-fundamental) representation ${\cal R}$, we denote the corresponding generators as $\tR^a$. In some sections of this review, for example in \sect{ZoologySection1}, we frequently use hatted indices for gauge-group indices of the gauge theories entering the double-copy construction to differentiate them from global indices.
\medskip

Finally, we conclude this appendix with a list of  acronyms commonly used throughout the review: \\

\begin{tabular}{l@{\hskip 1.5cm}l}
UV & Ultraviolet \\ 
IR & Infrared \\
KLT & Kawai-Lewellen-Tye (formula/relations) \\
CK duality  & Color/kinematics duality \\
BCJ  & Bern-Carrasco-Johansson (duality/relations)\\ 
YM  & Yang-Mills \\
SYM  & Super-Yang-Mills \\
NLSM & Nonlinear Sigma Model \\
DDM  & Del Duca-Dixon-Maltoni (amplitude representation)  \\
POP & Partially-ordered permutations  \\
BCFW & Britto-Cachazo-Feng-Witten (relations/recursion in field theory) \\
KK & Kleiss-Kuijf (amplitude relations)  \\
QCD & Quantum Chromodynamics \\
CPT & Charge-Parity-Time reversal (transformations)\\
MHV & Maximally-helicity-violating (amplitudes) \\
LSZ   & Lehmann-Symanzik-Zimmermann (reduction) \\
1PI  & One particle irreducible (effective action) \\
BMS   & Bondi-Metzner-Sachs (transformations) \\
SUSY & Supersymmetry (tables and figures only)\\
CSG & Conformal supergravity \\
BLG  & Bagger-Lambert-Gustavsson (theory) \\
ABJM & Aharony-Bergman-Jafferis-Maldacena (theory) \\
VEV & Vacuum expectation value \\
YM${}_\text{DR}$ &  Yang-Mills-scalar theory from dimensional reduction \\
YME & Yang-Mills-Einstein (theory) \\
DBI & Dirac-Born-Infeld (theories) \\
CHY & Cachazo-He-Yuan (formalism, also known as scattering equations) \\
SG & Supergravity (tables and figures only) \\
MZVs & Multiple zeta values \\
QFT & Quantum field theory \\
LIGO & Laser Interferometer Gravitational-Wave Observatory \\
\end{tabular}

\bigskip 

\bigskip

\section{Spinor helicity and on-shell superspaces}
\label{SpinorSuperspaceSection}

In explicit expressions for amplitudes, such as those in
\sect{ExamplesSection} or \app{GeneralizedUnitaritySection}, it is
very convenient to adopt a helicity (circular polarization) basis for the asymptotic states of
gluons or gravitons.  In this appendix, we summarize the
spinor-helicity formalism~\cite{SpinorHelicityBerends,
  SpinorHelicityKleissStirling, SpinorHelicityGunionKunszt, Xu:1984qe,
  SpinorHelicityXZC, ManganoParkeReview, TasiLance}, which offers a convenient
Lorentz covariant formalism for describing helicity, leading to
remarkably compact expressions for scattering amplitudes.  
The resulting states fit naturally into on-shell supermultiplets~\cite{Nair}.

\subsection{Basics of spinor helicity}
\label{SpinorHelicitySubsection}

The spinor-helicity formalism expresses the positive- and negative-helicity
polarizations of gluons (vectors) in terms of massless Weyl spinors 
\begin{equation}
\pol^{+}_\mu (k;q) =  \frac{{\langle{q}| {\sigma_\mu}}|k ]}
      { \sqrt2  \spa{q}.{k} } \, ,\hskip 1.5cm
\pol^{-}_\mu (k;q) =  \frac{[ q |  {\sigma_\mu} | k \rangle}
      { \sqrt{2} \spb{k}.{q}} \, ,
\label{PolarizationVector}
\end{equation}
where $q$ is an arbitrary null `reference' momentum which can be chosen independently for each
external state of amplitudes and, because of gauge invariance\footnote{Linearized gauge transformations, 
$\varepsilon_\mu(p)\rightarrow \varepsilon_\mu(p) + f(p) p_\mu$, is realized as shifts of the spinors associated to the reference vector, 
$|q\rangle \rightarrow |q\rangle + f(p) |p\rangle$, {\it etc}. }, drops out of the final expressions.  
We use the compact notation 
\begin{align}
& \langle ij \rangle  \equiv \frac{1}{2} {\bar u}(k_i)(1+\gamma_5) u(k_j) \,,
\hskip 2.2 cm
[ij] \equiv \frac{1}{2} {\bar u}(k_i) (1-\gamma_5) u(k_j) \,, \nn\\ 
& \langle{q}| {\sigma_\mu} |k ] \equiv \frac{1}{2} {\bar u}(q) \gamma^\mu (1-\gamma_5) u(k)  \,,  \hskip 1 cm 
  [q |  {\sigma_\mu} | k\rangle \equiv  \frac{1}{2} {\bar u}(q) \gamma^\mu  (1+\gamma_5) u(k) \,,
\label{spinorproducts}
\end{align}
with $u(k)$ following the standard textbook notation for solutions of the Dirac
equation~\cite{PeskinSchroeder}.  The spinors $|i\rangle$ and $|i]$
transform in the $({\bf 1},{\bf 2})$ and $({\bf 2},{\bf 1})$ representations of the
four-dimensional Lorentz group, respectively.  The spinor products~\eqref{spinorproducts} are
antisymmetric in their arguments. An important identity is the Schouten identity:
\begin{equation}
\spa{i}.{j}\spa{k}.{l} 
= \spa{i}.{l}\spa{k}.{j} + \spa{i}.{k}\spa{j}.{l} \, ,
\end{equation}
which is a consequence of the vanishing of all three-index antisymmetric tensor with each index taking two values.
The spinor products \eqref{spinorproducts} are related to the usual scalar products by
\begin{equation}
\spa{i}.j \spb{j}.i = 2 k_i \cdot k_j = s_{ij} \,,
\end{equation}
where the $k_i$ are null four momenta. 
The Fierz identity is in our conventions is
\be
\langle  i | \sigma^\mu | j ] \, \langle  k | \sigma_\mu | l ] = 2 \spa{i}.{k} \spb{l}.{j}\,.
\label{FierzId}
\ee

Helicity amplitudes (that is, amplitudes with polarization vectors or tensors in helicity notation) can be given a 
manifestly crossing symmetric representation. To this end it is necessary to assign all momenta to have the 
same orientation, either all outgoing or all incoming. When switching between the two different orientations the 
helicity label is reversed. This is, of course, natural: since the helicity measures the projection of the spin on 
the momentum, changing the orientation of the momentum reverses the helicity.
Using spinor helicity we can obtain exceptionally 
compact expressions for gauge-theory scattering amplitudes~\cite{ManganoParkeReview}.

The physical graviton polarization tensors in helicity notation are direct products of the gluon ones,
\begin{equation}
\pol^{+}_{\mu\nu} (k;q) = \pol^{+}_{\nu} (k;q)  \pol^{+}_{\nu} (k;q) \,, \hskip 1.5 cm 
\pol^{+}_{\mu\nu} (k;q) = \pol^{+}_{\nu} (k;q)  \pol^{+}_{\nu} (k;q) \,.
\label{pol_tensors}
\end{equation}
Their tracelessness , $\pol^{+\mu}_{\mu}=0$,  follows from the Fierz identity \eqref{FierzId} with the appropriate choice of spinors:
\begin{equation}
\langle{q}| {\sigma_\mu} |k]^2 = 0\,, \hskip 3 cm [{q}| {\sigma_\mu} |k\rangle^2 = 0\, .
\end{equation}
The relation \eqref{pol_tensors} between graviton and gluon
polarizations is the simplest manifestation of the double copy.  Of
course, the double copy holds for the full nonlinear theory, not just
for polarization tensors.

Loop calculations require regularization; we will not discuss details of this issue here except to note that 
to maximize the benefits of the spinor-helicity formalism, which is intrinsically four-dimensional, 
it is necessary to  choose a compatible version of dimensional regularization~\cite{FDH, FDH2}.

\begin{homework}
\label{HomeworkFourPoint}
Starting with Eqs.~\eqref{YMFourAmplitude}-\eqref{OtherChannels} apply \eqn{PolarizationVector} to
obtain four-gluon amplitudes for the various helicity configurations.  How clean can you make these expressions?
See, for example, Ref.~\cite{ManganoParkeReview}.
\end{homework}

\subsubsection{Massive spinor helicity \label{massiveSH}}

As some of the theories described in this review involve massive
fields, we will briefly outline how to adapt the spinor-helicity
formalism to this case.  A first possibility is to write a massive
momentum $k$ in terms of two massless momenta~\cite{Craig:2011ws}:
\begin{equation}
k = k_\perp + {m^2 \over 2 k \cdot q  } q  \, ,
\end{equation}
where $k_\perp$ is massless and we have also introduced a massless reference 
momentum $q$.  Polarizations for massive vectors are then written as
\begin{equation}
\pol_{+}^\mu (k;q) =  \frac{\langle q | \sigma^\mu  | k_\perp ]}
      {\sqrt2  \spa{q}.{k_{\perp}} }\, ,\hskip 1 cm 
\pol_{-}^\mu (k;q) =  \frac{[ q | \sigma^\mu | k_\perp \rangle}
      {\sqrt{2} \spb{k_{\perp}}.{q} } \, , \hskip 1 cm 
\pol_{0}^\mu (k;q) =  {1 \over m} \Big( k_\perp^\mu -  { q^\mu \over 2 k \cdot q  }  \Big) \, ,
\label{PolarizationVectorMassive}
\end{equation}
where the first two physical polarizations reproduce
(\ref{PolarizationVector}) in the massless limit and $\pol_0^\mu(k;q)$
gives the longitudinal polarization.  While this formalism allows us
to find relatively compact expressions for amplitudes with massive
fields, the reference momentum $q$ does not drop out from the final
expressions. This is to be expected: for a massive particle helicity is not a Lorentz-invariant 
quantity, so the decomposition \eqref{PolarizationVectorMassive} depends on the frame.

A more elegant approach involves a doublet of spinors $\lambda^a_\alpha, \tilde \lambda^a_{\dot \beta}$ which 
transform covariantly under the $SO(3)\cong SU(2)$ little group appropriate for describing massive particles 
in four dimensions. 
Massive momenta are then written as~\cite{Arkani-Hamed:2017jhn}:
\begin{equation}
k_{\alpha \dot \beta}= k_\mu \sigma^\mu_{\alpha \dot \beta} = \epsilon_{ab} |k^a \rangle_\alpha [ k^b|_{\dot \beta} 
= \epsilon_{ab} \lambda_\alpha^a \tilde \lambda^b_{\dot \beta} \, ,
\end{equation}
where $a,b$ are little group $SU(2)$ indices and $\alpha, \dot \beta$ are four-dimensional Weyl spinor indices. 
Massive vector polarizations are written n terms of the spinors $\lambda^a_\alpha, \tilde \lambda^a_{\dot \beta}$ as
\begin{equation}
\pol_\mu^{ab}(k) = {\langle k^{(a} | \sigma_\mu  |k^{b)}]  \over \sqrt{2} m} \,,
\end{equation}
where the little group indices are symmetrized. This formalism can be straightforwardly extended to construct polarization
tensors for higher-spin massive fields~\cite{Arkani-Hamed:2017jhn} and presents close analogies with massless 
spinor-helicity in six dimensions\cite{Cheung:2009dc}.

\subsection{On-shell superamplitudes}
\label{SuperAmplitudesSubsection}

For supersymmetric amplitudes, on-shell superspace provides a convenient organization 
of amplitudes according to their physical helicity states which also tracks the relationships between 
the different component amplitudes. 
The power of such an on-shell superspace follows from the
fact that, for generic momentum configurations, scattering amplitudes are insensitive to the nonlinear 
parts of (super)symmetry transformations (see \sect{GravitySymmetriesSection} for 
more details.). 
This greatly simplifies the evaluation of state sums in both the on-shell recursion~\cite{BCFW} and
generalized unitarity~\cite{UnitarityMethod, Fusing, BCFUnitarity} by allowing that all physical states 
be treated simultaneously.

To illustrate the ideas we use $\NeqFour$ SYM theory~\cite{Nair} as an example.  
Similar constructions exist in  cases with less than maximal supersymmetry~\cite{ElvangHuangLessSusy}
as well as supergravity theories~\cite{ElvangFreedmanN8superspace}. 
These superspaces are obtained by extending the usual momentum space (parametrized in terms of 
spinor variables) with unconstrained Grassmann variables, $\eta^I$ with $I=1, \dots , \mathcal{N}$, 
which transform in the fundamental representation of the $R$-symmetry group and carry unit little group weight. 
The bosonic spinor variables carry kinematic information, while the Grassmann 
variables carry information on the helicity and $R$-symmetry representation of the external states. On-shell 
superfields---\ie fields defined on these superspaces---have a finite expansion in the fermionic variables,
with each coefficient being a component fields of definite helicity and $R$-symmetry representation.
$\NeqFour$ SYM  has a simple structure because all states can be assembled into a single 
CPT-self-conjugate on-shell superfield:
\begin{equation}
\hskip -.5 cm 
\Phi(\eta)=g^++\eta^I f^+_I+\frac{1}{2}\eta^I\eta^J\phi_{IJ}
+\frac{1}{3!}\eta^I\eta^J\eta^Kf^{-}_{IJK}+\frac{1}{4!}\eta^I\eta^J\eta^K\eta^Lg_{IJKL}^-\,,
\label{Neq4multiplet}
\end{equation}   
where $g^+$ is the positive helicity gluon, $f_I^+$ four positive
helicity Majorana fermions, $\phi_{IJ}$ six real scalars, $f^{I -}\equiv \frac{1}{3!}\epsilon^{IJKL}f^-_{JKL}$
four negative helicity Majorana fermions and $g^-$ is the negative
helicity gluons, for a total of $8+8$ physical states (not including color degrees of freedom).  
The case of $\NeqEight$ supergravity is similar, with a four-dimensional CPT-self-conjugate on-shell 
superfield containing fields up to helicity $\pm 2$: 
\begin{eqnarray}
\hskip -.5 cm 
 \Phi(\eta)&=&
 h^{+}+\eta^I\psi^+_I+\frac{1}{2}\eta^I\eta^J A_{IJ} +\frac{1}{3!}\eta^I\eta^J\eta^K \chi^{+}_{IJK}+\frac{1}{4!}\eta^I\eta^J\eta^K\eta^L \phi_{IJKL}
\cr
&&
+\frac{1}{5!}\eta^I\eta^J\eta^K\eta^L \eta^M \chi^{-}_{IJKLM}
+\frac{1}{6!}\eta^I\eta^J\eta^K\eta^L \eta^M \eta^N A^{-}_{IJKLMN}
\label{Neq8multiplet}
\\
&& +\frac{1}{7!}\eta^I\eta^J\eta^K\eta^L \eta^M \eta^N \eta^O \psi^{-}_{IJKLMNO}
+\frac{1}{8!}\eta^I\eta^J\eta^K\eta^L \eta^M \eta^N \eta^O \eta^P h^{-}_{IJKLMNOP} \,.
\nonumber
\end{eqnarray}   
Each supergravity state is a double copy of the gauge-theory
states.\footnote{As we discussed in \sect{ZoologySection1},
  supergravity states can more generally be understood as being in
  one-to-one correspondence with gauge-invariant bilinears of
  gauge-theory states.}  The 256 physical states of $\NeqEight$
supergravity correspond to the $16 \times 16$ direct product of states
of two $\NeqFour$ SYM theories, as shown in
Table~\ref{table_N8States}.

\def\tI{{\tilde I}}
\def\tJ{{\tilde J}}
\def\tK{{\tilde K}}
\def\tL{{\tilde L}}
\begin{table}[t]
\begin{center}
\begin{tabular}{c||c|c|c|c|c}
& $\raisebox{-.3 cm}{ \vphantom{|}}  g_R{}^+$ & $f_R{}_\tI^+$ & $\phi_R{}_{\tI\tJ}$ & $f_R{}_{\tI\tJ\tK}^-$ & $g_R{}_{\tI\tJ\tK\tL}^-$     \cr
         \hline
$\raisebox{.3 cm}{ \vphantom{|}} g_L{}^+$     &   $ h^{+}$  &      $\psi^{+}_{\, \tI}$      & $A^+_{\,\tI\tJ}$  &   $\chi^{+}_{ \, \tI\tJ\tK}$   &  $\phi_{ \, \tI\tJ\tK\tL}$       \cr
$\raisebox{.3 cm}{ \vphantom{|}} f_L{}_{I}^+$  & $\psi^{+}_{I}$  &      $A^{+}_{I\, \tI}$  & $\chi^+_{I\, \tI\tJ} $ &  $\phi_{I\, \tI\tJ\tK}$    &  $\chi^{-}_{I \, \tI\tJ\tK\tL}$    \cr
$\raisebox{.3 cm}{ \vphantom{|}} \phi_L{}_{IJ}$     &   $A^{+}_{IJ}$ &  $\chi^+_{IJ\, \tI}$   & $\phi_{IJ\, \tI\tJ}$  &  $\chi^{-}_{IJ\, \tI\tJ\tK}$ &  $A^{-}_{IJ \, \tI\tJ\tK\tL }$     \cr
$\raisebox{.3 cm}{ \vphantom{|}} f_L{}^-_{IJKL}$   & $\chi^+_{IJKL}$ &      $\phi_{IJKL\, \tI}$      &   $\chi^-_{IJKL\, \tI\tJ }$ &   $A^{-}_{IJKL \, \tI\tJ\tK}$ &  $\psi^-_{IJKL \, \tI\tJ\tK\tL}$ \cr
$\raisebox{.3 cm}{ \vphantom{|}} g_L{}_{IJKL}^-$       &    $\phi_{IJKL}$  &  $\chi^{-}_{IJKL\, \tI}$   & $A^-_{IJKL\,\tI\tJ}$  &   $\psi^{-}_{IJKL \, \tI\tJ\tK}$        &  $h^{-}_{IJKL \, \tI\tJ\tK\tL}$
\end{tabular}
\caption{The states of $\NeqEight$ supergravity organized via the double copy. The $SU(8)$ representations are decomposed in
representations of the $SU(4)\times SU(4)$ subgroup which is manifest in the construction.
For gauge theory $(g^+, \lambda^{+}, \phi, \lambda^-, g^-)$ 
carry helicity $(+1, \frac{1}{2},  0, -\frac{1}{2}, -1)$, while the helicity
of the supergravity states are the sum of gauge-theory helicities. The $\pm$ decorating the entries represent the sign of the helicity 
of the corresponding state.
The double-copy states can be reorganized into the standard $\NeqEight$ multiplet
containing 256 physical states, cf. eq.~\eqref{Neq8multiplet}. 
}
\label{table_N8States} 
\end{center}
\end{table}

Supersymmetry transformations act on on-shell superfields (\ie single-particle supersymmetry transformations) as
\be
Q^{{\dot \alpha}}_I = {\tilde \lambda}^{\dot \alpha} \frac{\partial}{\partial \eta^I}
~~,\quad
Q_{{\dot \alpha}}^I = \eta^I  \frac{\partial}{\partial {\tilde \lambda}^{\dot \alpha}}
~~,\quad
Q^{\alpha I} = \lambda^\alpha \eta^I
~~,\quad
Q_{\alpha I} = \frac{\partial^2}{\partial \lambda^\alpha \partial\eta^I} \,;  
\label{susy_generators}
\ee
the (linearized) supersymmetry transformations of component fields are extracted by acting on superfields with these generators and reading off the
coefficient of the desired combination of Grassmann variables $\eta$. Multi-particle supersymmetry generators are obtained by summing the 
single-particle ones over all the particles; for example, $Q^{{\dot \alpha}}_I $ acts on a product of $n$ distinct fields as
\be
Q^{{\dot \alpha}}_I = \sum_{i=1}^nQ_i{}^{{\dot \alpha}}_I = \sum_{i=1}^n {\tilde \lambda}_i^{\dot \alpha} \frac{\partial}{\partial \eta_i^I} \, .
\ee
For supersymmetry algebras with less-than-maximal supersymmetry not all multiplets are CPT-self-conjugate; in such cases the fields of the theory form 
(perhaps several) CPT-conjugate pairs.

Scattering amplitudes in supersymmetric field theories can be combined into superamplitudes, defined as polynomials in Grassmann variables such that the
coefficient of each monomial is a component amplitude whose helicity configuration is dictated by the $\eta$ factors that multiply it and the structure of the 
superfields of the theory. The details depend on the amount of supersymmetry; as above, we illustrate these ideas for $\NeqFour$ SYM theory. See Refs.~\cite{ElvangHuangLessSusy,ElvangHuangReview}  for less supersymmetric cases.
On-shell supersymmetry Ward identities, relating component amplitudes with different external field configurations, can be derived by demanding that superamplitudes  are annihilated by the multi-particle supersymmetry generators. A detailed descriptions may be found in Refs.~\cite{ElvangFreedmanN8superspace, SWI3, BSTReview,ElvangHuangReview}.
The unconstrained nature of the Grassmann variables makes it
straightforward to translate summations of on-shell states needed in
unitarity cuts or on-shell recursion into Grassmann integrations,
which take care of the state bookkeeping. See Refs.~\cite{KorchemskyOneLoop,SuperSum} for details.
This procedure ensures that all generalized cuts are manifestly supersymmetric.

The minimum number of Grassmann variables in superamplitudes enforces the conservation of the polynomial supercharge $Q^{\alpha I}$ in 
Eq.~\eqref{susy_generators}, sometimes referred to as the ``supermomentum'':
\begin{equation}
\delta^{(8)}(Q)\equiv \delta^{(8)}\Biggl(\sum_{j=1}^n\lambda^{\alpha}_j\eta_j^I\Biggr)
= \prod_{I=1}^4\sum_{i < j}^n\langle ij\rangle\eta_i^I\eta_j^I\,.
\label{DeltaSpinor}
\end{equation}
The superamplitude with this minimal number of Grassmann variables are referred to as maximally-helicity-violating (MHV) superamplitude and
the corresponding component amplitudes are referred to in a similar manner. The name reflects the fact that
these amplitudes, with all incoming particles, exhibit the maximum imbalance between positive and negative helicities.\footnote{At loop 
level, in non-supersymmetric theories the imbalance can be even larger, as the all-plus and single-minus gluon amplitudes 
and their conjugates are nonvanishing~\cite{Cangemi:1996rx}.}
The $n$-point maximally-helicity-violating (MHV) tree amplitudes of $\NeqFour$ SYM theory are~\cite{Nair} 
\begin{equation}
\mathcal{A}^{\rm MHV}_n(1,2,\cdots,n)=\frac{i}{\prod_{j=1}^n\langle j(j+1)\rangle}\,
\delta^{(8)}\Biggl(\sum_{j=1}^n\lambda^{\alpha}_j\eta_j^I\Biggr)\,,
 \label{MHVSuperAmplitude}
\end{equation}   
where leg $n+1$ is to be identified with leg $1$. The coefficient of the supermomentum-conserving $\delta$-function is cyclically symmetric 
and can also be identified as the ratio 
\begin{equation}
A_{n}^\tree(1^-,2^-,3^+,\cdots,n^+)/{\langle 12\rangle^4}\,.
\end{equation}

General $n$-point  $\NeqFour$ SYM superamplitudes can be written as
\begin{eqnarray}
\mathcal{A}_n = \mathcal{A}^{\rm MHV}_n \left({\cal P}_n^{(0)} + {\cal P}_n^{(1)} + \dots + {\cal P}_n^{(n-4)}  \right) ,
\end{eqnarray}
where ${\cal P}_n^{(k)} $ is a polynomial of degree $4k$ in Grassmann
variables and ${\cal P}_n^{(0)}=1$.  Through $k_\text{mid} =
[(n-4)/2]$, the component amplitudes captured by these terms have a
smaller difference between the number of external states with positive and negative
helicity and are referred to as (next-to)$^k$-MHV amplitudes (or,
generically, non-MHV amplitudes). The component amplitudes with $k> k_\text{mid}$ can be obtained from 
those with $k<k_\text{mid}$ by conjugation.\footnote{Conjugation of
  superamplitudes exchanges $\eta$ with their conjugates and thus also 
  changes the type of on-shell superspace. The transformation to the original superspace 
  is given by the fermionic Fourier transform of all conjugate $\eta$ variables.. 
  For a discussion of various superspaces see Ref.~\cite{Huang:2011um}.}
We note that, while all component amplitudes of an MHV superamplitude are related to each other by supersymmetry transformations, several 
non-MHV component amplitudes are needed to generate entire superamplitude. The polynomials ${\cal P}_n^{(k\ge 1)} $ are expressed in terms 
of the $R$-invariants~\cite{Drummond:2008vq, DrummondAllTrees} which manifest the dual superconformal invariance of tree-level amplitudes 
of ${\cal N}=4$ SYM theory. This symmetry led to the derivation~\cite{DrummondAllTrees} 
of an explicit form of all tree-level amplitudes of this theory and, consequently, also of pure 
Yang-Mills theory, as superpartners do not contribute at tree level.

The MHV superamplitudes of $\mathcal{N}=8$ supergravity have a form similar to the one
of $\NeqFour$ SYM theory and are given by
 \begin{equation}
 \label{MHVsugra}
 \mathcal{M}_n^{\rm MHV}=\frac{M_{n}(1^-,2^-,3^+,\cdots,n^+)}{\langle 12\rangle^8}\,
  \delta^{(16)}\Biggl(\sum_{j=1}^n\lambda^{\alpha}_j\eta_j^I\Biggr)\,,
 \end{equation}
where $M_{n}(1^-,2^-,3^+,\cdots,n^+)$ is a tree-level MHV
pure-graviton amplitude.  The simplicity of this result follows
from the fact that, for MHV superamplitudes, the entire superspace content is contained in an overall 
supermomentum-conserving $\delta$-function. The supergravity one is the double-copy of the gauge 
theory one and it evaluates to
\begin{equation}
\delta^{(16)}\Biggl(\sum_{j=1}^n\lambda^{\alpha}_j\eta_j^a\Biggr)
= \prod_{I=1}^8\sum_{i<j}^n\langle ij\rangle\eta_i^I\eta_j^I\,.
\label{DeltaSpinoriN8}
\end{equation}
For more general amplitudes the results are more complicated, but
follow directly by applying the KLT or BCJ double copy to gauge-theory
superamplitudes to obtain superamplitudes in supergravity; the
manifestly-supersymmetric KLT relations were discussed
in~\cite{Feng:2010br}.  Since each gauge-theory amplitude exhibits a
factor of the supermomentum-conserving $\delta$-function, which is
symmetric under permutation of the external lines, the supergravity
amplitude inherits the complete supersymmetry of both gauge-theory
factors. Thus, the supergravity $R$-symmetry group is
$SU(N_L+N_R)$---see \sect{GravitySymmetriesSection} for more details.
Moreover, as we discuss in \sect{GravitySymmetriesSection},
double-copy supergravity exhibits an emergent $U(1)$ symmetry, which
is part of its U-duality group. Among its implications is the
vanishing of the double-copy of (super)amplitudes in different
N$^k$MHV sectors.

As discussed in Ref.~\cite{SuperSum}, we can obtain superamplitudes 
in theories with fewer supersymmetries by
appropriately grouping $R$-symmetry indices. This can be realized by truncating the supermultiplets 
\eqref{Neq4multiplet} and \eqref{Neq8multiplet} such that a certain subset of the $\eta$-variables 
always appear together; the corresponding component amplitudes are obtained from those of the 
maximally-supersymmetric theory by restricting them to the combinations of Grassmann variables that 
are allowed to appear for each external state.
The simplest example is that the tree amplitudes of pure non-supersymmetric Yang-Mills theory
are just the pure-gluon amplitudes of $\NeqFour$ SYM theory, only for these amplitudes the $\eta$
variables appear in the combination $\eta^1\eta^2\eta^3\eta^4$; all other states contain a subset of 
their corresponding Grassmann variables and thus decouple from these amplitudes.
In fact, using appropriate projections, one can even obtain QCD
tree-level amplitudes with quarks from $\NeqFour$ tree
amplitudes~\cite{DixonHenn}, leading to compact forms of QCD
amplitudes.\footnote{The necessary change in the color factor reflecting the change in gauge-group 
representation can easily be accounted for in tree-level amplitudes though a multiplicative factor.}

As an example, the MHV tree amplitudes for external gauge
supermultiplets in ${\cal N}$-extended SYM theory are given
by~\cite{SuperSum}
\begin{equation}
{\cal A}^{\rm MHV}_n(1, 2, \ldots, n)=
\frac{ \prod_{I=1}^{\cal N}\delta^{(2)}( Q^I) }
{\prod_{j=1}^n\langle j ~(j+1)\rangle}
\,\,
\biggl(\sum_{i<j}^n \spa{i}.{j}^{4-{\cal N}} 
\prod_{I={\cal N}+1}^{4} \eta_i^I  \eta_j^I\biggr) \,,
\label{MHVLessSuperAmplitude}
\end{equation}
with ${\cal N}$ counting the number of supersymmetries,
$Q^I=\sum_{i=1}^n \lambda_i \eta_i^I$, and $n\ge 3$. 

Using supersymmetric versions of MHV amplitudes~\cite{CSW} and on-shell
recursion~\cite{BCFW}, general tree superamplitudes in gauge and
gravity theories have been systematically constructed (see
\eg Refs.~\cite{GGK,ArkaniHamed2008gz, DrummondAllTrees, DixonHenn}).
As we briefly review in \app{GeneralizedUnitaritySection},
they can be used as input building blocks to construct the
integrands of loop superamplitudes. The cases of ${\cal N}<4$ superamplitudes, 
have been analyzed in Ref.~\cite{ElvangHuangLessSusy} in some detail.

\begin{homework}
\label{HomeworkFourPointSusy}
Using the supersymmetry generators \eqref{susy_generators} and the
on-shell superfield \eqref{Neq4multiplet}, construct the linearized
supersymmetry transformations of the component fields of ${\cal N}=4$
SYM theory. Derive the corresponding relations between the component
MHV amplitudes.
\end{homework}



\section{Generalized unitarity}
\label{GeneralizedUnitaritySection}

In this appendix we give a brief summary of the modern generalized
unitary method~\cite{UnitarityMethod, Fusing, TripleCuteeJets, BDDPR,
  BCFUnitarity, FiveLoop} used in multiloop calculations, focusing on
their applications in double-copy constructions.  This provides some of the 
necessary background for our review of the generalized double-copy 
construction in~\sect{GeneralizedDoubleCopySection}.  
We will present several examples illustrating the basic ideas and refer the 
reader to other reviews for further details~\cite{BDKUniarityReview, BernHuangReview, JJHenrikReview, BrittoReview}.  

The generalized-unitary method systematically builds
complete loop-level integrands using as input only on-shell tree-level amplitudes.
A central feature is that simplifications and features of the latter are directly 
imported into the former.  In particular, with this method we can use tree-level 
double-copy relations to construct gravity loop integrands.  We also briefly
review a variant of generalized unitarity, known as the maximal-cut
method~\cite{FiveLoop}, which meshes well with the generalized double
copy~\cite{GeneralizedDoubleCopy} discussed in
\sect{GeneralizedDoubleCopySection}.  A reorganization of the
generalized-unitarity method that has various advantages is found in
Ref.~\cite{Bourjaily:2017wjl}.

\begin{figure}[tb]
\centerline{\includegraphics[width=7cm]{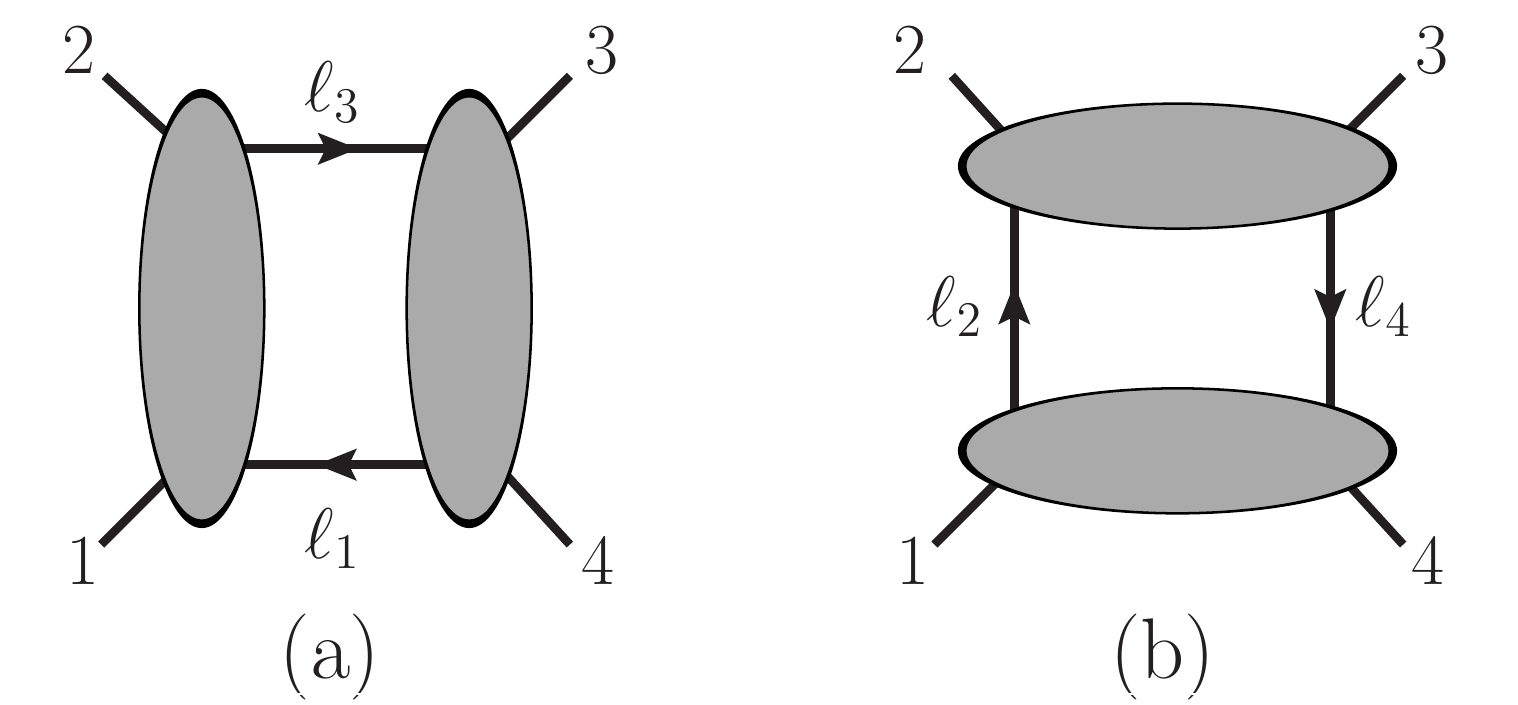}}
\caption[a]{\small The (a) $s$ and (b) $t$ channel two-particle cuts of a one-loop four-point 
amplitude. The exposed lines are all on-shell and the
  blobs represent tree amplitudes.}
\label{TwoParticleCutFigure}
\end{figure}

\begin{figure}[tb]
\centerline{\includegraphics[width=6 in]{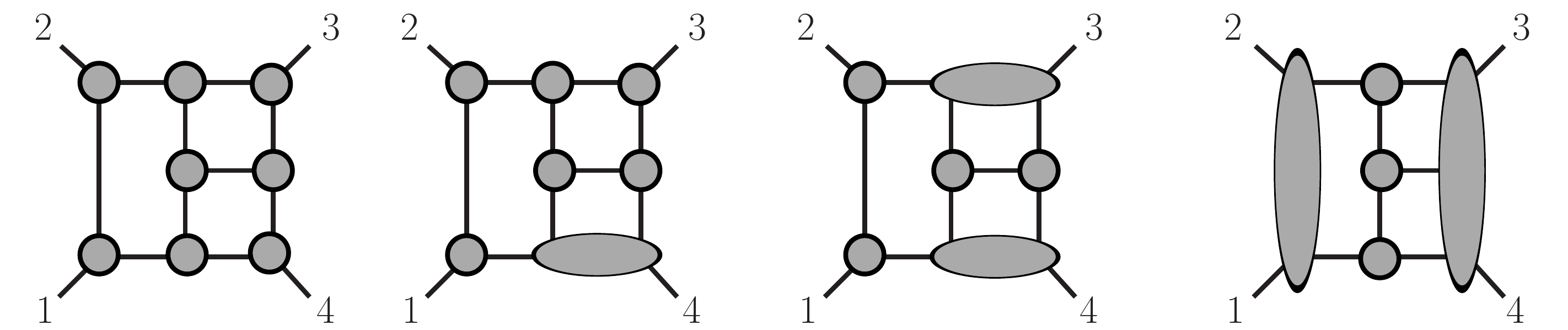}}
\caption[a]{\small Examples of generalized cuts for a three-loop
  four-point amplitude.  The exposed lines are all on-shell and the
  blobs represent tree amplitudes.
\label{ThreeLoopCutSampleFigure}
}
\end{figure}

Traditionally, unitarity of the scattering matrix is implemented at the integrated level 
via dispersion relations~\cite{AnalyticSMatrixBook}. For our purposes, however,
it is much more useful to use it at the integrand level.  We
introduce the concept of a generalized cut that reduces an integrand
to a sum of products of tree amplitudes $A^\tree_{(j)}$,
\begin{equation}
\sum_{\rm states} A^\tree_{(1)} A^\tree_{(2)} A^\tree_{(3)} \cdots  A^\tree_{(m)}\,.
\label{GenericCut}
\end{equation}
Each cut propagator is replaced with a delta function enforcing on-shell
constraint for the corresponding momentum.  
The sum runs over all intermediate physical states that
can contribute given the external states of the amplitude being constructed.  
Some generalized cuts of the one-loop four-point amplitude are shown in \fig{TwoParticleCutFigure} 
and of the three-loop four-point amplitude in \fig{ThreeLoopCutSampleFigure}.  In these
figures the exposed lines are all on-shell delta functions and the blobs represent
on-shell tree amplitudes.

Loop integrands are determined by  spanning set of generalized cuts, \ie a set of cuts which receive 
contributions from all the terms that could possibly be generated by the Feynman graphs of the theory.
Loop integrands are constructed by finding a single function whose cuts match all the products of tree amplitudes, 
summed over states corresponding to such a spanning set. 
Regardless of which set of cuts one uses to construct an integrand, one must always verify it on a minimal 
spanning set (\ie a spanning set that contains the minimal number of cuts).
To illustrate these ideas in practice we turn to a few simple unitarity cuts.

\subsection{One-loop example of unitarity cuts}

To illustrate the generalized unitarity method and how it meshes with
double-copy ideas consider the two-particle cuts of a one-loop
four-point color-ordered gauge-theory amplitude.  In these amplitudes
the color factors are stripped away and the external legs follow a
cyclic ordering~\cite{ManganoParkeReview,TasiLance}.  The two-particle
cuts of a one-loop four-point amplitude are obtained by putting two
intermediate lines on shell, as illustrated in
\fig{TwoParticleCutFigure}. For example, the $s$-channel cut in
\fig{TwoParticleCutFigure}(a) is given by
\begin{equation}
C_s^{\rm gauge} = \sum_{\rm states} \A(-\ell_1, 1,2, \ell_3) \,
           \A(-\ell_3, 3,4, \ell_1) \,,
\label{GeneralizedCutsChannel}
\end{equation}
where the sum runs over all physical states in the theory.
The cuts are evaluated using momenta that place all cut-line momenta on shell,
 \ie $\ell_i^2 = 0$ if the theory is massless.

A particularly simple example is the color-ordered one-loop four-point amplitude in $\NeqFour$ 
SYM theory, which is useful for illustrating how gauge-theory unitarity cuts can be converted to 
gravity ones.
For this theory, after summing over all physical states that cross the two-particle
cut, the result takes a remarkably compact form~\cite{BRY},
\begin{equation}
C_s^{\rm gauge} = 
= - ist \A (1, 2, 3, 4) 
   {1\over (\ell_1 - p_1)^2 } 
   {1\over (\ell_3 - p_3)^2 } \,.
\label{BasicYMCutting}
\end{equation}
All momenta are on shell. This expression is valid for any external states of the theory; the cut 
depends on them only through the overall factor $\A (1, 2, 3, 4)$. 
The $t$-channel cut in \fig{TwoParticleCutFigure}(b) is obtained by relabeling Eq.~\eqref{BasicYMCutting}.

\begin{figure}[tb]
\centerline{\includegraphics[width=4cm]{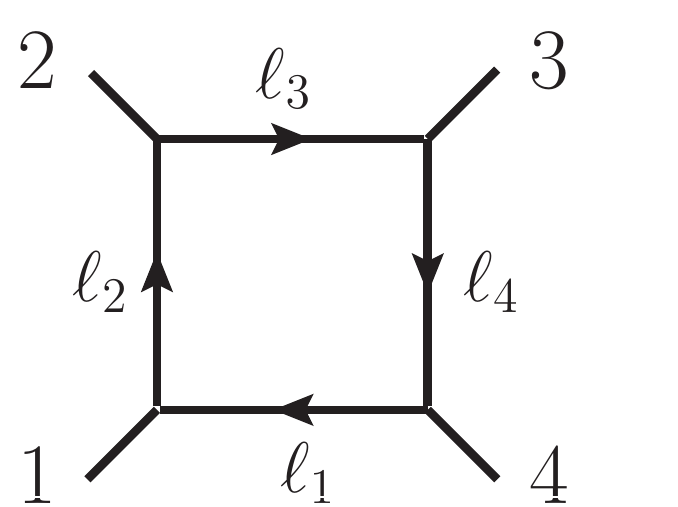}}
\caption[a]{\small The one-loop box integral and loop momentum labels used in the cut construction.}
\label{OneLoopBoxCutLabelsFigure}
\end{figure}

The most straightforward way to verify these equations is by using
four-dimensional helicity states and on-shell superspace, but they
hold in $D \le 10$ dimensions as well (where maximal supersymmetric
Yang-Mills theory is defined).  Details may be found in Ref.~\cite{BDDPR}.

Putting back the cut propagators and loop integration we find a function with 
the correct $s$-channel cut,
\begin{equation}
 i s t \, \A(1,2,3,4)  \, I_4(s,t)  \Bigr|_{\scut} 
  \,,
\label{SchannelCut}
\end{equation}
where $I_4(s,t)$ is the scalar box integral shown in \fig{OneLoopBoxCutLabelsFigure}
and given in \eqns{4PtIntegral}{ZeroMassBoxFullExpand}.
By the cut operation in the $s$ channel in \eqn{SchannelCut} we mean to remove 
the integration and to replace the two propagators $1/\ell^2$ and $1/(\ell - p_1 -p_2)^2$
with on-shell conditions, recovering \eqn{BasicYMCutting} after identifying $\ell = \ell_1$ 
and applying momentum conservation.  Similar considerations show that the $t$ channel
cut can be written as
\begin{equation}
 i s t \, \A(1,2,3,4)  \, I_4(s,t)  \Bigr|_{\tcut} 
  \, .
\label{TchannelCut}
\end{equation}

Once unitarity cuts are written in this way, as the cuts of a single function, it is easy 
to see that the one-loop four-point amplitude, with no cut conditions, is obtained simply 
by removing the cut conditions,
\begin{equation}
A^{\oneloop}_\NeqFour(1^-, 2^-, 3^+, 4^+)
= i \, s t \, \A(1,2,3,4) \, I_4(s,t) \, ,
\label{OneLoopN4}
\end{equation}
matching the result in \eqn{NeqFourYMFourPoint}.

These basic ideas generalize to any massless gauge theory
and underpin many theoretical studies, including those for collider
physics (see \eg Refs.~\cite{TripleCuteeJets,BlackHat, BlackHatW5}).

\subsection{Converting gauge-theory unitarity cuts to gravity ones}

As discussed in \sect{DualitySection}, the BCJ forms of loop-level gauge-theory 
integrands can be directly converted to gravity ones.  
Because the unitarity-based construction uses tree amplitudes as
input, we can also straightforwardly apply the KLT relations
\eqref{KLT456} to convert gravity unitarity cuts to sums of products of
gauge-theory cuts.
The BCJ form of the gauge-theory tree amplitudes \eqref{TreeAmpMpt}
may also be used to apply the double copy to convert gauge-theory cuts
to the corresponding gravity ones.  The KLT form is especially useful
when working with compact helicity amplitudes, while the BCJ form is
helpful in $D$ dimensions (i.e. when using dimensional
regularization) with formal polarization vectors.

Consider first the two-particle cut of a one-loop four-point 
amplitude show in \fig{TwoParticleCutFigure}(a) in a gravity theory. Using the KLT form 
of the double copy, it is given by
\begin{align}
C_{\rm GR} & =\sum_{\rm gravity \atop \rm states}
 \M(-\ell_1,  1, 2, \ell_3) \times
  \M(-\ell_3, 3, 4, \ell_1) \nn \\
& =
- s^2 \biggl(\sum_{\rm gauge \atop \rm states}
  \A(-\ell_1,  1, 2, \ell_3) \times
                \A(-\ell_3, 3, 4, \ell_1) \biggr)
   \nonumber  \\
& \hskip 2.0 truecm
\null \times
\biggl(\sum_{\rm gauge \atop \rm\ states}  \A(\ell_3,  1, 2, -\ell_1) \times
                \A(\ell_1, 3, 4, -\ell_3) \biggr) \,,
\label{TwoParticleCutGravity}
\end{align}
where we applied the KLT relation \eqref{KLTFourPt} to rewrite each
gravity tree amplitude in terms of a product of two gauge-theory
amplitudes.  
In this expression we have assumed that the gravity
theory of interest arises as a simple double copy, as it does for
$\NeqEight$ supergravity. This allows us to decompose each state in
the gravity theory into a ``left'' and a ``right'' gauge-theory state.
For the case of $\NeqEight$ supergravity, summing over the states in
the $\NeqEight$ multiplet is equivalent to summing independently over the left and
right $\NeqFour$ SYM gauge-theory multiplets.  
For theories which are not simple double copies, such as pure gravity, one must
remove unwanted states (i.e. dilaton and antisymmetric tensor) by
inserting explicit physical-state projectors into the cuts.  These
projectors have been used effectively at two-loops to study
ultraviolet properties of various theories, including pure
gravity~\cite{PureGravityTwoLoops} as well as for computing the third
post-Minkowskian correction to the conservative two-body Hamiltonian~\cite{3PM,3PMLong}. 
In some cases, it is sufficient to evaluate
the generalized unitarity cuts in four dimensions, where we can
simplify the input gauge-theory amplitudes enormously by using
helicity states.  In this case, a simple way to control which particles
circulate in the loops, is by correlating the state sum of the two
gauge theories, For example, if we want only gravitons to cross the cuts,
then for each term we should have identical helicity for the
corresponding gluons in the state sum~\cite{3PM,3PMLong}.
A similar procedure works well for supergravity theories which are obtained as orbifolds 
of  \eg  $\NeqEight$ supergravity where the orbifold action cannot be decomposed into 
independent actions on the left and right $\NeqFour$ SYM gauge theories~\cite{Chiodaroli2013upa}.

The one-loop four-point amplitude in $\NeqEight$ supergravity is an instructive 
illustration of how we can recycle gauge-theory unitarity cuts into gravity ones.  
For this case, \eqn{TwoParticleCutGravity} immediately collapses because, up to relabeling,  
each gauge-theory state sum is the $\NeqFour$ SYM state sum in
\eqn{BasicYMCutting}.  Inserting the simplified $\NeqFour$ SYM cut \eqref{BasicYMCutting} into 
\eqref{TwoParticleCutGravity} results in an equivalent relation for $\NeqEight$ supergravity,
\begin{align}
C_{\rm GR} & =s^2 (st)^2 \bigl[ \A(1, 2, 3, 4)\bigr]^2
   { 1\over (\ell_1 - p_1)^2  (\ell_3 - p_3)^2 
 (\ell_3 + p_1)^2  (\ell_1 + p_3)^2 }  \nn \\
& =  s^2\, [st\, A_4^{\rm tree} (1, 2, 3, 4)]^2
     { 1\over (\ell_1 - p_1)^2  (\ell_3 - p_3)^2 
   (\ell_1 - p_2)^2 (\ell_3 - p_4)^2 } \nn \\
& = i\, stu M_4^{\rm tree}(1, 2, 3, 4) 
 \biggl[{1\over (\ell_1 - p_1)^2 } + {1\over (\ell_1 - p_2)^2} \biggr]
\biggl[{1\over (\ell_3 - p_3)^2 } + {1\over (\ell_3 - p_4)^2} \biggr]\, .
\label{BasicGravityCutting}
\end{align}
To obtain this we partial fractioned the product of propagators and used the KLT relations
\eqref{4ptKLT} and the BCJ amplitude relations \eqref{4PtBCJAmplitude}.
As for gauge-theory cuts, the $\ell_1$ and $\ell_3$ are on shell.  The $t$- and $u$-channel formulae 
are obtained by relabeling the external legs in \eqn{BasicGravityCutting}.

Following the same strategy as for the reconstruction of the one-loop four-point $\NeqFour$ SYM 
amplitude, it is then straightforward to obtain the $\NeqEight$ one-loop four-point amplitude,
\begin{align}
{\cal M}_4^{ {\oneloop} }(1, 2, 3, 4)
& =  -i \Bigl( {\kappa \over 2}\Bigr)^4
s t u M_4^{\rm tree}(1,2,3,4)
 \Bigl(I_4(s,t) + I_4(s,u) + I_4(t,u) \Bigr) \,. 
\label{OneLoopN8GravResult}
\end{align}
Here $I_4(s,t)$ is the box integrals defined in \eqn{4PtIntegral}, while $I_4(s,u)$ and $I_4(t,u)$ 
are obtained by appropriate relabeling of external legs.
This agrees with the form obtained using the BCJ double copy in
\sect{ExamplesSection} and agrees with the result first obtained by
Brink, Green, Schwarz~\cite{GSB} in the field-theory limit of
superstring theory.

One can also use the BCJ double copy \eqref{DCformula} for the component
tree amplitudes of the gravity unitarity cuts.  This is especially efficient when working in $D$
dimensions (\eg when using dimensional regularization), because compact helicity-based expressions
for tree-level amplitudes are no longer available and, consequently, the result of the KLT relations 
will be cumbersome to use.
In $D$ dimensions, the BCJ form is a more natural form because it preserves the diagram structure when
converting from gauge theory to gravity.  
For example, the two-particle cut (a) in \fig{TwoParticleCutFigure}, can be evaluated using 
the form of the double copy in \eqn{cnrepl},
\begin{equation}
{\cal C}_{\rm GR}^{\rm (a)} = \sum_{\rm pols.} \M(-\ell_1,  1, 2, \ell_3) \times
  \M(-\ell_3, 3, 4, \ell_1) \ ,
\end{equation}
where the graviton tree amplitudes in the cut are obtained from the
double copy form in \eqn{cnrepl} by simple relabelings.  The
sum over polarizations gives the physical-state projector. For
gravitons in $D$ dimensions the projector is
\begin{align}
P^{\mu\nu \rho \sigma}(p,q)= \sum_{\rm pols.} \pol^{\mu\nu}(-p) \pol^{\rho\sigma}(p) 
=  &\, \frac{1}{2} \bigl( P^{\mu\rho} P^{\nu\sigma} + P^{\mu\sigma} P^{\nu\rho} \bigr)
- \frac{1}{D_s-2} P^{\mu\nu} P^{\rho\sigma} \,,
\end{align}
where $P^{\mu\nu}$ is the gluon physical-state projector 
\begin{equation}
P^{\mu\nu}(p,q) =  \sum_{\rm pols.} \pol^\mu(-p) \pol^\nu(p)
 = \eta^{\mu\nu} - { q^\mu p^\nu + p^\mu q^\nu \over q\cdot p } \,,
\label{YMProjector}
\end{equation}
with momentum $p$ and a null reference momentum $q$.  
In some cases, terms that vanish on-shell can be added to tree amplitudes so that  
the dependence on the reference momentum disappears because of the on-shell 
Ward identity for the gauge symmetry~\cite{3PM, KoemansCollado:2019ggb,3PMLong}.

\subsection{Method of Maximal Cuts}
\label{MaximalCutsSubsection}

A refinement of the unitarity method~\cite{UnitarityMethod,Fusing},
which is especially helpful at higher loop orders, is the method of
maximal cuts~\cite{FiveLoop}.  This method is not only a basic tool
for checking and building double-copy gravity integrands, but it also
plays a central role in the generalized-double-copy construction
described in \sect{GeneralizedDoubleCopySection}.  This construction was 
central to a computation determining the
ultraviolet properties of $\NeqEight$ supergravity at five
loops~\cite{GeneralizedDoubleCopy, GeneralizedDoubleCopyFiveLoops, UVFiveLoops}.

In the method of maximal cuts, the unitarity cuts
are clustered in levels according to the number $k$ of internal
propagators allowed to remain off shell,
\begin{equation}
{\cal C}^{{\rm N}^k{\rm MC}} = 
\sum_{\rm states} {\cal A}_{m(1)}^{\rm tree} \cdots {\cal A}_{m(p)}^{\rm tree}\,,
\label{GeneralizedCut}
\end{equation}
where ${\cal  A}_{m(i)}^{\rm tree}$ are tree-level $m(i)$-multiplicity amplitudes
corresponding to the blobs, illustrated for various cuts of a three-loop 
four-point amplitude illustrated in \fig{ThreeLoopCutSampleFigure}.  The level $k$
is related to the multiplicity of the various factors by
\begin{equation}
k = \sum_{i=1}^p (m(i)-3) \ .
\end{equation}
The cuts \eqref{GeneralizedCut} can be applied to either gauge or gravity
amplitudes.  As illustrated in the first diagram in
\fig{ThreeLoopCutSampleFigure}, at the maximal cut (MC) level the maximum
number of propagators are replaced by on-shell conditions and all tree
amplitudes appearing in \eqn{GeneralizedCut} are three-point
amplitudes.  At the next-to-maximal-cut (NMC) level, illustrated 
in the second cut of \fig{ThreeLoopCutSampleFigure}, a single
propagator is placed off shell and so forth.

With this organization of generalized cuts, the integrands for $L$-loop amplitudes
are obtained by first establishing an integrand whose maximal cuts are
correct, then adding to it terms so that NMCs are all correct and
systematically proceeding through the next$^k$ maximal cuts (\N{k}s),
until no further contributions are found.  Where this process
completes is dictated by the power counting of the theory and by
choices made at each level.  For example, if minimal power counting
is assigned to each contribution, for $\NeqFour$ SYM
four-point amplitudes, cuts through NMCs, \N2s and \N3s are sufficient
at three~\cite{BCJLoop}, four~\cite{SimplifyingBCJ} and five
loops~\cite{FiveLoopN4}, respectively.

\begin{figure}[t]
\begin{center}
  \includegraphics[clip,scale=0.45]{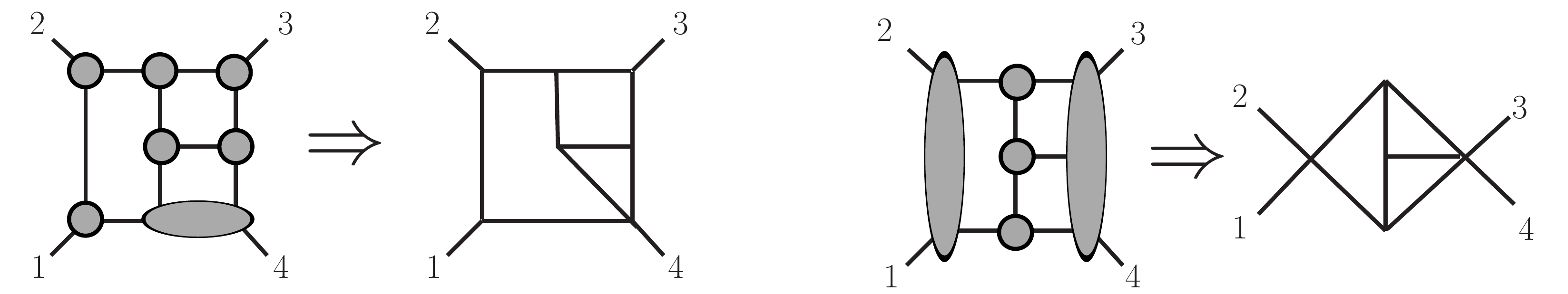}
\end{center}
\vskip -.2 cm
\caption[a]{New contribution found via the method of maximal cuts can
  be assigned directly to contact terms. In these diagrams all the exposed lines are 
on-shell, and propagators within a blob remain off shell. }
\label{CutToContactsFigure}
\end{figure}

Most calculations~(see \eg Refs.~\cite{GravityThree,CompactThree, SimplifyingBCJ, N4GravFourLoop,
  N5GravFourLoop,BCJDifficulty,FourLoopFormFactor,FiveLoopFormFactor}) find it
convenient to organize the integrands in terms of diagrams with purely cubic vertices, such as the 
three-loop ones illustrated in \fig{ThreeLoopDiagramsFigure}.  Representations with only cubic
diagrams have certain advantages: they are useful for establishing
minimal power counting in each diagram, and the number of diagrams used
to describe the result proliferate minimally with the loop order and
multiplicity.  
A disadvantage is that ans\"atze are required for imposing various
properties on each diagram, including the desired power counting,
symmetry, and the multiple unitarity cuts to which a given diagram
contributes.
As the loop order increases, it becomes cumbersome to
solve the requisite system of equations that imposes these
constraints. We can avoid this in
the generalized double-copy construction if we instead assign any new
information obtained in a \N{k} to a contact diagram, as discussed
in \sect{GeneralizedDoubleCopySection} and illustrated in
\fig{CutToContactsFigure}.  This is necessarily local because the
nonlocal contributions are accounted for at previous levels.

\subsubsection{Sewing superamplitudes}

We now briefly comment on the use of the on-shell superspace described in
\app{SpinorSuperspaceSection} for the evaluation of the sums over the states 
crossing a unitarity cut. It turns out~\cite{FreedmanUnitarity, KorchemskyOneLoop, SuperSum} 
that it can be conveniently expressed as an integration over the Grassmann parameters
of the cut legs.  The generalized supercut in ${\cal N}$-extended
SYM theory is then given by\footnote{In this formulation
  CPT-conjugate multiplets are both interpreted as being embedded in
  an ${\cal N}=4$ multiplet. This is equivalent to the formulation of
  Ref.~\cite{ElvangHuangLessSusy} up to Grassmann Fourier-transform
  with respect to four $\eta$ variables.}
\begin{equation}
{\cal C} = 
\int \Bigl[ \prod_{i=1}^{k} {d^{\cal N}\eta_i}\Bigr]
 {\cal A}^\tree_{(1)}
  {\cal A}^\tree_{(2)}
  {\cal A}^\tree_{(3)} \cdots 
   {\cal  A}^\tree_{(m)}\,,
\label{SuperCut}
\end{equation}
where ${\cal A}_{(j)}^{\rm tree}$ are the tree-level superamplitudes
connected by $k$ on-shell cut legs. For each cut leg, the integral over $\eta$ 
selects  all possible states on that leg and sums up their contribution to the cut.
These supercuts are functions on the on-shell superspace. For four and higher points 
the tree amplitudes
${\cal A}_{(j)}^{\rm tree}$ are always proportional to a supermomentum
delta function.  Using the simple identity
$\delta(A)\delta(B)=\delta(A+B)\delta(B)$, this implies that all such
cuts are proportional to an overall supermomentum
$\delta$-function~\cite{KorchemskyOneLoop}.  It turns out that such
supercuts are sufficient for determining massless superamplitudes.  This then
implies that the four-dimensional cuts of any loop amplitude with four
or more external legs must be proportional to an overall supermomentum
conservation $\delta$-function. Barring supersymmetry anomalies, this
will also be the case for the corresponding superamplitudes.

Fermionic integration provides one of the several different methods
for the evaluation of supersums in unitarity
cuts~\cite{KorchemskyOneLoop, ArkaniHamed2008gz, FreedmanUnitarity,
  SuperSum}.  There are two main approaches for organizing the
integration over the $\eta$ parameters.
In the first way, the fermionic $\delta$-functions can be used to
localize the integration, so that the evaluation of the supersum
amounts to solving a system of linear equations~\cite{KorchemskyOneLoop,SuperSum}. 
In a second complementary approach, ``index diagrams'' are used to track the
various contributions to the sum over states~\cite{SuperSum}. This approach leads to a simple
algorithm for reading off the contribution of the entire
supermultiplet from the purely gluonic ones and for reducing the number
of supersymmetries. It was used in the construction of the complete
four-loop four-point amplitude of $\NeqFour$ SYM
theory~\cite{Neq44np}.

The overall supermomentum-conserving $\delta$-function has
consequences on the ultraviolet properties of the theory akin to
those of off-shell superspaces. In particular, in a theory with ${\cal N}$-extended 
supersymmetry, it implies that at least ${\cal N}$ powers of momenta in the numerators 
of each diagram are external momenta. In turn, this implies that the superficial
degree of divergence of each diagram is improved by ${\cal N}$
compared to that of the non-supersymmetric theory.
For $\NeqFour$ SYM theory, this simple  power counting implies
the well known~\cite{MandelstamN4SYM, BrinkN4SYM, HoweStellN4SYM}
ultraviolet finiteness of all of its superamplitudes~\cite{SuperSum}. 
For $\NeqEight$ supergravity the BCJ
double copy appears to imply that individual diagrams generally have a poor
power count, because the kinematic numerators are products of
corresponding gauge-theory ones.  
However, it is difficult to draw any conclusions based on this observation
because of the existence of ``enhanced cancellations'' which are nontrivial (and 
not yet fully understood) cancellations between diagrams~\cite{N5GravFourLoop}.




\end{appendices}

\newpage

\setlength{\bibsep}{4pt plus 0.2ex}

{\small
\bibliography{bcjreview}
\bibliographystyle{JHEP}
}
\end{document}